%% file: thesis.tex
\documentclass[11pt,a4paper,twoside]{book}
\pdfoutput=1

\input{MyThesisPreamble.sty}

\author{Juan Bermejo Vega}

\title{Normalizer Circuits and Quantum Computation}

\def\biblio{\bibliographystyle{utphys2}\bibliography{database}}

\begin{document}
\def\biblio{}

\frontmatter

\input{title_page.tex}

\chapter*{}

\input{quote.tex}

\phantomsection
\addcontentsline{toc}{chapter}{Quote}

\chapter*{Abstract}
\phantomsection
\addcontentsline{toc}{chapter}{Abstract}

\input{abstract.tex}

\chapter*{Zusammenfassung}
\phantomsection
\addcontentsline{toc}{chapter}{Zusammenfassung}

\input{Zusammenfassung.tex}

\chapter*{Publications}
\phantomsection
\addcontentsline{toc}{chapter}{Publications}

\input{Publications.tex}

\chapter*{Acknowledgements}
\phantomsection
\addcontentsline{toc}{chapter}{Acknowledgements}

\input{Acknowledgements.tex}

\cleardoublepage

\input{tableofcontents.tex}

\mainmatter

\setcounter{chapter}{-1} 

\input{chapter0_Introduction.tex}

\input{chapterC.tex}

\input{chapterGT.tex}

\input{chapter1_finite}

\input{chapter2_infinite}

\input{chapter3_blackbox.tex}

\input{chapter4_hyperproject.tex}

\appendix

\part*{Appendices}

\input{appendix2_infinite_chapter.tex}


\input{appendix1_finite_chapter.tex}

\input{appendix3_blackbox.tex}


\input{appendix4_hypergroups.tex}

\input{appendix23_normalizer=gaussian}

\cleardoublepage
\phantomsection
\addcontentsline{toc}{chapter}{Bibliography}
\bibliographystyle{utphys2}
\bibliography{database}

\end{document}

%% file: title_page.tex
\begin{titlepage}

\begin{center}

\begin{minipage}{0.24\textwidth}
\centering
\includegraphics[height=1.5cm]{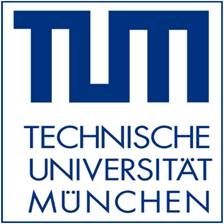}
\end{minipage}
\begin{minipage}{0.5\textwidth}
\centering
\begin{center}
Technische Universität München\\
\smallskip
Max-Planck-Institut für Quantenoptik\\
\end{center}

\end{minipage}
\begin{minipage}{0.24\textwidth}
\centering
\includegraphics[height=1.5cm]{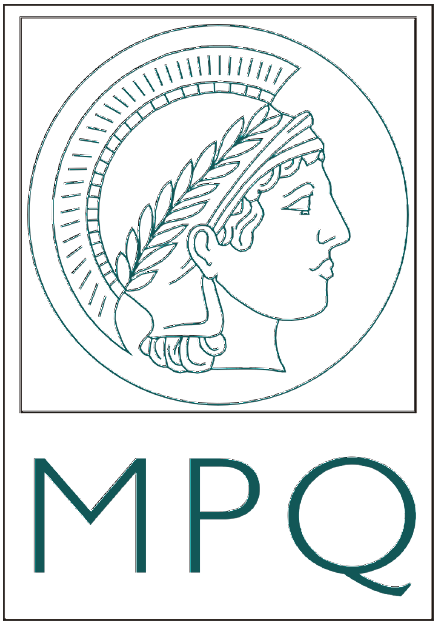}
\end{minipage}

\bigskip
\bigskip

\bigskip
\bigskip

\begin{minipage}{\textwidth}
\centering
\HRule \\[1cm]{\huge \textsf{ {Normalizer Circuits and Quantum Computation}\\ 
\LARGE \vspace{20pt} \huge Juan Bermejo-Vega} \LARGE
}\\[1cm]
\HRule 
\end{minipage}

\bigskip
\bigskip
\bigskip
\bigskip
\bigskip

\begin{center}
\includegraphics[height=3cm]{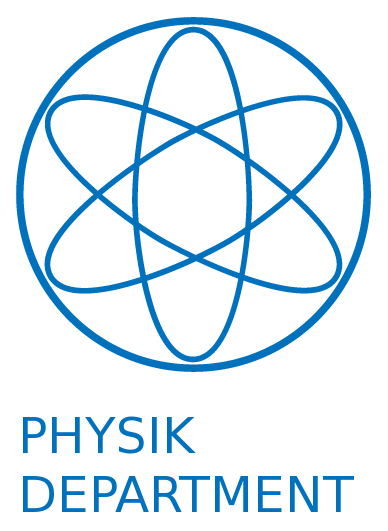}\end{center}

\bigskip
\bigskip
\bigskip
\bigskip
\bigskip
\bigskip

\begin{minipage}{0.85\textwidth}
Vollständiger Abdruck der von der Fakultät für Physik der Technischen Universität München zur Erlangung des akademischen Grades eines  Doktors der Naturwissenschaften (\emph{Dr.\ rer.\ nat.}) genehmigten Dissertation.
\end{minipage}

\bigskip
\bigskip

\begin{center}
\begin{minipage}[t]{0.4\textwidth}
\begin{flushleft}
\begin{enumerate}
\item[] Vorsitzender:\\
\item[] Prüfer der Dissertation:
\end{enumerate}
\end{flushleft}
\end{minipage}
\begin{minipage}[t]{0.55\textwidth}
\begin{flushleft}
\begin{enumerate}
\item[] Univ-Prof.\ Dr.\ Alexander Holleitner\\
\item Hon-Prof.\  Juan Ignacio Cirac, Ph.D.\\
\item Univ-Prof.\ Dr.\ Robert König\\
\item Univ.-Prof.\ Dr.\ Jens Eisert,\\ Freie Universität Berlin \\(nur schriftliche Beurteilung)
\end{enumerate}
\end{flushleft}
\end{minipage}
\end{center}

\bigskip
\bigskip

\begin{minipage}{0.85\textwidth}
Die Dissertation wurde am 26.11.2015 bei der Technischen Universität München eingereicht und durch die Fakultät für Physik am 01.02.2016 angenommen.
\end{minipage}
\end{center}

\end{titlepage}

%% file: quote.tex
\begin{quotation}
\em 
``This—'' He indicated his sword again, seeing Bellis begin to understand. ``—is a sword of possible strikes. A Possible Sword. It’s a conductor for a very rare kind of energy. It’s a node in a circuit, a possibility machine. This—''	He patted the little pack strapped to his waist. ``—is the power: a clockwork engine. These,” the wires stitched into his armor, ``draw the power up. And the sword completes the circuit. When I grip it, the engine’s whole.

If the clockwork is running, my arm and the sword mine possibilities. For every factual attack there are a thousand possibilities, nigh-sword ghosts, and all of them strike down together.''

Doul sheathed the blade and stared up into the trees’ pitch-black canopy.

“Some of the most likely are very nearly real. Some are fainter than mirages, and their power to cut...is faint. There are countless nigh-blades, of all probabilities, all striking together.''

\medskip
\raggedleft
\emph{China Miéville}, The Scar.
\end{quotation}

%% file: abstract.tex
In this thesis, we introduce new  models of quantum computation to study the  potential  and limitations  of quantum computer algorithms. Our models are based on algebraic  extensions of the  qubit Clifford gates (CNOT, Hadamard and $\uppi/4$-phase gates) and Gottesman's stabilizer formalism of quantum codes. We give two main kinds of technical contributions with applications in  quantum algorithm design, classical simulations and for the description of generalized stabilizer states and codes.

Our first main contribution is a formalism of restricted quantum operations, which we name the \emph{normalizer circuit formalism}, wherein the allowed gates are  quantum Fourier transforms (QFTs), automorphism gates and quadratic phase gates associated to a set $G$, which is either an abelian group or an abelian hypergroup.   These gates extend the qubit Clifford gates, which only have non-universal quantum computational power and can be efficiently simulated classically, to comprise additional powerful gates such as QFTs, which are central in Shor’s celebrated factoring algorithm. Using our formalism, we show that normalizer circuit models with different choices of $G$   encompass  famous quantum algorithms, including Shor's and those that solve abelian {Hidden Subgroup Problems} (HSP). Exploiting self-developed classical-simulation  techniques, we further characterize  under which scenarios normalizer circuits succeed or fail to  provide a quantum speed-up. In particular, we derive several no-go results for finding new quantum algorithms with the standard abelian Fourier sampling techniques. We also devise new quantum algorithms (with exponential speedups) for finding hidden commutative hyperstructures. These results offer new insights into the source of the quantum speed-up of the quantum algorithms for abelian and normal HSPs.

Our second main contribution  is a  framework for  describing quantum  many-body states, quantum codes and for the classical simulation of quantum circuits. Our framework comprises algebraic extensions of Gottesman’s Pauli
Stabilizer Formalism (PSF) \cite{Gottesman_PhD_Thesis}, in which quantum states/codes are
written as joint eigenspaces of stabilizer groups of commuting Pauli
operators. We use our framework to obtain various generalizations of the
seminal Gottesman-Knill theorem \cite{Gottesman99_HeisenbergRepresentation_of_Q_Computers,Gottesman98Fault_Tolerant_QC_HigherDimensions}, which asserts the classical
simulability of Clifford operations. Specifically, we use group and hypergroup theoretic methods to manipulate novel types of stabilizer groups and hypergroups, from infinite continuous ones, to  others that contain non-monomial non-unitary stabilizers and mimic reactions of physical particles. While the PSF is only valid for qubit and  (low dimensional) qudit systems, our formalism can be applied both to discrete and continuous-variable systems, hybrid settings, and  anyonic systems. These results enlarge the known families of quantum states/codes that can be efficiently described with classical methods.

This thesis also establishes the existence of a precise connection between the quantum algorithm of Shor and the stabilizer formalism, revealing a common mathematical structure in several quantum speed-ups and error-correcting codes. This connection  permits a beautiful transfer of ideas between the fields of quantum algorithms and codes, which lies at the roots of our methods and results.

%% file: Zusammenfassung.tex
\selectlanguage{ngerman}

In dieser Doktorarbeit führen wir neue \emph{quantum-computing}-Modelle ein, um das Potential und die Einschränkungen von Quantenalgorithmen zu untersuchen. Unsere Modelle basieren auf algebraischen Erweiterungen der Clifford-Gatter für Qubits (CNOT-, Hadamard-, und $\pi/4$-Phasen-Gatter) und auf Gottesmans Stabilisator-Formalismus für Quantencodes. Wir legen hier zwei Arten von zentralen technischen Beiträgen und Ergebnissen vor und zeigen deren Anwendung für das Design von Quantenalgorithmen, die klassische Simulation von Quantensystemen und
die Beschreibung verallgemeinerter Stabilisator-Zustände und -Codes auf.

Unser erstes Hauptergebnis ist ein Formalismus {zur Beschreibung} eingeschränkter Quantenoperationen, den wir als Normalisator Schaltkreis-Formalismus (\emph{normalizer circuit formalism}) bezeichnen. Darin sind die folgenden Quantengatter erlaubt: die {Quanten-Fourier\-trans\-for\-ma\-ti\-onen} (QFTs), Automorphismen-Gatter und quadratische Phasengatter, die {alle} zu einer
Menge $G$ assoziiert sind, die entweder eine abelsche Gruppe oder eine abelsche Hypergruppe darstellt. Diese Gatter erweitern die Qubit-Clifford-Gatter, die kein universelles
Quantencomputing erlauben und (mit einem klassischen Computer) effizient
simuliert werden können, um mächtige zusätzliche Gatter wie z.B. die QFTs,
die in Shors gefeiertem Algorithmus eine zentrale Rolle spielen. Mit unserem Formalismus zeigen wir, dass das Normalisator-Schaltkreis-Modell
mit geeigneter Wahl von $G$ wichtige Quantenalgorithmen, wie den Shor-Algorithmus und die Algorithmen zur Lösung abelscher \emph{Hidden-Subgroup}-Probleme (HSP) umfasst. Weiterhin charakterisieren wir unter Verwendung selbstentwickelter klassischer Simulationstechniken die Szenarien, unter denen mit Normalisator-Schaltkreise eine Quantenbeschleunigung erreicht werden kann bzw.\ wann das nicht möglich ist. Insbesondere beweisen wir eine Reihe von No-go-Resultaten bezüglich der Möglichkeit, mit herkömmlichen abelschen Fourier-Sampling-Techniken neue Quantenalgorithmen {(mit Quantenbeschleunigung)} zu finden. Außerdem konstruieren wir neue Quantenalgorithmen zum Auffinden verborgener kommutativer Hyperstrukturen. Diese Ergebnisse ermöglichen neue Einsichten in die Ursache der exponentiellen Quantenbeschleunigung, die die Quantenalgorithmen zur Lösung des abelschen und normalen HSPs bieten. 

Unser zweites Hauptergebnis ist ein Rahmen zur Beschreibung von Vielteilchen-Quan\-ten\-zuständen und Quantencodes und zur klassischen Simulation
von Quanten-Schaltkreise. Unser Rahmen umfasst algebraische Erweiterungen von Gottesmans
Pauli-Stabilisator-Formalismus (PSF) \cite{Gottesman_PhD_Thesis} -- in dem Quantenzustände/-codes als gemeinsame Eigenräume von Stabilisatorgruppen kommutierender Pauli-Operatoren geschrieben werden -- und wir benutzen ihn, um verschiedene Verallgemeinerungen des fruchtbaren Gottesman-Knill Theorems \cite{Gottesman99_HeisenbergRepresentation_of_Q_Computers,Gottesman98Fault_Tolerant_QC_HigherDimensions}, das die {effiziente} klassische Simulierbarkeit von Clifford-Operationen beweist, abzuleiten. Genauer gesagt, verwenden wir gruppen- und hypergruppentheoretische Methoden um neue Typen von Stabilisatorgruppen und -hypergruppen zu
behandeln, von unendlich-kontinuierlichen Gruppen zu solchen, die nicht-monomiale, nicht-unitäre Stabilisatoren enthalten und die Reaktionen physikalischer Teilchen nachbilden. Während der PSF nur für Qubit- (und niedrigdimensionale) Qudit-Systeme gültig ist, kann unser Formalismus sowohl auf diskrete wie kontinuierliche Systeme, auf hybride Fälle und auf anyonische Systeme angewendet werden. Diese Ergebnisse vergrößern die
bekannten Familien von Quantenzuständen/-codes, die mit klassischen Methoden effizient beschrieben werden können. 

Diese Arbeit zeigt außerdem eine präzise Verbindung zwischen Shors Quantenalgorithmus und dem Stabilisatorformalismus und enthüllt eine mathematische Struktur, die
zahlreichen Algorithmen mit Quantenbeschleunigung und fehlerkorrigierenden Codes gemeinsam ist. Diese Verbindung ermöglicht einen eleganten Transfer von
Ideen zwischen den Quantenalgorithmen und Quantencodes und stellt die Grundlage unserer Methoden und Resultate dar. 

\selectlanguage{english}

%% file: Publications.tex
\noindent Publications and preprints this thesis is based on:

\begin{enumerate}
\item Juan Bermejo-Vega and Kevin C.\ Zatloukal, \emph{Abelian hypergroups and quantum computation}, preprint (2015), \href{http://arxiv.org/abs/1509.05806}{arXiv:1509.05806 [quant-ph]}.
\item Juan Bermejo-Vega, Cedric Yen-Yu Lin, Maarten Van den Nest. \emph{The computational power of normalizer circuits over black-box groups},  preprint (2014), \href{http://arxiv.org/abs/1409.4800}{arXiv:1409.4800 [quant-ph]}.
\item Juan Bermejo-Vega, Cedric Yen-Yu Lin, Maarten Van den Nest. \emph{Normalizer circuits and a Gottesman-Knill theorem for infinite-dimensional systems}, Quantum Information and Computation 2016, Vol 16., No 5\&6 (2016), \href{http://arxiv.org/abs/1409.3208}{arXiv:1409.3208  [quant-ph]}.
\item Juan Bermejo-Vega, Maarten Van den Nest. \emph{Classical simulations of Abelian-group normalizer circuits with intermediate measurements},  Quantum Information and Computation, Vol 14, No 3\&4 (2014), \href{http://arxiv.org/abs/1210.3637}{arXiv:1201.4867  [quant-ph]}.
\end{enumerate}
Other publications/preprints I contributed to:
\begin{enumerate}
\item Robert Raussendorf, Daniel E. Browne, Nicolas Delfosse, Cihan Okay, Juan Bermejo-Vega, \emph{Contextuality and Wigner function negativity in qubit quantum computation}, preprint (2015), \href{http://arxiv.org/abs/1511.08506}{arXiv:1511.08506 [quant-ph]}.
\end{enumerate}

%% file: Acknowledgements.tex
My deep gratitude extends to my thesis advisors for their invaluable guidance.  I am grateful to Juan Ignacio Cirac for his generous support and  the outstanding research environment he provided. I thank  Maarten Van den Nest for long hours of academic counseling, unforgettable times working together and teaching me his way to do science.  I am grateful to Geza Giedke for his advice me during my last PhD years,  fascinating research discussions, and for his help to put together this thesis.

I thank the Quantum Computing Control and Communication (QCCC), International PhD Program of Excellence and its director, Thomas Schulte-Herbrüggen, for providing a unique interdisciplinary  environment to do a PhD.

This work benefited from  fruitful interactions with great scientific minds. I thank my collaborators Maarten Van den Nest, Cedric Yen-Yu Lin, Kevin C.\ Zatloukal, Geza Giedke, Robert Rau\-ssen\-dorf, Dan E.\ Browne, Nicolas Delfosse and Cihan Okay for the enthusiasm and brilliance they put in our projects together.  I  owe a big thanks to many quantum colleagues around the world with whom I had the great pleasure to discuss my resear chronologically, Earl T. Campbell, Mykhaylo (Mischa) Panchenko, Uri Vool, Pawel Wocjan,  Raúl García-Patrón, Mari Carmen Bañuls, Liang Jiang, Steven M. Girvin, Barbara M.\ Terhal, Hussain Anwar, Tobias J.\ Osborne,  Aram Harrow, Oliver Buerschaper, Martin Schwarz and Richard Jozsa.

My warmest thanks go to the members of Max Planck Institute of Quantum Optics and the MPQ Theory division, with whom I shared lovely years and experiences. I thank the MPQ Theory team as a whole for our great group atmosphere, our top-notch Wednesday Seminar series, our jolly annual retreats, the Wiesn Workshops and the heroic effort everyone made to follow my superluminal speech  velocity. I thank Vanessa Paulisch and András Molnár for the Kicker matches, the excursions to the Honghong-Marat-Flex facilities and their help to print this thesis. Raúl García-Patrón, for insightful conversations on quantum time-travel, Turing machines and men's clothing in the Philosophenweg. Román Orús and our visiting fellow Ondiz Aizpuru, for helpful comments on tensor networks and tortilla de patatas. Miguel Aguado, for his clear vision on topological order, spicy food and tropes. Anna Hackenbroick,  Eliška Greplová, Nayeli Azucena Rodríguez Briones, Thorsten Wahl,  Alexander Müller-Hermes, Xiaotong Ni and Christine Muschik, for fun times  at the office, the infamous Friedrichshafen shipwreck and math riddles.  My many other colleagues of the Quantum Information subggroup, Oliver Buerschaper, Fernando Pastawski, Mari Carmen Bañuls, Hong Hao Tu, Gemma de las Cuevas, Johannes Kofler, Stefan Kühn, Yimin Ge, Henrik Dreyer and Nicola Pancotti,  for our Quantum Coffee journal club series and  inspiring   discussions on science, pseudoscience and geopolitics. Hyung-Won Kim, Senaida Hernández Santana,  Martí Perarnau,  and Jordi Tura, for their memorable  visits to the group.  Veronika Lechner, Andrea Kluth, Lena Baumann and Karin Kügler, for countless hours of administrative help.

I am grateful to my colleagues from the MIT Center for Theoretical Physics for their  hospitality during my stay in 2013, a visit that allowed me to meet great people, broadened my views of quantum computation and enriched this thesis. My sincere thanks go to Aram Harrow, Eddie Farhi and Scott Morley for financial and administrative help to arrange the visit. Them, Cedric Yen-Yu Lin, Pawel Wocjan, Kristan Temme, Shelby Kimmel, Lina Necib, Thomas Vidick, Kamil Michnicki,
Han Hsuan Lin and David Rosenbaum, for introducing me to  MIT's vibrant research environment and graduate life.

During my PhD I had the privilege to meet great researchers and have enlightening discussions at numerous  workshops  and  scientific visits. I  thank  Reinhard Werner and Ciara Morgan (Leibniz University, Hannover) and Robert Raussendorf (University of British Columbia, Vancouver) for inviting me to their groups to present my work.

A warm thank goes to my  colleagues from the Max-Planck-Society's PhDnet (the codeword for our lovely doctoral society) with whom I had the pleasure to organize fantastic scientific events and activities. I am grateful to  Ahmed Omran, Alexander Prehn, Matt Holbran and Axel Beyer, for the energy they put into organizing the 2014 MPQ Student Condensates, the 2014 Student Colloquia and the 2014 MPQ Summer Symposium. To them and to Rosa Glöckner, for our joint efforts that led to the 2014 MPQ PhD Satisfaction Survey. To the next generation of PhD reps,  Julian Krauth, Dominik Ehberger, Vanessa Paulisch,  Matthias Körber and Johannes Zeiher, who continued these efforts and organized many get-together with other MPI PhD researchers in Munich.  To Eva-Regkina Symeonidou, Julia Hutenburg, Bj\o{}rt Kragesteen and Johanna Schulz for pushing forward the PhDnet Equal Opportunity group. To the directors of the MPQ (in particular, to Ignacio Cirac) for supporting the scientific activities of the doctoral researchers, which made life at the Max Planck Society a unique interdisciplinary experience.

I thank Abel Molina and Robert Raussendorf for  piquing my research interest  towards quantum computation early on when I was just a lil undergrad student. A special thank goes to Robert for welcoming me at his research group in 2008  during my undergrad exchange at UBC. They two convinced me that now is an exciting time to work on quantum computation.

Working on my thesis at MPQ Theory was not only an exciting time, but also a life-changing and enriching experience. Like the hobbit character of every epic, I could have not committed the extraordinary deed of completing this manuscript without overcoming some struggles and challenges, which  made me grow as a scientist and as a person. As my main MPQ collaborators and quantum information colleagues departed from the group in 2014, I rose up to become an independent researcher. When the last stages of my thesis writing  were slowed down by an episode of poor health, I endured it and taught myself to be patient. My deepest gratitude goes to  Esther Román García, my Munich friends, my family, Kevin, Robert and Albóndiga, who accompanied me during these times; and to Maarten and Geza, who were always available remotely. A special thank goes to Felix Ehrentraut for being a long-time supporter, flatmate and best friend during  my Munich years.

I am grateful to my university and high school teachers, who took an important part in my education and motivated me to pursue a career in  quantum physics. In particular, I thank Pastora Vega Cruz, for being both a great teacher and caring family member.

I thank  my parents, Juan José and Amparo, and my brother Andrés, who have always loved me and supported me unconditionally. I thank Esther for being my companion of daily adventures, and  Albóndiga, for being our squire.\\

\noindent Funding from the Max Planck Institute of Quantum Optics, QCCC, SIQS, ALG-I and AQuS is gratefully acknowledged.

%% file: tableofcontents.tex
\tableofcontents

%% file: chapter0_Introduction.tex
\chapter{Introduction}\label{chapter0}

This chapter summarizes the technical contributions of this thesis, the methodology we use and points out key connections to prior work. Our aim is to present the main ideas and methods to a non-expert readership and locate them in their historical background. Some technical material (including some results, techniques and connections) are not included in  our exposition and delegated to the technical chapters of the thesis. We refer the reader to chapters \ref{chapterC}-\ref{chapterH} for  full statements of results and extended discussions of their technical significance.

\section{Motivation}

\subsubsection*{Quantum algorithms: the quest and the challenges}

{What are the potentials and limitations of quantum computers?} Arguably, the most attractive feature of quantum computers is their ability to efficiently solve problems with no known  classical solution, as demonstrated by Shor's 1994 groundbreaking discovery of an efficient quantum algorithm for factoring numbers \cite{Shor}. To date, although a more than significant  number of quantum algorithms  has been discovered\footnote{At the time of writing the
``Quantum Algorithm Zoo" website \cite{Jordan_Quantum_Algorithm_Zoo} (one the best known online resources of this field) cites 262 papers on quantum algorithms.} \cite{Mosca09_Quantum_Algorithms_REVIEW,childs_vandam_10_qu_algorithms_algebraic_problems,Bacon10_Recent_Progres_Quantum_Algorithms,VanDamSasaki12_Q_algorithms_number_theory_REVIEW,MoscaSmith12_Algorithms_Quantum_Computers_REVIEW,Childs08_Quantum_Algorithms_Equation_Solving,Aaronson15_QA_MachineLearning_FinePrint,Jordan_Quantum_Algorithm_Zoo},  there is still great demand for finding new ones and applications of them \cite{Childs08_Quantum_Algorithms_Equation_Solving,Aaronson15_QA_MachineLearning_FinePrint}.  Consequently, one of the greatest challenges of the field of quantum computing as per today is to understand for which precise problems quantum algorithms can be \emph{exponentially} (or \emph{super-polynomially}) faster than their classical counterparts.

One lesson gleaned from more than 20 years of quantum algorithm research, is that  quantum computers can exponentially outperform classical ones at solving  certain ``structured'' problems \cite{nielsen_chuang,Aaronson14_Structure_Quantum_Speedups}. Yet, our understanding of what these ``structures'' are remains limited and, even today, the  search for quantum algorithms with exponential benefits remains more of an art than a science.

But what makes quantum algorithms with exponential advantages so hard to find? Although it is nearly impossible to give a mathematical answer to this question, a number of potential reasons have been pointed out in the literature. On the one hand, from the computer science perspective,  a 2004 list\footnote{The list (i-iv) was made in 2004 \cite{Shor_04_Progress_Quantum_Algorithms}, ten years after the factoring algorithm, but it is remarkably up to date.} due to Shor \cite{Shor_04_Progress_Quantum_Algorithms}, highlighted several major obstacles:
\begin{itemize}
\item[(i)] the lack of an analogue   \emph{classical} theory for deciding which problems can be solved efficiently on a \emph{classical} computer; 
\item[(ii)] the modest number of quantum algorithmic techniques discovered so far;
\item[(iii)] the constrains to find  interesting candidate problems to tackle with quantum computers, which should,  ideally, ``neither be in \textbf{P} nor  \textbf{NP}-hard\footnote{Strong evidence suggests that quantum computers cannot efficiently solve NP-complete problems \cite{Shor_04_Progress_Quantum_Algorithms,Aaronson05np-completeproblems}. It is standard  in quantum computing nowadays to assume that to be the case; and that $\textbf{P}\neq\textbf{NP}$,  $\textbf{BPP}\neq \textbf{BQP}$. We take all these assumptions in this thesis. We remind the reader that \textbf{P}  and \textbf{NP} are the classes of problems that can be \emph{solved} and \emph{verified} (respectively) in polynomial time on a deterministic classical computer, while \textbf{BPP} and \textbf{BQP} consist of problems that can be solved in polynomial time in probabilistic classical computers and quantum ones.}'' while most  problems of interest in computer science are in  either one of these two classes;  and also because 
\item[(iv)] candidate problems must remain unsolved after 60 years of classical algorithm research. 
\end{itemize}
One the other hand, from the physical side, one needs to add  the inherent difficulty of comprehending the \emph{emergent complexity} of quantum systems, which poses a barrier not only to understand  quantum speed-ups, but, more generally, quantum many-body phenomena. The same complexity that prevents us from simulating complex quantum dynamics on classical computers \cite{Manin,Manin1998-1999,Feynman} is a double edge possibility sword, which makes possible (in principle) the existence of a quantum speed-up, but does not easily let us unravel  the physical mechanisms that sustain them. In fact, the obstacle of complexity might be found (again and again)  in other formidable unsolved problems in many-body physics,  such as, e.g., deciding whether  topological order is stable at non-zero temperature \cite{Yoshida2011_feasability_selfcorrection}; whether  realizable self-correcting  quantum memories 
\cite{Brown14_QuantumMemories_Finite_Temper,Terhal15_QEC_for_quantumMemories} exist; in identifying the physical ingredients that sustain high $T_c$ superconductivity \cite{Crabtree2011_100_years_superconductivity}; and last but not least, in the principles that guide the complex quantum many-body quantum dynamical evolution of a quantum computer.

\section{This thesis in a nutshell}

A Holy Grail of quantum computation  would be to have a  \emph{theory of quantum speed-ups}\footnote{Throughout this thesis, ``quantum speed-up'' will be  synonymous of ``superpolynomial quantum speed-up''. We do not investigate  quantum algorithms that yield   polynomial advantages over classical computers.} that would tell from basic principles which problems can be efficiently solved with a quantum computer. Ideally, such a theory would delineate the  physical algorithmic mechanisms behind quantum speed-ups and be helpful in the design of new quantum algorithms. Although it is not a priori not clear  whether one can even hope for such a theory within the state of the art of quantum physics and complexity theory (because of the obstacles surveyed above), in this thesis we make progress towards this ambitious goal by developing a theory for a \emph{subclass} of quantum computational speed-ups. 

In the rest of this preliminary chapter we outline how the above program will be  implemented along the thesis. We begin with a discussion of why classical simulation methods and restricted gate models are central to our program, motivating the study of our first models of abelian-group normalizer circuits for gaining insight into quantum Fourier transforms (section \ref{sect:chapter1-chapter2}). After looking at limitations of our first models, we discuss more powerful ones based on the notion of black-box groups, which we prove are useful to describe  quantum algorithms and identifying no-go scenarios for finding them (section \ref{sect:chapter3}). Lastly, we describe our last  normalizer-circuit formalism based on abelian  hyper-structures, which we exploit to devise new quantum algorithms and infer  insights into the working mechanisms of existing ones (section \ref{sect:chapter4}). 

In parallel, we  explain how algebraic generalizations of the  stabilizer formalism \cite{Gottesman_PhD_Thesis} can be constructed and applied to  address the main questions of this thesis.

In section \ref{sect:RelationshipPastWork_c0}  we discuss a few  connections to previous work. 

In section \ref{sect:ReadingGuideThesis} we summarize the structure of the remaining chapters of this thesis.

\subsection{Classical simulations,  normalizer circuits and  quantum Fourier transforms}\label{sect:chapter1-chapter2}

A fruitful approach to understand the emergence and the structure of exponential quantum speed-ups  is to study \emph{restricted} models of quantum computation. Ideally, the latter should exhibit interesting quantum features  and, at the same time, have less power than universal quantum computers (up to reasonable computational complexity assumptions). To date, several  models studied in the literature seem to have these desirable properties, including Clifford circuits
\cite{Gottesman_PhD_Thesis,Gottesman99_HeisenbergRepresentation_of_Q_Computers,Knill96non-binaryunitary,Gottesman98Fault_Tolerant_QC_HigherDimensions}, nearest-neighbor matchgates \cite{Valiant02_matchgates,Knill01_Fermionic_Linear_Optics,Terhal02_Simulation_noninteracting_fermion_circuits,Jozsa08_Matchgates_classical_simulation},  Gaussian operations  \cite{Lloyd98_Analog_Error_Correction,LloydBraunstein99_QC_over_CVs,Bartlett02Continuous-Variable-GK-Theorem,BartlettSanders02Simulations_Optical_QI_Circuits},  the one-clean qubit (DQC1) model \cite{KnillLaflamme98_DQC1}, and commuting circuits \cite{Shepherd10_PhD_thesis,ShepherdBremner09_Temporally_Unstructured_QC,BremnerJozsaShepherd08,Ni13Commuting_Circuits} (a more complete list is given in section \ref{sect:RelationshipPastWork_c0}).

The first result concerning restricted gate models  in the history of quantum computation is the celebrated Gottesman-Knill theorem, which states that any quantum circuit built out of Clifford gates (Hadamards, CNOTs, $\uppi/2$-phase gates) and Pauli measurements can be \emph{efficiently} simulated on a classical computer \cite{Gottesman_PhD_Thesis,Gottesman99_HeisenbergRepresentation_of_Q_Computers, nielsen_chuang}; thus, a quantum computer that works exclusively with these operations cannot achieve \textit{exponential quantum speed-ups}.

The Gottesman-Knill theorem illustrates how subtle the frontier between classical and quantum computational power can be. For example, even though Clifford circuits can be simulated efficiently classically, replacing the $\uppi/2$-phase gates by a $\uppi/4$-phase gate immediately yields a quantum \emph{universal} gate set \cite{Boykin_etal_99_Clifford_pifourth_is_universal_OPEN,Boykin_etal_00_Clifford_pifourth_is_universal_ELSEVIER}. Another interesting feature is that, even though the computing power of Clifford circuits is not stronger than classical computation,  their behavior is genuinely quantum: they can be used, for instance, to prepare highly entangled states (such as cluster states \cite{raussen_briegel_01_Cluster_State, nest06Entanglement_in_Graph_States, raussen_briegel_onewayQC}), or to perform quantum teleportation \cite{Gottesman99_HeisenbergRepresentation_of_Q_Computers}. Yet, in spite of the high degrees of entanglement that may be involved, the evolution of a physical system under Clifford operations can be tracked efficiently using a Heisenberg picture: the \textit{stabilizer formalism},  backbone tool and basis of modern quantum error correction \cite{Terhal15_QEC_for_quantumMemories}.

\subsubsection{Normalizer circuits over abelian groups (setting in \textbf{chapters \ref{chapterC}}-\ref{chapterI})}

The fact that the Gottesman-Knill theorem yields a powerful  tool to identify  non-trivial families of quantum circuits that cannot lead to a quantum speed-up, motivates us to adopt it as the starting point of this thesis. Unfortunately, for our purposes, the theorem presents the major downside  that it can  only be applied to study Clifford gate circuits, which have no known  applications in quantum algorithm design\footnote{Note that Clifford circuits do provide a good setting to study which \emph{quantum states} are universal resources in quantum computation via state injection \cite{BravyiKitaev05MagicStateDistillation,Veitch12_Negative_QuasiProbability_Resource_QC,Howard14Contextuality_Magic,Veitch14_Resource_Theory_Stabilizer_QC,Howard14Contextuality_Magic,Delfosee14_Wigner_function_Rebits,Raussendorf15QubitQCSI}; yet, this thesis is not concerned with  universality but quantum speed-ups.}. To overcome this limitation, we dedicate  the first part of this thesis  (\textbf{chapters \ref{chapterC}-\ref{chapterI}}) to the study of  \emph{new} restricted models of quantum circuits that contain more types of quantum gates.

More precisely, in \textbf{chapters \ref{chapterC}-\ref{chapterI}} we introduce our first models of \emph{normalizer circuits}, which have the most interesting feature of containing \emph{quantum Fourier transforms} (QFT\footnote{Throughout the thesis, the acronym ``QFT'' will always stand for ``quantum Fourier transform'' and not for ``quantum field theory''.}),  quantum gates that are essential in  Shor's factoring algorithm \cite{Shor}   and  are sometimes pointed out to be root of its exponential quantum speed-up. Specifically, we define a \emph{normalizer circuit over an abelian group $G$} to be a quantum circuit consisting of three types of gates:
\begin{itemize*}
\item[---] Quantum Fourier transforms over $G$;
\item[---] Gates which compute automorphisms of $G$;
\item[---] Gates which compute quadratic functions on $G$.
\end{itemize*}
We introduce the above normalizer circuit models in full detail and give examples in chapter \ref{chapterC}, and in chapter \ref{chapterGT} we develop classical group theoretic and algorithmic tools to investigate them. In chapters \ref{chapterF}-\ref{chapterI} we present classical simulation results for normalizer circuits, which we summarize next.

\subsubsection{Chapter \ref{chapterF}: finite abelian group $G$} 

In chapter \ref{chapterF} we fix  $G$  to be a finite abelian group. When $G=\mathbb{Z}_2^{n}$ (the group of $n$-bit strings with addition modulo 2),  normalizer circuits coincide precisely with  the standard Clifford circuits. However, more exotic families of circuits can be obtained by simply modifying the parameter $G$. But there is more: for $G=\mathbb{Z}_{2^n}$, the associated normalizer circuit contain precisely the QFTs which are used in Shor's discrete-logarithm and factoring algorithms \cite{Shor}; for other choices of $G$, normalizer circuits contain highly entangling gates and QFTs associated to arbitrary abelian groups, which are central subroutines in Kitaev's ubiquitous quantum phase estimation algorithm \cite{kitaev_phase_estimation} and in quantum algorithms for  solving so-called abelian \textbf{Hidden Subgroup Problems} (HSPs) \cite{Shor,Deutsch85quantumtheory,Simon94onthe,Boneh95QCryptanalysis,Grigoriev97_testing_shift_equivalence_polynomials,kitaev_phase_estimation,Kitaev97_QCs:_algorithms_error_correction,Brassard_Hoyer97_Exact_Quantum_Algorithm_Simons_Problem,Hoyer99Conjugated_operators,MoscaEkert98_The_HSP_and_Eigenvalue_Estimation,Damgard_QIP_note_HSP_algorithm}: the latter comprise not only Shor's, but also Deutsch's \cite{Deutsch85quantumtheory}, Simon's \cite{Simon94onthe} quantum algorithms; furthermore, all famous quantum algorithms for breaking  widely used public-key cryptosystems (namely, RSA \cite{RSA}, Diffie-Hellman's \cite{DiffieHellman} and elliptic curve cryptopgraphy  \cite{Menezes96_cryptography_book,Buchmann00_cryptography_book}) belong to this class. Our motivation to investigate this abelian-group normalizer circuit model in  \textbf{chapter \ref{chapterF}} (see also \textbf{chapter \ref{chapterC}}) is to gain insight into the question of \textit{``When does the QFT serve as a resource for quantum computation?''} and, specifically,  of \emph{when does it lead  exponential quantum  speed-ups?}.

This chapter is based on \cite{BermejoVega_12_GKTheorem} (joint work with Maarten Van den Nest).

\paragraph{Main results and techniques.} Our first main result in \textbf{chapter \ref{chapterF}} (cf.\ \textbf{theorem \ref{thm_main}}) is a generalized Gottesman-Knill theorem, which states that  normalizer circuits over a group $G$ can be efficiently  simulated in a classical computer if $G$ is given to us in a canonically decomposed form. Specifically, when $G$ is given as a product of cyclic group factors
\begin{equation}\notag
G=\Z_{D_1}\times \cdots \times \Z_{D_m}
\end{equation}
----in which case normalizer gates over  $G$ act  on  a Hilbert space
\begin{equation}\label{eq:Hilbert Space abelian Group}
\mathcal{H}_G= \C^{D_1}\otimes\cdots \otimes \C^{D_m},
\end{equation}
i.e.\ $\mathcal{H}_G$  is a collection of $m$  finite systems of arbitrarily large dimensions $D_1,D_2,\ldots,D_m$---our result says that \emph{any} quantum circuit built of normalizer gates over $G$ can be classically simulated in time  at most \emph{polynomial} in the number of QFTs and gates present of the circuit, the number of factors $m$, and the logarithms $\log D_i$ of all local dimensions (hence, the simulation is efficient in the dimension of $\mathcal{H}_G$ even if $D_i=2^n$ is exponentially large). The significance of this result is that it identifies many non-trivial families of quantum computations that \emph{fail} to harness the power of QFTs in order to achieve achieve exponential quantum speed-ups. 

Our second main contribution in  \textbf{chapter \ref{chapterF}} (cf. \textbf{theorems \ref{thm structure test}, \ref{thm_Measurement_Update_rules}, \ref{thm Normal form of an stabilizer state}}) is a generalized
\textbf{stabilizer formalism over finite abelian groups}, i.e., for systems of the form (\ref{eq:Hilbert Space abelian Group}).  For a given finite abelian group $G$, our formalism lets us describe rich families of quantum states and codes within $\mathcal{H}_G$, which we name \emph{stabilizer states/codes} over $G$, as joint eigenspaces of stabilizer groups of  \emph{generalized Pauli operators} over $G$: for groups of the form $\Z_2^n$, we recover the  standard definitions of qubit Pauli operator and qubit stabilizer state/code, hence, our formalism extends Gottesman's PSF. We show that our formalism can be used to efficiently track the evolution of abelian-group stabilizer states  under arbitrarily-long normalizer circuits in a Heisenberg picture, by tracking a small-number of stabilizer group generators. Furthermore, we develop explicit analytic \emph{normal forms} for the evolved states in terms of subgroup cosets and quadratic functions. This techniques  are key to prove our main simulation result.

The main technical effort in chapter \ref{chapterF} goes into developing our generalized stabilizer formalism. Prior to our work, classical techniques to simulate Clifford circuits had been developed for \emph{qudit} systems of constant dimension $d$ (in our setting, this parameter grows unboundedly); most works further assumed $d$ to be prime, in which case $G$ is a vector space and exploited standard field-theoretic algorithms in the simulation  (e.g. Gaussian elimination). For our simulations we develop different techniques that involve representation theory of abelian groups, computational group theory,  and Smith normal forms.

\subsubsection{Chapter \ref{chapterI}: infinite abelian group $G$} 

In \textbf{chapter \ref{chapterI}} we introduce new generalized families of normalizer circuits over \emph{infinite} abelian groups that act on \emph{infinite dimensional} systems. Specifically, we define normalizer circuits over groups of the form\footnote{Our construction can be applied to define normalizer circuit models over arbitrary abelian groups, but we focus on these  types for the reasons given in the main text.}    $$G=\Z^a \times \T^b \times \Z_{D_1}\times \cdots \times \Z_{D_c},$$ extending our prior setting by allowing  new types of group factors, namely, integer lattices $\Z^a$ of arbitrary rank $a$, and  hypertori $\T^b$ of arbitrary dimension $b$. The motivation for adding $\Z$ is that several number theoretical problems of interest in quantum computation are naturally connected to problems over the integers (e.g., factoring is related to hidden subgroup over $\mathbb{Z}$); we further add $\mathbb{T}$ because it is connected to $\Z$ via the Fourier transform over this group.  

This chapter is based on \cite{BermejoLinVdN13_Infinite_Normalizers} (joint work with Cedric Yen-Yu Lin and Maarten Van den~Nest).

\paragraph{Main results and techniques.} Our main contributions in this chapter are  a generalized \emph{stabilizer formalism} and a \emph{Gottesman-Knill theorem} for infinite dimensional systems, which states that all normalizer circuits over infinite groups as above can be \emph{simulated} classically in polynomial time. These results extend those of chapter \ref{chapterF} to infinite dimensions. 

The simulation techniques in chapter \ref{chapterI} differ strongly from  previous work on stabilizer simulations because they can handle continuous infinite groups $G$ as well as and \emph{continuous-infinite stabilizer groups}. The groups under consideration are notoriously difficult to manipulate because they are neither finite, nor finitely generated, nor countable; they are not vector spaces and do not have bases; and they are not compact. Remarkably, in this setting, generalized stabilizer groups \emph{can no longer} be described with finite sets of generators. Instead, we develop a novel machinery for handling infinite stabilizer groups based on \emph{linear map} encodings,  \emph{normal forms} for quadratic functions and group morphisms. Combining this technology with novel \emph{$\varepsilon$-net techniques}, we devise the most powerful classical algorithm to date to sample the support of  infinite dimensional stabilizer states. This leads to our simulation result a la Gottesman-Knill for infinite dimensional systems.

\subsubsection{Discussion (chapters \ref{chapterF}-\ref{chapterI})}

We end this subsection discussing potential applications of the techniques developed in chapters \ref{chapterF}-\ref{chapterI} outside the scope of this thesis.

First, we recall that Gottesman's original Paul Stabilizer Formalism and the Gottesman-Knill theorem for qubits and qudits  has been used in a variety of settings. The PSF itself is a central tool in, e.g., 	measurement-based quantum computation  (with qubits \cite{raussen_briegel_onewayQC} and qudits \cite{ZhouZengXuSun03,Schlingemann04ClusterStates}), quantum error-correction    (qubits \cite{Gottesman_PhD_Thesis,Gottesman99_HeisenbergRepresentation_of_Q_Computers,BravyiKitaev05MagicStateDistillation}, qudits \cite{Gottesman98Fault_Tolerant_QC_HigherDimensions,CampbellAnwarBrowne12MagicStateDistillation_QUTRITS,CampbellAnwarBrowne12MagicStateDistillation_in_all_prime_dimensions,Anwar14_Decoders_Qudit_Topological_Codes,Campbell14_Enhance_FTQC_d_level_systems,Watson15_Qudit_Color_n_Gauge_Color_Codes}), secret-sharing (qubits \cite{Hillery99Quantum_Secret_Sharing}, qudits \cite{Cleve99Quantum_Secret_Sharing, Gottesman00Quantum_Secret_Sharing}); in the study of topologically-ordered systems (qubits \cite{kitaev_anyons}, qudits \cite{BombinDelgado07HomologicalQEC, BullockBrennen07QUDIT_surface_code,DuclosCianci_Poulin13ToricCode_QUDITS}) and  universal resources for  quantum computation via state injection (rebits \cite{Delfosee14_Wigner_function_Rebits}, qubits \cite{Raussendorf15QubitQCSI}, qudits \cite{Veitch12_Negative_QuasiProbability_Resource_QC}), among others. The standard Gottesman-Knill  is often  applied in fault-tolerant  quantum computation in order to simulate Pauli/Clifford noise channels \cite{Paler14ErrorTracking,Gutierrez87.030302} and delay recovery operations \cite{DiVincenzo07_effectiveFTQC_Slow_Measurements}, which indirectly reduces noise threshold requirements \cite{Steane03_Overhead_Threshold_FTQEC, Knill05_QComp_RealisticallyNoisy,Cross:2008:FQC:1559452,Aliferis:2006:QAT:2011665.2011666,Cross:2009:CCS:2011814.2011815}. 

It is plausible that the techniques developed in chapters \ref{chapterF}--\ref{chapterI} could find applications in  the fields mentioned above. An attractive feature of our work  is that it leads to the first known stabilizer formalism and  normalizer gate models for \emph{hybrid} systems  of asymmetric qudits $\mathcal{H}_{D_1}\otimes \cdots \otimes \mathcal{H}_{D_a}$, harmonic oscillators $\mathcal{H}_{\textrm{osc}}^{\otimes b}$  and quantum rotors $\mathcal{H}_{\textrm{rot}}^{\otimes c+d}$, which have Hilbert spaces labeled by groups of the form  $\DProd{D}{a}$, $\R^b$ and $\Z^c\times \T^d$, respectively: qudits and harmonic oscillators   have well-known applications in QIP over discrete and continuous-variables; the latter, quantum rotors, describe other QIP platforms like, e.g., \emph{Josephson tunneling junctions}, which are the basic constituent of all superconducting-qubit designs for building quantum computers, and  electromagnetic modes carrying angular momentum  \cite{Rigas11_OAM_in_phase_space,Rigas_Soto10_nonnegative_Wigner_OAM_states}.  Our normalizer gate models for these systems  extend the well-known families of qudit Clifford gates and continuous-variable  Gaussian unitaries \cite{Lloyd98_Analog_Error_Correction,LloydBraunstein99_QC_over_CVs,Bartlett02Continuous-Variable-GK-Theorem,BartlettSanders02Simulations_Optical_QI_Circuits}  and define ``superconducting'' analogues the latter, which might find uses in QEC and QIP\footnote{Normalizer gates over groups of the form $\Z_d^a$ and $\R^b$, respectively, yield the standard qudit Clifford gates and Gaussian unitaries (see chapter \ref{chapterC}, section  \ref{sect:Introduction2} and appendix \ref{aG}).  Within  our formalism, we  obtain more general models gates by either looking at systems whose  Hilbert spaces $\mathcal{H}_G$ are labeled by different groups and/or by combining registers $\mathcal{H}_{G_1}\otimes \mathcal{H}_{G_2}$ into a larger ``hybrid'' system.}. Our stabilizer formalism  yields  a framework for defining and analyzing stabilizer codes and states for all these platforms:  in finite dimensions, our techniques are  novel in that they can  handle qudit dimensions that differ, or are not prime numbers, or can be  large; for infinite dimensional systems labeled by $\Z$ and $\T$ groups, our methods could be  applied  to simulate charge/phase/flux noise or delay recovery operations in fault-tolerant quantum computing schemes based on ``rotating-variable'' superconducting codes (such as, e.g. Kitaev's  0-$\uppi$ codes \cite{Kitaev06_Protected_Qubit_Supercond_Mirror,Brooks13_Protected_gates_for_superconducting_qubits}).

\subsection{The computational power of normalizer circuits over black-box groups}\label{sect:chapter3}

The models of normalizer circuit over abelian groups  studied in chapters \ref{chapterF}-\ref{chapterI} let us identify scenarios where QFTs fail to achieve achieve exponential quantum speed-ups. In  chapter \ref{chapterB} of this thesis  we introduce  models of quantum computation based on extended normalizer circuits that have non-trivial quantum power. We further use these models and classical simulation techniques in order to characterize the computational power of a large family of  quantum algorithms.

Specifically, in chapter \ref{chapterB}, we consider \emph{black box normalizer circuits} that are  associated  to finite abelian groups that are \emph{black box groups} $\mathbf{B}$ (as introduced by Babai and Szemerédi in   \cite{BabaiSzmeredi_Complexity_MatrixGroup_Problems_I}) and allow $G$ to be of the  form. 
\begin{equation}\label{eq:Black Box Groups INTRO}
G=\left(\Z^a \times \T^b \times \Z_{D_1}\times \cdots \times \Z_{D_c}\right)_{\textrm{previous setting}}\times \left(\mathbf{B}_{}\right)_{\textrm{new setting}}.
\end{equation}
Note that the  difference between this and earlier settings is that the group $\mathbf{B}$ is no longer assumed to be given in a decomposed form $\Z_{D1}\times \cdots \times \Z_{D_d}$, which makes a distinction in terms of computational complexity: though every finite abelian group is isomorphic to some decomposed group $\Z_{D_1}\times \cdots \times \Z_{D_d}$, computing such a  decomposition is at least as hard as  factoring, which is polynomial-time reducible to the problem of decomposing \emph{multiplicative} groups  $\mathbb{Z}^{\times}_N$ of integers modulo $N$ \cite{Shoup08_A_Computational_Introducttion_to_Number_Theory_and_Algebra}. In chapter \ref{chapterB},  abelian groups for which such a direct product product decomposition is a priori unknown are modeled as \emph{black-box group} for which it is only known how elements can be \emph{efficiently} represented (as bit-strings) and multiplied/added.  

This chapter is based on \cite{BermejoLinVdN13_BlackBox_Normalizers} (joint work with Cedric Yen-Yu Lin and Maarten Van den~Nest).

\paragraph{Main results and techniques.} In contrast to our classical simulation result for decomposed abelian groups, we find that allowing black-box groups in our setting dramatically changes the computational power of normalizer circuits. In particular, we show that many of the most famous quantum algorithms are particular instances of normalizer circuits over black-box groups (\ref{eq:Black Box Groups INTRO}), thereby proving that normalizer circuits over black box groups can offer \emph{{exponential quantum speed-ups}} and break widely used public-key cryptographic systems.  Namely, in our generalized formalism, the following algorithms are  examples of black-box normalizer circuits over a group $G$ of form (\ref{eq:Black Box Groups INTRO})---or have equivalent normalizer circuit versions:	
\begin{itemize}
\item  Shor's algorithm for computing discrete logarithms \cite{Shor}:  $G= \Z_{p-1}\times\Z_p^\times$;
\item Shor's factoring algorithm \cite{Shor}: $G= \Z\times\Z_N^\times$;
\item The generalized Shor's algorithm for finding  discrete-logarithms over an elliptic curve $E$ \cite{ProosZalka03_Shors_DiscreteLog_Elliptic_Curves,Kaye05_optimized_Quantum_Elliptic_Curve,CheungMaslovMathew08_Design_QuantumAttack_Elliptic_CC}: $G=\Z^2\times E$;  
\item Simon's algorithm \cite{Simon94onthe} and other oracular abelian hidden subgroup problem algorithms  \cite{kitaev_phase_estimation,Boneh95QCryptanalysis}, are normalizer circuits over groups of the form $G\times\mathbf{B}$, where $G$ and $\mathbf{B}$ are a group and a black-box group determined by the input of the HSP;
\item Cheung-Mosca's algorithm for decomposing black-box finite abelian groups \cite{mosca_phd, cheung_mosca_01_decomp_abelian_groups} is a combination of several types of black-box normalizer circuits.
\end{itemize}
The above results establish a precise  connection between Clifford circuits and Shor-like quantum algorithms and, furthermore, imply that black-box normalizer circuits are powerful enough to break important cryptosystems such as RSA \cite{RSA}, Diffie-Hellman's \cite{DiffieHellman} and elliptic curve cryptopgraphy  \cite{Menezes96_cryptography_book,Buchmann00_cryptography_book}. In the rest of chapter \ref{chapterB}, we further exploit the \emph{abelian group stabilizer formalism} developed in earlier chapters to tightly \emph{characterize} the computational power of normalizer circuits, as outlined next.
\begin{itemize}
\item We show that the problem of decomposing black-box  groups is {\emph{complete}} for the class of computational problems solvable by black-box normalizer circuits:  once an oracle to solve that problem is provided, we show  (\textbf{theorem \ref{thm:Simulation}}) that our simulation techniques from earlier chapters \emph{render black-box normalizer circuits efficiently  classically simulable}.  For this result, we need to introduce a generalized version of the group decomposition problem considered by Cheung-Mosca \cite{mosca_phd, cheung_mosca_01_decomp_abelian_groups}, for which we give an efficient quantum algorithm based on normalizer circuits; extending, along the way,  the result of \cite{mosca_phd, cheung_mosca_01_decomp_abelian_groups}. These results demonstrate that the computational power of normalizer circuits is encapsulated precise in the classical hardness of decomposing black-box groups.
\item We give  a \emph{no-go theorem} (\textbf{theorem \ref{thm:No Go Theorem}})  for finding new quantum algorithms within the class of black-box normalizer circuits considered. This result has  immediate  implications for quantum algorithm design, for it imposes provable restrictions to quantum computing theorists for finding new quantum algorithms with the basic set of \emph{Fourier sampling techniques over finite abelian groups}, which are covered by our normalizer circuit model: specifically, our result shows that any (potentially sophisticated) quantum algorithm based on such techniques can be emulated by smartly using  our extended Cheung-Mosca quantum algorithm\footnote{Like all of our results, this no-go theorem is for  quantum algorithms with superpolynomial speed-ups.}.
\item Another consequence of theorem \ref{thm:No Go Theorem} is a \emph{universality result for short normalizer circuits} that explains a curious quantum computing mystery: although many quantum algorithms use quantum Fourier transforms, interestingly,  many of them  use only a small number of them and, in fact, two are often enough \cite{Shor,Brassard_Hoyer97_Exact_Quantum_Algorithm_Simons_Problem,Hoyer99Conjugated_operators,MoscaEkert98_The_HSP_and_Eigenvalue_Estimation,Damgard_QIP_note_HSP_algorithm,Deutsch85quantumtheory,Simon94onthe,Boneh95QCryptanalysis,Grigoriev97_testing_shift_equivalence_polynomials,kitaev_phase_estimation,Kitaev97_QCs:_algorithms_error_correction}. A corollary of theorem \ref{thm:No Go Theorem} is that all quantum algorithms based on normalizer circuits can be simulated by sequences of quantum circuits that  contain two QFTs and use intermediate classical processing. Hence, normalizer circuits cannot gain any significant  superpolynomial advantage from using more than two Fourier transforms.
\end{itemize}

\subsubsection{Discussion (chapter \ref{chapterB})}

The no-go theorem presented in  chapter \ref{chapterB} is not only a useful tool to identify  approaches for finding quantum algorithm that do not work, for, by carefully analyzing the conditions under which the theorem  holds, one may guess promising avenues for finding new ones (cf. section \ref{sect:Discussion}). In fact, this and other insights from chapter \ref{chapterB} helped us to find the new quantum algorithms that we present in our next chapter \ref{chapterH}.

\subsection{Abelian hypergroups and quantum computation}\label{sect:chapter4}

In chapters \ref{chapterF}-\ref{chapterB} we showed that the abelian group normalizer circuit framework and the abelian group  stabilizer formalism are helpful tools to understand the exponential quantum speed-ups of Shor's algorithm and the quantum algorithms  for solving abelian hidden subgroup problems (HSP). In the final  chapter of this thesis (\textbf{chapter \ref{chapterH}}) we attempt to extend this approach in order to gain insight into quantum algorithms for \emph{nonabelian groups} hidden subgroup problems, which have been object of intense research work  \cite{Hallgren00NormalSubgroups:HSP,EttingerHoyerKnill2004_Hidden_Subgroup,Kuperberg2005_Dihedral_Hidden_Subgroup,Regev2004_Dihedral_Hidden_Subgroup,Kuperberg2013_Hidden_Subgroup,RoettelerBeth1998_Hidden_Subgroup,IvanyosMagniezSantha2001_Hidden_Subgroup,MooreRockmoreRussellSchulman2004,InuiLeGall2007_Hidden_Subgroup,BaconChildsVDam2005_Hidden_Subgroup,ChiKimLee2006_Hidden_Subgroup,IvanyosSanselmeSantha2007_Hidden_Subgroup,MagnoCosmePortugal2007_Hidden_Subgroup,IvanyosSanselmeSantha2007_Nil2_Groups,FriedlIvanyosMagniezSanthaSen2003_Hidden_Translation,Gavinsky2004_Hidden_Subgroup,ChildsVDam2007_Hidden_Shift,DenneyMooreRussel2010_Conjugate_Stabilizer_Subgroups,Wallach2013_Hidden_Subgroup,lomont_HSP_review,childs_lecture_8,VanDamSasaki12_Q_algorithms_number_theory_REVIEW} in the last decades   of quantum computation. The motivation of the HSP research program followed the breakthrough discoveries that solving the  HSP over symmetric and dihedral groups would lead to  revolutionary efficient algorithms  for  Graph Isomorphism \cite{Ettinger99aquantum} and certain latticed-based problems  \cite{Regev:2004:QCL:976327.987177}. Despite much effort, no efficient quantum algorithm for dihedral or symmetric HSP has yet been found.

Specifically, our initial goal   in chapter \ref{chapterH} is to gain understanding into   a  seminal \emph{efficient} quantum algorithm of Hallgren, Russell, and Ta-Shma \cite{Hallgren00NormalSubgroups:HSP} for finding hidden \emph{normal} subgroups, which, remarkably, works efficiently for \emph{any} nonabelian group. Surprisingly, despite the fact that the HRT algorithm is also the basis of several sophisticated algorithms for nonabelian HSPs \cite{Gavinsky:2004:QSH:2011617.2011625,Ivanyos:2008:EQA:1792918.1792983}, its efficiency remains poorly explained. Given the   success of the normalizer circuit framework (chapters \ref{chapterF}-\ref{chapterB}) at understanding abelian HSP quantum algorithms, we address the question of whether a more sophisticated stabilizer formalism can shed light into this question and lead to new applications of quantum computation. Our main results, summarized below, answer this question in the affirmative.

This chapter is based on \cite{BermejoVegaZatloukal14Hypergroups} (joint work with Kevin C.\ Zatloukal).

\paragraph*{Main results and techniques.}

\begin{itemize}
\item Our first result in chapter \ref{chapterH} is a \emph{connection} between the hidden normal subgroup problem (HNSP) and \emph{abelian hypergroups}, which are algebraic objects that model collisions of physical particles and anti-particles and generalize abelian groups. Our result shows that  in many natural cases the HNSP  can be reduced to the commutative  problem of finding subhypergroups of abelian hypergroups. 
\end{itemize}
The above connection and the fact that abelian groups are particular instances of abelian hypergroups, motivates us to explore whether abelian hypergroups and the normalizer circuit framework can be combined to answer our initial main question.
Our findings are presented next.\begin{itemize}

\item \textbf{A hypergroup stabilizer formalism.} We present a generalized stabilizer formalism  based on commuting \emph{hypergroups} of generalized Pauli operators (whose multiplication mimics particle annihilation processes) as well as extended families of (Clifford-like)  normalizer circuits over abelian hypergroups. Using our formalism, we devise classical algorithms for simulating hypergroup normalizer circuits and develop analytic normal forms for describing quantum many-body quantum states (namely, hypergroup coset states) and analyzing the convergence of quantum algorithms.
\item \textbf{New quantum algorithms.} We devise  the first provably efficient quantum algorithms for finding hidden subhypergroups of abelian hypergroups and, exploiting our hypergroup-HNSP connection, also new quantum algorithms for the latter problem. Our algorithms are based on  hypergroup  normalizer gates, which let us apply our hypergroup stabilizer methods in our analysis. We show that our algorithms provably work for hypergroups that arise from nilpotent, dihedral and symmetric groups, which are the most interesting groups from the nonabelian HSP perspective. In contrast, no efficient quantum algorithm for nilpotent, dihedral or symmetric HSPs is known.
\end{itemize}

\subsubsection{Discussion (chapter \ref{chapterH})}

Our HNSP quantum algorithms are different from the one of Hallgren et al.\ in that they exploit commutative structures that are related to those present in Shor's  algorithm via a stabilizer formalism: this provides an important new insight into why the HNSP is much easier than the general nonabelian HSP. Furthermore, our quantum algorithm for finding abelian subhypergroups   provide strong evidence that the abelian Hidden Subhypergroup Problem \cite{Amini_hiddensub-hypergroup,Amini2011fourier} is a much easier problem for quantum computers  than the nonabelian HSP (perhaps even more natural one because of its elegant connection to a stabilizer picture). 
 
A main building block of the quantum algorithms in this chapter is a novel \emph{adaptive/recursive quantum Fourier sampling} technique of independent interest. This technique  overcomes the limitations of an earlier abelian HSHP quantum algorithm \cite{Amini_hiddensub-hypergroup,Amini2011fourier} based on Shor-Kitaev's quantum phase estimation \cite{kitaev_phase_estimation}, which we prove to be inefficient on easy instances.

Beyond the scope of this thesis, abelian hypergroups have important  applications in, e.g., convex   optimization  \cite{KlerkSDP_AssociationSchemes,anjos2011handbook}, classical classical error correction  \cite{corsini2003applications} and conformal field theory   \cite{Wildberger1994HypergroupsApplications}. In  topological quantum computation \cite{kitaev_anyons}, fusion-rule hypergroups   are indispensable  in the study of nonabelian anyons and topological order \cite{Kitaev2006_Anyons_Exactly_Solved_Model}. 

Our stabilizer formalism over abelian hypergroups provide the first generalization and alternative to  Gottesman's Pauli Stabilizer Formalism  where  stabilizer operators are not necessarily \emph{unitary}, nor \emph{monomial}, nor \emph{sparse} matrices. These techniques are likely to find applications in  quantum error correction and classical simulations, e.g., for probing the classical simulability of protected gates over topological quantum field theories \cite{Beverland14_ProtectedGates_Topological}. 

\subsection{Summary of complexity theoretic results}
 
\begin{figure}
\centering
\includegraphics[width=0.7\linewidth]{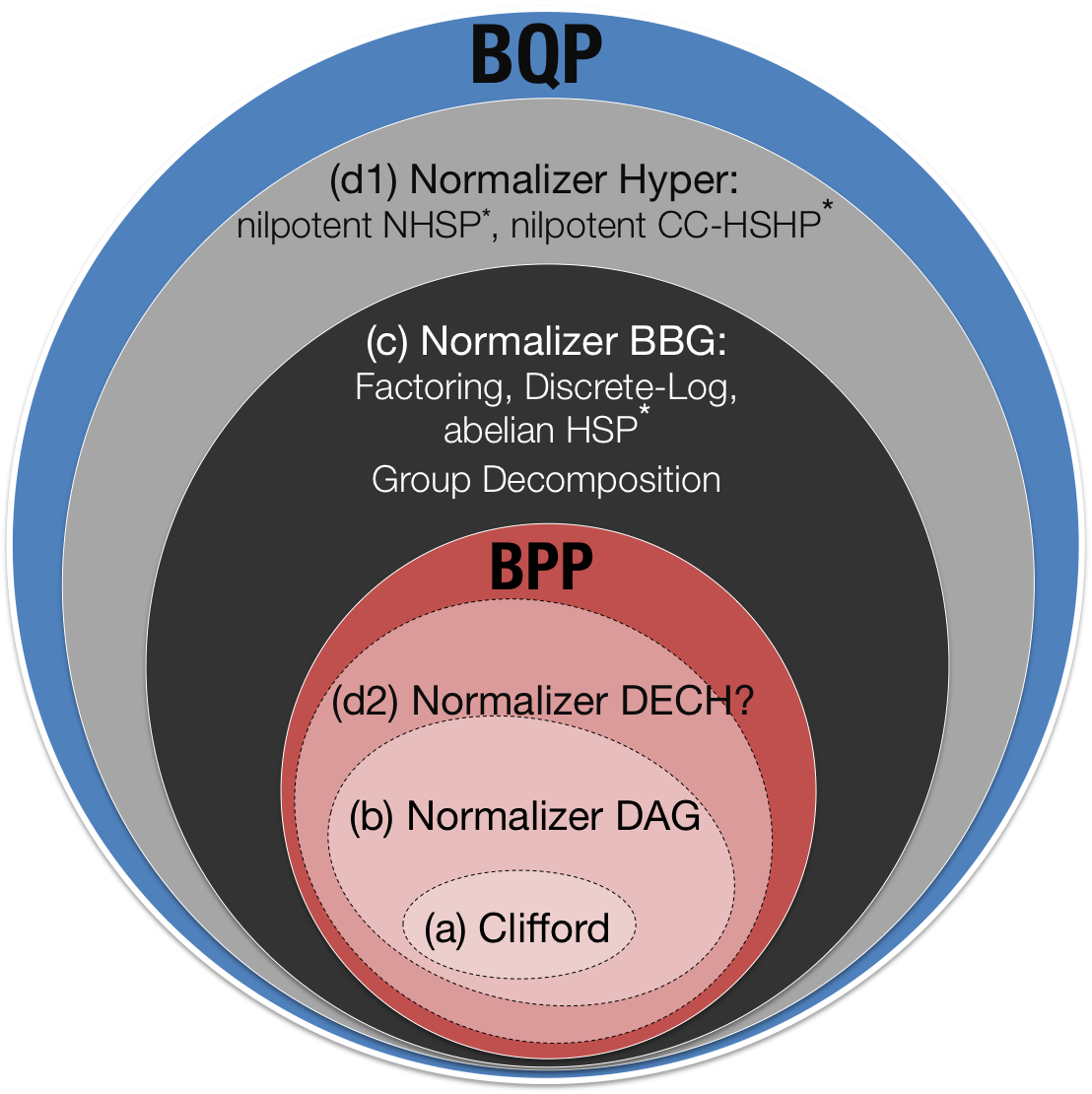}
\caption{Circles in this Venn diagram represent complexity classes of computational problems: a class of problems X is plotted inside Y if Y is known to include X. BPP and BQP are the classes of problems that can be solved by quantum and (probabilistic) classical computers in polynomial time \cite{nielsen_chuang}. The quantum circuits corresponding to classes outside BPP have access to classical computers that carry out classical post-processing tasks. Class (a) Clifford, which contains the problems solvable via qubit Clifford circuits, forms a  subset of BPP \cite{Gottesman99_HeisenbergRepresentation_of_Q_Computers}  (this containment is believed to be strict \cite{AaronsonGottesman04_Improved_Simul_stabilizer}). Class (b) Normalizer DAG represents the problems solvable via normalizer circuits over arbitrary decomposed abelian groups (chapters \ref{chapterF}-\ref{chapterI}). Class (c) Normalizer BBG represents the problems solvable via black-box-group normalizer circuits, e.g., factoring and discrete-log (chapter \ref{chapterB}). Class (d1) Normalizer Hyper---resp.\ class (d2) Normalizer DECH---contain problems solvable via normalizer circuits over efficiently---resp\ doubly-efficiently---computable hypergroups (chapter \ref{chapterH}).  The  containment of (a) and (b) inside (d2) (marked with ``?'') is conjectured in chapter \ref{chapterH} and only  proven for CSS-preserving operations. Problems marked by a superscript ``*'' are oracular and  proven to be contained in their respective classes if certain subroutines to carry out  some classical calculations are provided externally. We refer the reader to chapters \ref{chapterC}-\ref{chapterH} for details on the normalizer circuit models associated to each class.}
\label{fig:ComplexityClasses}
\end{figure}

In order to summarize our complexity theoretic results, we provide a Venn diagram (figure \ref{fig:ComplexityClasses}) that represents the known complexity classes associated to the different families of normalizer circuits investigated in this thesis and their relationships. 

\section{Relationship to previous works}\label{sect:RelationshipPastWork_c0}

We discuss some  (non-technical) connections between our thesis and other works on restricted models of quantum computations and/or classical simulations. We refer to chapters \ref{PreviousWork_c1}, \ref{PreviousWork_c2}, \ref{PreviousWork_c3} and \ref{PreviousWork_c4} for connections related to the main technical contributions of the thesis.

The normalizer circuit formalism presented in this thesis was developed by a sequence of various works. Normalizer circuits over \emph{finite abelian groups} with terminal measurements were  introduced by Van den Nest \cite{VDNest_12_QFTs} and Bermejo-Vega-VdN \cite{BermejoVega_12_GKTheorem}, over infinite and black-box groups by BV-Lin-VdN \cite{BermejoLinVdN13_Infinite_Normalizers,BermejoLinVdN13_BlackBox_Normalizers}, and over abelian hypergroups by BV-Zatloukal  \cite{BermejoVegaZatloukal14Hypergroups}. 

Clifford circuits over qubits and qudits, which can be understood as normalizer circuits over groups of the form $\Z_2^m$ and $\Z_d^m$ (section \ref{sect_examples}), and  Gaussian unitaries, which can be approximated by normalizer circuits over $\R^m$ to any degree of accuracy (chapter \ref{PreviousWork_c2}, appendix \ref{aG}), have been extensively investigated in the literature: see, e.g.,   \cite{Gottesman_PhD_Thesis,Gottesman99_HeisenbergRepresentation_of_Q_Computers,Gottesman98Fault_Tolerant_QC_HigherDimensions,dehaene_demoor_coefficients,AaronsonGottesman04_Improved_Simul_stabilizer,dehaene_demoor_hostens,deBeaudrap12_linearised_stabiliser_formalism,VdNest10_Classical_Simulation_GKT_SlightlyBeyond,JozsaVdNest14_Classical_Simulation_Extended_Clifford_Circuits} and \cite{Lloyd98_Analog_Error_Correction,LloydBraunstein99_QC_over_CVs,Bartlett02Continuous-Variable-GK-Theorem,BartlettSanders02Simulations_Optical_QI_Circuits,Gottesman01_Encoding_Qubit_inan_Oscillator,Barnes04StabilizerCodes_for_CV_WEC,BraunsteinLoock05QI_with_CV_REVIEW,GarciaPatron12_Gaussian_quantum_information,Aaronson11_Computational_Complexity_Linear_Optics} for Clifford and Gaussian references, respectively. 

Certain generalizations of Clifford circuits that are not normalizer circuits have also been studied: \cite{AaronsonGottesman04_Improved_Simul_stabilizer,BravyiKitaev05MagicStateDistillation,Jozsa08_Matchgates_classical_simulation,VdNest10_Classical_Simulation_GKT_SlightlyBeyond,JozsaVdNest14_Classical_Simulation_Extended_Clifford_Circuits} consider Clifford circuits supplemented with some non-Clifford ingredients;  a different form of  Clifford circuits based on projective normalizers of unitary groups
 was investigated in \cite{ClarkJozsaLinden08Generalized_Clifford_Groups}.

Aside from generalizations of Clifford circuits, many other classes of restricted quantum circuits have been studied in the literature (very often within the context of classical simulations). Some examples (by no means meant to be an exhaustive list) are nearest-neighbor matchgate circuits \cite{Valiant02_matchgates,Knill01_Fermionic_Linear_Optics,Terhal02_Simulation_noninteracting_fermion_circuits,Jozsa08_Matchgates_classical_simulation,Bravyi05_Lagrangian_Rep_Fermionic_Linear_Optics,JozsaKrausMiyakeWatrous10Matchgates,VdNest11_Matchgates,BravyiKoenig12_Simulation_Dissipative_Fermionic_Linear_Optics,deMeloCwiklinskiTerhal13_Noisy_Fermionic_Quantum_Computation}, the one-clean qubit model \cite{AmbainisShulmanVazirani06_Computing_highly_mixed_states,Pouline03_Integrability_DQC1,Poulin04_Fidelity_Decay_DQC1,Shepherd06_DQC1,ShorJordan08_jones_polynomial_Complete_DQC1,JordanWocjan09_DQC1_Jones_Homfly_polynomials,jordan2014approximating_Turaev_Viro_DQC1,MorimaeFujiiFitzsimons14_Hardness_Simulating_DQC1}, circuit models based on Gaussian and linear-optical operations
\cite{LloydBraunstein99_QC_over_CVs,Bartlett02Continuous-Variable-GK-Theorem,BartlettSanders02Simulations_Optical_QI_Circuits,Aaronson11_Computational_Complexity_Linear_Optics,Veitch12_Negative_QuasiProbability_Resource_QC,MariEisert12_Positive_Wigner_Functions_Quantum_Computation,VeitchWiebeFerrieEmerson13_Simulation_scheme_large_class_quantum_optics_experiments}, commuting circuits \cite{Shepherd10_PhD_thesis,ShepherdBremner09_Temporally_Unstructured_QC,BremnerJozsaShepherd08,Ni13Commuting_Circuits}, low-entangling\footnote{Here entanglement is measured with respect to the Schmidt-rank measure (low-entangling circuits with respect to continuous entanglement measures are universal for quantum computation \cite{VDNest2012_Little_Entanglement}).} circuits \cite{Jozsa03_Role_Entanglement_Quantum_Computational_SpeedUP,Vidal_03_Efficient_Simulation_Sligtly_Entangled} ,
low-depth circuits \cite{TerhalDiVincenzo02Adaptive_ConstantDepth_Quantum_Comp,MarkovShi08_Simulating_QuantumComp_TensorNetwork},
tree-like circuits \cite{MarkovShi08_Simulating_QuantumComp_TensorNetwork,aharonov_AQFT,yoran_short_QFT,browne_QFT,Yoran08_Contractable_circults_little_entanglement}, low-interference circuits \cite{nest_weak_simulations,Stahlke14_Interference_resource_speedup}  and a few others \cite{Jordan10_Permutational_Quantum_Computing,schwarz2013simulating}.

In this thesis,  classical simulation techniques play an important role in the development of the complexity theoretic hardness results and quantum algorithms we present. In this way, our results relate to other projects where classical simulations methods helped to find new quantum algorithms \cite{VdNest_Q_alg_spin_models_simulable_gate_sets,Ni13Commuting_Circuits} and/or  complexity theoretic hardness results \cite{Aaronson11_Computational_Complexity_Linear_Optics,BremnerJozsaShepherd08,MorimaeFujiiFitzsimons14_Hardness_Simulating_DQC1}.

Like Clifford circuits, the models we introduce are also unlikely to be universal: in chapters \ref{chapterF}-\ref{chapterB} we know they are not unless computational complexity classes that are believed to be distinct collapse; the universality of the model in chapter \ref{chapterH} was not fully investigated.

\section{Reading guide}\label{sect:ReadingGuideThesis}

In chapter \ref{chapterC} we  illustrate the many of the ideas developed in the thesis by introducing the simplest circuit families  we investigate, namely, our models  of normalizer circuits that arise from \emph{abelian groups}---and from non-black-box ones. Therein, we  give several examples of normalizer gates, and explain their connection with Clifford unitaries.

In chapter \ref{chapterGT} we introduce classical group-theoretic and algorithmic techniques that will be essential in  chapters \ref{chapterF}-\ref{chapterB}, including a theory of matrix representations for group morphisms, normal forms for quadratic functions and algorithms for solving systems of linear equations over groups.

The remaining chapters contain the quantum contributions of the thesis. In chapters \ref{chapterF} and \ref{chapterI} we develop techniques for simulating normalizer circuits over decomposed abelian groups, and develop their associated stabilizer formalism. In chapter  \ref{chapterB} we add black-box groups to these models, show how the resulting circuits can implement quantum algorithms and derive our first complexity-theoretic hardness results. 

Finally, in chapter \ref{chapterB}, we move above from the abelian-group setting allowing normalizer circuits to act on commutative hyper-structures. In this setting we investigate and devise new quantum algorithms for the normal HSP and the abelian HSHP, as well as a hypergroup stabilizer formalism.

%% file: chapterC.tex
\chapter{Normalizer circuits over abelian groups}\label{sect_Summary_of_Concepts}\label{sect:Summary}\label{chapterC}

Quantum Fourier transforms (QFT) lie among the most important quantum operations in quantum computation, being key components of many quintessential quantum algorithms \cite{childs_vandam_10_qu_algorithms_algebraic_problems} and often linked to the  exponential speed-up of, e.g., Shor's factoring algorithm. In this chapter we introduce quantum circuit models that contain QFTs and resemble the circuits employed in Shor-like quantum algorithms. Specifically, we propose \emph{normalizer circuits over abelian groups} \cite{VDNest_12_QFTs,BermejoVega_12_GKTheorem,BermejoLinVdN13_Infinite_Normalizers} as high-dimensional generalizations of the well-known Clifford circuits \cite{Gottesman_PhD_Thesis,Gottesman99_HeisenbergRepresentation_of_Q_Computers,Gottesman98Fault_Tolerant_QC_HigherDimensions} that contain group QFTs, automorphism gates and quadratic-phase gates. 

Normalizer circuit models provide a framework that we exploit to develop the program   of this thesis: in chapters \ref{chapterF}-\ref{chapterI} we show that the normalizer circuits in this chapter cannot provide quantum speed-ups despite the presence of QFTs in various settings; in chapters \ref{chapterB}-\ref{chapterH}, we propose  extended models of normalizer circuit model that lead to quantum algorithms. The purpose of this chapter is to introduce the simplest normalizer circuit models of the thesis (chapters \ref{chapterF}-\ref{chapterI})  and convey their key quantum features before moving to more involved (albeit powerful) ones (chapters \ref{chapterB}-\ref{chapterH}). To illustrate our definitions, we give several examples of normalizer gates and also present some other concepts that appear later in the thesis: namely, the notions of Clifford and Pauli operators.

Section \ref{sect:Normalizer Gates Finite Group} of this chapter is based on \cite{BermejoVega_12_GKTheorem} (joint work with Maarten Van den Nest). Section \ref{sect:Normalizer Gates Infinite Group} is based on \cite{BermejoLinVdN13_Infinite_Normalizers} (joint work with Cedric Yen-Yu Lin and Maarten Van den Nest).  Prior to us, normalizer circuits over finite abelian groups were considered in \cite{VDNest_12_QFTs} by Van den Nest. Connections to prior work are surveyed at the end of section \ref{sect:Introduction}.

\section{Introduction}\label{sect:Introduction}

Clifford gates are a winsome family of restricted quantum operations with a wide range o applications in quantum computation and information processing  and, at the same time, a beautiful mathematical theory to describe them: i.e. the stabilizer formalism \cite{Gottesman_PhD_Thesis,Gottesman99_HeisenbergRepresentation_of_Q_Computers}. By definition, an $n$-qubit \emph{Clifford circuit} $\mathcal{C}$ is any unitary gate that leaves invariant the $n$-qubit Pauli group\footnote{Ie.\ the group generated by the arbitrary $n$-fold tensor products of the Pauli matrices $\sigma_x, \sigma_y, \sigma_z$.} under conjugation; equivalently,  $\mathcal{C}$ is any circuit  built of sequences of Hadamard gates,  CNOTs,  CZ gates and  Phase gates $S=$ diag$(1, i)$ (acting on arbitrary qubits).

In this section we introduce \emph{normalizer circuits} associated to an \emph{abelian group} $G$, as group theoretic generalizations of the Clifford circuits containing abelian-group quantum Fourier transforms (QFTs), autormorphism and quadratic phase gates; the latter  generalize the  Hadamard, CNOT, CZ and $S$ gates, respectively.  Specifically, we will focus on groups of the form $G=\DProd{D}{a} \times \Z^b \times \T^c$, where $\mathbb{Z}_D=\{ 0, 1,\ldots, D-1\}$ is the additive group of integers modulo $N$, $\Z$ is the additive group of integers and $\T^c=\T\times \cdots\times \T$ is a $b$-dimensional hypertorus. Our motivation to consider these different types of group factors is twofold:
\begin{itemize}
\item On the one hand, our motivation to consider \emph{\textbf{finite}} abelian groups of form $\DProd{D}{a}$ is that the `standard'' quantum Fourier transform $\mathcal{F}_{2^n}$ used by Shor in its factoring and discrete-log quantum algorithms can be understood as a QFT over $\Z_{2^n}$.  The associated finite-group normalizer circuits will be investigated in \textbf{chapter \ref{chapterF}}. 

\item On the other hand, our interest in  the \emph{\textbf{infinite}} integer group $\Z$ is that several number theoretical problems are naturally connected to problems over the integers, a crucial example being the factoring problem, which is reducible to a hidden subgroup problem over $\mathbb{Z}$ \cite{Brassard_Hoyer97_Exact_Quantum_Algorithm_Simons_Problem,Hoyer99Conjugated_operators,MoscaEkert98_The_HSP_and_Eigenvalue_Estimation,Damgard_QIP_note_HSP_algorithm}. The motivation to consider hypertori $\T^m$ is that they are intrinsically connected to integer groups $\Z^m$ via a quantum Fourier transform. The associated normalizer circuits over the latter infinite groups will be investigated in \textbf{chapter \ref{chapterI}}. 
\end{itemize}
The families of normalizer circuits above are not the only ones we investigate along the thesis: in \textbf{chapters \ref{chapterB}} and \textbf{\ref{chapterH}}, in order to develop some of our main quantum algorithm and complexity theoretic results,  we will introduce more general (and powerful) models of normalizer gates that are related to \emph{\textbf{black-box groups}} and \emph{\textbf{abelian hypergroups}}. Our later models, which are slightly more abstract, will be much easier to understand after going through the simpler examples in this chapter. 

Also, in order to enrich the discussion in the introduction of the thesis (chapter \ref{chapter0}), we consider in \textbf{appendix \ref{aG}} a model of normalizer circuits over \textbf{\emph{real groups $\R^m$}}  and show that they realize the well-known families of (bosonic) \emph{Gaussian unitaries}, the latter being  central in \emph{continuous}-variable quantum information processing  \cite{Braunstein98_Error_Correction_Continuous_Quantum_Variables,Lloyd98_Analog_Error_Correction,Gottesman01_Encoding_Qubit_inan_Oscillator,Bartlett02Continuous-Variable-GK-Theorem,BartlettSanders02Simulations_Optical_QI_Circuits,Barnes04StabilizerCodes_for_CV_WEC,LloydBraunstein99_QC_over_CVs,BraunsteinLoock05QI_with_CV_REVIEW,GarciaPatron12_Gaussian_quantum_information}. Since Clifford gates are, in turn, a fundamental gate-set for QIP with \emph{discrete}-variables, this side result motivates the search of potential applications of normalizer circuit models over more general commutative algebraic structures in quantum information processing. We leave this potential research avenue open to future investigations (cf.\ chapter \ref{chapterI} for extended discussion)

\subsection{Chapter outline}

We split the discussion of this section as follows. In section \ref{section:NormalizerGatesGeneral}, we introduce normalizer gates over (fully) arbitrary abelian groups in a low level of detail. In section \ref{sect:Normalizer Gates Finite Group} we introduce our first quantum circuit model based on normalizer circuits over \emph{finite} abelian groups  $\DProd{D}{a}$. In section \ref{sect:Normalizer Gates Infinite Group}, we introduce the more involved  infinite-dimensional normalizer circuit model over groups  $\Z^a\times\T^b\times \DProd{D}{c}$. 

Along the section we illustrate our definitions with several examples. Section \ref{sect_examples} contains finite-dimensional ones, and explains the connection between normalizer circuits, the qubit and qudits Clifford gates used so-widely in quantum error correction \cite{Knill96non-binaryunitary,Gottesman98Fault_Tolerant_QC_HigherDimensions}, and Shor's quantum Fourier transform \cite{Shor}. Section \ref{sect:ExamplesInfinite} contains examples of infinite-dimensional normalizer gates.

\section{Normalizer gates}\label{section:NormalizerGatesGeneral}

In short, \emph{normalizer gates} are quantum gates that act on a Hilbert space $\mathcal{H}_G$ which has an orthonormal standard basis $\{\ket{g}\}_{g\in G}$ labeled by the elements of an abelian group $G$. The latter can be finite or infinite, but it must have a well-defined integration (or summation) rule (namely, a Haar measure) and a well-defined classical Fourier transform. Given these conditions, we define a \emph{normalizer gate over $G$} to be any gate of the following three types:
\begin{enumerate}
\item[(i)] \textbf{Quantum Fourier transforms}. These gates implement the (classical) Fourier transform of the group $\psi(x)\rightarrow\hat{\psi}(p)$ as a quantum operation $\int \psi(x)\ket{x}\rightarrow \int \hat{\psi}(p)\ket{p}$. Here, $\psi$ is a  complex function acting on the group and $\hat{\psi}$ is its Fourier transform.
\item[(ii)] \textbf{Group automorphism gates}. These implement  group automorphisms $\alpha:G \rightarrow G$, at the quantum level $\ket{g}\rightarrow\ket{\alpha(g)}$, $g\in G$. When $G$ is infinite, we require $\alpha$ to be continuous.
\item[(iii)] \textbf{Quadratic phase gates} are diagonal gates  that multiply standard basis states with \emph{quadratic} phases $\ket{g}\rightarrow \xi(g)\ket{g}$, where $|\xi(g)|=1$. ``Quadratic'' means that $g\rightarrow \xi(g)$ is an ``almost multiplicative''  function with the property $\xi(g+h)=\xi(g)\xi(h)B(g,h)$, and $B(g,h)$ is a  bi-character of $G$: i.e., a bi-multiplicative correcting term fulfilling
\begin{equation*}
B(x+y,g)=B(x,g)B(y,g), \qquad B(g,x+y)=B(g,x)B(g,y), \qquad\textnormal{for all } x,y,g\in G. 
\end{equation*}
Again, when $G$ is infinite, we require $\xi$, $B$ to be continuous in all arguments.
\end{enumerate}

\paragraph{Classical Fourier transforms.} In  the definition of QFT above (i), the \emph{classical} Fourier transform over an abelian group $G$ is defined canonically through the notion of \emph{character functions} of $G$: a complex function $\chi_p$ on $G$ is said to be a \emph{character} if $\chi_p(x+y)=\chi_p(x)\chi_p(y)$ and $|\chi_p(x)|=1$ holds for every $x,y\in G$; the set of all such functions is denoted $\widehat{G}$. Then, for all abelian groups $G$ with reasonable topologies\footnote{This holds for locally compact Hausdorff ones (i.e.\ the vast majority of groups used in quantum mechanics).}, the Fourier transform $$\psi(x) \qquad \xrightarrow{\textrm{QFT over $G$}} \qquad \hat{\psi}(p):=\sum_{\chi_p \in \widehat{G}} \, \chi_p(g)\psi(g)$$
defines a unitary transformation (up to normalization), which we will regard as a valid quantum circuit element. 

The properties of character functions will be reviewed in chapter \ref{chapterGT}. In the next sections,  we give examples of QFTs for various groups.

\paragraph{The groups.}Although normalizer circuits as above can be associated to almost\footnote{Our  circuit model is well-defined for any locally compact abelian group (cf.\ discussion in chapter \ref{chapterI}, appendix \ref{aG}), though sometimes a renormalization factor is required for the map $\ket{g}\rightarrow\ket{\alpha(g)}$ to be unitary. Within this thesis, this re-scaling only plays a role in appendix \ref{aG} (cf.\ discussion).} any abelian group, in this thesis we focus\footnote{In appendix \ref{aG} we briefly study normalizer circuits over $\R^m$ and show that they coincide with the well-known family of (bosonic) continuous-variable  Gaussian unitaries, widely used in the CV-QIP literature \cite{Lloyd98_Analog_Error_Correction,LloydBraunstein99_QC_over_CVs,BraunsteinLoock05QI_with_CV_REVIEW,GarciaPatron12_Gaussian_quantum_information}.} on abelian groups of the form 
\begin{equation}\label{group_hilbert_space} 
G=\DProd{D}{a} \times \Z^b \times \T^c
\end{equation}
where $\mathbb{Z}_D=\{ 0, 1,\ldots, D-1\}$ is the additive group of integers modulo $N$, $\Z$ is the additive group of integers and $\T^c=\T\times \cdots\times \T$ is a $b$-dimensional hypertorus. These particular groups are chosen for their connection with hidden subgroup problems (chapter \ref{chapterF}-\ref{chapterB}). Throughout the paper, the elements of $\T$ (the circle group\footnote{In our notation, the circle group  $\T$ is a one-dimensional torus and $\T^2$ is the usual two-dimensional one.}) are represented as real numbers in $[0,1)$ modulo 1 (these are angles measured in units of $2\uppi$).

It is important to note that the  finite abelian groups $\DProd{D}{a}$ in (\ref{group_hilbert_space}) are fully arbitrary  because of a well-known group-theoretic result.
\begin{theorem}[\textbf{Fundamental Theorem of Finite Abelian Groups} \cite{Humphrey96_Course_GroupTheory}]\label{thm:Fundamental Theorem FAGroups}
Any finite abelian group $G$ has a decomposition into a direct product of cyclic groups, i.e. 
\begin{equation}\label{GcC}
G=\DProd{D}{k}
\end{equation}
for some positive integers $D_1,\cdots,D_k$. Here, the elements of (\ref{GcC})  are $m$-tuples of the form $g=(g(1),\allowbreak \ldots,\allowbreak g(k))$ with $g(i)\in\Z_{D_i}$ and addition of two group elements is component-wise modulo $D_i$.  The order (or cardinality) of $G$ is denoted by $|G|$, and fulfills $|G|=D_1 D_2 \cdots D_k$.
\end{theorem}

\paragraph{\textbf{On complexity.}}
Although theorem \ref{thm:Fundamental Theorem FAGroups} states that any finite abelian group can be expressed as a product of the type (\ref{GcC}) via isomorphism, computing this decomposition is regarded as a difficult computational problem (at least as hard as factoring integers\footnote{Decomposing $G=\Z_N^{\times}$ yields an efficient algorithm to compute the Euler Totient function and this knowledge can be used to factorize in polynomial time  \cite[chapter 10]{Shoup08_A_Computational_Introducttion_to_Number_Theory_and_Algebra}. In turn, efficient \textit{quantum} algorithms to decompose abelian groups exist, at least for ``reasonably presented'' (black-box) groups (cf.\ chapter \ref{chapterB}).}). In this section and in chapters \ref{chapterF}-\ref{chapterI} a product decomposition (\ref{GcC}) of $G$ will always be explicitly given; however, this assumption will be removed in chapter  \ref{chapterB}.

\paragraph{The Hilbert space of a group.}

The Hilbert space $\mathcal{H}_G$ associated to any group of the form (\ref{group_hilbert_space}), inherits a natural tensor-product structure from the  factors of $G$
\begin{equation}
\label{eq:HilbertSpaceGeneral} \mathcal{H}_{G}=\mathcal{H}_{\Z_{D_1}}\otimes \dots\otimes \mathcal{H}_{\Z_{D_a}}\otimes \mathcal{H}_{\Z}^{\otimes b}\otimes\mathcal{H}_{\T}^{\otimes c}
\end{equation}
A normalizer circuit over $G$ performs a  quantum computation on the $m:=a+b+c$ computational registers of $\mathcal{H}_{G}$. The former $a$ registers form a finite-dimensional subspace of $D_i$-level systems $\mathcal{H}_{\Z_{D_1}}\otimes\cdots\otimes \mathcal{H}_{\Z_{D_m}}$, where  $\mathcal{H}_{\Z_{D_i}}\cong\C^{D_i}$. The latter, form a subspace of $(b+c)$  infinite-dimensional \emph{quantum rotors}\footnote{The rotors we consider are sometimes called \emph{quantum fixed-axis rigid rotors} \cite{Atiyah61_Characters_Cohomology}}, which may be regarded as quantum particles that can move in a circular orbit around a fixed axis, having angular position and integral momentum bases labeled by $\T$ and $\Z$: the position is given by a continuous angular coordinate and the angular momentum is quantized in $\pm 1$ units (the sign indicates the direction in which the particle rotates \cite{aruldhasquantum}). 
 A normalizer computation over $G$ will act on specific \emph{designated basis} of $\mathcal{H}_G$: the first of these bases is the  standard group-element basis $\mathcal{B}_G$ of product states labeled by elements of $G$
\begin{equation}\label{eq:Computational Basis EQUATION}
\ket{g}=\ket{g(1)}\otimes\cdots\otimes \ket{g(m)} \quad \textnormal{for all} \quad g\in G.
\end{equation}
The remaining bases can be obtained from (\ref{eq:Computational Basis EQUATION}) by performing single-register quantum Fourier transforms (QFT), which we introduce below.

\section{Normalizer circuits over finite $G$}\label{sect:Normalizer Gates Finite Group}

We introduce now our models of normalizer circuit models over finite abelian groups letting $G=\DProd{D}{m}$. These models will be investigated later in chapter \ref{chapterF}. The latter act on the Hilbert spaces of form $\mathcal{H}_G=\mathcal{H}_{\Z_{D_1}}\otimes \dots\otimes \mathcal{H}_{\Z_{D_m}}$ that are always \emph{finite}-dimensional. Restricting to this case  allows us to introduce our circuit models without technical complications that are only relevant in an infinite dimensional setting, such as in chapters \ref{chapterI}-\ref{chapterB}.

In short, a \emph{normalizer circuit over $G$} is a circuit composed of normalizer gates (i-ii-iii) acting on group-element states $\ket{g}$. The \textit{size} of a normalizer circuit is the number of normalizer gates it contains. To complete the definition of this model, we define QFTs over $G$ and give examples of automorphism gates and quadratic-phase gates.

\myparagraph{Input states:} The  allowed initial states of a normalizer circuit are group element states (\ref{eq:Computational Basis EQUATION}). At later steps the quantum state of the computation is of the form $\sum_{g\in G}\psi(g)\ket{g}$.  

\myparagraph{QFT over finite $G$:}The QFT over $\Z_D$ implements a unitary change of basis on $\mathcal{H}_{\Z_{D}}$: 
\begin{equation} \notag
\mathcal{F}_{\Z_N} := \sum_{x,y\in\Z_N} \langle{\tilde{y}}\,\ket{x}\ket{x}\bra{y}, \quad\textnormal{with} \quad 
|\tilde y\rangle:=\frac{1}{\sqrt{N}}\sum_{x\in \Z_N}\overline{e^{2\pi i \frac{xy}{N}}} |x\rangle\quad \mbox{ for every } y\in\Z_N.
\end{equation}
\emph{The global QFT} over the entire group $G$ acting on the entire space $\mathcal{H}_G$ is given by
\begin{equation}\label{QFT_cyclic}
\Fourier{G} = \Fourier{\mathbb{ Z}_{D_1}}\otimes \cdots \otimes \Fourier{\mathbb{ Z}_{D_m}}= \frac{1}{\sqrt{|G|}}\sum_{g,h\in G}\chi_g(h)\ket{g}\bra{h},
\end{equation}
where $\chi_g$ are the \emph{character functions} of the group $G$, which  fulfill $\chi_g(x+y)=\chi_g(x)\chi_g(y)=\chi_{g}(x+y)$ for any $x,y,g\in G$ and are defined as follows:
\begin{equation}\label{Character Functions DEFINITION}
 \chi_g(h)=\exp{\left(2\pii \sum_{i=1}^{m} \frac{g(i)h(i)}{D_i}\right)}.
\end{equation}
\emph{A partial QFT} is any operator obtained by replacing a subset of the gates $\Fourier{\mathbb{ Z}_{D_i}}$ in this tensor product by identity operators. The unitarity of all QFTs above follows from well-known character orthogonality relationships\footnote{For any $G$, these relationships say that $ \langle \chi_g, \chi_h\rangle = \frac{1}{|G|}\sum_{x\in G} \overline{\chi_g (x)}  \chi_h(x).$ }.

\myparagraph{Measurements:} 
Throughout the thesis,  measurements in the standard basis (\ref{group_labels_basis}) at the end of normalizer circuit are always allowed, although in  chapter \ref{chapterF} we  will also allow measurements of any \emph{generalized Pauli operator} $\sigma(a,g,h)$  over $G$, which we define to be of the form
\begin{equation}\label{Pauli Operators DEFINITION}
\sigma(a,g,h) := \gamma^a Z(g)X(h),\qquad 
X(g) := \sum_{h\in G} |h+g\rangle\langle h|, \qquad Z({g}) := \sum_{h\in G} \chi_{  g}({  h}) |{  h}\rangle\langle h|.
\end{equation}
All operators  $X(g)$, $Z(g)$ are {unitary} (the former just permute standard basis and the latter multiply by a complex phase). Generalized Pauli operators form a group $\mathcal{P}_G$ (cf.\ \cite{VDNest_12_QFTs} or section \ref{section Pauli and Clifford and Normalizer}), henceforth called the \emph{Pauli group over $G$}.

\myparagraph{Relationship to Clifford operations:}
A unitary operator $U$ on $\mathcal{H}_G$ is called a \textit{Clifford operator} over $G$   if it maps the Pauli group ${\cal P}_G$ onto itself under conjugation $\sigma\to U\sigma U^{\dagger}$. The set of all Clifford operators forms a group, henceforth called the \emph{Clifford group} ${\cal C}_G$. Formally, ${\cal C}_G$ is the (group theoretic) normalizer of the Pauli group in the full unitary group acting on $\mathcal{H}_G$.

It was proven in  (\cite{VDNest_12_QFTs} (see theorem \ref{thm:NormalizerCliffordC1}, chapter \ref{chapterF}) that every finite-$G$ normalizer circuit is a Clifford operator, but it is currently not known whether all possible Clifford operators can be implemented via normalizer gates. Such a question is of considerable relevance, since the finding of a non-normalizer Clifford operation could lead to a new quantum gate. However, in section \ref{section conjecture} we give  supporting evidence (\textbf{lemma \ref{lemma:PermutationNormalizer=Clifford}}) against the existence of such gates and further   \textit{conjecture} that any Clifford operator can be implemented as a poly-size normalizer circuit (\textbf{conjecture \ref{thm_conjecture}}).  Further evidence is given below, where we explain how  these notions are equivalent for regular Clifford circuits on qubits and qudits.

\subsection{Examples with finite $G$}\label{sect_examples}

 Here we give examples of Pauli and normalizer operations for several choices of finite abelian group $G$. We illustrate in particular how the definitions of the preceding section generalize existing notions of Pauli and Clifford operators for qubits and qudits.

\subsubsection{Qubit Clifford circuits: \texorpdfstring{$G=\Z_{2}^{m}$}{G = (Z2) to the m}}
 
Recall that qubit \emph{Clifford circuits} \cite{Gottesman_PhD_Thesis,Gottesman99_HeisenbergRepresentation_of_Q_Computers} are quantum circuits that normalize the qubit Pauli group and can be generated by sequences of  CNOTs,  CZ gates,  Hadamard gates and  Phase gates $S=$ diag$(1, i)$ (acting on arbitrary qubits). Below, we show that for $G=\Z_2^m$, qubit Clifford circuits become  examples of normalizer gates for $G=\Z_{2}^{m}$; this was first observed in \cite{VDNest_12_QFTs}. Note that, in this case, $\mathcal{H}_G = \mathbb{C}^2\otimes  \cdots \otimes\mathbb{C}^2$ is  a system of $m$ qubits and its group-element basis $\{\ket{x},x\in\Z_2^m\}$ is the standard basis labeled by $m$ bit-strings. 
\begin{enumerate}
\item \textbf{Hadamards}: Applying  (\ref{QFT_cyclic}) one finds that  the QFT over $\mathbb{Z}_2$ is simply the Hadamard gate $H$; the QFT over $\mathbb{Z}_2^m$ is   $H^{\otimes m}$;   and partial QFTs are obtained via combinations of single-qubit Hadamard action on qubit subsets.
\item \textbf{CNOT}, as a classical operation, implements the boolean map $(x_1,x_2)\to A(x_1,x_2)=(x_1,x_1+x_2\pmod 2)$ where $A$ denotes an invertible $2\times 2$ matrix over $\mathbb{  Z}_2^2$, hence, a $\mathbb{Z}_2^2$  automorphism. It follows that $\mathrm{CNOT}\ket{x_1,x_2}=\ket{A(x_1,x_2)}$ is a $\mathbb{Z}_2^2$ automorphism gate.
\item \textbf{S and CZ}. Let $A$ be an $m\times m$ matrix with entries in $\mathbb{Z}_2$ and let $a\in \mathbb{Z}_2^m$. Then,  the following functions are quadratic\footnote{Note that the exponent in $\xi_A$ is polynomial of degree 2 in $x$, whereas the exponent in $\xi_a$ has degree 1. Hence, the notion of quadratic functions we use differs from the usual notion of ``quadratic form'' used in, e.g., \cite{dehaene_demoor_coefficients,dehaene_demoor_hostens}} over $\Z^m$:
\be \xi_A: x\to (-1)^{x^TAx}\quad \mbox{ and }\quad \xi_a: x\to \imun^{(a^\transpose x)\bmod{2}},\qquad  x\in\mathbb{Z}_2^m.\ee 
The fact that these functions are quadratic follows from the following equations\footnote{Identity (\ref{i_quadratic}) can be proved by distinguishing between the 4 cases $a^Tx, a^Ty\in\{0, 1\}$.}:
\begin{eqnarray} \xi_A(x+y) &=& \xi_A(x)\xi_A(y) (-1)^{x^T(A+A^T)y} \\\label{i_quadratic} \xi_a(x+y)&=& \xi(x) \xi(y) (-1)^{q(x, y)} \quad \mbox{ with } q(x, y) =  (a^Tx)(a^Ty)\end{eqnarray}
As a particular case, we obtain that for $m=1$ and $m=2$ the functions $x\to i^x$  and $(x, y)\to (-1)^{xy}$ are quadratic. Finally, note  the Clifford gates $D$, $CZ$ are simply the quadratic functions associated to these gates, since:$$ D= \mbox{ diag}(1, i);\quad  CZ= \mbox{ diag}(1, 1, 1, -1).$$
\end{enumerate}
Moreover, all normalizer circuits over $\Z_2^m$ are also qubit Clifford circuits and, hence, these circuits families coincide. This follows from the fact that normalizer circuits leave the generalized Pauli group $\mathcal{P}_G$ invariant under conjugation and, moreover,  for $G=\Z^m$, $\mathcal{P}_G$  becomes the standard qubit Pauli group: to see this, let $\sigma_x$ and $\sigma_z$ denote the standard Pauli matrices and let $g\in \mathbb{Z}_2^m$ be an $m$-bit string; then, applying definition (\ref{Pauli Operators DEFINITION}) one finds that
 \be X(g) = \sigma_x^{g(1)}\otimes \cdots \otimes \sigma_x^{g(m)},\qquad  Z(g) = \sigma_z^{g(1)}\otimes \cdots \otimes \sigma_z^{g(m)}, \qquad \textnormal{for }g\in \mathbb{Z}_2^m\ee
 In short, $X(g)$ is a tensor product of $\sigma_x$-matrices and identities, and $Z(g)$ is a tensor product of $\sigma_z$-matrices and identities. Therefore, every Pauli operator (\ref{Pauli Operators DEFINITION}) has the form $\sigma \propto U_1\otimes \cdots  \otimes U_m$ where each $U_i$ is a single-qubit operator of the form $\sigma_x^u \sigma_z^v$ for some $u, v\in \mathbb{Z}_2$. This recovers the usual notion of a Pauli operator on $m$ qubits \cite{Gottesman_PhD_Thesis,Gottesman99_HeisenbergRepresentation_of_Q_Computers} .

\subsubsection{Qudit Clifford circuits: \texorpdfstring{$G=\Z_{d}^{m}$}{G = (Zd) to the m}}\label{Examples Paulis/Normalizer for qudits}

In this case the Hilbert space $\mathcal{H}_G = \mathbb{C}^d\otimes \cdots \otimes\mathbb{C}^d$ is a system of $m$ $d$-level systems (qudits) and Pauli operators have the form $\sigma \propto U_1\otimes \cdots  \otimes U_m$, where each $U_i$ is a single-qudit operator of the form $X_d^u Z_d^v$ for some $u, v\in \mathbb{Z}_d$, where  $X_d$ and $Z_d$ are the usual generalizations of $\sigma_x$ and $\sigma_z$ for $d$-level systems: \be\label{X_Z_qudits} X_d = \sum_{x\in\Z_d} |x+1\rangle\langle x| \quad \mbox{ and }\quad Z_d = \sum_{x\in\Z_d} \euler^{2\pii x/d}|x\rangle\langle x| \ee 
Examples of normalizer gates over $\mathbb{Z}_d^m$  are the standard Clifford operations for qubits,
\begin{align}\label{gates_qudits}
\mbox{SUM}_d = \sum_{x\in\Z_d} |x, x+ y\rangle\langle x, y|, \quad \mbox{CZ}_d = \sum \omega_d^{xy} |x, y\rangle\langle x, y|, \quad \omega_d:= \euler^{2\pii /d} \\ \Fourier{\mathbb{ Z}_{d}} =\frac{1}{\sqrt{d}} \sum \euler^{2\pii xy/d} |x\rangle\langle y|, \qquad  S_d = \sum  \xi_d^{x(x + d)}|x\rangle\langle x|, \quad \xi_d:= \euler^{\pii/d}.\end{align}
To show that SUM$_d$ is a normalizer gate, note that $(x, y)\to(x, x+y)$ is indeed an automorphism of $\mathbb{Z}_d\times \mathbb{Z}_d$. The gates $\mbox{CZ}_d$ and $S_d$ are quadratic phase gates \cite[section 11]{VDNest_12_QFTs}. In addition, the ``multiplication gate'' $M_{d, a} = \sum |ax\rangle \langle x|$ is also a normalizer gate, for every $a\in \mathbb{Z}_d$ which is coprime to $d$. Indeed, for such $a$ the map $x\to ax$ is known to be an automorphism of $\mathbb{Z}_d$. Furthermore, it is known that the \textit{entire} Clifford group for qudits (for arbitrary $d$) is generated by the gates SUM$_d$, $\Fourier{\mathbb{ Z}_{d}}$, $S_d$ and $M_a$ \cite{dehaene_demoor_hostens}; hence, for $G=\Z_2^m$ normalizer circuits become the qudit Clifford circuits.

Lastly, the diagonal gates associated to the  functions below are quadratic phase gates   \cite{VDNest_12_QFTs}: \be z\to \omega^{{  b}{  z}^2 + cz} \quad \mbox{ and } z\to \gamma^{  b  z(z + d)}; \quad \omega := \euler^{2\uppi \imun/d}, \ \gamma := \omega^{1/2}.\ee 

\subsubsection{Shor's discrete quantum Fourier transform: \texorpdfstring{$G = \mathbb{Z}_{2^m}$}{G = Z mod (2 to the m)}}

In our last example, we consider $G$ to be the single cyclic group  $G= \mathbb{  Z}_{2^m}$. In this case, $\mathcal{H}_G$ is a $2^m$-dimensional Hilbert space with standard basis $\{|0\rangle,\ldots, |2^m-1\rangle\}$. Note that, in contrast with previous examples  (e.g.\ $G=\mathbb{Z}_2^m$),  the structure of $\mathbb{Z}_{2^m}$ does not naturally induce a factorization of the Hilbert space into $m$ single-qubit systems. As a consequence, normalizer gates over $\mathbb{Z}_{2^m}$ act \emph{globally} on $\mathcal{H}_{G}$.

Examples  of normalizer gates are now given by $\Fourier{\mathbb{ Z}_{2^m}}$, $S_{2^m}$ and $M_{2^m,\, a}$, following the definitions of the previous example with $d=2^m$. Crucially, here the gate   $\Fourier{\mathbb{ Z}_{2^m}}$ is the ``standard'' $F_{2^m}$ QFT used in e.g Shor's algorithm and the phase estimation quantum algorithm \cite{kitaev_phase_estimation}.

\section{Normalizer circuits over infinite $G$}

\label{sect:Normalizer Gates Infinite Group}

We now extend the circuit model from previous section introducing  normalizer gates over arbitrary infinite abelian groups $G=\DProd{D}{a} \times \Z^b \times \T^c$ with associated  infinite-dimensional  Hilbert spaces $\mathcal{H}_G=\mathcal{H}_{\Z_{D_1}}\otimes \dots\otimes \mathcal{H}_{\Z_{D_m}}\otimes \mathcal{H}_{\Z}^{\otimes m}\otimes \mathcal{H}_{\T}^{\otimes m}$. We investigate these circuits in detail in chapter \ref{chapterI} and use them to understand Shor's factoring algorithm in chapter \ref{chapterB}.

\subsection{Infinite-dimensional aspects of infinite-group normalizer gates}
\label{sect:Quantum states over infinite abelian groups}

We now introduce some idiosyncratic features of infinite-dimensional Hilbert spaces that, as explained next, will affect our treatment of infinite-dimensional quantum states, quantum Fourier transforms and allowed measurement bases. These aspects will be important to  construct well-defined computational models based on infinite-group normalizer gates  (section \ref{sect:Designated basis}).

\subsubsection*{Infinite-dimensional quantum states}\label{sect_Hilbert_space_Z}

For  $\mathbb{Z}$ (resp.\ $\mathbb{T}$), a quantum state in $\mathcal{H}_{\Z}$  (resp. $\mathcal{H}_\T$) is associated to any normalized \emph{sequence} of complex numbers $\{\psi(x): x\in\mathbb{Z}\}$ with $\sum |\psi(x)|^2=1$ (resp.\ normalized  complex \emph{function} $\{\phi(p):p\in \T\}$  with $\int_{\T}\mathrm{d}p\, |\phi(p)|^2=1$):
 \begin{equation}\label{psi_Z} 
|\psi\rangle = \sum_{x\in\Z} \psi(x)|x\rangle; \qquad \qquad \ket{\phi}=   \int_{\mathbb{T}} \mbox{d}p \ \phi(p) |p\rangle,
\end{equation}
where d$p$ denotes the standard Haar/Lebesgue measure on $\T$ and  we introduced the  \emph{plane-wave states}
\begin{equation}\label{eq:Fourier basis state over Z}
|p\rangle:=\sum_{z\in \Z} \overline{e^{2\pi i zp}}|z\rangle \quad p\in \T = [0, 1).
\end{equation}
Plane-wave states, as well as those $\ket{\psi}$ states whose squared sums are not finite, are \emph{non-normalizable} unphysical states that  do not belong to $\mathcal{H}_{\Z}$. Nonetheless, it will be  convenient in our formalism to consider them.

\subsubsection*{Infinite-dimensional Quantum Fourier transforms}

\textbf{The QFT over $\Z$:} Though non-normalizable, the plane-wave states  $|p\rangle$  (\ref{eq:Fourier basis state over Z}) labeled by torus elements define a dual ``orthonormal basis''\footnote{Although we use this terminology, the $|p\rangle$ states do not form a basis in the usual sense since they lie outside of  $\mathcal{H}_{\Z}$.  Rigorously, the $|p\rangle$ kets should be understood as Dirac-delta measures, or as Schwartz-Bruhat tempered distributions \cite{Bruhat61_Schwatz-Bruhat-functions,Osborne75_Schwartz_Bruhat}. The theory of rigged Hilbert spaces \cite{delaMadrid05_roleofthe_riggedHilbert, Antoine98_QM_beyond_Hilber_space, Gadella02_unified_Dirac_formalism, gadella12_Riggings_LCA_Groups} (often used to study observables with continuous spectrum) establishes that the $|p\rangle$ kets  can be ``used as a  basis'' for all practical purposes.} of  $\mathcal{H}_\Z$, in the sense that  the map $\psi \rightarrow\hat{\psi}$:
\begin{equation}
\label{eq:QFT over Z}
\ket{\psi} \quad \xrightarrow{\text{QFT over $\Z$}} \quad  |\hat{\psi}\rangle= \int_{\mathbb{T}}\mbox{d}p\  \hat\psi(p) |p\rangle \quad\mbox{ with }\quad \hat\psi(p):=\langle p |\psi \rangle=  \sum_{x\in\mathbb{Z}} {\euler^{2\pi i px}} \psi(x),
\end{equation}
is a well-defined \emph{unitary} transformation and $\bra{p}p'\rangle = \delta(p-p')$ is a normalized Dirac delta.  The \emph{{QFT over $\Z$}} (denoted $\mathcal{F}_\Z$) is defined as the unitary transformation that implements the change of basis  (\ref{eq:QFT over Z}). It is crucial to note that, strictly speaking, the QFT over $\Z$ is an isomorphism from the Hilbert space $\mathcal{H}_\Z$ onto $\mathcal{H}_\T$, since it changes the underlying integral basis into a continuous one labeled by angles. Therefore, it is natural to identify $\mathcal{H}_\Z=\mathcal{H}_\T$ as spaces associated to  two different canonical bases of quantum states of a  single \emph{physical} system (i.e., a \emph{quantum rotor} with angular position and integral momentum \cite{aruldhasquantum}). Henceforth, we adopt this convention and use the index group $\Z$, $\T$ to denote  in which basis we work.

We immediately observe, that, because the $\Z$ group-element and Fourier basis  have different cardinality, their associated infinite-dimensional QFTs must have two unique exotic features with no finite-dimensional counterpart.
\begin{itemize}
\item [(a)]  \textbf{{\bf QFTs are not gates:}} Since the standard basis $\{|x\rangle: x\in \Z\}$ and Fourier basis $\{|p\rangle: p\in \T\}$ have different cardinality they cannot be rotated onto each other. Instead, the QFT  is a change of basis between two orthonormal basis, but it does not define a  unitary rotation as in the standard (finite dimensional) circuit model\footnote{Mathematically, this Fourier transform is a unitary transformation between two different functional spaces, $L^2(\Z)$ and $L^2(\T)$. The latter two define one quantum mechanical system with two possible bases (of Dirac-delta measures) labeled by $\Z$ and $\T$. In the finite dimensional case, the picture is simpler because the QFT is a unitary transformation of $L^2(\Z_N)$ onto itself. (These facts are consequences of the Plancherel theorem for locally compact abelian groups \cite{rudin62_Fourier_Analysis_on_groups,HofmannMorris06The_Structure_of_Compact_Groups}.)}. This is in strong contrast, with the finite group case where the QFT could be implemented either as a change of basis or as a gate: e.g. the QFT over $\Z_N$ implemented the change of basis $\ket{y}\rightarrow\ket{\hat{y}}$.
\item[(b)] \textbf{QFT over $\T$.} A technical obstacle   to construct a well-defined infinite-dimensional normalizer circuit model is that such models cannot be based on QFTs over $\Z$ only because they cannot be \emph{concatenated} one after another: this happens because the QFT over $\Z$ changes the underlying group labeling the basis from $\Z$ to $\T$, and a QFT over $\Z$ is only a well-defined normalizer gate in the $\Z$ basis. To cope with this issue,  we need to allow  a normalizer circuit over $\Z$ to contain not only QFTs over $\Z$ but also a \emph{distinct type} of \emph{\textbf{QFT over $\T$}} that re-expresses a state $|\phi\rangle= \int_{\mathbb{T}}\mbox{d}p\  \psi(p) |p\rangle$ back in the integer basis of $\mathcal{H}_\Z$
 \begin{equation}\label{eq:Fourier coefficients}
 \ket{\phi} \quad \xrightarrow{\text{QFT over $\T$}} \quad   |\hat{\phi}\rangle= \sum_{x\in\mathbb{Z}} \hat\phi(x) |x\rangle \quad\mbox{ with }\quad  \hat\phi(x):=  \overline{\langle x |\phi \rangle} = \int_{\mathbb{T}}\mbox{d}p\ {\euler^{2\pi i px} }\phi(p).
    \end{equation}
\end{itemize}
We stress that, in our circuit model, the QFT over $\Z$ (resp.\ over $\T$) may only be applied if we work in the group-element basis labeled by $\Z$ elements (resp.\ $\T$ elements). Also, some readers may  note, at this point, that the latter QFT over $\T$ implements the well-known classical \emph{Fourier series} of a \emph{periodic} real function as a quantum gate. Conversely, the QFT over $\Z$  is nothing but the quantum version of the (also well-known) discrete-time Fourier transform \cite{oppenheim_Signals_and_Systems}, which turns a discretized signal into a periodic function.

\subsubsection*{Designated bases} 
 
Above, we saw the action on QFTs  on the Hilbert space $\mathcal{H}_\Z=\mathcal{H}_\T$ is two perform a change between \emph{two} distinct natural bases. As a consequence of this feature, it follows that  computational models based on normalizer gates over $\Z$ and $\T$ do not have a unique preferred ``standard basis'',  as opposed to the finite-dimensional setting of section \ref{sect:Normalizer Gates Finite Group}. Instead, we will let a  normalizer circuit over an infinite group $G=\DProd{D}{a} \times \Z^b \times \T^c$ act on a \emph{time-dependent} \emph{designated basis}: the latter is a ``standard basis''  that  is subject to change along the computation.
\begin{definition}[\textbf{Designated basis}]
At every time step $t$ in a normalizer circuit there is a \emph{designated  basis} ${\cal B}_{G_t}$ of the Hilbert space $\mathcal{H}_G$, which is the group-element basis of a group $G_t$ picked from a family of size $2^{b+c}$ constructed below (\ref{group_labels_basis}). The pair ($G_t$, ${\cal B}_{G_t}$) determines the allowed normalizer gates at time $t$ as well as the basis in which measurements are performed.

Specifically,  each \emph{designated basis} $\mathcal{B}_{G'}$ is the group-element basis of a group  $G'$  of form
\begin{align}\label{group_labels_basis} G'&:=\mathbb{Z}_{D1}\otimes \dots\mathbb{Z}_{Da} \times G_1'\times\cdots \times G_{b+c}', \quad \textnormal{where each }\quad G_i \in\{\mathbb{Z}, \mathbb{T}\}.\\
\label{basis} {\cal B}_{G'}&:=\{ |g\rangle=|g(1)\rangle\otimes \dots\otimes|g(m)\rangle; \quad g=(g(1), \dots, g(m))\in G'\}.
 \end{align}
\end{definition}

The notation $G_i = \T$ indicates that  $\ket{g(i)}$ is a Fourier state of $\Z$ (\ref{eq:Fourier basis state over Z}). The states $\ket{g}$ are product-states with respect to the tensor-product decomposition of $\mathcal{H}_G$. There are $2^b$ possible choices of groups in (\ref{group_labels_basis}) (which are, in fact,  related via Pontryagin duality\footnote{From a mathematical point of view, all groups (\ref{group_labels_basis}) form a family (in fact, a category) which is generated by replacing the factors $G_i$ of the  group $\DProd{D}{a}\times \Z^{b}$ with their character groups  $G_i^*$ (cf.\ chapter \ref{chapterGT}), and identifying isomorphic groups. Pontryagin duality \cite{Morris77_Pontryagin_Duality_and_LCA_groups,Stroppel06_Locally_Compact_Groups,Dikranjan11_IntroTopologicalGroups,rudin62_Fourier_Analysis_on_groups,HofmannMorris06The_Structure_of_Compact_Groups,Armacost81_Structure_LCA_Groups,Baez08LCA_groups_Blog_Post} then tells us that there are $2^b$ different groups and bases. Note that this  multiplicity is a purely \emph{infinite-dimensional feature}, since  all finite groups are isomorphic to their character groups; consequently, this feature does not play a role in the study of finite-dimensional normalizer circuits.}) and $2^b$ inequivalent group-element basis of the Hilbert space.

\myparagraph{Example 1:} The designated basis $\mathcal{B}_G$ is the group-element basis labeled by $G$ elements:
\begin{equation}
\label{standard_basis_all} |x(1)\rangle\otimes \cdots\otimes |x(a)\rangle\otimes |y(1)\rangle\otimes\dots\otimes |y(b)\rangle \otimes |z(1)\rangle\otimes\dots\otimes |z(c)\rangle, \quad x(i)\in\mathbb{Z}_{Di}; \ (y,z)\in\Z^b\times\T^c.
\end{equation}
\myparagraph{Example 2:} In turn,  choosing the Fourier basis in the $(a+b)$-th space we obtain in turn
\begin{equation}
|x\rangle\otimes \left(|y'(1)\rangle\otimes \dots\otimes|y'(b-1)\rangle\otimes |p\rangle \right)\otimes \ket{z'}\quad, (y',p,z')\in\left( \Z^{b-1}\times \T\right)\times\T^{c},
\end{equation}
which is labeled by the elements of $\DProd{D}{a}\times \Z^{b-1}\times \T^{c+1}$.

\myparagraph{Update rules:} The action of a QFT over $\Z$ or $\T$ in a normalizer circuit will be \emph{precisely} to change the designated basis of the computation, as explained next:
\begin{enumerate}
\item Precisely, when we say that the \emph{QFT over $\Z$} (\ref{eq:Fourier basis state over Z}) is applied to $|\psi\rangle$, we mean that the designated basis is changed from the $\Z$ group-element to its  Fourier $\T$-element basis: here, the state does not actually change (no gate is physically applied), but the normalizer gates acting after the QFT  will be associated with $\T$ (\emph{not} $\Z$), and measurements will be performed in the $\T$ basis (cf.\ next section \ref{sect:Designated basis}). Correspondingly,  the wavefunction of the state $\ket{\psi}$ ought to be re-expressed  in the Fourier basis (\ref{eq:QFT over Z}).
\item Respectively, when we say that the \emph{QFT over $\T$} is applied to $|\psi\rangle$, we mean that the designated basis is changed from the $\T$-element basis to the $\Z$ group-element basis. Like in the previous case, we must re-express the state $|\psi\rangle$ in the new designated basis (\ref{eq:Fourier coefficients}).
\end{enumerate}

\subsection{The full infinite-dimensional normalizer circuit model}\label{sect:Designated basis}\label{sect_normalizer_circuits_sub}

We now  present our infinite-group normalizer circuit model in precise terms. Below, we fix   $$G=\DProd{D}{a}\times \Z^{\otimes b}\times \T^c, \qquad\qquad \mathcal{H}_{G}=\mathcal{H}_{\Z_{D_1}}\otimes \dots\otimes \mathcal{H}_{\Z_{D_a}}\otimes \mathcal{H}_{\Z}^{\otimes b}\otimes\mathcal{H}_{\T}^{\otimes c},$$ and let $m:=a+b+c$ be the number of total registers of the computation. In this decomposition, the parameters $a$, $b$, $c$, $D_i$  can be chosen arbitrarily.

Roughly speaking, a \emph{\textbf{normalizer circuit over $G$}} of size $T$ is a quantum circuit ${\cal C}=U_T\cdots U_1$ composed of $T$ \emph{normalizer gates} $U_i$ as in section \ref{sect:Normalizer Gates Finite Group}. However, in contrast with  finite-group setting, now not all gates  $U_i$ need to be normalizer gates  over the group $G$, but over any group $G'$ (\ref{group_labels_basis}) that labels one of the allowed designated basis of $G$. Specifically,  a {normalizer circuit over $G$} is  any quantum circuit generated by the following rules:
\begin{itemize}
\item \textsf{\textbf{Input states:}} The input states of a normalizer computation are  elements of some designated group basis $\mathcal{B}_{G_0}$ at time zero. For instance,  if we choose  $G_0=\DProd{D}{a}\times \Z^{\otimes b}\times \T^c$, in our notation, then the registers  $\mathcal{H}_\Z^{\otimes b}$ and $\mathcal{H}_\T^{\otimes c}$ are fed, respectively, with standard-basis  $\ket{n}$, $n\in \Z$ and Fourier-basis states  $\ket{p}$, $p\in \T$.

\item \textsf{\textbf{Structure of the circuit:}}
\begin{itemize}
\item[$\circ$] \emph{At time $t=1$}, the gate $U_1$ is applied, which is either an automorphism gate, quadratic phase gate over $G_0$ (see section \ref{sect:ExamplesInfinite}) or a QFT. As earlier, we allow  the application of \emph{partial QFTs} on any subset of the individual registers $\mathcal{H}_{\Z_{N_i}}$, $\mathcal{H}_{\Z}$, $\mathcal{H}_{\T}$ and the full \emph{QFT over $G$}  is the combination of all partial QFTs acting on the smaller  registers.

\item[$\circ$] \emph{At time $t=1$}, after the action of $U_1$, the designated basis is changed from ${\cal B}_{G_0}$ to ${\cal B}_{G_1}$, for some group $G_1$ in the family (\ref{group_labels_basis}), which may only differ from $G_0$ if a QFT was applied. Specifically: whenever if $G_0(i)=\Z$ (respectively, if  $G_0(i)=\T$) and a QFT acts  on the $i$th register, then the group $G_1$ is chosen so that $G_1(i)=\T$ (resp. $G_1(i)=\Z$); in all other case $G_1(i)=G_0(i)$ is left unchanged.

\item[$\circ$] \emph{At time $t=2$}, the gate $U_1$ is applied, which is, again, either an automorphism gate, quadratic phase gate or a QFT over $G_1$. The designated basis is changed from ${\cal B}_{G_1}$ to ${\cal B}_{G_2}$, for some group $G_2$, following the rules of the previous step.

\item[$\circ$] The gates $U_3, \dots, U_t$ are considered similarly. We denote by ${\cal B}_{G_t}$ the designated basis after application of $U_t$ (for some group $G_t$ in the family (\ref{group_labels_basis})), for all $t=3, \dots, T$. Thus, after all gates have been applied, the designated basis is $G_T$.

\item[$\circ$] After the circuit, a measurement in the designated basis $G_T$ is performed.
\end{itemize}
\end{itemize}

\subsection{Examples of infinite-dimensional normalizer gates} \label{sect:ExamplesInfinite}

Finally, we illustrate  the above definitions giving examples of  normalizer gates over  $\Z^m$ and $\T^m$.

\subsubsection*{The infinite case $G=\Z^m$}

First, the formulas below show how the QFT over $\Z$ acts on  quantum states:\footnote{The transformations we depict can be found in standard signal processing textbooks  \cite{oppenheim_Signals_and_Systems}.}
\begin{center}
\begin{minipage}{0.45\linewidth}
\centering
{\it State before QFT over $\Z$}
\begin{equation*} \ket{x}, x\in \Z \end{equation*}
\begin{equation*} \sum_{x\in\Z} \overline{\euler^{2\pii\, px}}\ket{x} \end{equation*}
\begin{equation*} \sum_{\substack{x\in\Z\\\:}} \ket{rx} \end{equation*}
\end{minipage}
\begin{minipage}{0.45\linewidth}
\centering
{\it State after QFT over $\Z$}
\begin{equation*} \int_{\mbb{T}} dp\, {\euler^{2\pii\, px}}\ket{p} \end{equation*}
\begin{equation*} \ket{p}, p\in \T\end{equation*}
\begin{equation*} \frac{1}{r}\sum_{\substack{k\in\Z :\\ k/r\in \mbb{T}}}\ket{k/r}  \end{equation*}
\end{minipage}
\end{center}
Second, the  gates below are examples of automorphism  and quadratic phase gates, respectively:
\begin{equation}\nonumber \text{SUM}_{\Z,a} = \sum_{x,y\in\Z} \ket{x,x+ay} \bra{x,y},\qquad S_p = \sum_{x\in\Z} \exp{(\pii p x^2)} \ket{x} \bra{x} \end{equation}
where $a$ is an arbitrary integer and $p$ is an arbitrary real number. The fact that these gates are indeed normalizer gates follows from  general normal forms for matrix representations group homomorphisms (lemma \ref{lemma:Normal form of a matrix representation}) and quadratic functions (theorem \ref{thm:Normal form of a quadratic function}) that we introduced in chapter \ref{chapterGT}.

\subsubsection*{The infinite case $G=\T^m$}

We now take a look at the effect of the quantum Fourier transform over $\T$ over some states.\footnote{These examples also illustrate that the QFT over $\T$ is the quantum version of the Fourier series \cite{oppenheim_Signals_and_Systems}.}
\begin{center}
\begin{minipage}{0.45\linewidth}
\centering
\vspace{+5pt}{ \em State before QFT over $\T$}
\begin{equation*} \ket{p},p\in \T \end{equation*}
\begin{equation*} \int_{\mbb{T}} dp\, \euler^{2\pii\, px}\ket{p} \end{equation*}
\begin{equation*} \frac{1}{r}\sum_{\substack{k\in\Z :\\ k/r\in \mbb{T}}}\ket{k/r} \end{equation*}
\end{minipage}
\begin{minipage}{0.45\linewidth}
\centering
{\it State after  QFT over $\T$}
\begin{equation*} \sum_{x\in\Z} \euler^{2\pii\, px}\ket{x},x\in\Z \end{equation*}
\begin{equation*} \ket{-x} \end{equation*}
\begin{equation*} \sum_{\substack{x\in\Z\\\:}} \ket{rx} = \sum_{\substack{x\in\Z\\\:}} \ket{-rx} \end{equation*}
\end{minipage}
\end{center}
Comparing the effect of the QFT on $\mathcal{H}_{\Z}$ and the QFT on $\mathcal{H}_{\T}$, we see that the former is the ``inverse'' of the latter up to a change of sign of the group elements labeling the basis; concatenating the two of them yields the transformation $|x\rangle$ to $|-x\rangle$. This is a general phenomenon, which we shall observe throughout the thesis.

Examples of automorphism gates over $\T^m$ are the sum and sign-flip gates:
\begin{equation} \nonumber \text{SUM}_{\T,b} = \iint_{\T} \mathrm{d}p\mathrm{d}q\,\ket{p,q+bp} \bra{p,q},\qquad M_{\T,s} =\int_{\T} \mathrm{d}p\,\ket{sp} \bra{p} \end{equation}
where $b$ is any\ integer and $s=\pm1$ (the correctness of these formulas comes from lemma \ref{lemma:Normal form of a matrix representation}). 

Unlike the previous examples we have considered, any quadratic phase gate over $G$  is \emph{purely multiplicative} (i.e., the bi-multiplicative function $B(g,h)$ is always trivial\footnote{This fact can be understood in the light of a later result, theorem \ref{thm:Normal form of a quadratic function} and it is related to  nonexistence of nontrivial group homomorphisms from $\T^m$ to $\Z^m$, the latter being the character group of $\T^m$ up to isomorphism.}). In the case $m=1$, this is equivalent to saying that any such gate is of the form
\begin{equation}\nonumber \int_{\T}\mathrm{d}p\, \exp{(2\pi i b p)}\ket{p} \bra{p} \end{equation}
with $b$  an arbitrary integer.

%% file: chapterGT.tex
\chapter{Classical group theoretic and algorithmic techniques}\label{chapterGT}

In this chapter we develop a series of \emph{classical} \textbf{group-theoretic} and \textbf{algorithmic techniques}. These tools will provide a basic  language in this thesis to attack quantum computing problems in chapters \ref{chapterF}-\ref{chapterB}. The main contributions in this chapter are threefold: 
\begin{itemize}
\item[I.] {\textbf{A theory of matrix representations for abelian-group homomorphisms}}. We show that, similarly to real linear maps, homomorphisms  between {elementary} abelian  groups of form $\R^a \times \Z^b \times \T^c \times \DProd{N}{d}$ admit concise classical descriptions in terms of matrix representations with well-behaved algebraic properties (lemmas  \ref{lemma:properties of matrix representations}, \ref{lemma:existence of matrix representations}). We give a \emph{normal form} that fully characterizes the structure of such matrices (lemma \ref{lemma:Normal form of a matrix representation}). 
\item[II.]  {\textbf{Normal forms for quadratic and bi-character functions}} over abelian-groups (theorem \ref{thm:Normal form of a quadratic function}, lemma \ref{lemma symmetric matrix representation of the bicharacter homomorphism}), based on matrix representations of group homomorphisms.
\item[III.] \textbf{Classical algorithms for  group theoretic problems.} We give \emph{efficient} classical algorithms for solving \emph{linear systems of equations over abelian groups} $\alpha(x)=b$ where $\alpha:G\rightarrow H$ is an abelian-group homomorphism, $x\in G,b\in H$ (theorem \ref{thm:General Solution of systems of linear equations over elementary LCA groups}): our algorithms decide the existence of and find  \emph{general solution} (definition \ref{def:General Solution of a system}) for any such system if a matrix representation $A$ of $\alpha$ is provided; our technique is based on a reduction to mixed real-integer systems of equations (\ref{eq:System of Mixed Integer linear equations}) and the Smith normal form.
\end{itemize}
We highlight that in chapters \ref{chapterF}-\ref{chapterB}  we will identify a rich variety of \emph{quantum applications} for  results I-II-III. For this reason, we regard the latter as main contributions of the thesis. To illustrate the versatility of these methods, we anticipate some of these applications:
\begin{itemize}
\item \emph{Matrix representations} (result I) and \emph{our normal form for quadratic functions}  (result II) will be applied to define efficient \emph{classical encodings} for abelian-group normalizer gates (s.\ \ref{section Standard encodings of normalizer circuits}, \ref{sect:Main Result}) and  infinite-dimensional stabilizer states (s.\ \ref{sect:Tracking normalizer evolutions with stabilizer groups}); as well as to derive  our classical simulation results (theorems \ref{thm_main}, \ref{thm:Main Result}, \ref{thm:Normalizer gates are Clifford}) and  our complexity theoretic hardness results (theorem \ref{thm:Simulation}, \ref{thm:No Go Theorem}).
\item   \emph{Our normal form for quadratic functions}  can also be applied to characterize the wave-functions of abelian-group\footnote{This result is proven only for finite-dimensional stabilizer states but it can be easily extended to the infinite dimensional settings of chapter \ref{chapterI}, appendix \ref{aG}.} \emph{stabilizer states} in combination with another normal form for the latter (theorem \ref{thm Normal form of an stabilizer state}). These results  partially\footnote{The theorem in \cite{Gross06_discrete_Hudson_theorem,Gross_PhD_Thesis} also says that such states are precise those with a positive Wigner representation; this fact does not easily extend to even dimensions due to certain  nonlocal features (cf.\ discussion in chapter \ref{chapterI}).} extend Gross' discrete Hudson theorem, which describes odd-dimensional pure qudit stabilizer states via quadratic forms \cite{Gross06_discrete_Hudson_theorem,Gross_PhD_Thesis}.
\item \emph{Our  group theoretic algorithms} (result III) are used ubiquitously in chapters \ref{chapterF}-\ref{chapterB}, more importantly, to manipulate  abelian-group stabilizer states, stabilizer groups and generalized abelian-group Pauli operators. As examples of key results where these techniques play a key role, the reader might look at  theorems \ref{thm structure test}, \ref{thm Normal form of an stabilizer state}, \ref{thm_Measurement_Update_rules}, \ref{thm:Main Result}, \ref{thm:Algorithm to sample subgroups} and our extended Cheung-Mosca quantum algorithm \ref{alg:group decomposition} for the group decomposition problem.
\end{itemize}
Last, we point out that the concept of quadratic function explored by \cite{VDNest_12_QFTs} and  this thesis has a great theoretical value for understanding the algebraic structure of the stabilizer formalism. For instance,  we saw in  the examples of chapter \ref{chapterC} that this notion yields a one-line unified definition for all diagonal Clifford gates for qubits and qudits, as well as a (previously unknown) common group-theoretic operational interpretation for these gates (a generalization of this result for finite abelian groups is given by lemma \ref{lemma:PermutationNormalizer=Clifford}). Furthermore, our later result (theorem \ref{thm Normal form of an stabilizer state}) further shows that these functions yield a description of all phases of stabilizer states.  It would be interesting to  explore if quadratic functions have applications beyond this thesis in the area of fault-tolerant quantum computation, e.g., to  understand better which quantum error correcting codes have transversal cubic (non-Clifford) diagonal gates \cite{Campbell14_Enhance_FTQC_d_level_systems}. We propose these questions to the reader as motivation for further research.

\subsection{Relationship to previous work}

The author makes no claim about the novelty of the methods in this chapter  for  solving non-quantum problems: it is quite possible that some of the results I-II-III might be known, e.g.,  by group theorists and/or computer scientists working on (classical) algorithms for algebraic problems, even if we did not find  explicit proofs for them in the literature. The connections to existing classical works that we are aware of are pointed out throughout the chapter.  

In our view, the value of the techniques in this chapter comes from  their  applications to solve problems in quantum information and computation. In this sense, we regard our results I-II-III as novel and our new and the techniques employed in their proofs of interest to the general quantum audience.

To the best of our knowledge, Van den Nest \cite[section 6]{VDNest_12_QFTs} and us \cite{BermejoVega_12_GKTheorem}, were the first to point out and exploit the notions of quadratic functions and abelian-group-homomorphism matrix representations in quantum computation theory.  Quadratic functions over finite abelian groups were  introduced in  \cite{VDNest_12_QFTs,BermejoVega_12_GKTheorem}, and over infinite groups in  \cite{BermejoLinVdN13_Infinite_Normalizers}. Prior to these works, quadratic \emph{forms} (which are instances of quadratic functions) were used, e.g., in \cite{dehaene_demoor_coefficients,dehaene_demoor_hostens} to study  the qubit/qudit stabilizer formalism (see also section  \ref{sect:ExistingNormalForms}).  VdN and us are also among the first to introduce classical algorithms for solving linear systems of equations. Prior to us, some quantum  applications of  classical algorithms for solving linear systems of equations were known (though the concept had not been introduced). Implicitly, methods for solving certain instances of these systems (of the type given in lemma \ref{lemma:Algorithms_FA_Groups}.(e)) were  employed in the classical post-processing of quantum algorithms for abelian hidden subgroup problems \cite{lomont_HSP_review} and in the classical simulation algorithm of \cite[theorems 3,4]{VDNest_12_QFTs}. This technique was formalized in a group-theoretic language and generalized to the full extent of theorem \ref{thm:General Solution of systems of linear equations over elementary LCA groups}  by  us in \cite{BermejoVega_12_GKTheorem,BermejoLinVdN13_Infinite_Normalizers}, as part of this thesis.

Our account in this chapter is based on \cite{BermejoLinVdN13_Infinite_Normalizers} (joint work with Cedric Yen-Yu Lin and Maarten Van den Nest), which contains our most general algorithms for infinite-abelian-group problems.

\subsection{Chapter outline}

The proofs of the theorem in this chapter have been moved to appendices \ref{appendix:Supplement to section Homomorphisms}-\ref{app:Supplement Quadratic Functions}, in order to give more attention  to the quantum contributions of the thesis. Section \ref{sect:elementary_Abelian_groups} surveys some necessary notions of group and character theory. In section \ref{sect:Homomorphisms and matrix representations} we develop our  theory of matrix representations of group homomorphisms.  In section \ref{sect:quadratic_functions} we present normal forms for quadratic functions. In section \ref{sect_CompComp_FiniteAbelianGroups} we study computational aspects of the abelian groups in this thesis, including the computational complexity of solving systems of linear equations over groups (s.  \ref{sect:Systems of linear equations over groups}), for which we give polynomial-time deterministic classical algorithms.

\section{Introduction to abelian group theory}\label{sect:GTDefinitions}

\subsection{Definitions}\label{sect:elementary_Abelian_groups} 

\paragraph{Elementary abelian groups:} A commutative group $G$ is called \emph{elementary} if it is of the form
\be 
G=\Z^a \times \mathbb{R}^b\times  \DProd{N}{c} \times \T^d
\ee  
Below, we often let $F:= \DProd{N}{c}$ be the finite subgroup in the above decomposition. Though the main results in this thesis are not for normalizer circuits over real-number groups $\mathbb{R}^b$ (see appendix \ref{aG}), we consider these groups in this chapter to develop of our classical methods.

An elementary abelian group of the form $\Z$, $\R$, $\T$ or $\Z_N$ is said to be \emph{primitive}. Thus every elementary abelian group can be written as $G=G_1\times\dots \times G_m$ with each $G_i$ primitive; we will often use this notation. We will also use the notation $G_{\Z}$, $G_{\R}$, $G_{F}$, $G_{\T}$ to denote elementary abelian groups that are, respectively, integer lattices $\Z^a$, real lattices $\R^b$, finite groups $F$ and tori $\T^d$. We will also assume that the  factors $G_i$ of $G$ are arranged so that $G=G_{\Z}\times G_{\R}\times G_{F}\times G_{\T}$.

\myparagraph{Characteristic:} The group characteristic $\charac{G}$ of a primitive group is a number defined as
\begin{equation}\label{eq:Characteristic}
\charac{\Z}:=0, \quad \charac{\R}:=0,\quad \charac{\Z_N}:=N, \quad \charac{\T}:=1.
\end{equation}
Group theoretically,  $\charac{G}$ can be equivalently defined as (a) the \emph{order} of $1$ in $G$ if 1 has finite order (which is the case for $\Z_N$ and $\T$); (b) zero, if $1$ has infinite order in $G$ (which is the case for $\Z$ and $\R$).

\myparagraph{Group element encodings:} Consider an elementary abelian group $G= G_1\times\dots \times G_m$ where $c_i$ is the characteristic of $G_i$. Each element $g\in G$ can be represented as an $m$-tuple $g=(g_1, \dots, g_m)$ of real numbers. If $x=(x_1, \dots, x_m)$ is an arbitrary $m$-tuple of real numbers, we say that $x$ is congruent to $g$, denoted by  $x\equiv g$ $(\mbox{mod } G)$, if \be x_i \equiv  g_i\  (\mbox{mod } c_i) \quad \mbox{ for every } i=1, \dots, m.\ee
For example, every string of the form $x=(\lambda_1c_1, \dots, \lambda_mc_m)$ with $\lambda_i\in \Z$ is congruent to $0\in G$.

\subsection{Character functions and character duality}\label{sect:characters}

\begin{definition}[\textbf{Character} \cite{Morris77_Pontryagin_Duality_and_LCA_groups,Moreno05_Analytic_Number_Theory_L_Functions}] \label{def:Characters} Let $G$ be an elementary abelian group. A character of $G$ is  a complex function $\chi_\mu$ on $G$  that fulfills two properties:
\begin{equation*}
\textbf{(i)} \quad \chi(g+h)=\chi(g)\chi(h),  \quad \textnormal{ for every $g, h\in G$,} \qquad  \textbf{(ii)} \quad  |\chi(g)|=1, \quad \textnormal{ for every $g\in G$.}
\end{equation*}
\textbf{Properties:} For any two characters $\chi_1,\chi_2$ the function $\chi_1\chi_2$ is a new character. Furthermore,  character functions   form a new elementary abelian group under the functional point-wise product called the \textit{character group} or \emph{dual group} $\widehat{G}$. Finally, the character group of a direct product group  $G=G_1\times \cdots \times G_m$ is the product of character groups $\widehat{G}=\widehat{G}_1\times \cdots \times \widehat{G}_m$.
\end{definition}

\subsubsection*{Examples of groups and their character groups} Let $G=G_1\times \cdots \times G_m$ be an elementary abelian group. Then $\widehat{G}$ is isomorphic to another elementary abelian group $G^*$ obtained via the following map:
\begin{equation}\label{eq:Elementary Character Group}
G=\R^a\times\T^b \times \Z^c \times \DProd{N}{d} \quad \rightarrow\quad   G^*:=\left(\R^a\times \Z^b \times \T^c \times \DProd{N}{d}\right) \cong \widehat{G}.
\end{equation}
Thus, in particular, $\widehat{\mathbb{R}}$ is isomorphic to $\mathbb{R}$ itself and similarly $\widehat{\Z_{N_i}}$ is isomorphic to ${\Z_{N_i}}$ itself; these groups are called autodual. On the other hand, $\widehat{\mathbb{Z}}$ is isomorphic to $\mathbb{T}$ and, conversely, $\widehat{\mathbb{T}}$ is isomorphic to $\mathbb{Z}$. We also note from the rule (\ref{eq:Elementary Character Group}) that the dual group of $\widehat{G}$ is isomorphic to $G$ itself. This is a manifestation of the Pontryagin-Van Kampen duality \cite{Morris77_Pontryagin_Duality_and_LCA_groups,Stroppel06_Locally_Compact_Groups,Dikranjan11_IntroTopologicalGroups} (cf.\ lemma~\ref{lemma:Pontryagin duality for characters}).

We now give explicit formulas for the characters of any primitive abelian group.
\begin{itemize}
\item The characters of $\mathbb{R}$ are
\begin{equation}\label{char_R}
 \chi_x(y):=\exp{\left(2\pi i xy\right)}, \quad\text{for every $x$, $y\in\R$}.
\end{equation}
Thus each character is labeled by a real number. Note that $\chi_x\chi_{x'}= \chi_{x+x'}$ for all $x, x'\in \mathbb{R}$. The map $x\to \chi_x$ is an isomorphism from $\R$ to $\widehat{\R}$, so that $\R$ is autodual.
\item The characters of $\mathbb{Z}_N$ are
\begin{equation}\label{char_F}
\chi_x(y):=\exp{\left(\frac{2\pi i}{N}\ xy\right)}, \quad\text{for every $x$, $y\in\mathbb{Z}_N$}.
\end{equation}
Thus each character is labeled by an element of $\mathbb{Z}_N$. As above, we have $\chi_x\chi_{x'}= \chi_{x+x'}$ for all $x, x'\in\mathbb{Z}_N$. The map $x\to \chi_x$ is an isomorphism from $\mathbb{Z}_N$ to $\widehat{\mathbb{Z}}_N$, so that $\mathbb{Z}_N$ is autodual.
\item  The characters of $\Z$ are
\begin{equation}\label{char_Z}
\chi_p(m):=\exp{(2\pi i pm)}, \quad\text{for every $p\in\T$, $m\in\Z$},
\end{equation}
Each character is labeled by an element of $\mathbb{T}$. Again we have $\chi_p\chi_{p'}= \chi_{p+p'}$ for all $p, p'\in\mathbb{T}$ and the map $p\to \chi_p$ is an isomorphism from $\mathbb{T}$ to $\widehat{\mathbb{Z}}$.
\item The characters of $\T$ are
\begin{equation}\label{char_T}
\chi_m(p):=\exp{(2\pi i pm)}, \quad\text{for every $p\in\T$, $m\in\Z$};
\end{equation}
Each character is labeled by an element of $\mathbb{Z}$. Again we have $\chi_m\chi_{m'}= \chi_{m+m'}$ for all $m, m'\in\mathbb{Z}$ and the map $m\to \chi_m$ is an isomorphism from $\mathbb{Z}$ to $\widehat{\mathbb{T}}$.
\end{itemize}
If $G$ is a general elementary abelian group, its characters are obtained by taking products of the characters described above. More precisely, if $A$ and $B$ are two elementary abelian groups, the character group of $A\times B$ consists of all products $\chi_A\chi_B$ with $\chi_A\in \widehat{A}$ and $\chi_B\in \widehat{B}$, and where $\chi_A\chi_B(a, b):= \chi_A(a)\chi_B(b)$ for every $(a, b)\in A\times B$. To obtain all characters of a group $G$ having the form (\ref{eq:Elementary Character Group}), we denote
\be G^*:=\R^a\times \Z^b \times \T^c \times F.\ee Considering an arbitrary element \be\label{mu} \mu= (r_1, \dots, r_a ,z_1, \dots, z_b, t_1, \dots, t_c, f_1, \dots, f_d)\in G^*,\ee the associated character is given by the product \be \label{mu_character} \chi_{\mu}:=\chi_{r_1}\dots\chi_{r_a}\ \chi_{z_1}\dots\chi_{z_b}\ \chi_{t_1}\dots\chi_{t_c}\ \chi_{f_1}\dots\chi_{f_d}\ee where the individual characters $\chi_{r_i}, \chi_{z_j}, \chi_{t_k}, \chi_{f_l}\dots$ of $\mathbb{R}, \mathbb{Z}, \mathbb{T}$ and $\mathbb{Z}_{N_l}$  are defined above. The character group of $G$ is given by \be \widehat{G}= \{\chi_{\mu}: \mu\in G^*\}.\ee

\subsection{Duality theory of abelian groups} Note that rule (\ref{eq:Elementary Character Group}) immediately implies that $(G^*)^* = G$, i.e.,  the character group of $G^*$ is $\{\chi_{g}: g\in G\}$, where $\chi_g$ is defined in full analogy with (\ref{mu_character}). Furthermore, these equations illustrate  two fundamental features of elementary abelian groups and their character functions. 
\begin{lemma}[\textbf{The character group is elementary}]\label{lemma:Character Multiplication}
For every $\mu,\nu\in G^*, g\in G$ it holds that
\begin{equation}
\chi_{\mu+\nu}(g)=\chi_{\mu}(g)\chi_{\nu}(g).
\end{equation}
As a consequence,  the map $\mu\to \chi_\mu$ realizes the group isomorphism between $G^*$ and $\widehat{G}$.
\end{lemma}
\begin{lemma}[\textbf{Group-character duality}]\label{lemma:Pontryagin duality for characters}
For every $g\in G$ and $\mu\in G^*$ we have
\be \chi_{\mu}(g)= \chi_g(\mu).\ee
This identity implies that the map $g \to \chi_g$ defines a group isomorphism between $G$ and the character group of $\widehat{G}$, establishing a  duality between groups and their associated characters\footnote{This is a manifestation of the   Pontryagin-Van Kampen duality \cite{Morris77_Pontryagin_Duality_and_LCA_groups,Stroppel06_Locally_Compact_Groups,Dikranjan11_IntroTopologicalGroups,rudin62_Fourier_Analysis_on_groups,HofmannMorris06The_Structure_of_Compact_Groups,Armacost81_Structure_LCA_Groups,Baez08LCA_groups_Blog_Post}, which says that any locally compact abelian group $G$ is isomorphic to $\doublewidehat{G}$ via the map $g\to \widetilde{\chi}_g$ where $\widetilde{\chi}_g(\chi_\mu)=\chi_\mu(g)$.  
}
\end{lemma}
Both lemmas \ref{lemma:Pontryagin duality for characters} follow from inspection of the characters of $\mathbb{R}, \mathbb{Z}, \mathbb{T}$ and $\mathbb{Z}_{N}$ defined in (\ref{char_R})-(\ref{char_T}). The lemmas also reflect the strong duality between $G$ and $G^*$.

 Finally, the definition of every character function $\chi_{a}(b)$ as given in (\ref{char_R})-(\ref{char_T}), which is in principle defined for $a$ in  $\R, \Z_N, \Z, \T$  and $b$ in $\R, \Z_N, \T, \Z$, respectively, can be readily extended to the entire domain of real numbers, yielding functions $\chi_x(y)$ with $x ,y\in \R$.  
Consequently, the character functions (\ref{mu_character}) of general elementary abelian groups $G=G_1\times \dots\times G_m$ can also be extended to a larger domain, giving rise to functions $\chi_x(y)$ where $x, y\in \R^m$. With this extended notion, we have the following basic property:

\begin{lemma}\label{thm_extended_characters}
Let $g\in G$ and $\mu\in G^*$. For every  $x, y\in \R^m$ such that $x\equiv g$ $(\mbox{mod } G)$ and $y\equiv \mu$ $(\mbox{mod } G^*)$, we have
\be
\chi_y(x)=\chi_{\mu}(g).
\ee
\end{lemma}
The proof is easily given for primitive groups, and then extended to   elementary abelian groups.

\subsection{Duality of subgroups and morphisms}\label{sect:Annihilators}

Character functions  give rise to set-theoretical dualities among abelian group subgroups and morphisms, via the notions of \emph{dual morphisms} and \emph{subgroup annihilators}\footnote{Annihilator subgroups have been called ``orthogonal subgroups'' in some quantum computing works \cite{lomont_HSP_review,VDNest_12_QFTs,BermejoVega_12_GKTheorem}. We avoid using this term because it is hardly ever used in group theory and because ``subgroup orthogonality'' differs from the usual ``orthogonality'' of vector spaces.}. We review these concepts next.

\myparagraph{Dual morphism:}Let $\alpha:G\rightarrow H$ be a continuous group homomorphism between two elementary abelian groups $G$ and $H$. Then,  there exists a unique continuous group homomorphism $\alpha^*:H^*\rightarrow G^*$, which we call the \textbf{dual homomorphism} of $\alpha$ \cite[prop.\ 30]{Morris77_Pontryagin_Duality_and_LCA_groups},  defined as
\begin{equation}\label{eq:Dual Automorphism DEF}
\chi_{\alpha^*(\mu)}(g)=\chi_\mu(\alpha(g)).
\end{equation}
Again, we have $\alpha^{**}=\alpha$ by duality.

\myparagraph{Annihilator subgroup:} Let $G$ be an elementary abelian group and $X$ be any subset of $G$. The \emph{annihilator}\footnote{As mentioned earlier, in the quantum computation literature (see e.g. \cite{Brassard_Hoyer97_Exact_Quantum_Algorithm_Simons_Problem,lomont_HSP_review,VDNest_12_QFTs}) the annihilator $H^\perp$ of a subgroup $H$ is sometimes known as the \emph{orthogonal subgroup} of $H$.} $X^\perp$ is the subset
\begin{equation}\label{orthogonal group EQUATION}
X^\perp:=\{\mu\in G^* :\: \chi_\mu(x)=1\textnormal{ for every }x\in X\}.
\end{equation}
We can define the annihilator $Y^\perp$ of a subset $Y\subseteq G^*$ analogously as \be Y^\perp:=\{x\in G :\: \chi_\mu(x)=1\textnormal{ for every }\mu\in Y\}.\ee By combining the two definitions it is possible to define  double annihilator sets $X^\Perp:= (X^{\perp})^{\perp}$, which is a subset of the initial group $G$, for every set $X\subseteq G$. Similarly,   $Y^{\Perp}\subseteq G^*$  for every $Y\subseteq G^*$. The following lemma states that $X$ and $X^\Perp$ are related to each other and, in fact, identical sets \emph{iff} $X$ is a closed subgroup.
\begin{lemma}[\textbf{Annihilator properties} \cite{Stroppel06_Locally_Compact_Groups}]\label{lemma:Annihilator properties}
Let $X,Y$ and $H,K$ be, respectively, two arbitrary \emph{subsets} and two  \emph{closed subgroups} of an elementary abelian group $G$. Then the following  holds.\\

\noindent \emph{For subsets:}
\begin{enumerate*}
\item[(a)] $X^\perp$  is a closed subgroup of $G^*$ (and  $X^\Perp$ is a closed subgroup of $G$).
\item[(b)] $X^\Perp$  is the smallest closed subgroup of $G$ containing $X$.
\item[(c)] If $Y$ is a subset of $G$ such that $X\subseteq Y$ then $X^\perp \supseteq Y^\perp$ and $X^\Perp\subseteq Y^\Perp$.
\end{enumerate*}
\emph{For closed subgroups:}
\begin{enumerate*}
\item[(a)] $H^\Perp=H$.
\item[(c)] $H^\perp$ is isomorphic to $(G/H)^*$. 
\item[(d)] $|H^{\perp}| = |G/H| = |G|/|H|$ if $G$ is \emph{finite}.
\item[(e)] $(H\cap K)^{\perp} = \langle H^{\perp},\,  K^{\perp}\rangle$.
\end{enumerate*}
\end{lemma}

\subsection{Final note on notation: simplifying characters via the bullet group}

In order to simplify calculations with characters in the next sections, it will be convenient to renormalize the elements of the  group $G^*$ with a map $\mu \rightarrow \mu^\bullet$ which is defined so that the following equation holds for any  $g\in G$ and $\mu\in G^*$:
\be\label{eq:definition of bullet map}
\chi_{\mu}(g)=\exp{\left(2\pii\, \sum_{i=1}^m \mu_i^\bullet g_i \right)}.
\ee
For this reason, we introduce a new abelian group $G^{\bullet}$, called the bullet group of $G$, which is isomorphic to $G^\bullet= G_1^{\bullet}\times \dots\times G_m^{\bullet}$ and $\widehat{G}$, defined as
\begin{align}
\Z_N^{\bullet} &:= \left\{ 0, \frac{1}{N}, \frac{2}{N}, \dots, \frac{N-1}{N}  \bmod{1}\right\},\notag\\
\R^{\bullet} &:= \R^*= \R; \quad \Z^{\bullet} := \Z^* = \T; \quad
\T^{\bullet} := \T^* = \Z.\label{eq:Bullet Group}
\end{align}
Thus the only difference between the groups $G^*$ and $G^\bullet$ is in the $\Z_{N_i}$ components. The groups $G^*$ and $G^\bullet$ are manifestly isomorphic via the ``bullet map''
\begin{equation}
\mu\in G^*\to \mu^\bullet:= (\mu_1^\bullet, \dots, \mu_m^\bullet) \in G^\bullet,
\end{equation}
where $\mu_i^\bullet:= \mu_i/N$ if $\mu_i\in \Z_N$ and $\mu_i^\bullet = \mu_i$ if $\mu_i$ belongs to either $\R, \Z$ or $\T$.

\section{Homomorphisms and matrix representations}\label{sect:Homomorphisms and matrix representations}\label{section homo iso auto morphisms}

Given two elementary abelian groups $H$ and $G$, a  group homomorphism from $G$ to $H$ is a  map $\alpha: H \rightarrow G$ that fulfills $\alpha(g+h) = \alpha(g) + \alpha(h)$ for every $g, h\in H$. (In other words, $\alpha$ is the group-theoretic analogue of a \textit{linear} map.) An isomorphism from $G$ to $H$ is an invertible group homomorphism. An automorphism of $G$ is an isomorphism of the form $\alpha:G\to G$, i.e. from a group onto itself. The set of all automorphisms of $G$ forms a group, called the automorphism group.

Throughout this thesis, continuous group homomorphisms between abelian groups are to be described in terms of \emph{matrix representations}. In this section we introduce and develop these techniques.

\subsection{Normal form of a homomorphisms}\label{sect:Homomorphisms}
 
Let $G=G_1\times \ldots \times G_m$ and $H=H_1\times \ldots \times H_n$ be two elementary finite abelian groups, where $G_i$, $H_j$ are primitive subgroups. As discussed in section \ref{sect:elementary_Abelian_groups}, we assume that the $G_i$ and $H_j$ are ordered so that $G=G_{\Z}\times G_{\R}\times G_{F}\times G_{\T}$
and $H=H_{\Z}\times H_{\R}\times H_{F}\times H_{\T}$.

Consider a continuous group homomorphism $\alpha: G\to H$. Let $\alpha_{\mathbb{Z}\mathbb{Z}}: {G_{\Z}}\to {G_{\Z}}$ be the map obtained by restricting the input and output of $\alpha$ to ${H_{\Z}}$. More precisely, for $g\in G_{\Z}$ consider the map \be (g, 0, 0, 0)\in G \to \alpha(g, 0, 0, 0)\in H\ee and define $\alpha_{\mathbb{Z}\mathbb{Z}}(g)$ to be the $G_{\Z}$-component of $\alpha(g, 0, 0, 0)$. The resulting map $\alpha_{\mathbb{Z}\mathbb{Z}}$ is a continuous homomorphism from $\Z$ to $\Z$. Analogously, we define the continuous group homomorphisms $\alpha_{XY}: G_Y\rightarrow H_X$ with $X, Y = \Z, \R, \T, F$. It follows that, for any $g=(z, r, f, t)\in G$, we have
\be\label{eq:block decomp Homomorphism}
\alpha(g) = \begin{pmatrix}
      \alpha_{\Z\Z}(z) + \alpha_{\Z\R}(r) + \alpha_{\Z F}(f) + \alpha_{\Z\T}(t) \\
      \alpha_{\R\Z}(z) + \alpha_{\R\R}(r) + \alpha_{\R F}(f) + \alpha_{\R\T}(t) \\
      \alpha_{F\Z}(z) + \alpha_{F\R}(r) + \alpha_{FF}(f) + \alpha_{F\T}(t) \\
      \alpha_{\T\Z}(z) + \alpha_{\T\R}(r) + \alpha_{\T F}(f) + \alpha_{\T\T}(t)
    \end{pmatrix} \leftrightarrow \begin{pmatrix}
      \alpha_{\Z\Z} & \alpha_{\Z\R} & \alpha_{\Z F} & \alpha_{\Z\T} \\
      \alpha_{\R\Z} & \alpha_{\R\R} & \alpha_{\R F} & \alpha_{\R\T} \\
      \alpha_{F\Z} & \alpha_{F\R} & \alpha_{FF} & \alpha_{F\T} \\
      \alpha_{\T\Z} & \alpha_{\T\R} & \alpha_{\T F} & \alpha_{\T\T}
    \end{pmatrix}
    \begin{pmatrix}
    z\\ r\\f\\t
    \end{pmatrix}
\ee
 
$\alpha$ is therefore naturally identified with the $4\times 4$ ``matrix of maps'' given in the r.h.s of (\ref{eq:block decomp Homomorphism}).

The following lemma (see e.g. \cite{PrasadVemuri08_classification_heisenberg_groupss} for a proof) shows that homomorphisms between elementary abelian groups must have a particular block structure.
\begin{lemma}[\textbf{Homomorphism normal form}]\label{corollary:block-structure of Group Homomorphisms} Let $\alpha: G\to H$ be a continuous group homomorphism. Then $\alpha$ has the following block structure
\begin{equation}\label{eq:block decomposition of a homomorphism}
    \alpha \leftrightarrow
    \begin{pmatrix}
      \alpha_{\Z\Z} & \mathpzc{0}  &\mathpzc{0}  &  \mathpzc{0} \\
      \alpha_{\R\Z} & \alpha_{\R\R} &  \mathpzc{0} &  \mathpzc{0} \\
      \alpha_{F\Z} &  \mathpzc{0} & \alpha_{FF} &  \mathpzc{0} \\
      \alpha_{\T\Z} & \alpha_{\T\R} & \alpha_{\T F} & \alpha_{\T\T}
    \end{pmatrix}
\end{equation}
where $\mathpzc{0}$ denotes the trivial group homomorphism.
\end{lemma}
 
The lemma shows, in particular, that there are no non-trivial continuous group homomorphisms between certain pairs of primitive groups: for instance, continuous groups cannot be mapped into discrete ones, nor can finite groups be mapped into zero-characteristic groups.

\newpage
\subsection{Matrix representations}\label{sect:Matrix Representations}

\begin{definition}
[\textbf{Matrix representation}]\label{def:Matrix representation}
Consider elementary abelian groups $G= G_1\times\dots \times G_m$ and $H= H_1\times\dots \times H_n$ and a group homomorphism $\alpha:G\rightarrow H$. A \emph{matrix representation of $\alpha$} is an $n\times m$ real matrix  $A$ satisfying the following property:
 \be\label{def_matrix_rep} \alpha (g)\equiv Ax \ (\mbox{mod } H) \quad \mbox{ for every } g\in G \mbox{ and } x\in \R^m \mbox{ satisfying } x\equiv g \ (\mbox{mod } G)\ee
Conversely, a real $n\times m$ matrix $A$ is said to define a group homomorphism if there exists a group homomorphism $\alpha$ satisfying (\ref{def_matrix_rep}).
\end{definition}
It is important to highlight that in the definition of matrix representation we impose that the identity $\alpha(g)= Ax \pmod{H}$ holds in a very general sense: the output of the map must be equal for inputs $x,\,x'$ that are \emph{different} as strings of real numbers but correspond to the \emph{same} group element $g$ in the group $G$. In particular, all strings that are congruent to zero in $G$ must be mapped to strings congruent to zero in $H$. Though these requirements are (of course) irrelevant when we only consider groups of zero characteristic (like $\Z$ or $\R$), they are crucial when quotient groups are involved (such as $\Z_N$ or $\T$).

As a simple example of a matrix representation, we consider the bullet map\footnote{Strictly speaking, definition \ref{def:Matrix representation} cannot be applied to the bullet map, since $G^\bullet$ is not an elementary abelian group. However the definition is straightforwardly extended to remedy this.}, which is an isomorphism from $G^*$ to $G^\bullet$ . Define the diagonal $m\times m$ matrix $\Upsilon$ with diagonal entries defined as
\begin{equation}\label{eq:matrix representation of bullet isomorphism}
\Upsilon(i,i)=
\begin{cases}
    1/N_i &\textrm{if ${G_i}= \Z_{N_i}$ for some $N_i$}, \\
   1  &\textrm{otherwise}.
\end{cases}
\end{equation}
It is easily verified that $\Upsilon$ satisfies the following property: for every $\mu\in G^*$ and $x\in \R^m$ satisfying $x\equiv \mu $ $(\mbox{mod } G^*)$, we have
\begin{equation}\label{eq:Upsilon is a Mat Rep of Bullet}
\mu^\bullet\equiv\Upsilon x \pmod{G^\bullet}.
\end{equation}
Note that, with the definition of $\Upsilon$, equation (\ref{eq:definition of bullet map}) implies
\begin{equation}\label{eq:definition of bullet map2}
\chi_{\mu}(g)=\exp{\left(2\pii\, \sum_{i=1}^m \mu^\bullet(i) g(i) \right)} = \exp{\left(2\pii\, \mu^T \Upsilon g\right)}.
\end{equation}
Looking at equation  (\ref{eq:Upsilon is a Mat Rep of Bullet}) coefficient-wise, we obtain a relationship  
$\mu^\bullet(i)\equiv\frac{x(i)}{N_i}\pmod{1}$ for each factor $G_i$ of the form $\Z_{N_i}$; other factors are left unaffected by the bullet map. From this expression it is easy to derive that  $\Upsilon^{-1}$ is a matrix representation of the inverse of the bullet map\footnote{We ought to highlight that the latter is by no means a general property of matrix representations. In fact, in many cases, the matrix-inverse $A^{-1}$ (if it exists) of a matrix representation $A$ of  a group isomorphism is not a valid matrix representation of a group homomorphism. (This happens, for instance, for all group automorphisms of the group $\Z_N$ that are different from the identity.)  In lemma \ref{lemma:Normal form of a matrix representation} we characterize which matrices are valid matrix representations. Also, in section \ref{sect:Computin Inverses} we discuss the problem of computing  matrix representations of  group automorphisms.}, i.e.\ the group isomorphism $\mu^\bullet\rightarrow\mu \pmod{G^*}$.

The next lemma (see appendix \ref{appendix:Supplement to section Homomorphisms} for a proof) summarizes some useful properties of matrix representations.
\begin{lemma}[\textbf{Properties of matrix representations}]\label{lemma:properties of matrix representations}

Let $G$, $H$, $J$ be elementary  abelian groups, and $\alpha :G\rightarrow H$ and $\beta: H\rightarrow J$ be group homomorphisms with matrix representations $A,\, B$, respectively. Then it holds that
\begin{itemize*}
\item[(a)] $BA$ is a matrix representation of the composed homomorphism $\beta\circ\alpha$;
\item[(b)] The matrix $A^*:=\Upsilon_{G}^{- 1} A^\transpose \, \Upsilon_H $ is a matrix representation of the dual homomorphism  $\alpha^*$, where  $\Upsilon_X$ denotes the matrix representation of the bullet map $X^*\rightarrow X^\bullet$.
\end{itemize*}
\end{lemma}

As before, let $G=G_1\times\dots\times G_m$ be an elementary abelian group with each $G_i$ of primitive type. Let \be e_i= (0, \dots, 0, 1, 0, \dots, 0)\ee denote the $i$-th canonical basis vector  of $\mathbb{R}^m$. If we regard $g\in G$ as an element of $\R^m$, we may write $g= \sum g(i) e_i$. Note however that $e_i$ may not belong to $G$ itself. In particular, if $G_i = \T$ then $e_i\notin \T$ (since $1\notin \T$ in the representation we use, i.e.\ $\T= [0, 1)$).

\begin{lemma}[\textbf{Existence of matrix representations}]\label{lemma:existence of matrix representations} Every group homomorphism $\alpha: G\rightarrow H$ has a matrix representation $A$. As a direct consequence, we have $\alpha(g)\equiv \sum_i g(i)Ae_i$ $(\mbox{mod }H)$, for every $g=\sum_i g(i)e_i\in G$.
\end{lemma}
The last property of lemma \ref{lemma:existence of matrix representations} is remarkable, since the coefficients $g(i)$ are real numbers when $G_i$ is of the types $\R$ and $\T$. We give a proof of the lemma in appendix \ref{appendix:Supplement to section Homomorphisms}.

\label{section Quadratic functions}

We finish this section by giving a normal form for matrix representations and characterizing  which types of matrices constitute valid matrix representations as in definition \ref{def:Matrix representation}.
\begin{lemma}[\textbf{Normal form of a matrix representation}] \label{lemma:Normal form of a matrix representation}
Let $G = G_1\times\dots\times G_m$ and $H=H_1\times \dots\times H_n$ be elementary abelian groups. Let $c_j, c_j^*, d_i$ and $d_i^*$  denote the characteristic of $G_j, G_j^*, H_i$ and $H_i^*$, respectively. Define $\mathbf{Rep}$ to be the subgroup of all $n\times m$   real matrices  that have integer coefficients in those rows $i$ for which $H_i$ has the form  $\Z$ or $\Z_{d_i}$. A real $n\times m$ matrix $A$ is a valid matrix representation of some group homomorphism $\alpha: G\rightarrow H$ iff  $A$ is an element of $\mathbf{Rep}$ fulfilling two (dual) sets of consistency conditions:
\begin{align}\label{eq:Consistency Conditions Homomorphism}
c_j A(i,j) = 0 \mod d_i,\qquad {d}_i^* A^*(i,j) = 0 \mod c^*_j, 
\end{align}
for every $i=1,\ldots,n$, $j=1,\ldots, m$, and being $A^*$  the $m\times n$ matrix defined in lemma \ref{lemma:properties of matrix representations}(b). 
Equivalently, $A$ must be of the form 
\begin{equation}\label{eq:block-structure of matrix representations}
A:= \begin{pmatrix}
      A_{\Z\Z} & 0 & 0 & 0 \\
      A_{\R\Z} & A_{\R\R} & 0 & 0 \\
      A_{F\Z} & 0 & A_{FF} & 0 \\
      A_{\T\Z} & A_{\T\R} & A_{\T F} & A_{\T\T}.
    \end{pmatrix}
\end{equation}
with the following restrictions:
  \begin{enumerate}
  \item $A_{\Z\Z}$ and $A_{\T\T}$ are arbitrary  integer matrices.
  \item $A_{\R\Z}$, $A_{\R\R}$ are arbitrary  real matrices.
  \item $A_{F\Z}$, $A_{FF}$ are integer matrices: the first can be arbitrary; the coefficients of the second must be of the form \begin{equation}\label{eq:coefficients of Matrix Rep for nonzero characteristic groups}
  A(i,j)= \alpha_{i,j}\, \frac{d_i}{\gcd{(d_i, c_j)}}
  \end{equation}
  where  $\alpha_{i,j}$ can be arbitrary integers\footnote{Since $A_{F\Z}$, $A_{FF}$ multiply integer tuples and output integer tuples modulo $F=\DProd{N}{c}$, for some $N_i$s, the coefficients of their $i$th rows can be chosen w.l.o.g. to lie in the range $[0,N_i)$ (by taking remainders).}.
  \item $A_{\T\Z}$, $A_{\T\R}$ and $A_{\T F}$ are real matrices: the first two are arbitrary; the coefficients of the third are of the form  $A(i,j)= \alpha_{i,j}/c_j$ where  $\alpha_{i,j}$ can be arbitrary integers\footnote{Due to the periodicity of the torus, the coefficients of $A_{\T\Z}$, $A_{\T F}$ can be chosen to lie in the range  $[0,1)$.}.
  \end{enumerate}
\end{lemma}
The result is proven in appendix \ref{appendix:Supplement to section Homomorphisms}.

\section{Quadratic functions}\label{sect:quadratic_functions}

In this section we study the properties of quadratic functions over arbitrary elementary groups of the form $G=\R^a\times \T^a\times \Z^b \times \DProd{N}{c}$. Most importantly, we give normal forms for quadratic functions and bicharacters. We list results without proof, since all techniques used throughout the section are classical. Yet, we highlight that the normal form in theorem \ref{thm:Main Result} has quantum applications, since we will show in chapter \ref{chapterF} that quadratic functions can be used to give a powerful normal form for stabilizer states over elementary groups.

All results in this section are proven in appendix \ref{app:Supplement Quadratic Functions}. 

\subsection{Definitions}

Let $G$ be an elementary abelian group. Recall from chapter \ref{chapterC} that a bicharacter of $G$ is a continuous complex function $B:G\times G\to U(1)$ such that the restriction of $B$ to either one of its arguments is a character of $G$.  Recall that a quadratic function $\xi:G\to U(1)$ is a continuous function for which there exists a bicharacter  $B$ such that  
\begin{equation} 
\xi(g+h) = \xi(g)\xi(h)B(g, h) \quad \mbox{ for all } g,h \in G.
\end{equation}
In this section, we call $\xi$  a \emph{$B$-representation} if the above equation holds. A bicharacter $B$ is said to be \emph{symmetric} if $B(g, h)= B(h, g)$ for all $g, h\in G$. Symmetric bicharacters are natural objects to consider in the context of quadratic functions: if $\xi$ is a $B$-representation then $B$ is symmetric since 
\begin{equation} B(g, h)= \xi(g+h)\overline{\xi(g)}\overline{\xi(h)}= \xi(h+g)\overline{\xi(h)}\overline{\xi(g)}=B(h,g).
\end{equation}

\subsection{Normal form of  bicharacters}

The next lemmas characterize bicharacter functions.
\begin{lemma}[\textbf{Normal form of a bicharacter}]\label{lemma:Normal form of a bicharacter 1}
Given an elementary abelian group $G$, then a function $B:G\times G\rightarrow U(1)$ is a bi-character iff it  can be written in the normal form
\begin{equation}\label{eq:first normal fomr of a bicharacter}
B(g,h)=\chi_{\beta(g)}(h)
\end{equation}
where $\beta$ is some group homomorphism from $G$ into $G^*$.
\end{lemma}
 
This result generalizes lemma 5(a) in \cite{VDNest_12_QFTs}.The next lemma gives a explicit characterization of symmetric bicharacter functions.
\begin{lemma}[\textbf{Normal form of a symmetric bicharacter}]\label{lemma symmetric matrix representation of the bicharacter homomorphism}
Let $B$ be a symmetric bicharacter of $G$ in the form (\ref{eq:first normal fomr of a bicharacter}) and let $A$ be a matrix representation of the homomorphism $\beta$. Let $\Upsilon$ denote the default matrix representation of the bullet map $G^*\rightarrow G^\bullet$ as in (\ref{eq:matrix representation of bullet isomorphism}), and $M=\Upsilon A$. Then
\begin{enumerate*}
\item[(a)] $B(g,h)=\exp{\left(2\pii \, g^\transpose M h\right)}$ for all $g,h\in G$.
\item[(b)] $M$ is a matrix representation of the homomorphism $G\stackrel{\beta}{\rightarrow} G^*\stackrel{\bullet}{\rightarrow}G^\bullet$.
\item[(c)] If $x, y\in\R^m$ and $g, h\in G$ are such that $x\equiv g$ (mod $G$) and $y\equiv h$ (mod $G$), then
\be
B(g, h) = \exp{\left(2\pii \, x^\transpose M y\right)}.
\ee

\item[(d)] The matrix $M$ is symmetric modulo integer factors, i.e.\ $M = M^{T} \bmod{\Z}$.
\item[(e)] The matrix $M$ can be efficiently symmetrized: i.e.\ one can compute in classical polynomial time a symmetric matrix $M'=M^{'T}$ that also fulfills (a)-(b)-(c).
\end{enumerate*}
\end{lemma}

\subsection{Normal form of quadratic functions}

Our final goal is to characterize all quadratic functions. This is achieved in theorem \ref{thm:Normal form of a quadratic function}. To show this result a few lemmas are needed.
\begin{lemma}\label{lemma quadratic B-representations differ by a character}
Two quadratic functions $\xi_1$,$\xi_2$ that are $B$-representations of the same bicharacter  $B$ must be equal up to multiplication by a character of $G$, i.e. there exists $\mu\in G^*$ such that
\begin{equation}
\xi_1(g)=\chi_{\mu}(g)\xi_2(g), \quad \text{for every $g\in G$.}
\end{equation}
\end{lemma}
\begin{proof}
This lemma can be proven using projective representation theory \cite{BackBrad71_projective_representations_of_Abelian_groups}. Here, we give a simple alternative proof. 
We prove that the function $f(g):=\xi_1(g)/\xi_2(g)$ is a character, implying that there exists $\mu\in G^*$ such that $\chi_{\mu}=f$:
\begin{equation}\label{eq proof of quadratic B-representations differ by character}
f(g+h):=\frac{\xi_1(g)}{\xi_2(g)}\frac{\xi_1(h)}{\xi_2(h)}\frac{B(g,h)}{B(g,h)}=f(g)f(h).
\end{equation}
\end{proof}

Our approach now will be to find a method to construct a quadratic function that is a $B$-representation for any given bicharacter $B$. Given one $B$-representation,  lemma \ref{lemma quadratic B-representations differ by a character} tells us how all other $B$-representation look like. We can exploit this to characterize all possible quadratic functions, since we know how symmetric bicharacters look (lemma \ref{lemma symmetric matrix representation of the bicharacter homomorphism}).

The next lemma shows how to construct $B$-representations canonically.
\begin{lemma}\label{lemma:B-representations can always be constructed}
Let $B$ be a bicharacter $B$ of $G$. Consider  a symmetric real matrix $M$ such that $B(g,h)=\exp{\left(2\pii \, g^{\transpose} M h\right)}$. Then the following function is quadratic and  a $B$-representation:
\begin{equation}
Q(g):=\euler^{\pii \, \left(  g^{\transpose} M g + C^\transpose g \right)},
\end{equation}
where $C$ is an integer vector dependent on $M$, defined component-wise as $C(i)=M(i,i)c_i$, where $c_i$ denotes the characteristic of the group $G_i$.
\end{lemma}
Finally, we arrive at the main result of this section.
\begin{theorem}[\textbf{Normal form of a quadratic function}]\label{thm:Normal form of a quadratic function}
Let $G$ be an elementary abelian group. Then a function $\xi: G\rightarrow U(1)$ is quadratic if and only if
\begin{equation}\label{eq:Normal Form Quadratic Function}
\xi(g)=\euler^{\pii \,\left(g^{\transpose} M g \: +  \: C^{\transpose} g  \: +  \: 2v^\transpose g\right)}
\end{equation}
where $C$, $v$, $M$ are, respectively, two vectors and a matrix that satisfy the following:
\begin{itemize}
\item $v$ is an element of the bullet group $G^\bullet$;
\item $M$ is the matrix representation of a group homomorphism from $G$ to $G^\bullet$; and
\item $C$ is an integer vector dependent on $M$, defined component-wise as $C(i)=M(i,i)c_i$, where $c_i$ is the characteristic of the group $G_i$.
\end{itemize}
\end{theorem}
The normal form in theorem \ref{thm:Normal form of a quadratic function} can be very useful to perform certain calculations within the space of quadratic functions, as illustrated by the following lemma.
\begin{lemma}\label{lemma:Quadratic Function composed with Automorphism}
Let $\xi_{M,v}$ be the quadratic function (\ref{eq:Normal Form Quadratic Function}) over $G$. Let $A$ be the matrix representation of a continuous group homomorphism $\alpha:G\rightarrow G$. Then the composed function $\xi_{M,v}\circ \alpha$ is also quadratic and can be written in the normal form
(\ref{eq:Normal Form Quadratic Function})  as $\xi_{M',v'}$, with
\begin{equation}
M':=A^\transpose M A, \qquad v':= A^\transpose v + v_{A,M},\qquad v_{A,M}:= A^\transpose C_M - C_{A^\transpose M A},
\end{equation}
 
 where $C_M$ is the vector $C$ associated with $M$ in (\ref{eq:Normal Form Quadratic Function}).
\end{lemma}

\section{Computational group theory}\label{sect_CompComp_FiniteAbelianGroups}

Computational aspects of finite abelian groups are now discussed; our discourse focuses on a selected catalog of  computational problems relevant to this chapter and efficient classical algorithms to solve them. Since this section concerns only classical computational complexity, we will tend to omit the epithet \textit{classical} all the way throughout it.

\subsection{Basic group operations}

We begin  recalling that basic arithmetical computations within groups of the form 
\begin{equation}\label{eq:GroupElementaryLCA}
G=\R^a \times \mathbb{T}^b\times\cdots  \times \Z^c \times F, \qquad F:= \DProd{N}{d} .
\end{equation}
 can be efficiently performed in a classical computer \cite{brent_zimmerman10CompArithmetic}. From now on,  the  \emph{size} $\|N \|_\mathbf{b}$ of an integer $N$ is the number of bits in its binary expansion (recall that $\|N_\mathrm{max}\|_\mathbf{b}$ is roughly $\log |N|$ times the absolute value of $N$).  Throughout this thesis the elements of (\ref{eq:GroupElementaryLCA}) will always be represented as $m:=a+b+c+d$ vectors of fractions $g=(g(1),\ldots,g(m))\in G$, which can be efficiently stored in a computer: when $G$ is finite,  $O(\polylog{|G|})$ bits of memory are enough\footnote{This follows from the inequalities $2^{m}\leq \Inputsize$ and $N_i \leq \Inputsize$.}, in general, $O(m\, \poly{\|N_\mathrm{max}\|_\mathbf{b}})$ bits are enough where $N_\mathrm{max}$ denotes the largest numerator/denominator in $g$ that we need to store. The bit-size scaling of these descriptions is efficient in the size of the input. Similarly,  matrix representations as in lemma \ref{lemma:Normal form of a matrix representation} can be efficiently described in terms of rational \emph{matrices}, instead of vectors.

We discuss now how to perform some basic operations efficiently within any finite abelian group (\ref{eq:GroupElementaryLCA}). First, given two fractions $a$ and $b$ with numerators and denominators of size at most $l$, common arithmetic operations can be computed in \ppoly{l} time with elementary algorithms: such as their sum, product, the quotient of $a$ divided by $b$, and the remainder $a\bmod{b}$ \cite{brent_zimmerman10CompArithmetic}. Therefore, given $g, h\in G$, the sum $g+h$  can be obtained in $O(m\, \poly{\|N_\mathrm{max}\|_\mathbf{b}})$ time  by computing the $m$ remainders $g(i) + h(i) \bmod{\mathrm{char}(G_i)}$, where $\mathrm{char}(G_i)$ is the characteristic function (\ref{eq:Characteristic}). Similarly, given an integer $n$, the element $ng$  can be obtained in $ \ppolylog{m\|N_\mathrm{max}\|_\mathbf{b},\|n\|_\mathbf{b} }$ time  by computing the remainders $ng(i)$ mod $d_i$.

In connection with section \ref{sect:Homomorphisms and matrix representations}, it follows from the properties just introduced that matrix representations  can be stored using only a polynomial amount of memory, and, moreover, that given the matrix representation $A$ we can efficiently compute $Ah\pmod{G}$. Specifically,  given a matrix representation $A$ of the homomorphism $\alpha: G_1\times \cdots \times G_m \rightarrow H_1\times\cdots  \times H_n$, we need $\ppolylog{mn , \|N_\mathrm{max}\|_\mathbf{b}}$ space to store its columns $a_i$ as tuples of integers, and $\ppolylog{mn , \|N_\mathrm{max}\|_\mathbf{b}}$ time to compute the function $\alpha(h)$. 

\subsection{Systems of linear equations over abelian groups}\label{sect:Systems of linear equations over groups}

Let $\alpha:G\rightarrow H$ be a  continuous group homomorphism between  elementary abelian groups $G$, $H$ and let $A$ be a rational matrix representation of $\alpha$. We consider systems of equations of the form
\begin{equation}\label{eq:Systems of linear equations over groups}
\alpha(x)\equiv Ax\equiv b\pmod{H},\quad \textnormal{where } x\in G,
\end{equation}
which we dub \emph{systems of linear equations over (elementary) abelian groups}.  In this section we develop algorithms to find solutions of such systems. 

Systems of linear equations over  abelian groups form a large class of problems, containing, as particular instances, standard systems of linear equations over real vectors spaces,
\begin{equation}\label{eq:Linear System Ax=b over R}
\mathbf{A}\mathbf{x}=\mathbf{b},\quad \mathbf{A}\in\R^{n\times m},\, \mathbf{x}\in \R^m , \mathbf{b}\in \R^n,
\end{equation}
as well as systems of linear equations over other types of vector spaces, such as $\Z_2^n$, e.g.\
\begin{equation}\label{eq:Linear System Ax=b over Z_2}
\mathbf{B}\mathbf{y}=\mathbf{c},\quad \mathbf{B}\in\Z_2^{n\times m},\, \mathbf{y}\in \Z_2^m , \mathbf{c}\in \Z_2^n.
\end{equation}
In (\ref{eq:Linear System Ax=b over R}) the matrix $\mathbf{A}$ defines a linear map from $\R^m$ to $\R^n$, i.e.\ a map that fulfills $\mathbf{A}( a \mathbf{x}+b\mathbf{y})=\mathbf{A}( a \mathbf{x})+\mathbf{A}( b\mathbf{y}),$ for every $a,b\in\R,\,\mathbf{x,y}\in\R^n$ and is, hence, compatible with the vector space operations; analogously,  $\mathbf{B}$ in (\ref{eq:Linear System Ax=b over Z_2}) is a linear map between  $\Z_2$ vector spaces.

We dub systems (\ref{eq:Systems of linear equations over groups}) ``linear'' to highlight this resemblance. Yet the reader must beware that, in general, the groups $G$ and $H$ in problem (\ref{eq:Systems of linear equations over groups}) are \emph{not} vector spaces (primitive factors of the form $\Z$ or $\Z_d$, with non-prime $d$, are rings yet \emph{not} fields; the circle $\T$ is not even a ring, as it lacks a well-defined multiplication operation\footnote{Note that $\T=\R/\Z$ is a quotient group of $\R$ and that the addition in $\T$ is  well-defined  group operation between equivalence classes. It is, however, not possible to define a multiplication $a b$ for $a,b\in \T$ operation between equivalence classes: different choices of class representatives yield different results.}), and that the map $A$ is a group homomorphism between groups, but \emph{not} a linear map between vector spaces. 

Indeed, there are interesting classes of problems that fit in the class (\ref{eq:Systems of linear equations over groups}) and that are not systems of linear equations over vectors spaces. For infinite groups, an example are  systems of mixed real-integer linear equations \cite{BowmanBurget74_systems-Mixed-Integer_Linear_equations, HurtWaid70_Integral_Generalized_Inverse}, which we introduce  in equation (\ref{eq:System of Mixed Integer linear equations}) in this section. Furthermore, in the next chapters, we will encounter a wide range of computational problems directly related to simulating normalizer circuits that can be reduced to linear systems over abelian groups (cf.\ lemma \ref{lemma:Algorithms_FA_Groups} in chapter \ref{chapterF}): hence, the techniques developed in this section will be useful throughout the thesis.

\paragraph{Input of the problem.}

In this thesis, we only consider systems of the form (\ref{eq:Systems of linear equations over groups}) where the matrix $A$ is \emph{rational}. In other words,  we always assume that the group homomorphism $\alpha$ has a rational matrix representation $A$; the latter is  given to us in the input of our problem. Exact integer arithmetic will be used to store the rational coefficients of $A$; floating point arithmetic will never  
be needed in our proofs.\footnote{Of course, not all group homomorphisms have rational matrix representations (cf.\ lemma \ref{lemma:Normal form of a matrix representation}). However, for the applications we are interested  (cf.\  chapter \ref{chapter0}-\ref{chapterB}) it is enough to study this subclass.} 

\paragraph*{General solutions of system (\ref{eq:Linear System Ax=b over R})}

Since $A$ is a homomorphism, it follows that the set $G_{\textnormal{sol}}$ of all solutions of (\ref{eq:Systems of linear equations over groups}) is either empty or a coset of the kernel of $A$:
\begin{equation}\label{eq:Systems of linear equations over groups: Solution Space}
G_{\textnormal{sol}}=x_0+\ker{A}
\end{equation}
The main purpose of this section is to devise efficient algorithms to solve system (\ref{eq:Systems of linear equations over groups}) when $A$, $b$ are given as  
input, in the following sense: we say that we have \emph{solved} system (\ref{eq:Systems of linear equations over groups}) if we manage to find a \emph{general solution} of (\ref{eq:Systems of linear equations over groups}) as defined next.
\begin{definition}[\textbf{General solution of system (\ref{eq:Systems of linear equations over groups})}]\label{def:General Solution of a system} A {\emph{general solution}} of a system of equations $Ax\equiv b\pmod{H}$ as in $(\ref{eq:Systems of linear equations over groups})$ is a pair $(x_0, \mathcal{E})$ where $x_0$ is a particular solution of the system and $\mathcal{E}$ is a continuous group homomorphism (given as a matrix representation) from an auxiliary group $\mathcal{X}:=\R^{\alpha} \times \Z^{\beta}$ into   $G$, whose image $\textnormal{im}\,\mathcal{E}$ is the kernel of ${A}$.
\end{definition}
Although it is not straightforward to prove, general solutions of solvable systems of the form (\ref{eq:Systems of linear equations over groups}) \emph{always} exist. This is shown in appendix \ref{appendix:Closed subgroups of LCA groups / existence of general-solutions of linear systems}.

A  main contribution of this chapter is a deterministic classical algorithm that finds a general solution of any system of the form (\ref{eq:Systems of linear equations over groups}) in polynomial time.  
This is the content of the next theorem, which is one of our main technical results.
\begin{theorem}[\textbf{Classical algorithms for linear systems over groups (\ref{eq:Systems of linear equations over groups})}]\label{thm:General Solution of systems of linear equations over elementary LCA groups}
Let $A$, $b$  define a system of linear equations (over elementary abelian groups) of form (\ref{eq:Systems of linear equations over groups}), with the group $G$ as solution space and image group $H$. Let $m$ and $n$ denote the number of direct-product factors of $G$ and $H$ respectively and let $c_i$, $d_j$ denote the characteristics of $G_i$ and $H_j$.   
Then there exist efficient, deterministic, exact classical algorithms to solve the following tasks in  $O\left(\ppoly{m,n,\|A\|_{\mathbf{b}}, \|b\|_{\mathbf{b}}, \log{c_i},\log{d_j}}\right)$ time:
\begin{enumerate}
\item Deciding whether system (\ref{eq:Systems of linear equations over groups}) admits a solution.
\item Finding a general solution $(x_0,\mathcal{E})$ of (\ref{eq:Systems of linear equations over groups}).
\item Simplifying ``discrete'' solutions: given a  finitely generated  $G$ and a  solution $(x_0,\mathcal{E})$ where $\mathcal{E}$ acts on $\mathcal{X}=\Z^{\alpha+\beta}$\footnote{Note that $\mathcal{X}$ needs to be of form $\mathcal{X}=\Z^{\alpha+\beta}$ for discrete $G$ (lemma \ref{lemma:Normal form of a matrix representation}).}; find  $\{Q,\mathcal{E}_\mathrm{iso}\}$ such that (a) $Q$ is an elementary group isomorphic to the quotient $\mathcal{X}/\ker{\mathcal{E}}$ and (b) $\mathcal{E}_\mathrm{iso}$ is a matrix representation of the isomorphism $Q\xrightarrow{\mathcal{E}}\mathrm{im}\mathcal{E}$.
\item  If $G$ is finite,   counting the number of solutions of (\ref{eq:Systems of linear equations over groups}) and finding $x_1, \ldots, x_r\in G$ such that all solutions of the system are linear combinations of the form $x_0+\sum k_i x_i$. 
\end{enumerate}
\end{theorem}
A rigorous proof of this theorem is given in appendix \ref{appendix:Systems of Linear Equations over Groups}. Below, we sketch the key ideas behind our algorithms for task (1-2); our algorithms for tasks (3-4) crucially combine the former ones  with fast classical methods to compute  Smith normal forms \cite{Storjohann10_Phd_Thesis}.

In short, for tasks (1-2), we show that the problem of finding a general solution of a system of the form (\ref{eq:Systems of linear equations over groups}) reduces in polynomial time to the problem of finding a general solution of a so-called \emph{system of mixed real-integer linear equations} \cite{BowmanBurget74_systems-Mixed-Integer_Linear_equations}.
\begin{equation}\label{eq:System of Mixed Integer linear equations}
A'x'+B'y'=c,\quad  \textnormal{where } x'\in \Z^a, y' \in\R^b,
\end{equation}
 where $A'$ and $B'$  are rational matrices and $c$ is a rational vector. Denoting by $\R^{b'}$ the given space in which $c$ lives, we see that, in our notation, $\begin{pmatrix}
A & B
\end{pmatrix}w=c$, where $w\in \Z^a\times \R^b$ is a particular instance of a system of linear equations over elementary locally compact abelian groups that are products of $\Z$ and $\R$. Systems (\ref{eq:System of Mixed Integer linear equations}) play an important role within the class of problems (\ref{eq:Systems of linear equations over groups}), since any efficient algorithm to solve the former  can be adapted to solve the latter in polynomial time.

The second main idea in the proof of theorem \ref{thm:General Solution of systems of linear equations over elementary LCA groups} is to apply an existing (deterministic) algorithm by Bowman and Burdet \cite{BowmanBurget74_systems-Mixed-Integer_Linear_equations} that computes a  general solution to a system of the form (\ref{eq:System of Mixed Integer linear equations}). Although Bowman and Burdet did not prove the efficiency of their algorithm in \cite{BowmanBurget74_systems-Mixed-Integer_Linear_equations}, we  show in appendix \ref{appendix:Efficiency of Bowman Burdet} that it can be implemented in polynomial-time, completing the proof of the theorem.

\subsubsection*{Application of theorem \ref{thm:General Solution of systems of linear equations over elementary LCA groups}: computing inverses of group automorphisms}\label{sect:Computin Inverses}

In section \ref{sect:Matrix Representations} we discussed that computing a matrix representation of the inverse $\alpha^{-1}$ of a group automorphism $\alpha$ cannot be done by simply inverting a (given) matrix representation $A$ of $\alpha$. However, the  algorithm given in theorem \ref{thm:General Solution of systems of linear equations over elementary LCA groups} can be adapted to solve this problem.
\begin{lemma}\label{lemma:Computing Inverses}
Let $\alpha:G\rightarrow G$ be a continuous group automorphism. Given any  matrix representation $A$ of $\alpha$, there exists efficient classical algorithms that compute a matrix representation $X$ of the inverse group automorphism $\alpha^{-1}$.
\end{lemma}
A proof (and an algorithm) is given in appendix \ref{appendix:Computing inverses}.

%% file: chapter1_finite.tex
\chapter{Classical simulations of normalizer circuits over finite abelian groups}\label{chapterF}

In chapter \ref{chapterC} we introduced \textit{normalizer circuits over finite abelian groups}  as group theoretic generalizations of Clifford circuits composed of quantum Fourier transforms, automorphism gates and quadratic-phase gates (cf.\ chapters \ref{chapterC}-\ref{sect:Normalizer Gates Finite Group}). An interesting feature of this circuit families is the presence of QFTs over finite abelian groups, which are central in Shor's  factoring algorithm \cite{Shor} and in  quantum algorithms for abelian  hidden subgroups problems \cite{lomont_HSP_review,childs_lecture_8, childs_vandam_10_qu_algorithms_algebraic_problems}. 

Normalizer circuits over finite abelian groups were first studied by Van den Nest in \cite{VDNest_12_QFTs}, who proved that the action of such circuits on  computational basis states and followed by computational basis measurements can be simulated classically efficiently. In this section, we generalize the result in  \cite{VDNest_12_QFTs} in several ways. Most importantly, we show that normalizer circuits supplemented with \emph{intermediate measurements}  of arbitrary (generalized) Pauli operators can also be simulated efficiently classically, even when the computation proceeds \emph{adaptively}.  This yields a generalization of the Gottesman-Knill theorem (valid for $n$-qubit Clifford operations \cite{Gottesman_PhD_Thesis,Gottesman99_HeisenbergRepresentation_of_Q_Computers}) to quantum circuits described by arbitrary finite abelian groups. Moreover, our simulations are twofold: we present efficient classical algorithms to (a) sample the measurement probability distribution of any adaptive-normalizer computation, as well as (b) to compute the amplitudes of the state vector in every step of it.

Finally we develop a generalization of the \emph{stabilizer formalism} \cite{Gottesman_PhD_Thesis,Gottesman99_HeisenbergRepresentation_of_Q_Computers} relative to arbitrary finite abelian groups: for example we characterize how to update stabilizers under generalized Pauli measurements and provide a normal form of the amplitudes of generalized stabilizer states using quadratic functions and subgroup cosets.

The results in this chapter, together with \cite{VDNest_12_QFTs}'s identify a large family of (arbitrarily long) quantum computations  that \emph{cannot} yield exponential speed-ups in spite of usage of the QFT. In chapter \ref{chapterI} we will show that many of our results can even be generalized to an infinite-dimensional setting. In the second part of this thesis (cf.\ chapter \ref{chapterB}), the techniques developed in these first chapters will help us to identify  and analyze more powerful models of normalizer gates that achieve exponential quantum speedups

This chapter is based on \cite{BermejoVega_12_GKTheorem} (joint work with Maarten Van den Nest).

\section{Introduction}

In the circuit model considered in \cite{VDNest_12_QFTs} the allowed operations are normalizer gates over a finite abelian group $G=\DProd{D}{m}$ (cf.\ chapter \ref{sect:Normalizer Gates Finite Group}) supplemented with standard basis states and terminal measurements in the standard basis. The main result in \cite{VDNest_12_QFTs} states that any such circuit is  efficiently classically simulable for any group $G$.  The main contribution of this chapter is a generalization of the result of \cite{VDNest_12_QFTs}  where \emph{intermediate measurements} are allowed at arbitrary times in the computation. This extension recovers a missing feature that was present in the original \cite{Gottesman_PhD_Thesis,Gottesman99_HeisenbergRepresentation_of_Q_Computers} Gottesman-Knill theorem, which states that  intermediate measurements of Pauli operators interspersed along a Clifford circuit can also be classically simulated, and even if operations are chosen adaptively.

Specifically, in this work, we define \textit{adaptive normalizer circuits} over $G$ to comprise the following three fundamental ingredients:
\begin{itemize}
\item \textbf{Normalizer gates over $G$}, i.e. QFTs, automorphism gates, quadratic phase gates.
\item \textbf{Measurements} of generalized Pauli operators over $G$ at any time of  the computation.
\item \textbf{Adaptiveness:} the choice of normalizer gate at any time may depend (in a polynomial-time computable way) on the outcomes obtained in all previous measurement rounds.
\end{itemize}
If $G$ is chosen to be $\mathbb{Z}_2^{n}$, the corresponding class of adaptive normalizer circuits precisely corresponds to the class of adaptive Clifford circuits allowed in the original Gottesman-Knill theorem.

\subsection{Main results}

This chapter contains several results, summarized as follows:
\begin{itemize}
\item[I.] \textbf{A Gottesman-Knill theorem for all finite abelian groups} (Theorem \ref{thm_main}). Given any abelian group $G$, every poly-size adaptive normalizer circuit over $G$, acting on any standard basis input, can be efficiently simulated by a classical computer. That is, we show that the conditional probability distribution arising at each measurement (given the outcomes of the previous ones) can be sampled in classical polynomial time.

\item[II.]  \textbf{A stabilizer formalism for finite abelian groups}. Generalizing the well-known stabilizer formalism for qubits, we develop a stabilizer formalism for arbitrary abelian groups. This framework is a key ingredient to efficiently track the evolution of quantum states under normalizer circuits. In particular, our results are:
    \begin{itemize}
    \item We provide an analytic formula, as well as an efficient algorithm,  to compute the dimension of any stabilizer code over a finite abelian group (Theorem \ref{thm structure test}).
    \item We provide an analytic formula, as well as an efficient algorithm, to compute the update of any stabilizer group under Pauli measurements over arbitrary finite abelian groups (Theorem \ref{thm_Measurement_Update_rules}).
    \end{itemize}
\item[III.] \textbf{A normal form for stabilizer states} (Theorem \ref{thm Normal form of an stabilizer state}). We give an analytic formula to characterize the amplitudes of stabilizer states over abelian groups and show how to compute these amplitudes efficiently. It follows that all stabilizer states over abelian groups belong to the class of Computationally Tractable (CT) states, introduced in \cite{nest_weak_simulations}. The interest in this property is that all CT states can be simulated classically in various contexts well beyond the setting of the present work---cf. \cite{nest_weak_simulations} for a discussion.
\end{itemize}
In all the results above the term \textit{efficient} is used as synonym of ``in polynomial time in $\log{|G|}$'' (where $|G|$ denotes the cardinality of the group $G$). All algorithms presented show good performance regarding computational errors: the sampling algorithm given in theorem \ref{thm_main} is \textit{exact} (i.e.\ it samples the output probability of the adaptive normalizer circuit \emph{exactly} in polynomial time\footnote{In our model, for simplicity we assume availability of a subroutine which allows to generate, with zero error, a uniformly random integer in the interval $[0, N]$ in polylog$(N)$ time, for any integer $N$. Under this assumption, our classical sampling algorithm for simulating normalizer circuits also has perfect accuracy i.e. no additional errors are introduced.}), whereas the algorithms in theorem \ref{thm Normal form of an stabilizer state} yield \textit{exponentially} accurate estimates of state amplitudes and normalization constants.

\subsection{Discussion}

\subsubsection*{Technical aspects of the results.}
 
An important technical difference (and difficulty) of our setting compared to the  Gottesman-Knill theorem qubit one is that in the context of arbitrary finite abelian groups (such as $G=\mathbb{Z}_{2^n}$) arithmetic is generally over large integers. This is in contrast to $\mathbb{Z}_2^n$ where arithmetic is simply over $\mathbb{Z}_2$ i.e.\ modulo 2. The difference is in fact twofold:\\

\noindent
$\bullet$\hspace{1mm} First, $\mathbb{Z}_2$ is a \textbf{field}. As a result, it is possible to describe the ``standard'' stabilizer formalism for qubits with vector space techniques over $\mathbb{Z}_2$. In this context methods like Gaussian elimination have straightforward analogues, which can be exploited in the design of classical algorithms. General abelian groups are however no longer fields. This complicates both the analytic and algorithmic aspects of our abelian-group stabilizer formalism due to, for instance, the presence of zero divisors.\\

\noindent $\bullet$\hspace{1mm} Second, in $\mathbb{Z}_2$ arithmetic is with small numbers (namely 0s and 1s), whereas in general finite abelian groups arithmetic is with \textbf{large integers}. For example, this is the case with $G=\mathbb{Z}_{2^n}$. Of course, one must beware that some problems in number theory are widely believed to be \textit{intractable} for classical computers: consider, for instance, the integer factorization problem or computing discrete logarithms. One of the main challenges in our scenario is to show that the ``integer arithmetic'' used in our classical simulation algorithms can be carried out efficiently. For this purpose, a significant technical portion of our work is dedicated to solving \textit{systems of linear equations modulo a finite abelian group}, defined as follows:  given a pair of finite abelian groups  $G_{sol}$ and  $G$ (both of which are given as a direct product of cyclic groups), and a homomorphism $\alpha$ between them, we look at systems of the form $\alpha(x)=b$ where $x\in G_{sol}$ and $b\in G$. We present polynomial-time deterministic classical algorithms for counting and finding solutions of these systems. These efficient algorithms lie at the core of our classical simulations of normalizer circuits.

\subsubsection*{The power of adaptiveness.}

Another interesting feature in our work compared to the qubit setting, is that abelian-group normalizer circuits with intermediate measurements  are  more powerful than those with only terminal measurements for quantum state preparation:  in {section \ref{sect:adaptiveness}}, we show that certain families of abelian group stabilizer states (namely, abelian-group coset states) can only be prepared if intermediate measurement is allowed\footnote{This is analogous to a known feature of stabilizer states in composite qudit dimensions, some of which cannot be prepared without measuring Paulis, \emph{even though} mere terminal measurements are enough in the qubit setting  (cf.\ discussion in section \ref{sect:adaptiveness}).}. Albeit, despite displaying superior QIP features, our main simulation result says that enhanced normalizer circuits with intermediate measurements can still not outperform classical computers.

\subsubsection*{Applications.} Finally, we recall that the stabilizer formalism has been used in a variety of settings (both for qubits and $d$-level systems) beyond the context of the Gottesman-Knill theorem. This includes e.g.\ 	measurement-based quantum computation \cite{raussen_briegel_onewayQC,ZhouZengXuSun03,Schlingemann04ClusterStates}, quantum error-correction  and fault-tolerance \cite{Gottesman98Fault_Tolerant_QC_HigherDimensions, BravyiKitaev05MagicStateDistillation,CampbellAnwarBrowne12MagicStateDistillation_QUTRITS,CampbellAnwarBrowne12MagicStateDistillation_in_all_prime_dimensions,Anwar14_Decoders_Qudit_Topological_Codes,Campbell14_Enhance_FTQC_d_level_systems,Watson15_Qudit_Color_n_Gauge_Color_Codes}, secret-sharing \cite{Hillery99Quantum_Secret_Sharing, Cleve99Quantum_Secret_Sharing, Gottesman00Quantum_Secret_Sharing}, topological systems \cite{kitaev_anyons, BombinDelgado07HomologicalQEC, BullockBrennen07QUDIT_surface_code,DuclosCianci_Poulin13ToricCode_QUDITS},  quantum computation via state injection (rebits \cite{Delfosee14_Wigner_function_Rebits,Raussendorf15QubitQCSI,Veitch12_Negative_QuasiProbability_Resource_QC}) and other applications. The mathematical tools developed in the present work may therefore also have applications outside the realm of classical simulations of quantum circuits.

\subsection{Relationship to previous work}\label{PreviousWork_c1}

In \cite{VDNest_12_QFTs} it was proven that one can sample classically in poly-time the output distribution of any \emph{non-adaptive} normalizer circuit followed by a terminal measurement in the standard basis. Our work extends  this result in various ways, as outlined above in I-II-III. Main differences are the fact that here we consider adaptive normalizer circuits, and two different types of simulations: sampling output distributions and computation of amplitudes.

To our knowledge, ref.\ \cite{VDNest_12_QFTs} and the present work are the first studies to investigate normalizer circuits over arbitrary finite abelian groups, including those of the form $G=\mathbb{Z}_{d}^{m}$ where $d$ can be an \textit{exponentially} large number, such as $d=2^n$; they are also the first to consider normalizer operations that act on high-dimensional physical systems without a natural tensor product decomposition (such as \ $\C^{p}$ where $p> 2^{n}$ is an exponentially big prime number), or clusters of heterogeneous qudits (e.g. $\C^{a}\times \C^{b}\times\C^{c}$ when $a, b, c$ are different, as opposed to $\C^{d^{\otimes n}}$).

Restricting to groups of the form $G=\mathbb{Z}_{d}^{m}$ where $d$ is \textit{constant}, our work recovers previous results regarding classical simulations of Clifford circuits for \textit{qudits}. We emphasize that in this second scenario $d$ is a \textit{fixed} parameter that does not scale; this is in contrast with the cases studied in \cite{VDNest_12_QFTs} and in the present paper. We briefly summarize prior work on qudits.
 \begin{itemize}
\item Results when $d$ is a \textbf{constant prime} number: if $d = 2$, the ability to \textit{sample} classically efficiently follows from the Gottesman-Knill theorem \cite{Gottesman_PhD_Thesis,Gottesman99_HeisenbergRepresentation_of_Q_Computers}, whereas the computation of \textit{amplitudes} from \cite{dehaene_demoor_coefficients};  for prime values of $d$ larger than 2, techniques given in \cite{Gottesman98Fault_Tolerant_QC_HigherDimensions} yield efficient sampling simulations also for adaptive Clifford circuits.
\item Results when $d$ is an \textbf{arbitrary constant}: techniques given in  \cite{dehaene_demoor_hostens} can be used to simulate \textit{non}-adaptive Clifford circuits  followed by a terminal standard basis measurement (sampling output distributions and computation of amplitudes); tools developed in  \cite{deBeaudrap12_linearised_stabiliser_formalism} can be used to sample in the adaptive case.
\end{itemize}

Finally, our work also connects to previous studies on the simulability of abelian quantum Fourier transforms (QFTs) \cite{aharonov_AQFT,yoran_short_QFT,browne_QFT,VDNest_12_QFTs}. In \cite{aharonov_AQFT} it was shown that the action of the approximate QFT over $\mathbb{Z}_{2^n}$ on product states, followed by a terminal measurement in a product basis can be classically simulated  in \emph{quasi}-polynomial $O(n^{\log n})$ time. This result was improved in \cite{yoran_short_QFT}  where fully efficient classical simulation algorithms were given for this setting and, more generally, for constant-depth circuits of bounded interaction range, interspersed with a constant amount of approximate QFTs. In \cite{browne_QFT} it was shown that the ``semi-classical'' QFT  acting on a class of entangled input states can be efficiently classically simulated.  Finally, Van den Nest \cite{VDNest_12_QFTs} gave efficient classical algorithms for circuits of arbitrary size containing QFTs and normalizer gates.

A common ingredient in works  \cite{aharonov_AQFT,yoran_short_QFT,browne_QFT} is that they all employ  tensor contraction schemes in their simulations, which crucially depend  on the geometric \emph{structure} of the quantum circuit: in particular, all these methods can only be efficient if the graph representing the gate structure of the circuit has a strong tree-like structure (measured by the tree-width \cite{MarkovShi08_Simulating_QuantumComp_TensorNetwork}). Unlike the simulations given in \cite{VDNest_12_QFTs} and in the present work, the circuits in  \cite{aharonov_AQFT,yoran_short_QFT,browne_QFT} can only generate limited amounts of entanglement \cite{Yoran08_Contractable_circults_little_entanglement}. Also, the simulations in \cite{VDNest_12_QFTs} and in the present work  are fully \emph{independent} of the structure of the circuit. This generality  comes at the cost that we have to restrict our allowed gates, similarly to the original  Gottesman-Knill theorem.

\subsection{Chapter outline}

We refer the reader to chapter \ref{chapterC}-\ref{sect:Normalizer Gates Finite Group} for an introduction to the normalizer circuit model of this chapter, its relationship to the standard Pauli and Clifford operations and definitions of character, quadratic and bicharacter functions. The rest of the chapter is organized as follows.

 Sections \ref{sect_prelimin_fab} and \ref{section Pauli and Clifford and Normalizer} contain technical preliminaries. Section \ref{sect_prelimin_fab} presents a number of efficient classical algorithms to solve algebraic computational problems based on the classical techniques we developed in chapter \ref{chapterGT}. Section \ref{section Pauli and Clifford and Normalizer} gives a detailed account of the mathematical properties of Pauli, Clifford and (unitary) normalizer operations.

The remaining sections contain the main results of our work. In section \ref{section abelian group stabilizer formalism}, a theory of abelian-group stabilizer codes is developed. In section \ref{section Normal form of stabilizer states} we give normal forms for stabilizer states. Section \ref{section  Pauli measurements Implementation} explains how intermediate (generalized) Pauli operator measurements can be implemented, and how they transform abelian-group stabilizer states. In section \ref{section Gottesman-Knill theorem} we show how to simulate adaptive normalizer circuits classically and discuss the power of these operations for state preparation.

\section{Preliminaries on finite abelian groups \label{sect_prelimin_fab}}

\paragraph{Conventions and methodology:}Throughout this section we fix the group $G$ to be of the form
 \begin{equation}\label{G}
 G=\DProd{d}{m},
 \end{equation} with parameters $d_i$, $m$ chosen arbitrarily. The   elements  and \textit{canonical generators}\footnote{These elements play a similar role as the canonical basis vectors of vector spaces like $\R^{m}$ or $\C^{m}$ (though $G$ is \textit{not} a vector space).} of $G$ are denoted by $g=(g(1),\ldots, g(m)), g(i)\in\Z_{d_i}$, resp., $e_i=(0,\ldots,1_i,\ldots,0)$. The $m$ elements ${  e}_i$ generate $G$ and  for any   ${ g}\in G$ can be naturally written as ${  g} = \sum g(i) {  e}_i $. Throughout the section, we use the shorthand ($\bmod{G}$) as synonym of $(\bmod d_1, \ldots, \bmod d_m)$.

 The classical simulation and stabilizer formalism methods in this chapter exploit several of the classical group theoretic techniques that we develop in chapter \ref{chapterGT}. In particular, we will apply the notions  and main properties of  annihilator subgroup $H^\perp$ (section \ref{sect:Annihilators}), and character $\chi_g(h)$, quadratic $\xi(h)$ and bicharacter $B(g,h)$ functions (chapters \ref{chapterC}, \ref{sect:GTDefinitions},
\ref{sect:quadratic_functions})  to devise analytic tools to describe stabilizer states and codes. Furthermore, the theory of matrix representations for group homomorphisms that we developed in chapter will be useful to characterize normalizer gates and various linear structures present in the problems we study.

\paragraph{Computational group theory:} 
 Computational aspects of finite abelian groups are now discussed; our discourse focuses on a selected catalog of  computational problems relevant to this chapter and efficient classical algorithms to solve them.

In computational complexity theory a (classical or quantum) algorithm is said to be \textit{efficient} if it solves a given computational problem of input-size $n$ in (classical or quantum) \poly(n) time: when  one looks at problems related to finite abelian groups, this will be synonym of ``in \ppolylog{|G_1|,|G_2|,\ldots,|G_n|} time'', being $G_1,\ldots,G_n$ the groups involved in a problem of interest. 
 Since this section concerns only classical computational complexity, we will tend to omit the epithet \textit{classical} all the way throughout it.

Periodically, and at crucial stages of this chapter, some advanced algebraic computational problems are bound to arise. The following lemma compiles a list of group theoretical problems that will be relevant to us and can be solved efficiently by classical computers.
\begin{lemma}[\textbf{Algorithms for finite abelian groups}]\label{lemma:Algorithms_FA_Groups}
Given $H$, $K$, two subgroups of $G$, and $\{h_i\}$, $\{k_j\}$, polynomial-size generating-sets of them, there exist efficient classical algorithms to solve the following problems deterministically.
\begin{itemize*}
\item[(a)] Decide whether  $b\in G$ belongs to $H$;  if so, find integers $w_i$ such that $b=\sum w_ih_i$.
\item[(b)] Count the number of elements of $H$.
\item[(c)] Find a generating-set of the intersection $H\cap K$.
\item[(d)] Find a generating-set of the annihilator subgroup $H^{\perp}$ (cf.\ definition in chapter \ref{sect:Annihilators}).
\item[(e)] Given the system of equations $\chi_{h_i}(g)=\gamma^{a_i}$, find elements $(g_0, g_1, \ldots, g_s)$ such that all solutions can be written as linear combinations of the form $g_0+\sum v_i g_i$.
\item[(h)] Find a $r\times m$ matrix representation $\Omega$ of a homomorphism $\varpi:G\rightarrow \Z_{d}^r$ such that $H$ coincides with the kernel of $\varpi$ and $r, d$ have polynomial bit-size.
\end{itemize*}
\end{lemma} 
The  proof of the lemma is given in  appendix \ref{Appendix:A1}, where we prove the following statement.
\begin{lemma}\label{LEMMA REDUCTION TO SYSTEMS OF LINEAR EQUATIONS}
Problems (a-e) in lemma \ref{lemma:Algorithms_FA_Groups} are polynomial-time reducible to either counting or finding solutions of systems of equations of the form $\alpha(x)=Ax=b$; where $\alpha$ is a group homomorphism between two (canonically-decomposed) finite abelian groups, $\mbf{G}_{{sol}}$ and $\mbf{G}$, to which $x$, $b$ respectively belong and $A$ is a matrix representation of $\alpha$.
\end{lemma}
We recall that the system of equation in lemma \ref{LEMMA REDUCTION TO SYSTEMS OF LINEAR EQUATIONS} is a \emph{linear system over groups} in the sense of (chapter \ref{sect:Systems of linear equations over groups}), which can be solved efficiently with our classical algorithm in theorem \ref{thm:General Solution of systems of linear equations over elementary LCA groups}: the latter may be applied to count solutions and/or output an element $x_0$ and a poly-size generating set of $\ker{\alpha}$  such that $X_{sol}=x_0+\ker{\alpha}$ is the total number of solutions of the system.

\section{Pauli operators and normalizer circuits over abelian groups}\label{section Pauli and Clifford and Normalizer}

\subsection{Definitions and terminology}

We recall (section \ref{sect:Normalizer Gates Finite Group}) that a generalized Pauli operator over $G$ (hereafter often simply denoted \emph{Pauli operator}) is any  unitary operator of the form
\begin{equation}\label{eq:Pauli Operators DEFINITION}
\sigma(a,g,h) := \gamma^a Z(g)X(h), \quad X(g) := \sum_{h\in G} |h+g\rangle\langle h|, \quad  Z({g}) := \sum_{h\in G} \chi_{  g}({  h}) |{  h}\rangle\langle h|
\end{equation}
where $\chi_g$ is a character, $\gamma:= \euler^{\imun\uppi /|G|}$ is a primitive root of unity, and $a\in\mathbb{Z}_{2|G|}$.  Throughout this chapter, the triple $(a,g,h)$ describing the Pauli operator is called the \textit{label} of $\sigma$. It is important to observe that, although $\sigma$ is a $|G|\times|G|$ matrix, its label $(a,g,h)$ is  an \textit{efficient}  description of itself comprising $O(\log|G|)$ bits; from now on, we will specify Pauli operators in terms of their labels, and refer to the latter as the \textit{standard encoding} of these operators.

\subsection{Manipulation of Pauli operators}\label{sect:Manipulation_Paulis}

First, note that every Pauli operator factorizes as a tensor product relative to the tensor decomposition of $\mathcal{H}_G$ i.e.\ $\sigma$ can be written as $\sigma = U_1\otimes\cdots\otimes U_m$ where $U_i$ acts on $\mathbb{C}^{d_i}$. This property simplifies several proofs; it can be verified straightforwardly by applying (\ref{eq:Pauli Operators DEFINITION}) and the definition (\ref{Character Functions DEFINITION}) of the characters of $G$.

Basic manipulations of Pauli operators can be carried out transparently by translating them into transformations of their labels: we review now some of these rules. First, the Pauli matrices (\ref{eq:Pauli Operators DEFINITION}) obey the following commutation rules:
\begin{align}
X({  g})X({  h}) &= X({  g}+{  h}) = X({  h})X({  g}) \nonumber\\  \label{commutation relations Pauli operators} Z({  g})Z({  h})&=Z({  g}+{  h})=Z({  h})Z({  g}) \\
Z({  g}) X({  h}) &= \chi_{  g}({  h})X({  h})Z({  g}).\nonumber\end{align}
Combinations of these rules straightforwardly lead to the next two lemmas.
\begin{lemma}[\textbf{Products and powers of Pauli operators \cite{VDNest_12_QFTs}}]\label{pauli_products_powers PROPERTY}
Consider Pauli operators $\sigma$ and $\tau$ and a positive integer $n$. Then  $\sigma\tau$, $\sigma^n$ and $\sigma^{\dagger}$ are also Pauli operators, the labels of which can be computed in \ppolylog{|G|, n} time on input of $n$ and the labels of $\sigma$ and $\tau$. Moreover,  $\sigma^{\dagger}= \sigma^{2|G|-1}$.
\end{lemma}
\begin{lemma}[\textbf{Commutativity}]\label{Commutativity of Pauli Stabilizer Groups PROPOSITION}
Consider two Pauli operators $\sigma(a_1,g_1,h_1)=\sigma_1$ and $\sigma(a_2,g_2,h_2)=\sigma_2$. Then the following statements are equivalent:
\begin{itemize*}
\item[(i)] $\sigma_1$ and $\sigma_2$ commute;
\item[(ii)] $\chi_{g_1}(h_2)=\chi_{g_2}(h_1)$;
\item[(iii)] $x:=(g_1,h_1)$ and $y:=(h_2,-g_2)$ annihilate each other as elements of $G\times G$: i.e.\ $\chi_x(y)=1$.
\end{itemize*}
\end{lemma}
Lemma  \ref{pauli_products_powers PROPERTY} implies that the set of all Pauli operators ${\cal P}_G$ over $G$ forms a (finite) group, called the Pauli group (over $G$).

\subsection{Normalizer quantum circuits\label{section Standard encodings of normalizer circuits}}

 Hitherto we have not considered technical aspects of normalizer circuits, such as how to describe normalizer circuits efficiently, or how to compute their action on Pauli operators; we address these questions in this section.

\subsubsection{Describing normalizer operations}

In this chapter we will be interested in classical simulations of \textit{normalizer circuits}. To make meaningful statements about classical simulations one must first specify which \textit{classical descriptions} of normalizer circuits are considered to be available.  In the case of Pauli operators over $G$, we saw in the previous section that it is possible to describe them using few (\polylog{|G|}) memory resources, by choosing their labels $(a,g,h)$ as standard encodings; this property holds for \textit{all} normalizer gates and---hence---circuits \cite{VDNest_12_QFTs}: all of them admit efficient classical descriptions. This is discussed next.

\begin{itemize}
\item First, a partial quantum Fourier transform  is described by the set of systems $\mathcal{H}_{\Z_{d_i}}$ on which it acts non-trivially
\item Second, an automorphism gate is described by the \textit{matrix representation} of the associated automorphism (cf.\ chapter \ref{sect:Matrix Representations}).
\item Third, let $\xi$ be an arbitrary quadratic function. Then, it follows from our normal form for quadratic functions (theorem \ref{thm:Normal form of a quadratic function}) that there exists $n(g)\in\mathbb{Z}_{2|G|}$ such that $\xi(g) = \euler^{\pii n(g)/|G|}$ for every $g\in G$;  furthermore,  the $O(m^2)$ integers $n(e_i)$ and $n(e_i+e_j)$ comprise an efficient description of $\xi$ and, thus, of the associated quadratic phase gate.\footnote{Given these integers one can efficiently compute $M$ and  vector $v$ as in (\ref{eq:Normal Form Quadratic Function}) (cf.\ appendix \ref{sect:QuadraticPhaseMatrices}): it follows that $\xi$  can be efficiently computed given these numbers (see also \cite{VDNest_12_QFTs} for an earlier proof of this fact).}
\end{itemize}
Henceforth we will assume that all normalizer gates are specified in terms of the descriptions given above, which will be called their \emph{standard encodings}. The standard encoding of each type of gate comprises polylog$(|G|)$ bits. The standard encoding of a normalizer circuits is the sequence of classical descriptions of its gates.

\subsubsection{Normalizer  vs Clifford \label{section conjecture}}

The following theorem from \cite{VDNest_12_QFTs} states that every normalizer gate belongs to the Clifford group, and the action of any normalizer gate on a Pauli operator via conjugation can be described efficiently classically.
\begin{theorem}[{\bf Normalizer gates are Clifford \cite{VDNest_12_QFTs}}]\label{thm_G_circuit_fundamental}
Every normalizer gate is a Clifford operator. Furthermore let $U$ be a normalizer gate specified in terms of its standard classical encoding as above, and let $\sigma$ be a Pauli operator specified in terms of its label; then the label of $U \sigma U^{\dagger}$ can be computed in \polylog{\Inputsize} time.
\end{theorem}
\begin{proof} We do not reproduce the original proof of this theorem  since we present an infinite-dimensional generalization of it in chapter \ref{chapterI} (theorem \ref{thm:Normalizer gates are Clifford}). However, we illustrate here how the main types of normalizer gates $\mathcal{F}_{G}$, $U_{\alpha}$,  $D_{\xi} $ act  on Pauli operators $X(g)$, $Z(g)$ under conjugation:
\begin{eqnarray}\label{thm:NormalizerCliffordC1} \begin{array}{ccll}
\mathcal{F}_{G}&: & X({  g})\to Z({  g}); & Z({  g})\to X({  -g})\vspace{1mm}\\
U_{\alpha}&: & X({  g})\to X(\alpha({  g})); & Z({  g})\to Z(\alpha^{-*}({  g}))\vspace{1mm}\\
D_{\xi}& : & X({  g})\to \xi({  g}) X({  g}) Z(\beta({  g})) ; \ \  & Z({  g})\to Z({  g})
\end{array} \end{eqnarray}
Above,  $\beta:G\to G$ denotes the homomorphism in lemma  \ref{lemma:Normal form of a bicharacter 1},  $\alpha^*$ is the \emph{dual group automorphism} of $\alpha$ (\ref{eq:Dual Automorphism DEF});  and $\alpha^{-*}$ denotes the inverse of $\alpha^*$.
\end{proof}
It is unknown whether the entire Clifford group can be generated (up to global phase factors) by normalizer gates in full generality. However, it was proven in  \cite{dehaene_demoor_hostens} (see also the examples in section \ref{sect_examples}) that this is indeed the case for groups of the form $G=\mathbb{Z}_{d}^m$ (i.e.\ $m$ qudit systems); more strongly, every Clifford group element (over $\mathbb{Z}_{d}^m$) can be written as a product of at most polylog$(|G|)$ such operators. We \textit{conjecture} that this feature holds true for Clifford operators over \textit{arbitrary} finite abelian groups.
\begin{conjecture}\label{thm_conjecture} Let $G$ be an arbitrary (canonically decomposed) finite abelian group. Then, up to a global phase, every Clifford operator over $G$ can be written as a product of \polylog{\Inputsize} normalizer gates.
\end{conjecture}
Finally, in the following lemma we provide some partial support for this conjecture. We show that  both  automorphism gates and quadratic phase gates have a distinguished role within the Clifford group, characterized  as follows:
\begin{lemma}\label{lemma:PermutationNormalizer=Clifford}
Up to a global phase, every Clifford operator which acts on the standard basis as a permutation has the form $X(g)U_{\alpha}$ for some $g\in G$ and some automorphism gate $U_{\alpha}$. Every diagonal Clifford operator is, up to a global phase, a quadratic phase gate.
\end{lemma}
\begin{proof}
The first statement was proved in \cite{VDNest_12_QFTs}. We prove the second statement. Let $D = \sum \xi(g) |g\rangle\langle g|$ be a diagonal unitary operator (so that $|\xi(g)|=1$ for all $g\in G$) in the Clifford group. Without loss of generality we may set $\xi(0)=1$, which can always be ensured by choosing a suitable (irrelevant) overall phase. Then for every $h\in G$, $D$ sends $X(h)$ to a Pauli operator under conjugation. This implies that there exists a complex phase $\gamma(h)$ and group elements  $f_1(h), f_2(h)\in G$ such that \be\label{condition_diag_clifford} D X(h) D^{\dagger} = \gamma(h) X(f_1(h))Z(f_2(h)). \ee Since $D$ is diagonal, it is easy to verify that we must have $f_1(h)=h$ for every $h\in G$. Now consider an arbitrary $g\in G$. Then
\begin{eqnarray}
D X(h) D^{\dagger}|g\rangle &=& \overline\xi(g) \xi(g+h) |g+h\rangle;\label{quadratic_1}\\
\gamma(h) X(h)Z(f_2(h))|g\rangle &=&  \gamma(h) \chi_g(f_2(h)) |g+h\rangle \label{quadratic_2}.
\end{eqnarray}
Condition (\ref{condition_diag_clifford}) implies that (\ref{quadratic_1}) is identical to (\ref{quadratic_2}) for every $g, h\in G$. Choosing $g=0$ and using that $\xi(0)=1$ and $\chi_0(x)=1$ for every $x\in G$ it follows that $\gamma(h)=\xi(h)$. We thus find that \be\xi(g+h) = \xi(g)\xi(h) \chi_g(f_2(h)). \label{condition_diag_clifford2}\ee The function  $B(g, h):=\xi(g+h)\overline \xi(g)\overline\xi(h)$ is manifestly linear in $g$, since $B(g, h) = \chi_g(f_2(h))$. Furthermore by definition $B$ is symmetric in $g$ and $h$. Thus $B$ is also linear in $h$.
\end{proof}

\section{An abelian Group Stabilizer Formalism}\label{section abelian group stabilizer formalism}

 In this section we develop further the stabilizer formalism for finite abelian groups as started in \cite{VDNest_12_QFTs}. We provide new analytic and algorithmic tools to describe them and analyze their properties. Throughout this section we consider an arbitrary abelian  group of the form $G=\DProd{d}{m}$.

\subsection{Stabilizer states and codes}

 Let ${\cal S}$ be a subgroup of the Pauli group ${\cal P}_G$. Then ${\cal S}$ is said to be a stabilizer group (over $G$) if there exists a non-zero vector $|\psi\rangle\in \mathcal{H}_G$ which is invariant under all elements in ${\cal S}$ i.e.\ $\sigma|\psi\rangle = |\psi\rangle$ for every $\sigma\in{\cal S}$. The linear subspace ${\cal V}:= \{|\psi\rangle : \sigma|\psi\rangle = |\psi\rangle \mbox{ for all } \sigma\in {\cal S}\}$ is called the stabilizer code associated with ${\cal S}$. If ${\cal V}$ is one-dimensional, its unique element (up to a multiplicative constant) is called the stabilizer state associated with ${\cal S}$.  In this chapter we will mainly be interested in stabilizer states. Occasionally, however, it will be useful to consider the general setting of stabilizer codes (cf. e.g.\  theorem \ref{thm structure test}).

Note that every stabilizer group ${\cal S}$ is abelian. To see this, consider a state $|\psi\rangle\neq 0$ which is invariant under the action of all elements in ${\cal S}$ and consider two arbitrary $\sigma, \tau\in {\cal S}$. Then (\ref{commutation relations Pauli operators}) implies that there exists a complex phase $\alpha$ such that $\sigma\tau = \alpha \tau\sigma$.  It follows that $|\psi\rangle = \sigma\tau |\psi\rangle = \alpha\tau\sigma|\psi\rangle = \alpha|\psi\rangle$, where we have used that $\sigma|\psi\rangle= |\psi\rangle = \tau|\psi\rangle$. We thus find that $|\psi\rangle = \alpha|\psi\rangle$ so that $\alpha = 1$  (i.e.\ $\sigma $ and $\tau$ commute).

On the other hand, not every abelian subgroup of the Pauli group is a stabilizer group. A simple counterexample is the group $\{I, -I\}$ where $I$ is the identity operator acting on $\mathcal{H}_G$.

The support of a stabilizer code ${\cal V}$ is the set of all $g\in G$ for which $|g\rangle$ has a  nonzero overlap with ${\cal V}$ i.e.\ there exists $|\psi\rangle\in {\cal V}$ such that $\langle g|\psi\rangle\neq 0$. The support of a stabilizer state $|\phi\rangle$ is simply the set of all $g\in G$ for which $\langle g|\phi\rangle\neq 0$.

\subsection{Label groups\label{sect_label_groups}}

 Let $\mc{S}$ be a stabilizer group over $G$. The diagonal subgroup $\mc{D}$ is the subgroup of $\mc{S}$ formed by its diagonal operators i.e.\ it  consists of all operators in ${\cal S}$ of the form $\gamma^a Z(g)$.
Second, we introduce two subgroups $\mathbb{H}$ and  $\mathbb{D}$ of $G$ called the label groups of $\mc{S}$:
\begin{align}
\label{Label group H}\mathbb{H} &= \{ h\in G \:\colon\: \textnormal{there exists $\gamma^aZ(g)X(h)\in\mc{S}$} \},\\
\label{Label group D}\mathbb{D} &= \{ g\in G \:\colon\: \textnormal{there exists $\gamma^aZ(g)\in\mc{D}$} \},
\end{align}
Using  (\ref{commutation relations Pauli operators}) it is straightforward to verify that $\mathbb{D}$ is indeed a subgroup of $G$. To prove that $\mathbb{H}$ is a subgroup as well, one argues as follows. Let $\sigma$ be a Pauli operator with label $(a, {  g}, {  h})$. We call $g$ the ``$Z$-component'' and $h$ the ``$X$-component'' of $\sigma$. Denote the $X$-component formally by $\varphi(\sigma):= {  h}$.  Then $\mathbb{H}$ is the image of ${\cal S}$ under the map $\varphi$.  The commutation relations (\ref{commutation relations Pauli operators})   yield \be\label{X_component} \varphi(\sigma\tau)= \varphi(\sigma) + \varphi(\tau)\quad\mbox{ for all } \sigma, \tau \in {\cal S}.\ee
This implies that $\varphi$ is a homomorphism from ${\cal S}$ to $G$. It follows that $\mathbb{H}$ is a subgroup of $G$.
\begin{lemma}[\textbf{Label groups}]\label{label groups of an stabilizer DEFINITION}
Let $\mc{S}$ be a stabilizer group and assume that the labels of $k=$ \polylog{\Inputsize} generators of $\mc{S}$ are given as an input. Then the label groups of ${\cal S}$ fulfill:
\begin{enumerate*}
\item[(i)] $\mathbb{H} \subseteq \mathbb{D}^{\perp}$, where $\mathbb{D}^\perp$ denotes the annihilator of $\mathbb{D}$ (section \ref{sect:Annihilators});
\item[(ii)] Generating sets of $\,\mathbb{H}$, $\mathbb{D}$ can be efficiently computed classically;
\item[(iii)] The labels of a generating set of $\mc{D}$  can be efficiently computed classically.
\end{enumerate*}
\end{lemma}
\begin{proof}
Property (i) is a straightforward consequence of the commutation relations given in lemma  \ref{Commutativity of Pauli Stabilizer Groups PROPOSITION} and the definition of annihilator subgroup (\ref{orthogonal group EQUATION}).
To show property (ii), recall that the map $\varphi$ defined above is a homomorphism from ${\cal S}$ to $G$ with $\mathbb{H}= $ Im$(\varphi)$. Suppose that ${\cal S}$ is generated by $\{\sigma_1, \ldots, \sigma_k\}$.  Then $\mathbb{H}$ is generated by $\{\varphi(\sigma_1), \ldots, \sigma(\sigma_k)\}$: this yields an efficient method to compute generators of $\mathbb{H}$. To prove the second statement of (ii) as well as (iii) requires more work. The argument is a direct generalization of the proof of lemma 9 in \cite{VDNest_12_QFTs} and the reader is referred to this work.
\end{proof}

\subsection{Certificates}

 The main purpose of this section is to provide a criterion to verify when a stabilizer group gives rise to a one-dimensional stabilizer code i.e.\ a stabilizer state. This is accomplished in corollary \ref{corollary_Uniqueness_Test}. To arrive at this statement we first analyze how the dimension of a general stabilizer code is related the structure of its stabilizer group.
\begin{theorem}[\textbf{Structure Test}]\label{thm structure test}
Let ${\cal S}$ be a stabilizer group with stabilizer code $\mc{V}$ and  $\mathbb{D}^\perp$ be the annihilator of the label subgroup $\mathbb{D}$ (section \ref{sect:Annihilators}). Then, there exists $g_0\in G$ such that
\begin{equation}
(i) \textnormal{ supp}(\mc{V})= g_0 + \mathbb{D}^{\perp}, \qquad\qquad (ii) \textnormal{ dim}(\mc{V})  = \frac{|\mathbb{D}^{\perp}|}{|\mathbb{H}|},
\end{equation}
where $\mathbb{H}$, $\mathbb{D}$ are the label subgroups of ${\cal S}$. Furthermore, there exist efficient classical algorithms to compute a representative $g_0$ of the support, a generating set of $\mathbb{D}^{\perp}$ and the dimension $\textnormal{dim}(\mc{V})$.
\end{theorem}
Before proving theorem \ref{thm structure test}, we note that combining property (ii) together with lemma \ref{label groups of an stabilizer DEFINITION}(i) immediately yield:
\begin{corollary}[\textbf{Uniqueness Test}]\label{corollary_Uniqueness_Test}
Let ${\cal S}$ be a stabilizer group with stabilizer code ${\cal V}$. Then  $\mc{V}$ is one-dimensional if and only if  $\mathbb{H}$ and $\mathbb{D}$ annihilate each other: i.e.\ iff  $\mathbb{H}=\mathbb{D}^{\perp}$.
\end{corollary}
Theorem \ref{thm structure test}(ii) also leads to an alternative formula for the dimension of a stabilizer code:
\begin{corollary}\label{corollary dimension of stabilizer code}
The dimension of ${\cal V}$ equals $|G|/|{\cal S}|$.
\end{corollary}
The result in corollary \ref{corollary dimension of stabilizer code} is well known for stabilizer codes over qubits \cite{Gottesman_PhD_Thesis, nielsen_chuang} (i.e.\ where $G = \mathbb{Z}_2^m$ so that $\Inputsize=2^m$) and qudits (where $G=\mathbb{Z}_d^m$) \cite{Gottesman_PhD_Thesis,gheorghiu11Qudit_Stabilisers}.
\begin{proof}{[of corollary \ref{corollary dimension of stabilizer code}]}
Consider the  map $\varphi: {\cal S}\to G$, defined in section \ref{sect_label_groups}, which is a group homomorphism with image $\mathbb{H}$. Furthermore the kernel of $\varphi$ is precisely the diagonal subgroup ${\cal D}$ of $G$. Since $|\mbox{Im } \varphi| = |{\cal S}|/|\mbox{ker }\varphi|$ it follows that $|\mathbb{H}| = |{\cal S}|/|{\cal D}|$. Finally we claim that ${\cal D}$ and $\mathbb{D}$ are isomorphic groups so that $|{\cal D}| = |\mathbb{D}|$. To prove this,  consider the map $\delta: {\cal D}\to \mathbb{D}$ that sends $\sigma = \gamma^a Z(g)$ to $\delta(\sigma)= g$. Using (\ref{commutation relations Pauli operators}) it follows that this map is a homomorphism; furthermore, it is a surjective one by definition of $\mathbb{D}$, and thus $\textnormal{im}\delta =\mathbb{D}$. The kernel of $\delta$ is the set of all $\sigma\in {\cal S}$ having the form $\sigma = \gamma^a I$. But the only operator in ${\cal S} $ proportional to the identity is the identity itself, since otherwise ${\cal S}$ cannot have a common $+1$ eigenstate. This shows that the kernel of $\delta$ is trivial, so that ${\cal D}$ and $\mathbb{D}$ are isomorphic, as claimed. The resulting identity $|\mathbb{H}| = |{\cal S}|/|\mathbb{D}|$ together with $|\mathbb{D}^{\perp}| = |G|/|\mathbb{D}|$ (recall lemma \ref{lemma:Annihilator properties}) and theorem \ref{thm structure test}(ii) proves the result.
\end{proof}
We now prove theorem \ref{thm structure test} using  techniques developed in \cite{nest_MMS} where the properties of so-called M-spaces were studied. We briefly recall basic concepts and results.

A unitary operator acting on $\mathcal{H}_G$ is said to be \emph{monomial} if it can be written as a product $U = DP$ where $D$ is diagonal and $P$ is a permutation matrix. A subspace ${\cal M}$ of $\mathcal{H}_G$  is called an \emph{M-space} if there exists a group of monomial unitary operators ${\cal G}$ such that $|\varphi\rangle\in {\cal M}$ iff $U|\varphi\rangle = |\varphi\rangle$ for every $U\in {
\cal G}$. The group ${\cal G}$ is called a stabilizer group of ${\cal M}$. If ${\cal M}$ is one-dimensional, its unique (up to a multiplicative factor) element $|\psi\rangle$ is called an M-state.  The support of ${\cal M}$ is defined analogously to the support of a stabilizer code i.e.\ it is the set of all $g\in G$ such that $|g\rangle$ has a nontrivial overlap with ${\cal M}$. With this terminology, every stabilizer code is an instance of an M-space and every stabilizer state is an M-state. To see this, note that every Pauli operator $\sigma(a, g, h)$ is a monomial unitary operator. Indeed, $\sigma$ can be written as a product $\sigma = DP$ where $D = \gamma^aZ(g)$ is diagonal and $P = X(h)$ is a permutation matrix.

We introduce some further terminology. Let ${\cal G}$ be an arbitrary monomial stabilizer group. For every $g\in G$, let $\mc{G}_{g}$ be the subset of $\mc{G}$ consisting of all $U\in \mc{G}$ satisfying $U|g\rangle\propto |g\rangle$ i.e.\ $U$ acts trivially on $g$, up to an overall phase. This subset is easily seen to be a subgroup of $\mc{G}$. Also, we define the orbit $\mc{O}_g$ of $g$ as:
\begin{equation}\label{orbit DEFINITION} \mc{O}_g = \{h: \exists U\in\mc{G}\mbox{ s.t. } U|g\rangle\propto|h\rangle\}
\end{equation}
In the following  result the support of any M-space is characterized in terms of the orbits ${\cal O}_g$ and the subgroups ${\cal G}_g$.
\begin{theorem}[\textbf{Support of M-space \cite{nest_MMS}}]\label{thm_support_M_space}
Consider an M-space ${\cal M}$ with monomial stabilizer group $\mc{G}$. Then the following statements hold:
\begin{itemize*}
\item[(i)] There exist orbits $\mc{O}_{g_1},\ldots,\mc{O}_{g_\mbf{d}}$ such that $\mbf{d}= \dim({\cal M})$ and \be \textnormal{supp}({\cal M})= {\cal O}_{g_1}\cup\cdots\cup {\cal O}_{g_\mbf{d}}.\ee
\item[(ii)] Consider $g\in G$ and an arbitrary set of generators $\{V_1, \ldots,V_r\}$ of $\mc{G}_g$. Then  $g\in$ supp(${\cal M}$) if and only if $V_{ i}|g\rangle = |g\rangle$ for every $i$.
\end{itemize*}
\end{theorem}
Using this result, we can now prove theorem \ref{thm structure test}.
\begin{proof}\textbf{[of theorem \ref{thm structure test}]}
We apply theorem \ref{thm_support_M_space} to the Pauli stabilizer group $\mc{S}$. In this case, the group $\mc{S}_g$ and the orbit $\mc{O}_g$ fulfill
\begin{equation}\label{Action Orbits&Stabilizer for Pauli Stabilizer Groups}
\mc{O}_g=g+\mathbb{H}, \qquad\qquad \qquad \mc{S}_g = \mc{D}.
\end{equation}
To demonstrate the first identity in  (\ref{Action Orbits&Stabilizer for Pauli Stabilizer Groups}), we use  (\ref{eq:Pauli Operators DEFINITION}) which implies $\sigma(a,x,y)\ket{g}\propto\ket{g+y}$ for every $\sigma(a,x,y)\in\mc{S}$. To show the second identity, first note that $D|g\rangle\propto|g\rangle$ for every diagonal operator $D\in {\cal D}$, showing that ${\cal D}\subseteq {\cal S}_g$. Conversely, if $\sigma\in{\cal S}_g$  has label $(a, x, y)$ then $\sigma |g\rangle\propto |g+y\rangle$. Since $\sigma\in {\cal S}_g$ the state $|g\rangle$ is an eigenvector of $\sigma$; this can only be true if $y=0$, showing that $\sigma\in {\cal D}$.

Using lemma \ref{label groups of an stabilizer DEFINITION}, we can efficiently compute  the labels of a generating set $\{\sigma_1,\ldots,\sigma_r\}$  of $\mc{S}_g=\mc{D}$, where $\sigma_i = \gamma^{a_i} Z(g_i)$ for some $a_i\in\mathbb{Z}_{2|G|}$ and $g_i\in G$. Owing to theorem \ref{thm_support_M_space}(ii), any $g\in G$  belongs to the support of ${\cal V}$ if and only if $\sigma_i\ket{g} = \ket{g}$ for every $i=1,\ldots, r$. Equivalently, $g$ satisfies
\begin{equation}\label{Support of a Stabilizer Code Characteristic EQUATIONS}
\gamma^{a_i}\chi_{g_i}(g)=1 \quad \textnormal{for all} \; i =1,\ldots, r.
\end{equation}
Since the elements $g_i$ generate the label group $\mathbb{D}$, the solutions of the system are easily seen---use the multiplicativity of characters and (\ref{orthogonal group EQUATION})---to form a coset of the form $\textnormal{supp}(\mc{V})=g_0+\mathbb{D}^{\perp}$ for some particular solution $g_0$. Moreover, the classical algorithm in lemma \ref{LEMMA REDUCTION TO SYSTEMS OF LINEAR EQUATIONS}.(e), returns a valid $g_0$ and a generating set of $\mathbb{D}^\perp$, showing  (i).

Further, we combine (i) with theorem \ref{thm_support_M_space}(i) to get a short proof of (ii): the equation
\begin{equation*}
\textnormal{supp}(\mc{V})= \mc{O}_{g_1}\cup\cdots\cup \mc{O}_{g_{\mbf{d}}}=\left(g_1+\mathbb{H}\right)\cup\cdots\cup (g_{\mbf{d}}+\mathbb{H})= g_0+\mathbb{D}^{\perp}
\end{equation*}
implies, computing the cardinalities of the sets involved, that $\textbf{d}|\mathbb{H}|=\dim{\mc{V}}|\mathbb{H}|=|\mathbb{D}^{\perp}|$.

Finally, the ability to compute $g_0$ and to find generators of $\mathbb{D}^{\perp}$ efficiently classically follows by applying theorem \ref{thm:General Solution of systems of linear equations over elementary LCA groups} to a linear system  described by a $r\times m$ matrix  $\Omega$ that defines a  homomorphism from $G$ to $\Z_{|G|}^{r}$, with  $r\in O(\polylog{\Inputsize})$. Furthermore, we can compute $\dim \mc{V}$ directly using formula (ii) together with lemma \ref{label groups of an stabilizer DEFINITION} and the algorithms of lemma \ref{lemma:Algorithms_FA_Groups}.
\end{proof}

\section{Normal form of a stabilizer state \label{section Normal form of stabilizer states}}

We now apply the stabilizer formalism of section \ref{section abelian group stabilizer formalism} to develop an analytic characterization of the amplitudes of arbitrary stabilizer states over finite abelian groups. In addition, we show that the wavefunction of any stabilizer state can always be efficiently computed and sampled, which we use later to study the classical simulability of normalizer circuits.
\begin{theorem}[\textbf{Normal form of stabilizer states}]\label{thm Normal form of an stabilizer state}
Every stabilizer state $\ket{\phi}$ over a finite abelian group $G$ with stabilizer group $\mc{S}$  has the form
\begin{equation}\label{Normal form of an stabilizer state EQUATION}
\ket{\phi}=\alpha \frac{1}{\sqrt{|\mathbb{H}|}}\sum_{h\in \mathbb{H}}\xi(h)\ket{s+h}.
\end{equation}
Here $\alpha$ is a global phase, $\mathbb{H}$ is the label group (\ref{Label group H}), $s\in G$, and relative phases are described by a quadratic function $\xi$ on the group $\mathbb{H}$.  Furthermore, if a generating set $\{\sigma_1,\ldots,\sigma_r\}$ of $\mc{S}$ is specified, the following tasks can be carried out efficiently:

(a) Compute $s$;

(b) Given $g\in G$, determine if $g\in s + \mathbb{H}$;

(c) Given $h\in \mathbb{H}$, compute $\xi(h)$ up to $n$ bits in $\ppoly{n, \log |G|}$ time;

(d) Compute $\sqrt{|\mathbb{H}|}$.
\end{theorem}
\begin{proof}
Corollary \ref{corollary_Uniqueness_Test} implies that $\mathbb{D}^{\perp} = \mathbb{H}$. Using this identity together with theorem \ref{thm structure test}(i), we find that supp($|\phi\rangle) = s+ \mathbb{H}$ for some $s\in G$. By definition of $\mathbb{H}$, for every $h\in\mathbb{H}$ there exists some element $\sigma(a,g,h)\in\mc{S}$. Using that $\sigma(a,g,h)|\phi\rangle = |\phi\rangle$ we then have
\begin{equation}\label{Quadratic Relative Phases DEFINITION}
\langle s + h | {\phi} \rangle=\langle s + h |\sigma(a,g,h)|{\phi} \rangle= {\gamma^{a}\chi_{s+h}(g)}\langle s |{\phi} \rangle
\end{equation}
This implies that $|\langle s + h | {\phi} \rangle| = |\langle s| {\phi} \rangle|$ for all $h\in \mathbb{H}$. Together with the property that supp($|\phi\rangle) = s+ \mathbb{H}$, it follows that $|\phi\rangle$ can be written as
\begin{equation}\label{uniform supperposition}
|\phi\rangle = \frac{1}{\sqrt{|\mathbb{H}|}} \sum_{h\in\mathbb{H}} \xi(h) |s+h\rangle
\end{equation}  for some complex phases $\xi(h)$. By suitably choosing an (irrelevant) global phase, w.l.o.g. we can assume that $\xi(0)=1$.

We now show that the function $ h\in H \to \xi(h)$ is  quadratic. Using (\ref{Quadratic Relative Phases DEFINITION}, \ref{uniform supperposition}) we derive \be\label{proof_relphases_are_quadratic0}\xi(h) = \sqrt{|\mathbb{H}|} \langle s+h|\phi\rangle = \sqrt{|\mathbb{H}|}{\gamma^{a}\chi_{s+h}(g)}\langle s |{\phi} \rangle = \gamma^{a}\chi_{s+h}(g)\xi(0)=\gamma^{a}\chi_{s+h}(g).\ee Since $\xi(h)$ by definition only depends on $h$, the quantity $\gamma^{a}\chi_{s+h}(g)$ only depends on $h$ as well: i.e.\ it is independent of $a$ and $g$. Now select $h_1, h_2\in \mathbb{H}$ and two associated stabilizer operators $\sigma_1=\sigma(a_1,g_1,h_1)$, $\sigma_2=\sigma(a_2,g_2,h_2)\in {\cal S}$. Then
\begin{align}
\xi(h_1 + h_2)&= \sqrt{|\mathbb{H}|}\langle s + h_1+h_2 | {\phi} \rangle\\
&= \sqrt{|\mathbb{H}|}\langle s + h_1+h_2 | \sigma_1 \sigma_2 | {\phi} \rangle \label{proof_relphases_are_quadratic1}\\
&= \sqrt{|\mathbb{H}|} \gamma^{a_1}\chi_{s+h_1+h_2}(g_1)\: \langle s + h_2 |  \sigma_2 | {\phi} \rangle \label{proof_relphases_are_quadratic2}\\
&= \left[ \gamma^{a_1}\chi_{s+h_1}(g_1) \right] \: \left[ \gamma^{a_2}\chi_{s+h_2}(g_2)\right]  \: \chi_{g_1}(h_2) \: \xi(0) \label{proof_relphases_are_quadratic3} \\ &= \xi(h_1) \xi(h_2) \chi_{g_1}(h_2)  \label{proof_relphases_are_quadratic4}
\end{align}
In (\ref{proof_relphases_are_quadratic1}) we used that $\sigma_1\sigma_2|\phi\rangle = |\phi\rangle$; in (\ref{proof_relphases_are_quadratic2}-\ref{proof_relphases_are_quadratic3}) we used the definitions of Pauli operators and the fact that $\sqrt{|\mathbb{H}|}\langle s|\phi\rangle = \xi(0)$; finally in (\ref{proof_relphases_are_quadratic4}) we used identity (\ref{proof_relphases_are_quadratic0}) and the fact that $\xi(0)=1$.  Now define $B(h_1,h_2)= \xi(h_1+h_2)\overline \xi(h_1) \overline \xi(h_2)$. We  claim that $B$ is a bicharacter function of $\mathbb{H}$.  To see this, note that the derivation above shows that $B(h_1, h_2)=\chi_{g_1}(h_2)$ for any $\sigma(a_1,g_1,h_1)\in\mathcal{S}$. Linearity in the second argument $h_2$ is immediate. Furthermore, by definition $B$ is a symmetric function i.e.\ $B(h_1,h_2) = B(h_2,h_1)$.  This shows that $B$ is bicharacter, as desired.

We now address (a)-(d). As for (a) recall that  $s+\mathbb{H}$ is the support of a stabilizer state  $|\phi\rangle$; theorem \ref{thm structure test} then provides an efficient method to compute a suitable representative $s$. Note also that a generating set of $\mathbb{H}$ can be computed efficiently owing to lemma \ref{label groups of an stabilizer DEFINITION}. Statement (b) follows from lemma \ref{lemma:Algorithms_FA_Groups}(a). Statement (d) follows from lemma \ref{lemma:Algorithms_FA_Groups}(b). Finally we prove (c), by showing that the following procedure to compute $\xi(h)$ is efficient, given any $h\in \mathbb{H}$:

\vspace{2mm}

(i) determine some element $\sigma \in {\cal S}$ such that $\sigma|s\rangle \propto |s+h\rangle$;

(ii) compute $\langle s+h|\sigma|s\rangle = \xi(h)$.

\vspace{2mm}

\noindent To achieve (i), it suffices to determine an arbitrary stabilizer element of the form $\sigma = \sigma(a,g,h)\in\mc{S}$. Assume that generators $\sigma_1=\sigma(a_1,g_1,h_1),\ldots,\sigma_r=\sigma(a_r,g_r,h_r)$ are given to us. We can then use algorithm (a) in lemma \ref{lemma:Algorithms_FA_Groups} to find integers $w_i$ such that $h=\sum w_i h_i$, for which $\sigma = \prod \sigma_i^{w_i}$ is an operator of form $\sigma(a,g,h)$ for some values of $a,g$---use (\ref{commutation relations Pauli operators}). Moreover, given the $w_i$ the label $(a,g,h)$ of $\sigma$ can be computed efficiently; this accomplishes (i). Finally, it is straightforward that (ii) can be carried out efficiently: using formula $\xi(h)=\gamma^{a}\chi_{s+h}(g)$ and standard algorithms to compute elementary functions \cite{brent_zimmerman10CompArithmetic}.
\end{proof}

\subsection{Reproduction of existing normal forms}\label{sect:ExistingNormalForms}

Theorem \ref{thm Normal form of an stabilizer state} generalizes result from \cite{dehaene_demoor_coefficients,dehaene_demoor_hostens,Gross06_discrete_Hudson_theorem} where analogous characterizations were given for qubits and qudits, although those works do not consider the notion of quadratic functions used here (furthermore their methods are completely different from ours). For example, in ref.\ \cite{dehaene_demoor_coefficients} it was shown that every Pauli stabilizer state for qubits (corresponding to the group $\mathbb{Z}_2^m$) can be written as \be |\phi\rangle \propto \frac{1}{\sqrt{|S|}} \sum_{x\in S} (-1)^{q(x)} i^{l(x)} |x+s\rangle.\ee Here $S$ is a linear subspace of $\mathbb{Z}_2^m$, $q(x) = x^T Ax\mod{2}$ is a quadratic form over $\mathbb{Z}_2$, and $l(x)\mod{2}$ is a linear form. This characterization indeed conforms with theorem \ref{thm Normal form of an stabilizer state}: the set $S$ is a subgroup of $\mathbb{Z}_2^m$ and the function \be x\in \mathbb{Z}_2^m \to \xi(x):= (-1)^{q(x)} i^{l(x)}\ee is quadratic (chapter \ref{sect_examples}).

\subsection*{Application: computational tractability of stabilizer states}

Theorem \ref{thm Normal form of an stabilizer state} also implies that every stabilizer state belongs to the family of Computationally Tractable states (CT states) \cite{nest_weak_simulations}. A state $|\psi\rangle = \sum \psi_g|g\rangle \in \mathcal{H}_G$ is said to be CT (relative to its classical description) if the following properties are satisfied:
\begin{itemize}
\item[(a)] there exists an efficient randomized classical algorithm to sample the distribution $\{|\psi_g|^2\}$;
\item[(b)] given $g\in G$, the coefficient $\psi_g$ can be computed efficiently with exponential precision.
\end{itemize}
CT states form a basic component in a general class of quantum computations that can be simulated efficiently classically using probabilistic simulation methods. For example consider a quantum circuit ${\cal C}$ acting on a CT state and followed by a final standard basis measurement on one of the qubits. Then, regardless of which CT state is considered, such computation can be efficiently simulated classically when ${\cal C}$ is e.g.\ an arbitrary Clifford circuit, matchgate circuit, constant-depth circuit or sparse unitary. See \cite{nest_weak_simulations} for an extensive discussion of classical simulations with CT states.

Here we show that every stabilizer state $|\psi\rangle\in \mathcal{H}_G$ over a finite abelian group $G$ is CT. To be precise, we prove that such states are CT \textit{up to a global phase}. That is, instead of (b) we prove a slightly weaker statement which takes into account the fact that any stabilizer state specified in terms of its stabilizer is only determined up to an overall phase. Formally, we consider the property
\begin{itemize}
\item[(b')] there exists an efficient classical algorithm that, on input of $g\in G$, computes a coefficient $\psi_g'$, where the collection of coefficients $\{\psi_g':g\in G\}$ is such that $|\psi\rangle = \alpha \sum \psi_g'|g\rangle$ for some complex phase $\alpha$.
\end{itemize}
\begin{corollary}\label{corollary:CT}
Let $|\psi\rangle$ be a stabilizer state over an abelian group $G$, specified in terms of a generating set of \polylog{\Inputsize} stabilizers. Then $|\psi\rangle$ is CT in the sense (a)-(b').
\end{corollary}
\begin{proof}
Property (a) was proved in \cite{VDNest_12_QFTs}. To prove (b'), note that theorem  \ref{thm Normal form of an stabilizer state} implies there exists a global phase $\alpha$ such that \be \langle g|\psi\rangle = \left\{ \begin{array}{cl}\alpha \cdot \frac{1}{\sqrt{|\mathbb{H}|}}\cdot  \xi(h) & \mbox{ if } g=s+h \mbox{ for some } h\in \mathbb{H} \\ 0 & \mbox{ if } g\notin \mathbb{H}+ s.  \end{array}\right.\ee
Using theorem \ref{thm Normal form of an stabilizer state}(b) it can be efficiently determined whether $g$ belongs to $\mathbb{H}+s$. If not, then $\langle g|\psi\rangle=0$. If yes, then compute $h=g-s$; then $\xi(h)$ can be computed owing to theorem  \ref{thm Normal form of an stabilizer state}(c). Finally, $\sqrt{|\mathbb{H}|}$ can be computed owing to theorem  \ref{thm Normal form of an stabilizer state}(d).
\end{proof}

\section{Pauli measurements in the stabilizer formalism}\label{sect:PauliMeauserementsC1}

The rest of this chapter is investigates normalizer circuits that contain intermediate measurements of generalized Pauli operators. In this section we show how generalized Pauli measurements can be implemented and neatly described within our stabilizer formalism over abelian groups (cf.\ section \ref{section abelian group stabilizer formalism}). The main result of this section (\textbf{theorem \ref{thm_Measurement_Update_rules}}) is an update-rule for describing the output state after measuring any generalized Pauli operator on an abelian-group stabilizer state. We  use these tools to probe the classical simulability of adaptive normalizer circuits in section \ref{section Gottesman-Knill theorem}.

\subsection{Definition}

 Associated with every Pauli operator $\sigma$ 
(\ref{eq:Pauli Operators DEFINITION}) we will consider a quantum measurement in the eigenbasis\footnote{Recall that $\sigma$ is not hermitian but still unitary, hence, diagonalizable.} of $\sigma$.  Consider the spectral decomposition $\sigma = \sum \lambda P_{\lambda}$ where $\lambda\in\C$ are the distinct eigenvalues of $\sigma$ and $P_{\lambda}$ is the projector on the eigenspace associated with eigenvalue $\lambda$.
Given a state $|\psi\rangle\in \mathcal{H}_G$, the measurement associated with $\sigma$ is now defined as follows: the possible outcomes of the measurement are labeled by the eigenvalues $\{{\lambda}\}$ where each $\lambda$ occurs with probability $\|P_{\lambda}|\psi\rangle\|^2$; furthermore, if the outcome $\lambda$ occurs, the state after the measurement  equals to $P_{\lambda}|\psi\rangle$ up to normalization.

Consider a group $G$ of the form (1), with associated physical system $\mathcal{H}^{G}=\C^{d_1}\otimes\cdots\otimes\C^{d_m}$. We remark that a measurement of the $i$-th system $\mathbb{C}^{d_i}$ in the standard basis $\{|0\rangle, \ldots, |d_i-1\rangle\}$ can be realized as a measurement of a suitable Pauli operator, for every $i$ ranging from 1 to $m$. To keep notation simple, we demonstrate this statement for the special case $G=\Z_{d}^{m}$, yet the argument generalizes straightforwardly to arbitrary $G$. Denote by ${e}_i\in G$ the group element which has $1\in\mathbb{Z}_{d}$ in its $i$-th component and zeroes elsewhere. Then definition (\ref{eq:Pauli Operators DEFINITION}) implies that the Pauli operator $Z(e_i)$ acts as $Z_{d}$ on the $i$-th qudit and as the identity elsewhere, where $Z_d$ was defined in (\ref{X_Z_qudits}). Note that $Z_d$ has $d$ distinct eigenvalues, each having a rank-one eigenprojector $|x\rangle\langle x|$ with $x\in\mathbb{Z}_d$. It follows straightforwardly that measurement of $Z(e_i)$ corresponds to measurement of the $i$-th qudit in the standard basis.

\subsection{Implementation\label{section  Pauli measurements Implementation}}

 It is easily verified that every Pauli operator $\sigma$ can be realized as a poly-size (unitary) quantum circuit \cite{VDNest_12_QFTs}. Therefore, measurement of $\sigma$ can be implemented efficiently on a quantum computer using standard phase estimation methods \cite{nielsen_chuang}.  Here we provide an alternate method. In particular we show that every Pauli measurement can be implemented using \emph{only} normalizer circuits and measurements in the standard basis, which will be a useful ingredient in our proof of theorem \ref{thm_main}. To this end, we will use the following result from \cite{Nielsen02UniversalSimulations}.
\begin{lemma}[\textbf{\cite{Nielsen02UniversalSimulations}}] \label{lem_Nielsen02UniversalSimulations} For any dimension $d$ and for integers $j$ and $k$ such that $j,k \in \Z_d$, there exists a poly-size normalizer circuit $\mc{C}$ over the group $\Z_d$ that transforms $Z(j) X(k)$ into a diagonal Pauli operator of the form $\gamma^{a} Z(\gcd(j,k))$. Furthermore, there are efficient classical algorithms to compute a description of $\mc{C}$.
\end{lemma}
\begin{corollary}\label{PEG lemma for abelian groups}
Consider a Pauli operator $\sigma$ over an arbitrary finite abelian group $G$. Then there exists a poly-size normalizer circuit $\mc{C}$ over $G$ such that ${\cal C}\sigma{\cal C}^{\dagger} =\gamma^{a}Z(g)$. Furthermore, there are efficient classical algorithms to compute a description of $\mc{C}$ as well as  $\gamma^a$, a and $g$.
\end{corollary}
\begin{proof}
To compute $\mc{C}$ note that every Pauli operator over $G$ has the form $\sigma \propto U_1\otimes\cdots\otimes U_m$ where $U_i$ is a Pauli operator over $\mathbb{Z}_{d_i}$ and apply lemma \ref{lem_Nielsen02UniversalSimulations} to each factor. The rest follows by applying theorem \ref{thm_G_circuit_fundamental} to compute the label of $\mc{C}\sigma{\cal C}^{\dagger}$ and, in the case of $\gamma^{a}$, by using standard algorithms to compute scalar exponentials.
\end{proof}
Lemma \ref{lem_Nielsen02UniversalSimulations} and corollary \ref{PEG lemma for abelian groups} reduce the problem of measuring general Pauli operators to that of implementing measurements of $Z(g)$. Indeed,  given an arbitrary $\sigma$ to be measured, we can always compute a poly-size normalizer circuit that transforms it into a diagonal operator $\gamma^{a}Z(g)$, using corollary \ref{PEG lemma for abelian groups}.  Then, the measurement of $\sigma$ is equivalent to the procedure (a) apply ${\cal C}$; (b) measure $\gamma^{a}Z(g)$; (c) apply ${\cal C}^{\dagger}$. Finally, Pauli operators that are proportional to each other define the \textit{same} quantum measurement, up to a simple relabeling of the outcomes. Therefore it suffices to focus on the problem of measuring an operator of the form $Z(g)$.

Note now that, by definition, the eigenvalues of $Z(g)$ have the form $\chi_{g}(h)$. Define the following function $\omega$ from $G$ to $\Z_d$, where  $d=\textnormal{lcm}(d_1,\ldots,d_m)$:

\begin{equation}\label{Pauli Measurement - Labelling Function}
\omega(h)= \sum_{i}\, \frac{d}{d_i}\, g(i) h(i) \mod d.
\end{equation}
With this definition one has $\chi_{g}(h) = \euler^{2\pii \omega(h)/d}$. Given any $y\in\mathbb{Z}_d$, the eigenspace of $Z(g)$ belonging to the eigenvalue $\lambda= \euler^{2\pii y/d}$ is spanned by all standard basis states $|h\rangle$ with $\omega(h)=y$.

Next, note that $\omega$ is a group homomorphism from $G$ to $\Z_d$ due to lemma \ref{lemma:Normal form of a matrix representation}. As a result, the controlled operation $f(h,a)= (h,a+\omega(h))$ is a group automorphism of $G\times \Z_d$ and it can be implemented by a normalizer gate $U_f\ket{h,a}=\ket{h,a+\omega(h)}$.

The gate  $U_f$ can now be used to measure $Z(g)$, with a routine inspired by the coset-state preparation method used in the standard quantum algorithm to solve the abelian hidden subgroup problem \cite{lomont_HSP_review, childs_lecture_8}: first, add an auxiliary  $d$-dimensional system $\C^d$ in the state $\ket{0}$ to $\mathcal{H}^{G}$, the latter being in some arbitrary state $\ket{\psi}$; second, apply the global interaction $U_f$; third, measure the ancilla in the standard basis. The global evolution of the system along this process is
\begin{equation*}\label{implementing Pauli measurement Equation}
\ket{\psi}\ket{0}=\sum_{h\in G}\psi(h)\ket{h}\ket{0}\xrightarrow{U_f}\sum_{h\in G}\psi(h)\ket{h}\ket{\omega(h)} \xrightarrow{\textnormal{Measure} } \: \frac{1}{\sqrt{p_y}}\: \left(\sum_{h: \omega(h)=y}\psi(h)\ket{h}\ket{y}\right)
\end{equation*}
The measurement yields an outcome $y\in\mathbb{Z}_d$ with probability $p_y=\sum_{h: \omega(h)=y}|\psi(h)|^2$. The latter precisely coincides with $\|P_{\lambda}\ket{\psi}\|^{2}$ where $P_\lambda$ is the eigenprojector associated with the eigenvalue $\lambda = \euler^{2\pii y/d}$ and, therefore, we have implemented the desired measurement.

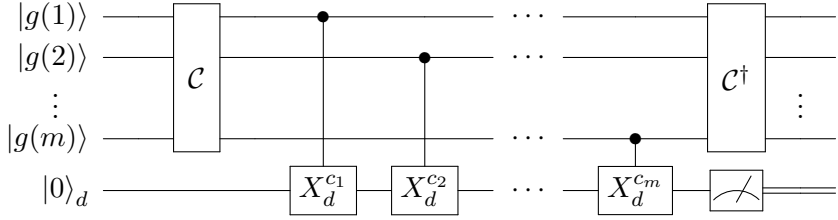
\begin{figure}[htbp]
\begin{equation*}
 \Qcircuit @C=1.2em @R=.5em {
  \lstick{\ket{g(1)}}    & \qw & \multigate{3}{\mc{C}}&  \qw & \ctrl{4} & \qw        & \qw & \cdots & & \qw                & \multigate{3}{\mc{C}^{\dagger}} & \qw    & \qw  \\
  \lstick{\ket{g(2)}}    & \qw & \ghost{\mc{C}}  &  \qw & \qw      & \ctrl{3}   & \qw & \cdots & & \qw                & \ghost{\mc{C}^{\dagger}}     & \qw & \qw   \\
  \lstick{\vdots \quad } &    &   \pureghost{\mc{C}}&   &          &            &     &        & &                    & \pureghost{\mc{C}^{\dagger}}    & \vdots &  &     \\
  \lstick{\ket{g(m)}}    & \qw & \ghost{\mc{C}}  & \qw & \qw      & \qw        & \qw & \cdots & & \ctrl{1}           & \ghost{\mc{C}^{\dagger}}     & \qw  &\qw\\	
  \lstick{\ket{0}_d} & \qw &\qw  & \qw & \gate{X_d^{c_1}} & \gate{X_d^{c_2}} & \qw & \cdots & & \gate{X_d^{c_m}} & \meter & \cw & \cw
 }
\end{equation*}
\caption{Quantum circuit implementing measurement of operator $\sigma=\mathcal{C} Z(g) \mathcal{C}^{\dagger}$ }
\label{figure Quantum circuit for Pauli measurements}
\end{figure}

In figure \ref{figure Quantum circuit for Pauli measurements} we show a poly-size quantum circuit that implements the measurement of the Pauli operator $\sigma=\mc{C}Z(g)\mc{C}^{\dagger}$ in the way just described. In the picture, the $m+1$ horizontal lines represent the $m$ physical subsystems that form $\mathcal{H}^{G}={\C^{d_1}\otimes\cdots\otimes\C^{d_m}}$ and the ancillary system ${\C^{d}}$; the numbers $c_i:=d/d_i\, g(i)$ are chosen to compute the function (\ref{Pauli Measurement - Labelling Function}) in the ancillary system. For merely pictorial reasons, the depicted measurement acts on a standard-basis state.
We now make two remarks. First, the state of the ancilla could be reset (with Pauli gates) to its original value once the measurement outcome $\omega(x)$ is recorded; this could be used to implement a series of measurements using only one ancilla. Second, the value $\omega(x)$ can be used to compute $\lambda=\chi_{g}(x)=\euler^{2\pii \omega(x)/d }$.

Finally we mention that a procedure given in \cite{deBeaudrap12_linearised_stabiliser_formalism} to implement measurements of qudit Pauli operators (as presented in section \ref{Examples Paulis/Normalizer for qudits}) can be recovered from ours by choosing $G=\Z_d^{m}$.

\subsection{Measurement update rule}

In this section we show that  Pauli measurements transform stabilizer states into new stabilizer states. We give an analytic formula to update their description. Moreover we show that the update can be carried out efficiently.
\begin{theorem}[\textbf{Measurement update-rule}]\label{thm_Measurement_Update_rules}
Consider a stabilizer state $\ket{\phi}$ over $G$ with stabilizer group $\mc{S}$ and let $\sigma$ be a Pauli operator. Perform a measurement of $\sigma$ on $|\phi\rangle$, let the measurement outcome be labeled by an eigenvalue $\lambda$ of $\sigma$, and let $|\phi_m\rangle$ denote the post-measurement state.  Then the following statements hold:
\begin{itemize}
\item[(i)] The state $\ket{\phi_m}$  is a stabilizer state, with stabilizer group
\begin{equation}
\mc{S}_m=\langle\, \overline{\lambda}\sigma,\, C_{\mc{S}}(\sigma)\rangle.
\end{equation}
Here $C_{\mc{S}}(\sigma)$ denotes the centralizer of $\sigma$ inside $\mc{S}$, i.e.\ the group containing all elements of $\mc{\mc{S}}$ that commute with $\sigma$.
\item[(ii)] The labels of a generating set of ${\cal S}_m$ can be computed efficiently classically, given the labels of a generating set of ${\cal S}$.
\end{itemize}
\end{theorem}
\begin{proof}
First we show that $\ket{\phi_m}$ is stabilized by $\mc{S}_m$. To see this, first note that $\ket{\phi_m}$ is trivially stabilized by $\overline{\lambda}\sigma$. Furthermore, $\sigma$ commutes with every $\tau\in C_{\mc{S}}(\sigma)$. It follows that the projector $P$ onto the $\lambda$-eigenspace of $\sigma$ commutes with $\tau$ as well (this can easily be shown by considering $\sigma$ and $\tau$ in their joint eigenbasis). Hence $\tau P|\phi\rangle = P\tau |\phi\rangle = P|\phi\rangle$. Using that $\ket{\phi_m}\propto P|\phi\rangle$, we find that $\tau|\phi_m\rangle = |\phi_m\rangle$ for every $\tau\in C_{\mc{S}}(\sigma)$.

Second, we prove that $\ket{\phi_m}$ is the unique state stabilized by the group $\mc{S}_m$.  Without loss of generality we  restrict to the case $\sigma=Z(g_m)$. This is sufficient since, first, every Pauli operator can be transformed into an operator of the form $\alpha Z(g)$ with a suitable normalizer circuit $\mathcal{C}$ (cf. the discussion in section \ref{section  Pauli measurements Implementation}); and, second,  for any normalizer circuit $\mc{C}$, the quantum state $\ket{\phi_m}$ is a stabilizer state with stabilizer group $\mc{S}_m$ if and only if $\mc{C}\ket{\psi_m}$ is a stabilizer state with stabilizer group $\mc{C}\mc{S}_m\mc{C}^{\dagger}$.

Working with the assumption $\sigma=Z(g_m)$, we write the label subgroups $\mathbb{H}_{m}$ and $\mathbb{D}_{m}$ of $\mc{S}_m$ in terms of the label groups  $\mathbb{H}$ and  $\mathbb{D}$ of ${\cal S}$.  We have $\mc{S}_m=\langle\, \overline{\lambda}\sigma,\, C_{\mc{S}}(\sigma)\,\rangle$ where $\sigma=Z(g_m)$ for some $g_m\in G$. This implies that only the labels of  $C_{\mc{S}}(\sigma)$ contribute to $\mathbb{H}_{m}$. The centralizer $C_{\mc{S}}(\sigma)$ can be written as $C_{\mc{S}}(\sigma)=\mc{S}\cap C_{\mc{P}}(\sigma)$,  where $C_{\mc{P}}(\sigma)$ is  the subgroup of all Pauli operators that commute with $\sigma$. Thence, using lemma \ref{Commutativity of Pauli Stabilizer Groups PROPOSITION} we see that $C_{\mc{P}}(\sigma)$ consists of all $\gamma^{a}Z(g)X(h)$ with labels $h\in \langle g_m\rangle^{\perp}$. Hence,
\begin{equation}\label{H_m}
\mathbb{H}_{m}=\mathbb{H}\cap \langle g_m\rangle^{\perp}
\end{equation}
Due to the commutativity of $Z(g_m)$ and $C_{\mc{S}}(\sigma)$,  any element in $\mc{S}_m$  can be reordered as $\tau\, Z(g_m)^{i}$, with $\tau\in C_{\mc{S}}(\sigma)$.  Therefore, the diagonal group of ${\mc{S}_m}$ can be written as $\mc{D}_{m}= \langle\,\mc{D}',\, Z(g_m) \rangle $ where ${\cal D}'$ is the diagonal subgroup of $C_{\mc{S}}(\sigma)$. We now claim that ${\cal D}' ={\cal D}$ where ${\cal D}$ is the diagonal subgroup of ${\cal S}$. To see this, first note that trivially ${\cal D}' \subseteq {\cal D}$ since $C_{\mc{S}}(\sigma)$ is a subgroup of ${\cal S}$. Conversely, ${\cal D} \subseteq {\cal D}'$: as every diagonal element of ${\cal S}$ commutes with $Z(g_m)$, we have ${\cal D}\subseteq C_{\mc{S}}(\sigma)$; but this implies ${\cal D} \subseteq {\cal D}'$.

Putting everything together, we thus find $\mc{D}_{m}= \langle\,\mc{D},\, Z(g_m) \rangle $. It follows:
\begin{equation}\label{D_mH_m}
\mathbb{D}_{m}=\langle \mathbb{D}, \langle g_m\rangle \rangle \quad\Longrightarrow\quad \mathbb{D}_{m}^{\perp}=\left\langle \mathbb{D}, \langle g_m\rangle\right\rangle^{\perp}=\mathbb{D}^{\perp}\cap \langle g_m\rangle^{\perp},
\end{equation}
where we used lemma \ref{lemma:Annihilator properties}. Since ${\cal S}$ uniquely stabilizes $|\phi\rangle$, we have $\mathbb{H}=\mathbb{D}^{\perp}$ owing to corollary \ref{corollary_Uniqueness_Test}. With (\ref{D_mH_m}) and (\ref{H_m}) this implies that $\mathbb{H}_{m}=\mathbb{D}_{m}^{\perp}$. Again using corollary \ref{corollary_Uniqueness_Test}, it follows that ${\cal S}_m$ uniquely stabilizes $|\phi_m\rangle$.\\

To complete the proof of the theorem, we give an efficient classical algorithm to find a generating set of the centralizer $C_{\mc{S}}(\sigma)$; our approach is to reduce this task to a certain problem over the group $G\times G$ that can be efficiently solved using lemma \ref{lemma:Algorithms_FA_Groups}. Let $K\subset G\times G$  be the group of tuples $(g,h)$ such that there exists a stabilizer operator $\sigma(a,g,h)\in C_{\mc{S}}(\sigma)$; we prove that  $C_{\mc{S}}(\sigma)$ is isomorphic to $K$ via the map $\kappa : \sigma(a,g,h) \rightarrow (g,h)$, and that $\kappa$ is efficiently classically invertible: this  reduces the problem to finding a generating set of $K$ and applying the map $\kappa^{-1}$ to all its elements.

First, it is straightforward to verify that $\kappa$ is an isomorphism. Equations (\ref{commutation relations Pauli operators}) imply that the map is indeed linear. Surjectivity is granted by definition. Invertibility follows then from the fact that only  elements of the type $\gamma^{a}I\in C_\mathcal{S}(\sigma)$, for some $a$,  belong to $\ker\kappa$ (where $I$ denotes the identity): the latter are invalid stabilizer operators unless $\gamma^a=1$.

Second, we show how to compute $\kappa^{-1}$. Let the operator to measure $\sigma$ be of the (general) form  $\sigma=\gamma^{a_m}Z(x)X(y)$, and let $(a',g',h')$ be the label of an arbitrary stabilizer $\tau\in \mathcal{S}$. Given a set of (mutually commuting) generators $\sigma_1, \ldots, \sigma_r$ of $\mc{S}$, with corresponding labels $(a_i,g_i,h_i)$, the element $\tau$ can be written in terms of them as \be \label{group isomorphic to centralizer1}\tau=\prod \sigma_{i}^{v_i}= \gamma^{a'} X\left(\sum v_i g_i \right) Z\left(\sum v_i h_i \right)\ee for some integers $v_i$. From this equation it follows that
$K\subset \langle (g_1, h_1),\ldots,(g_r, h_r) \rangle$, which leads us to the following algorithm to compute $\kappa^{-1}$: given $(g,h)\in K$,  use the algorithm of lemma \ref{lemma:Algorithms_FA_Groups}(a) to compute $r$ integers $w_i$ such that $({g},h)=\sum w_i(g_i,h_i)$; due to (\ref{commutation relations Pauli operators}), the stabilizer operator defined as $\varsigma=\prod \sigma_i^{w_i}$ (whose label can be efficiently computed) is proportional to $X(g)Z({h})$; it follows that $\kappa(\varsigma)=(g,h)$ and, hence, $\varsigma$ equals $\kappa^{-1}(g,h)$.

Finally, combining (\ref{group isomorphic to centralizer1}) with formula (iii) in lemma  \ref{Commutativity of Pauli Stabilizer Groups PROPOSITION} we obtain
\begin{equation}\label{group isomorphic to centralizer2}
K=\langle y,-x\rangle^{\perp}\cap \langle (g_1, h_1),\ldots,(g_r, h_r) \rangle.
\end{equation}
Using eq.\ (\ref{group isomorphic to centralizer2}) together with algorithms (c-d) of lemma \ref{lemma:Algorithms_FA_Groups}, we can efficiently compute $s=\polylog{\Inputsize}$ elements $(x_1,y_1),\ldots,(x_s,y_s)$ that generate $K$; applying $\kappa^{-1}$ to these, we end up with a set of stabilizer operators $\kappa^{-1}(x_i,y_i)$ that generates $C_{\mc{S}}(\sigma)$.
\end{proof}

\section{Classical simulation of adaptive normalizer Circuits \label{section Gottesman-Knill theorem}}

In this section we prove our main result, i.e., a classical simulation theorem for adaptive normalizer circuits a la Gottesman-Knill (\textbf{theorem \ref{thm_main}}). We conclude the chapter discussing some significant differences between the power of adaptiveness for quantum state preparation in our formalism compared to previous qubit and  prime qudit works (\textbf{section \ref{sect:adaptiveness}}). 

\subsection{Simulation result}

Recall that in \cite{VDNest_12_QFTs} the following classical simulation result was shown:
\begin{theorem}\label{thm_maarten}
Let  $G=\mathbb{ Z}_{d_1}\times\cdots \times \mathbb{ Z}_{d_m}$ be a finite abelian group. Consider a polynomial size unitary normalizer circuit over $G$ acting on a standard basis input state. Both circuit and input are specified in terms of their standard encodings as described above. The circuit is followed by a measurement in the standard basis. Then there exists an efficient classical algorithm to sample the corresponding output distribution.
\end{theorem}
In the theorem, the \textit{standard encoding} of a normalizer circuit is defined as in section \ref{section Pauli and Clifford and Normalizer} in this chapter; the \textit{standard encoding} of a standard basis input state $\ket{g}$ is simply the tuple $g$, i.e. a collection of $m$ integers. Recall also that ``efficient'' is synonymous to ``in polynomial time in $\log |G|$''. 
\begin{proof}[Proof of theorem \ref{thm_maarten}]
For the sake of completeness, we will prove the result using the techniques of this chapter and refer the reader to \cite{VDNest_12_QFTs} for the original proof. Let ${\cal C}$ denote the normalizer circuit. Without loss of generality we assume that the input state is $|0\rangle$. Indeed, any standard basis state $|g\rangle$ can be written as $|g\rangle = X(g)|0\rangle$. The Pauli operator $X(g)$ can be realized as a polynomial-size normalizer circuit because of corollary \ref{PEG lemma for abelian groups} and because  diagonal  $Z(g)$ gates can be implemented with standard phase kickback tricks \cite{VDNest_12_QFTs,KLM_QC_07}. Hence, $X(g)$ can  be absorbed in the overall adaptive normalizer circuit. 

Next, let $e_i\in G$ denote the $i$-th ``canonical basis vector'' and note that the state $|0\rangle$ is a stabilizer state with stabilizer generators $Z(e_1),\ldots, Z(e_m)$: clearly, all $Z(e_i)$s stabilize $|0\rangle$; furthermore, the label subgroups associated to the stabilizer code $\mathcal{S}:=\langle Z(e_1),\ldots, Z(e_m)\rangle$ fulfill\footnote{This identity follows from lemma \ref{lemma:Annihilator properties}(a-b), which yields $G=(G^\perp)^\perp=\{0\}^\perp$.} $\mathbb{D}=G=\{0\}^\perp=\mathbb{H}^\perp$,  hence, $|0\rangle$ is uniquely stabilized because of corollary \ref{corollary_Uniqueness_Test}. It follows that the state $\ket{\psi}=\mathcal{C}\ket{0}$ is a stabilizer state uniquely stabilized by $\mathcal{C}\mathcal{S}\mathcal{C}^\dagger$ with generators $\{\mathcal{C} Z(e_i) \mathcal{C}^\dagger\}_{i=0}^m$, the labels of which can be efficiently computed due to theorem  \ref{thm_G_circuit_fundamental}. Finally, since simulating a terminal measurement in the standard basis is equivalent to sampling the probability distribution $|\psi(x)|^2$, the claim follows from the fact that stabilizer states described in terms of poly-size sets of stabilizer generators are computationally tractable (corollary \ref{corollary:CT}).\end{proof}

The main classical simulation result of this chapter (theorem \ref{thm_main} below) is a generalization of the above result. Rather than unitary normalizer circuits, the family of quantum circuits  considered here is that of the \emph{adaptive normalizer circuits}. A polynomial-size adaptive normalizer circuit consists of \polylog{\Inputsize} elementary steps, each of which is either a unitary normalizer gate $U$ or a Pauli measurement $M$. Furthermore, the choice of which $U$ or $M$ to apply in any given step may depend, in a (classical) polynomial-time computable way, on the collection of outcomes obtained in all previous measurements. The notion of adaptive normalizer circuits is thus a direct generalization of the adaptive Clifford circuits considered in the original Gottesman-Knill theorem \cite{Gottesman_PhD_Thesis,Gottesman99_HeisenbergRepresentation_of_Q_Computers}. Note that, compared to theorem \ref{thm_maarten}, two elements are added. First, measurements are no longer restricted to be standard basis measurements but arbitrary Pauli measurements. Second, the circuits are adaptive.

Before stating our classical simulation result, we make precise what is meant by an efficient classical simulation of an adaptive normalizer circuit. First, recall that the outcomes of any Pauli measurement are labeled by the eigenvalues of the associated Pauli operator. Since $\sigma^{2|G|}=I$ (recall lemma \ref{pauli_products_powers PROPERTY}) it follows that each Pauli operator eigenvalue is a $2|G|$-th root of unity i.e.\ it has the form $\lambda = \euler^{\pii k/|G|}$ for some $k\in \{0,\ldots, 2|G|-1\}$. This implies that any Pauli measurement gives rise to a probability distribution over the set of $2|G|$-th roots of unity; we denote the latter set by $S_{2|G|}$. Now consider an adaptive normalizer circuit ${\cal C}$. Let $P_i(\lambda | \lambda_1 \cdots \lambda_{i-1})$ denote the conditional probability of obtaining the outcome $\lambda\in S_{2|G|}$ in the $i$-th measurement, given that in previous measurements the outcomes $\lambda_1 \cdots \lambda_{i-1}\in S_{2|G|}$ were measured. We now say that ${\cal C}$ can be simulated efficiently classically if \textit{for every $i$} the $i$-th conditional probability distribution $P_i(\lambda | \lambda_1 \cdots \lambda_{i-1})$ can be sampled efficiently on a classical computer,  given the description of all gates and measurement operators in the circuit.
\begin{theorem}[\textbf{Classical simulation of adaptive normalizer circuits}]\label{thm_main}
Consider a polynomial size \emph{adaptive} normalizer circuit over $G$, specified as a list of normalizer gates in their  standard encoding, which  acts on an arbitrary standard basis input state. Then any such circuit can be efficiently simulated classically.
\end{theorem}
\begin{proof}
Let ${\cal C}$ denote the adaptive normalizer circuit. Without loss of generality we assume that the input state is $|0\rangle$. Indeed, any standard basis state $|g\rangle$ can be written as $|g\rangle = X(g)|0\rangle$; the Pauli operator $X(g)$ can be realized as a polynomial-size normalizer circuit \cite{VDNest_12_QFTs} and can thus be absorbed in the overall adaptive normalizer circuit. Letting $e_i\in G$ denote the $i$-th ``canonical basis vector'', the state $|0\rangle$ is a stabilizer state with stabilizer generators $Z(e_1),\ldots, Z(e_m)$ \cite{VDNest_12_QFTs}.
 We now recall the following facts, proved above:
\begin{itemize}
\item[(a)] Given any normalizer gate $U$ and any stabilizer state $|\psi\rangle $ specified in terms of a generating set of \polylog{\Inputsize} generators, the state $U|\psi\rangle$ is again a stabilizer state; moreover a set of generators can be determined efficiently (see theorem \ref{thm_G_circuit_fundamental}).
\item[(b)] Given any Pauli operator $\sigma$ and any stabilizer state $|\psi\rangle $ specified in terms of a generating set of polylog$(|G|)$ generators, the state $|\psi_{\lambda}\rangle$ obtained after measurement of $\sigma$, for any outcome $\lambda$, is again a stabilizer state; moreover a set of generators can be determined efficiently (cf. theorem \ref{thm_Measurement_Update_rules}).

Furthermore, the measurement probability distribution can be sampled efficiently in polynomial time on a classical computer. The latter is argued as follows. First, it follows from the discussion in section \ref{section  Pauli measurements Implementation} that the simulation of any Pauli measurement, on some input stabilizer state $\ket{\psi}$, reduces to simulating a unitary normalizer circuit (the description of which can be computed efficiently) followed by a \textit{standard basis} measurements (acting on the same input $\ket{\psi}$ and a suitable ancillary stabilizer state $\ket{0}$). Second, normalizer circuits acting on stabilizer state inputs and followed by standard basis measurements on stabilizer states can be simulated efficiently: this was shown in the proof of theorem \ref{thm_maarten} for an input stabilizer state with stabilizer generators $Z(e_1),\ldots, Z(e_m)$; the  argument, however, carries over immediately to the general case.
\end{itemize}
The proof of the result is now straightforward. Given any tuple $\lambda_1,\ldots, \lambda_{i-1}$, a generating set of stabilizers can be computed efficiently for the state of the quantum register obtained immediately before the $i$-th measurement, given that the previous measurement outcomes were $\lambda_1 \cdots \lambda_{i-1}$. Furthermore, given this stabilizer description, the distribution $P_i(\lambda | \lambda_1 \cdots \lambda_{i-1})$ can be sampled efficiently on a classical computer, as argued in (b).
\end{proof}

\subsection{The role of adaptiveness}\label{sect:adaptiveness}

To conclude this section we comment on an interesting difference between normalizer circuits and the ``standard'' qubit Clifford circuits, concerning the role of adaptiveness as a \textbf{tool for state preparation}. For qubits, adaptiveness adds no new state preparation power to the unitary Clifford operations. Indeed for any $n$-qubit stabilizer state $\ket{\psi}$ there exists a (poly-size) \textit{unitary} Clifford circuit $\mathcal{C}$ such that $\ket{\psi}=\gamma\mathcal{C}\ket{0}^{\otimes n}$, for some global phase $\gamma$ \cite{dehaene_demoor_coefficients}. In contrast, over general abelian groups $G$ this feature is no longer true. The associated adaptive normalizer circuits allow to prepare a \textit{strictly} larger class of stabilizer states compared to unitary normalizer circuits alone.

To demonstrate this claim, we provide a simple example of a stabilizer state over $G=\Z_{4}$  that \textit{cannot} be prepared from standard basis input states via unitary normalizer transformations over $G$, even in exponential time. However, the same state can be prepared efficiently  \textit{deterministically} if one considers adaptive normalizer schemes.
We consider
\begin{equation}\label{Peculiar stabilizer state}
\ket{{\psi}}=\frac{1}{\sqrt{2}}\left(\ket{0}+\ket{2}\right)
\end{equation}
Suppose that there existed a unitary Clifford operator $U\in\mc{C}^{G}$ which generates $\ket{{\psi}}$ from $|0\rangle$. Since the stabilizer group of $\ket{0}$ is  generated by $Z(1)$, the stabilizer group of $\ket{\psi}$ would be generated by  $U Z_d U^{\dagger}$. However it was shown in \cite{dehaene_demoor_hostens} that the stabilizer group of $\ket{{\psi}}$  cannot be generated by {one} single Pauli operator (i.e. at least two generators are needed), thus leading to a contradiction.

On the other hand, we now provide an efficient adaptive normalizer scheme to prepare,  not only the example $\ket{\psi}$, but in fact any coset state \cite{lomont_HSP_review,childs_lecture_8,childs_vandam_10_qu_algorithms_algebraic_problems}  of any finite abelian group $G$. This refers to any state of the form
\begin{equation}
|H+x\rangle:=\frac{1}{\sqrt{|H|}}\sum_{h\in H}\ket{h+x},
\end{equation}
where $H$ is a subgroup of $G$ and $x\in G$. Note that $|\psi\rangle$ is a coset state of the group $G=\Z_4$ with $H:=\langle 2\rangle$ and $x:=0$.

Our algorithm to efficiently prepare general coset states $|H+x\rangle$ receives the element $x$ and a polynomial number of generators of $H$. Here, we use the classical  algorithm in lemma \ref{lemma:Algorithms_FA_Groups}.(h) to  efficiently compute the matrix representation of a group homomorphism $\varpi:G\rightarrow \Z_d^{s}$ such that $\ker{\varpi}=H$, where $s$ and $d$ have polynomial bit-size. Given $\varpi$, we define a group automorphism $\alpha$ of the group $G\times \Z_d^{s}$ by $\alpha(g,h):=(g,h+\varpi(g))$. We now consider the following procedure\footnote{Observe that $\varpi$ can be considered as a function that \textit{hides} the subgroup $H$ in the sense of the hidden subgroup problem (HSP) \cite{lomont_HSP_review,childs_lecture_8,childs_vandam_10_qu_algorithms_algebraic_problems}. That is, for every $g, g' \in G$ we have $\varpi(g)=\varpi(g')$ iff $g-g'\in H$. Procedure (\ref{coset_preparation}) is essentially the routine used in the quantum algorithm for HSP to prepare random coset states.}:
\begin{equation}\label{coset_preparation}
\ket{0}\ket{0} \xrightarrow{\Fourier{G}\otimes I} \sum_{h\in G} \ket{h}\ket{0} \xrightarrow{U_{\alpha}} \sum_{h\in G} \ket{h}\ket{\varpi(h)} \xrightarrow{{M}} \: \frac{1}{\sqrt{|H|}}\: \sum_{h\in H}\ket{g+h}\ket{b} = \ket{g+H}\ket{b},
\end{equation}
where $\Fourier{G}$ denotes the QFT over $G$, the unitary $U_{\alpha}$ is the automorphism gate sending $\ket{g,h}$ to $\ket{\alpha(g,h)}$, and $M$ is a measurement of the second register in the standard basis. If the measurement outcome is $b$, then the post-measurement state is $\ket{g+H}\ket{b}$ where $g$ is a solution of the equation $\varpi(g)= b$. It can be verified that each coset state of $H$ (and thus also the desired coset state $\ket{x+H}$) occurs equally likely, i.e. with  probability $p=|H|/|G|$: in general, $p$ can be exponentially small. However, if we apply adaptive operations, we can always prepare $\ket{x+H}$ with probability 1, as follows. First, given the measurement outcome $b$ we efficiently compute an element $g'\in G$ satisfying  $\varpi(g')=b$ using theorem \ref{thm:General Solution of systems of linear equations over elementary LCA groups}. Then we apply a ``correcting'' Pauli operation $X(x-g')$ to the first register state, yielding $X(x-g')\ket{g+H}=\ket{x +(g-g')+H}=\ket{x+H}$ as desired (we implicitly used $g-g'\in H$).

%% file: chapter2_infinite.tex
\chapter{Normalizer circuits  and a Gottesman-Knill theorem for infinite-dimensional systems}\label{chapterI}

In chapter \ref{chapterF} we studied models of {normalizer circuits over finite abelian groups} that act on  arbitrary finite-dimensional systems  $\mathcal{H}_{D_1}\otimes \cdots \otimes \mathcal{H}_{D_n}$. The latter constituted group theoretic generalized Clifford operations that implemented quantum  Fourier transforms, group automorphism gates and quadratic phase gates over a group of the form $G=\mathbb{Z}_{D_1}\times\cdots \times \mathbb{Z}_{D_n}$. In this chapter, we extend the normalizer-circuit formalism to \emph{infinite dimensions}, by allowing normalizer gates to act on systems of the form  $\mathcal{H}_\mathbb{Z}^{\otimes a}$:  each factor $\mathcal{H}_\mathbb{Z}$ has a  standard basis labeled by \emph{integers} $\mathbb{Z}$, and a  Fourier basis  labeled by \emph{angles}, elements of the \emph{circle group} $\mathbb{T}$. As discussed in chapter \ref{chapterC}, in this setting, normalizer circuits become hybrid quantum circuits acting both on continuous- and discrete-variable systems. Here, we show that infinite-dimensional normalizer circuits can be efficiently simulated classically with a generalized \textbf{\emph{stabilizer formalism}} for Hilbert spaces associated with groups of the form $\mathbb{Z}^a\times \mathbb{T}^b \times \mathbb{Z}_{D_1}\times\cdots\times \mathbb{Z}_{D_n}$. We develop new techniques to track stabilizer-groups based on  the \emph{normal forms} for group automorphisms and quadratic functions we developed in chapter \ref{chapterGT}: we use the latter classical techniques to reduce  the problem of simulating these extended normalizer circuits to that of finding general solutions of  systems of mixed real-integer linear equations  \cite{BowmanBurget74_systems-Mixed-Integer_Linear_equations} and exploit this fact to devise a robust simulation algorithm: the latter remains efficient even in pathological cases where stabilizer groups become \emph{infinite}, \emph{uncountable} and \emph{non-compact}.  The techniques developed in this chapter might find applications  in the study of fault-tolerant quantum computation with  superconducting qubits \cite{Kitaev06_Protected_Qubit_Supercond_Mirror,Brooks13_Protected_gates_for_superconducting_qubits}.

This chapter is based on \cite{BermejoLinVdN13_Infinite_Normalizers} (joint work with Cedric Yen-Yu Lin and Maarten Van den Nest).

\section{Introduction}\label{sect:Introduction2}

In this chapter we investigate our  generalized infinite-dimensional normalizer circuit model (chapter \ref{chapterC}) where normalizer gates were associated to  abelian groups $G$  that can be \emph{infinite}. Specifically,  our interest is to focus on groups of the form  $G=F\times \Z^a$,  where $F=\DProd{d}{n}$ is a finite abelian group (decomposed into cyclic groups) as in the normalizer circuit setting considered in chapter \ref{chapterF}, and where  $\Z$ denotes the additive group of integers---the latter being an infinite group. The motivation for adding $\Z$ is that several number theoretical problems are naturally connected to problems over the integers, a crucial example being the \emph{factoring problem}, which is reducible to a hidden subgroup problem over $\mathbb{Z}$ \cite{Brassard_Hoyer97_Exact_Quantum_Algorithm_Simons_Problem,Hoyer99Conjugated_operators,MoscaEkert98_The_HSP_and_Eigenvalue_Estimation,Damgard_QIP_note_HSP_algorithm}.  The main result of this paper is a proof that all  normalizer circuits over infinite groups $G$ can be simulated classically in polynomial time, thereby extending the classical simulation results obtained in chapter \ref{chapterF} for normalizer circuits over finite abelian groups.

Similarly to the finite group case, normalizer circuits over an infinite group $G$  are composed of automorphism gates, quadratic phase gates and quantum Fourier transforms (chapter \ref{chapterC}, \ref{sect:Normalizer Gates Infinite Group}). However, as discussed in  chapter \ref{chapterC}-\ref{sect:Quantum states over infinite abelian groups}, several issues  that are not present in finite dimensions arise in extending normalizer circuits  to infinite groups $G$: 
\begin{itemize}
\item[(i)] First, because the physical system associated with $G$ has standard basis vectors $|g\rangle$ labeled by elements in $G$, the Hilbert space of the computation  is infinite-dimensional.
\item[(ii)] Second, infinite-dimensional quantum Fourier transforms (QFT)  perform  changes of basis between the group element basis $\{|g\rangle\}$ and a \emph{Fourier basis}, but  are no longer  gates: this was true in chapter \ref{chapterF} because Fourier basis elements were labeled by elements of the character group $\widehat{G}$ that was  isomorphic to $G$; however,  the infinite-yet-discrete group  $\Z$ has a \emph{continuous-variable} Fourier basis labeled by  elements  of the  circle group $\T$\footnote{We recall that $\T=[0, 1)$ (resp.\ $\T^n$) are the group of angles (given in $2\uppi$ units) with the addition and   the $n$-dimensional hypertorus: as discussed in chapter \ref{sect:characters},  $\T^n$ is isomorphic to the character group of ${\Z^n}$ for any $n$.}.  
\end{itemize}
In chapter \ref{sect:Quantum states over infinite abelian groups}, we saw that (i-ii)  have important consequences for the treatment of normalizer gates over $G$. In particular, in order to construct a closed normalizer formalism over groups $F\times \Z^a$ in this chapter, we will need to consider continuous ones $F\times \Z^a\times\T^b$, let the  computational basis change along the computation via the action of QFTs (the latter changed the underlying group $G$ indexing the basis) and allow initial state preparations and measurements in all group-element and Fourier bases associated to the Hilbert space $\mathcal{H}_G$.

\subsection{Main results}

To achieve an efficient classical simulation of normalizer circuits over $F\times \Z^a\times\T^b$ (\textbf{theorem \ref{thm:Main Result}}),  we develop new \emph{\textbf{stabilizer formalism}} techniques which extend the stabilizer formalism for finite abelian groups of  chapter \ref{chapterF} (which, in turn, extended the well-known stabilizer formalism for qubit/qudit systems  \cite{Gottesman_PhD_Thesis,Gottesman99_HeisenbergRepresentation_of_Q_Computers,Knill96non-binaryunitary,dehaene_demoor_coefficients,dehaene_demoor_hostens,AaronsonGottesman04_Improved_Simul_stabilizer,deBeaudrap12_linearised_stabiliser_formalism}). 

In generalizing the {stabilizer formalism} to describe infinite-dimensional normalizer circuits, several complications arise from the presence of non-finite nor-finitely-generated associated groups. One immediate consequence is that the associated Pauli stabilizer groups are  \emph{no longer finitely generated} either, which is  a \emph{unique} infinite-dimensional feature. This fact requires the development of new simulation techniques, since in our (and all known) finite-dimensional stabilizer  formalisms, a central  common feature is that quantum  states can be \emph{efficiently} described   as eigenstates of stabilizer groups of Pauli operators that are finite and fully determined by \emph{small lists} of group generators. Furthermore, if a Clifford gate is applied to the state, the list of generators transforms in a transparent way which can be efficiently updated. Performing such updates throughout the computation yields a stabilizer description of the output state, from which   final measurement statistics can be efficiently reproduced classically by suitably manipulating the stabilizer generators of the final state.

For the above reasons, in this chapter we abandon the standard method to track stabilizer groups. Instead, we devise a new simulation method based on the existence of certain concise \textbf{\emph{normal forms}} for quadratic functions and homomorphisms on $G$, which we presented as independent (classical) results in chapter \ref{chapterGT}. Thereby, we demonstrate that these purely group-theoretic contributions of the thesis have  interesting applications  in quantum information, far beyond those discussed in chapter \ref{chapterF}: therein,  we exploited  matrix representations in simulations and used  quadratic functions to develop normal forms for  \emph{stabilizer states}\footnote{This is due to our  normal form for stabilizer states in theorem \ref{thm Normal form of an stabilizer state}; we mention that this result can also be easily extended  to infinite dimensional stabilizer states as considered  later in this chapter (section \ref{sect:Stabilizer States}).}; in this chapter, we find new uses of these classical tools, e.g., to develop  \emph{efficient classical encodings} for infinite-dimensional stabilizer states that cannot be described  with any other available method.

A crucial ingredient in the last step of our simulation is a  polynomial-time classical algorithm that computes the \textbf{\emph{support}} of a stabilizer state, given a stabilizer group that describes it. This algorithm exploits a classical reduction of this problem to solving systems of \textbf{\emph{linear equations over infinite groups}}, which we showed how to solve in chapter \ref{chapterGT}. To find this reduction, we make crucial use of  the afore-mentioned normal forms and our infinite-group stabilizer formalism.

Lastly, we mention a technical issue that arises in the simulation of the final measurement of a normalizer computation: the basis in which the measurement is performed may be \emph{continuous} (stemming again from the fact that $G$ contains factors of $\T$). As a result, accuracy issues need to be taken into account in the simulation. For this purpose, we develop \textbf{\emph{$\boldsymbol{\varepsilon}$-net techniques}} to sample the support of stabilizer states.

\subsection{Relationship to previous work}\label{PreviousWork_c2}

In the particular case when $G$ is finite, our results completely generalize the results in \cite{VDNest_12_QFTs} and some of the results of chapter \ref{chapterF} (ref.\ \cite{BermejoVega_12_GKTheorem}): here, we fully characterize the support of stabilizer states in infinite dimensions, but, for simplicity, we will no longer allow adaptive Pauli measurements in the middle of a normalizer computation.\footnote{We believe it should be possible to fully extend the simulation techniques in chapter \ref{chapterF} to infinite dimensions.}

Prior to our work, an  infinite-dimensional stabilizer formalism best-known as ``the \emph{continuous variable} (CV) \emph{stabilizer formalism}'' was developed for systems that can be described in terms of harmonic oscillators \cite{Braunstein98_Error_Correction_Continuous_Quantum_Variables,Lloyd98_Analog_Error_Correction,Gottesman01_Encoding_Qubit_inan_Oscillator,Bartlett02Continuous-Variable-GK-Theorem,BartlettSanders02Simulations_Optical_QI_Circuits,Barnes04StabilizerCodes_for_CV_WEC}, which can be used as ``continuous variable'' carriers of quantum information. The CV stabilizer formalism is  used in the field of  continuous-variable  quantum information processing \cite{Braunstein98_Error_Correction_Continuous_Quantum_Variables,Lloyd98_Analog_Error_Correction,Gottesman01_Encoding_Qubit_inan_Oscillator,Bartlett02Continuous-Variable-GK-Theorem,BartlettSanders02Simulations_Optical_QI_Circuits,Barnes04StabilizerCodes_for_CV_WEC,LloydBraunstein99_QC_over_CVs,BraunsteinLoock05QI_with_CV_REVIEW,GarciaPatron12_Gaussian_quantum_information}, being a key ingredient in  current schemes for  CV quantum error correction \cite{Gottesman01_Encoding_Qubit_inan_Oscillator,Menicucci14FaultTolertant_MBQC_CV_ClusterState} and CV measurement-based quantum computation with CV cluster states \cite{ZhangBraunstein13_CV_Gaussian_Cluster_States,Menicucci06_UQC_with_CV_cluster_states,GuMileWeedbrook09_QC_with_CV_cluster,Menicucci14FaultTolertant_MBQC_CV_ClusterState}. A CV version of the Gottesman-Knill theorem \cite{Bartlett02Continuous-Variable-GK-Theorem,BartlettSanders02Simulations_Optical_QI_Circuits} for simulations of Gaussian unitaries (acting on Gaussian states) has been derived in this framework. 

We  stress that, although our infinite-group stabilizer formalism in this chapter and  the CV stabilizer formalism share some similarities, they are physically and mathematically \emph{inequivalent} and should not be confused with each other. The results in this chapter are for Hilbert spaces of the form $\mathcal{H}_{\Z}^{\otimes a}\otimes\mathcal{H}_{\T}^{\otimes b}\otimes\mathcal{H}_{\Z_{N_1}} \otimes \cdots \otimes \mathcal{H}_{\Z_{N_c}}$  with a basis  $\ket{g}$ labeled by the elements of $\T^a\times \Z^b\times \DProd{N}{c}$: the last $c$ registers correspond to finite-dimensional ``discrete variable'' systems;  the first $a+b$ registers can be thought of infinite-dimensional ``\textbf{rotating-variable}'' systems that are best described in terms of \textbf{quantum rotors}\footnote{\noindent The quantum states of a \textbf{quantum fixed-axis rigid rotor} (a quantum particle that can move in a circular orbit around a fixed axis) live in a Hilbert space with position and momentum bases labeled by $\T$ and $\Z$: the position is given by a continuous angular coordinate and the angular momentum is quantized in $\pm 1$ units (the sign indicates the direction in which the particle rotates \cite{aruldhasquantum}).}. In the CV formalism \cite{Bartlett02Continuous-Variable-GK-Theorem}, in contrast, the Hilbert space is $\mathcal{H}_\R^m$ with a standard basis labeled by $\R^m$ (explicitly constructed as a product basis of    position and momentum eigenstates of $m$ harmonic oscillators). Due to these differences, the available families of normalizer gates and Pauli operators in each framework (see sections \ref{sect_normalizer_circuits_sub}, \ref{sect:Pauli operators over abelian groups} and \cite{Bartlett02Continuous-Variable-GK-Theorem} for examples) are simply inequivalent.

Furthermore, dealing with continuous-variable stabilizer groups as in \cite{Gottesman01_Encoding_Qubit_inan_Oscillator, Bartlett02Continuous-Variable-GK-Theorem, Barnes04StabilizerCodes_for_CV_WEC} is sometimes simpler, from the simulation point of view, because the group $\R^m$ is also a finite-dimensional \textbf{vector space} with  
a \emph{finite basis}. In our setting, in turn, $G$ is \emph{no longer} a vector space but a \textbf{group} that may  well be \emph{uncountable} yet having \emph{neither a basis} \emph{nor a  finite generating set}; on top of that, our groups contain \emph{zero divisors}. These differences require new techniques to track stabilizer groups as they \emph{inherit} all these rich properties. For further reading on these issues we refer to the discussion in chapter \ref{chapterF}, where the differences between prime-qudit stabilizer codes \cite{Gottesman_PhD_Thesis,Gottesman99_HeisenbergRepresentation_of_Q_Computers,Gottesman98Fault_Tolerant_QC_HigherDimensions} (which can described in terms of fields and vector spaces) and  stabilizer codes over arbitrary spaces $\mathcal{H}_{d_1}\otimes\cdots\otimes\mathcal{H}_{d_n}$ (which are associated to a finite abelian group) are explained in detail. Last, we mention that a similarity with the $\R^m$ case is that our groups are not \emph{compact}.

Finally, we mention some related work on the classical simulability of Clifford circuits based on different techniques other than stabilizer groups: see \cite{VdNest10_Classical_Simulation_GKT_SlightlyBeyond} for simulations of  qubit non-adaptive Clifford circuits in the Schrödinger picture based on the stabilizer-state normal form of \cite{dehaene_demoor_coefficients}; see \cite{Veitch12_Negative_QuasiProbability_Resource_QC,MariEisert12_Positive_Wigner_Functions_Quantum_Computation} for phase-space simulations of odd-dimensional qudit Clifford operation exploiting  a local hidden variable theory based on the discrete Wigner function of \cite{Gibbons04_Discrete_Phase_Space_Finite_Fields,Gross06_discrete_Hudson_theorem,Gross_PhD_Thesis};  see \cite{Delfosee14_Wigner_function_Rebits} and  \cite{Raussendorf15QubitQCSI}  for phase-space simulations of restricted types of  Clifford operations based on sampling rebit and qubit Wigner functions. 

It should be insightful at this point to discuss briefly  whether the  latter results may extend to our set-up.  In this regard, it seems plausible to the authors that efficient simulation schemes for normalizer circuits analog to those in \cite{VdNest10_Classical_Simulation_GKT_SlightlyBeyond} might exist and may even benefit from the techniques  developed in the present work (specifically, our normal forms, as well as those given in chapter \ref{chapterF}). Within certain limitations, it might also be possible to extend the results in \cite{Veitch12_Negative_QuasiProbability_Resource_QC,MariEisert12_Positive_Wigner_Functions_Quantum_Computation,Delfosee14_Wigner_function_Rebits} and \cite{Raussendorf15QubitQCSI} to our setting. One fundamental limitation is that local hidden variable models  for the full-fledged stabilizer formalism on \emph{qubits} (which we generalize here) cannot exist due to the existence of certain stabilizer-type Bell inequalities \cite{Mermin90_extreme_quantum_entanglement,Scarani05_Nonlocality_Cluster_states,OtfriedToth05_Bell_Ineq_for_Graph_States}. Consequently, in order to find a hypothetical non-negative quasi-probability representation of normalizer circuits with  properties analogue to those of the standard discrete Wigner function as in  \cite{Gross06_discrete_Hudson_theorem,Gross_PhD_Thesis,Delfosee14_Wigner_function_Rebits}  (which leads to local HVMs), one would necessarily need to specialize to restricted normalizer-circuit models\footnote{Note that this might not be true for  quasi-probability representations that do not lead to non-local HVMs. The locality of the hidden variable models given in \cite{Gibbons04_Discrete_Phase_Space_Finite_Fields,Gross_PhD_Thesis,Veitch12_Negative_QuasiProbability_Resource_QC,Delfosee14_Wigner_function_Rebits} comes both from the positivity of the Wigner function and an additional factorizability property (cf. \cite{Gross_PhD_Thesis} and \cite{Delfosee14_Wigner_function_Rebits} page 5, property 4): in principle, classical simulation approaches that sample non-negative quasi-probability distributions without the factorizability property are well-defined and could also work, even if they do not  lead to local hidden variable models.} with, e.g., fewer types of gates, input states or measurements; this, in fact, is part of the approach followed in \cite{Delfosee14_Wigner_function_Rebits,Raussendorf15QubitQCSI}. The case for \cite{Raussendorf15QubitQCSI} is more subtle, since the classical simulation method therein is based on more  general \emph{non-contextual} hidden variable models: it is presently a subject of ongoing research  whether the standard (qubit) Gottesman-Knill can be recovered by sampling non-contextual HVMs; this program deals with some  interesting challenges related to  the phenomenon of state-independent quantum contextuality  with Pauli observables (we refer the reader to \cite{Raussendorf15QubitQCSI} for discussion).

Currently, there are no good candidate Wigner functions for extending the results of \cite{Veitch12_Negative_QuasiProbability_Resource_QC,MariEisert12_Positive_Wigner_Functions_Quantum_Computation} or \cite{Delfosee14_Wigner_function_Rebits} to systems of the form $\mathcal{H}_\Z^{\otimes a}$: the proposed ones (see  \cite{Rigas_Soto10_nonnegative_Wigner_OAM_states,Rigas11_OAM_in_phase_space,HinarejosPerezBanuls12_Wigner_function_Lattices} and references therein)  associate negative Wigner values to  Fourier basis states (which are allowed inputs in our formalism and also in \cite{Veitch12_Negative_QuasiProbability_Resource_QC,MariEisert12_Positive_Wigner_Functions_Quantum_Computation,Delfosee14_Wigner_function_Rebits}) that we introduce in section \ref{sect:Quantum states over infinite abelian groups}; for one qubit, these  are the usual $\ket{+}$, $\ket{-}$ states. The existence of a  non-negative Wigner representation for this individual case has not been ruled out by Bell inequalities or contextuality arguments, up to our best knowledge.

\subsection{Discussion and outlook}

Finally, we discuss open questions suggested by the work in this chapter as well as a few potential avenues for future research.

First,  we anticipate that the techniques developed in this chapter will play a key role in the  following  chapter \ref{chapterB}, where we will draw a rigorous connection between the normalizer circuit framework developed here and a large family of quantum algorithms, including Shor's factoring algorithm. (In particular, normalizer circuits over infinite groups of the form $\Z$ will be essential to understand this celebrated quantum algorithm.) 

\myparagraph{Connections with quantum error correction.} We also point out that, due to the presence of Hilbert spaces of the form $\mathcal{H}_\Z$, our stabilizer formalism over infinite groups yields a natural framework to study continuous-``rotating''-variable error correcting codes for quantum computing architectures based on superconducting qubits. Consider, for instance, the so-called  \emph{$0$-$\uppi$ qubits} \cite{Kitaev06_Protected_Qubit_Supercond_Mirror,Brooks13_Protected_gates_for_superconducting_qubits}. These are encoded qubits that, in our formalism, can be written as eigenspaces of groups of (commuting) generalized Pauli operators associated to  $\Z$ and $\T$ (cf.\ sections \ref{sect:Pauli operators over abelian groups}-\ref{sect:Stabilizer States} and also the definitions in \cite{Kitaev06_Protected_Qubit_Supercond_Mirror,Brooks13_Protected_gates_for_superconducting_qubits}). Hence, we can interpret them as instances of generalized \emph{stabilizer codes}\footnote{In this chapter we only discuss stabilizer states but it is easy to adapt our techniques in chapter \ref{section abelian group stabilizer formalism} to study codes.} over the groups $\Z$ and $\T$. The authors believe that it should be possible to apply the simulation techniques in this paper (e.g., our generalized Gottesman-Knill theorem) in the study of fault-tolerant quantum-computing schemes  that employ this form of generalized stabilizer codes: we remind the reader that the standard Gottesman-Knill theorem \cite{Gottesman_PhD_Thesis,Gottesman99_HeisenbergRepresentation_of_Q_Computers} is often  applied in fault-tolerant schemes for quantum computing with  traditional qubits, in order to   delay recovery operations and track the evolution of Pauli errors (see, for instance, \cite{Steane03_Overhead_Threshold_FTQEC, Knill05_QComp_RealisticallyNoisy,DiVincenzo07_effectiveFTQC_Slow_Measurements,Paler14ErrorTracking}).

Also in relation with quantum error correction, it would interesting to improve the stabilizer formalism in this chapter in order to describe \emph{adaptive Pauli measurements}; this would fully extend our simulation  results from chapter \ref{chapterF}.

\paragraph{Connection with bosonic Gaussian unitaries.} In relation to works on continuous-variable and hybrid quantum information processing, it would be  interesting to investigate normalizer circuits over more general types of infinite groups. After completion of this work, we became aware that \emph{normalizer circuits over $\R^m$} groups and Gaussian unitaries can be used to efficiently approximate each other to any accuracy (this result is presented in appendix \ref{aG}, theorem \ref{thm:Gaussian=Normalizer}), hence, define identical gate models for all practical purposes.  Because of the importance of the Clifford and Gaussian formalisms in discrete- and continuous-variable QIP, a natural next question (that we leave to future investigations) would be to analyze the potential QIP applications of normalizer gates over hybrid systems $\mathcal{H}_{\R}^{\otimes a}\otimes\mathcal{H}_{\Z}^{\otimes b}\otimes\mathcal{H}_{\T}^{\otimes c}\otimes\mathcal{H}_{\Z_{N_1}} \otimes \cdots \otimes \mathcal{H}_{\Z_{N_d}}$. We  highlight that  results developed in this chapter can be  extended to study such hybrid system context with only tiny modifications (cf.\ discussion in appendix  \ref{aG}).

\paragraph{Connection with duality theory.} Lastly, we  mention that an important ingredient underlying the consistency of our normalizer/stabilizer formalism is the fact that the groups associated to the Hilbert space fulfill the so-called \textbf{Pontryagin-Van Kampen duality}\footnote{Aspects of this duality and a generalization of it  feature also in the circuit models studied in chapters \ref{chapterB}-\ref{chapterH}.} \cite{Morris77_Pontryagin_Duality_and_LCA_groups,Stroppel06_Locally_Compact_Groups,Dikranjan11_IntroTopologicalGroups,rudin62_Fourier_Analysis_on_groups,HofmannMorris06The_Structure_of_Compact_Groups,Armacost81_Structure_LCA_Groups,Baez08LCA_groups_Blog_Post}. From a mathematical point of view, it is possible to associate a family of normalizer gates to every group in such class, which accounts for all possible abelian groups that are locally compact Hausdorff (often called LCA groups). Some LCA groups are notoriously complex objects and remain unclassified to date. Hilbert spaces associated to them can exhibit exotic properties, such as   \emph{non-separability}, and may not always be in correspondence with natural quantum mechanical systems. In order to construct a physically relevant model of quantum circuits, we have restricted ourselves to groups of the form $\Z^a \times \T^b \times \DProd{N}{c}$, which can be naturally associated to known quantum mechanical systems. As aforementioned, we believe that the results presented in this paper can easily  be extended to all groups of the form $\Z^a \times \T^b \times \DProd{N}{c}\times \R^d$, which we call ``\emph{elementary}'', and form a well-studied class of groups known as ``\emph{compactly generated abelian Lie groups}'' \cite{Stroppel06_Locally_Compact_Groups}. Some examples of LCA groups that are not elementary are the $p$-adic numbers $\mathbb{Q}_p$ and the adele ring $\mathbb{A}_F$ of an algebraic number field $F$ \cite{Dikranjan11_IntroTopologicalGroups}.

\subsection{Chapter outline}
We refer the reader to chapter \ref{chapterC} for a detailed introduction to the \textbf{normalizer circuit model} over finite and infinite abelian  groups: specifically, see   section \ref{sect:Normalizer Gates Infinite Group} for specific details on  the infinite-group case, including a discussion of the most technical infinite-dimensional aspects of these circuits compared to the finite-group setting (s.\ \ref{sect:Quantum states over infinite abelian groups}), a full description of the circuit model (s.\ \ref{sect:Designated basis}) and examples \ref{sect:ExamplesInfinite}.

In section \ref{sect:Main Result} we state the \textbf{main result} (theorem \ref{thm:Main Result}) of this chapter.  In section \ref{sect:Pauli operators over abelian groups} we study the properties of Pauli operators over abelian groups. In section \ref{sect:Stabilizer States} we present stabilizer group techniques based on these operators. Finally, in \ref{sect:Proof of theorem 1}, we prove our main result.

We refer the reader to chapter \ref{chapterGT} for a description of the \textbf{classical techniques} that we exploit in the classical simulations of this chapter: namely, see sections  \ref{sect:Homomorphisms and matrix representations}, \ref{sect:quadratic_functions} for details on our   matrix representations and normal forms  for group homomorphisms and  quadratic functions, and  section \ref{sect:Systems of linear equations over groups} for  our classical algorithms for solving linear equations over groups.

\section{Main result}\label{sect:Main Result}

In our main result (theorem \ref{thm:Main Result} below) we show that any polynomial-size normalizer circuit (see c.\ \ref{sect_normalizer_circuits_sub} for definitions and details on our computational model) over  any group of  form
\begin{equation}\label{eq:GroupChapterI}
G= \Z^{\otimes a}\times \T^b\times\DProd{N}{c}, \quad\textnormal{ with }\quad \mathcal{H}_{G}=\mathcal{H}_{\Z}^{\otimes a}\otimes\mathcal{H}_{\T}^{\otimes b}\otimes \mathcal{H}_{\Z_{N_1}}\otimes \dots\otimes \mathcal{H}_{\Z_{N_c}}
\end{equation}   can be simulated efficiently classically. Before stating the result, we will rigorously state what it is meant in our work by an efficient classical simulation of a normalizer circuit, in terms of computational complexity.

In short, the  \textbf{computational problem} we consider is the  following: given a classical description of a normalizer quantum circuit, its input quantum state and the measurement performed at the end of the computation, our task is to sample the probability  distribution of final measurement outcomes with a classical algorithm. Any classical algorithm to solve this problem in polynomial time (in the bit-size of the input) is said to be \emph{efficient}.

We specify next how an instance of the computational problem is presented to us.

First, we introduce \textbf{standard encodings} that we use to describe normalizer gates. Our encodings  are  \emph{efficient}, in the sense that the number of bits needed to store a description of a normalizer gate scales as $O(\poly{m},\polylog{N_i})$, where $m$ is the total number of {registers} of the Hilbert space (\ref{eq:GroupChapterI}) and $N_i$ are the local dimensions of the finite dimensional registers (the memory size of each normalizer gate in these encodings is given in table \ref{table:Main Result INPUT SIZE}). This polynomial (as opposed to exponential) scaling in $m$ is crucial in our setting, since normalizer gates may act non-trivially on all $m$ registers of the Hilbert space (\ref{eq:GroupChapterI})---this is an important difference between our computational model (based on normalizer gates) and the standard quantum circuit model \cite{nielsen_chuang}, where a  quantum circuit is always given as a sequence of one- and two-qubit gates. 
\begin{enumerate}
\item[(i)] A partial quantum Fourier transform $\Fourier{G_i}$ over $G_i$ (the $i$th factor of $G$) is described by the index $i$ indicating the register where the gate acts non-trivially.
\item[(ii)] An automorphism gate $U_\alpha$ is described by  what we call a \emph{matrix representation $A$}  of the automorphism $\alpha$ (definition \ref{def:Matrix representation}): an $m\times m$ real matrix $A$ that specifies the action of the map $\alpha$.
\item[(iii)] A quadratic phase gate $D_\xi$ is described by an $m\times m$ real matrix $M$ and an $m$-dimensional real vector $v$. The pair $(M,v)$ specifies the action of the quadratic function $\xi$ associated to $D_\xi$. Here we exploit the normal form for quadratic functions that we gave  in theorem~\ref{thm:Normal form of a quadratic function}. 
\end{enumerate}
 In this chapter, we assume that all maps $\alpha$ and $\xi$ can be represented \emph{exactly} by rational matrices and vectors $A$, $M$, $v$, which are explicitly given to us\footnote{Some automorphisms and quadratic functions can only be represented by matrices with irrational entries (cf.\ the normal forms in sections \ref{sect:Homomorphisms and matrix representations},\ref{sect:quadratic_functions}). Restricting ourselves to study the rational ones allows us to develop \emph{exact simulation algorithms}. We believe irrational matrices (even with transcendental entries) could also be handled by taking into account floating-point errors. We highlight that the stabilizer formalism in this paper and all of our normal forms are developed \emph{\textbf{analytically}}, and hold even if transcendental numbers appear in the matrix representations of $\alpha$ and $\xi$. (It is an good question to explore whether an exact simulation results may hold for matrices with algebraic coefficients.)}.
 
Second, a normalizer circuit is specified as a list of normalizer gates given to us in their standard encodings.

The existence and efficiency of our standard encodings is guaranteed by  the  (classical) \textbf{theory of matrix representations} and \textbf{quadratic functions} that we  developed in chapter \ref{chapterGT}: specifically, because of our  existence lemma  \ref{lemma:existence of matrix representations} for group-homomorphism matrix representations; our  normal form that characterizes the structure of these matrices (lemma \ref{lemma:Normal form of a matrix representation}); and  our analytic normal forms for bicharacter functions (lemmas \ref{lemma:Normal form of a bicharacter 1}, \ref{lemma symmetric matrix representation of the bicharacter homomorphism}) and quadratic functions (theorem \ref{thm:Normal form of a quadratic function}). Because of their quantum applications, these earlier classical results are also main contributions of this thesis.

Lastly, we mention that, in this chapter, we allow the matrices $A$, $M$ and the vector $v$ in (i-iii) to contain \emph{arbitrarily large \emph{and} arbitrarily small} coefficients. This degree of generality is necessary in the setting we consider, since we allow \emph{all} normalizer gates to be valid components of a normalizer circuit. However,  the presence of infinite groups in (\ref{eq:GroupChapterI}) implies that there exists an infinite number of normalizer gates (namely, of automorphism and quadratic gates, which follows from the our  analysis in sections  \ref{sect:Homomorphisms and matrix representations} and \ref{sect:quadratic_functions}). This is in contrast with the settings considered in chapter \ref{chapterF}, where both the group (\ref{eq:GroupChapterI}) and the associated set of  normalizer gates are finite. As a result, the arithmetic precision needed to store the coefficients of $A$, $M$, $v$ in our standard encodings becomes a variable of the model (just like in the standard problem of multiplying two integer matrices).

We state now our main result.
\begin{theorem}[\textbf{Main result}]\label{thm:Main Result} Let $\mathcal{C}$ be any normalizer circuit over any group $G=\Z^a \times \T^b \times \DProd{N}{c}$ as defined in section \ref{sect_normalizer_circuits_sub}. Let $\mathcal{C}$ act on a input state $\ket{g},g\in G_0$ in the designated standard basis $\mathcal{B}_{G_0}$ at time zero, and be followed by a final  measurement in  the  designated basis $\mathcal{B}_{G_T}$ at time $T$. Then the output probability distribution can be sampled classically  efficiently  using an infinite-dimensional stabilizer formalism and epsilon-net methods.
\end{theorem}
We remind the reader that in theorem \ref{thm:Main Result} both standard and Fourier basis states of $\mathcal{H}_{\Z}$ are allowed inputs (cf.\ section \ref{sect_normalizer_circuits_sub}). 

In theorem \ref{thm:Main Result}, the state $\ket{g}$ is described by the group element $g$, which is encoded as a tuple of $m$ rational\footnote{In this work we do not use floating point arithmetic.} numbers of varying size (see table \ref{table:Main Result INPUT SIZE}, row 1).  The memory needed to store the normalizer gates comprising $\mathcal{C}$ is summarized in table \ref{table:Main Result INPUT SIZE}, row 2. By ``\emph{classically efficiently}'' it is meant that there exists a classical algorithm (\textbf{theorem \ref{thm:Algorithm to sample subgroups}}) to perform the given task whose worst-time running time scales \emph{polynomially} in the input-size (namely, in the number of subsystems $m$, the number of normalizer gates of $\mathcal{C}$) and of all other variables listed in the ``bits needed'' column of table \ref{table:Main Result INPUT SIZE}), and \emph{polylogarithmically}  in the parameters ${\frac{1}{\varepsilon}}$, ${\Delta}$ that specify the number of points in a \emph{$(\Delta, \varepsilon)$-net} (which we introduce below) and their geometrical arrangement. 
\begin{table}[ht]
\begin{center}   
  \begin{tabular}{| c | c |  c|}
           \hline 
           Input element & Description needs to  & Bits needed\\ 
           \hline \multirow{2}{*}{Input state $\ket{g}$}  & Specify element $g(i)$ of infinite group $\Z$, $\T$ & variable\\
          & Specify element $g(j)$ of finite group $\Z_{N_j}$ & $\log N_j$   \\ 
          \hline \multirow{3}{*}{Normalizer circuit $\mathcal{C}$}  & Specify  quantum Fourier transform $\Fourier{G_i}$ & $\log{m}$ \\ 
          & Specify automorphism gate $U_\alpha$ via $A$ & $m^2 \|A\|_{\mathbf{b}}$  \\
          & Specify quadratic phase gate $D_\xi$ via $M,v$ & $ m^2 \|M\|_{\mathbf{b}}+m \|v\|_{\mathbf{b}}$  \\ \hline     
    \end{tabular}
  \end{center}
   \caption{The input-size in theorem \ref{thm:Main Result}. Above, $m=a+b+c+d$ denotes the number of primitive factors of $G$. $\|X\|_{\mathbf{b}}$ denotes the number of bits used to store one single coefficient of $X$, which is always assumed to be a rational matrix/vector. Formulas in column 3 are written in Big Theta $\Theta$ notation and do not  include constant factors.} \label{table:Main Result INPUT SIZE}
\end{table}

\subsubsection*{Sampling techniques}

We finish this section by saying a few words about the $(\Delta, \epsilon)$-net methods used in the proof of theorem \ref{thm:Main Result}. These techniques are fully developed in sections \ref{sect:Proof of theorem 1} and \ref{sect:Sampling the support of a state}.

We shall show later (lemma \ref{lemma:StabStates_are_Uniform_Superpositions}) that the final quantum state $\ket{\psi}$ generated by a normalizer circuit is always a uniform quantum superposition in any  designated  basis ${\cal B}_{G_t}$,  (\ref{basis}) at any time step $t$: i.e., given the most general form of state $\ket{\psi}$, which is
\begin{equation}
\ket{\psi}=\int_X\mathrm{d}g\, \psi(g)\ket{g},
\end{equation} where $\mathrm{d}g$ denotes the Haar measure\footnote{If $X$ is a discrete set, this Haar integral is simply  a sum over all elements in $G_t$.} on $G_t$, $\psi(g)$ are the amplitudes of a normalized wavefunction  and  $X$ is the support of $\psi$ (i.e., the set of $x\in G_t$ such that $\psi (x)\neq 0$); we show that $|\psi(x)|=|\psi(y)|$ for all $x,y\in X$ at any time step.  As a result the final distribution of measurement outcomes of a normalizer circuit is always a flat distribution over some set $X$. 

Moreover, we show in section  \ref{sect:Sampling the support of a state} that $X$ is always isomorphic to a group of the form $K \times \Z^\mathbf{r}$ where $K$ is compact, and that an isomorphism can be efficiently computed: as a result, we see that, although $X$ is not compact, the non-compact component of $X$ inherits a simple Euclidean geometry from $\mathbb{R}^\mathbf{r}$. Our sampling algorithms are based on this fact: to sample $X$ in an approximate sense, we construct a subset $\mathcal{N}_{\Delta,\varepsilon} \subset X$ of the form
\begin{equation}
\mathcal{N}_{\Delta,\varepsilon} =\mathcal{N}_{\varepsilon}\oplus\mathcal{P}_\Delta,
\end{equation}
where $\mathcal{N}_{\varepsilon}$ is an $\varepsilon$-net (definition \ref{def:Epsilon Net}) of the compact component $K$ of $X$ and $\mathcal{P}_\Delta$ is a $\mathbf{r}$-dimensional parallelotope contained in the Euclidean component $\Z^\mathbf{r}$, centered at 0, with edges of length $2\Delta_1,\ldots, 2\Delta_\mathbf{r}$. We call $\mathcal{N}_{\Delta,\varepsilon}$ a ($\Delta, \varepsilon$)-net  (definition \ref{def:Delta Epsilon Net}). The algorithm in theorem \ref{thm:Main Result} can efficiently construct and sample uniformly from such sets for any $\varepsilon$ and $\Delta:=\Delta_1,\ldots, \Delta_\mathbf{r}$ of our choice: its worst-case running-time is $O(\polylog{\tfrac{1}{\varepsilon}},\polylog{\Delta_i}) $,  as a function of these parameters. We refer the reader to section \ref{sect:Proof of theorem 1} and theorem \ref{thm:Algorithm to sample subgroups} for more details.

\subsubsection*{Treatment of finite-squeezing errors}

It follows from the facts that we have just discussed that when $G$ is not a compact group (i.e.\, $G$ contains $\Z$ primitive factors) the support $X$ of the quantum state $\ket{\psi}$ can be an unbounded set. In such case, it follows from the fact that $\ket{\psi}$ is a uniform superposition that the quantum state is \emph{unphysical} and that the physical preparation of such a state requires infinite energy; in the continuous-variable quantum information community, states like $\ket{\psi}$ are often called \emph{infinitely squeezed states} \cite{KokLovett10Intro_to_Optical_Quantum_Information_Processing}. In a physical implementation (cf.\ chapter \ref{chapterB}), these states can be replaced by physical finitely-squeezed states, whose amplitudes will decay towards the infinite ends\footnote{The particular form of the damping depends on the implementation. These effects vanish in the limit of infinite squeezing.}  of the support set $X$. This leads to finite-squeezing errors, compared to the ideal scenario.

In this chapter, we consider normalizer circuits to work perfectly in the ideal infinite-squeezing scenario. Our simulation algorithm in theorem \ref{thm:Main Result} samples the ideal distribution that one would obtain in the infinite precision limit, neglecting the presence of  finite-squeezing errors. This is achieved with the  $(\Delta, \epsilon)$-net methods described above, which we use to discretize and sample the  manifold $X$ that supports the ideal output state $\ket{\psi}$;  the output of this procedure  reveals the information  encoded in the  wavefunction of the state.

In the following chapter \ref{chapterB}, we will make use of this simulation algorithm to study quantum algorithms based on normalizer circuits. We will also study  how  information can be represented with finitely-squeezed states in a computation.

\section{Pauli operators over abelian groups}\label{sect:Pauli operators over abelian groups}

In this section we introduce Pauli operators over groups of the form $G= \Z^a\times \T^b\times F$ (note that we no longer include factors of $\R^d$ because these groups are not related to the Hilbert spaces that we study in this paper), discuss some of their basic properties and finally show that normalizer gates map any Pauli operator to another Pauli operator. The latter property is a generalization of a well known property for qubit systems, namely that Clifford operations map the Pauli group to itself.

\textbf{Note on terminology.} Throughout the rest of the paper,  we sometimes use the symbol $\mathcal{H}_\T$ as a second name for the Hilbert space $\mathcal{H}_\Z$. Whenever this notation is used, we make implicit that we are working on the Fourier basis of  $\mathcal{H}_\Z$, which is labeled by the circle group $\T$. Sometimes, this basis will be called the $\T$ standard basis or just $\T$ basis. From now on, $G^*= \T^a\times \Z^b\times F$ will always denote the uniquely-define dual elementary group  that is isomorphic to the character group $\widehat{G}$ of $G$ (cf.\ chapter \ref{sect:characters} for details about group characters).

\subsection{Definition and basic properties}\label{sect:pauli_def}

Consider an abelian group of the form $G= \Z^a\times \T^b\times F$ and the associated Hilbert space $\mathcal{H}_{G}$ with the associated group-element basis $\{|g\rangle: g\in G\}$ as defined in section \ref{sect:Quantum states over infinite abelian groups} . We define two types of unitary gates acting on $\mathcal{H}_G$, which we call the \emph{Pauli operators of} $G$. The first type of Pauli operators are the X-type operators $X_G(g)$ (often called \emph{shift operators} in generalized harmonic analysis):
\be\label{eq:Pauli operator type X, over G, shift definition}
X_G(g)\psi(h) := \psi(h-g), \quad \text{for every $g, h\in G$},
\ee
where the $\psi(h)$ are the coefficients of some quantum state $|\psi\rangle$ in $\mathcal{H}_{G}$. These operators can also be written via their action on the standard basis, which yields a more familiar definition:
\begin{equation}\label{eq:Pauli operator type X, over G, eigenket definition}
X_{{G}}(g)\ket{h} = \ket{g+h}, \quad  \text{for every $g,h\in G$}.
\end{equation}
In representation theory, the map $g\rightarrow X_G(g)$ is called the \emph{regular representation} of the group $G$. The second type of Pauli operators are the Z-type operators $Z_G(\mu)$:
\begin{equation}\label{eq:Pauli operator type Z}
Z_{G}(\mu)\ket{g} :=\chi_{\mu}(g)\ket{g}, \quad \text{for every $g\in G,\, \mu\in G^*$}.
\end{equation}
We define a \emph{ generalized Pauli operator} of $G$ to be any unitary operator of the form
\begin{equation}\label{eq definition of Pauli operator}
 \sigma:=\gamma Z_G(\mu) X_G(g)
\end{equation}
where $\gamma$ is a complex number with unit modulus. We will call the pair $(\mu, g)$ and the complex number $\gamma$, respectively, the \emph{label} and the \emph{phase} of the Pauli operator $\sigma$. Furthermore we will regard the label  $(\mu, g)$ as an element of the abelian group $G^*\times G$. The above definition of Pauli operators is a generalization of the notion of Pauli operators over finite abelian groups as considered in chapter \ref{chapterF}, which was in turn a generalization of the standard notion of Pauli operators for qubit systems. An important distinction between Pauli operators for finite abelian groups and the current setting is that the $Z_G(\mu)$ are labeled by $\mu\in G^*$. For finite abelian groups, we have  $G^* = G$ and consequently the Z-type operators are also labeled by elements of $G$.

Using the definition of Pauli operators, it is straightforward to verify the following commutation
relations, which hold for all $g\in G$ and $\mu\in G^*$:
\begin{align}
X_{G}(g)X_{G}(h) &=  X_G(g+h)=X_G(h)X_G(g)\nonumber \\
Z_{G}(\mu)Z_{G}(\nu) &=  Z_{G}(\mu+\nu)=Z_{G}(\nu)Z_{G}(\mu)\label{eq:pauli_commutation}\\
Z_{G}(\mu)X_{G}(g) &=  \chi_{\mu}(g)X_{G}(g)Z_{G}(\mu)\nonumber
\end{align}
It follows that the set of generalized Pauli operators of $G$
form a group, which we shall call the \emph{Pauli group} of $G$.

\subsection{Evolution of Pauli operators} \label{sect:Evolution Pauli Operators}

The connection between normalizer gates and the Pauli group is that the former ``preserve'' the latter  under conjugation, as we will show in this section. This property will be a generalization of the well known fact that the Pauli group for $n$ qubits is mapped to itself under the conjugation map $\sigma\to U\sigma U^{\dagger}$, where $U$ is either a Hadamard gate, CNOT gate or $(\pi/2)$-phase gate \cite{Gottesman_PhD_Thesis,Gottesman99_HeisenbergRepresentation_of_Q_Computers}. More generally, as we know from \cite{VDNest_12_QFTs} and chapter \ref{chapterF}, normalizer gates over any finite abelian group $G$ also map the corresponding Pauli group over $G$ to itself under conjugation. In generalizing the latter result to abelian groups of the form $G= \Z^a\times \T^b\times F$, we will however note an important distinction. Namely, normalizer gates over $G$ will map Pauli operators over $G$ to Pauli operators over a group $G'$ which is, in general, \emph{different} from the initial group $G$. This feature is a consequence of the fact that the groups $\Z^a$ and $\T^b$ are no longer autodual (whereas all finite abelian groups are). Consequently, as we have seen in chapter \ref{sect:Normalizer Gates Infinite Group}, the QFT over $G$ (or any partial QFT) will change the  group that labels the designated basis of $\mathcal{H}_{G}$ from $G$ to $G'$. We will therefore find that the QFT maps Pauli operators over $G$ to Pauli operators over $G'$. In contrast, such a situation does not occur for automorphism gates and quadratic phase gates, which do not change the group $G$ that labels the designated basis.

Before describing the action of normalizer gates on Pauli operators (theorem \ref{thm:Normalizer gates are Clifford}), we provide two properties of QFTs.
\begin{lemma}[\textbf{Fourier transforms diagonalize shift operators}]\label{lemma:Fourier transform diagonalizes the Regular Representation}
Consider a group of the form $G= \Z^a\times \T^b\times F$. Then the X-type Pauli operators of $G$ and the Z-type operator of $G^*$ are related via the quantum Fourier transform $\Fourier{G}$ over $G$:
\be
Z_{G^*}(g)=\Fourier{G} X_G(g) \Fourier{G}^\dagger.
\ee
\end{lemma}
\begin{proof}
We show this by direct evaluation of the operator $X_G(h)$  on the Fourier basis states. Using the definitions introduced in section \ref{sect:characters} we can write the vectors in the Fourier basis of $G$ (chapter \ref{sect:Normalizer Gates Infinite Group}) in terms of character functions (definition \ref{def:Characters}): letting $\ket{\mu}$ be the state
\begin{equation}\label{eq:Fourier Basis in terms of Characters}
\ket{\mu}= \int_{G} \mathrm{d} h \, \overline{\chi_\mu(h)}\ket{h} = \int_{G} \mathrm{d} h \, \chi_{-\mu}(h)\ket{h},
\end{equation}
then the Fourier basis of $G$ is just the set $\{\ket{\mu}, \,\mu \in G^*\}$. Now it is easy to derive
\begin{align}
X_G(g)\ket{\mu} &= X_G(g) \left(\int_{G} \mathrm{d} h \overline{\chi_\mu(h)}\ket{h}\right) =  \int_{G} \mathrm{d} h \overline{\chi_\mu(h)}\ket{g+h} =  \int_{G} \mathrm{d} h' \overline{\chi_\mu(h'-g)}\ket{h'}\notag\\
&= \overline{\chi_\mu(-g)}\left(\int_{G} \mathrm{d} h' \chi_\mu(h')\ket{h'} \right)= \chi_{g}(\mu)\ket{\mu} = Z_{G^*}(g) \ket{\mu}.
\end{align}
In the derivation we use  lemmas \ref{lemma:Pontryagin duality for characters}, \ref{lemma:Character Multiplication} and equation (\ref{eq:Pauli operator type Z}) applied to the group $G^*$.
\end{proof}
The next theorem shows that normalizer gates are generalized Clifford operations, i.e.\ they transform Pauli operators into Pauli operators under conjugation and, therefore, they \emph{normalize}  the group of all Pauli operators within the group of all unitary gates\footnote{It is usual in quantum information theory to call the normalizer group of the $n$-qubit Pauli group ``the Clifford group'' because of a ``tenuous relationship'' \cite[Gottesman]{Gottesman09Intro_QEC_and_FTQC}  to Clifford algebras.}. 
\begin{theorem}[\textbf{Normalizer gates are Clifford}]\label{thm:Normalizer gates are Clifford}
Consider a group of the form $G= \Z^a\times \T^b\times F$. Let $U$ be a normalizer gate of the group $G$. Then $U$ corresponds to an isometry from $\mathcal{H}_{G}$ to $\mathcal{H}_{G'}$ for some suitable group $G'$, as discussed in section  
\ref{sect_normalizer_circuits_sub}.  Then the conjugation
map $\sigma\rightarrow U\sigma U^{\dagger}$ sends Pauli operators
of $G$ to Pauli operators of $G'$, hence, $U$ is a generalized Clifford operator.
\end{theorem}
Our result generalizes Van den Nest's theorem for finite abelian group normalizer gates (theorem \ref{thm_G_circuit_fundamental}), which we reviewed in chapter \ref{chapterF}.
\begin{proof}
We provide an explicit proof for Pauli operators of type $X_{G}(g)$  and  $Z_{G}(\mu)$. This is enough to prove the lemma due to  (\ref{eq:pauli_commutation}). As before,  $G=G_{1}\times\cdots\times G_{m}$ where the $G_{i}$ are groups of primitive type.

We break the proof into three cases.
\begin{itemize}
\item If $U$ is an automorphism gate $U_{\alpha}:\:|h\rangle\rightarrow|\alpha(h)\rangle$ then
\begin{equation}\label{eq:Automorphism Gate on X}
U_{\alpha}X_{G}(g)U_{\alpha}^{\dagger}|h\rangle=|\alpha(\alpha^{-1}(h)+g)\rangle=|h+\alpha(g)\rangle=X_{G}(\alpha(g))|h\rangle,
\end{equation}
\begin{equation}\label{eq:Automorphism Gate on Z}
U_{\alpha}Z_{G}(\mu)U_{\alpha}^{\dagger}|h\rangle=\chi_{\mu}(\alpha^{-1}(h))|h\rangle=\chi_{\alpha^{*^{\inverse}}(\mu)}(h)|h\rangle=Z_{G}\left(\alpha^{*^{\inverse}}(\mu)\right)|h\rangle,
\end{equation}
where $\alpha^*$ is the dual group automorphism (\ref{eq:Dual Automorphism DEF}).
\item If $U$ is a quadratic phase gate $D_{\xi}$ associated  
with a quadratic function $\xi$  then
\begin{align}\label{eq:QuadraticPhase Gate on X}
D_{\xi}X_{G}(g)D_{\xi}^{\dagger}|h\rangle &= \xi(g+h)\overline{\xi(h)}|g+h\rangle=\xi(g)B(g,h)|g+h\rangle \notag\\
 & =  \xi(g)\chi_{\beta(g)}(h)|g+h\rangle=\xi(g)X(g)Z(\beta(g))|h\rangle,
\end{align}
where, in the second line, $\beta$ is the group homomorphism in the bi-character normal form of lemma \ref{lemma:Normal form of a bicharacter 1}. Moreover $D_{\xi}Z_{G}(\mu)D_{\xi}^{\dagger}= Z_{G}(\mu)$ since diagonal gates commute.
\item If $U$ is the  Fourier transform $\Fourier{G}$ on the  $\mathcal{H}_{G}$ then
\begin{equation}\label{eq:Fourier transform on X and Z}
\Fourier{G} X_{G}(g)\Fourier{G}^{\dagger}=Z_{G^*}(g),\qquad\quad\quad \Fourier{G} Z_{G}(\mu)\Fourier{G}^{\dagger}=X_{G^*}(-\mu).
\end{equation}
The first identity is the content of lemma \ref{lemma:Fourier transform diagonalizes the Regular Representation}. The second is proved in a similar way:
\begin{align}
Z_G(\mu)\ket{\nu} &= Z_G(\mu) \left(\int_{G} \mathrm{d} h \chi_{-\nu}(h)\ket{h}\right) =  \int_{G} \mathrm{d} h \chi_{-\nu}(h)\chi_\mu(h)\ket{h} =  \int_{G} \mathrm{d} h \chi_{-(\nu-\mu)}(h)\ket{h}\notag\\
&= \ket{\nu-\mu} = X_{G^*}(-\mu) \ket{\nu},
\end{align}
where we apply (\ref{eq:Fourier Basis in terms of Characters}), lemma \ref{lemma:Character Multiplication} and (\ref{eq:Pauli operator type X, over G, eigenket definition},\ref{eq:Pauli operator type Z}). These formula also apply to partial Fourier transforms $\Fourier{G_i}$, since Pauli operators decompose as tensor products.\qedhere
\end{itemize}
\end{proof}

\section{Stabilizer states}\label{sect:Stabilizer States}

In this section we develop a stabilizer framework to simulate normalizer circuits over infinite abelian groups of the form $G=\Z^a\times\T^b\times \DProd{N}{c}$. As explained in section \ref{sect:Introduction2}, our techniques generalize methods given in chapter \ref{chapterF} (which apply to groups of the form $F=\DProd{N}{c}$) and are closely related to the (more general) monomial stabilizer formalism~\cite{nest_MMS}.

 \subsection{Definition and basic properties}

A \emph{stabilizer group $\mathcal{S}$}  over  $G$ is any group of \emph{commuting} Pauli operators of $G$ with a nontrivial +1 common eigenspace.  
Here we are interested in stabilizer groups where the +1 common eigenspace is one-dimensional, i.e. there exists a state $|\psi\rangle$ such that $\sigma|\psi\rangle = |\psi\rangle$ for all $\sigma\in {\cal S}$, and moreover $|\psi\rangle$ is the unique state (up to normalization) with this property. Such states are called stabilizer states (over $G$). This terminology is an extension of the already established stabilizer formalism for finite-dimensional systems (chapter \ref{chapterF},  \cite{Gottesman_PhD_Thesis,Gottesman99_HeisenbergRepresentation_of_Q_Computers,Knill96non-binaryunitary,Gottesman98Fault_Tolerant_QC_HigherDimensions}).

We stress here  that stabilizer states $|\psi\rangle$ are allowed to be unnormalizable states;  
in other words, we do not require $|\psi\rangle$ to belong to the physical Hilbert space $\mathcal{H}_{G}$. In a more precise language, stabilizer states may be tempered distributions in the Schwartz-Bruhat space $\mathcal{S}_G^\times$ \cite{Bruhat61_Schwatz-Bruhat-functions,Osborne75_Schwartz_Bruhat}. This issue arises only when considering infinite groups, i.e. groups containing  $\Z$ or $\T$. An example of a non-physical stabilizer state is the  Fourier basis state $|p\rangle$ (\ref{eq:Fourier basis state over Z})  (we argue below that this is indeed a stabilizer state). Note that not all stabilizer states for $G=\T$ must be unphysical; an example of a physical stabilizer state within $\mathcal{H}_{G}$ is 
\begin{equation}
\int_{\T} dp |p\rangle \quad \xrightarrow{\textnormal{QFT over } \T}\quad \ket{0},0\in\Z.
\end{equation} 
The stabilizer group of this state is $\{X_\T(p): p\in \T\}$, which can be alternatively written as $\{Z_\Z(p): p\in \T\}$ (lemma \ref{lemma:Fourier transform diagonalizes the Regular Representation}). Similar examples of stabilizer states within and outside $\mathcal{H}_{G}$ can be given for $G=\Z$. Note, however, that in this case the standard basis states $|x\rangle$ with $x\in \Z$ (which are again stabilizer states) \emph{do} belong to $\mathcal{H}_{\Z}$.

Next we show that all standard basis states are stabilizer states. 
\begin{lemma}\label{lemma:Stabilizer Group for Standard Basis State}
Consider $G= \Z^a\times \T^b\times F$ with associated Hilbert space $\mathcal{H}_{G}$ and standard basis states $\{|g\rangle:g\in G\}$. Then every standard basis state $|g\rangle$ is a stabilizer state. Its stabilizer group is \be\label{stabilizer_g}
\{\overline{\chi_{\mu}}(g)Z_G(\mu): \mu\in G^*\}.\ee
\end{lemma}
The lemma implies that the Fourier basis states and, in general,
any of the allowed group-element basis states (11) are stabilizer states.
 
\begin{proof} Let us first prove the theorem for $g=0$, and show that $\ket{0}$ is the unique state that is stabilized by $\mathcal{S}=\{Z_G(\mu): \mu\in G^*\}$. It is easy to check that a standard-basis state $\ket{h}$ with $h\in G$ is a common +1-eigenstate of $\mathcal{S}$ if and only if $\chi_\mu(h)=1$ for all $\mu\in G^*$ or, equivalently, iff $h$ belongs to $G^\perp$, the annihilator of $G$. It is known that $G^\perp$ coincides with the trivial subgroup $\{0\}$ of $G^*$ \cite[corollary 20.22]{Stroppel06_Locally_Compact_Groups}, and therefore $\ket{0}$ is the unique standard-basis state that is also a common +1 eigenstate of $\mathcal{S}$. Since all unitary operators of $\mathcal{S}$ are diagonal in the standard basis, $\ket{0}$ is the unique common +1 eigenstate of $\mathcal{S}$.

For arbitrary $\ket{g}=X_G(g)\ket{0}$, the stabilizer group of $\ket{g}$ is $X_G(g)\mathcal{S}X_G(g)^\dagger$, which equals $\{\overline{\chi_{\mu}}(g)Z_G(\mu): \mu\in G^*\}$ (see equation  (\ref{eq:pauli_commutation})).

\end{proof}
Let $|\psi\rangle$ be a stabilizer state with stabilizer group ${\cal S}$. We define the following sets, all of which are easily  
verified to be abelian groups:
\begin{align}\label{label_groups}
\mathbb{L}&:=\{(\mu,g)\in G^*\times G\ : \ \mathcal{S} \textnormal{ contains a Pauli operator of the form }\gamma Z(\mu)X(g)\};\nonumber\\
\mathbb{H}&:=\{g\in G\ : \ \mathcal{S} \textnormal{ contains a Pauli operator of the form }\gamma Z(\mu)X(g)\};\nonumber\\
\mathbb{D}&:=\{\mu\in G^*\ : \ \mathcal{S} \textnormal{ contains a Pauli operator of the form }\gamma Z(\mu)\}
\end{align}
The groups $\mathbb{L}$, $\mathbb{D}$ and $\mathbb{H}$ contain information about the labels of the operators in ${\cal S}$. We highlight that, although $\mathbb{D}$ and $\mathbb{H}$ are subsets of very different groups (namely $G$ and $G^*$, respectively),  they are actually closely related to each other by the relation
\begin{equation}
\mathbb{H}\subseteq \mathbb{D}^\perp \quad (\mbox{or, equivalently, } \mathbb{D}\subseteq \mathbb{H}^\perp),
\end{equation}
which follows from the commutativity of the elements in $\mathcal{S}$ and the definition of annihilator (recall sections \ref{sect:pauli_def} and \ref{sect:Annihilators}). 

Finally, let $\mathcal{D}$ be the subgroup of all diagonal Pauli operators of $\mathcal{S}$. It is easy to see that, by definition, $\mathcal{D}$ and $\mathbb{D}$ are isomorphic to each other.

\subsection{Support of a stabilizer state}

We show that the support of a stabilizer state $\ket{\psi}$ (the manifold of points where the wavefunction $\psi(x)$ is not zero) can be fully characterized in terms of the  label groups $\mathbb{H}$, $\mathbb{D}$.

Our next result characterizes the structure of this wavefunction.
\begin{lemma}
Every stabilizer state \label{lemma:StabStates_are_Uniform_Superpositions}$\ket{\psi}$ over $G$ is a uniform quantum superposition over some subset of the form $s+\mathbb{H}$, where $\mathbb{H}\subset G$ is the label subgroup  defined as in (\ref{label_groups}). Equivalently, any stabilizer state  $\ket{\psi}$ can be writte in the form
\begin{equation}\label{eq:Stabilizer_States_r_Uniform_Superpositions}
\ket{\psi}=\int_{\mathbb{H}} \mathrm{d}h \, \psi(h) \ket{s+h}
\end{equation}
where $\mathrm{d}h$ is the Haar measure over $\mathbb{H}$ and all amplitudes have equal absolute value $|\psi(h)|=1$. We call the $s+\mathbb{H}$ the \emph{support} of $\ket{\psi}$.
\end{lemma}
 
This lemma generalizes corollary 1 in \cite{VDNest_12_QFTs}.
\begin{proof}
 Let $\ket{\psi}$ be an arbitrary quantum state $\ket{\psi}=\int_X\mathrm{d}g\, \psi(g)\ket{g}$. The action of an arbitrary Pauli operator $U=\gamma Z_G(\mu)X_G(h)\in \mathcal{S}$ on the state is
\begin{equation}\label{eq:inproof_Support_Stabilizer_State 0}
U\ket{\psi}= \gamma \int_X\mathrm{d}g\, \chi_{\mu}(g+h)\psi(g)\ket{g+h} = \int_X\mathrm{d}g\, \psi(g)\ket{g} =\ket{\psi}.
\end{equation}
Recall the definition of $\mathbb{H}$ in (\ref{label_groups}). Comparing the two  integrals in (\ref{eq:inproof_Support_Stabilizer_State 0}), and knowing that $|\chi_\mu(x)|=1$ for every $x\in G$, we find that the absolute value of $\psi$ cannot change if we shift this function  by an element of $\mathbb{H}$; in other words,
\begin{equation}\label{eq:inproof_Support_Stabilizer_States 1}
\textit{for every $g\in X$ it holds $|\psi(g)|=|\psi(g+h)|$ for every $h\in\mathbb{H}$.}
\end{equation}
Now let $Y\subset X$  denote the subset of points $y\in X$ for which $\psi(y)\neq0$. Eq.\ (\ref{eq:inproof_Support_Stabilizer_States 1}) implies that $Y$ is a disjoint union of cosets of $\mathbb{H}$, i.e.\
\begin{equation}
Y=\bigcup_{\iota\in I} s_\iota + \mathbb{H},
\end{equation}
where $I$ is a (potentially uncountable) index set, and that $\ket{\psi}$ is of the form
\begin{equation}\label{eq:inproof_Support_Stabilizer_States 2}
\ket{\psi} = \int_Y\mathrm{d}y\, \psi(y)\ket{y}  = \int_I \mathrm{d} \iota\, \alpha(\iota) \ket{\phi_\iota},
\end{equation}
where the states $\ket{\phi_\iota}$ are \emph{non-zero} linearly-independent uniform superpositions over  the cosets $x_\iota+\mathbb{H}$:
\begin{equation}\label{eq:inproof_Support_Stabilizer_States_3}
\ket{\phi_\iota}=\int_{\mathbb{H}}\mathrm{d}h\, \phi_\iota(h)\ket{s_\iota + h}
\end{equation}
and $|\phi_\iota(h)|=1$ for every $h$. Putting together (\ref{eq:inproof_Support_Stabilizer_States 2}) and (\ref{eq:inproof_Support_Stabilizer_States_3}) we conclude that, for any $U\in \mathcal{S}$, the condition $U\ket{\psi}=\ket{\psi}$ is fulfilled  if and only if $U\ket{\phi_\iota}=\ket{\phi_\iota}$ for every $\ket{\phi_\iota}$:  this holds because $U$ leaves invariant the mutually-orthogonal vector spaces  $\mathcal{V}_\iota:=\mathrm{span}\{\ket{s_\iota + h}:\: h\in \mathbb{H}\}$. Consequently,  every state $\ket{\phi_\iota}$ is a (non-zero) common +1 eigenstate of all operators in $\mathcal{S}$.  
Finally, since we know that  $\ket{\psi}$ is the \emph{unique} +1 common eigenstate of $\mathcal{S}$,  it follows from  (\ref{eq:inproof_Support_Stabilizer_States 2}, \ref{eq:inproof_Support_Stabilizer_States_3}) that  $I$ has exactly one element and $Y=s+\mathbb{H}$; as a result,  $\ket{\psi}$ is a uniform superposition of the form (\ref{eq:inproof_Support_Stabilizer_States_3}). This proves the lemma.
\end{proof}
\begin{lemma}\label{lemma:Support of StabStates}
An element $x\in G$  belongs to the support $s+\mathbb{H}$ of a stabilizer state $\ket{\psi}$  iff
\begin{equation}\label{eq:Support of Stabilizer State 1}
D\ket{x}=\ket{x} \quad \textrm{for all $D\in \mathcal{D}$}.
\end{equation}
Equivalently, using that   $D=\gamma_\mu Z_{G}(\mu)$ for some $\mu\in\mathbb{D}$ and that $\gamma_\mu$ is determined by $\mu$, we get
\begin{equation}\label{eq:Support of Stabilizer State 2}
\mathrm{supp}(\ket{\psi}) =\{ x\in G\: : \: \chi_\mu(x)=\overline{\gamma_\mu}\textrm{ for all $\mu\in \mathbb{D}$}\}.
\end{equation}
\end{lemma}
Lemma \ref{lemma:Support of StabStates} was proven for finite groups in \cite{VDNest_12_QFTs} and by us in chapter \ref{chapterF}, partially exploiting the monomial stabilizer formalism (MSF) developed in \cite{nest_MMS}. Since the MSF framework has not been generalized to infinite dimensional Hilbert spaces, the techniques in \cite{nest_MMS,VDNest_12_QFTs} and chapter \ref{chapterF} can no longer be applied in our present setting\footnote{The authors believe that the MSF formalism in \cite{nest_MMS} should be easy to extend to infinite dimensional systems if one looks at monomial stabilizer groups with normalizable eigenstates. However, dealing with monomial operators with unnormalizable eigenstates---which can be the case for (\ref{eq:Pauli operator type X, over G, shift definition},\ref{eq:Pauli operator type Z})---seems to be notoriously harder.}. Our proof works in infinite dimensions and even in the case when the Pauli operators (\ref{eq:Pauli operator type X, over G, shift definition},\ref{eq:Pauli operator type Z}) have unnormalizable eigenstates.
\begin{proof}
Write  $\ket{\psi}$ as in (\ref{eq:Stabilizer_States_r_Uniform_Superpositions}) integrating over  $X:=s+\mathbb{H}$. Then, the ``if'' condition follows easily by evaluating the action of an arbitrary diagonal stabilizer operator $D=\gamma_\mu Z_{G}(\mu)$  on a the stabilizer state $\ket{\psi}$: indeed, the condition
\begin{equation}
D\ket{\psi}=\ket{\psi} \quad \iff\quad  \int_{X}\mathrm{d}x\, \left(\gamma_\mu \chi_{\mu}(x)\right)\psi(x)\ket{x} = \int_X\mathrm{d}x\, \psi(x)\ket{x},
\end{equation}
holds only if $\gamma_\mu \chi_\mu (x) =1$, which is equivalent to $D\ket{x}=\ket{x}$ (here, we use implicitly that $\psi(x)\neq0$ for all integration points).

Now we prove the reverse implication. Take $x\in G$ such that  $D\ket{x}=\ket{x}$ for all $D\in \mathcal{D}$. We want to show that  $\ket{x}$ belongs to the set $s+\mathbb{H}$. We argue by contradiction, showing that $x\notin s+\mathbb{H}$ implies that there exists a nonzero  common +1 eigenstate $\ket{\phi}$ of all $\mathcal{S}$ that is not proportional to $\ket{\psi}$, which cannot happen.

We now show how to  construct such a $\ket{\phi}$.

Let   $Y:=\{\xi(\mu, g)Z_{G}(\mu)X_G(g)\}$ be  a system of representatives of the factor group $\mathcal{S}/\mathcal{D}$. For every $h\in \mathbb{H}$, we use the notation $V_h$ to denote a Pauli operator of the form $\xi(\nu_h, h)Z_{G^*}(\nu_h)X_G(h)$. It is easy to see that the set of all such $V_h$ forms an equivalence class in $\mathcal{S/D}$, so that there is a one-to-one correspondence between $\mathbb{H}$ and $\mathcal{S/D}$. Therefore, if to every $h\in\mathbb{H}$ we associate a $U_h\in Y$ (in a unique way), written as $U_h:=\xi(\nu_h,h) Z_{G*}(\nu_h)X(h)$, then we have that:
\begin{itemize}
\item[(a)] any Pauli operator  $V\in\mathcal{S}$ can be written as $V=U_x D$ for some $U_x\in Y$ and $D\in\mathcal{D}$;
\item[(b)]   $U_g U_h =U_{g+h} D_{g,h}$ for every $U_g$, $U_h\in Y$ and some $D_{g,h}\in \mathcal{D}$.
\end{itemize}
With this conventions, we take $\phi$ to be the state
\begin{equation}\label{eq:inproof_Support of Stab States 1}
\ket{\phi}:=\left( \int_{Y} \mathrm{d}U \, U\right ) \ket{x} = \left( \int_{\mathbb{H}} \mathrm{d}h \, U_h\right ) \ket{x} = \int_{\mbb{H}} \mathrm{d}h \,  \xi(\nu_h, h) \chi_{\nu_h}(x+h)  \ket{x+h} \textrm{d}{h}.
\end{equation}
The last equality in (\ref{eq:inproof_Support of Stab States 1}) shows that $\ket{\phi}$ is a uniform superposition over $x+\mathbb{H}$. As a result, $\ket{\phi}$ is non-zero. Moreover, $\ket{\phi}$ linearly independent from $\ket{\psi}$ if we assume $x\notin \mathrm{supp}(\psi)$, since this  implies that $\mathrm{supp}(\phi) = x+\mathbb{H}$ and $\mathrm{supp}(\psi)= s+\mathbb{H}$  are disjoint.   
Lastly, we prove that $\ket{\phi}$ is  stabilized by all Pauli operators in $\mathcal{S}$. First, for any diagonal stabilizer $D$ we get
\begin{equation}
D\ket{\phi}=D\left( \int_{Y} \mathrm{d}U \, U\right ) \ket{x} = \left( \int_{Y} \mathrm{d}U \, U\right ) D \ket{x} = \left( \int_{Y} \mathrm{d}U \, U\right ) \ket{x},
\end{equation}
due to commutativity and the promise that $D\ket{x}=\ket{x}$. Also, any stabilizer of the form $U_x$ from the set of representatives $Y$ fulfills
\begin{align}
U_x \ket{\phi}&=U_x\left( \int_{\mathbb{H}} \mathrm{d}h\, U_h\right)\ket{x} =\left( \int_{\mathbb{H}} \mathrm{d}h\, U_x U_h\right ) \ket{x} = \left( \int_{\mathbb{H}} \mathrm{d}h\, U_{x+h} \right) D_{x,h} \ket{x}\\
&= \left( \int_{\mathbb{H}} \mathrm{d}h'\, U_{h'} \right) \ket{x} = \ket{\phi}
\end{align}
Hence, using property (a) above, it follows that any arbitrary stabilizer $V$ stabilizes $\ket{\phi}$ as well.
\end{proof}

\begin{corollary}\label{corollary:StabStates have Closed Support}
The sets $\mathbb{H}$ and $\mathrm{supp}(\ket{\psi})=s+\mathbb{H}$ are closed.
\end{corollary}
\begin{proof}
It follows from $(\ref{eq:Support of Stabilizer State 2})$ that supp($\ket{\psi}$) is of the form $x_0+\mathbb{D}^\perp$.  Putting this together with (\ref{eq:Stabilizer_States_r_Uniform_Superpositions}) in lemma \ref{lemma:StabStates_are_Uniform_Superpositions} it follows that $\mathbb{H}=\mathbb{D}^\perp$. Since any annihilator is closed (lemma \ref{lemma:Annihilator properties}),  $\mathbb{H}$ is closed. Since the group operation of $G$ is a continuous map\footnote{This is a fundamental property of topological groups. Consult e.g.\ \cite{Stroppel06_Locally_Compact_Groups,Dikranjan11_IntroTopologicalGroups} for details.},  $s+\mathbb{H}$ is closed too.
\end{proof}

\section{Proof of theorem \ref{thm:Main Result} }\label{sect:Proof of theorem 1}
 
In this section we prove our main result (theorem \ref{thm:Main Result}). As anticipated, we divide the proof in three parts. In section \ref{sect:Tracking normalizer evolutions with stabilizer groups}, we show that the evolution of the quantum state during a normalizer computation can be tracked efficiently using \textbf{stabilizer groups} (which we introduced in the previous section). In section \ref{sect:Computing the support of the final state} we show how to compute the support of the final quantum state by reducing the problem to solving systems of \textbf{{linear equations over an abelian group}}, which can be reduced to systems of mixed real-integer linear equations \cite{BowmanBurget74_systems-Mixed-Integer_Linear_equations} and solved with the  classical algorithms presented in section  \ref{sect_CompComp_FiniteAbelianGroups}. Finally, in section \ref{sect:Sampling the support of a state}, we show how to simulate the final measurement of a normalizer computation by developing \textbf{net techniques} (based, again, on techniques of section \ref{sect_CompComp_FiniteAbelianGroups}) to sample from the support of the final state.

\subsection{Tracking normalizer evolutions with stabilizer groups}\label{sect:Tracking normalizer evolutions with stabilizer groups}

As in the celebrated Gottesman-Knill theorem \cite{Gottesman_PhD_Thesis,Gottesman99_HeisenbergRepresentation_of_Q_Computers}  and  its existing generalizations (cf.\ chapter \ref{chapterF}), our approach will be to track the evolution of the system in a stabilizer picture. Since we know that the initial state $|0\rangle$ is a stabilizer state (lemma \ref{lemma:Stabilizer Group for Standard Basis State}) and that normalizer gates are Clifford operations (lemma \ref{lemma:Evolution of Pauli labels}), it follows that the quantum state at every time step of a normalizer computation is  a stabilizer state. It is thus tempting to use stabilizer groups of abelian-group Pauli operators to classically describe the evolution of the system during the computation; this is the approach we used in  chapter \ref{chapterF} to simulate normalizer circuits over finite abelian groups. 

However, complications arise compared to all previous cases where normalizer circuits are associated to a finite group $G$. We discuss these issues next.\\

\textbf{Stabilizer groups are infinitely generated.} A common ingredient in all previously known methods to simulate Clifford circuits and normalizer circuits over finite abelian groups can no longer be used in our setting: traditionally\footnote{As discussed in section ``Relationship to previous work'', there are a few simulation methods \cite{VdNest10_Classical_Simulation_GKT_SlightlyBeyond,Veitch12_Negative_QuasiProbability_Resource_QC,MariEisert12_Positive_Wigner_Functions_Quantum_Computation} for Clifford circuits that are not based on stabilizer-groups, but they are more limited than stabilizer-group methods: the Schrödinger-picture simulation in \cite{VdNest10_Classical_Simulation_GKT_SlightlyBeyond} is  for non-adaptive qubit Clifford circuits; the Wigner-function simulation in \cite{Veitch12_Negative_QuasiProbability_Resource_QC,MariEisert12_Positive_Wigner_Functions_Quantum_Computation} is for odd-dimensional qudit Clifford circuits (cf.\ also section \ref{sect:Introduction2}).}, simulation algorithms based on stabilizer groups keep track of a list of (polynomially  many) generators of a stabilizer group, which can be updated to reflect the action of Clifford/normalizer gates. 
In our set-up,  this is a \emph{futile approach} because \emph{stabilizer groups over infinite abelian groups can have an \textbf{infinite} number of generators}. Consider for example the state $|0\rangle$ with $G=\Z$, which has a continuous stabilizer group  $\{Z_{G}(p)|p\in\mathbb{T}\}$ (lemma \ref{lemma:Stabilizer Group for Standard Basis State}); the group  that describes the labels of the Pauli operators is the circle group $\T$, which cannot be generated by a finite number of elements (since it is uncountable).\\

\textbf{Fourier transforms change the group $G$}. In chapter \ref{chapterF}, the group $G$ associated to a normalizer circuits was a parameter that does not change during the computation. In section \ref{sect:Evolution Pauli Operators} we discussed that our setting is now different, as Fourier transforms can change the group that labels the designated basis (theorem \ref{thm:Normalizer gates are Clifford}, eq.\ \ref{eq:Fourier transform on X and Z}); this reflects that groups (\ref{eq:GroupChapterI}) are not autodual. \\

In this section we will develop new methods to track the evolution of stabilizer groups, that deal with the issues mentioned above. 

From now on, unless stated otherwise, we consider a normalizer circuit ${\cal C}$ comprising $T$ gates. The input is the $\ket{0}$ state of a group $G$, which we denote by $G(0)$ to indicate that this group occurs at time $t=0$.  The stabilizer group of $|0\rangle $ is $\{ Z_G(\mu):  \mu \in\ G(0)^*\}$. The quantum state at any time $t$ during the computation will have the form $\ket{\psi(t)}=\mathcal{C}_t\ket{0}$ where ${\cal C}_t$ is the normalizer circuit containing the first $t$ gates of ${\cal C}$. This state is a stabilizer state over a group $G(t)$. The stabilizer group of $\ket{\psi(t)}$ is  ${\cal S}(t):=\{\mathcal{C}_t Z_{G}(\mu) \mathcal{C}_t^\dagger,\:  \mu\in G(0)^* \}$.

Throughout this section, we always assume that normalizer gates are given in the standard encodings defined in section \ref{sect:Main Result}.

\subsubsection*{Tracking the change of  group $G$}

 First, we show how to keep track of how the group $G$ that labels the designated basis changes along the computation. Let $G=G_1\times\dots\times G_m$ with each $G_i$ of primitive type. Define now the larger group $\Gamma:=G^*\times G$. Note that the labels $(\mu, g)$ of a Pauli operator $\gamma Z_G(\mu)X_G(g)$ can be regarded as an element of $\Gamma$, so that the transformations of these labels  in theorem \ref{thm:Normalizer gates are Clifford} can be understood as transformations of this group. We show next that the transformations induced on this group by normalizer gates are \emph{continuous group isomorphisms}, that can be stored in terms of matrix representations. This will give us a method to keep track of $G$ and $G^*$ at the same time. Studying the transformation of $\Gamma$ as a whole (instead of just $G$)  will be useful in the next section, where we consider the evolution of Pauli operators.

First, note that both automorphism gates and quadratic phase gates leave $G$ (and thus $\Gamma$) unchanged (theorem \ref{thm:Normalizer gates are Clifford}). We can keep track of this effect by storing the $2m\times 2m$ identity matrix $I_{2m}$ (the matrix clearly defines a group automorphism of $\Gamma$). Moreover, (\ref{eq:Fourier transform on X and Z}) shows that Fourier transforms just induce a signed-swap operation on the factors of $\Gamma$. We can associate a $2m\times 2m$ matrix $S_i$ to this operation, defined as follows: $S_i$ acts  non-trivially (under multiplication) only on the factors $G_i^*$ and $G_i$; in the  subgroup  $G_i^*\times G_i$ formed by these factors $S_i$ acts as 
 \begin{equation}\label{eq:Signed Swap}
 (\mu(i), g(i)) \in G_i^* \times G_i \quad \longrightarrow \quad (g(i), -\mu(i)) \in G_i \times G_i^*.
 \end{equation}
By construction, $\Gamma'=S_i\Gamma$. Manifestly, $S_i$ defines a group isomorphism $S_i:\Gamma\rightarrow\Gamma'$.

Lastly, let $G(t)$ denote the underlying group at time step $t$ of the computation. Define $\Gamma(t):=G^*(t)\times G(t)$ and let $V_1,\ldots,V_t$ be the matrices associated to the first $t$ gates describing the transformations of $\Gamma$. Then, we have $\Gamma(t)=V_t V_{t-1}\cdots V_1 \Gamma(0)$, so that it is enough to store the matrix $V_t V_{t-1}\cdots V_1 $ to keep track of the group $\Gamma(t)$. 

\subsubsection*{Tracking Pauli operators}

We deal next with the fact that we can no longer store the ``generators'' of a stabilizer group.  
We will exploit a crucial mathematical property of our stabilizer groups:  for any stabilizer group ${\cal S}$ arising along the course of a normalizer circuit, we will show that there always exists a classical description for $\mathcal{S}$ consisting of a triple $(\Lambda, M, v)$ where $\Lambda$ and $M$ are real matrices and $v$ is a real vector. If we have $G=\T^a\times \Z^b \times \DProd{N}{c}$ with $m=a+b+c$, then all elements of the triple  $(\Lambda,M,v)$ will  have $O(\poly{m})$ entries. As a result, we can use these triples to describe the stabilizer state $\ket{\psi}$ associated to ${\cal S}$ efficiently classically. Moreover, we shall show (lemmas \ref{lemma:Evolution of Pauli labels}, \ref{lemma:Evolution of Pauli Phases}) that the description  $(\Lambda, M, v)$ can be \emph{efficiently transformed} to track the evolution of $\ket{\psi}$ under the action of a normalizer circuit.

Let $\Gamma(t)$ be the group $G^*(t)\times G(t)$. Recalling the definition of the group $\mathbb{L}$ in (\ref{label_groups}), we denote by $\mathbb{L}(t)\subseteq \Gamma(t)$ this group at time $t$. We want to keep track of this group in a way that does not involve storing an infinite number of generators. As a first step, we consider the initial standard basis state  $|0\rangle$, where \be \mathbb{L}(0)= \{(\mu,0):\mu\in G(0)^*\}.\ee  
A key observation is that this group can be written as the image of a continuous group homomorphism
\be \Lambda_0: (\mu, g)\in \Gamma(0)\rightarrow (\mu, 0)\in \Gamma(0);\ee it is easy to verify  $\mathbb{L}(0) = \mathrm{im}\,\Lambda_0$. Therefore, in order to keep track of the (potentially uncountable) set $\mathbb{L}(0)$ it is enough to store a $2m\times 2m$ matrix representation of $\Lambda_0$ (which we  denote by the same symbol):
 \begin{equation}\label{Lambda_0}
\Lambda_0  =\begin{pmatrix}
I & 0 \\ 0 & 0
\end{pmatrix}
 \end{equation}
Motivated by this property,  we will track the evolution of the group $\mathbb{L}(t)$ of Pauli-operator labels by means of a matrix representation of a group homomorphism: $\Lambda_t: \Gamma(0)\rightarrow\Gamma(t)$ whose image is precisely $\mathbb{L}(t)$. The following lemma states that this approach works.
\begin{lemma}[\textbf{Evolution of Pauli labels}]\label{lemma:Evolution of Pauli labels}
There exists a group homomorphism $\Lambda_t$
from $\Gamma(0)$ to $\Gamma(t)$ satisfying
\begin{equation}
\mathbb{L}(t)=\mathrm{im}\, \Lambda_t.
\end{equation}
Moreover,  a  matrix representation of $\Lambda_t$ can be computed in classical polynomial time,  using $O(\ppoly{m,t})$ basic arithmetic operations.
\end{lemma}
\begin{proof}
We show this by induction. As discussed above, at $t=0$
we choose  $\Lambda_0$ as in (\ref{Lambda_0}). Now, given the homomorphism $\Lambda_t$ at time $t$, we show how to compute $\Lambda_{t+1}$ for every type of normalizer gate. The proof relies heavily on the identities in the proof of theorem \ref{thm:Normalizer gates are Clifford}. We also note that the equations below are for groups of commuting Pauli operators but they can be readily applied to any single Pauli operator just by considering the stabilizer group it generates.
\begin{itemize}
\item Automorphism gate $U_{\alpha}$: Let $A$ be a matrix representation of $\alpha$; then equations (\ref{eq:Automorphism Gate on X})-
(\ref{eq:Automorphism Gate on Z}) imply

\begin{equation}\label{inproof:LabelUpdateAutomorphismGate}
\Lambda_{t+1}=\left(\begin{array}{cc}
A^{*^{\inverse}} & 0\\
0 & A
\end{array}\right)\Lambda_t.
\end{equation}
The matrix $A^{*^{\inverse}}$ can be computed efficiently due to lemmas  \ref{lemma:Computing Inverses} and  \ref{lemma:properties of matrix representations}.(b).
\item Quadratic phase gate $D_{\xi}$: suppose that $\xi$ is a $B$-representation for some bicharacter $B$ (recall section \ref{sect:quadratic_functions}). Let $M$ be a matrix representation of the homomorphism $\beta$ that appears in lemma \ref{lemma:Normal form of a bicharacter 1}.  Then  (\ref{eq:QuadraticPhase Gate on X}) implies
\begin{equation}
\Lambda_{t+1}=\left(\begin{array}{cc}
I & M \\
0 & I
\end{array}\right)\Lambda_t.
\end{equation}
\item Partial Fourier transform $\Fourier{G_i}$: recalling (\ref{eq:Fourier transform on X and Z}), we  simply have
\begin{equation}
\Lambda(t+1)=S_i\Lambda(t),
\end{equation}
with
\begin{equation}
S_i = 
\begin{pmatrix}
\begin{matrix}
\textbf{ \Large 1} & & \\
& 0 & \\
& & \textbf{ \Large 1}
\end{matrix}  &\vline &
\begin{matrix}
\textbf{ \Large 0} & & \\
& 1 & \\
& & \textbf{ \Large 0}
\end{matrix} \\
\hline
\begin{matrix}
\textbf{ \Large 0} & & \\
& -1 & \\
& & \textbf{ \Large 0}
\end{matrix} &\vline & \begin{matrix}
\textbf{ \Large 1} & & \\
& 0 & \\
& & \textbf{ \Large 1}
\end{matrix}
\end{pmatrix}
\end{equation}
where the $ \begin{pmatrix}
0 & 1 \\ -1 & 0
\end{pmatrix}$ subblock in $S_i$ corresponds to the $i$-th entries of $G^*$ and $G$. \qedhere
\end{itemize}
\end{proof}

We now show how the phases of the Pauli operators in ${\cal S}(t)$ can be tracked.  

Suppose that there exists $(\mu, g)\in \mathbb{L}$ and complex phases $\gamma$ and $\beta$ such that both \be \sigma:=\gamma Z(\mu)X(g)\quad \mbox{ and } \tau:=\beta Z(\mu)X(g)\ee belong to ${\cal S}$.   
Then $\sigma^{\dagger}\tau$ must also belong to ${\cal S}$, where $\sigma^{\dagger}\tau= \overline{\gamma}\beta I $ with $I$ the identity operator. But this implies that $\overline{\gamma}\beta|\psi\rangle = |\psi\rangle$, so that $\gamma = \beta$. This shows that the phase of $\sigma$ is uniquely determined by the couple $(\mu, g)\in\mathbb{L}$.
We may thus define a function $\gamma: \mathbb{L}\to U(1)$ such that \be {\cal S}= \{\gamma(\mu, g)Z(\mu)X(g):\ (\mu, g)\in \mathbb{L}\}.\ee
\begin{lemma}\label{thm_gamma_quadratic}
The function $\gamma$ is a quadratic function on $\mathbb{L}$.
\end{lemma}
\begin{proof}
By comparing the phases of two stabilizer operators $\sigma_1=\gamma(\mu_1,g_1) Z_{G}(\mu_1)X_{G}(g_1)$ and $\sigma_2=\gamma(\mu_2,g_2)\allowbreak Z_{G}(\mu_2)X_{G}(g_2)$ to the phase $ \gamma((\mu_1,g_1)+(\mu_2,g_2))$ of their product operator $\sigma_2\sigma_1$, we obtain
 \begin{equation}
 \gamma((\mu_1,g_1)+(\mu_2,g_2))=\gamma(\mu_1,g_1)\gamma(\mu_2,g_2)\overline{\chi_{\mu_2}(g_1)},
 \end{equation}
 which implies that $\gamma$ is quadratic.
 \end{proof}
Although it does not follow from lemma \ref{thm_gamma_quadratic}, in our setting, the quadratic function $\gamma$ will always be \emph{continuous}. As a result, we can apply the normal form given in theorem \ref{thm:Normal form of a quadratic function} to describe  the phases of the Pauli operators of a stabilizer group. Intuitively, $\gamma$ must be continuous in our setting, since  this is the case for the allowed family of input states (lemma \ref{lemma:Stabilizer Group for Standard Basis State}) and normalizer gates continuously transform Pauli operators under conjugation; this is rigorously shown using induction in the proof of  theorem \ref{thm_evolution_phases}.

We will use that these phases of Pauli operators are described by quadratic functions on $\mathbb{L}(t)$ (recall lemma \ref{thm_gamma_quadratic}).  In particular, theorem \ref{thm:Normal form of a quadratic function} shows that every quadratic function can be described by means of an $m\times m$  matrix $M$ and a $m$-dimensional vector $v$. For the initial state $\ket{0}$, we  simply set both $M$, $v$ to be zero. The next lemma shows that $M$, $v$ can be efficiently updated during any normalizer computation.

\begin{lemma}[\textbf{Evolution of Pauli phases}]\label{lemma:Evolution of Pauli Phases} \label{thm_evolution_phases}
At every time step $t$ of a normalizer circuit, there exists a $2m\times 2m$ rational matrix $M_t$ and a $m$-dimensional rational vector $v_t$ such that the quadratic function describing the phases of the Pauli operators in $\mathcal{S}(t)$ is $\xi_{M_t,v_t}$ (as in theorem \ref{thm:Normal form of a quadratic function}). Moreover, $M_t$ and $v_t$ can be efficiently computed classically with $O(\ppoly{m,n})$ basic arithmetic operations.
\end{lemma}
\begin{proof} The proof is similar to the proof of lemma \ref{lemma:Evolution of Pauli labels}. We act by induction. At $t=0$ we just take $M_0$ to be the zero matrix and $v_0$ to be the zero vector. Then, given $M_t$ and $v_t$ at time $t$, we show how to compute $M_{t+1}$, $v_{t+1}$. In the following, we denote by $\mathbf{A}$ the matrix that fulfills $\Lambda_{t+1}=\mathbf{A}\Lambda_{t}$ in each case of lemma \ref{lemma:Evolution of Pauli labels} and write $(\mu',g')=\mathbf{A}(\mu,g)$ for every $(\mu, g)\in \Gamma_{t}$. Finally, let $\xi_t$ and $\xi_{t+1}$ denote the quadratic phase functions for ${\cal S}(t)$ and ${\cal S}(t+1)$, respectively.
\begin{itemize}
\item \textbf{Automorphism gate $U_\alpha$.} Let $A$, $A^{*^{\inverse}}$ be matrix representations of $\alpha$, $\alpha^{*^{\inverse}}$. Using (\ref{eq:Automorphism Gate on X}, \ref{eq:Automorphism Gate on Z}) we have
\begin{align}
\xi_t(\mu,g)Z_G(\mu)X_{G}(g)\:\xrightarrow{U_\alpha} \:\xi_t(\mu,g)Z_G(\mu')X_{G}(g')
\end{align}
with $(\mu',g')=\mathbf{A}(\mu, g)$ and $\mathbf{A}= \begin{pmatrix}
A^{*^{\inverse}} & 0 \\ 0 & A
\end{pmatrix}$. The matrix $A^{*^{\inverse}}$ can be computed using lemmas \ref{lemma:Computing Inverses} and \ref{lemma:properties of matrix representations}.(b). The phase $\xi_t(\mu,g)$ of the Pauli operator can be written now as a function  $\xi_{t+1}$ of $(\mu',g')$ defined as
\begin{equation}
\xi_{t+1}(\mu',g'):=\xi_{t}(\mathbf{A}^{-1}(\mu', g')) = \xi_{t}(\mu,g).
\end{equation}
The function is manifestly quadratic. By applying lemma \ref{lemma:Quadratic Function composed with Automorphism} we obtain
\begin{equation}
 M_{t+1}= \mathbf{A}^{-\transpose} M_t \mathbf{A}^{-1}, \qquad v_{t+1} =\mathbf{A}^{-\transpose} v_t + v_{\mathbf{A}^{-1},M_t},
\end{equation}
where $v_{\mathbf{A}^{-1},M_t}$ is defined as $v_{A,M}$ in lemma \ref{lemma:Quadratic Function composed with Automorphism}.

\item \textbf{Partial Fourier transform $\Fourier{G_i}$.}  The proof is analogous using that  $\mathbf{A}= S_i$. Since the Fourier transform at the register $i$th exchanges the order of the X and Z Pauli operators acting on the subsystem $\mathcal{H}_{G_i}$ (\ref{eq:Fourier transform on X and Z}), we locally exchange the operators using (\ref{eq:pauli_commutation}), gaining an extra phase. Assume for simplicity that $i=1$ and re-write $G=G_1\times \cdots\times G_m$ as $G= A \times B$; let $g=(a,b)$ and $\mu=(\alpha, \beta)$. Then  $\Fourier{G_1}$ acts trivially on $\mathcal{H}_{G'}$ and we get
\begin{align}\notag
\xi_t(\mu,g)Z_{G_1}(\alpha)X_{G_1}(a)\otimes U \:\xrightarrow{\Fourier{G_1}+\mathrm{reorder}} \:\left(\xi_{t}(\mu,g)\chi_{\left(\alpha,0\right)}(a,0)\right) Z_{G_1^*}(a)X_{G_1^*}(-\alpha)\otimes U.
\end{align}
In general, for arbitrary $i$, we gain a phase factor $\overline{\chi_{\left( 0,\ldots,\mu(i),\ldots,0\right)}{\left( (0,\ldots, g(i),\ldots,0)\right)}}$. Using the change of variables $(\mu',g')=\mathbf{A}(\mu, g)=S_i(\mu, g)$, we define $\xi_{t+1}$ to be function that carries on the accumulated phase of the operator. For arbitrary $i$ we obtain
\begin{align}
\xi_{t+1}(\mu',g')&:=\xi_{t}(\mu,g)\,\chi_{\left( 0,\ldots,\mu(i),\ldots,0\right)}{\left( (0,\ldots, g(i),\ldots,0)\right)}.
\end{align}
The character $\chi_{\left( 0,\ldots,\mu(i),\ldots,0\right)}{\left( (0,\ldots, g(i),\ldots,0)\right)}$ can be written as a quadratic function $\xi_{M_F,v_F}(\mu, g)$ with $v_F=0$ and 
\begin{equation}
M_{F}:=\begin{pmatrix}
\begin{matrix}
 & & \\
& \textbf{ {\Huge 0}} & \\
&&
\end{matrix}  &\vline &
\begin{matrix}
\textbf{ \Large 0} & & \\
& \Upsilon_G(i,i) & \\
& & \textbf{ \Large 0}
\end{matrix} \\
\hline
\begin{matrix}
\textbf{ \Large 0} & & \\
& \Upsilon_G(i,i) & \\
& & \textbf{ \Large 0}
\end{matrix} &\vline & \begin{matrix}
& & \\
& \textbf{ {\Huge 0}} & \\
& &
\end{matrix}
\end{pmatrix},
\end{equation}
where $\Upsilon_G(i,i)$ is the $i$th diagonal element of  $\Upsilon_G$ (\ref{eq:definition of bullet map}). Applying lemma \ref{lemma:Quadratic Function composed with Automorphism}  we obtain
\begin{equation}
 M_{t+1}= \mathbf{A}^{-\transpose}\left( M_t +M_{F}\right) \mathbf{A}^{-1}, \qquad v_{t+1} =\mathbf{A}^{-\transpose} v_t +  v_{\mathbf{A}^{-1},M_t+M_F}.
\end{equation}
\item \textbf{Quadratic phase gate $D_\xi$.} Let $\xi=\xi_{M_Q,v_Q}$ be the quadratic function implemented by the gate and $M_\beta$ be the matrix representation of $\beta$ as in (\ref{lemma:Normal form of a bicharacter 1}). We know from lemma \ref{lemma symmetric matrix representation of the bicharacter homomorphism} that $M_Q=\Upsilon_G M_\beta$. Using (\ref{eq:QuadraticPhase Gate on X}) and reordering Pauli gates (similarly to the previous case) we get
\begin{align}\notag
\xi_t(\mu,g)Z_{G}(\mu)X_{G}(g) \:\xrightarrow{D_\xi+\mathrm{reorder}} \:\left( \xi_{t}(\mu,g)\xi_{M_Q,v_Q}(g)\overline{\chi_{\beta(g)}(g)}\right) Z_{G}(\mu+\beta(g))X_{G}(g)
\end{align}
The accumulated phase can be written as a quadratic function $\xi_{M',v'}$ with  \begin{equation}
M':=  M_t + \begin{pmatrix}
 0 & 0 \\ 0 & M_Q
 \end{pmatrix} -\begin{pmatrix}
 0 & 0 \\ 0 & 2M_Q
 \end{pmatrix}, \qquad v' := v + \begin{pmatrix}
 0\\ v_Q
 \end{pmatrix}
 \end{equation}
 Using lemma \ref{lemma:Quadratic Function composed with Automorphism} and  $\mathbf{A}= \begin{pmatrix}
I & M_\beta \\ 0 & I
\end{pmatrix}$  (from the proof of lemma \ref{lemma:Evolution of Pauli labels}) we arrive at:
\begin{align}
 M_{t+1}&=\mathbf{A}^{-\transpose} M' \mathbf{A}^{-1}, \qquad
v' = \mathbf{A}^{-\transpose} v' +  v'_{\mathbf{A}^{-1}, M'}
 \end{align}\qedhere
\end{itemize}
\end{proof}
Combining lemmas \ref{lemma:Evolution of Pauli labels} and \ref{thm_evolution_phases}, we find that 
the triple $(\Lambda_t, M_t, v_t)$, which  constitutes a classical description of the stabilizer state $|\psi(t)\rangle$, can be efficiently computed for all $t$. This yields a poly-time algorithm to compute the description $(\Lambda_T, M_T, v_T)$ of the output state $|\psi_T\rangle$ of the circuit.  Henceforth we continue to work with this final state and drop the reference to $T$ throughout. That is, the final state is denoted by $|\psi\rangle$, which is a stabilizer state over $G$ with stabilizer  ${\cal S}$. The latter is described by the triple $(\Lambda, M, v)$, the map from $\Gamma(0)$ to $\Gamma$ is described by $\Lambda$, etc.

\subsection{Computing the support of the final state}\label{sect:Computing the support of the final state}

Given the triple $(\Lambda, M, v)$ describing the final state $|\psi\rangle$ of the computation,  
we now consider the problem of determining the support of $\ket{\psi}$. Recall that the latter has the form $x+\mathbb{H}$ where the label group $\mathbb{H}$ was defined in (\ref{label_groups}) and $x\in G$ is any element satisfying conditions (\ref{eq:Support of Stabilizer State 1}). Since $\mathbb{L}= \Lambda \Gamma(0)$ and $\Lambda$ is given, a description of $\mathbb{H}$ is readily obtained:  the $m\times 2m$ matrix $P = (0 \ I)$
is a matrix representation of the homomorphism  $(\mu, g)\in \Gamma\to g\in G$. It easily follows that $\mathbb{H}= P\Lambda\Gamma(0)$. Thus the matrix $P\Lambda$ yields an efficient description for $\mathbb{H}$. To compute an $x$ in the support of $|\psi\rangle$, we need to solve the equations (\ref{eq:Support of Stabilizer State 1}).  In the case of finite groups $G$, treated in chapter \ref{chapterF}, the approach consisted of first computing a (finite) set of generators $\{D_1, \dots, D_r\}$ of $\mathcal{D}$. Note that $x\in G$ satisfies (\ref{eq:Support of Stabilizer State 1}) if and only if $D_i|x\rangle = |x\rangle$ for all $i$. This gives rise to a finite number of equations. In chapter \ref{chapterF} we showed how such equations can be solved efficiently.  In contrast with such a finite group setting, here the group $G$, and hence also the group $\mathcal{D}$, can be continuous, so that $\mathcal{D}$ can in general not be described by a finite list of generators. Consequently, the approach followed for finite groups does no longer work. Next we provide an alternative approach to compute an $x$ in the support of $|\psi\rangle$ in polynomial time.

\subsubsection{Computing $\mathcal{D}$}

We want to solve the system of equations (\ref{eq:Support of Stabilizer State 2}). Our approach will be to reduce  
this problem to a system of linear equations over a group of the form (\ref{eq:Systems of linear equations over groups}) and apply the algorithm in theorem \ref{thm:General Solution of systems of linear equations over elementary LCA groups} to solve it. To compute $\mathcal{D}$ it is enough to find a compact way to represent $\mathbb{D}$, since we can compute the phases of the diagonal operators using the classical description $(\Lambda, M,v )$ of the stabilizer group. To compute $\mathbb{D}$ we argue as follows. An arbitrary element of $\mathbb{L}$ has the form $\Lambda u$ with $u\in \Gamma(0)$. Write $\Lambda$ in a block form
\be
\Lambda =
\begin{pmatrix}
\Lambda_1\\ \Lambda_2
\end{pmatrix}
\ee
so that $\Lambda u = (\Lambda_1 u, \Lambda_2 u)$ with $\Lambda_1 u\in G^*$ and $\Lambda_2 u\in G$. Then
\begin{equation}\notag
\mathbb{D} =\left\{\Lambda_1 u:  u \mbox{ satisfies } \Lambda_2 u \equiv 0 \mbox{ mod } G.
\right\}
\end{equation}
The equation $\Lambda_2 u \equiv 0 \mbox{ mod } G$ defines a linear system of constrains over a group as in  (\ref{eq:Systems of linear equations over groups}). This means (because of our algorithm in theorem \ref{thm:General Solution of systems of linear equations over elementary LCA groups}) that we can compute in polynomial time a \emph{general solution} of it and, in particular,  a   matrix representation  $\mathcal{E}_\mathbb{D}$ of a group homomorphism $\mathcal{E}_\mathbb{D}:\R^a\times \Z^b\rightarrow G^*$ whose image is precisely $\mathbb{D}$, i.e.: $$ \mathbb{D}=\{\mathcal{E}_\mathbb{D}w:\ w \in \R^a\times \Z^b\}.$$
In particular, this means that we can efficiently compute a classical description $\mathcal{E}_\mathbb{D}$ of $\mathbb{D}$.
\subsubsection{Computing the support $x_0+\mathbb{H}$}

Recalling the support equations (\ref{eq:Support of Stabilizer State 2}) and the fact that $|\psi\rangle$ is described by the triple $(\Lambda, M, v)$, we find that $x_0$ belongs to the support of $|\psi\rangle$ if and only if
\begin{equation}\label{eq_support_final}
\xi_{M,v}(\mu,0)\chi_{\mu}(x_0)=1,\quad\textrm{for all }\mu\in \mathbb{D}.
\end{equation} We will now write the elements $\mu\in \mathbb{D}$ in the form $\mu=\mathcal{E}_{\mathbb{D}}w$ where $w$ is an arbitrary element in $\R^a\times \Z^b$. We further denote $\mathcal{E}_{\mathbb{D},\mathrm{pad}}:=\begin{pmatrix}
\mathcal{E}_\mathbb{D} \\ 0
\end{pmatrix}$. We now realize that
\begin{itemize}
\item $\xi_{M,v}(\mathcal{E}_\mathbb{D}w,0)$, as a function of $w$ only, is a quadratic function of $\R^a\times\Z^{b}$, since $\xi_{M,v}$ is quadratic and $\mathcal{E}_\mathbb{D}$ is a homomorphism. Furthermore \be \xi_{M,v}(\mathcal{E}_\mathbb{D}w,0)= \xi_{M',v'}(w)\qquad \mbox{ with } M':=\mathcal{E}_{\mathbb{D},\mathrm{pad}}^\transpose M\mathcal{E}_{\mathbb{D},\mathrm{pad}},\ v':=\mathcal{E}_{\mathbb{D},\mathrm{pad}}^\transpose v.
    \ee
\item $\chi_{\mathcal{E}_{\mathbb{D} }w } (x_0) $, as a function of $w$ only, is a character function of $\R^a\times\Z^{b}$ which can be written as $\chi_\varpi$ with $\varpi:={\mathcal{E}_\mathbb{D}}^*(x_0)$.
\end{itemize}
It follows that $x_0$ satisfies (\ref{eq_support_final}) if and only if the quadratic function $\xi_{M',v'}$ is a character and coincides with $\chi_\varpi$. Using lemma \ref{lemma symmetric matrix representation of the bicharacter homomorphism} and theorem \ref{thm:Normal form of a quadratic function}, we can write these two conditions equivalently as:
\begin{align}
 w_1^\transpose M' w_2 &= 0 \pmod{\Z}, \quad \textnormal{for all $w_1$, $w_2\in  \R^a\times \Z^b$}\\
\mathcal{E}_\mathbb{D}^*(x_0)&=\mathcal{E}_{\mathbb{D},\mathrm{pad}}^\transpose v \pmod{\R^a\times \T^b}.\label{eq:FinalSystemEquations}
\end{align}
The first equation does not depend on $x_0$ and  it must hold by promise:  we are guaranteed that the support is not empty, so that the above equations must admit a solution. Furthermore, and crucially, the second equation is again a \emph{system of linear equations over groups} (section \ref{sect:Systems of linear equations over groups}),  a general solution $(x_0,\mathcal{E})$ of which can be efficiently computed  with our classical algorithm in theorem \ref{thm:General Solution of systems of linear equations over elementary LCA groups}.

\subsection{Sampling the support of a state}\label{sect:Sampling the support of a state}

In the last section we showed how to efficiently compute a classical description of the  (uniform) support of the final state   supp($\ket{\psi}$), re-expressing it as the set of solutions  $G_{\textnormal{sol}}=x_0+\textnormal{im}\,\mathcal{E}$ of a linear system over groups (\ref{eq:FinalSystemEquations}) and using our classical algorithm (theorem \ref{thm:General Solution of systems of linear equations over elementary LCA groups}) to find a general solution  $(x_0,\mathcal{E})$. In this section, we complete our classical simulation algorithm by devising a classical subroutine to uniformly sample from the solution space $G_{\textnormal{sol}}$ of any linear system $\alpha(x)=b$ with variables $x$ in a group 

\begin{equation}\label{eq:Elementary TZF Groups 1}
G=\T^a \times \Z^b \times \DProd{N}{c}.
\end{equation}
More generally, our main result (\textbf{theorem \ref{thm:Algorithm to sample subgroups}}) is an algorithm to uniformly sample elements from  any coset of form $x_0+\textnormal{im}\,\mathcal{E}$, provided that  $(x_0,\mathcal{E})$ is given to us as an input: here,  $\mathcal{E}$ denotes   an arbitrary matrix representation of a group homomorphism from  $\R^\alpha\times\Z^\beta$ to $G$ with known  $\alpha$, $\beta$. Throughout the section, we let $H:=\mathrm{im{\mathcal{E}}}$ be the image of $\mathcal{E}$ and denote $m:=a+b+c$. Because stabilizer states have uniform support (lemma \ref{lemma:StabStates_are_Uniform_Superpositions}), this yields our final classical algorithm to simulate final measurement statistics.

\myparagraph{Input of the problem and assumptions:} For our algorithm to work, $H$ needs to  be a \emph{closed} subgroup (in the topological sense), which is enough since the subgroup $\mathbb{H}$ that defines the support of a stabilizer state (and we aim to sample) is always closed (corollary \ref{corollary:StabStates have Closed Support}). Below, we use the word ``subgroup'' as a synonym of ``closed subgroup''.

\myparagraph{A simple heuristic:} The coset structure of supp($\ket{\psi}$)=$G_\mathrm{sol}$ immediately suggests us a \emph{simple heuristic} to sample this set, which will be the first step towards our algorithm:
\begin{enumerate}
\item[(a)] Choose a random element $v\in \R^\alpha\times\Z^\beta$ using some efficient classical procedure. This step should be easy since this group  is a simple product of a conventional real Euclidean space $\R^\alpha$ and an integer lattice $\Z^\beta$.
\item[(b)] Apply the map $v\rightarrow x_0 + \mathcal{E}(v)$ to obtain a probability distribution on $G_{\textnormal{sol}}$.
\end{enumerate}
Unfortunately, this straightforward strategy has important caveats and does not yet yield an algorithm  to sample $G_\mathrm{sol}$. In the first place, the heuristic neglects two delicate mathematical properties of the groups under consideration, namely, that they are \emph{continuous} and \emph{unbounded}. Moreover, the second step of the heuristic involves the transformation of a given probability distribution  on a space $\R^\alpha\times\Z^\beta$  by the application of a \emph{non-injective} map $\mathcal{E}:\R^\alpha\times\Z^\beta\rightarrow G$; this step is prone to create a wild number of \emph{collisions} among samples, about which the heuristic gives no information.

\subsubsection*{Norms}
In the rest of this section we show how to tackle problems (a-b) with a strategy that uses epsilon-net methods. To this end, our first step is to introduce a  suitable notion of 2-norm  for any group of form $G:=\Z^a\times \R^b \times \DProd{N}{c}\times \T^d$ analogous to the standard 2-norm $\|\cdot\|_2$ of a real Euclidean space (we denote the group 2-norm simply by $\|\cdot \|_{G}$):  given  $g=(g_\Z, g_\R, g_F, g_\T)\in G$, 
\begin{equation}\label{eq:Norms}
\|g\|_G:= \left\| \left( g_\Z, g_\R,  g_F^{\circlearrowleft}, g_\T^{\circlearrowleft} \right) 	\right\|_2
\end{equation}
where  $g_F^{\circlearrowleft}$ (resp.\ $g_\T^{\circlearrowleft}$) stands for any integer  tuple $x\in \Z^a$  (resp.\ real tuple $y\in \R^d$)  that is congruent to $g_F$ (resp.\ $g_\T$) and has minimal two norm $\|\cdot\|_2$. The reader should note that, although $g_F^{\circlearrowleft}$, $g_\T^{\circlearrowleft}$ may not be uniquely defined, the value of $\|g\|_G$ is always unique.

The following relationship between norms will later be useful:
\begin{equation}\label{eq:Relationship between norms}
\textnormal{if}\quad\|g\|_{2}\leq \tfrac{1}{2}\quad \textnormal{then}\quad \|g\|_{G}=\|g\|_{2}\leq \tfrac{1}{2};
\end{equation}
or, in other words, if an element $g\in G$ has small $\|\cdot\|_2$ norm as a tuple of real numbers, then its norm $\|g\|_{G}$ as a group element of $G$ is also small and equal to $\|g\|_2$.

\subsubsection*{Net techniques}

Groups of the form (\ref{eq:Elementary TZF Groups 1}) contain subgroups that are \emph{continuous} and/or \emph{unbounded} as sets. These properties must  be taken into account in the design of algorithms to sample subgroups.  We briefly discuss the technical issues---absent from the case of finite $G$ as in chapter \ref{chapterF} ---that arise, and present net techniques to tackle them.
 
The first issue to confront, related to continuity, is the presence of discretization errors due to finite precision limitations, for no realistic algorithm can sample from a continuous subgroup $H$ exactly. Instead, we will  sample from some distinguished discrete subset $\mathcal{N}_\varepsilon$  of $H$  that, informally, ``discretizes''  $H$ and that can be efficiently represented in a computer. More precisely, we choose $\mathcal{N}_\epsilon$ to be a certain type of $\varepsilon$-net:
 \begin{definition}[$\boldsymbol{\varepsilon}$\textbf{-net}]\label{def:Epsilon Net} An $\varepsilon$-net\footnote{Our definition of  $\varepsilon$-net is based on the ones used in \cite{HaydenLeungShorWinter04Randomizing_Quantum_States,YaoyunXiaodi@12_E_Net_Method_Optimizations_Separable,Ni13Commuting_Circuits,Deza13Encyclopedia_of_Distances}. We adopt an additional non-standard convention, that $\mathcal{N}$ must be a subgroup, because it is convenient for our purposes.} $\mathcal{N}$ of a subgroup $H$ is a finitely generated subgroup of $H$ such that for every $h\in H$ there exists  $\mathpzc{n}\in \mathcal{N}$ with $\| h -  \mathpzc{n} \|_{G}\leq \varepsilon$.
 \end{definition}
The second issue in our setting is the unboundedness of certain subgroups of $G$ \emph{by themselves}. We must carefully define  a notion of sampling for such sets that suits our needs, dealing with the fact that  uniform distributions over unbounded sets (like $\R$ or $\Z$) cannot be interpreted as well-defined probability distributions; as a consequence, one cannot simply ``sample'' from $\mathcal{N}$ or $H$ uniformly. However, in order to simulate the distribution of measurement outcomes of a \emph{physical} normalizer quantum computation (where the initial states $\ket{g}$ can only be prepared approximately) it is enough to sample uniformly from some bounded compact region of $H$ with finite volume $V$. We can approach the infinite-precision limit by choosing $V$ to be larger and larger, and in the $V\rightarrow \infty$ limit we will approach an exact quantum normalizer computation. 

We will slightly modify the definition of $\epsilon$-net so that we can sample from $H$ in the sense described above. For this, we need to review some structural properties of the subgroups of groups of the form (\ref{eq:Elementary TZF Groups 1})

It is known that any arbitrary closed subgroup $H$ of an elementary group $G$ of the form (\ref{eq:Elementary TZF Groups 1}) is isomorphic to an elementary group also of the form  (\ref{eq:Elementary TZF Groups 1})  (see \cite{Stroppel06_Locally_Compact_Groups} theorem 21.19 and proposition 21.13). As a result, any subgroup $H$ is of the form $H=H_{\mathrm{comp}}\oplus H_{\mathrm{free}}$ where $H_{\mathrm{comp}}$ is a \emph{compact} abelian subgroup of $H$ and $H_{\mathrm{free}}$ is either the trivial subgroup or an \emph{unbounded} subgroup that does not contain non-zero finite-order elements (it is \emph{torsion-free}, in group theoretical jargon). By the same argument, any $\varepsilon$-net $\mathcal{N}_\varepsilon$ of $H$ decomposes in the same way
 \begin{equation}\label{eq:eNet_compact-free_decomposition}
 \mathcal{N}_\varepsilon:=\mathcal{N}_{\varepsilon,\mathrm{comp}}\oplus \mathcal{N}_{\varepsilon,\mathrm{free}}.
 \end{equation}
 where $\mathcal{N}_{\varepsilon,\mathrm{comp}}$ is a finite subgroup of $H_{\mathrm{comp}}$ and $\mathcal{N}_{\varepsilon,\mathrm{free}}$ is a finitely generated torsion-free subgroup of $H_{\mathrm{free}}$. The fundamental theorem of finitely generated abelian groups tells us that $\mathcal{N}_{\varepsilon,\mathrm{free}}$ is  isomorphic to a group of the form $\Z^\mathbf{r}$ (a \emph{lattice} of rank $\mathbf{r}$) and, therefore, it has a \emph{$\Z$-basis} \cite{Mollin09Advanced_Number_Theory_with_Applications}: i.e.\ a set  $\{\mathpzc{b}_1, \ldots, 	\mathpzc{b}_\mathbf{r}\}$ of  elements  such that every $\mathpzc{n}\in \mathcal{N}_{\varepsilon,\mathrm{free}}$ can be written in one and only one way as a linear combination of basis elements with \emph{integer} coefficients:
  \begin{equation}
   \mathcal{N}_{\varepsilon,\mathrm{free}}=\left\{\mathpzc{n} = \sum_{i=1}^{\mathbf{r}} \mathpzc{n}_i  \mathpzc{b}_i, \,\textnormal{ for some } \mathpzc{n}_i \in \Z\right\}.
   \end{equation}
In view of equation (\ref{eq:Elementary TZF Groups 1}) we introduce a more general notion of nets that is adequate for sampling from this type of set.
  \begin{definition}\label{def:Delta Epsilon Net} Let $\mathcal{N}_\varepsilon$ be an $\varepsilon$-net of $H$ and let $\{\mathpzc{b}_1, \ldots, 	\mathpzc{b}_\mathbf{r}\}$ be a prescribed  basis of $\mathcal{N}_{\varepsilon,\mathrm{free}}$. Then, we call a $\boldsymbol{(\Delta, \varepsilon )}$\textbf{\em-net} any  finite subset $\mathcal{N}_{\Delta,\varepsilon}$ of $\mathcal{N}_\varepsilon$ of the form
  \begin{equation}
   \mathcal{N}_{\Delta,\varepsilon}= \mathcal{N}_{\varepsilon,\mathrm{comp}}\oplus \mathcal{P}_\Delta,
   \end{equation}
  where  $\mathcal{P}_\Delta$ denotes the  parallelotope contained in $\mathcal{N}_{\varepsilon,\mathrm{free}}$ with vertices $\pm\Delta_1 \mathpzc{b}_1, \ldots, 	\pm\Delta_\mathbf{r} \mathpzc{b}_\mathbf{r}$,
    \begin{equation}\label{eq:Parallelotope DEFINITION}
    \mathcal{P}_\Delta := \left\{\mathpzc{n} = \sum_{i=1}^{\mathbf{r}} \mathpzc{n}_i  \mathpzc{b}_i, \,\textnormal{ where } \mathpzc{n}_i \in \{0,\pm 1, \pm 2,\ldots , \pm \Delta_i\}\right\}.
    \end{equation}
    The index of $\mathcal{P}_\Delta$ is a  tuple of positive integers $\Delta:=(\Delta_1,\ldots,\Delta_\mathbf{r})$ that specifies the lengths of the edges of $\mathcal{P}_\Delta$.
 \end{definition}
 Notice that $\mathcal{N}_{\Delta,\varepsilon}\rightarrow \mathcal{N}_\varepsilon$ in the limit where the edges  $\Delta_i$ of $\mathcal{P}_\Delta$ become infinitely long and that the volume covered by $\mathcal{N}_{\Delta,\varepsilon}$ increases monotonically as a function of the edge-lengths. Hence, any algorithm to construct and sample $(\Delta, \varepsilon)$-nets of $H$ can be used to sample from  $H$ in the sense we want. Moreover, the next theorem (a  main contribution of this chapter) states that there exist  classical algorithms to sample the subgroup $H$ through $(\Delta,\varepsilon)$-nets \emph{efficiently}.
 \begin{theorem}\label{thm:Algorithm to sample subgroups}
Let  $H$ be an arbitrary closed subgroup of an elementary group $G=\T^a\times\Z^b\times \DProd{N}{c}$. Assume we are given a matrix-representation $\mathcal{E}$ of a group homomorphism $\mathcal{E}:\R^\alpha\times \Z^\beta\rightarrow G$ such that $H$ is the image of $\mathcal{E}$. Then, there exist  classical algorithms to sample $H$ through $(\Delta, \varepsilon)$-nets using $O(\ppoly{m, \alpha,\beta, \log{N_i},\|\mathcal{E}\|_\mathbf{b}, \log\frac{1}{\varepsilon}, \log\Delta_i})$ time and bits of memory.
 \end{theorem}
 Again, $\|\mathcal{E}\|_\mathbf{b}$ denotes the maximal number of \emph{bits} needed to store a coefficient of $\mathcal{E}$ as a fraction. The proof is the content of the next section, where we devise a classical algorithm with the advertised properties.

\subsubsection*{Proof of theorem \ref{thm:Algorithm to sample subgroups}: an algorithm to sample subgroups}

We  denote by $\mathcal{E}_{\T\R}$ the block of $\mathcal{E}$ with image contained in $\T^a$ and with domain $\mathbb{R}^\alpha$. Define  a new set  $\mathcal{L}:= (\varepsilon_1 \Z) ^\alpha \times \Z^\beta$, which is a subgroup of $\R^\alpha\times \Z^\beta$, and let   $\mathcal{N}:= \mathcal{E}(\mathcal{L})$ be the image of $\mathcal{L}$ under the action of the homomorphism $\mathcal{E}$.  

In first place, we show that  by setting $\varepsilon_1$ to be  smaller than $2\varepsilon/(\alpha \sqrt{a}|\mathcal{E}|)$, we can ensure that $\mathcal{N}$ is an $\varepsilon$-net of $H$ for any $\varepsilon$ of our choice. We will use that $\mathcal{L}$ is, by definition,  a $(\frac{\varepsilon_1\sqrt{\alpha}}{2})$-net of $\R^\alpha\times \Z^\beta$. (This follows from the fact that, for every   $x\in\R^\alpha$ there exists $x'\in (\varepsilon_1 \Z)^\alpha$ such that $|x(i)-x'(i)|\leq \varepsilon_1/2$, so that $\|x-x'\|_2\leq \varepsilon_1\sqrt{\alpha}/2$).  Of course, we must have that   $\mathcal{N}$ must be an $\varepsilon$-net of $H$ for some value of $\varepsilon$. To bound this $\varepsilon$ we will use the following bound for the operator norm of the matrix $\mathcal{E}_{\T\R}$:
\begin{equation}\label{eq:Error propagation in e-nets}
\|\mathcal{E}_{\T\R}\|_{\mathrm{op}}^2\leq \alpha a  |\mathcal{E}_{\T\R}|^2 \leq \alpha a |\mathcal{E}|^2 .
\end{equation}
The first inequality in (\ref{eq:Error propagation in e-nets})  follows from Schur's bound on the maximal singular value of a real matrix. This bound implies that, if two elements $\mathpzc{x}:=(x,z)\in\mathcal{L}$ and $\mathpzc{x}':=(x',z)\in\mathcal{L}$ are $\varepsilon_1\sqrt{\alpha}/2$-close to each other, then
\begin{equation}\label{eq:Error propagation in e-nets 2}
 \|\mathcal{E}\mathpzc{x}-\mathcal{E}\mathpzc{x}'\|_2\leq \|\mathcal{E}_{\T\R}\|_{\mathrm{op}} \|x-x'\|_2 \leq \tfrac{1}{2}\alpha \sqrt{a} |\mathcal{E}|\varepsilon_1
\end{equation}
 (In the first inequality, we apply the normal form in lemma \ref{lemma:Normal form of a matrix representation}.) Finally, by   imposing $\frac{ \alpha \sqrt{a}}{2}|\mathcal{E}|\varepsilon_1\leq \varepsilon\leq \tfrac{1}{2}$, we get  that $\|\mathcal{E}(\mathpzc{x}-\mathpzc{x'})\|_G\leq \varepsilon$ due to property (\ref{eq:Relationship between norms}); it follows that $\mathcal{N}$ is an $\varepsilon$-net of $H$  if $\varepsilon_1\leq 2\varepsilon/(\alpha \sqrt{a}|\mathcal{E}|)$ for every $\varepsilon\leq \tfrac{1}{2}$.

Assuming that $\varepsilon_1$ is chosen so that $\mathcal{N}$ is an $\varepsilon$-net, our next step will be to devise an algorithm to construct and sample an $(\Delta,\varepsilon)$-net $\mathcal{N}_\Delta\subset\mathcal{N}$. The key step of our algorithm will be a subroutine that computes a nicely-behaved classical representation of the quotient group $Q\cong\mathcal{L}/\ker{\mathcal{E}}$ and a matrix representation  of the group isomorphism\footnote{$Q$ and $\mathcal{N}$   are isomorphic due to the first isomorphism theorem \cite{Stroppel06_Locally_Compact_Groups}.} $\mathcal{E}_{\mathrm{iso}}:Q\rightarrow \mathcal{N}$. We will use the computed representation of $Q$ to construct  a $(\Delta,\varepsilon)$-net $Q_\Delta\subset Q$ and sample elements form it; then,  by applying the map $\mathcal{E}_{\mathrm{iso}}$ to the sampled elements, we will effectively sample a  $(\Delta,\varepsilon)$-net $\mathcal{N}_\Delta\subset \mathcal{N}$; and, moreover, in a clean \emph{collision free} fashion. The  first steps of our subroutine are described next.
\begin{enumerate}
\item \textbf{Set precision.} Choose $\varepsilon_1$ so that $\mathcal{N}$ is an $\varepsilon$-net using the above bounds.
\item  \textbf{Put $\mathcal{L}$ in standard form.} Note that $\mathcal{L}$ is isomorphic to the discrete finite-generated abelian group  $\mathcal{L}':=\Z^{\alpha+\beta}$ via an isomorphism $\phi:\mathcal{L}'\rightarrow\mathcal{L}$ with matrix representation $\varepsilon_1 I_\alpha\oplus I_\beta$. In order to apply the algorithms in theorem \ref{thm:General Solution of systems of linear equations over elementary LCA groups}, we  ``absorb'' the $\varepsilon_1$ parameter into the map $\phi$ and replace   $\mathcal{L}$ with $\mathcal{L'}$, and $\mathcal{E}$ with the map $\mathcal{E'}:= \mathcal{E}(\varepsilon_1 I_\alpha\oplus I_\beta)$. 
\item \textbf{Take quotient.} Apply algorithm 3 in  theorem \ref{thm:General Solution of systems of linear equations over elementary LCA groups} to compute an efficient  decomposition $Q'= \DProd{\sigma}{\mathbf{a}}\times \Z^{\mathbf{b}}\cong$ of  the quotient $\mathcal{L}'/\ker{\mathcal{E}'}\cong Q'$ and  a new matrix representation $\mathcal{E}_{\mathrm{iso}}'$ of the isomorphism $\mathcal{E}_{\mathrm{iso}}':Q'\rightarrow \mathcal{N}$. Our earlier result says that this step can be implemented in time at most polynomial in the variables  $m$, $\alpha$, $\beta$, $ \log{N_i}$, $\|\mathcal{E}\|_{\mathbf{b}}$, and $\log{\frac{1}{\varepsilon}}$.
\end{enumerate}
The above steps procedure outputs a new classical representation $(x_0, \mathcal{E}_{\mathrm{iso}}')$ of the support $\mathrm{supp}(\ket{\psi})=x_0+\mathcal{E}_{\mathrm{iso}}'$ with the remarkable property that now $\mathcal{E}_{\mathrm{iso}}'$ is invertible., 
Moreover, since the matrix $\mathcal{E}_{\mathrm{iso}}'$ acts isomorphically on $Q'$, it follows that the epsilon-net subgroup $\mathcal{N}$ is a direct sum of cyclic subgroups generated by the  columns of  $\mathcal{E}_{\mathrm{iso}}$:
\begin{equation}
\mathcal{N}=  \langle \mathpzc{f}_1 \rangle\oplus\cdots \oplus \langle  \mathpzc{f}_\mathbf{a} \rangle\oplus \langle \mathpzc{b}_1 \rangle \oplus \cdots \oplus \langle \mathpzc{b}_\mathbf{b}\rangle,
\end{equation} 
where $\mathpzc{f}_i$, $\mathpzc{b}_j$ stand for the  $(i)$th and the $(\mathbf{a}+j)$th column of $\mathcal{E}_{\mathrm{iso}}$; via isomorphism, it must also hold that the elements $\mathpzc{f}_i$s (resp.\ $\mathpzc{b}_j$)   generate the compact subgroup $\mathcal{N}_{\mathrm{comp}}$  (resp.\  form a $\Z$-basis of $\mathcal{N}_{\mathrm{free}}$). Finally, taking $\{\mathpzc{f}_i\}$ (resp.\  $\{\mathpzc{b}_j\}$) as default generating-set (resp.\ default basis) of $\mathcal{N}_{\mathrm{comp}}$ and $\mathcal{N}_{\mathrm{free}}$, we  select a parallelotope $\mathcal{P}_\Delta$ of the form (\ref{eq:Parallelotope DEFINITION}) with some desired $\Delta=(\Delta_1,\ldots,\Delta_\mathbf{b})$. This procedures specifies a net $\mathcal{N}_\Delta=\mathcal{N}_{\mathrm{comp}}\oplus\mathcal{P}_\Delta$ that can be efficiently represented with $O(\ppoly{m,\alpha,\beta,\log{N_i}, \|\mathcal{E}_{\mathrm{iso}}'\|_{\mathbf{b}}, \log{\frac{1}{\varepsilon}} \log{\Delta_i}})$ bits of memory (by keeping track of the generating-sets of $\mathcal{N}$ and the numbers $\Delta_i$). Moreover, we can efficiently sample from  $\mathcal{N}_\Delta$ uniformly and collision-freely by generating random strings of the form
\begin{equation}
\sum_{i=1}^{\mathbf{a}} \mathpzc{x}_i \mathpzc{f}_i + \sum_{j=1}^{\mathbf{b}} \mathpzc{y}_j \mathpzc{b}_j,
\end{equation}
where $\mathpzc{x}_i\in\Z_{\sigma_i}$ and $\mathpzc{y}_j\in\{0,\pm1,\ldots,\pm\Delta_j\}$. This completes the proof.

%% file: chapter3_blackbox.tex
\chapter{The computational power of normalizer circuits over black box groups}\label{chapterB}

In this chapter we present a precise connection between Clifford circuits,  Shor's factoring algorithm and several other famous quantum algorithms with exponential quantum speed-ups for solving abelian hidden subgroup problems. We show that all these different forms of quantum computation belong to a common new restricted model of quantum operations that we call \emph{black-box normalizer circuits}. To define these, we extend the  model of normalizer circuits  of chapters \ref{chapterF}-\ref{chapterI} where normalizer gates could be quantum Fourier transforms, group automorphism and quadratic phase gates associated with a (finite or infinite) abelian group $G$. While earlier $G$ was always given in an explicitly decomposed form, in this chapter we remove this assumption and allow $G$ to be a black-box group \cite{BabaiSzmeredi_Complexity_MatrixGroup_Problems_I}. In contrast with standard normalizer circuits, which we showed to be efficiently classically simulable, we find that black-box normalizer circuits are powerful enough to factorize and solve classically-hard problems in the black-box setting. We further set upper limits to their computational power by showing that decomposing finite abelian groups is complete for the associated complexity class. In particular, solving this problem renders black-box normalizer circuits efficiently classically simulable by exploiting the generalized stabilizer formalism  of chapters \ref{chapterF}-\ref{chapterI}. Lastly, we employ our connection to draw a few practical implications for quantum algorithm design: namely,  we give a no-go theorem for finding new quantum algorithms with black-box normalizer circuits, a universality result for low-depth normalizer circuits, and identify two other complete problems.

This chapter is based on \cite{BermejoLinVdN13_BlackBox_Normalizers} (joint work with Cedric Yen-Yu Lin and Maarten Van den Nest).

\section{Introduction}\label{sect:Introduction3}

In this chapter, we introduce \emph{\textbf{black-box normalizer circuits}}, a \emph{new restricted family} of quantum operations, and characterize their computational power. Our new model extends the classes of {normalizer circuits over abelian groups} of chapters \ref{chapterF}-\ref{chapterI} as explained next. In previous chapters,  normalizer circuits acted in  high and infinite dimensional systems associated with an abelian group $G$: in our construction, we associated $G$ with a Hilbert space $\mathcal{H}_G$ with a standard basis $\{\ket{g}\}_{g\in G}$ labeled by $G$ elements. Furthermore, previously, the group $G$ was assumed to be given in an explicit factorized form, which endows the Hilbert space of the computation  with a tensor-product structure:
\begin{equation}\label{eq:Elementary Group}
G=\Z^a\times \T^b\times \DProd{N}{c}\quad \longleftrightarrow \quad \mathcal{H}_{G}=\mathcal{H}_{\Z}^{\otimes (a+b)}\otimes \mathcal{H}_{\Z_{N_1}}\otimes \cdots \otimes \mathcal{H}_{\Z_{N_c}};
\end{equation} 
above,  $\Z$ is the group of \emph{integers}, $\Z_N$ the group of integers modulo $N$,  and $\T$ is the \emph{circle group}, consisting of angles from 0 to 1 (in  units of $2\uppi$) with the addition modulo 1. The Hilbert space $\mathcal{H}_\Z$ has a standard basis labeled by integers ($\Z$ basis) and a Fourier-basis labeled by angles ($\T$ basis). A \emph{normalizer circuit over $G$}  was a circuit built of three types of normalizer gates: Quantum Fourier transforms over $G$, group automorphism gates and quadratic phase gates. 

With these definitions, we saw (section \ref{sect_examples}) that $n$-qubit Clifford circuits are examples of normalizer circuits over the group $\Z_2^n$.

In chapters \ref{chapterF}-\ref{chapterI} we showed that, despite containing arbitrary numbers of QFTs, which play an important role in Shor's algorithms \cite{Shor}, and entangling gates (automorphism, quadratic phase gates),  normalizer circuits can be \emph{efficiently simulated} by classical computers. For this, we exploited an  extended \emph{stabilizer formalism} over groups to track the evolution of normalizer circuits, thereby generalizing the celebrated Gottesman-Knill theorem \cite{Gottesman_PhD_Thesis,Gottesman99_HeisenbergRepresentation_of_Q_Computers}.

The key new element in the present chapter  are normalizer circuits that can be associated with abelian \emph{\textbf{black-box groups}} \cite{BabaiSzmeredi_Complexity_MatrixGroup_Problems_I}, which we may simply call ``black-box normalizer circuits''. A group $\mathbf{B}$ (always abelian in this work) is a black-box group if it is \emph{finite}, its elements are uniquely encoded by strings of some length $n$ and the group operations are performed by a black-box (the \emph{group black box}) in one time-step. We define  \emph{black-box normalizer circuits} to be a normalizer circuits associated with groups of the form $G=G_\textrm{prev}\times \mathbf{B}$, with   $G_\textrm{prev}$ is of form~(\ref{eq:Elementary Group}). 

The \textbf{\emph{key new feature}} in this chapter is that the black-box  group $\mathbf{B}$ is \emph{not} given to us in a factorized form. This is a subtle yet tremendously important difference: although such a  decomposition \emph{always} exists  for any finite abelian group (chapter \ref{chapterC}, theorem \ref{thm:Fundamental Theorem FAGroups}), finding just one is regarded as a \emph{hard computational problem}; indeed, it is provably at least as hard as \emph{factoring}\footnote{\label{footnote:Hardness Group Decomposition}Knowing $\mathbf{B}\cong \DProd{d}{m}$ implies that the order of the group is $|G|=d_1 d_2\cdots d_m$. Hardness results for computing orders \cite{BabaiSzmeredi_Complexity_MatrixGroup_Problems_I,Babai98apolynomial_time_theory_of_Black_Box_Groups} imply that the problem is provably hard for classical computers in the black-box setting. For groups  $\Z_N^\times$, computing $\varphi(N):=|\Z_N^\times|$ (the Euler totient function) is equivalent to factoring \cite{Shoup08_A_Computational_Introducttion_to_Number_Theory_and_Algebra}.}. Our \textbf{\emph{motivation}} to adopt the notion of black-box group is to study abelian groups for which the group multiplication can be performed in classical polynomial-time while no efficient classical algorithm to decompose them is known. A key example\textsuperscript{\ref{footnote:Hardness Group Decomposition}} is $\mathbb{Z}^{\times}_N$, the multiplicative group of integers modulo $N$, which  plays an important role in Shor's factoring algorithm \cite{Shor}. With some abuse of notation, we  call any such group also a ``black-box group''\footnote{It will always be clear from context whether the group multiplication is performed by an oracle at unit cost or by some well-known polynomial-time classical algorithm; most results will be stated in the black-box setting.}.

\subsection{Main results}

This chapter focuses on understanding the potential uses and limitations of black-box normalizer circuits. Our results (listed below) give a precise characterization of their \textbf{computational power}. On one hand, we show that several famous quantum algorithms, including Shor's celebrated \emph{factoring algorithm}, can be implemented with black-box normalizer circuits. On the other hand, we apply  our former simulation results  (chapters  \ref{chapterF}- \ref{chapterI})  to set upper limits to the class of problems that these circuits can solve, as well as to draw practical implications for quantum algorithm design.

Our main results are now summarized:

\begin{enumerate}
\item \textbf{Quantum algorithms.} We show that many of the best known quantum algorithms  are particular instances of normalizer circuits over black-box groups, including  Shor's celebrated factoring and discrete-log algorithms; it follows that black-box normalizer circuits can achieve \emph{\textbf{exponential quantum speed-ups}} over all known classical algorithms. Namely, the following algorithms are examples of black-box normalizer circuits.
\begin{itemize}
\item \textbf{Discrete logarithm.} Shor's discrete-log quantum algorithm  \cite{Shor} is a normalizer circuit over $\Z_{p-1}^2\times\Z_p^\times$ (theorem \ref{thm:Discrete Log}, section \ref{sect:Discrete Log}).
\item \textbf{Factoring.} We show that  a hybrid infinite-finite dimensional version of Shor's factoring algorithm \cite{Shor}  can be implemented with normalizer circuit over $\Z\times\Z_N^\times$. We prove that there is a close relationship between \emph{Shor's original algorithm} and our version: Shor's  can be understood as a discretized qubit implementation of ours (theorems \ref{thm:Order Finding}, \ref{thm:Shor normalizer}).  We also discuss that the \emph{\textbf{infinite group}} $\Z$ plays a key role in our ``infinite Shor's algorithm'', by showing that it is impossible to implement Shor's modular-exponentiation gate efficiently, {even approximately}, with finite-dimensional normalizer circuits (theorem \ref{thm:ModExp requires Z}). Last,  we further \emph{conjecture} that only normalizer circuits over {infinite groups} can factorize (conjecture \ref{conj: Factoring not over finite groups}).
\item \textbf{Elliptic curves.} The generalized Shor's algorithm for computing discrete logarithms over an elliptic curve \cite{ProosZalka03_Shors_DiscreteLog_Elliptic_Curves,Kaye05_optimized_Quantum_Elliptic_Curve,CheungMaslovMathew08_Design_QuantumAttack_Elliptic_CC} can be implemented with black-box normalizer circuits (section \ref{sect:Elliptic Curve}); in this case, the black-box group is the group of integral points $E$ of the elliptic curve instead of $\Z_p^\times$.
\item \textbf{Group decomposition.}  Cheung-Mosca's algorithm for decomposing black-box finite abelian groups \cite{mosca_phd, cheung_mosca_01_decomp_abelian_groups} is a combination of several types of black-box normalizer circuits. Furthermore, we present a new \emph{extended} quantum  algorithm building upon Cheung-Mosca's that finds even more information about the structure of the group and is also normalizer-circuit based (section \ref{sect:Group Decomposition}).
\item \textbf{Hidden subgroup problem.} Deutsch's \cite{Deutsch85quantumtheory}, Simon's  \cite{Simon94onthe} and, in fact, all quantum algorithms that solve abelian hidden subgroup problems \cite{Boneh95QCryptanalysis,Grigoriev97_testing_shift_equivalence_polynomials,kitaev_phase_estimation,Kitaev97_QCs:_algorithms_error_correction,Brassard_Hoyer97_Exact_Quantum_Algorithm_Simons_Problem,Hoyer99Conjugated_operators,MoscaEkert98_The_HSP_and_Eigenvalue_Estimation,Damgard_QIP_note_HSP_algorithm}, are normalizer circuits over groups of the form $G\times\mathcal{O}$, where $G$ is the group that contains the hidden subgroup $H$ and $\mathcal{O}$ is a group isomorphic to $G/H$ (section \ref{sect:Abelian HSPs}). The group $\mathcal{O}$, however, is not a black-box group due to a small technical difference between our  oracle model we use and the oracle setting in the HSP.
 
\item \textbf{Hidden kernel problem.}  The group $\mathcal{O} \cong G/H$ in the previous section becomes a black-box group if the oracle function in the HSP is a homomorphism between black-box groups: we call this subcase the {\emph{hidden kernel problem}} (HKP). The difference does not seem to be very significant, and can be eliminated by choosing different oracle models (section \ref{sect:Abelian HSPs}). However, we will never refer to Simon's or to general abelian HSP algorithms  as ``black-box normalizer circuits'', in order to be consistent with our and pre-existing terminology. 
\end{itemize}
Note that it follows from the above that black-box normalizer circuits can render insecure widespread public-key cryptosystems, namely,  Diffie-Hellman key-exchange  \cite{DiffieHellman}, RSA \cite{RSA} and elliptic curve cryptography  \cite{Menezes96_cryptography_book,Buchmann00_cryptography_book}.

\item \textbf{Group decomposition is \emph{as hard as} simulating normalizer circuits.} Another main contribution of this work is to show that the group decomposition problem (suitably formalized) is, in fact, \textbf{\emph{complete}} for the complexity class \textbf{Black-Box Normalizer}, of problems efficiently solvable by probabilistic classical computers with oracular access to black-box normalizer circuits. Since normalizer circuits over decomposed groups are efficiently classically simulable (chapters  \ref{chapterF}- \ref{chapterI}), this result suggests that the computational power of normalizer circuits  originates \emph{precisely} in the classical hardness of learning the structure of a black-box group. 

We obtain this last  result by proving a significantly \textbf{\emph{stronger theorem}} (theorem \ref{thm:Simulation}), which states that any black-box normalizer circuit  can be efficiently simulated \emph{step by step} by a classical computer if  an efficient subroutine for decomposing finite abelian groups is provided.

\item \textbf{A no-go theorem for new quantum algorithms.} In this work, we provide an negative answer to the question ``\emph{can new quantum algorithms based on normalizer circuits be found?}'': by applying the latter simulation result, we conclude that any new algorithm not in our list can be efficiently simulated step-by-step using our extended Cheung-Mosca algorithm and classical post-processing. This implies (theorem \ref{thm:No Go Theorem}) that new \emph{exponential} speed-ups cannot be found without changing our setting (we discuss how the setting might be changed in the discussion, section \ref{sect:Discussion}). This result says nothing about polynomial speed-ups.

\item \textbf{Universality of short normalizer circuits.} A practical consequence of our no-go theorem is that all problems in the class \textbf{Black Box Normalizer} can be solved using short normalizer circuits with a \emph{constant} number of normalizer gates. (We may still need polynomially many runs of such circuits, along with classical processing in between, but each individual normalizer circuit is short.) We find this observation interesting, in that it explains a very curious feature present in all the quantum algorithms that we study \cite{Shor,ProosZalka03_Shors_DiscreteLog_Elliptic_Curves,Kaye05_optimized_Quantum_Elliptic_Curve,CheungMaslovMathew08_Design_QuantumAttack_Elliptic_CC,mosca_phd, cheung_mosca_01_decomp_abelian_groups,Deutsch85quantumtheory,Simon94onthe,Boneh95QCryptanalysis,Grigoriev97_testing_shift_equivalence_polynomials,kitaev_phase_estimation,Kitaev97_QCs:_algorithms_error_correction,Brassard_Hoyer97_Exact_Quantum_Algorithm_Simons_Problem,Hoyer99Conjugated_operators,MoscaEkert98_The_HSP_and_Eigenvalue_Estimation,Damgard_QIP_note_HSP_algorithm} (section \ref{sect:Quantum Algorithms3}): they all contain at most a constant number of \emph{quantum Fourier transforms} (actually at most two).
\item \textbf{Other complete problems.} As our last contribution in this series, we identify another two complete problems for the class \textbf{Black Box Normalizer} (section \ref{sect:Complete Problems}): these are the (afore-mentioned) abelian \emph{hidden kernel problem}, and  the problem of finding a general-solution to a \emph{system of linear equations over black-box groups} (the latter are related to the systems of linear equations over groups  studied in chapters \ref{chapterGT}, \ref{chapterF},  \ref{chapterI}).
\end{enumerate}

\subsection{The link between Clifford circuits and Shor's  algorithm}

The results in this chapter together with those previously obtained in  chapters \ref{chapterF}-\ref{chapterI} (see also \cite{VDNest_12_QFTs}) demonstrate the existence of a precise connection between Clifford circuits and  Shor's factoring algorithm. At first glance, it might be hard to digest that two types of quantum circuits that seem to be so far away from each other might be related at all. Indeed, classically simulating Shor's algorithm is widely believed to be an intractable problem (at least as hard as factoring), while a zoo of classical techniques and efficient classical algorithms exist for simulating and computing properties of Clifford circuits \cite{Gottesman_PhD_Thesis,Gottesman99_HeisenbergRepresentation_of_Q_Computers,Knill96non-binaryunitary,Gottesman98Fault_Tolerant_QC_HigherDimensions,dehaene_demoor_coefficients,AaronsonGottesman04_Improved_Simul_stabilizer,dehaene_demoor_hostens,AndersBriegel06_Simulation_Stabilizer_GraphStates,VdNest10_Classical_Simulation_GKT_SlightlyBeyond,deBeaudrap12_linearised_stabiliser_formalism,JozsaVdNest14_Classical_Simulation_Extended_Clifford_Circuits}. However, from the point of view of this chapter, both turn out to be \emph{intimately related} in that they  both are just different types of normalizer circuits. In other words, they are both \emph{members of a common family of quantum operations}.

Remarkably, this correspondence between Clifford and Shor, rather than being just a mere mathematical curiosity, has also some  sensible consequences for the theory of quantum computing. One that follows from theorem \ref{thm:Simulation}, our simulation result, is that all algorithms studied in this chapter (Shor's factoring and discrete-log algorithms,  Cheung-Mosca's, etc.) have a \emph{rich hidden structure} which enables simulating them classically  with a stabilizer picture approach ``à la Gottesman-Knill'' \cite{Gottesman_PhD_Thesis,Gottesman99_HeisenbergRepresentation_of_Q_Computers}. This structure lets us track the evolution of the quantum state of the computation \emph{step by step} with a very special algorithm, which, despite being inefficient, exploits \emph{completely different} algorithmic principles than the naive brute-force approa i.e.,\ writing down the coefficients of the initial quantum state and tracking their quantum mechanical evolution through the gates of the circuit\footnote{Note that throughout this chapter we always work at a high-level of abstraction (algorithmically speaking), and that the ``steps'' in a normalizer-based quantum algorithm are always counted at the logic level of normalizer gates, disregarding smaller gates needed to implement them. In spite of this, we find the above  simulability property of black-box normalizer circuits to be truly fascinating. To get a better grasp of its significance, we may 	perform the following  thought experiment. Imagine, we would repeatedly concatenate black-box normalizer circuits in some intentionally complex geometric arrangement, in order to form a gargantuan, intricate ``Shor's algorithm'' of monstrous size. Even in this case, our simulation result states that if we can decompose abelian groups (say, with an oracle), then we can efficiently simulate the evolution of the circuit, normalizer-gate after normalizer-gate, independently of the number of Fourier transforms, automorphism and quadratic-phase gates involved in the computation (the overhead of our classical simulation is  at most polynomial in the input-size).}. Although the stabilizer-picture simulation is \emph{inefficient}  when black-box groups are present (i.e., it does not yield an efficient classical algorithm for simulating Shor's algorithm), the mere existence of such an algorithm reveals how much mathematical structure these quantum algorithms have in common with Clifford and normalizer circuits.

In retrospect, and from an applied point of view, it is also rather satisfactory that one can gracefully exploit the above connection to draw practical implications for quantum algorithm design:  in this chapter, we have actively used our knowledge of the hidden ``Clifford-ish'' mathematical features of the abelian hidden subgroup problem algorithms in deriving results 2, 3, 4 and 5 (in the list given in the previous section).
 
As a side remark, we regard it a memorable curiosity that replacing decomposed groups with black-box groups not only renders the simulation methods in chapters \ref{chapterF}-\ref{chapterI}   inefficient (this is, in fact, something to be expected, due to the existence of hard computational problems related to black-box groups), but it is also precisely this modification that suddenly bridges the gap between Clifford/normalizer circuits, Shor's algorithms, Simon's and so on. 

Finally, it is  mathematically elegant to note that all normalizer circuits we have studied are related through the so-called \textbf{Pontryagin-Van Kampen duality} \cite{Morris77_Pontryagin_Duality_and_LCA_groups,Stroppel06_Locally_Compact_Groups,Dikranjan11_IntroTopologicalGroups,rudin62_Fourier_Analysis_on_groups,HofmannMorris06The_Structure_of_Compact_Groups,Armacost81_Structure_LCA_Groups,Baez08LCA_groups_Blog_Post}, which states that all locally-compact abelian (LCA) groups are dual to their character groups. The role of this duality in the normalizer circuit model was discussed in chapter \ref{chapterI}.

\subsection{Relationship to previous work}\label{sect:Relationship Previous Work}\label{PreviousWork_c3}

Up to our best knowledge,  neither normalizer circuits over black-box groups, nor their relationship with Shor's algorithm or the abelian hidden subgroup problem,  have been investigated before this thesis.

The hidden subgroup problem (HSP) has played a central role in the history of quantum algorithms and has been extensively studied before our thesis. The abelian HSP, which is also a central subject of this chapter, is related to most of the best known quantum algorithms that were found in the early days of the field \cite{Deutsch85quantumtheory,Simon94onthe,Boneh95QCryptanalysis,Grigoriev97_testing_shift_equivalence_polynomials,kitaev_phase_estimation,Kitaev97_QCs:_algorithms_error_correction,Brassard_Hoyer97_Exact_Quantum_Algorithm_Simons_Problem,Hoyer99Conjugated_operators,MoscaEkert98_The_HSP_and_Eigenvalue_Estimation,Damgard_QIP_note_HSP_algorithm}. Its best-known generalization, the non-abelian HSP (which we investigate in chapter \ref{chapterH}), has also been heavily investigated due to its relationship to the graph isomorphism problem and certain shortest-vector-lattice problems \cite{EttingerHoyerKnill2004_Hidden_Subgroup,HallgrenRusselTaShma2003_Normal_Subgroup,Kuperberg2005_Dihedral_Hidden_Subgroup,Regev2004_Dihedral_Hidden_Subgroup,Kuperberg2013_Hidden_Subgroup,RoettelerBeth1998_Hidden_Subgroup,IvanyosMagniezSantha2001_Hidden_Subgroup,MooreRockmoreRussellSchulman2004,InuiLeGall2007_Hidden_Subgroup,BaconChildsVDam2005_Hidden_Subgroup,ChiKimLee2006_Hidden_Subgroup,IvanyosSanselmeSantha2007_Hidden_Subgroup,MagnoCosmePortugal2007_Hidden_Subgroup,IvanyosSanselmeSantha2007_Nil2_Groups,FriedlIvanyosMagniezSanthaSen2003_Hidden_Translation,Gavinsky2004_Hidden_Subgroup,ChildsVDam2007_Hidden_Shift,DenneyMooreRussel2010_Conjugate_Stabilizer_Subgroups,Wallach2013_Hidden_Subgroup} (see also the reviews \cite{lomont_HSP_review,childs_lecture_8,VanDamSasaki12_Q_algorithms_number_theory_REVIEW} and references therein).
  
The notion of black-box group, which is a key concept in our setting, was first considered by Babai and Szemerédi in  \cite{BabaiSzmeredi_Complexity_MatrixGroup_Problems_I} and has since been extensively studied in classical complexity theory  \cite{Arvind97solvableblack-box,Babai1991_Vertex_Transive_Graphs_Random_Generation_Finite_Groups,Babai1992_Bounded_Round_Interactive_Proofs_Finite_Groups,Babai97_Randomization_group_algorithms,Babai98apolynomial_time_theory_of_Black_Box_Groups}. In general, black-box groups may not be abelian and do not need to have uniquely represented elements \cite{BabaiSzmeredi_Complexity_MatrixGroup_Problems_I}; in the present work, we only consider abelian uniquely-encoded black-box groups. 

In quantum computing, black-box groups were previously investigated in the context of quantum algorithms, both in the abelian \cite{mosca_phd,cheung_mosca_01_decomp_abelian_groups,Zhang11Decomposing} and the non-abelian group setting \cite{watrous00_quantumAlgorithms_solvableGroups,IvanyosMagniezSantha2001_Hidden_Subgroup,FriedlIvanyosMagniezSanthaSen2003_Hidden_Translation,MagniezNayak2005_Group_Commutativity,Fenner05_QAlg_Group_Theoretic_Problems,IvanyosSanselmeSantha2007_Nil2_Groups,LeGall2010_Group_Isomorphism,Zatloukal2013_Equivalent_Group_Extensions}. Except for a few exceptions (cf.\ \cite{watrous00_quantumAlgorithms_solvableGroups,Zhang11Decomposing}) most quantum results have been obtained for uniquely-encoded black-box groups.

\subsection{Discussion and outlook}\label{sect:Discussion}

We finish our introduction by discussing a few potential avenues for finding new quantum algorithms as well as some open questions suggested by the work in this chapter.

In this work, we provide a strict no-go theorem for finding new quantum algorithms with black-box normalizer circuits, as we define them. There are, however, a few possible ways to  modify our setting leading to scenarios where one could bypass these results and, indeed, find new interesting quantum algorithms. We now discuss some.

One interesting possibility would be to consider more general types of normalizer circuits than ours, by \textbf{\emph{extending the class of abelian groups}} they can be associated with. However, looking at more general  \emph{decomposed} groups does not look particularly promising: we believe that our methods in  chapters \ref{chapterI}-\ref{chapterB} can be extended, e.g., to efficiently simulate normalizer circuits over groups of the form $\R^a\times \Z^b \times \T^c \times \DProd{N}{d} \times \mathbf{B}$, with additional $\R$ factors, once we know how to decompose $\mathbf{B}$ (see also our discussion in chapter \ref{chapterI}). On the other hand, allowing more general types of groups to act as \emph{black-boxes} looks rather promising to us: one may, for instance, attempt to extend the notion of normalizer circuits to act on Hilbert spaces associated with multi-dimensional infrastructures \cite{Sarvepalli14_1D_infrastructures,FonteinWocjan11_Q_Alg_Period_Lattice_Infrastructure}, which may, informally, be understood as ``infinite black-box groups''\footnote{An $n$-dimensional infrastructure $\mathcal{I}$ provides a classical presentation for an $n$-dimensional hypertorus group $\R^n/\Lambda\cong \T^n$, where $\Lambda$ is an (unknown) period lattice $\Lambda$. The elements of this continuous group are represented with some classical structures known as \emph{$f$-representations}, which are  endowed with an operation that allows us to compute within the torus. Although one must deal carefully with non-trivial technical aspects of infinite groups in order to properly define and compute with $f$-representations (cf.\ \cite{Sarvepalli14_1D_infrastructures,FonteinWocjan11_Q_Alg_Period_Lattice_Infrastructure} and references therein), one may intuitively understand infrastructures as ``generalized black-box hypertoruses''. We stress, though, that it is not standard terminology to call ``black-box group'' to an infinite group.} We expect, in fact, that  known quantum algorithms for finding hidden periods and hidden lattices within real vector spaces  \cite{Hallgren07_Pells_equation,Jozsa03_Hallgrens_Algorithm,Schmidt05_Q_Algorithm_Computation_Unit_Group,Hallgren2005_Unit_Group_Class_Group} and/or or infrastructures \cite{Sarvepalli14_1D_infrastructures,FonteinWocjan11_Q_Alg_Period_Lattice_Infrastructure} (e.g.,\ Hallgren's algorithm for solving Pell's equation \cite{Hallgren07_Pells_equation,Jozsa03_Hallgrens_Algorithm}) could be at least partially interpreted as generalized normalizer circuits in this sense. Addressing this question would  require a careful treatment of precision errors that appear in such algorithms due to the presence of transcendental numbers, which play no role in the present chapter\footnote{No such treatment is needed in this work, since we study quantum algorithms for finding hidden structures in \emph{discrete} groups.}. Some  open questions in this quantum algorithm subfield have been discussed in \cite{FonteinWocjan11_Q_Alg_Period_Lattice_Infrastructure}.

A second enticing possibility would be to study possible extensions of the normalizer circuit framework to \emph{non-abelian groups}, in connection with non-abelian hidden subgroup problems \cite{EttingerHoyerKnill2004_Hidden_Subgroup,HallgrenRusselTaShma2003_Normal_Subgroup,Kuperberg2005_Dihedral_Hidden_Subgroup,Regev2004_Dihedral_Hidden_Subgroup,Kuperberg2013_Hidden_Subgroup,RoettelerBeth1998_Hidden_Subgroup,IvanyosMagniezSantha2001_Hidden_Subgroup,MooreRockmoreRussellSchulman2004,InuiLeGall2007_Hidden_Subgroup,BaconChildsVDam2005_Hidden_Subgroup,ChiKimLee2006_Hidden_Subgroup,IvanyosSanselmeSantha2007_Hidden_Subgroup,MagnoCosmePortugal2007_Hidden_Subgroup,IvanyosSanselmeSantha2007_Nil2_Groups,FriedlIvanyosMagniezSanthaSen2003_Hidden_Translation,Gavinsky2004_Hidden_Subgroup,ChildsVDam2007_Hidden_Shift,DenneyMooreRussel2010_Conjugate_Stabilizer_Subgroups,Wallach2013_Hidden_Subgroup}. This direction will be explored in the last chapter of this thesis (chapter \ref{chapterH}) where we    develop a possible nonabelian model of normalizer circuits and use it to devise new efficient quantum algorithms for the so called normal Hidden Subgroup Problem \cite{HallgrenRusselTaShma2003_Normal_Subgroup}.

A  third possible direction to investigate would be whether different models of normalizer circuits could be constructed over \textbf{\emph{algebraic structures that are not groups}}.

Our results in chapter \ref{chapterH} will also make significant progress in this direction: therein, we consider (in general) normalizer circuit models over so-called \emph{abelian hypergroups}, which generalize abelian groups, and used them to develop the first provably-efficient quantum  algorithms for a hypergroup extension of the hidden subgroup problem.

Furthermore, one could, for instance, consider sets with \emph{less algebraic structure} than groups, like semi-groups. In this regard, we highlight that a quantum algorithm for finding discrete logarithms over finite semigroups was recently given in   \cite{Childs14_Discrete_Log_Semigroups}. Alternatively, one could study also \emph{sets} with more structure than groups, such as \emph{fields}, whose study is relevant to Van Dam-Seroussi's  quantum algorithm for estimating Gauss sums  \cite{VanDamSeroussi02_Gauss_Sums_QALG}. 

Lastly, we mention some open questions suggested by the work of this chapter.

In this work, we have not investigated the computational complexity of black-box normalizer circuits \emph{without} classical post-processing. There are two facts which suggest  that power of black-box normalizer circuits alone might, in fact, be significantly smaller. The first is the fact that the complexity class of problems solvable  by non-adaptive Clifford circuits with standard basis inputs and measurements is $\oplus \mathbf{L}$ \cite{AaronsonGottesman04_Improved_Simul_stabilizer}, which is believed to be a strict subclass\footnote{This is the class of problems solvable by classical poly-size  circuits of NOT and CNOT gates \cite{AaronsonGottesman04_Improved_Simul_stabilizer}.} of $\mathbf{P}$. The second is that finite-dimensional normalizer circuits  are unable of implementing classical boolean functions coherently  in various settings (see \cite{VDNest_12_QFTs} and lemma \ref{lemma:PermutationNormalizer=Clifford} in  chapter \ref{chapterF}).

Finally, one may study whether considering more general types of inputs,  measurements or adaptive operations might change the power of black-box normalizer circuits. Allowing, for instance, input product states has potential to increase the power of these circuits, since this already occurs for standard Clifford circuits \cite{BravyiKitaev05MagicStateDistillation,JozsaVdNest14_Classical_Simulation_Extended_Clifford_Circuits}. Concerning measurements, the authors believe that allowing, e.g.\, adaptive Pauli operator measurements (in the sense of chapter \ref{chapterF}) is unlikely to give any additional computational power to black-box normalizer circuits: in the best scenario, this could only happen  in infinite dimensions, since we showed (chapter \ref{chapterF}) that adaptive normalizer circuits over finite abelian groups are also efficiently classically simulable with stabilizer techniques. With more general types of measurements, it should be possible to recover full quantum universality, given that qubit cluster-states (which can be generated by Clifford circuits) are a universal resource for measurement-based quantum computation \cite{raussen_briegel_01_Cluster_State,raussen_briegel_onewayQC}. The possibility of obtaining  intermediate hardness results if non-adaptive yet also non-Pauli measurements are allowed (in the lines of \cite{Aaronson11_Computational_Complexity_Linear_Optics}  or \cite[theorem 7]{JozsaVdNest14_Classical_Simulation_Extended_Clifford_Circuits}) remains also open. 

\subsection{Chapter outline}

In section \ref{sect:Groups} we introduce our normalizer-circuit models over black-box groups.
In section \ref{sect:Quantum Algorithms3} we show how the  quantum algorithms in result 1 above are examples of black-box normalizer circuits\footnote{For the sake of conciseness, our results about  quantum algorithms to compute discrete-logarithms over elliptic-curves (which require a brief introduction to the latter abstract groups) are given in appendix \ref{sect:Elliptic Curve}.}. In section \ref{sect:Simulation_c3} we give our first completeness result and our no-go theorem (results 2-3). In section \ref{sect:Universality} we present our universality result. Finally, in section  \ref{sect:Complete Problems} we study additional complete problems (result 5).

\section{Black-box groups and black-box normalizer circuits}\label{sect:Groups}

 In this section we introduce abelian black-box groups and present  models of black-box group normalizer circuits; the latter generalize our earlier models in chapters \ref{chapterC}, \ref{chapterF}-\ref{chapterI}.

\subsection{Decomposed groups and black-box groups}

The most general groups we consider in this chapter are abelian groups of the form
\begin{equation}
\label{general_group}
G= \Z^a \times \T^b \times \DProd{N}{c} \times \mathbf{B}.
\end{equation}
where   the parameters $a$, $b$, $N_1,\cdots,N_c$ are arbitrary integers of unbounded size and $\mathbf{B}$ is an arbitrary (finite) \emph{abelian} black-box group. Following the nomenclature of chapters \ref{chapterF}-\ref{chapterI},  $\Z$ denotes the (infinite, discrete) group of integers under addition, $\T$ is the (infinite, continuous) group of angles in the interval $[0,1)$ under addition modulo $1$ and $\Z_{N_i}$ is the (finite) group of integers modulo $N_i$. 

 Note that the key difference with earlier chapters is the presence of a black-box group  $\mathbf{B}$. In terms of computational complexity, there is an stark separation between  decomposed abelian groups and black-box groups, which is discussed next.

\myparagraph{Decomposed abelian groups:} A finite abelian group $G$ is \emph{decomposed}  if it is of the form
\be\label{eq:Group Decomposition}
G = \Z_{N_1} \times \Z_{N_2} \times \cdots \times \Z_{N_k}
\ee
and the  positive integers $k, N_1,\cdots,N_k$ are given to us as a description of $G$. By the fundamental theorem of finite abelian groups (chapter \ref{chapter0}, theorem \ref{thm:Fundamental Theorem FAGroups}), any finite group can be put in the form (\ref{eq:Group Decomposition}) via isomorphism. Yet finding such a decomposition for a group $G$ may be difficult in practice. As a key example, there is currently not known efficient classical algorithm to decompose  the multiplicative group\footnote{Recall that  $\Z_N^\times$ is formed of integers relatively prime to $N$ multiplied modulo $N$.} $\Z_N^\times$ of integers modulo $N$ into  cyclic subgroups. In fact, the latter problem has long been believed to be classically hard even in the simple case $N=pq$ for $p$, $q$ prime: in this case,  $\Z_{pq}^\times \cong \Z_{p-1} \times \Z_{q-1}$ and  decomposing $\Z_{pq}^\times$ becomes at least as hard as \emph{factoring} $pq$ and, hence,  breaking the ubiquitous RSA cryptosystem \cite{RSA}. More generally, decomposing $\Z_N^\times$ is known to be polynomial time equivalent to factoring \cite{Shoup08_A_Computational_Introducttion_to_Number_Theory_and_Algebra}. In the quantum case, however, Cheung and Mosca gave an algorithm \cite{mosca_phd,cheung_mosca_01_decomp_abelian_groups} to decompose any finite abelian group.

\myparagraph{Black box groups:}\label{sect:Black Box Groups} In equation (\ref{general_group}), the factors $\DProd{N}{c}$ represent an arbitrary finite abelian group for which {the group decomposition is known}. The case where the decomposition is unknown will be covered by the black box group $\mathbf{B}$.

In this chapter, we define a \emph{black-box group} $\mathbf{B}$ \cite{BabaiSzmeredi_Complexity_MatrixGroup_Problems_I}  to be a finite  group whose elements are uniquely encoded by binary strings of a certain size $n$, which is the length of the encoding. The elements of the black-box group can be multiplied and inverted  at unit cost by querying a black-bock, or \emph{group oracle}, which computes these operations for us.  The order of a  black-box group with encoding length $n$ is bounded above by $2^n$: the precise order   $|\mathbf{B}|$ may not be given to us, but it is assumed that the group black box can identify which  strings in the group encoding correspond to elements of the group. When we say that a particular black-box group (or subgroup) is given (as the input to some algorithm), it is meant that a \emph{list of generators} of the group or subgroup is explicitly provided.

From now on, all  black-box groups in this work will be assumed to be \emph{abelian}. Although we only consider finite abelian black-box groups, we stress now, that  the only (albeit subtle) difference between these groups and the explicitly decomposed finite abelian groups in chapter \ref{chapterF} is that, for black-box groups, we assume no knowledge of a decomposition (\ref{eq:Group Decomposition}).  Our motivation to introduce black-box groups in our setting is precisely to model those abelian groups that cannot be efficiently decomposed with  known classical algorithms that have, nevertheless, efficiently classically computable group operations. With some abuse of notation, we shall call all  such groups also ``black-box groups'', even if no oracle is needed to define them; in such cases, oracle calls will be replaced by \ppoly{n}-size classical circuits for computing group multiplications and inversions.

As an  example, let us consider again the group $\Z_N^\times$. This group can be naturally modeled as a black-box group in the above sense: on one hand, for any $x,y \in \Z_N^\times$, $xy$ and $x^{-1}$ can be efficiently  computed using Euclid's algorithm \cite{brent_zimmerman10CompArithmetic}; the same algorithm  tells us whether a given integer $z\in\Z$ belongs to $\Z_N^\times$ (i.e., whether $z$ is coprime to $N$); on the other hand, decomposing $\Z_N^\times$ is as hard as factoring \cite{Shoup08_A_Computational_Introducttion_to_Number_Theory_and_Algebra}. Last, note that a generating set of $\Z_N^\times$ can be efficiently found by uniformly sampling random integers from $\{1,\ldots,N\}$ because $|\Z_N^\times|/|\Z_N|\in \Omega(1/\log\log{N})$ \cite{rosser1962} and due to the following lemma.
\begin{lemma}\label{lemma:SamplingGeneratingSets}
For any uniquely-encoded black-box group $\mathbf{B}$ with encoding length $n$, if it holds that $|\mathbf{B} |/2^n\in \Omega(1/\mathrm{poly}(n))$ (i.e., if the encoding used does not incur  into superpolynomial overhead), then a generating set of $\mathbf{B}$ can be found in probabilistic polynomial-time by sampling bit-strings in $\{0,1\}^n$ uniformly at random and rejecting those not that are not elements of $\mathbf{B}$.
\end{lemma}
\begin{proof}
Because $|\mathbf{B} |/2^n\in \Omega(1/\mathrm{poly}(n))$, we obtain an uniformly-random element of $\mathbf{B}$  after $T\in  O(\poly{N})$ trials with  $\Omega(1-c^{T})$ probability for some constant $c\in(0,1)$ (via the Chernoff-Hoeffding bound~\cite{Hoeffdingdoi:10.1080/01621459.1963.10500830}). Furthermore, by uniformly sampling $t\in\Theta(\log |G|)$ elements $g_1,\ldots, g_t$ from any finite group $G$, we  obtain a generating-set with probability exponentially close to 1. To see this, note that  if $G_i:=\langle g_1,\ldots,g_i\rangle$ is a proper subgroup of $G$, then $g_{i+1}\in G$ belongs to $G_i$ with a small probability $|G_i|/|G|\leq 1/2$. Further if $g_{i+1}\notin G_i$, then $|G_{i+1}|/|G_i|\geq 2$. Hence,   the cardinality $|G_t|$ converges exponentially fast to $|G|$ in  $t$. 
\end{proof}

\subsection{Black box normalizer circuits}\label{sect:Normalizer circuits over blackbox groups}\label{sect:Hilbert space}

We now define families of \emph{normalizer circuits over any group $G$} of form (\ref{general_group}) with Hilbert space
\begin{equation*}
\mathcal{H}_{G}=\mathcal{H}_\Z^{a}\otimes\mathcal{H}_\T^{b}\otimes \left(\mathcal{H}_{\Z_{N_1}}\otimes \cdots \otimes\mathcal{H}_{\Z_{N_c}}\right)\otimes \mathcal{H}_\mathbf{B},
\end{equation*}
which we view as the physical system of $m:=a+b+c+1$ computational registers. Because, mathematically speaking, $\mathbf{B}$ is equivalent  to some group $\DProd{D}{m}$ (via some isomorphism), the definitions of \emph{normalizer gates} over $G$ in this chapter will be identical to previous ones (chapter \ref{chapterC}). Consequently, in this chapter we adopt the model of normalizer circuit over infinite groups from  chapters \ref{chapterC}-\ref{chapterI}  with only a few minor \emph{modifications}---cf.\  (a-b-c-d) below---, which are needed to take some aspects of our new setting into consideration. In short, modifications below are related to, mainly, two facts:  (i) now   less information is given to us about the structure of the black-box group $\textbf{B}$;  (ii) in this chapter, precision errors need to be handled with more care than before since we aim at characterizing the power of black-box normalizer circuit model rigorously in terms of computational complexity classes.

\myparagraph{(a) Black-box designated bases:} As in chapter \ref{sect:Normalizer Gates Infinite Group}, all \emph{designated bases} (ditto for input states and final measurements\footnote{As in chapter \ref{sect:Normalizer Gates Infinite Group}, we only consider terminal measurements performed at the end of the computation.}) of a normalizer computation on a Hilbert space $\mathcal{H}_G$ are labeled by a group that is either fixed along the computation (when $G$ is a finite group) or may change via the action of infinite-group QFTs. Each designated basis ${\cal B}_{G'}$ is parametrized by a group $G'$ from the following family:
\begin{align}\label{group_labels_basis_c3} G'= G_1'\times\cdots \times G_{a+b}'\times \mathbb{Z}_{N_1}\otimes \dots\mathbb{Z}_{N_c}\times \mathbf{B} \quad\textnormal{where } G_i' \in\{\mathbb{Z}, \mathbb{T}\},\\
{\cal B}_{G'}:=\left\lbrace |g\rangle:=|g(1)\rangle\otimes \dots|g(m)\rangle, \quad g=(g(1), \dots, g(m))\in G'\right\rbrace.
\end{align}
where we simply adapted earlier formulas to consider $\mathbf{B}$. Note that, in the case of the (finite) black box group $\mathbf{B}$,  the Hilbert space $\mathcal{H}_\mathbf{B}$ has a (unique) standard  basis  $\{|\mathbf{b}\rangle\}$ where $\mathbf{b}$ ranges over all elements of $\mathbf{B}$. (It follows that $\mathcal{H}_\mathbf{B}$ is $|\mathbf{B}|$-dimensional.) Hence, the existence of multiple designated bases is, again, an infinite-dimensional feature as in chapter  \ref{sect:Normalizer Gates Infinite Group} and \emph{does not} come from the black-box. In fact, though multiple designated bases will  play a role in some of the quantum algorithms we consider (see, e.g.,  the factoring algorithm in section~\ref{sect:Factoring}) they will now show up e.g.\ in Shor's discrete log quantum algorithm (section~\ref{sect:Discrete Log}).

\myparagraph{(b) Black-box normalizer gates:} As in chapter \ref{chapterC}, a normalizer circuit over $G$ is a sequence of automorphism gates, quadratic phase functions and  QFTs over any group $G'$. However, in this chapter we will \emph{not allow} QFTs to act  on the black box subspace $\mathcal{H}_{\mathbf{B}}$ of the total system $\mathcal{H}_G$. This   restriction is natural because,  although  $\mathcal{H}_\mathbf{B}$ has a mathematically well-defined Fourier basis, it is not currently known how to implement its associated QFT without decomposing the black-box group first\footnote{To our best knowledge, all existing QFT-based quantum algorithms exploit only those abelian-group QFTs that act  on (explicitly) factorized systems of the form $\mathcal{H}_{\Z_{N_1}}\otimes \dots\otimes \mathcal{H}_{\Z_{N_c}}$ for this reason.}.

\subsubsection*{(c) Classical encodings for black-box normalizer gates}

Because of the presence of black box groups in $G=\Z^a \times \T^b\times \DProd{N}{c}\times \mathbf{B}$, it is no longer possible in this chapter to use the classical encodings of chapters \ref{chapterF}-\ref{chapterI} to represent automorphism and quadratic-phase gates since the latter crucially exploited that $G$ was given in a fully decomposed form\footnote{Note that, by construction, one cannot use, our earlier matrix representations encoding to represent group automorphisms over a non-decomposed  $\mathbf{B}$; similar issues affect our prior encodings for quadratic-phase gates.}. (This issue does not affect QFTs since we allow only decomposed-group ones.) 

For the above reason, throughout this chapter, we assume that any \emph{automorphism gate} $U_\alpha$ and \emph{quadratic phase gate} $D_\xi$ over $G$ can be given to us in a more general format, namely,  as a black-box quantum gates (i.e.,\ an \emph{oracle}) that can be implemented either at unit cost or by a poly-size quantum algorithm (that is explicitly given to us). Furthermore, we  make some additional simplifying assumptions about the associated classical functions $\alpha, \xi$, which we will assume \emph{efficiently computable rational functions}, which have the following restrictions.
\begin{enumerate}
\item \textbf{Rational.\footnote{We expect this assumption not to be essential, but it simplifies our proofs by allowing us to use exact arithmetic operations. Our stabilizer formalism in chapter \ref{chapterI} can still be applied if the functions $\alpha$, $\xi$ are not rational, and we expect some version of the  simulation result (theorem \ref{thm:Simulation}) to hold even when transcendental numbers are involved (taking carefully into account precision errors). It is a good question to explore whether an exact simulation result may hold for algebraic numbers \cite{Cohen:1995Course_Computational_Algebraic_Number_Theory}.}} An  automorphism/function $\alpha:G\rightarrow G$ is rational if it returns rational outputs for all rational inputs. A quadratic function $\xi$ is rational if it can be written in the form $\xi(g)=\exp\left(2\pii\, q(g) \right)$ where $q$ is a rational function from $G$ into $\R$ modulo $2\Z$.
\item\textbf{Efficiently computable.}  $\alpha$ and  $q$ can be computed by  polynomial-time uniform family of classical circuits $\{\alpha_i\}$, $\{q_i\}$. All   $\alpha_i$, $q_i$ are $\ppoly{m,i}$ size classical circuits that  query the \emph{group black box} at most $\ppoly{m,i}$ times: their inputs are strings of rational numbers whose numerators and denominators are represented by $i$ classical bits (their size is $O(2^i)$). For any rational element $g\in G$ that can be represented with so many bits (if $G$ contains factors of the form $\T$ these are approximated by fractions), it holds that $\alpha_i(g)=\alpha(g)$
 and $q_i(g)=q(g)$. 
 
In certain cases (see section \ref{sect:Quantum Algorithms3}) we will consider groups like $\Z_N^\times$  which, strictly speaking, are not black-box groups (because polynomial time algorithms for group multiplication for them are available  and there is no need to introduce oracles). In those cases, the queries to the group black box (in the above model) are substituted by some efficient subroutine.
\end{enumerate}
We add a third restriction to the above.
\begin{itemize}
\item[3.] \textbf{Precision bound.} For any  $q$ or $\alpha$ that acts on an \emph{infinite} group  a bound $n_\textrm{out}$ is given so that for every $i$, the number of bits needed to specify the numerators and denominators in the output of $q_i$ or $\alpha_i$ exactly is at most $i+n_{\mathrm{out}}$. The bound $n_\textrm{out}$ is independent of $i$ and indicates how much the input of each function may grow or shrink along the computation of the output\footnote{For infinite groups there is no fundamental limit to how much the output of $\alpha$ or $q$ may grow/shrink with respect to the input (this follows from the normal forms in chapter \ref{chapterGT}). The number $n_\textrm{out}$ parametrizes the precision needed to compute the function. This assumption might be weakened if a treatment for precision errors is incorporated in the model.}. This bound is used to correctly store the output of  maps $\alpha:\Z^a\rightarrow \Z^a$, $\alpha':\Z^a\rightarrow \T^a$ and to detect whether the output  of a function $\alpha'':\T^b \rightarrow \T^b$ might get truncated modulo 1.
\end{itemize}
The allowed automorphism gates $U_\alpha$ and quadratic phase gates $D_\xi$ are those associated with efficiently computable rational functions $\alpha$, $\xi$. We  ask these  unitaries to be efficiently implementable as well\footnote{Recall that, in finite dimensions, the gate cost of implementing a classical function $\alpha$ as a quantum gate is at most the classical cost \cite{nielsen_chuang} and that computing $q$ efficiently is enough to implement $\xi$ using  phase kick-back tricks \cite{Kaye05_optimized_Quantum_Elliptic_Curve}. We expect these results to extend to  infinite dimensional systems of the form $\mathcal{H}_\Z$.}, by \ppoly{m,i,n_\textrm{out}}-size quantum circuits comprising at most \ppoly{m,i,n_\textrm{out}} quantum queries of the group black box. The variable $i$ denotes the bit size used to store the labels $g$ of the inputs $\ket{g}$ and bounds the precision level $d$ of the normalizer computation, which we set to fulfill $\log d  \in O(i+n_\textrm{out})$. The complexity of a normalizer gate is measured by the number of gates and (quantum) oracle queries needed to implement them.

In the next section \ref{sect:Quantum Algorithms3}, we will see particular examples of efficiently computable normalizer gates. We will repeatedly make use of automorphism gates of the form \begin{equation}\nonumber
U_\alpha \ket{k_1,\ldots,k_m,x}\longrightarrow \ket{k_1,\ldots,k_m,b_1^{k_1}\cdots  b_m^{k_m}x}
\end{equation} where $k_i$ are integers and $b_j$, $x$ are elements of some black-box group $\mathbf{B}$. These gates are allowed in our model, since there exist well-known efficient classical circuits for modular exponentiation given access to a group multiplication oracle \cite{brent_zimmerman10CompArithmetic}. In this case, a precision bound can be easily computed: since the infinite elements $k_i$ do not change in size and all the elements of $\mathbf{B}$ are specified with strings of the same size, the  output of $\alpha$  can be represented with as many bits as the input and we can simply take $n_\textrm{out}= 0$ (no extra bits are needed).

Many examples of efficiently computable normalizer gates were given in chapter \ref{chapterC}; for decomposed finite group $\DProd{N}{c}$. It was also shown in  \cite{VDNest_12_QFTs} that all normalizer gates over such groups can be efficiently implemented.

\subsubsection*{(d) Precision requirements}

Finally, recall that in the model of quantum circuits we use, input states and final measurements in the Fourier-basis $\{\ket{p}, p\in\T\}$ of $\mathcal{H}_\Z$  can never be implemented with perfect accuracy, a limitation that stems from the fact that the $\ket{p}$ states are \emph{unphysical}. This can be quickly seen in two ways: first, in the $\Z$ basis, these states are infinitely-spread plane-waves  $\ket{p}=\sum \overline{\euler^{2\pii zp}} \ket{z}$; second, in the $\T$
basis, they are infinitely-localized Dirac-delta pulses. Physically, preparing Fourier-basis states or measuring in this basis \emph{perfectly} would require infinite energy and lead to infinite precision issues in our computational model.

In the algorithms we study in this work (namely, the order-finding algorithm in theorem \ref{thm:Order Finding}), Fourier states over $\Z$ can be substituted with \emph{\textbf{realistic physical approximations}}. The degree of \emph{precision} used in the process of  Fourier state preparation is treated as a \emph{computational resource}. We model the precision used in a computation as follows.

Since  our goal is  to use the Fourier basis $\ket{p}$, $p\in \T$, to represent information in a computation, we require the ability to store and retrieve information in this  continuous-variable basis. Our assumption is that for any finite set $X$ with cardinality $d=|X|$, we can divide the continuous circle-group $\T$ spectrum into $d$ equally sized sectors of length $1/d$ and use them to represent the elements of $X$. More precisely, to each element of $X$ we assign a number in $\Z_d$. The element $x_i\in X$ with index $i\in\Z_d$ is then represented by any state of the subspace $V_{i,d}=\mathrm{span}\{\ket{\tfrac{i}{d}+\Delta} \textnormal{ with }|\Delta|  <\tfrac{1}{2d}\}$. We call the latter states  \emph{$d$-approximate Fourier states} and refer to $d$ as the \emph{precision level}  of the computation. We assume that these states can be prepared and distinguished to any desired precision $d$ in the following way:
\begin{itemize}
\item[1.] \textbf{State preparation assumption.} Inputs $\ket{\psi_i}$ with at least $\tfrac{2}{3}$ fidelity to some element of  $V_{i,d}$ can be prepared for any $i\in \Z_d$.
\item[2.] \textbf{Distinguishability assumption.}  The subspaces $\{V_{i,d}\}_i$ can be reliably distinguished.
\end{itemize}
Note that $d$ determines how much information is stored in the Fourier basis.

\begin{definition}[\textbf{Efficient use of precision}] A \emph{quantum algorithm} that uses $d$-approximate Fourier states to solve  a computational problem with input size $n$ is said to use an \emph{efficient} amount of precision if and  only if $\log{d}$ is upper bounded by some polynomial of $n$. Analogously,  an algorithm that stores information  in the standard basis $\{\ket{m},\,m\in\Z\}$ is said to be \emph{efficient} if the states with $m$ larger than some threshold $\log{( m_\textrm{max})}\in O(\poly{n})$ do not play a role in the computation.\footnote{Note that this definition is not necessary to define normalizer circuits but to discuss the physicality of the model. We point out that there might be better ways to model precision than ours (which may, e.g., lead to tighter bounds or more efficient algorithms), but our simple model is enough to derive our main results. We advance that, even if these precision requirements turned out to be high in practice, there exist efficient discretized \emph{qubit} implementations  of all the infinite-dimensional quantum algorithms that we study later in the chapter (cf.\ theorem  \ref{thm:Shor normalizer}).}
\end{definition}

\section{Quantum algorithms}\label{sect:Quantum Algorithms3}

\subsection{The discrete logarithm problem over $\Z_p^\times$}\label{sect:Discrete Log}

In this section we consider the discrete-logarithm problem studied by Shor \cite{Shor}. For any prime number $p$, let $\Z_p^\times$  be the multiplicative group of non-zero integers modulo $p$. An instance of the discrete-log  problem over $\Z_p^\times$ is determined by two  elements $a$, $b\in \Z_p^\times$, such that $a$ generates the group $\Z_p^\times$. Our task is to find the smallest non-negative integer $s$ that is a solution to the equation $a^s=b \bmod{p}$; the number is called the discrete logarithm $s=\log_a b$.

We now review Shor's algorithm \cite{Shor,childs_vandam_10_qu_algorithms_algebraic_problems} for this problem and prove our first result.
\begin{theorem}
[\textbf{Discrete logarithm}]\label{thm:Discrete Log} Shor's quantum algorithm for the discrete logarithm problem over $\Z_p^\times$ is a black-box normalizer circuit over the group $\Z_{p-1}^2\times \Z_p^\times$.
\end{theorem}
Theorem \ref{thm:Discrete Log} shows that black box normalizer circuits over \emph{finite} abelian groups can efficiently solve a problem for which no efficient classical algorithm is known. In addition, it tells us that black-box normalizer circuits can render widespread public-key cryptosystems vulnerable: namely, they break the Diffie-Helman key-exchange protocol \cite{DiffieHellman}, whose security  relies in the assumed classical intractability of the discrete-log problem.
\begin{proof}
Let us first recall the main steps in Shor's discrete log algorithm.
\begin{algorithm}[Shor's algorithm for the discrete logarithm]\label{alg:Discrete Log}$\,$
\begin{alg_in} Positive integers $a$, $b$, where $\Z_p^\times = \langle a \rangle $.
\end{alg_in}
\begin{alg_out} The least nonnegative integer $s$ such that $a^s\equiv b \pmod{p}$.
\end{alg_out}
We will use three registers indexed by integers, the first two modulo $p-1$ and the last modulo $p$. The first two registers will correspond to the additive group $\Z_{p-1}$, while the third register will correspond to the multiplicative group $\Z_p^\times$. Two important ingredients of the algorithm will be the unitary gates $U_a:\ket{s}\rightarrow\ket{sa}$ and $U_b:\ket{s}\rightarrow\ket{sb}$.
\begin{enumerate}
\item \textbf{Initialization:} Start in the state $\ket{0}\ket{0}\ket{1}$.

\item Create the superposition state $\frac{1}{p-1} \sum_{x,y=0}^{p-1} \ket{x} \ket{y} \ket{1} $, by applying the standard quantum Fourier transform on the first two registers.

\item Apply the unitary $U$ defined by $U\ket{x}\ket{y} \ket{z} = \ket{x} \ket{y} \ket{za^xb^y}$, to obtain the state
\[ 
\frac{1}{p-1}\sum_{x,y=0}^{p-1} \ket{x} \ket{y} \ket{a^xb^y}
\]
This is equivalent to applying the controlled-$U_a^x$ gate between the first and third registers, and the controlled-$U_b^y$ between the second and third registers.

\item Measure and discard the third register. This step generates a so-called coset  state
\[ 
\frac{1}{\sqrt{p-1}}\sum_{k=0}^{p-1} \ket{\gamma  +ks,-k},
\]
where $\gamma$ is some uniformly random element of $\Z_{p-1}$ and $s$ is the discrete logarithm.

\item Apply the quantum Fourier transform over $\Z_{p-1}$ to the first two registers, to obtain
\[ 
\frac{1}{\sqrt{p-1}}\sum_{k'=0}^{p-1} \euler^{2\pii \frac{k'\gamma}{p-1}}\ket{k',k's},
\]
\item Measure the system in the standard basis to obtain a pair of the form $(k',k's)\bmod{p}$ uniformly at random.

\item Classical post-processing. By repeating the above process $n$ times, one can extract the discrete logarithm $s$ from these pairs with exponentially high probability (at least $1- 2^{-n}$), in classical polynomial time.
\end{enumerate}
\end{algorithm}
Note that the Hilbert space of the third register precisely corresponds to  $\mathcal{H}_{\mathbf{B}}$ if we choose the black-box group to be $\mathbf{B}=\Z_p^\times$. It is now easy to realize that Shor's algorithm for discrete log is a normalizer circuit over $\Z_{p-1}\times\Z_{p-1}\times\Z_{p}^\times$: steps 2 and 4 correspond to applying partial QFTs over $\Z_{p-1}$, and the gate $U$ applied in state 3 is a group automorphism over $\Z_{p-1}\times\Z_{p-1}\times\Z_{p}^\times$. 
\end{proof}
We stress that, in the proof above, there is no known efficient classical algorithm for solving the group decomposition problem for the group $\Z_p^\times$  (as we define it in section \ref{sect:Group Decomposition}): although, by assumption, we know that $\Z_p^\times=\langle a \rangle \cong \Z_{p-1}$, this information does not allow us to convert elements from one representation to the other, since this requires solving the discrete-logarithm problem itself. In other words, we are unable to compute classically the \emph{group isomorphism} $\Z_p^\times \cong \Z_{p-1}$. In our version of the group decomposition problem, we require the ability to \emph{compute} this group isomorphism. For this reason, we treat the group  $\Z_p^\times$ as a \emph{black-box group}.

\subsection{Shor's factoring algorithm	}\label{sect:Factoring}

In this section we will show  that normalizer circuits can efficiently compute the order of elements of (suitably encoded) abelian groups. Specifically, we show how to efficiently solve the {order finding problem}  for every  (finite) abelian black-box  group $\mathbf{B}$  \cite{BabaiSzmeredi_Complexity_MatrixGroup_Problems_I} with normalizer circuits. Due to the well-known classical reduction of the factoring problem to the problem of computing orders of elements of the group $\Z_N^\times$, our result implies that black-box normalizer circuits can efficiently factor large composite numbers, and thus break the widely used RSA public-key cryptosystem~\cite{RSA}.

We briefly introduce the \textbf{order finding problem} over a  black-box group $\mathbf{B}$, that we always assume to be finite and abelian. In addition, we assume that the elements of the black-box group can be uniquely encoded with  $n$-bit strings, for some known $n$. The task we consider is the following: given an  element $a$ of $\mathbf{B}$, we want to compute the order $|a|$ of $a$ (the  smallest positive integer $r$ with the property\footnote{Since $\mathbf{B}$ is finite, the order $|a|$ is  a well-defined number.} $a^r=1$).  Our next theorem states that this version of the order finding problem can be efficiently solved by a quantum computer based on normalizer circuits.
\begin{theorem}[\textbf{Order finding over $\mathbf{B}$}]\label{thm:Order Finding}  Let $\mathbf{B}$ be a finite abelian black-box group  and $\mathcal{H}_\mathbf{B}$ its associated Hilbert space. Let $V_a$ be the unitary that performs the group multiplication operation on $\mathcal{H}_\mathbf{B}$: $V_a\ket{x}=\ket{ax}$. We denote by $c\mbox{\,-}V_a$  the unitary that performs $V_a$ on $\mathcal{H}_\mathbf{B}$ controlled on the value of an ancillary register $\mathcal{H}_\Z $:
 \begin{equation}\notag
 \ket{m,x}\quad\xrightarrow{\quad c-V_{a}\quad}\quad\ket{m,a^mx},\qquad\textnormal{for any $m$ in } \Z.
 \end{equation}
Assume that we can query an oracle that implements $c\mbox{\,-}V_a$ in one time step for any  $a\in \mathbf{B}$. Then, there exists a hybrid version of Shor's order-finding algorithm, which can compute the order $|a|$ of any $a\in\mathbf{B}$ efficiently, using  normalizer circuits over the group $\Z\times \mathbf{B}$ and classical post-processing. The algorithm runs in polynomial-time, uses an efficient amount of precision and succeeds with high probability.
\end{theorem}
In theorem \ref{thm:Order Finding},  by ``efficient amount of precision'' we mean that instead of preparing Fourier basis states of $\mathcal{H}_{\Z}$ or measuring on this (unphysical) basis, it is enough to use realistic physical approximations of these states (cf. section \ref{sect:Normalizer circuits over blackbox groups}).
\begin{proof}
We divide the proof into two steps. In the first part, we give an infinite-precision  quantum algorithm to randomly sample elements from the set $\mathrm{Out}_a=\{\frac{k}{|a|}: k\in \Z\}$ that uses normalizer circuits over the group $\Z\times\mathbf{B}$ in polynomially many steps. In this first algorithm, we assume that Fourier basis states of $\mathcal{H}_\Z$ can be prepared perfectly and that there are no physical limits in measurement precision; the outcomes $k/|a|$ will be stored with floating point arithmetic and with finite precision. The algorithm allows one to extract the period $|a|$ efficiently by sampling fractions $k/|a|$ (quantumly) and then using a continued fraction expansion (classically).

 In the second part of the proof, we will remove the infinite precision assumption.

Our first algorithm is essentially a variation of Shor's algorithm for order finding \cite{Shor} with one key modification: whereas Shor's algorithm uses a large $n$-qubit register $\mathcal{H}_2^n$ to estimate the eigenvalues of the unitary $V_a$, we will replace this multiqubit register with a single \emph{infinite} dimensional Hilbert space $\mathcal{H}_{\Z}$. The algorithm is \emph{hybrid} in the sense that it involves both continuous- and discrete-variable registers. The key feature of this algorithm is that, at every time step, the implemented gates are \emph{normalizer gates}, associated with the groups $\Z\times \Z_N^\times$ and  $\T\times \Z_N^\times$ (which are, themselves, related via the partial Fourier transforms $\Fourier{\Z}$ and $\Fourier{\T}$). The algorithm succeeds with constant probability.
\begin{algorithm}[\textbf{Hybrid order finding with infinite precision}]\label{alg:Order Finding infinite precision}
 $\,$
 \begin{alg_in}
A black box (finite abelian) group $\mathbf{B}$, and an element $a \in \mathbf{B}$.
 \end{alg_in}
 \begin{alg_out}
 The order $r:=|a|$ of $a$ in $\mathbf{B}$, i.e. the least positive integer $r$ such that $a^r = 1$.
 \end{alg_out}
 We will use multiplicative notation for the black box group $\mathbf{B}$, and additive notation for all other subgroups.
\begin{enumerate}
\item \textbf{Initialization:} Initialize $\mathcal{H}_\Z$ on the Fourier basis state $\ket{0}$ with $0\in \T$, and $\mathcal{H}_{\mathbf{B}}$ on the state $\ket{1}$, with $1\in \mathbf{B}$. In our formalism, we will regard $\ket{0,1}$ as a standard-basis state of the basis labeled by $\T\times\mathbf{B}$.
\item  Apply the Fourier transform $\Fourier{\T}$ to the register $\mathcal{H}_\Z$.  This changes the designated basis of this register to be  the one labeled by the group $\Z$. The state $\ket{0}$ in the new basis is an infinitely-spread comb of the form $\sum_{m\in\Z}\ket{m}$.
\item Let the oracle $V_a$ act jointly on $\mathcal{H_\Z}\times \mathcal{H}_\mathbf{B}$; then the state is mapped in the following manner:
\begin{equation}
\sum_{m\in\Z}\ket{m}\ket{1}\quad\xrightarrow{\quad c\mbox{\,-}V_a\quad }\quad\sum_{m\in \Z} \ket{m, a^m}.
\end{equation}
Note that, in our formalism,  the oracle $c\mbox{\,-}V_a$  can be regarded as an automorphism gate $U_\alpha$. Indeed, the gate implements a classical invertible function on the group $\alpha(m,x)=(m,a^mx)$. The function is, in addition, a continuous\footnote{This is vacuously true: since the group $G:=\Z\times\mathbf{B}$ is discrete, \emph{any} function $f:G\rightarrow G$ is continuous.} group automorphism, since
\begin{align}	\notag
\alpha\left((m,x)(n,y)\right)=\alpha(m+n,xy)&=(m+n,(a^{m+n})(xy))\\&=(m+n,(a^{m}x)(a^{n}y))=(m,a^{m}x)(n,a^{n}y)\label{eq:ModExp is Automorphism}\\
&=\alpha(m,x)\alpha(n,y).\notag
\end{align}
\item Measure and discard the register $\mathcal{H}_{\mathbf{B}}$. Say we obtain $a^s$ as the measurement outcome. Note that the function $a^m$ is periodic with period $r=|a|$, the order of the element. Due to periodicity, the state after measuring $a^s$ will be of the form
\begin{equation}\label{eq:Periodic State}
\left(\sum_{j\in \Z} \ket{s+jr}\right) \ket{a^s}.
\end{equation}	
After discarding $\mathcal{H}_{\mathbf{B}}$ we end up in a periodic state $\sum \ket{s+jr}$ which encodes $r=|a|$.
\item  Apply  the Fourier transform $\Fourier{\Z}$ to the  register $\mathcal{H}_\Z$.  We work again in the Fourier basis of $\mathcal{H}_\Z$, which is labeled by the circle group $\T$. The periodic state $\sum\ket{s+jr}$ in the dual $\T$ basis reads \cite{oppenheim_Signals_and_Systems}
\begin{equation}\label{eq:Periodic state after Fourier Transform}
\sum_{k=0}^{r-1} \euler^{2\pii \frac{sk}{r}} \ket{\tfrac{k}{r}}  
\end{equation}
\item Measure $\mathcal{H}_{\Z}$ in the Fourier basis (the basis labeled by $\T$). Since  we assume that the initial state of the computation can be as close to $\ket{0}$ as we wish, the wavefunction of the final state  (\ref{eq:Periodic state after Fourier Transform}) is \emph{sharply peaked} around values $p\in\T$ of the form $k/r$. As a result, a high resolution measurement will let us sample these numbers (within some floating-point precision window $\Delta$) nearly uniformly at random.
\item \textbf{Classical postprocessing:} Repeat steps 1-7 a few times and use a (classical) continued-fraction expansion algorithm \cite{nielsen_chuang,KLM_QC_07}  to extract the order $r$ from the randomly sampled multiples $\{k_i/r\}_i$. This can be done, for instance, with an  algorithm from \cite{Knill95onshors} that obtains $r$ with \emph{constant} probability after sampling two numbers $\tfrac{k_1}{r}$, $\tfrac{k_2}{r}$, if the measurement resolution is high enough:  $\Delta \leq 1/2r^2$ is enough for our purposes.
\end{enumerate}\vspace*{-10pt}
\end{algorithm}
Manifestly,  there is a strong similarity between algorithm \ref{alg:Order Finding infinite precision} and Shor's factoring algorithm:  the quantum Fourier transforms $\Fourier{\T}$ in our algorithm $\Fourier{\Z}$ plays the role of the discrete Fourier transform $\Fourier{2^n}$, and $c\mbox{\,-}V_a$ acts as the modular exponentiation gate \cite{Shor}.  In fact, one can regard algorithm \ref{alg:Order Finding infinite precision} as a ``hybrid'' version of Shor's algorithm combining both  continuous and discrete variable registers. The remarkable feature of this version of Shor's algorithm is that the quantum part of the algorithm 1-6 is a normalizer computation.

Algorithm \ref{alg:Order Finding infinite precision} is efficient if we just look at the number of gates it uses. However, the algorithm is \emph{inefficient} in that it uses infinitely-spread Fourier states $\ket{p}=\sum_{m\in\Z}\euler^{-2\pii pm}\ket{m}$ (which are unphysical and cannot be prepared with finite computational resources) and arbitrarily precise measurements. We finish the proof of theorem \ref{thm:Order Finding} by giving an improved algorithm that does not rely on unphysical requirements.

\newpage

\begin{algorithm}[\textbf{Hybrid order finding with finite precision}] \label{alg:Order Finding Finite precision}$\,$
\begin{enumerate}
 \item[1-2] \textbf{Initialization:} Initialize $\mathcal{H}_{\mathbf{B}}$ to $\ket{1}$. The register  $\mathcal{H}_\Z$ will begin in an \emph{approximate} Fourier basis state $\ket{\widetilde{0}}= \tfrac{1}{\sqrt{2M+1}}\sum_{-M}^{+M}\ket{m}$,  i.e.\ a square pulse of length $2M+1$ in the integer basis, centered at 0. This step simulates steps 1-2 in algorithm \ref{alg:Order Finding infinite precision}.
 
 \item[3-4] Repeat steps 3-4 of algorithm \ref{alg:Order Finding infinite precision}. The state after obtaining the measurement outcome $a^s$ is now different due to the finite ``length'' of the comb $\sum_{m=0}^{M} \ket{m}$; we obtain
 \begin{equation}\label{eq:Periodic state finite}
|\psi\rangle =\frac{1}{\sqrt{L}}\sum_{-L_a}^{L_b}\ket{s+jr},
 \end{equation}
 where $L=L_a+L_b+1$ and $s$ is obtained nearly uniformly at random from $\{ 0,\ldots, r-1\}$. The values $L_a$, $L_b$ are positive integers of of the form  $\lfloor M/r\rfloor - \epsilon $ with $-2\leq \epsilon\leq 0$  (the particular value of $\epsilon$ depends on $s$, but it is irrelevant in our analysis). Consequently,  we have  $L=2\lfloor M/r\rfloor - (\epsilon_a +\epsilon_b)$.
\item[5]  Apply  the Fourier transform $\Fourier{\Z}$ to the  register $\mathcal{H}_\Z$ . The wavefunction of the final state  $\hat{\psi}$  is the Fourier transform of  the wavefunction $\psi$ of  (\ref{eq:Periodic state finite}). We  compute $\hat{\psi}$ using formula~(\ref{eq:QFT over Z}):
\begin{align}
 \hat{\psi}(p)&=\sum_{x\in\Z} \euler^{2\pii p x}\psi(x)=\frac{1}{\sqrt{L}}\sum_{-L_a}^{L_b} \euler^{2\pii p (s+jr)}=\frac{1}{\sqrt{L}} \left(\euler^{2\pii p s}\right)\frac{\euler^{2\pii p r (L_b+1)}-\euler^{-2\pii p r L_a}}{\euler^{2\pii p r }-1} \notag\\
 &= \frac{\euler^{2\pii p\left( s+\tfrac{L_b-L_a}{2}\right)}}{\sqrt{L}}  \frac{ \sin{ \left( \pi L pr  \right)} } { \sin{\left( \pi pr  \right)}} = \frac{\euler^{2\pii p\left(s +\tfrac{L_b-L_a}{2}\right) }}{\sqrt{L}} D_{L,r}({p}) 	\label{eq:Dirichlet State}
\end{align}
(to derive the equation, we apply the summation formula of the geometric series and  re-express the result in terms of the Dirichlet kernel \cite{rudin62_Fourier_Analysis_on_groups}
\begin{equation}\label{eq:Dirichlet Kernel}
D_{L,r}(p)= \frac{\sin{ \left( \pi L pr \right)}}{\sin{\left( \pi pr  \right)}}.
\end{equation} 
\item[6] \textbf{Measure $\mathcal{H}_{\Z}$ } in the Fourier basis. We show now that, if  the resolution is high enough, then the probability distribution of measurement outcomes will be ``polynomially close'' to the one obtained in the infinite precision case (\ref{eq:Periodic state after Fourier Transform}). Intuitively, this is a consequence of the fact that in the limit $M\rightarrow \infty$  (when the initial state becomes an infinitely-spread comb), we have also $L\rightarrow\infty$ and that the function $D_{L},r(p)$  converges  to a train  $\sum_{k=0}^{r-1} \delta_{k/r}(p)$ of Dirac measures \cite{rudin62_Fourier_Analysis_on_groups}. In addition, for a high finite value of $M$, we  find that the probability of obtaining some outcome  $p$ within a $\Delta=\tfrac{1}{Lr}$ window of a fraction  $\tfrac{k}{r}$ is also high.
\begin{equation}\label{eq:Probability in Shor Infinite}
\mathrm{Pr}(|p-\tfrac{k}{r}|\leq \tfrac{\Delta}{2})=\frac{1}{L} \int_{-\tfrac{\Delta}{2}}^{+\tfrac{\Delta}{2}} \frac{\sin^2{ \left( \pi L pr \right)}}{\sin^2{\left( \pi pr  \right)}}  \,\mathrm{d}p  \geq \tfrac{\Delta}{L}\frac{\sin^2{ \left( \tfrac{\uppi}{2} \right)}}{\sin^2{\left( \tfrac{\uppi}{2L}   \right)}} \geq \frac{4}{\uppi^2r},
\end{equation}
where we use the mean value theorem  and the bound $\sin(x)^2\leq x^2$.  It follows that with  \emph{constant} probability (larger than $4/\pi^2\approx0.41$)  the measurement will output some outcome $\tfrac{\Delta}{2}$-close to a number of the form $k/r$. (A tighter lower bound of $2/3$ for the success probability can be obtained by evaluating the integral numerically.)

Lastly, note that although the derivation of (\ref{eq:Probability in Shor Infinite}) implicitly assumes that the finial measurement is infinitely precise, it is enough to implement measurements with resolution close to $\Delta$. Due to the peaked shape of the final distribution (\ref{eq:Probability in Shor Infinite}), it follows that $\Theta(\tfrac{1}{M})$ resolution is enough if our task is to sample $\tfrac{\Delta}{2}$-estimates of these fractions nearly uniformly at random; this scaling  is \emph{efficient} as a function of   $M$ (cf. section \ref{sect:Normalizer circuits over blackbox groups}).

\item[7] \textbf{Classical postprocessing:} We now set $M$ (the length of the initial comb state) to be large enough so that $\tfrac{\Delta}{2}=\tfrac{1}{2Lr}\leq \tfrac{1}{2r^2}$; since $r\leq |\mathbf{B}|$, taking $\log M \in O(\poly{n})$, where $n$ denotes the encoding length of $\mathbf{B}$, is enough for our purposes. With such an $M$, the measurement step 6 will output a number $p$ that is $\tfrac{1}{2r^2}$ close to a $\tfrac{k}{r}$ with  high probability, which can be increased to be arbitrarily close to 1 with a few repetitions. We then proceed as in step 7 of algorithm \ref{alg:Order Finding infinite precision} to compute the order $r$.\qedhere
\end{enumerate}\vspace*{-10pt}
\end{algorithm}
\end{proof}

\subsubsection*{Shor's algorithm as a normalizer circuit}

Our discussion in the previous section reveals strong a resemblance between our hybrid normalizer quantum algorithm for order finding and Shor's original quantum algorithm for this problem \cite{Shor}: indeed, both quantum algorithms employ remarkably similar circuitry. In this section we show that this resemblance is actually more than a mere fortuitous analogy, and that, in fact, one can understand Shor's original order-finding algorithm as a discretized version of our finite-precision hybrid algorithm for order finding \ref{alg:Order Finding infinite precision}.

\begin{theorem}[\textbf{Shor's algorithm as a normalizer circuit}] \label{thm:Shor normalizer} Shor's order-finding algorithm \cite{Shor} provides an efficient discretized implementation of our hybrid normalizer algorithm \ref{alg:Order Finding Finite precision}. 
\end{theorem}
Note that the theorem does not imply that all possible quantum algorithms for order finding are normalizer circuits (or discretized versions of some normalizer circuit). What it shows is that the one first found by Shor in \cite{Shor} does exhibit such a structure.
\begin{proof}
Our approach will be to show explicitly that the evolution of the initial quantum state  in Shor's algorithm is analogous to that of the initial state in algorithm \ref{alg:Order Finding Finite precision} if we discretize the computation. Recall that Shor's algorithm implements a quantum phase estimation \cite{kitaev_phase_estimation} for the unitary $V_a$. Let $D$ be the dimension of the Hilbert space used to record such phase. We assume $D$ to be odd\footnote{This choice is not essential, neither in Shor's algorithm nor in algorithm \ref{alg:Order Finding Finite precision}, but it simplifies the proof.} and write $D=2M+1$. Then Shor's algorithm can be written as follows:
\begin{enumerate}
\item Initialize the state $\ket{0,1}$ on the Hilbert space $\mathcal{H}_D\times \mathcal{H}_{\Z_N^\times}$.
\item Apply the discrete Fourier transform $\Fourier{\Z_D}$ on $\mathcal{H}_D$ to obtain
\begin{equation}
\sum_{m=0}^{D-1} \ket{m}\ket{1}=\sum_{-M}^{M} \ket{m}\ket{1}.
\end{equation}
So far, we have simulated  step 1 in algorithm \ref{alg:Order Finding Finite precision}  by constructing the same periodic state. These first two steps are also clearly analogous to  steps 1-2 in algorithm \ref{alg:Order Finding infinite precision}.
\item[3-4] Apply the modular exponentiation gate $U_{\mathrm{me}}$, which is the following unitary  \cite{Shor}
\begin{equation}\label{eq:Modular Exponentiation Gate}
U_{\mathrm{me}}\ket{m,x}=\ket{m,a^mx },
\end{equation}
to the state. Measure the register $\mathcal{H}_{\Z_N^\times}$ in the standard basis. We obtain, again, a quantum state of the form (\ref{eq:Periodic state finite}), with $L\leq D$. 

\item[6] We apply the discrete Fourier transform $\Fourier{\Z_D}$ to the register $\mathcal{H}_{\Z_D}$ again. We claim now that the output state will be a discretized version of (\ref{eq:Dirichlet State}) due to a remarkable \textbf{mathematical correspondence} between  Fourier transforms. Note that any quantum state $\ket{\psi}$ of the infinite-dimensional Hilbert space $\mathcal{H}_{\Z}$ can be regarded as a quantum state of $\mathcal{H}_{D}$ given that the support of $\ket{\psi}$ is limited to the standard basis states $\ket{0}, \ket{\pm 1}, \ldots, \ket{\pm M}$. Let us denote the latter state $\ket{\psi_D}$ to distinguish both. Then, we observe a correspondence between letting $\Fourier{\Z}$ act on $\ket{\psi}$ and letting $\Fourier{\Z_D}$ act on $\ket{\psi_D}$.
\begin{equation}\label{eq:Fourier Correspondence}
 \displaystyle\hat{\psi}(p)= \sum_{x=-M}^{x=+M} \euler^{2\pii p x}\psi(x)\qquad \longleftrightarrow\qquad\displaystyle\hat{\psi}_D(k)= \sum_{x=-M}^{x=+M} \euler^{2\pii \tfrac{kx}{D}}\psi_D(x)
\end{equation}
The correspondence (equation \ref{eq:Fourier Correspondence}) tells us that, since we have $\psi(x)=\psi_D(x)$, it follows that the Fourier transformed function $\hat{\psi}_D(k)$ is precisely the function $\hat{\psi}(p)$ evaluated at points of the form $p=\tfrac{k}{D}$. The final state can be written as 
\begin{equation}\label{eq:Discretized Shor Output}
\sum_{k=0}^{D-1} \hat{\psi}\left(\tfrac{k}{D}\right)\ket{k}. 
\end{equation}
which is, indeed, a discretized version of  (\ref{eq:Dirichlet State}). 
\item[7-8] The last steps of Shor's algorithm are identical to 7-8 in algorithm \ref{alg:Order Finding Finite precision}, with the only difference being that the wavefunction (\ref{eq:Discretized Shor Output}) is now a discretization of (\ref{eq:Dirichlet State}). The probability of measuring a number $k$ such that $\tfrac{k}{D}$ is close to a multiple of the form $\tfrac{k'}{r}$ will again be high, due to the properties of the Dirichlet kernel (\ref{eq:Dirichlet Kernel}). Indeed, one can show (see, e.g.\, \cite{childs_vandam_10_qu_algorithms_algebraic_problems}) with an argument similar to (\ref{eq:Probability in Shor Infinite}) that, by setting $D=N^2$, the algorithm outputs with constant probability and almost uniformly a fraction $\tfrac{k}{D}$ among the two closest fraction to some value of the form $k/r$ (see e.g. \cite{Shor} for details). The period $r$ can be recovered, again, with a continued fraction expansion.\qedhere
\end{enumerate}
 \end{proof}

 \subsubsection*{Normalizer gates over $\infty$ groups are necessary to factorize}
 
 At this point, it is a natural question to ask whether it is necessary at all to replace the Hilbert space $\mathcal{H}_2^n$ with an infinite-dimensional space $\mathcal{H}_\Z$ with an integer basis in order to be able to factorize with normalizer circuits. We discuss in this section that, in the view of the authors, this is a \textbf{key indispensable ingredient} of our proof. 
 
We begin our discussion  pointing out obstacles for finding quantum factoring algorithm  based on \emph{modular exponentiation} gates (controlled $V_a$ rotations), showing that implementing the latter by normalizer circuits over \emph{finite} groups $\Z_M\times \mathbf{B}$ is not possible without solving a computational problem at least as hard as factoring. 
 \begin{theorem} \label{thm:ModExp requires Z} Let $\mathcal{H}_{\Z_M}=$ be the Hilbert space with basis $\{\ket{0},\ldots,\ket{M-1}\}$ and dimension $M$.  Let $\mathbf{B}$ be an abelian black-box group with associated Hilbert space $\mathcal{H}_{\mathbf{B}}$. Consider the composite Hilbert space $\mathcal{H}=\mathcal{H}_{\Z_M}\times \mathcal{H}_{\mathbf{B}}$ and define  $U_\mathrm{me}$ to be the unitary gate on $\mathcal{H}$ defined as $U_{\mathrm{me}}\ket{m,x}=\ket{m,a^mx }$, where $a,x \in \mathbf{B}$ and  $m\in\Z_M$. Then, unless $M$ is a multiple of the order of $a$, there does not exist any normalizer circuit over $\mathcal{H}$ (even of exponential size) satisfying $\|\mathcal{C}-U_\mathrm{me} \|_{\mathrm{op}}\leq1-{2}^{-1/2}$.
 \end{theorem}
We prove the theorem in appendix \ref{app:ModExp requires Z}. We highlight that a similar result was proven in  \cite[theorem 2]{VDNest_12_QFTs}: that normalizer circuits over groups of the form $\Z_{2^n}\times\Z_N$ also fail to approximate the modular exponentiation.  Also, we point out that it is easy to see that the converse of theorem \ref{thm:ModExp requires Z} is also true:  if $|a|$ divides $M$, then an argument similar to (\ref{inproof:Ume is no Clifford})  shows that $(m,x)\rightarrow(m,a^mx)$ is a group automorphism of $\Z_M\times \mathbf{B}$, and the gate  $U_\mathrm{me}$ automatically becomes a normalizer automorphism gate.
 
The main implication of theorem \ref{thm:ModExp requires Z}  is that, although  finite-group normalizer circuits over $\Z_N \times \mathbf{B}$ can easily implement the quantum Fourier transforms needed for Shor's factoring algorithm, they \emph{cannot} implement nor approximate the quantum modular exponentiation gate between $\mathcal{H}_\mathbf{B}$, playing the role of the target system, and some ancillary control system, \emph{unless} a multiple $M=\lambda |a|$ of the order  of $a$ is known in advance. Yet the problem of finding  multiples of orders is \emph{at least as hard as factoring and order-finding}: for $\mathbf{B}=\Z_N^\times$, a subroutine to find multiples of orders  can be used to efficiently compute classically a multiple of the order of the group $\varphi(N)$, where $\varphi$ is the Euler totient function, and  it is known that factoring is  polynomial-time reducible to the problem of finding a single multiple of the form  $\lambda\varphi(N)$  \cite{Shoup08_A_Computational_Introducttion_to_Number_Theory_and_Algebra}. 

The above no-go result highlights a deep reason why  normalizer gates over $\Z\times \mathbf{B}$ (where we may view $\Z$ as the limit of $\Z_M$ when $M\rightarrow \infty$) are needed in theorem \ref{thm:Shor normalizer} for implementing a modular exponentiation gate. We further conjecture that the obstacles displayed above are a general feature of finite-group normalizer gates, and that no finite-dimensional black-box normalizer circuit can implement an efficient factoring algorithm.
 \begin{conjecture} \label{conj: Factoring not over finite groups}
 Unless factoring is contained in \textnormal{BPP}, there is no efficient quantum algorithm to solve the factoring problem using only normalizer circuits over finite abelian groups (even when these are allowed to be black-box groups) and classical pre- and  post-processing.
 \end{conjecture}
 We back up our conjecture with two facts. On one hand, Shor's algorithm for factoring \cite{Shor} (to our knowledge, the only quantum algorithm for factoring thus far) uses a modular exponentiation gate to estimate the phases of the unitary $V_a$, and these gates are hard to implement with finite-group normalizer circuits due to theorem \ref{thm:ModExp requires Z}. On the other hand,  the reason why this does works for the group $\Z$ seems to be, in the view of the authors,   intimately related to the fact that the order-finding problem can be naturally cast as an instance of the abelian \textbf{\emph{hidden subgroup problem}} over $\Z$  (see also section \ref{sect:Abelian HSPs}).  Note that, although one can always cast the order-finding problem  as an HSP over any finite group $\Z_{\lambda \varphi(N)}$ for an integer $\lambda$, this formulation of the problem is unnatural in our setting, as it requires (again) the prior knowledge of a multiple of $\varphi(N)$, which we could use to factorize and find orders classically without the need of a quantum computer \cite{Shoup08_A_Computational_Introducttion_to_Number_Theory_and_Algebra}.

\subsubsection{Elliptic curves}\label{sect:Elliptic Curve_NOTE}

We finish our discussion of Shor's algorithms for discrete-log and factoring by highlighting that the techniques in sections \ref{sect:Discrete Log}-\ref{sect:Factoring} can be combined to show that existing generalized quantum algorithms for computing discrete-logarithms \cite{ProosZalka03_Shors_DiscreteLog_Elliptic_Curves,Kaye05_optimized_Quantum_Elliptic_Curve,CheungMaslovMathew08_Design_QuantumAttack_Elliptic_CC} over \emph{elliptic curves}\footnote{This last result extends easily even to arbitrary  black-box groups.} can also be implemented with black-box normalizer circuits (over the infinite group $\Z^2 \times E$): here, the black-box group is  the group of points $E$ of an elliptic curve; despite the latter groups being relatively more abstract that those in sections above,  they are finite, abelian and efficient (unique) encodings and fast multiplication algorithms for them are known (hence, they can be modeled as black-box groups).

This last result, which we give in \textbf{appendix \ref{sect:Elliptic Curve}} implies that normalizer circuits can also render \emph{elliptic curve cryptography (ECC)} vulnerable, as discussed in the chapter introduction.

\subsection{The hidden subgroup problem}\label{sect:Abelian HSPs}

All problems we have considered this far---finding discrete logarithms and orders of abelian group elements---fit inside  a general class of problems known as hidden subgroup problems over abelian groups \cite{Brassard_Hoyer97_Exact_Quantum_Algorithm_Simons_Problem,Hoyer99Conjugated_operators,MoscaEkert98_The_HSP_and_Eigenvalue_Estimation,Damgard_QIP_note_HSP_algorithm}. Most quantum algorithms discovered in the early days of quantum computation solve problems that can be recast as abelian HSPs, including Deutsch's problem \cite{Deutsch85quantumtheory}, Simon's \cite{Simon94onthe}, order finding and discrete logarithms \cite{Shor}, finding hidden linear functions \cite{Boneh95QCryptanalysis}, testing shift-equivalence of polynomials \cite{Grigoriev97_testing_shift_equivalence_polynomials}, and Kitaev's abelian stabilizer problem \cite{kitaev_phase_estimation,Kitaev97_QCs:_algorithms_error_correction}.

In view of our previous results, it is natural to ask how many of these problems can be solved within the normalizer framework. In this section we show that a well-known quantum algorithm  that solves the abelian HSPs (in full generality) can be modeled as a normalizer circuit over an abelian group $\mathcal{O}$. Unlike previous cases, the group involved in this computation cannot be regarded as a black-box group, as it will not be clear how to perform group multiplications of its elements. This fact reflects the presence of oracular functions with unknown structure are present in the algorithm, to which the group $\mathcal{O}$ is associated; thus,  we  call $\mathcal{O}$ an \emph{oracular group}. We will discuss, however, that this latter difference does not seem to be very substantial, and that the abelian HSP algorithm can be naturally regarded as a normalizer computation.

\subsubsection*{The quantum algorithm for the abelian HSP}

In the \emph{abelian hidden subgroup} problem we are given a function $f:G\rightarrow X$ from an abelian finite\footnote{In this section we assume $G$ to be finite for simplicity. For a case where $G$ is infinite, we refer the reader back to section \ref{sect:Factoring}, where we studied the order finding problem (which is a HSP over $\Z$).} group $G$ to a finite set $X$. The function $f$ is constant on cosets of the form $g+H$, where $H$ is a subgroup ``hidden'' by the function; moreover, $f$ is different between different cosets. Given $f$ as a black-box, our task is to find such a subgroup $H$.

The abelian HSP is a hard problem for classical computers, which need to query the oracle $f$ a superpolynomial amount of times in order to identify $H$ \cite{childs_vandam_10_qu_algorithms_algebraic_problems}. In contrast, a quantum computer can determine $H$ in polynomial time $O(\polylog{|G|})$, and using the same amount of queries to the oracle. We describe next a celebrated quantum algorithm for this task \cite{Brassard_Hoyer97_Exact_Quantum_Algorithm_Simons_Problem,Hoyer99Conjugated_operators,mosca_phd}. The algorithm is efficient given that the group $G$ is explicitly given\footnote{If the group $G$ is not given in a factorized form, the abelian HSP may still be solved by  applying  Cheung-Mosca's algorithm to decompose $G$ (see next section).} in the form $G=\DProd{d}{m}$ \cite{mosca_phd,cheung_mosca_01_decomp_abelian_groups,Damgard_QIP_note_HSP_algorithm}.
\begin{algorithm}[\textbf{Abelian HSP}]\label{alg:HSP} $ $
\begin{alg_in}
An explicitly decomposed finite abelian group $G=\DProd{d}{m}$, and oracular access to a function $f:\:G\rightarrow X$ for some set $X$. $f$ satisfies the promise that $f(g_1) = f(g_2)$ iff $g_1 = g_2+h$ for some $h \in H$, where $H\subseteq G$ is some fixed but unknown subgroup of $G$.
\end{alg_in}
\begin{alg_out}
A generating set for $H$.
\end{alg_out}
\begin{enumerate}
\item Apply the QFT over the group $G$ to an initial state $\ket{0}$ in order to obtain a uniform superposition over the elements of the group $\sum_{g\in G}\ket{g}$.
\item Query the oracle $f$ in an ancilla register, creating the state
\begin{equation}
\frac{1}{\sqrt{|G|}}\sum_{g\in G}\ket{g,f(g)}
\end{equation}
\item Apply the QFT over $G$ to the first register.
\item Measure the first register in the standard basis.
\item After repeating 1-3 polynomially many times, the obtained outcomes can be postprocessed classically to obtain a generating set of $H$ with exponentially high probability (we refer the reader to \cite{lomont_HSP_review} for details on this classical part).
\end{enumerate}
\end{algorithm}
We now claim that the quantum part of algorithm \ref{alg:HSP} is a \emph{normalizer circuit}, of a slightly more general kind than the ones we have already studied. The normalizer structure of the HSP-solving quantum circuit is, however, remarkably well-hidden compared to the other quantum algorithms that we have already studied. It is indeed a   surprising fact that there is \emph{any} normalizer structure in the circuit, due to the presence of an oracular function, whose inner structure appears to be completely unknown to us!
\begin{theorem}[\textbf{The abelian HSP algorithm is a normalizer circuit.}]\label{thm:HSP} In any abelian HSP, the subgroup-hiding property of the oracle function $f$ induces a group structure $\mathcal{O}$ in the set $X$. With respect to this hidden ``linear structure'', the function $f$ is a group homomorphism, and the HSP-solving quantum circuit is a normalizer circuit over $G\times \mathcal{O}$.
\end{theorem}
The proof is the content of the next two sections.

\subsubsection*{Unweaving the hidden-subgroup oracle}

The key ingredient in the proof of the theorem (which is the content of the next section) is to realize that the oracle $f$ cannot fulfill the subgroup-hiding property without having a hidden homomorphism structure, which is also present in the quantum algorithm. 

First, we show that $f$ induces a \emph{\textbf{group structure}} on $X$. Without loss of generality, we  assume that the function $f$ is surjective, so that $\mathrm{im}f = X$. (If this is not true, we can redefine $X$ to be the image of $f$.) Thus, for every element  $x\in X$, the preimage $f^{-1}(x)$ is contained in $G$, and is a coset of the form  $f^{-1}(x)= g_x+H$, where $H$ is the hidden subgroup and $f(g_x)=x$. With these observations in mind, we can define a group operation in $X$  as follows:
\begin{equation}\label{eq:Oracular Group Operation}
x\cdot y = \tilde{f}\left(f^{-1}(x)+f^{-1}(y)\right).
\end{equation}
In (\ref{eq:Oracular Group Operation}) we denote by $\tilde{f}$ the function $\tilde{f}(x+H)=f(x)$ that sends cosets $x+H$ to elements of $X$. The subgroup-hiding property guarantees that this function is well-defined; moreover, $f$ and $\tilde{f}$ are related via $f(x)=\tilde{f}(x+H)$. The addition operation on cosets $f^{-1}(x)=g_x+H$ and  $f^{-1}(y)= g_y+H$ is just the usual group operation of the quotient group $G/H$ \cite{Humphrey96_Course_GroupTheory}:
\begin{equation}\label{eq:Factor Group DEfinition}
f^{-1}(x)+f^{-1}(y)=\left(g_x+H\right)+\left(g_y+H\right)=(g_x+g_y)+ H.
\end{equation}
By combining the two expressions, we get an explicit formula for the group multiplication in terms of coset representatives: $x\cdot y = f(g_x+g_y)$. It is routine to check that this operation is associative and invertible, turning $X$ into a group, which we denote by $\mathcal{O}$. The neutral element of the group is the string $e$ in $X$ such that $e=f(0)=f(H)$, which we show explicitly:
\begin{equation}
x\cdot e = e\cdot x= \tilde{f}\left(f^{-1}(x)+f^{-1}(e)\right) = \tilde{f}\left(f^{-1}(x)+ H \right) = x
\end{equation}
The group $\mathcal{O}$ is manifestly finite and abelian---the latter property is due to the fact that  the addition  (\ref{eq:Factor Group DEfinition}) is commutative. 

Lastly, it is straightforward to check that the oracle $f$  \textbf{\emph{is a group homomorphism}} from $G$ to $\mathcal{O}$: for any $g$, $h\in G$ let $x:=f(g)$ and $y:=f(h)$, we have 
\begin{align}\label{eq:HSP oracle is group homomorphism}
f(g+h)&=\tilde{f} \left(g+h+H\right)=\tilde{f} \left(\left(g+H\right)+\left(h+H\right)\right)=\tilde{f} \left(f^{-1}\left(x\right)+f^{-1}\left(y\right)\right)\\
&=x\cdot y = f(g)\cdot f(h).
\end{align}
It follows from the first isomorphism theorem in group theory \cite{Humphrey96_Course_GroupTheory} that $\mathcal{O}$ is isomorphic to the quotient group  $G/H$  via the map $\tilde{f}$.

\subsubsection*{The HSP quantum algorithm is a normalizer circuit}

We will now analyze the role of the different quantum gates used in algorithm \ref{alg:HSP} and see that they are examples of normalizer gates over the group $G\times \mathcal{O}$, where $\mathcal{O}$ is the oracular group that we have just introduced.  

The Hilbert space underlying the computation can be written as $\mathcal{H}_G\otimes \mathcal{H}_\mathcal{O}$ with the standard basis  $\left\{\ket{g,x}:g\in G,\,x\in\mathcal{O}\right\}$ associated with this group. We will initialize the ancillary  registers to the state $\ket{e}$, where $e=f(0)$ is the neutral element of the group; the total state at step 1 will be $\ket{0,e}$. The Fourier transforms in steps 1 and 3 are just partial QFTs over the group $G$, which are normalizer gates. The quantum state at the end of step 1 is $\sum_{g\in G} \ket{g,e}$. 

Next, we look now at step 2 of the computation: 
\begin{equation}
\frac{1}{\sqrt{|G|}}\sum_{g\in G} \ket{g,e} \quad \longrightarrow \quad  \frac{1}{\sqrt{|G|}}\sum_{g\in G}\ket{g,f(g)}.
\end{equation}
This step can be implemented by a normalizer automorphism gate defined as follows. Let $\alpha:G\times \mathcal{O}\rightarrow G\times \mathcal{O}$ be the function $\alpha(g, x)= (g, f(g)\cdot x)$. Using the fact that $f:G\rightarrow \mathcal{O}$ is a group homomorphism (\ref{eq:HSP oracle is group homomorphism}), it is easy to check that $\alpha$  is a group automorphism of $G\times \mathcal{O}$. Then the evolution at step 2 corresponds to the action of the automorphism gate  $U_\alpha$:
\begin{equation}
U_\alpha \sum_g \ket{g,e}= \sum_g \ket{\alpha(g,e)} = \sum_g \ket{g,f(g)\cdot e} = \sum_g \ket{g,f(g)}.
\end{equation}
Finally, note that in the last two steps of the algorithm  we  measure the register $\mathcal{H}_G$ in the standard basis and post-process the information classically like in a normalizer computation. Hence, we have shown that every step in the quantum algorithm \ref{alg:HSP} can be implemented by a normalizer gate over $G\times \mathcal{O}$. This finishes the proof of theorem \ref{thm:HSP}.

\subsubsection*{The oracular group $\mathcal{O}$ is not a black-box group (but almost)}\label{sect:Oracular Group is not Black Box}

We ought to stress, at this point, that although theorem \ref{thm:HSP} shows that the abelian HSP quantum algorithm is a normalizer computation over an abelian group $G\times \mathcal{O}$, the oracular group $\mathcal{O}$ is not a black-box group (as defined in section \ref{sect:Black Box Groups}), since it is not clear how to compute the group operation (\ref{eq:Oracular Group Operation}), due to our lack of knowledge about the oracular function which defines the multiplication rule. Yet, even in the absence of an efficiently computable group operation, we regard it natural to call the abelian HSP quantum algorithm a normalizer circuit over $G\times \mathcal{O}$. Our reasons are multi-fold. 

First,  there is a manifest strong similarity between the quantum circuit in algorithm \ref{alg:HSP} and the other normalizer circuits that we have studied in previous sections, which suggests that normalizer operations naturally capture the logic of the abelian HSP quantum algorithm. 

Second, it is in fact possible to argue that, although $\mathcal{O}$ is not a black-box group, it behaves \emph{effectively} as a black-box group in the quantum algorithm. Observe that, although it is true that one cannot generally compute $x\cdot y$ for arbitrary $x,\,y\in \mathcal{O}$, it is indeed always possible to multiply any element $x$  by the neutral element $e$, since the computation is trivial in this case: $x\cdot e = e\cdot x = x$. Similarly, in the previous section, it is not clear at all how to implement the unitary transformation $U_\alpha\ket{g,x}=\ket{g,f(g)\cdot x}$ for arbitrary inputs. However, for the restricted set of inputs that we need in the quantum algorithm (which is just the state $\ket{e}$), it is trivial to implement the unitary, for in this case $U_\alpha\ket{g,e}=\ket{g,f(g)}$; since quantum queries to the oracle function are allowed (as in step 2 of the algorithm), the unitary can be simulated by such process, regardless of how it is implemented. Consequently, the circuit \emph{effectively} behaves as a normalizer circuit over a black-box group.

Third,  although the oracular model in the black-box normalizer circuit setting is slightly different from the one used in the abelian HSP they are still \emph{remarkably close} to each other. To see this, let $x_i$ be the elements of $X$ defined as $x_i:=f(e_i)$ where $e_i$ is the bit string containing a 1 in the $i$th position and zeroes elsewhere. Since the $e_i$s form a generating set of $G$, the $x_i$s generate the group $\mathcal{O}$. Moreover, the value of the function $f$ evaluated on  an element $g=\sum g(i)e_i$ is $f(g)=x_1^{g(1)} x_2^{g(2)}\cdots  x_m^{g(m)}$, since $f$ is a group homomorphism. It follows from this expression that the group homomorphism is \emph{implicitly multiplying} elements of the group $\mathcal{O}$. We cannot use this property to multiply elements of $\mathcal{O}$ ourselves, since everything happens at the hidden level. However, this observation shows that the assuming that $f$ is computable is \emph{tightly related} to the assumption that we can multiply in $\mathcal{O}$, although slightly weaker. (See also the next section.)

Finally, we mention that this very last feature can be exploited to extend several of our main results, which we derive in the black-box setting, to the more-general ``HSP oracular group setting'' (although proofs become more technical). For details, we refer the reader to sections \ref{sect:Simulation_c3}-\ref{sect:Complete Problems} and appendix  \ref{app:Extending}.

\subsubsection*{A connection to a result by Mosca and Ekert}

Prior to our work,  it was observed by Mosca and Ekert \cite{MoscaEkert98_The_HSP_and_Eigenvalue_Estimation,mosca_phd} that $f$ must have a hidden homomorphism structure, i.e. that $f$ can be decomposed as $\mathcal{E}\circ\alpha$ where $\alpha$ is a group homomorphism between $G$ and another abelian group $Q\cong G/H$, and $\mathcal{E}$ is a one-to-one hiding function from $Q$ to the set $X$. In this decomposition, $\mathcal{E}$ hides the homomorphism structure of the oracle.

Our result differs from Mosca-Ekert's in that we show that $X$ \emph{itself} can always be viewed as a group, with a group operation that is induced by the oracle, with no need to know the decomposition $\mathcal{E}\circ\alpha$. 

It is possible to relate both results as follows.  Since both $Q$ and $\mathcal{O}$  are isomorphic to  $G/H$,  they are also mutually isomorphic. Explicitly, if  $\beta$ is an isomorphism from $Q$ to $G/H$ (this map depends on the particular decomposition  $f=\mathcal{E}\circ\alpha$), then  $Q$ and $\mathcal{O}$ are isomorphic via the map $\tilde{f}\circ \beta$.

\subsection{Decomposing finite abelian groups}\label{sect:Group Decomposition}

As mentioned earlier, there is a quantum algorithm for decomposing abelian groups, due to Cheung and Mosca \cite{mosca_phd,cheung_mosca_01_decomp_abelian_groups}. In this section, we will introduce this problem, and present a quantum algorithm that solves it, which uses only  black-box normalizer circuits supplemented with classical computation.  The algorithm we give is based on Cheung-Mosca's, but it reveals some additional  information about the structure of the black-box group. We will refer to it as our \emph{\textbf{extended Cheung-Mosca's algorithm}}.

\subsubsection*{The group decomposition problem}

In this work, we define the    \textbf{group decomposition} problem as follows. The input of the problem is a list  of generators $\alpha=( \alpha_1,\cdots,\alpha_k)$ of some abelian black-box group $\mathbf{B}$. Our task is to return a \emph{group-decomposition table} for $\mathbf{B}$. A group-decomposition table is a tuple $(\alpha, \beta, A, B, c)$ consisting of the original string $\alpha$ and four additional elements:
\begin{enumerate}
\item[(a)] A new generating set $\beta={\beta_1,\ldots,\beta_\ell}$ with the property $\mathbf{B}=\langle \beta_1\rangle\oplus\cdots\oplus\langle \beta_\ell\rangle$. We will say that these new generators are \emph{linearly independent}.
\item[(b)] An integer vector $c$ containing the orders of the linearly independent generators $\beta_i$.
\item[(c)]  Two integer matrices $A$, $B$ that relate the old and new generators as follows:
\begin{equation}\label{eq:Matrix of Relationships}
\begin{pmatrix}
 \beta_1,\ldots,\beta_\ell
\end{pmatrix}=
\begin{pmatrix}
\alpha_1, 
\ldots
\alpha_k
\end{pmatrix} A,\qquad\quad \begin{pmatrix}
\alpha_1, 
\ldots
\alpha_k
\end{pmatrix}= \begin{pmatrix}
 \beta_1,\ldots,\beta_\ell
\end{pmatrix}B.
\end{equation}
This last equation should be read in multiplicative notation (as in e.g.\ \cite{Cohen00_Advanced_Topics_ComputationalNumber_Theory}), where ``vectors'' of group elements are right-multiplied by matrices as follows: given  the $i$th column $a_i$  of $A$ (for the left hand case), we have $\beta_{i}=(\alpha_1,\ldots,\alpha_k) a_i=\alpha_1^{a_i(1)}\cdots \alpha_k^{a_i(k)}$.
\end{enumerate}
Our definition of the group decomposition is more general than the one given in \cite{mosca_phd,cheung_mosca_01_decomp_abelian_groups}. In Cheung and Mosca's formulation, the task is to find just $\beta$ and $c$. The algorithm they give also computes the matrix $A$ in order to find the generators $\beta_i$ (cf.\ the next section). What is completely new in our formulation is that we ask in addition for the matrix $B$.

Note that a {\textbf{group-decomposition table}} $(\alpha,\beta, A, B, c)$ contains a lot of information about the group structure of  $\mathbf{B}$. First of all, the tuple elements (a-b)  tell us  that  $\mathbf{B}$ is isomorphic to a decomposed group $G=\DProd{c}{k}$. In addition, the matrices $A$ and $B$ provide us with an efficient method to re-write linear combinations of the original generators $\alpha_i$ as linear combinations of the new generators $\beta_j$ (and vice-versa). Indeed, equation (\ref{eq:Matrix of Relationships})  implies
\begin{align}\alpha_{1}^{x_1}\cdots \alpha_{k}^{x_k}&=  \begin{pmatrix}
\alpha_1, 
\ldots
\alpha_k
\end{pmatrix}x= \begin{pmatrix}
 \beta_1,\ldots,\beta_\ell
\end{pmatrix}(Bx) ,\quad\textnormal{  for any  $x\in\Z^k$,}\notag\\
\beta_{1}^{y_1}\cdots \beta_{1}^{y_\ell} &= \begin{pmatrix}
 \beta_1,\ldots,\beta_\ell
\end{pmatrix}y=
\begin{pmatrix}
\alpha_1, 
\ldots
\alpha_k
\end{pmatrix} (Ay),\quad\textnormal{  for any $y\in\Z^\ell$}.\notag
\end{align}
It follows that, for any given $x$, the integer string $y=Bx$ (which can be efficiently computed classically)  fulfills the condition $\alpha_{1}^{x_1}\cdots \alpha_{1}^{x_k}=\beta_{1}^{y_1}\cdots \beta_{1}^{y_\ell}$. (A symmetric argument proves the opposite direction.)

As we  discussed earlier in the introduction, the group decomposition problem is \emph{provably hard} for classical computers within the black-box setting, and it is at least \emph{as hard as} Factoring (or Order Finding) for matrix groups of the form $\Z_N^\times$ (the latter being polynomial-time reducible to group decomposition).\ It can be also shown that group decomposition is also at least as hard as computing discrete logarithms, a fact that we will use in the proof of theorems \ref{thm:Simulation}, \ref{thm:No Go Theorem}:
\begin{lemma}[\textbf{Multivariate discrete logarithms}]\label{lemma:Multivariate Discrete Log} Let $\beta_1,\ldots,\beta_\ell$ be  generators of some abelian black-box group $\mathbf{B}$ with the  property
$\mathbf{B}=\langle \beta_1\rangle\oplus\cdots\oplus\langle \beta_\ell\rangle$. Then, the following generalized version of the discrete-logarithm problem is polynomial time reducible to group decomposition: for a given $\beta\in\mathbf{B}$, find an integer string $x$ such that $\beta_{1}^{x_1}\cdots \beta_{\ell}^{x_\ell}=\beta$. 
\end{lemma}
\begin{proof}
Define a new set of generators  for $\mathbf{B}$  by  adding the element $\beta_{\ell+1}=\beta$ to the given set $\{\beta_i\}$. The array $\alpha':=(\beta_1,\ldots,\beta_{\ell+1})$ defines an instance of Group Decomposition. Assume that a group decomposition table $(\alpha', (\beta'_1,\ldots,\beta_m'), A', B', c')$ for this instance of the problem is given to us. We can now use the columns $b_i'$ of the matrix $B'$ to re-write the previous generators $\beta_i$ in terms of the new ones:
\begin{equation}\label{inproof:MultiVariate Discrete Log}
\beta_i = (\beta_1,\ldots,\beta_{\ell+1}) e_i = \begin{pmatrix}
 \beta_1',\ldots,\beta_m'
\end{pmatrix} (B'e_i) =  \begin{pmatrix}
 \beta_1',\ldots,\beta_m'
\end{pmatrix} b_i'=
 \beta_1'^{b_i'(1)}\cdots\beta_m'^{b_i'(m)}.
\end{equation}
Here, $e_i$ denotes the integer vector with $e(i)=1$ and $e(j)=0$ elsewhere. Conditions (a-b) imply that the columns $b_i'$ can be treated as elements of the group $G=\DProd{c'}{m}$. Using this identification, the original discrete logarithm problem reduces to finding an integer string $x\in G$ such that $b_{\ell+1}'=(b_1',\ldots,b_\ell')x=\sum x(i)b_i'$ (now in additive notation). The existence of such an $x$ can be easily proven using  that the elements $\beta_1,\ldots,\beta_\ell$ generate $\mathbf{B}$: the latter guarantees the existence of an $x$ such that
\begin{equation}
\beta_{\ell+1}=(\beta_1,\ldots,\beta_{\ell}) x = \begin{pmatrix}
 \beta_1',\ldots,\beta_m'
\end{pmatrix} (b_1',\ldots,b_\ell') x =   \begin{pmatrix}
 \beta_1',\ldots,\beta_m'
\end{pmatrix} b_{\ell+1}',
\end{equation}
which implies $(b_1',\ldots,b_\ell') x \equiv    b_{\ell+1}' \bmod{(c_1',\ldots,c_m')}$.
By finding such an  $x$, we can  solve the multivariate discrete problem, since $\beta_{1}^{x_1}\cdots \beta_{\ell}^{x_\ell}=\beta_1'^{b_{\ell+1}'(1)}\cdots\beta_m'^{b_{\ell+1}'(m)}=\beta_{\ell+1}=\beta$, due to    (\ref{inproof:MultiVariate Discrete Log}). Finally, note that we can find $x$  efficiently with existing our deterministic classical algorithms for Group Membership in finite abelian groups (lemma \ref{lemma:Algorithms_FA_Groups}).
\end{proof}
We highlight that, in order for the latter result to hold, it seems critical to use our formulation of group decomposition instead of Cheung-Mosca's.  Consider again the discrete-log problem over the group $\Z_p^\times$ (recall section \ref{sect:Discrete Log}). This group  $\Z_p^\times$ is cyclic of order $p-1$ and a generating element $a$ is given to us as part of the input of the discrete-log problem. Although  it is not known how to solve this problem efficiently, Cheung-Mosca's group decomposition problem (find some linearly independent generators and their orders) can be solved effortlessly in this case, by simply returning $a$ and $p-1$, since  $\langle a \rangle = \Z_p^\times \cong \Z_{p-1}$. The crucial difference is that Cheung-Mosca's algorithm returns a factorization $ \DProd{c}{\ell}$ of $\mathbf{B}$, but it cannot be used to convert elements between the two representations efficiently (one direction is easy; the other requires computing discrete logarithms). In our formulation, the matrices $A$, $B$ provide such a method.

\subsubsection*{Quantum algorithm for group decomposition}

We now present a quantum algorithm that solves the group decomposition problem. The first 3 steps of our algorithm mimic Cheung and Mosca's\footnote{Cheung-Mosca's original presentation first applied Shor's algorithm to decompose $B$ into Sylow $p$-subgroups and subsequently performed the group decomposition on these subgroups \cite{cheung_mosca_01_decomp_abelian_groups}. It is discussed in \cite{cheung_mosca_01_decomp_abelian_groups} that the first of these steps is not necessary. In our presentation we bypass this step.}. Our novel contribution here is step 4, which  computes the $B$ matrix in (\ref{eq:Matrix of Relationships}).

\begin{algorithm}[\textbf{Extended Cheung-Mosca's algorithm}]
\label{alg:group decomposition}
$ $ 

\begin{alg_in} A list of generators $\alpha=(\alpha_1,\ldots,\alpha_k)$ of an abelian black-box group $\mathbf{B}$.

\end{alg_in}

\begin{alg_out} A group decomposition table $(\alpha, \beta, A, B, c)$.

\end{alg_out}

\begin{enumerate}

\item  Use the order finding algorithm (comprising normalizer circuits over $\Z\times \mathbf{B}$ and classical postprocessing) to obtain the orders $d_i$ of the generators $\alpha_i$. Then, compute (classically) and store their least common multiplier $d=\mathrm{lcm}(d_1,\ldots,d_k)$.

\item  Define the function $f : \Z_{d}^k \rightarrow \mathbf{B}$ as $f(x) = \alpha_1^{x(1)} \cdots \alpha_k^{x(k)}$, which is a group homomorphism and hides the subgroup $\ker f$ (its own kernel). Apply the abelian HSP algorithm to compute a set  of generators $h_1,\ldots,h_m$ of $\ker f$.  This round uses normalizer circuits over $\Z_d^k\times \mathbf{B}$ and classical post-processing (cf.\ section \ref{sect:Abelian HSPs}).

\item Given the generators $h_i$ of $\ker f$ one can classically compute a $k\times \ell$ matrix $A$ (for some $\ell$) such that $(\beta_1,\ldots,\beta_\ell)=(\alpha_1,\ldots,\alpha_k) A$ is a system of linearly independent generators   \cite[theorem 7]{cheung_mosca_01_decomp_abelian_groups}.  $\beta$, $A$ and the orders $c_i$ of the $\beta_i$s (computed again via an order-finding subroutine) will form part of the output.
\item Finally, we show how to classically compute a valid relationship matrix $B$. (This step is not part of Cheung-Mosca's original algorithm.) The problem reduces to finding a $k\times \ell $ integer matrix $X$ with two properties: 
\begin{itemize}
\item[(a)] $X$ is a solution to the equation $(\alpha_1,\ldots,\alpha_k)X=(\alpha_1,\ldots,\alpha_k)$. Equivalently, every column $x_i$ of $X$ is equal (modulo $d$) to some element of the coset $e_i+\ker{f}\subset\Z_d^k$.
\item[(b)] Every column $x_i$ is an element of the image of the matrix $A$.
\end{itemize}
It is easy to see  that a matrix $X$ fulfilling (a-b) always exists, since for any $\alpha_i$, there exists some $y_i$ such that $\alpha_i=(\beta_1,\ldots,\beta_\ell)y_i$ (because the $\beta_i$s generate the group). It follows that $\alpha_i=(\alpha_1,\ldots,\alpha_k) (Ay_i)$. Then, the matrix with columns $x_i=Ay_i$ has the desired properties.

$\quad$ Our existence proof for $X$ is constructive, and tells us that  $X$ can be computed in quantum polynomial time by solving a multivariate discrete logarithm problem (lemma \ref{lemma:Multivariate Discrete Log}). However,  we will use a more subtle efficient \emph{classical} approach to obtain $X$, by reducing the problem to a {\textbf{system of linear equations over abelian groups}} as in chapter \ref{chapterGT}. Let $H$ be a matrix formed column-wise by the generators $h_i$ of $\ker f$. By construction, the image of the map $H:\Z_d^m \rightarrow \Z_d^k$ fulfills $\mathrm{im} H =\ker{f}$. Properties (a-b) imply that the $i$th column $x_i$ of $X$ must be a particular solution to the equations $x_i= A y_i$ with $y_i\in \Z^\ell$ and $x_i = e_i + Hz_i \bmod{d}$, with $z_i\in \Z_d^m$. These equations can be equivalently written as a system of linear equations over $\Z^{m+\ell}$:
\begin{equation}
\begin{pmatrix}
A & -H
\end{pmatrix}\begin{pmatrix}
y_i \\ z_i
\end{pmatrix} = e_i\mod{d},\qquad (y_i,z_i)\in \Z_d^{m}\times \Z^\ell,
\end{equation}
which can be  solved in classical polynomial time using the algorithms from chapter \ref{chapterGT}. Then, the matrix $X$ can be constructed column wise taking $x_i=Ay_i$.

$\quad$Finally, given such an $X$, it is easy to find a valid  $B$ by computing a Hurt-Waid integral pseudo-inverse $A^\#$ of $A$ \cite{HurtWaid70_Integral_Generalized_Inverse,BowmanBurget74_systems-Mixed-Integer_Linear_equations}:
\begin{equation}
\alpha= \alpha X = \alpha (A A^\#) X = (\alpha A ) (A^\#X) =(\beta_1,\ldots,\beta_\ell) (A^\# X).
\end{equation}
In the third step, we used that $A^\#$ acts as the inverse of $A$ on inputs $x\in \Z^k$ that live in the image of $A$ \cite{HurtWaid70_Integral_Generalized_Inverse}. Since integral pseudo-inverses can be computed efficiently using the Smith normal form (see appendix \ref{appendix:Efficiency of Bowman Burdet}), we finally set $B:=A^\# X$. 
\end{enumerate}
\end{algorithm}

\section{Simulation of  black-box normalizer circuits}\label{sect:Simulation_c3}

Our results so far show that the computational power of normalizer circuits over black-box groups (supplemented with classical pre- and post- processing) is \emph{strikingly high}: they can solve several problems believed to be classically intractable and render the RSA, Diffie-Hellman, and elliptic curve public-key cryptosystems vulnerable. In contrast,   normalizer circuits associated with abelian groups that are \emph{explicitly decomposed}, can be efficiently simulated classically, by exploiting the generalized stabilizer formalism of chapters chapters \ref{chapterF}-\ref{chapterI}.

It is natural to wonder at this point where the computational  power of black-box normalizer circuits originates. In this section, we will argue that the hardness of simulating black-box normalizer circuits resides \emph{precisely} in the hardness of decomposing black-box abelian groups. An equivalence is suggested by the fact that we can use these circuits to solve the group decomposition problem and, in turn, when the group is decomposed, the techniques of chapters \ref{chapterF}-\ref{chapterI}  render these circuits classically simulable. In this sense, then, the \emph{quantum speedup }of such circuits appears to be completely encapsulated in the group decomposition algorithm. This intuition can be made precise and be stated as a theorem.
\begin{theorem}[\textbf{Simulation of black-box normalizer circuits}]\label{thm:Simulation} Black-box normalizer circuits can be efficiently simulated classically using the stabilizer formalism over abelian groups  of chapters \ref{chapterF}-\ref{chapterI} if a subroutine for solving the group-decomposition problem is provided as an oracle.
\end{theorem}
The proof of this theorem is the subject of section \ref{app:Simulation proof} in the appendix.

Since normalizer circuits can solve the group decomposition problem (section \ref{sect:Group Decomposition}), we obtain that this problem is complete for the associated normalizer-circuit complexity class, which we now define.
\begin{definition}[\textbf{Black-Box Normalizer}] The complexity class \textbf{Black-Box Normalizer} is the set of oracle problems that can be solved with bounded error by at most polynomially many rounds of efficient black-box normalizer circuits (section \ref{sect:Normalizer circuits over blackbox groups}), with polynomial-sized classical computation interspersed between. In other words, if $N$ is an oracle that given an efficient (poly-size) black-box normalizer circuit as input, samples from its output distribution, then
\begin{equation}
\text{\textbf{Black-Box Normalizer}} = BPP^{N}.
\end{equation}
\end{definition}

\begin{corollary}[\textbf{Group decomposition is complete}]\label{corollary:Group Decomposition is complete}
Group decomposition is a complete problem for the complexity class \textnormal{\textbf{Black-Box Normalizer}} under classical polynomial-time Turing reductions.
\end{corollary}
We stress that theorem \ref{thm:Simulation}  tells us even more than the completeness of group decomposition. As we discussed in the introduction, an oracle for group decomposition gives us an efficient classical algorithm to simulate Shor's factoring and discrete-log algorithm (and all the others) \emph{step-by-step} with a stabilizer-picture approach ``à la Gottesman-Knill''.

We also highlight that theorem \ref{thm:Simulation} can be restated as a \emph{no-go theorem} for finding new quantum algorithms based on black-box normalizer circuits.
\begin{theorem}[\textbf{No-go theorem for new quantum algorithms}]\label{thm:No Go Theorem} It is not possible to find ``fundamentally new'' quantum algorithms within the class of black-box normalizer circuits studied in this work, in the sense that any new algorithm would be efficiently simulable  using the extended Cheung-Mosca algorithm and classical post-processing.
\end{theorem}
This theorem tells us that  black-box normalizer circuits cannot give exponential speedups over classical circuits that are not already covered by our extended Cheung-Mosca algorithm; the theorem may thus have applications to algorithm design. 

Note, however, that this no-go theorem says nothing about other possible \emph{polynomial} speed-ups for black-box normalizer circuits; there may well be other normalizer circuits that are polynomially faster, conceptually simpler, or easier to implement than the extended Cheung-Mosca algorithm. Our theorem neither denies that investigating black-box normalizer could be of pedagogical or practical value if, e.g., this led  to new interesting complete problems for the class \textbf{Black-Box Normalizer}.

Finally, we note that theorem \ref{thm:Simulation}  can be extended to  the general abelian hidden subgroup problem to show that the quantum algorithm for the abelian HSP becomes efficiently classically simulable if an algorithm for decomposing the oracular group $\mathcal{O}$ is given to us (cf.\ section \ref{sect:Abelian HSPs} and refer to appendix \ref{app:Extending} for a proof). We discuss some implications of this fact in the next sections.

\section{Universality of short quantum circuits}\label{sect:Universality}

Since all problems in \textbf{Black-Box Normalizer} are solvable by our extended Cheung-Mosca quantum algorithm (supplemented with classical processing), the structure of said quantum algorithm allows us to state the following:
\begin{theorem}[\textbf{Universality of short normalizer circuits}]\label{thm:Universality Short Circuits}
Any problem in the class \textbf{\emph{Black-Box Normalizer}} can be solved by a quantum algorithm composed of polynomially-many rounds of \emph{short} normalizer circuits, each with at most a \textbf{constant} number of normalizer gates, and additional classical computation. More precisely, in every round, normalizer circuits containing two quantum Fourier transforms and one automorphism gate (and no quadratic phase gate) are already sufficient.
\end{theorem}
\begin{proof}
This result follows immediately form the fact that group decomposition is complete for this class (theorem \ref{corollary:Group Decomposition is complete}) and from the structure of the extended Cheung-Mosca quantum algorithm with this problem, which has precisely this structure. 
\end{proof}
Similarly to theorem \ref{thm:Simulation}, theorem \ref{thm:Universality Short Circuits} can be extended to the general abelian HSP  setting. For details, we refer the reader to appendix \ref{app:Extending}.

We find the latter result is insightful, in that it actually explains a somewhat intriguing feature present in Deutsch's, Simon's, Shor's and virtually all known quantum algorithms for solving abelian hidden subgroup problems: they all contain at most two quantum Fourier transforms! Clearly, it follows from this theorem than no more than two are enough. 

Also, theorem \ref{thm:Universality Short Circuits} tells us that it is actually pretty \emph{useless} to use logarithmically or polynomially long  sequences of quantum Fourier transforms for solving abelian hidden subgroup problems, since just two of them suffice\footnote{This last comment does not imply that building up sequences of Fourier transforms is useless in general. On the contrary, this can be actually be useful, e.g., in QMA amplification  \cite{Nagaj2009_Fast_Amplification_QMA}.}. In this sense, the abelian HSP quantum algorithm uses an \emph{asymptotically optimal} (constant) number of quantum Fourier transforms. Furthermore, the normalizer-gate depth of this algorithm is optimal in general.
 
\section{Other Complete problems}\label{sect:Complete Problems}

We end this chapter by giving two other complete problems for the complexity class \textbf{Black Box Normalizer}.
\begin{theorem}[\textbf{Hidden kernel problem is complete}]\label{thm:Hidden Kernel Problem} Let the abelian hidden kernel problem (abelian HKP) be the subcase of the hidden subgroup problem where the oracle function $f$ is a group homomorphism from a group of the form  $G=\Z^a\times\DProd{N}{b}$ into a black-box group $\mathbf{B}$. This problem is complete for \textbf{\emph{Black Box Normalizer}} under polynomial-time Turing reductions.
\end{theorem}
\begin{proof}
Clearly group decomposition reduces to this problem, since the quantum steps of the extended Cheung-Mosca algorithm algorithm (steps 1 and 3) are solving instances of the abelian kernel problem. Therefore, the abelian HKP problem is hard for \textbf{Black Box Normalizer}.

Moreover, abelian HKP can be solved with the general abelian HSP quantum algorithm, which manifestly becomes a black-box normalizer circuit for oracle functions $f$ that are group homomorphisms onto black-box groups. This implies that abelian HKP is inside \textbf{Black Box Normalizer}, and therefore, it is complete.

\textbf{Note.} Although we originally stated the abelian HSP for finite groups, one can first apply the order-finding algorithm to compute a multiple $d$ of the orders of the elements $f(e_i)$, where $e_i$ are the canonical generators of $G$. This can be used to reduce the original HKP problem to a simplified HKP over the group $\Z_{d}^a\times \DProd{N}{b}$
\end{proof}
The latter result can be extended to  show that any abelian hidden subgroup problem is polynomial-time equivalent to decomposing groups of form $\mathcal{O}$ (cf.\ appendix \ref{app:Extending}).
\begin{theorem}[\textbf{System of linear equations over groups}] Let $\alpha$ be a group homomorphism from a group $G=\Z^a\times\DProd{N}{b}$ onto a black-box group $\mathbf{B}$. An instance of a linear system of equations over $G$ and $\mathbf{B}$  is given by a homomorphism  $\alpha$ and an element $\mathbf{b}\in \mathbf{B}$. Our task is to find a general $(x_0, K)$ solution to the equation
\begin{equation}\nonumber
\alpha(x)=\mathbf{b},\quad x\in G,
\end{equation}
where $x_0$ is any particular solution and  $K$ is a generating set of the kernel of $\alpha$. This problem is  complete for \textbf{\emph{Black Box Normalizer}} under polynomial-time Turing reductions.
\end{theorem}
\begin{proof}
Clearly, this problem is hard for our complexity class, since the abelian hidden kernel problem reduces to finding $K$.

Moreover, this problem can be solved with black-box normalizer circuits and classical computation, proving its completeness. First, we find a decomposition $\mathbf{B}=\bigoplus\langle \beta_i \rangle \cong H=\DProd{c}{\ell}$ with black-box normalizer circuits. Second, we recycle the ``de-black-boxing'' idea from the proof of theorem \ref{thm:Simulation} to compute a matrix representation of $\alpha$, and   solve the multivariate discrete logarithm problem $\mathbf{b}=\beta_1^{b(1)}\cdots\beta_\ell^{b(\ell)}$, $b\in H$, either with black-box normalizer circuits or classically (recall section \ref{sect:Group Decomposition}). The original system of equations can now be equivalently written as $A x = b \pmod{H}$. A general solution of this system can be computed with classical algorithms given in chapter \ref{chapterGT}.
\end{proof}

%% file: chapter4_hyperproject.tex
\chapter{Abelian hypergroups and quantum computation}\label{chapterH}
 
Motivated by a connection, described here for the first time, between the hidden normal subgroup problem (HNSP) and abelian hypergroups (algebraic objects that model collisions of physical particles), we develop a stabilizer formalism using abelian hypergroups and an associated classical simulation theorem (a la Gottesman-Knill). Using these tools, we develop the first provably efficient quantum algorithm for finding hidden subhypergroups of nilpotent abelian hypergroups and, via the aforementioned connection, a new, hypergroup-based algorithm for the HNSP on nilpotent groups. We also give efficient methods for manipulating non-unitary, non-monomial stabilizers and an adaptive Fourier sampling technique of general interest. 

This chapter is based on \cite{BermejoVegaZatloukal14Hypergroups} (joint work with Kevin C.\ Zatloukal).

\section{Introduction}

Ever since Shor's groundbreaking discovery of an efficient quantum algorithm
for factoring \cite{Shor}, researchers have striven to understand the source of its quantum speed up and  find new applications for quantum computers. An era of breakthroughs followed, in which researchers  found that factoring and discrete log are instances of the so-called  \emph{Hidden Subgroup Problem} (HSP), a more general  problem about \emph{finite groups}\footnote{In the HSP, the task is to find a subgroup $H$ of a finite group $G$ by evaluating a function $f:G\rightarrow X$, which is given to us and is promised to \emph{hide} $H$ in the sense  that $f(x)=f(y)$ iff $x=yh$ for some $h\in H$.};   developed efficient quantum algorithms for the abelian group HSP \cite{kitaev_phase_estimation,Kitaev97_QCs:_algorithms_error_correction,Brassard_Hoyer97_Exact_Quantum_Algorithm_Simons_Problem,Hoyer99Conjugated_operators,MoscaEkert98_The_HSP_and_Eigenvalue_Estimation,mosca_phd,cheung_mosca_01_decomp_abelian_groups,Damgard_QIP_note_HSP_algorithm}; and discovered that solving the nonabelian group HSP over symmetric and dihedral groups would lead to a revolutionary algorithm  for  Graph Isomorphism \cite{Ettinger99aquantum} and break lattice-based cryptography \cite{Regev:2004:QCL:976327.987177}. 

Motivated by these breakthroughs, there has been a great deal of research work over the last decade aimed at finding efficient quantum algorithms for nonabelian HSPs, leading to many  successes
\cite{Hallgren00NormalSubgroups:HSP,EttingerHoyerKnill2004_Hidden_Subgroup,Kuperberg2005_Dihedral_Hidden_Subgroup,Regev2004_Dihedral_Hidden_Subgroup,Kuperberg2013_Hidden_Subgroup,RoettelerBeth1998_Hidden_Subgroup,IvanyosMagniezSantha2001_Hidden_Subgroup,MooreRockmoreRussellSchulman2004,InuiLeGall2007_Hidden_Subgroup,BaconChildsVDam2005_Hidden_Subgroup,ChiKimLee2006_Hidden_Subgroup,IvanyosSanselmeSantha2007_Hidden_Subgroup,MagnoCosmePortugal2007_Hidden_Subgroup,IvanyosSanselmeSantha2007_Nil2_Groups,FriedlIvanyosMagniezSanthaSen2003_Hidden_Translation,Gavinsky2004_Hidden_Subgroup,ChildsVDam2007_Hidden_Shift,DenneyMooreRussel2010_Conjugate_Stabilizer_Subgroups,Wallach2013_Hidden_Subgroup,lomont_HSP_review,childs_lecture_8,VanDamSasaki12_Q_algorithms_number_theory_REVIEW}, though  efficient quantum algorithms for dihedral and symmetric HSP have still not been found.

Thus far, the foundation of nearly all known  quantum algorithms for nonabelian HSPs has been the seminal work of Hallgren, Russell, and Ta-Shma \cite{Hallgren00NormalSubgroups:HSP}, which showed that hidden \emph{normal} subgroups can be found efficiently  for \emph{any} nonabelian group. For example, the algorithms for (near) Hamiltonian groups \cite{Gavinsky:2004:QSH:2011617.2011625} work because all subgroups of such groups are (nearly) normal. Likewise, the sophisticated algorithm of Ivanyos et al. for 2-nilpotent groups \cite{Ivanyos:2008:EQA:1792918.1792983} cleverly reduces the problem of finding a hidden non-normal subgroup to
two problems of finding hidden normal subgroups. 

Surprisingly, given the importance of the nonabelian HSP program in the  history of quantum computing, the success of the quantum algorithm for the hidden \emph{normal} subgroup problem (HNSP) \cite{Hallgren00NormalSubgroups:HSP} remains poorly explained. The initial motivation for this work was to improve our understanding of the quantum algorithm for the HNSP up to the same level as those for abelian HSPs.

Our approach is inspired by the connection  between Shor's algorithm, Gottesman's \emph{Pauli stabilizer formalism} (PSF) \cite{Gottesman_PhD_Thesis},  and the Gottesman-Knill theorem \cite{Gottesman_PhD_Thesis,Gottesman99_HeisenbergRepresentation_of_Q_Computers,Gottesman98Fault_Tolerant_QC_HigherDimensions} of chapter \ref{chapterB}. In short, there we showed that all most-famous quantum algorithms for abelian Hidden Subgroup Problems  are generalized types of Clifford operations over groups; this connection, combined with  the generalized Group Stabilizer Formalism (GSF) for simulating normalizer circuits  (chapters \ref{chapterF}-\ref{chapterI}),  let us derive  a sharp \emph{no-go theorem} for finding new quantum algorithms with the standard abelian group Fourier sampling techniques. 

Given the success of our previous techniques  at understanding abelian HSP quantum algorithms, our aim in this chapter is to gain a deeper understanding of the algorithm for HNSPs on nonabelian groups using a more sophisticated stabilizer formalism. Furthermore, because the PSF (and generalizations)  have  seminal applications in fault tolerance \cite{Gottesman98Fault_Tolerant_QC_HigherDimensions,BravyiKitaev05MagicStateDistillation}, measurement based quantum computation \cite{raussen_briegel_onewayQC}, and condensed matter theory \cite{kitaev_anyons}, we expect a new stabilizer formalism to  find new uses outside of  quantum algorithm analysis.

\subsection{Main results}

While it would be natural to generalize the abelian group stabilizer formalisms into a nonabelian group stabilizer formalism, we find that the proper way to understand the quantum algorithm for the HNSP is not to generalize the ``abelian'' property but rather the ``group'' property. In particular, we will work with \emph{abelian hypergroups}. These are generalizations of groups and can be thought of as collections of particles (and anti-particles) with a ``collision'' operation that creates new particles. A group is a special case of a hypergroup where each collision produces exactly one resulting particle. 

Our first result is a formal connection between the HNSP and abelian hypergroups which will be helpful to understand why quantum computers can solve this problem:
\begin{enumerate}
\item[I.] \textbf{Connecting the HNSP to the abelian HSHP.} We demonstrate (\textbf{section \ref{sect:HNSP}})	that, in many natural cases, the HNSP can be reduced to a problem on abelian hypergroups, called the hidden \emph{subhypergroup} problem (HSHP) \cite{Amini_hiddensub-hypergroup,Amini2011fourier}. This occurs because all of the information about the normal subgroups of a nonabelian group is captured in its hypergroup of conjugacy classes. Even in a nonabelian group, there is a multiplication operation on conjugacy classes that remains abelian. Our results show that, in many natural cases, finding hidden normal subgroups remains a problem about an abelian algebraic structure even when the group is nonabelian.
\end{enumerate}
Our next results show that the tools that proved successful for understanding quantum algorithms for abelian group HSPs (as well as many other problems) can be generalized to the setting of abelian hypergroups:
\begin{enumerate}

\item[II.] \textbf{A hypergroup stabilizer formalism.} We extend the PSF \cite{Gottesman_PhD_Thesis,Gottesman99_HeisenbergRepresentation_of_Q_Computers,Gottesman98Fault_Tolerant_QC_HigherDimensions} and our earlier abelian-group extension (chapters \ref{chapterF}-\ref{chapterI}) into a stabilizer formalism that uses commuting \emph{hypergroups} (instead of \emph{groups}) of generalized Pauli operators. The latter are no longer unitary nor monomial but still exhibit rich Pauli-like features that let us manipulate them with (new) hypergroup techniques and are normalized by associated Clifford-like gates. We also provide a normal form for  hypergroup stabilizer states (\textbf{theorem \ref{thm:Normal Form CSS States}}) that are CSS-like \cite{CalderbankShor_good_QEC_exist,Steane1996_Multiple_Particle_Interference_QuantumErrorCorrection,Calderbank97_QEC_Orthogonal_Geometry} in our setting.

\item[III.] \textbf{A hypergroup Gottesman-Knill theorem.} We introduce models of  \emph{normalizer circuits over abelian hypergroups}, which contain hypergroup quantum Fourier transforms (QFTs) and other entangling gates.  These models provide a major generalization of the (finite\footnote{For simplicity, we do not consider infinite groups nor infinite dimensional Hilbert spaces in this chapter.}) abelian group normalizer circuit models of chapters \ref{chapterF}-\ref{chapterB}. We show (\textbf{theorem \ref{thm:Evolution of Stabilizer States}}) that the dynamical evolution of such circuits can be tracked in our hypergroup stabilizer picture and,  furthermore, that for large hypergroup families (including  products $\mathcal{T}^m$ of  constant size hypergroups), many hypergroup  normalizer circuits can be efficiently simulated classically (\textbf{theorem \ref{thm:simulation}}).\footnote{Here, we rely  on computability assumptions (section \ref{sect:Assumptions on Hypergroups})  that are always fulfilled in chapter \ref{chapterF}.}

\end{enumerate}
We complete our analysis of the HNSP, which we reduced to the abelian HSHP (result I.),  showing that our normalizer circuit model encompasses an earlier HSHP quantum algorithm based on a variant of Shor-Kitaev's quantum phase estimation, which was proposed by but not fully analyzed by  Amini-Kalantar-Roozbehani in \cite{Amini_hiddensub-hypergroup,Amini2011fourier}. Using our stabilizer formalism,  we prove the latter to be  inefficient on easy instances, and, thereby, point out the  abelian HSHP as the \emph{first} known commutative hidden substructure problem in quantum computing that \emph{cannot} be solved by standard phase estimation\footnote{Note that, in our context and in any HSP setting, phase estimation is used to extract information from a \emph{fixed} unitary oracle $U$. This should not be confused with  the settings where $U$ is not fixed and phase estimation becomes a BQP-complete problem \cite{WocjanZhang06SomeBQPcompleteproblems}.}. In spite of this  no-go result, we also show, in our last main contribution, that in the interesting cases from the nonabelian HSP perspective, the abelian HSHP can actually be solved with a novel \emph{adaptive/recursive} quantum Fourier sampling approa
\begin{enumerate}
\item[IV.] \textbf{New quantum algorithms.} We present the first provably efficient quantum algorithm for finding hidden subhypergroups of \emph{nilpotent}\footnote{These are conjugacy class hypergroups associated to \emph{nilpotent groups} \cite{Humphrey96_Course_GroupTheory}. The latter form a \emph{large} group class that includes abelian groups, Pauli/Heisenberg groups over $\Z_{p^r}$ with prime $p$, dihedral groups $D_{2N}$ with $N=2^n$, groups of prime-power order and their direct products.} abelian hypergroups, provided we have efficient circuits for the required QFTs. This algorithm also leads, via the connection above (result I.), to a new efficient quantum algorithm for the HNSP over nilpotent groups that directly exploits the abelian hypergroup structure and is fundamentally different from the algorithm of Hallgren et al. \cite{Hallgren00NormalSubgroups:HSP}.
\end{enumerate}
Our correctness proofs for these last quantum algorithms  can further be extended to  crucial non-nilpotent groups\footnote{We give another algorithm that works for all groups  under some additional mild assumptions.} (and their associated class hypergroups) such as  the dihedral and symmetric groups.\footnote{For dihedral groups/hypergroups we give a quantum algorithm; for symmetric ones, a \emph{classical} one already does the job  because symmetric groups/hypergroups have few normal subgroups/subhypergroups. } In contrast, no efficient quantum algorithm for the nilpotent, dihedral and symmetric HSPs is known. This provides strong evidence that abelian HSHP  is a much \emph{easier} problem for quantum computers than  nonabelian HSP, and, because of its Shor-like  connection with a stabilizer formalism, perhaps even a more \emph{natural} one.

\subsection{Applications} Though  lesser known than nonabelian groups, abelian hypergroups  have a wide range of applications in  convex   optimization (cf.\ association schemes \cite{KlerkSDP_AssociationSchemes,anjos2011handbook}), classical cryptography, coding theory \cite{corsini2003applications} and conformal field theory   \cite{Wildberger1994HypergroupsApplications}. In  topological quantum computation \cite{kitaev_anyons}, fusion-rule hypergroups  \cite{Kitaev2006_Anyons_Exactly_Solved_Model}  are indispensable  in the study of nonabelian anyons  \cite{Kitaev2006_Anyons_Exactly_Solved_Model}. Our stabilizer formalism over the latter hypergroups likely has applications for quantum error correction and for the simulation of protected gates over topological quantum field theories \cite{Beverland14_ProtectedGates_Topological}. 

The stabilizer formalism and classical simulation techniques presented in this chapter are unique in that they are the first and only available methods  to manipulate stabilizer operators that neither \emph{unitary}, nor \emph{ monomial}, nor \emph{sparse} that we are aware of \cite{nest_MMS}. Furthermore, our stabilizer formalism yields the first known families of qubit/qudit stabilizer operators for any arbitrary finite dimension $d$ that are not the standard Weyl-Heisenberg operators \cite{Gottesman98Fault_Tolerant_QC_HigherDimensions}, with associated normalizer gates that are \emph{not} the standard qudit Clifford gates. Additionally, our methods  allow great flexibility to construct new codes because the stabilizer families can be chosen over any hypergroup of interest.

\subsection{Relationship to prior work}
\label{PreviousWork_c4}

Because finite groups are particular examples of finite abelian groups (see section \ref{sect:Hypergroups}), the hypergroup normalizer circuits extend (finite) abelian group models of \cite{VDNest_12_QFTs}, chapters \ref{chapterF}-\ref{chapterH}. Though we will not consider infinite hypergroups, many of our results should extend to locally compact abelian hypergroups  (cf.\ \cite{BloomHeyer95_Harmonic_analysis,Amini2011fourier} and the discussion in chapter \ref{chapterI} on locally compact abelian groups).

Our efficient classical simulation result (\textbf{theorem \ref{thm:simulation}}) is not a full generalization of the Gottesman-Knill theorem \cite{Gottesman_PhD_Thesis,Gottesman99_HeisenbergRepresentation_of_Q_Computers}, 
 but of its CSS-preserving variant \cite{Delfosee14_Wigner_function_Rebits} and without intermediate measurements; in terms of gates and compared to \cite{VDNest_12_QFTs}, chapter \ref{chapterF}-\ref{chapterI}, this means that we can only simulate hypergroup normalizer circuits built of automorphism gates, global QFTs and generalized Pauli gates. In this chapter, we dedicate most effort to cope with the highly nontrivial difficulty that our Pauli operators are \emph{non-monomial} and \emph{non-unitary}, which renders \emph{all} existing stabilizer formalism techniques \cite{Gottesman_PhD_Thesis,Gottesman99_HeisenbergRepresentation_of_Q_Computers,Gottesman98Fault_Tolerant_QC_HigherDimensions,VDNest_12_QFTs,BermejoVega_12_GKTheorem,BermejoLinVdN13_Infinite_Normalizers,BermejoLinVdN13_BlackBox_Normalizers, AaronsonGottesman04_Improved_Simul_stabilizer,dehaene_demoor_coefficients,dehaene_demoor_hostens,VdNest10_Classical_Simulation_GKT_SlightlyBeyond,deBeaudrap12_linearised_stabiliser_formalism,nest_MMS,NiBuerschaperVdNest14_XS_Stabilizers} inapplicable (including those developed thus far within the thesis). To tackle this issue, we develop new simulation techniques based on hypergroup methods, up to a fairly mature state, though further improvement remains possible (see\ \textbf{section \ref{sect:Simulation}} for a discussion and a related conjecture).

The quantum algorithm results of this chapter solve a  question left open  in chapter \ref{chapterB} (cf.\ discussion) where we gave our no-go theorem for finding new quantum algorithms  based on black-box group normalizer circuits. Therein, we  raised the  question of whether normalizer circuits over \emph{different algebraic structures} could be found and be used to bypass our no-go theorem. Our quantum algorithms for abelian HSHPs answer their question in the affirmative: the circuits we use to solve that problem are instances of normalizer circuits over nonabelian groups/hypergroups (\textbf{section \ref{sect:Quantum Algorithms}}).

The  hidden subhypergroup problem (HSHP) we discuss was first considered by Amini, Kalantar and Roozbehani in \cite{Amini_hiddensub-hypergroup,Amini2011fourier}, yet (to the best of our knowledge) no  provably efficient quantum algorithm for this problem has been given before. We show that an earlier quantum algorithm proposed in \cite{Amini2011fourier} for solving the problem using a variant of Shor-Kitaev's quantum phase estimation \cite{kitaev_phase_estimation} is inefficient on easy instances (section \ref{sect:Quantum Algorithms}). Interestingly, this means that abelian HSHP is the first known \emph{commutative}  hidden substructure problem that cannot be solved by standard phase estimation. Instead, our quantum algorithm is based on  a novel \emph{adaptive/recursive Fourier sampling} quantum approach.

For any non-abelian group $G$, the simulation results we present lead  to efficient  classical algorithms for simulating quantum Fourier transforms over $G$ (specifically,  as employed in weak Fourier sampling routines) acting on coset states $\ket{aH}$, $a\in G$, $H\subset G$ such that $aH$ is invariant under conjugation\footnote{This happens, e.g., if $aH=N$ for normal $N$ or if the subgroup $H$ contains the derived subgroup $[G,G]$.}. In this sense, our work connects with \cite{bermejo2011classical}, where efficient classical algorithms  were given for simulating weak and strong quantum Fourier sampling on arbitrary coset states of  semi-direct products group $\Z_p\ltimes A$, where $p$ is  prime  and $A$ is an  abelian group given in a canonical  form $\Z_{N_1}\times \cdots \times \Z_{N_m}$.

Finally, we mention  it is not contradictory that some hypergroup normalizer circuits are efficiently classically simulable  while others lead to valuable quantum algorithms. Analogously to earlier chapters (see discussion in chapter \ref{chapterB}), this difference arises from the existence of hypergroups with weaker and stronger computability properties, which now play the role of ``decomposed abelian groups'' and ``abelian black-box groups'' (see section \ref{sect:Assumptions on Hypergroups} and appendix \ref{app:Discret Log}). 

\subsection{Chapter outline}

We give a non-technical introduction to the theory of hypergroups in \textbf{section \ref{sect:Hypergroups}}. We then re-introduce the hidden normal subgroup problem (HNSP) and prove its connection to the hidden subhypergroup problem (HSHP) in \textbf{section \ref{sect:HNSP}}. We present our models of hypergroup normalizer circuits, our hypergroup stabilizer formalism, and our simulation results in \textbf{sections \ref{sect: Circuit Model}-\ref{sect:Simulation}} and describe these on some examples. Finally, we use these tools to develop new quantum algorithms for abelian HSHP and HNSP in \textbf{section \ref{sect:Quantum Algorithms}}.

While our motivation for developing our hypergroup stabilizer formalism was to understand more about the HNSP, we note that the results of sections \ref{sect: Circuit Model}--\ref{sect:Simulation} are more general, as they apply to arbitrary hypergroups. We expect that these tools will have applications outside of the analysis of quantum algorithms such as to the development of new error correcting codes.

\section{Abelian hypergroups and hypergroup duality}\label{sect:Hypergroups}

This section is an introduction for quantum computer scientists to the beautiful theory of  \emph{finite abelian hypergroups}\footnote{The hypergroups we consider are frequently called ``finite commutative hypergroups'' in mathematics. We call them ``abelian'' because of the focus of this work on abelian and nonabelian HSPs. In some of our references \cite{Roth75_Character_Conjugacy_Hypergroups,McMullen79_Algebraic_Theory_Hypergroups,McMullen_Duality_abelian_Hypergroups}, the hypergroups in this work are called ``\emph{reversible abelian hypergroups}''.}, whose origin dates back to works by Dunkl \cite{Dunkl1973}, Jewett \cite{Jewett19751}, Spector \cite{Spector1978} in the  70s. Our account is based on \cite{Roth75_Character_Conjugacy_Hypergroups,McMullen79_Algebraic_Theory_Hypergroups,McMullen_Duality_abelian_Hypergroups,Wildberger97_Duality_Hypergroups,Wildberger_Lagrange,Wildberger1994HypergroupsApplications,Wildberger2001algebraic,BloomHeyer95_Harmonic_analysis} and borrows most notation and terminology from \cite{Wildberger1994HypergroupsApplications,Wildberger97_Duality_Hypergroups,Wildberger_Lagrange,Wildberger2001algebraic}. Throughout the chapter, hypergroups and groups are assumed to be \emph{finite} unless said otherwise.

In brief, abelian hypergroups are algebraic structures that generalize  abelian  groups, although   in a  \emph{different} way than nonabelian groups. Despite being less known than the latter, abelian hypergroups   have a wide number of applications in multiple fields, including  combinatorics, convex optimization \cite{KlerkSDP_AssociationSchemes,anjos2011handbook}; cryptography, classical error correction \cite{corsini2003applications}; classical information theory \cite{Wildberger1994HypergroupsApplications};  and conformal field theory  \cite{Wildberger1994HypergroupsApplications}. In  topological quantum computation \cite{kitaev_anyons}, certain hypergroups known by the names of ``fusion theories or categories'' \cite{Wildberger1994HypergroupsApplications,Kitaev2006_Anyons_Exactly_Solved_Model}  are invaluable in the study of topological order and nonabelian anyons  \cite{Kitaev2006_Anyons_Exactly_Solved_Model}. 

On top of their versatility, abelian hypergroups also admit a simple and intuitive \emph{physical} definition, which we give now before going into the full mathematical details of their theory. In simple terms, a \emph{finite abelian hypergroup} $\mathcal{T}$ is a set of particle types $\{x_0, x_1,
\dots, x_n\}$ that can collide. When $x_i$ collides with $x_j$ a particle $x_k$ is created with probability $n_{ij}^k$. Furthermore, a non reactive \emph{vacuum} particle $x_0$ will be created with non-zero probability by such process iff $x_i$  is the antiparticle of $x_k$ (which always exists).

\subsection{Definition}\label{sect:DefinitionsHypergroup}

\newcommand{\B}{\mathcal{B}}

We now turn the  intuitive definition of hypergroup above into a precise mathematical one.

A finite abelian hypergroup $\mathcal{T} = \{x_0, x_1,
\dots, x_n\}$ is a basis of a commutative complex $C^*$ algebra $\mathcal{A}(\mathcal{T})=\C \mathcal{T}$, called the \emph{hypergroup algebra} of $\mathcal{T}$, with a particular structure. $\mathcal{A}(\mathcal{T})$ is endowed with an associative commutative \emph{hyper}operation 
\begin{equation}
\label{eq:Hyperoperation}
x_i x_j = \sum_{k=0}^n n_{ij}^k x_k \ \ \ \forall x_i, x_j \in \mathcal{T},
\end{equation}
which returns a superposition of outcomes in $\mathcal{T}$ (we write ``$x_k\in x_ix_j$'' when $x_k$ is a possible outcome of $x_ix_j$, with $n_{ij}^k\neq 0$); a multiplicative identity $x_0=1$; and an involution $x_i \rightarrow \overline{x}_i$. Note that commutativity and the presence of the involution imply that  $n_{\overline{a},\overline{b}}^{\overline{c}}=n_{ba}^c=n_{ab}^c $ holds for any $a,b,c\in\mathcal{T}$.

Furthermore, the ``structure constants'' $n_{ij}^k \ge 0$ are \emph{real} numbers with three properties:
\begin{itemize}
\item[(i)] \textbf{Anti-element property.}  For every $x_i$ and any $x_j$, the identity $x_0=1$ can be an outcome of $x_ix_j$ if and only if $x_j = \overline{x}_i$. We call  $\overline{x}_i$  the \emph{anti-element} of $x_i$.
\item[(ii)] \textbf{Normalization property.} For all values of  $k=0,\ldots, n$ we have $\sum_{k=0}^n
  n_{i,j}^k = 1$;  in other words, $n_{ij}^k$ is a probability distribution (of outcomes) over $k$.

\item[(iii)]\textbf{Reversibility.}\footnote{This last property (iii) and (\ref{eq:Reversibility Property}) can both be derived from the previous axioms \cite{McMullen79_Algebraic_Theory_Hypergroups}.} For every $x,y,z\in\mathcal{T}$, it holds that $z\in xy$ if and only if $y\in\overline{x}z$. Moreover, if the \emph{weight of $x$} is defined as $\w{x}:=1/n_{x\overline{x}}^0$, the following identity holds:
\begin{equation}\label{eq:Reversibility Property}
\frac{n_{xy}^z}{\w{z}}=\frac{n_{\overline{x}z}^y}{\w{y}}=\frac{n_{z\overline{y}}^x}{\w{x}}
\end{equation}
\end{itemize} 

As a simple example,  any finite abelian group $G$ is an abelian hypergroup. The elements of $G$ define the basis of the group algebra $\Complex G$ and the
involution is $\overline{x} := x^{-1}$. In the case of a group, though, for any
$i, j \in \Integer_{n+1}$, there is only a single nonzero $n_{i,j}^k$ since
$x_i x_j = x_k$ for some $k$; though hypergroups have a more complicated
multiplication than groups,  they preserve the property that the product of
$x$ and  $\overline{x}$ includes the identity.

\myparagraph{Hypergroups in this work.} Though nonabelian hypergroups exist\footnote{In fact, every nonabelian group $G$ is also a kind of nonabelian hypergroup.}, this chapter focuses on  \emph{abelian} ones because they  fulfill certain useful dualities (see below). In sections \ref{sect:HNSP} and \ref{sect:Quantum Algorithms},  we further focus on specific abelian hypergroups that arise from \emph{finite groups} (section \ref{sect:Examples Hypergroups}). 

\subsection{Glossary}\label{sect:Glossary}
\label{sect:Orthogonality and QFT}\label{sect:Subhypergroups, Quotients}

We now give a glossary of hypergroup theoretic concepts for future reference. In all definitions below $\mathcal{T}$ is fixed to be an arbitrary \emph{abelian} finite hypergroup. 

\myparagraph{Weight functions.}Every subset $X\subset\mathcal{T}$ has a \emph{weight} $\varpi_X:=\sum_{x\in X}w_x$, with $w_x$ as in (\ref{eq:Reversibility Property}).

\myparagraph{Subhypergroup.}{A \emph{subhypergroup $\mathcal{N}$}}   is a subset of $\mathcal{T}$ that is also a hypergroup with the same identity, involution, structure constants and weights.

\myparagraph{Quotient hypergroup.}For any  subhypergroup $\mathcal{N}$ the {\emph{quotient hypergroup}} $\mathcal{T/N}$  is an abelian hypergroup whose elements are the cosets $a\mathcal{N}:=\{x\in\mathcal{T}:x\in a b \text{ for some }b\in \mathcal{N}
\}$. Its  hyperoperation is  defined \cite{McMullen79_Algebraic_Theory_Hypergroups,Wildberger_Lagrange} by, first, identifying each $a\mathcal{N}$ with an element of the  $\mathcal{A}(\mathcal{T})$ algebra\footnote{See \cite{Roth75_Character_Conjugacy_Hypergroups} for a set theoretic definition} via $a\mathcal{N} :=\sum_{x\in a\mathcal{N}}\w{x} x /\varpi_{a\mathcal{N}}$. Then, $\mathcal{T/N}$ inherits a hyperoperation with  structure constants $r_{a\mathcal{N},b\mathcal{N}}^{c\mathcal{N}} =\sum_{d\in c\mathcal{N}}n_{ab}^{d}$ and weights $\ws{a\mathcal{N}}=1/(\sum_{b\in\mathcal{N}}n_{a,\overline{a}}^{b})=\varw{a\mathcal{N}}/\varw{\mathcal{N}}$.

\myparagraph{Morphisms.} A map between two hypergroups $f:\mathcal{T}\rightarrow\mathcal{T'}$ is a \emph{homomorphism} if $f(ab)=f(a)f(b)=\sum_{c} n_{ab}^c f(c)$ and $f(\overline{a})=\overline{f(a)}$. An invertible homomorphism is an \emph{isomorphism}. An isomorphism from $\mathcal{T}$ to $\mathcal{T}$ is an \emph{automorphism}. As with groups, isomorphic hypergroups have identical hypergroup-theoretic properties (weights, subhypergroups, etc.).

\myparagraph{Character hypergroup $\mathcal{T^*}$.} A complex function $\mathcal{X}_\mu : \mathcal{T} \rightarrow \Complex$ is a \emph{character of $\mathcal{T}$} if it is not identically zero and satisfies the identity\footnote{If characters are linearly extended to act on the hypergroup algebra $\mathcal{A(T)}$, condition (\ref{eq:Character DEFINITION}) becomes  $\mathcal{X}_\mu(ab) =
\mathcal{X}_\mu(a) \mathcal{X}_\mu(b)$, $\forall a,b \in \mathcal{A(T)}$; in other words, the characters of $\mathcal{T}$ are also the characters of  $\mathcal{A(T)}$.}
\begin{equation}\label{eq:Character DEFINITION}
\mathcal{X}_\mu(ab)=\mathcal{X}_\mu(a)\mathcal{X}_\mu(b)= \sum_{c}n_{ab}^c \mathcal
X(c)\quad\textnormal{and}\quad \mathcal{X}_\mu(\overline{a})=\overline{\mathcal{X}_\mu}(a)\quad\textnormal{for all $a,b\in\mathcal{T}$}.
\end{equation}
For any abelian hypergroup $\mathcal{T}$, its set  $\mathcal{T^*}$ of character functions defines an  abelian \emph{signed} hypergroup with  the point-wise functional product as hyperoperation, the trivial character $\Hchi_1(a)=1$ as identity and  the complex conjugate map $\mathcal{X}_\mu\rightarrow \overline{\mathcal{X}_\mu}$ as involution: here, ``signed''  means that $\mathcal{T^*}$ fulfills (i-ii-iii) but may have some negative structure constants $m_{\mu\nu}^\gamma$, which represent negative probabilities. If all constants $m_{\mu\nu}^\gamma$ are non-negative, $\mathcal{T}^*$ is a hypergroup called the \emph{character hypergroup of $\mathcal{T}$}, and $\mathcal{T}$ is said to be \emph{strong} \cite{BloomHeyer95_Harmonic_analysis}. Throughout the chapter, we assume all hypergroups  to be strong (without notice) so that the associated character hypergroups $\mathcal{T}^*$ define  new ``dual theories'' of particle collisions\footnote{Many of the hypergroup concepts and properties presented in this section as well as our results in sections \ref{sect: Circuit Model}, \ref{sect:Simulation} can be effortlessly extended to the setting where $\mathcal{T}$ is an {abelian signed hypergroup}, in which case $\mathcal{T}^*$ is also an abelian signed hypergroup  \cite{McMullen79_Algebraic_Theory_Hypergroups,McMullen_Duality_abelian_Hypergroups,BloomHeyer95_Harmonic_analysis,Ichihara_thesis_Hypergroup_Extensions,Yamanaka2013_Thesis}). Though it seems plausible, we have not investigated whether our results  in section \ref{sect:Stabilizer Formailsm} can be extended to signed hypergroups. In the remaining sections, we focus on class and character hypergroups that arise from finite groups, which are always strong.}.

\myparagraph{Weight-order duality.} The hypergroups $\mathcal{T}$ and $\mathcal{T^*}$ have the same cardinalities and weights:
\begin{equation}\label{eq:Weight Duality}
\varw{\mathcal{T}}=\sum_{a\in\mathcal{T}}\w{a} = \sum_{\mathcal{X}_\mu\in\mathcal{T}^*}\w{\mathcal{X}_\mu}=\varw{\mathcal{T}^*}.
\end{equation}

\myparagraph{Abelian hypergroup duality.}\label{sect:Duality} The  hypergroup  $\mathcal{T}^{**}$ of characters of $\mathcal{T}^*$ is \emph{isomorphic} to the original hypergroup $\mathcal{T}$. This isomorphism is constructed canonically by sending $a \in \mathcal{T}$ to a character 
\begin{equation}\label{eq:Hypergroup Duality}
\widetilde{\mathcal{X}_a}(\mathcal{X}_\mu)=\overline{\Hchi_{\mu}}(a).
\end{equation} In particular, this shows that  the hypergroups $\mathcal{T}$, $\mathcal{T}^*$  have the same number of elements.

\begin{remark}[\textbf{Notation}] Throughout the text,  we identify dual characters $\widetilde{\mathcal{X}}_{a}\in\mathcal{T}^{**}$ with elements $a\in\mathcal{T}$ via the isomorphism (\ref{eq:Hypergroup Duality}). We write the hyperoperation of $\mathcal{T}^*$ compactly as $\Hchi_\mu \Hchi_\nu = \sum_{\gamma} m_{\mu\nu}^{\gamma} \Hchi_\gamma$ and, occasionally, use the expression $\Hchi_{\overline{\mu}}$ as a shorthand for $\overline{\Hchi_\mu}$.
\end{remark}
The notions of character and duality lead to a family of related concepts that are extremely valuable in hypergroup theory and in the present work:

\myparagraph{Annihilators.} The \emph{annihilator $\mathcal{N^\perp}$} of a subhypergroup $\mathcal{N\subset T}$  is a  subhypergroup  of $\mathcal{T^*}$
\begin{equation}\label{eq:annhilator}
\mathcal{N}^\perp := \{\mathcal{X}_\mu \in \mathcal{T}^*: \mathcal{X}_\mu(a)=1 \textnormal{ for all } a\in \mathcal{N}\}.
\end{equation}
\myparagraph{Subhypergroup duality.} A stronger form of hypergroup duality relates the notions of annihilator, subhypergroup and quotient: the annihilator $\mathcal{N}^\perp$ is \emph{isomorphic} to the characters $(\mathcal{T/N})^*$ of  $\mathcal{T/N}$; moreover, the character hypergroup $\mathcal{N^*}$ is isomorphic to $\mathcal{T^*/\mathcal{N}^\perp}$. 

\myparagraph{Character orthogonality.} Character functions are orthogonal with the inner product
\begin{equation}\label{eq:Character Orthogonality}
\langle \Hchi_{\mu},\Hchi_\nu\rangle =
\sum_{a\in\mathcal{T}} \frac{\ws{\mathcal{X}_\nu}\w{a}}{\varpi_{\mathcal{T}}} \overline{\Hchi_{\mu}}(a)\Hchi_\nu(a)=\delta_{\mu,\nu}.
\end{equation}
Moreover, due to hypergroup and subhypergroup duality, for any subhypergroup $\mathcal{N}\subset\mathcal{T}$, any two cosets $a\mathcal{N},b\mathcal{N}\in\mathcal{T/N}$ and any $\mathcal{X_\mu,X_\nu\in \mathcal{N}^\perp}$, the following generalized orthogonality relationships are always fulfilled
\begin{equation}\label{eq:Character Orthogonality Subhypergroups}
\sum_{a\in \mathcal{N}} \frac{\ws{\Hchi_\nu\mathcal{N}^\perp}\w{a} }{\varpi_\mathcal{N}} \overline{\Hchi_{\mu}}(a) \mathcal{X}_\nu(a) =\delta_{\mathcal{X}_\mu \mathcal{N}^\perp ,\mathcal{X}_\nu \mathcal{N}^\perp} ,\quad   \sum_{\mathcal{X}_\mu \in \mathcal{N}^\perp} \frac{\ws{b\mathcal{N}} \w{\mathcal{X}_\mu}}{{\varpi_{\mathcal{N}^\perp}}} \overline{\Hchi_{\mu}}(a) \mathcal{X}_\mu(b) ={\delta_{a\mathcal{N},b\mathcal{N} }},
\end{equation}

\subsection{Examples from group theory}
\label{sect:Examples Hypergroups}

We now introduce two  examples of hypergroups that play a central role in our work (namely, in sections \ref{sect:HNSP} and \ref{sect:Quantum Algorithms}).  For an arbitrary finite  group, these are the hypergroups of conjugacy classes and of characters, which are dual to each other in the sense of (\ref{eq:Hypergroup Duality}). The existence of these hypergroups linked to arbitrary groups lets us apply our hypergroup normalizer circuit and stabilizer formalisms (sections \ref{sect: Circuit Model}--\ref{sect:Stabilizer Formailsm}) to nonabelian groups.

\subsubsection{The hypergroup of  conjugacy classes of $G$}\label{sect:ConjClassHypergroup}

Let $G$ be any finite group. For any $g \in G$, we let $C_g := \set{g^a}{a
\in G}$, where $g^a := a^{-1}ga$ denotes the conjugacy class of $g$. We let
$\Conj{G}$ be the set of distinct conjugacy classes of $G$.

Let $C = \{g_1, g_2, \dots\}$ and $D = \{h_1, h_2, \dots\}$ be two conjugacy
classes.  Then, for any product, $g_i h_j$, its conjugate $(g_i h_j)^a = g_i^a
h_j^a$ is a product of conjugates, so it can be written as $g_k h_\ell$ for
some $k$ and $\ell$. Furthermore, if there are $M$ distinct products $g_{i_1}
h_{j_1}, \dots, g_{i_M} h_{j_M}$ producing some element $x$, then the distinct
products $g_{i_1}^a h_{j_1}^a, \dots, g_{i_M}^a h_{j_M}^a$ all produce $x^a$.
Thus, for each conjugacy class $E$ arising in the product of elements of $C$
and $D$, we get a well defined number of ``how many times'' that class arises,
which we denote $M_{C,D}^E$.

We will denote by $\A(\Conj{G})$ the complex vector space with the distinct
conjugacy classes as a basis, which make into a $\Complex$-algebra by defining
the product $C D := \sum_{E \in \Conj{G}} M_{C,D}^E E$.

We take the map $C_g \mapsto C_{g^{-1}}$, extended to all of $\A(\Conj{G})$ by
linearity, as our involution.

It is easy to see that $C_e$ arises in a product $C_g C_h$ iff $C_h$ contains
$g^{-1}$, which occurs iff $C_h = C_{g^{-1}}$. Thus, we can see that the first
of the two required properties holds for the product with structure constants
$M_{C,D}^E$.

To get the normalization property to hold, though, we must make a minor change.
For each $C_g \in \Conj{G}$, define $c_g$ to be the vector
$(1/\abs{C_g})\,C_g$. Then we will take $\set{c_g}{C_g \in \Conj{G}}$
to be a new basis. The structure constants become $m_{C,D}^E :=
M_{C,D}^E \abs{E} / \abs{C} \abs{D}$. Since the total number of products of
elements formed multiplying $C$ by $D$ is $\abs{C} \abs{D}$, we can see that
$\sum_{E \in \Conj{G}} M_{C,D}^E \abs{E} = \abs{C} \abs{D}$, which means that
these new structure constants, $m_{C,D}^E$, are properly normalized. Thus,  conjugacy classes define a hypergroup, up to this normalization, which we call the \textbf{class hypergroup}~$\overline{G}$. 

Finally, we note that this hypergroup is abelian, even if the underlying group
is not abelian. To see this, we calculate $gh = h h^{-1} g h = h g^h = (hg)^h$
(since $h^h = h$), which shows that $gh$ and $hg$ are in the same conjugacy
class. Hence, if we are multiplying conjugacy classes instead of elements, we
do not distinguish between $gh$ and $hg$, and we get an abelian structure.

\subsubsection{The hypergroup of characters}
\label{sect:CharacterHypergroup}

Let $\Repr{G}$ denote the set of irreducible characters of the finite group  of $G$, $\A(\Repr{G})$ the complex vector space with basis $\widehat{G}$, and $\chi_\mu$ the character of the irreducible representation $\mu$. As we explain next, $\Repr{G}$ has a natural hypergroup structure.

First, the  involution of $\widehat{G}$ will be the linear extension of the map  $\chi_\mu \mapsto
\overline{\chi}_\mu$, for $\chi_\mu \in \Repr{G}$, the image also being an
irreducible character.

Second, for any two characters, $\chi_\mu$ and $\chi_\sigma$, the pointwise product  $\chi_\mu \chi_\sigma$ is also a character, though it is not
necessarily irreducible. However, as is well known, any representation
can be written as a linear combination of irreducible characters: $\chi_\mu
\chi_\sigma = \sum_{\tau \in \Repr{G}} N_{\mu,\sigma}^\tau \chi_\tau$ for some
non-negative integers $N_{\mu,\sigma}^\tau$. Using this as our product,
$\A(\Repr{G})$ becomes a $C^*$-algebra, where the identity element is the
trivial irreducible representation, $\chi_1$, given by ${\chi_1}(g) \equiv 1$.

Third, the coefficient $N_{\mu,\sigma}^\tau$, as is also well known from
representation theory, is given by $\langle{\chi_\mu \chi_\sigma},{\chi_\tau}\rangle$,
where $\langle{\cdot},{\cdot}\rangle$ is the inner product 
 $\langle{\chi_\mu},{\chi_\sigma}\rangle = \abs{G}^{-1} \sum_{g \in G}
\chi_\mu(g) \overline{\chi_\sigma(g)}$.  From this, we can see that
$N_{\mu,\sigma}^{1} = \langle{\chi_\mu \chi_\sigma},{{\chi_1}}\rangle =
\langle{\chi_\mu},{\overline{\chi_\sigma}\,{\chi_1}}\rangle =
\langle{\chi_\mu},{\overline{\chi_\sigma}}\rangle$.  Since $\chi_\mu$ and
$\overline{\chi_\sigma}$ are both irreducible, this is 1 if $\chi_\sigma =
\overline{\chi_\mu}$ and 0 otherwise.  Hence, we can see that the structure
constants $N_{\mu,\sigma}^\tau$ have the first required property.

Finally, we will normalize the characters, as in section \ref{sect:ConjClassHypergroup}, in order to have (ii). For this, we define  $\widehat{G}$, the \textbf{character hypergroup of $G$}, to be the hypergroup with elements
\begin{equation}\label{eq:Normalized Character}
\mathcal{X}_\mu:=\frac{\chi_\mu}{d_\mu}
\end{equation}
where $d_\mu$ is the dimension of the irrep $\mu$. The  structure constants now become $n_{\mu,\sigma}^\tau
:= N_{\mu,\sigma}^\tau d_\tau / d_\mu d_\sigma$. Since $\chi_\mu
\chi_\sigma$ is actually the character of the representation $\mu \otimes
\sigma$, which splits into a direct sum of irreducible
representations (as described  above), we must have $\sum_{\tau
\in \Repr{G}} N_{\mu,\sigma}^\tau d_\tau = d_\mu d_\sigma$ as the latter is the
dimension of the tensor product. This implies that (ii) is fulfilled and that  $\widehat{G}$ (now suitably normalized) is indeed a hypergroup.

Finally, we note that, in this case,  our product is manifestly
abelian since $\chi_\mu \chi_\sigma$ denotes the element-wise product of these
functions, which takes place in the abelian group $\Complex$.

\subsubsection{The relationship between $\overline{G}$ and $\widehat{G}$}

Crucially, the characters of the hypergroup $\Conj{G}$ turn out to be the  normalized characters $\mathcal{X}_\mu=\chi_\mu/d_\mu$ of $G$ and, due to duality (\ref{eq:Hypergroups}),  conjugacy classes are the characters of $\widehat{G}$. Classes and characters have weights $\ws{C_g}=|C_g|$ and $\ws{\mathcal{X}_\mu}=d_\mu^2$, respectively, where $d_\mu$ is the dimension of the irrep $\mu$. This fantastic connection between groups and hypergroups lets one easily derive many well known results in nonabelian group character theory \cite{Humphrey96_Course_GroupTheory,Isaacs1994character} using the properties of section \ref{sect:DefinitionsHypergroup}, including the usual character orthogonality relationships and the famous $|G|=\sum_{\Hchi_\mu\in \widehat{G}} d_\mu^2$ identity: the latter can be derived from (\ref{eq:Weight Duality}), which leads to $\varw{\overline{G}}=\sum_{C_g\in\overline{G}} |C_g| =|\overline{G}|= \sum_{\Hchi_\mu\in \widehat{G}} d_\mu^2$ and also implies, $\varw{\overline{G}}=|\overline{G}|$ and $\varw{\widehat{G}}= |\widehat{G}|$.

\section{Understanding the Hidden Normal Subgroup Problem}
\label{sec:hnsp}
\label{sect:HNSP}

In this section, we demonstrate a formal connection between the hidden \emph{normal} subgroup problem (HNSP) and a problem on abelian hypergroups, defined below, which we call the CC-HSHP. Specifically, we show that, in many cases, we can efficiently reduce the HNSP to the CC-HSHP, classically. This reduction tells us that, even though the HNSP is defined in terms of nonabelian groups, it can be translated into a problem about an algebraic structure that is abelian, albeit one that is more complex than a group (a hypergroup).

In the remainder of the chapter, we will see the effects of moving from nonabelian groups to abelian hypergroups. While the switch from groups to hypergroups creates some new difficulties, we also gain a great deal by working with an abelian structure. In particular, we will see that the mathematical structure of abelian hypergroups leads to a beautiful stabilizer formalism and to new quantum algorithms. Here, we explain how abelian hypergroups arise specifically when looking for hidden normal subgroups  before moving to the more general setting.

In the first subsection, we formally define the two problems mentioned above, the HNSP and the CC-HSHP. Afterwards, we show how to reduce the former to the latter.

\subsection{The HNSP and the CC-HSHP}

In the HNSP, we are given an oracle $f:G\rightarrow \{0,1\}^*$, assigning
labels to group elements, that is promised to \emph{hide} some normal subgroup
$N \lhd G$. The latter means that we have $f(x) = f(x')$ for $x, x' \in G$ if
and only if $x' = xn$ for some $n \in N$. An algorithm solves the HNSP if it
can use this oracle and other quantum computation in order to determine the
subgroup $N$ with high probability.

The algorithm of Hallgren et al. for the HNSP finds the hidden subgroup $N$
using exclusively information provided by characters of the group. They showed
that this works only for normal subgroups as it cannot distinguish a non-normal
subgroup $H \le G$ from a conjugate subgroup $H^a \not= H$.

If we are only examining the characters of the group $G$, then it stands to
reason that we can get the same information from the hypergroup of characters
$\Repr{G}$ or, equivalently, from the hypergroup of conjugacy classes
$\Conj{G}$ since these two hypergroups contain the same
information.\footnote{After all, each can be recovered from the other as its
dual hypergroup.} Hence, we may expect that the HNSP on $G$ is related to some
problem on the abelian hypergroup $\Conj{G}$.

A natural question for abelian hypergroups like these is the hidden
subhypergroup problem \cite{Amini_hiddensub-hypergroup}. For our
abelian hypergroup of conjugacy classes, we will refer to this problem as the
conjugacy class hidden subhypergroup problem or CC-HSHP.  Here, we are given an
oracle $f : \Conj{G} \rightarrow \{0,1\}^*$, assigning labels to conjugacy
classes, that hides some subhypergroup, and we are
asked to determine that subhypergroup via oracle queries and quantum
computation. We will see next how this is related to the HNSP.

\subsection{Reducing the HNSP to the CC-HSHP}

Since a normal subgroup $N$ is (the union of) a set of conjugacy classes that
is closed under multiplication and taking inverses, it also defines a
subhypergroup of $\Conj{G}$, which we denote by $\subhyp{N}{G}$.\footnote{We
distinguish this from $N$, which is a set of group elements, because
$\subhyp{N}{G}$ is a set of conjugacy classes.} Hence, any subgroup that can be
found as the solution of the HNSP can also be found as the solution of the
CC-HSHP. Indeed, as we will see next, in many cases, we can directly reduce
the HNSP to the CC-HSHP.

In order to perform this reduction, we need to provide a CC-HSHP oracle. Our proofs will show how to translate an oracle for the HNSP into an oracle for the CC-HSHP. These translations assume that we can perform certain computations with conjugacy classes, described in detail in appendix~\ref{sect:class basis operations}, which we refer to as ``computing efficiently with conjugacy classes''. (While formally an assumption, we know of no group for which these calculations cannot be performed efficiently.)

\begin{theorem}[HNSP $\le$ CC-HSHP, I]
\label{thm:reduction1}
Let $G$ be a group. Suppose that we are given a hiding function $f : G \rightarrow \{0,1\}^*$ that is also a class function\footnote{This means that $f$ is constant on conjugacy classes. This will occur iff $G / N$ is abelian, where $N$ is the hidden subgroup.}. If we can compute efficiently with conjugacy classes, then we can efficiently reduce this HNSP to the CC-HSHP.
\end{theorem}
\begin{proof}
The assumptions about computing efficiently with conjugacy classes imply that, given a conjugacy class $C_g$, we can efficiently find an element $x \in C_g$ and apply $f$ to get a label. (Since $f$ is a class function, the label is the same for any $x' \in C_g$.) Let $N$ be the hidden subgroup.  Since $f$ hides $N$ and $N$ is normal, we can see that $f(xn) = f((xn)^a) = f(x^a n^a) = f(x^a) = f(x^a n')$ for any $n, n' \in N$. This shows that $f$ is constant on $C_g\subhyp{N}{G}$, which corresponds to a coset of the subhypergroup $\subhyp{N}{G} \le \Conj{G}$. It follows immediately that $f$ has distinct values on distinct cosets of $\subhyp{N}{G}$, so we can see that $f$ is a hiding function for this subhypergroup corresponding (uniquely) to $N$.
\end{proof}

\begin{theorem}[HNSP $\le$ CC-HSHP, II]
\label{thm:reduction2}
Let $G$ be a group. Suppose that we are given a hiding function $f : G \rightarrow H$ that is also a homomorphism.  If we can efficiently compute with conjugacy classes of $G$ and $H$, then we can efficiently reduce this HNSP to the CC-HSHP.
\end{theorem}

\begin{proof}
Consider any element $x \in G$. For any conjugate $x^a$, for some $a \in G$, we
see that $f(x^a) = f(a^{-1} x a) = f(a^{-1}) f(x) f(a)$ since $f$ is a
homomorphism. Furthermore, since $a^{-1} a = e$, we see that $f(a^{-1}) f(a) =
f(e) = e$, which shows that $f(a^{-1}) = f(a)^{-1}$. Putting these together, we
have $f(x^a) = f(a)^{-1} f(x) f(a) = f(x)^{f(a)}$. This means that the function
$\tilde{f}$ taking $x$ to the conjugacy class label of $f(x)$ is a class
function, which we can compute efficiently by assumption.\footnote{Also note
that, since $e$ is the only element in its conjugacy class, $\tilde{f}$ hides
the same subgroup as $f$.} Thus, by the same proof as in previous theorem, we can reduce this to
the CC-HSHP.
\end{proof}
This latter theorem applies to many of the important examples of HSPs. This
includes the oracles used for factoring, discrete logarithm over cyclic groups
and elliptic curves, and abelian group decomposition (cf.\ chapter \ref{chapterB}).

While all of these examples are abelian groups, it is true in general that, for
any normal subgroup of any group, there is always some hiding function that is a group
homomorphism.\footnote{If $H \trianglelefteq G$ is the hidden subgroup, then one example is
the canonical oracle $G \rightarrow G /H$ given by $x \mapsto xH$.}

From these proofs, we can see that the essential difference between the HNSP
and the CC-HSHP is the slightly differing requirements for their oracles.  We
have seen that, whenever we can convert an oracle for the former into one for
the latter, we can reduce the HNSP to the CC-HSHP.\footnote{This also assumes
the relatively minor assumption that we can compute with conjugacy classes.}
Above, we showed this can be done in the case that the two sets of requirements
are actually the same (\textbf{theorem~\ref{thm:reduction1}}) and the case where the
labels produced by the oracle are not opaque but rather come with enough
information to compute with their conjugacy classes
(\textbf{theorem~\ref{thm:reduction2}}).

Apart from this, it is worth reflecting on which of the types of oracle is the
most sensible for the problem of finding hidden normal subgroups. With this in
mind, we note that the oracle in the HNSP is not specific to normal
subgroups: the same type of oracle can hide non-normal subgroups as well --- we
are simply promised that, in these cases, the hidden subgroup happens to be
normal. In contrast, the oracle in the CC-HSHP can \emph{only} hide normal
subgroups because it is required to be constant on conjugacy classes. Hence,
even though we came upon the oracle definition from the CC-HSHP by looking at
hypergroups, it is arguable that this is actually a \emph{better} definition of 
hiding function for  normal subgroups.  Our proofs above demonstrate that,
whenever we are given an oracle of this type, we can reduce finding the hidden normal subgroup to the CC-HSHP.

We will return to the HNSP in section~\ref{sect:Quantum Algorithms}. There, we will show that the CC-HSHP can be efficiently solved on a quantum computer under reasonable assumptions. This, together with the theorems above, show that the HNSP is easy because the CC-HSHP is easy, which gives an explanation for why the HNSP is easy in terms of the presence of an \emph{abelian} algebraic structure.

Before we can do that, however, we need to first develop some tools for analyzing quantum algorithms using abelian hypergroups. These tools will be of independent interest.

\section{Normalizer circuits over abelian hypergroups}\label{sect: Circuit Model}

In section \ref{sect:HNSP}, we described our motivating example (the hidden normal subgroup problem) for considering how abelian hypergroups can be used to understand quantum computation. There, the abelian hypergroups arose from nonabelian groups. However, there are a vast number of interesting hypergroups with applications in physics and mathematics \cite{Wildberger1994HypergroupsApplications}, including many of the ones used in topological quantum computation \cite{kitaev_anyons,kitaev_anyons}, that do \emph{not} arise from groups. So in the next three sections, we will work with a general abelian hypergroup $\mathcal{T}$, which could come from any of these settings.

Our plan in these next few sections is to extend  the abelian-group normalizer circuit model that we applied to successfully  understand quantum algorithms in chapters \ref{chapterF}-\ref{chapterB}. We start, in this section, by defining a model of normalizer circuits over hypergroups that we will analyze. Definitions are given in section \ref{sect: Circuit Model Definitions}. In section \ref{sect:Examples Normalizer Circuits}, we go through a few examples of what these models consist of for different hypergroups. In later sections, we develop a stabilizer formalism and a Gottesman-Knill-type theorem that applies to these circuits.

\subsection{Circuit model}\label{sect: Circuit Model Definitions}

Fix $\mathcal{T}$  to be an arbitrary \emph{finite abelian hypergroup}. We now define a circuit model, which we call \emph{normalizer circuits} over $\mathcal{T}$. The gates of these circuits are called \emph{normalizer gates}.

\myparagraph{The Hilbert space:} Normalizer gates over  $\mathcal{T}$  act on a Hilbert space  $\Comp_\mathcal{T}$ with two orthonormal bases, $\mathcal{B}_{\mathcal{T}}=\{\ket{a}, a \in \mathcal{T}\}$ and  $\mathcal{B}_{\mathcal{T}^*}=\{\ket{\mathcal{X}_\mu}, \mathcal{X}_\mu \in \mathcal{T}^*\}$, labeled by elements and characters of $\mathcal{T}$,\footnote{Note that duality (\ref{eq:Hypergroup Duality}) implies $\dim \Comp_\mathcal{T} = \dim \Comp_\mathcal{T^*}$.} that  are related via the \emph{quantum Fourier transform} (QFT) of $\mathcal{T}$:
\begin{equation}\label{eq:Quantum Fourier Transform over Hypergroup T}
\Fourier{\mathcal{T}}\ket{a}=\sum_{\mathcal{X_\mu}\in\mathcal{T^*}}\sqrt{\frac{\w{\mathcal{X}_\mu} \w{a}}{\varpi_{\mathcal{T^*} }}} \mathcal{X}_\mu(a)\ket{\mathcal{X}_\mu},\qquad \Fourier{\mathcal{T}}^\dagger \ket{\mathcal{X_\mu}} = \sum_{a\in\mathcal{T}}\sqrt{\frac{\w{a}\w{\overline{\mathcal{X}_\mu}}}{\varpi_{\mathcal{T} }}} \overline{\mathcal{X}_\mu}(a)\ket{a}.
\end{equation}
Character orthogonality  (\ref{eq:Character Orthogonality}) implies that (\ref{eq:Quantum Fourier Transform over Hypergroup T}) is a unitary transformation.

\myparagraph{Registers:} Because in many settings it is important  to split a   quantum computation in multiple registers, we let   $\mathcal{T}$ and $\mathcal{H_T}$  have a general direct product and tensor product form
\begin{equation}\label{eq:Product Hypergroup and Hilbert Space}
\mathcal{T} = \mathcal{T}_1 \times \dots \times \mathcal{T}_m \qquad\longleftrightarrow \qquad
\Comp_{\mathcal{T}} \cong \Comp_{\mathcal{T}_1} \otimes \dots \otimes \Comp_{\mathcal{T}_m}.
\end{equation}
In this case, the QFT over $\mathcal{T}$ is the tensor product  of the QFTs over the $\mathcal{T}_i$'s:
\begin{equation}\label{eq:partial QFT}
\Fourier{\mathcal{T}}=\Fourier{\mathcal{T}_1}\otimes \cdots\otimes  \Fourier{\mathcal{T}_m}.
\end{equation}
\myparagraph{Input states:} Each register $\mathcal{H}_{\mathcal{T}_i}$ is initialized to be in either an \emph{element state} $\ket{{x_i}},{x_i}\in \mathcal{T}_i$ or in a \emph{character state} $\ket{\Hchi_\mu}, \Hchi_\mu\in\mathcal{T}_i^*$.

\myparagraph{Gates:}The allowed \emph{normalizer gates} at step $t$ of a normalizer circuit depend on a parameter $\mathcal{T}(t)$, which is a hypergroup, related to $\mathcal{T}$, of the form
\begin{equation}\label{eq:Hypergroups}
\mathcal{T}(t)=\mathcal{T}(t)_{1}\times\cdots \times \mathcal{T}(t)_{m} \quad \textnormal{with}\quad  \mathcal{T}(t)_i \in \{{\mathcal{T}}_i, {\mathcal{T}}_i^*\}.
\end{equation}
The role of $\mathcal{T}(t)$ is to indicate whether the operations carried out by circuit at time $t$ will be on the element or character basis.  At step $0$, $\mathcal{T}(0)$ is chosen so that  $\mathcal{T}_{i}(0)\in \{\mathcal{T}_i, \mathcal{T}_i^*\}$ indicates whether $\mathcal{H}_{\mathcal{T}_i}$ begins on an element or character state.  At any steps $t> 0$, $\mathcal{T}(t)$  depends on the gates that have been applied at earlier steps, following  rules given below.

Normalizer gates at time $t$  can be of four types:
\begin{enumerate}
\item \textbf{Pauli gates.}  Pauli gates of type X\footnote{In previous chapters we did not consider X-type Pauli gates to be normalizer gates explicitly but we proved that they can be implemented via normalizer circuits comprising 3 gates (lemma \ref{lemma:Fourier transform diagonalizes the Regular Representation}). This property is shared with hypergroup normalizer circuits, as shown in  (\ref{eq:QFTs are Clifford}), theorem \ref{thm:Evolution of Stabilizer States} below. For the sake of generality, we add them to the basic set of normalizer gates in this final chapter.}  implement the $\mathcal{T}(t)$ hyperoperation $\PX{\mathcal{T}(t)}(a)\ket{b}=\ket{ab}$ for invertible elements $a\in\mathcal{T}(t)$. Pauli gates of type Z multiply by phases $\PZ{\mathcal{T}(t)}(\mathcal{X}_\mu)\ket{b}=\mathcal{X}_\mu(b)\ket{b}$ which correspond to invertible characters in $\mathcal{T}(t)^*$. 
\item \textbf{Automorphism Gates.} Let $\alpha : \mathcal{T}(t) \rightarrow \mathcal{T}(t)$ be an
automorphism of the hypergroup $\mathcal{T}(t)$. Then the automorphism gate $U_\alpha$ taking $\ket{g}\mapsto \ket{\alpha(g)}$ is a valid normalizer gate. 

\item \textbf{Quadratic Phase Gates} A complex function  $\xi : \mathcal{T}(t) \rightarrow U(1)$ is
called ``quadratic'' if the map $B:\mathcal{T}(t)\times
\mathcal{T}(t)\rightarrow U(1)$ defined by
$\xi(gh)=\xi(g)\xi(h)B(g,h)$ is a bi-character, i.e., a character of
the hypergroup in either argument. A quadratic phase gate is a diagonal map
$D_\xi$ taking $\ket{g} \mapsto \xi(g) \ket{g}$ for some quadratic function
$\xi$.
\item \textbf{Quantum Fourier Transforms}. A \emph{global QFT} implements the  gate $\mathcal{F}_{\mathcal{T}(t)}$ over $\mathcal{T}(t)$  (\ref{eq:Quantum Fourier Transform over Hypergroup T}). \emph{Partial QFTs}  implement the gates $\mathcal{F}_{{\mathcal{T}(t)}_i}$ on single registers $\mathcal{H}_{\mathcal{T}(t)_i}$ (while the other registers remain unchanged).
\end{enumerate}
\myparagraph{Update rule:} Because QFTs change the hypergroup that labels the standard basis (\ref{eq:Quantum Fourier Transform over Hypergroup T}), the rules above do not specify which  normalizer gates should be applied on the second step. For this reason, in our gate model, we  \emph{update} the value of $\mathcal{T}(t+1)$ at time $t+1$ so that $\mathcal{T}(t+1)_i=\mathcal{T}(t)_i^*$ if a QFT acts on $\mathcal{H}_{\mathcal{T}(t)_i}$ and $\mathcal{T}(t+1)_i=\mathcal{T}(t)_i$ otherwise.

\myparagraph{Measurements:} At the final step $T$, every register $\mathcal{H}_{\mathcal{T}_i}$ is measured in either the element or the character basis depending on the configuration of the QFTs in the circuit: specifically,  $\mathcal{H}_{\mathcal{T}_{i}}$ is measured in basis $\mathcal{B}_{\mathcal{T}_i}$ labeled by elements of $\mathcal{T}_i$  when  $\mathcal{T}(T)_i=\mathcal{T}_i$, and in the character basis $\mathcal{B}_{\mathcal{T}_i^*}$ when  $\mathcal{T}(T)_i=\mathcal{T}_i^*$. In the end, the final string of measurement outcomes identifies an element of the hypergroup $\mathcal{T}(T)$.

\subsection{Examples from group theory}\label{sect:Examples Normalizer Circuits}

We now give examples of normalizer gates over conjugacy class and character hypergroups with the aim to illustrate our definitions and, furthermore, show how our results can be applied to define models of \emph{normalizer circuits over nonabelian groups}.

\subsubsection*{Example 1: Clifford and abelian-group normalizer circuits}

For an abelian group $G$, all conjugacy classes contain a single group element. Consequently, the class hypergroup $\overline{G}$ is always a \emph{group} and it is equal to $G$. In this scenario, our gate model coincides with the finite abelian-group normalizer-circuit model of earlier chapters, which contain numerous examples of normalizer gates, including the standard Clifford circuits for qubits/qudits \cite{Gottesman99_HeisenbergRepresentation_of_Q_Computers,Gottesman98Fault_Tolerant_QC_HigherDimensions} and circuits that contain abelian-group QFTs. We refer the reader to chapter  \ref{chapterC} for  examples of normalizer circuits that are efficiently classically simulable, and to chapter \ref{chapterB} for hard instances  that can implement famous quantum algorithms such as  Shor's \cite{Shor}.

\subsubsection*{Example 2: normalizer circuits over nonabelian groups}
\label{sect:Circuits nonabelian group}

We now apply our circuit formalism to introduce (new) models of normalizer gates over any finite nonabelian group $G$. For this, we  associate a Hilbert space $\mathcal{H}_{G}$ to $G$ with basis $\{\ket{g},g \in G\}$ and restrict the computation to act on its  (nontrivial) subspace  $\mathcal{I}_{\overline{G}}$ of  \emph{conjugation invariant} wavefunctions.\footnote{That is, wave functions $\psi(x)$ such that $\psi(x^g) = \psi(x)$ for all $x,g \in G$.}

As is well-known from representation theory \cite{Isaacs1994character}, the Dirac delta measures $\delta_{C_g}$ over conjugacy classes $C_g\in\overline{G}$ and the  character functions $\chi_\mu$ of the irreducible representations $\mu\in\mathrm{Irr}(G)$ form two dual orthonormal bases of $\mathcal{I}_\Conj{G}$. In our circuit model, recalling the definitions of class hypergroup $\overline{G}$ and character hypergroup $\widehat{G}$ (see section \ref{sect:Examples Hypergroups}), this means that $\mathcal{I}_{\overline{G}}$ can be viewed as the Hilbert space of the conjugacy class hypergroup $\mathcal{H}_{\overline{G}}$ with a \emph{conjugacy-class basis} $\mathcal{B}_{\overline{G}}=\{\ket{C_{g}},C_{g}\in G\}$ and a \emph{character basis} $\mathcal{B}_{\widehat{G}}=\{\ket{\mathcal{X}_{\mu}},\mathcal{X}_{\mu}\in\widehat{G}\}$ if we define these bases within $\mathcal{H}_G$ as the vectors
\begin{equation}\label{eq:Bases Nonabelian Group}
\ket{C_{g}}=\frac{1}{\sqrt{|C_{g}|}}\sum_{aga^{-1} \in C_{g}} \ket{aga^{-1}}\ \text{ and }\ \ket{\mathcal{X}_{{\mu}}}={\sqrt{\frac{d_\mu^2}{|G|}}}\sum_{g\in G}  \overline{\Hchi_{\mu}} (g)\ket{g}.
\end{equation}

With these identifications, we can now define a \emph{normalizer circuit over $G$} to be a normalizer circuit over the hypergroup $\overline{G}$: the latter acts on the conjugation invariant subspace, admits conjugacy class and character state inputs, and applies QFTs, group automorphisms, and quadratic phase functions associated to $\overline{G}$ and $\widehat{G}$. Furthermore, if we have a direct product $G=G_1\times \cdots \times G_m$, then $\overline{G}=\overline{G}_1\times \cdots \times \overline{G}_m$, $\widehat{G}=\widehat{G}_1\times \cdots \times \widehat{G}_m$ and $\mathcal{H}_{\Conj{G}}=\mathcal{H}_{\Conj{G}_1}\otimes \cdots \otimes \mathcal{H}_{\Conj{G}_m}$. In this setting, normalizer gates such as partial QFTs and entangling gates over different registers are allowed.

It is straightforward to check, using the identities $\ws{C_g}=|C_g|$, $\ws{\Hchi_\mu}=d_\mu^2$ and $\varw{\Conj{G}}=\varw{\widehat{G}}=|G|$ (section \ref{sect:Examples Hypergroups}), that the QFT defined with the bases from (\ref{eq:Bases Nonabelian Group}) is actually the identity map. Even so, the QFT performs a useful purpose in these circuits as it changes the basis used for subsequent gates, $\mathcal{T}(t+1)$. In particular, the QFT can change the final basis to the character basis, which means that the final measurement is performed in the character basis rather than the element basis.

As we show in appendix \ref{app:CC implementation details}, we can perform a final measurement in the character basis provided that we have an efficient QFT circuit for the group $G$. The same techniques also allow us to prepare initial states and perform all the gate types (Pauli gates, automorphisms, and quadratic phases) in the character basis efficiently.

Performing the gate types in the conjugacy class basis is straightforward if we make some modest assumptions about our ability to compute with conjugacy classes of the group. For example, we need a way to map an element $g \in G$ to a label of its conjugacy class $C_g$. These details are discussed in appendix~\ref{sect:class basis operations}, where we explain why these assumptions are easily satisfied for typical classes of groups.

\subsubsection*{Example 3: quaternionic circuits}\label{sect:Quaternionic circuits}

Lastly, we give concrete examples of normalizer gates over nonabelian groups  for systems of the form $Q_8^n$ where  $Q_8$ is the quaternion group with  presentation \begin{equation}
Q_8=\langle -1,i,j,k | (-1)^2=1,i^2=j^2=k^2=ijk=-1\rangle
\end{equation} Note that $Q_8$ is nonabelian and that, although it has  eight elements $\pm 1, \pm i, \pm j, \pm k$, it has only five conjugacy classes $\{1\}$, $\{-1\}$, $\{\pm i\}$, $\{\pm j\}$, $\{\pm k\}$. Hence, although the Hilbert space  $\mathcal{H}_{Q_8}^{\otimes n}=\{\ket{g}: g\in Q_8\}$ is $8^n$-dimensional, our quantum computation based on normalizer gates will never leave the $5^n$-dimensional  conjugation invariant subspace  $\Comp_{\Conj{Q_8}}^{\otimes n}$, which can be viewed as a system of 5-dimensional qudits. Using the group character table \cite{Atiyah61_Characters_Cohomology}, it is easy to write down the conjugacy class and character basis states of $\Comp_{\Conj{Q_8}}$:
\begin{itemize}
\item \textbf{Conjugacy-class states:}
\begin{equation*}
\ket{C_1}=\ket{1},\quad \ket{C_{-1}}=\ket{-1},\quad\ket{C_{i}}=\tfrac{\ket{i}+\ket{-i}}{\sqrt{2}},\quad\ket{C_{j}}=\tfrac{\ket{j}+\ket{-j}}{\sqrt{2}},\quad\ket{C_{k}}=\tfrac{\ket{k}+\ket{-k}}{\sqrt{2}}\vspace{-5pt}
\end{equation*}

\item \textbf{Character states:} 
\begin{align*}
\ket{\mathcal{X}_1} &= \tfrac{1}{\sqrt{8}} \left(\ket{C_1} +\ket{C_{-1}}+\sqrt{2}\ket{C_{i}}+\sqrt{2}\ket{C_{j}}+\sqrt{2}\ket{C_{k}}\right),\\
\ket{\mathcal{X}_i} &= \tfrac{1}{\sqrt{8}} \left(\ket{C_1} +\ket{C_{-1}}+\sqrt{2}\ket{C_{i}}-\sqrt{2}\ket{C_{j}}-\sqrt{2}\ket{C_{k}}\right),\\
\ket{\mathcal{X}_j} &= \tfrac{1}{\sqrt{8}} \left(\ket{C_1} +\ket{C_{-1}}-\sqrt{2}\ket{C_{i}}+\sqrt{2}\ket{C_{j}}-\sqrt{2}\ket{C_{k}}\right),\\
\ket{\mathcal{X}_k} &= \tfrac{1}{\sqrt{8}} \left(\ket{C_1} +\ket{C_{-1}}-\sqrt{2}\ket{C_{i}}-\sqrt{2}\ket{C_{j}}+\sqrt{2}\ket{C_{k}}\right),\\
\ket{\mathcal{X}_2} &= \tfrac{2}{\sqrt{8}} \left(\ket{C_1} -\ket{C_{-1}}\right),
\end{align*}
\end{itemize}
We now give a list of nontrivial normalizer gates (not intended to be exhaustive), which we obtain directly from the definitions in section \ref{sect: Circuit Model} applying basic properties of the quaternion group \cite{Humphrey96_Course_GroupTheory,Atiyah61_Characters_Cohomology}. For the sake of conciseness, the elementary group-theoretic derivations are omitted.
\begin{itemize}
\item \textbf{Quantum Fourier transform.} For one qudit, the QFT implements the change of basis between the conjugacy-class and character bases written above. For $n$-qudits, the total QFT implements this change of bases on all qudits. Partial QFTs, instead, implement the QFT on single qudits.
\vspace{5pt}
\item \textbf{Pauli gates:} $\PX{\overline{Q}_8}(-1)\ket{C_x}=\ket{-C_x},\quad \PZ{\overline{Q}_8}(\mathcal{X}_\ell)\ket{C_x}=\mathcal{X}_\ell(C_x)\ket{C_x}\quad$for $\ell = i, j, k$.
\vspace{5pt}
\item \textbf{Automorphism gates:} All automorphisms of the class-hypergroup can be obtained by composing  functions $\alpha_{xy}$ that swap pairs of conjugacy classes $C_x$, $C_y$ with $x, y\in \{i,j,k\}$. The corresponding \emph{swap gates} $U_{\alpha_{xy}}\ket{C_z}=\ket{\alpha_{xy}(C_z)}$ are instances of one-qudit quaternionic automorphism gates.

\item \textbf{Quadratic phase gate.} Next, we give examples of non-linear quadratic phase gates. For one qudit, quadratic phase gates $D_{\xi_{i}}$, $D_{\xi_{j}}$, $D_{\xi_{k}}$ defined as 
\begin{equation*}
D_{\xi_{x}}\ket{C_y}=\ket{C_y}, \textnormal{ if } y\in\langle x\rangle=\{\pm 1, \pm x\}  \quad\textnormal{and} \quad
D_{\xi_{x}}\ket{C_z}=i\ket{C_z}   \textnormal{ otherwise},
\end{equation*} 
  provide quaternionic analogues of the one-qubit $P=\mathrm{diag}(1,i)$ Clifford gate. 
  
  For two qudits, there is also a ``\emph{quaternionic controlled-Z gate}'' $D_\xi$, which implements a quadratic function  $\xi(C_x,C_y)=f_{C_x}(C_y)$, with $f_{C_x}$ being a linear character specified by the following rules: $f_{C_{\pm1}}=\mathcal{X}_1$ and $f_{C_{x}}=\mathcal{X}_x$ for $x=i,j,k$. We refer the reader to  appendix \ref{app:Quadratic functions} for a proof that the above functions are quadratic.
  
\end{itemize}

Most of the above gates act on a single copy of $\mathcal{H}_{Q_8}$ and, thus, cannot generate entanglement. Entangling normalizer gates can be found by considering two copies of $\mathcal{H}_{Q_8}$. The allowed  normalizer gates are now those associated to the group $Q_8\times Q_8$.

We give next three examples of two-qudit  automorphism gates $U_{\alpha_i}$, $U_{\alpha_j}$, $U_{\alpha_k}$, that can generate \textbf{quantum entanglement} and provide quaternionic analogues of the qubit CNOT \cite{Gottesman99_HeisenbergRepresentation_of_Q_Computers} and the qudit CSUM gates \cite{Gottesman98Fault_Tolerant_QC_HigherDimensions}. The three are defined as
\begin{equation}
U_{\alpha_x}\ket{C_1,C_2}=\ket{\alpha_x(C_1,C_2)}=\ket{C_1,f_x(C_1)C_2}
\end{equation}   where  $f_x(C_y)=C_1$ if $C_y$ is contained in the subgroup $\langle x \rangle = \{\pm 1,\pm x\}$ generated by $x$ and $f_{x}(C_y)=C_{-1}$ otherwise\footnote{The function   $f_x$ defines a group homomorphism from $\overline{Q}_8$ into its center $Z(Q_8)$. Using this fact, it is easy to show that $\alpha_x$ is a group automorphism.}. 
The action of any of these gates on  the product state
  $\ket{\mathcal{X}_1}\ket{C_1}$ generates an entangled  state; we show this explicitly for  $U_{\alpha_i}$:
\begin{equation*}
U_{\alpha_i}\ket{\mathcal{X}_1}\ket{C_1}= \tfrac{1}{2} \left( \tfrac{\ket{C_1} +\ket{C_{-1}}}{\sqrt{2}}+\ket{C_{i}}\right)\ket{C_1}+ \tfrac{1}{2}\left(\ket{C_{j}}+\ket{C_{k}}\right)\ket{C_{-1}}.
\end{equation*}
  Quaternionic quadratic-phase gates can also generate \textbf{highly entangled states}.
  For instance, the action of $D_\xi$ on a product state $\ket{\mathcal{X}_1}\ket{\mathcal{X}_1}$ creates an entangled bi-partite state with Schmidt rank 4, which is close to the maximal value of 5 achievable for a state in $\Comp_{\Conj{Q_8}}\otimes\Comp_{\Conj{Q_8}}$:
\begin{equation}\label{eq:Entangled State}
D_\xi\ket{\mathcal{X}_1}\ket{\mathcal{X}_1}= \tfrac{1}{4} \left( \left(\tfrac{\ket{C_1}+\ket{C_{-1}}}{\sqrt{2}}\right)\ket{\mathcal{X}_1}+\ket{C_{i}}\ket{\mathcal{X}_i}+\ket{C_{j}}\ket{\mathcal{X}_j}+\ket{C_{k}}\ket{\mathcal{X}_k}\right).
\end{equation}
A  quaternionic analogue of the ($d=4$) qudit \emph{cluster state} \cite{ZhouZengXuSun03} displaying  multi-partite entanglement can prepared by repeatedly  applying $D_\xi$ to all pairs of neighboring qudits on a lattice, chosen to be initially in the state $\ket{\mathcal{X}_1}$.

As this example shows, while normalizer circuits have fairly simple algebraic
properties, they can produce states that are very complicated and often highly
entangled. Thus, as in the abelian case, it comes as a surprise that these
circuits can often be classically simulated efficiently, as we will see in
section 6.

\section{A Hypergroup Stabilizer Formalism}\label{sect:Stabilizer Formailsm}

In this section we develop a stabilizer formalism based on  abelian hypergroups that extends Gottesman's PSF \cite{Gottesman_PhD_Thesis,Gottesman99_HeisenbergRepresentation_of_Q_Computers,Gottesman98Fault_Tolerant_QC_HigherDimensions} and the abelian group extension of chapters \ref{chapterF}-\ref{chapterB}. We  apply our formalism to the description of new types of  quantum many-body states,  including hypergroup coset states and  those that appear at intermediate steps of quantum computations by normalizer circuits over hypergroups.

This section is organized as follows. In section \ref{sect:Pauli Operators}, we introduce new  types of Pauli operators based on hypergroups that have richer properties than  those of chapters \ref{chapterF}-\ref{chapterB}: most remarkably, they can be  non-monomial and non-unitary matrices. In section (section \ref{sect:Stabilizer States over Hypergroups}) we show that commuting \emph{stabilizer hypergroups} built of the latter Paulis can be used to describe interesting families of quantum states, which we call \emph{hypergroup stabilizer states}, as well as track the dynamical evolution of hypergroup normalizer circuits (\textbf{theorem \ref{thm:Evolution of Stabilizer States}}). In section \ref{sect:Normal Form CSS hypergroup Stablizier States}, we give a  powerful \emph{normal form} (\textbf{theorem \ref{thm:Normal Form CSS States}}) for hypergroup stabilizer states that are CSS-like \cite{CalderbankShor_good_QEC_exist,Steane1996_Multiple_Particle_Interference_QuantumErrorCorrection}. The latter will be an invaluable tool in this chapter, which we later use to describe hypergroup coset states (equation \ref{eq:Hypergroup Coset State}, corollary \ref{corollary:Coset State Preparations}) and analyze the quantum algorithms of  section \ref{sect:Quantum Algorithms}. The techniques in this section will also be the basis of the classical simulation methods developed in section \ref{sect:Simulation}.

The fact that our hypergroup stabilizer formalism is based on non-monomial, non-unitary stabilizers introduces nontrivial technical difficulties that are discussed in detail in sections \ref{sect:Pauli Operators}-\ref{sect:Stabilizer States over Hypergroups}. The techniques we develop to cope with the issues are unique in the stabilizer formalism literature since both the original PSF and all of its previously known extensions \cite{Gottesman_PhD_Thesis,Gottesman99_HeisenbergRepresentation_of_Q_Computers,Gottesman98Fault_Tolerant_QC_HigherDimensions,VDNest_12_QFTs,BermejoVega_12_GKTheorem,BermejoLinVdN13_Infinite_Normalizers,BermejoLinVdN13_BlackBox_Normalizers, AaronsonGottesman04_Improved_Simul_stabilizer,dehaene_demoor_coefficients,dehaene_demoor_hostens,VdNest10_Classical_Simulation_GKT_SlightlyBeyond,deBeaudrap12_linearised_stabiliser_formalism,nest_MMS,NiBuerschaperVdNest14_XS_Stabilizers} were tailored to handle  unitary monomial stabilizer matrices. For this reason, we regard them as a main contribution of this chapter.

Like  in  previous section, we develop our stabilizer formalism over arbitrary abelian hypergroups. Throughout the section, we fix $\mathcal{T} = \mathcal{T}_1 \times \dots \times \mathcal{T}_m$ to be an arbitrary finite abelian hypergroup with Hilbert space $\mathcal{H}_{\mathcal{T}} \cong \Comp_{\mathcal{T}_1} \otimes \dots \otimes \Comp_{\mathcal{T}_m}$.

\subsection{Hypergroup Pauli operators}\label{sect:Pauli Operators}

We introduce generalized Pauli operators over $\mathcal{T}$ acting on $\Comp_{\mathcal{T}}$ with analogous properties to the qubit and abelian-group Pauli matrices (chapters \ref{chapterF}-\ref{chapterI}). In our formalism Pauli operators perform operations associated to the abelian hypergroups of section \ref{sect: Circuit Model}. First, \emph{Pauli operators over a hypergroup $\mathcal{T}$} (\ref{eq:Pauli operators DEFINITION})  implement multiplications by hypergroup characters as well as the hypergroup operation of $\mathcal{T}$. For the character group $\mathcal{T^*}$, \emph{Pauli operators over $\mathcal{T^*}$} are defined analogously (\ref{eq:Pauli operators DUAL BASIS}).  More precisely, for all $x,y \in\mathcal{T}$, $\mathcal{X}_\mu, \mathcal{X}_\nu\in\mathcal{T^*}$, we define
\begin{align}\label{eq:Pauli operators DEFINITION}
 \PZ{\mathcal{T}}(\mathcal{X}_\mu)\ket{x}:=\Hchi_\mu(x)\ket{x},\qquad\:  & \PX{\mathcal{T}}(x)\ket{y}:= \sum_{z\in\mathcal{T}} \sqrt{\tfrac{\w{y}}{\w{z}}}\,  n_{x,y}^{z} \ket{z},\\
\label{eq:Pauli operators DUAL BASIS}
 \PZ{\mathcal{T^*}}(x)\ket{\mathcal{X}_\mu}:= \Hchi_\mu(x) \ket{\mathcal{X}_\mu}, \qquad\: & \PX{\mathcal{T^*}}(\mathcal{X}_\mu)\ket{\mathcal{X}_\nu}:= \sum_{\Hchi_\gamma\in\mathcal{T^*}} \sqrt{\tfrac{\w{\nu}}{\w{\gamma}}}\,  m_{\mu,\nu}^{\gamma} \ket{\mathcal{X}_\gamma}.
\end{align}
With this definition, Pauli operators over a product $\mathcal{T}=\mathcal{T}_1\times\cdots \times \mathcal{T}_m$ inherit a tensor product form $\PX{\mathcal{T}}(a):=\PX{\mathcal{T}_1}(a_1)\otimes \cdots \otimes \PX{\mathcal{T}_m}(a_m)$ and $\PZ{\mathcal{T}}(\mathcal{X}_\mu)=\PZ{\mathcal{T}_1}(\mathcal{X}_{\mu_1})\otimes \cdots \otimes \PZ{\mathcal{T}_m}(\mathcal{X}_{\mu_m})$. Any operator  that can be written as a product of operators of type $\PX{\mathcal{T}}(a)$ and $\PZ{\mathcal{T}}(\chi_\mu)$ will be called a generalized \emph{hypergroup Pauli operator}.

We  state a few \emph{main properties} of hypergroup Pauli operators. First, it follows from (\ref{eq:Pauli operators DEFINITION})  that the  Pauli operators $\PZ{\mathcal{T}}(\mathcal{X}_\mu)$ commute and form a hypergroup isomorphic to $\mathcal{T}^*$: 
\begin{equation}\label{eq:Pauli Z Properties}
\PZ{\mathcal{T}}(\mathcal{X}_\mu) \PZ{\mathcal{T}}(\mathcal{X}_\nu)= \PZ{\mathcal{T}}(\mathcal{X}_\nu) \PZ{\mathcal{T}}(\mathcal{X}_\mu)=\sum_{\gamma} m_{\mu\nu}^{\gamma} \PZ{\mathcal{T}}(\mathcal{X}_\gamma),  \quad  \textnormal{for any $\mathcal{X}_\mu,\mathcal{X}_\nu\in \mathcal{T}^*$}.
\end{equation}
Although it is not obvious from the definitions, we show later (theorem \ref{thm:Evolution of Stabilizer States}, eq.\ \ref{eq:QFTs are Clifford}) that the X-Paulis $\PX{\mathcal{T}}(a)$ are also pair-wise commuting normal operators, which are diagonal in the the character basis  $\{ \ket{\mathcal{X}_\mu}, \mathcal{X}_\mu \in \mathcal{T}^* \}$, and form a faithful representation of the conjugacy-class hypergroup. Precisely, for any $a,b\in \mathcal{T}$, $\mathcal{X}_\mu\in\mathcal{T}^*$, we have
\begin{equation}\label{eq:Pauli X Properties}
\PX{\mathcal{T}}(a)\ket{ \mathcal{X}_\mu }=\mathcal{X}_{\mu}(a)\ket{\mathcal{X}_\mu}, \qquad \PX{\mathcal{T}}(a)\PX{\mathcal{T}}(b)=\PX{\mathcal{T}}(b)\PX{\mathcal{T}}(a)=\sum_{c} n_{ab}^c \PX{\mathcal{T}}(c).
\end{equation}

The following lemma characterizes when Pauli operators of different type commute.

\begin{lemma}[\textbf{Commutativity}]\label{lemma:Commutativity of Paulis}
The  operators $\PX{\mathcal{T}}(a)$, $\PZ{\mathcal{T}}(\mathcal{X}_\mu)$ commute iff $\mathcal{X}_\mu(a)=1$.
\end{lemma}
\begin{proof}
For any state $\ket{b}$ in the basis $\mathcal{B}_\mathcal{T}$, compare $\PZ{\mathcal{T}}(\mathcal{X}_\mu)\PX{\mathcal{T}}(a)\ket{b}=\sum_{c}\sqrt{\tfrac{\w{b}}{\w{c}}}n_{ab}^c \mathcal{X}_\mu(c)\ket{c}$ with $ \PX{\mathcal{T}}(a)\PZ{\mathcal{T}}(\mathcal{X}_\mu)\ket{b}=\sum_{c}\sqrt{\tfrac{\w{b}}{\w{c}}}n_{ab}^c \mathcal{X}_\mu(b)\ket{c}$. Then,  the ``if'' holds  because  $\mathcal{X}_\mu(a)=1$  implies  $\mathcal{X}_\mu(b)=\mathcal{X}_\mu(c)$ for any $b\in\mathcal{T}$, $c\in ab$ \cite[proposition 2.4.15]{BloomHeyer95_Harmonic_analysis}, and the two expressions coincide. Conversely, letting $b=e$ in these two expressions yields  the ``only if''.
\end{proof}
Note that the above properties are always fulfilled by qubit \cite{Gottesman99_HeisenbergRepresentation_of_Q_Computers}, qudit \cite{Gottesman98Fault_Tolerant_QC_HigherDimensions} and group Pauli operators (chapter \ref{chapterF}). In this sense, hypergroup Pauli operators provide a generalization of these concepts. Yet, as we will see in the next section, there are some remarkable properties of group Pauli operators that are fully shared by their hypergroup counterparts.

\subsection{Hypergroup stabilizer states}\label{sect:Stabilizer States over Hypergroups}

We will now extend the notion of stabilizer state from groups to hypergroups.

\begin{definition}[\textbf{Stabilizer hypergroup and stabilizer state}]\label{def:Stabilizer Hypergroup} \label{def:Stabilizer State}

A \emph{stabilizer hypergroup} $\mathcal{S}^\lambda$ is a hypergroup of commuting  hypergroup Pauli operators over $\mathcal{T}$ (\ref{eq:Pauli operators DEFINITION}-\ref{eq:Pauli operators DUAL BASIS}) with an associated \emph{stabilizer function} $\lambda$ that selects an eigenvalue $\lambda(U)$ for every  $U$ in $\mathcal{S}^\lambda$.

Let $\{\mathcal{S}_i^{\lambda_i}\}_{i=1}^{r}$ be a \emph{collection} of $r$ mutually commuting stabilizer hypergroups. Then, a  quantum state $\ket{\psi}$ is called a \emph{stabilizer state} stabilized by   $\{\mathcal{S}_i^{\lambda_i}\}_{i=1}^{r}$ if, up to normalization and global phases, it is the unique non-zero solution to the system of spectral equations
\begin{equation}\label{eq:Stabilizer State Condition}
U\ket{\psi}= \lambda_i(U) \ket{\psi},\quad\textnormal{for all }U\in\mathcal{S}_i^{\lambda_i},\, i=1,\ldots, r.
\end{equation} 
By definition, every function $\lambda_i$ is further constrained to be a \emph{character} of $\mathcal{S}_i^{\lambda_i}$. This is necessary for the system  (\ref{eq:Stabilizer State Condition}) to admit nontrivial solutions.\footnote{Let $UV=\sum_{W}s_{UV}^{W} W$ for any $U,V\in\mathcal{S}_i^{\lambda_i}$ with   structure constants  $s_{U,V}^{W}$. Then (\ref{eq:Stabilizer State Condition}) implies  $UV\ket{\psi}=\lambda(U)\lambda(V)\ket{\psi}=\sum_{W}s_{UV}^{W} \lambda(W)\ket{\psi}$. Since $\ket{\psi}\neq 0$, every non-zero $\lambda$ must be a character (by definition).}
\end{definition}
In this work we focus on \emph{pure} stabilizer states and do not discuss mixed ones. 

Though, in the PSF and in the group GSF (chapters \ref{chapterF}-\ref{chapterB}), stabilizer groups can always be described efficiently in terms of generators or matrix representations of morphisms, the existence of efficient descriptions is hard to prove  in the hypergroup setting. The stabilizer hypergroups $\mathcal{S}^\lambda$ in this chapter (see section \ref{sect:Normal Form CSS hypergroup Stablizier States}) have efficient  $\polylog{|\mathcal{T}|}$-size classical descriptions (where $|\mathcal{T}|$ is  the  dimension of the  Hilbert space $\mathcal{H_T}$) if poly-size descriptions for the subhypergroups of $\mathcal{T}$ and $\mathcal{T^*}$ are promised to exist. We highlight that this latter condition is fulfilled for many hypergroups of interest, including  conjugacy class and character hypergroups, and anyonic fusion rule theories\footnote{All examples mentioned belong to a class of so-called \emph{resonance} hypergroups $\mathcal{T}$ that  have integral weights, $\w{a}$, and fulfill a Lagrange theorem \cite{Wildberger_Lagrange}, which says that $\varpi_\mathcal{N}$ for any subhypergroup $\mathcal{N\subset T}$ always divides $\varpi_\mathcal{T}$. As in the finite group case (lemma \ref{lemma:SamplingGeneratingSets}), this implies that a random set $\{a_1,\ldots,a_m\}\subset\mathcal{T}$ with $m\in \Theta(\log \varpi_{\mathcal{T}})$ generates $\mathcal{T}$ via hyperoperations with  high probability $\Omega(1-c^m)$ form some constant $c\in(0,1)$.}. The stabilizer hypergroups obtained from all those cases provide a powerful means to describe quantum many-body states that are uniquely defined via an equation of the form (\ref{eq:Stabilizer State Condition}).

The definition of stabilizer hypergroup and state generalizes the standard notions used in the PSF and the Group Stabilizer Formalism. At the same time, our hypergroup Paulis  also have novel interesting  properties that are  explained next.

\begin{comparison}[\textbf{Relationship to group stabilizer states.}]
All qubit, qudit, and abelian-group  stabilizer states (chapter \ref{chapterF}) are instances of hypergroup stabilizer states over an abelian hypergroup $\overline{G}$, where $G$ chosen to be a group of the form $\Integers_2^m$, $\Integers_{d}^m$, and $\Integers_{N_1}\times \cdots\times \Integers_{N_m}$ (respectively) with $\lambda(U)=+1$. Hypergroup Pauli operators and stabilizer hypergroups over $\overline{G}$ (\ref{eq:Pauli operators DEFINITION}) become standard Pauli operators and stabilizer groups over $G$ (in the notation of chapter~\ref{chapterF}).
\end{comparison}

\begin{comparison}[\textbf{Subtleties of hypergroup Pauli operators}] Interestingly, in spite of having some  Pauli-like mathematical properties (section \ref{sect:Pauli Operators}),  hypergroup Pauli operators are \emph{not as simple} as group Pauli operators (chapter \ref{chapterF}): namely, they are not necessarily \emph{unitary} and no longer \emph{monomial} (1-sparse) matrices because the hyperoperations in (\ref{eq:Pauli operators DEFINITION})  are non-invertible and return multiple outcomes. The absence of these two properties is reflected in definition \ref{def:Stabilizer Hypergroup}. In our formalism,  stabilizer \emph{hypergroups} $\mathcal{S}^\lambda$ of commuting Paulis do not  necessarily form \emph{groups}. Stabilizer states are no longer restricted to be $+1$-eigenstates of hypergroup Pauli operators $U\in\mathcal{S}^\lambda$, as in chapter \ref{chapterF}, for Pauli operators can  now have zero eigenvalues\footnote{This follows from (\ref{eq:Pauli operators DEFINITION}) because nonabelian group  and hypergroup characters can take zero values \cite{Wildberger97_Duality_Hypergroups,Amini2011fourier}.}. This allows us to include more states in the formalism.
\end{comparison}
\begin{comparison}[\textbf{Non-commutativity up to a phase}]
Hypergroup Paulis do not satisfy an identity of the form $\PZ{\mathcal{T}}(\mathcal{X}_\mu)\PX{\mathcal{T}}(a)= \mathcal{X}_\mu(a) \PX{\mathcal{T}}(a)\PZ{\mathcal{T}}(\mathcal{X}_\mu)$ in general, although this is the case when $\mathcal{T}$ is a group (chapter \ref{chapterF}). In our setting, this is fulfilled only in some special cases, e.g., when either $a$ or $\mathcal{X}_\mu$ is an invertible hypergroup element  (theorem \ref{thm:Evolution of Stabilizer States}, eq.\ \ref{eq:Pauli gates are Clifford}) or when $\mathcal{X}_\mu(a){=}1$ (where $\PZ{\mathcal{T}}(\mathcal{X}_\mu)$ and $\PX{\mathcal{T}}(a)$ \emph{commute} due to lemma \ref{lemma:Commutativity of Paulis}). 
\end{comparison}
\begin{comparison}[\textbf{Multiple stabilizer hypergroups}] The reason why we use multiple stabilizer hypergroups $\{\mathcal{S}_i^{\lambda_i}\}$ instead of merging them into a single commutative algebra is that finding a basis with hypergroup structure for the latter object is not a simple  task\footnote{We have not investigated this problem nor whether a hypergroup basis can always be found.}. On the other hand, we can easily keep track and exploit the available  hypergroup structures by simply storing a poly-size list of pairwise commuting stabilizer hypergroups.\footnote{Throughout the chapter $r$ will always be poly-sized and all  examples  we give (section \ref{sect:Normal Form CSS hypergroup Stablizier States}) have $r\leq 2$.}
\end{comparison}
The next two results imply that any intermediate quantum state of a hypergroup normalizer circuit (section \ref{sect: Circuit Model}) computation is a hypergroup stabilizer state and, hence, can be characterized concisely as a joint-eigenstate of some commuting hypergroup Pauli operators. This observation motivates our further development of these concepts.

\begin{claim}[\textbf{Standard basis states}]\label{lemma:Standard Basis States are Stabilizer States}  Conjugacy-class and character states (\ref{eq:Quantum Fourier Transform over Hypergroup T}) are instances of hypergroup stabilizer states stabilized by \emph{single} stabilizer hypergroups.
\end{claim} 

\begin{theorem}[\textbf{Evolution of stabilizer states}]\label{thm:Evolution of Stabilizer States}
Normalizer gates map hypergroup Pauli operators to new hypergroup Pauli operators under conjugation and, therefore, transform hypergroup stabilizer states into stabilizer states. It follows from this and the previous claim that the quantum state of a normalizer circuit is always a stabilizer state. 
\end{theorem}
Note that theorem \ref{thm:Evolution of Stabilizer States} extends Van den Nest's theorem \ref{thm_G_circuit_fundamental} for abelian group stabilizer states.

In order to prove claim \ref{lemma:Standard Basis States are Stabilizer States}, theorem \ref{thm:Evolution of Stabilizer States}, and  many of the main results  in the next sections, we will develop a new kind of hypergroup  stabilizer formalism techniques that can cope with the \emph{\textbf{non-monomiality}} and the \emph{\textbf{non-unitarity}} of hypergroup Pauli operators. A central part of the chapter will be dedicated exclusively to this end. We stress the necessity to develop such techniques since currently available stabilizer-formalism methods---including the PSF \cite{Gottesman_PhD_Thesis,Gottesman99_HeisenbergRepresentation_of_Q_Computers,Gottesman98Fault_Tolerant_QC_HigherDimensions,AaronsonGottesman04_Improved_Simul_stabilizer,dehaene_demoor_coefficients,dehaene_demoor_hostens,VdNest10_Classical_Simulation_GKT_SlightlyBeyond,deBeaudrap12_linearised_stabiliser_formalism}, the Group Stabilizer Formalism (chapters \ref{chapterF}-\ref{chapterB}),  the general  Monomial Stabilizer Formalism of Van den Nest \cite{nest_MMS}, and the recent  XS stabilizer formalism \cite{NiBuerschaperVdNest14_XS_Stabilizers}---can not be applied in our setting as they \emph{critically exploit  the monomiality/unitarity of stabilizer operators} for central tasks such as simulating Clifford/Normalizer operations, analyzing stabilizer state and code properties (e.g., code dimension,  code support), and giving normal forms for stabilizer states\footnote{The role of monomiality and unitarity in the PSF has been extensively discussed in \cite{nest_MMS}.}. The lack of these beneficial properties requires a change of paradigm in our setting.

To prove  claim \ref{lemma:Standard Basis States are Stabilizer States}, we  show a stronger result (\textbf{theorem \ref{thm:Normal Form CSS States}} below), which gives a \emph{normal form} for  hypergroup stabilizer states; we outline the proof of theorem \ref{thm:Evolution of Stabilizer States} below, with details referred to   appendix \ref{app:A}.

\begin{proof}[Proof of theorem \ref{thm:Evolution of Stabilizer States}, part I] We show that normalizer gates  transform X- and Z-type Pauli operators  over $\mathcal{T}$  into new  Pauli operators (which may involve products of X, Z Paulis) under conjugation. This result extends to arbitrary products of these operators. 

Specifically, for any invertible element $s\in\mathcal{T}$, invertible character $\mathcal{X}_\varsigma\in\mathcal{T^*}$, automorphism $\alpha$, quadratic function $\xi$, we can calculate this action for the normalizer gates $Z_\mathcal{T}(\mathcal{X}_\mu)X_\mathcal{T}(a)$, $U_\alpha$, $D_\xi$, and the hypergroup QFT:
\begin{align}
\PX{\mathcal{T}}(a)& \xrightarrow{\:Z_\mathcal{T}(\mathcal{X}_\varsigma)X_\mathcal{T}(s)\:} \mathcal{X}_{{\varsigma}}(a) \PX{\mathcal{T}}(a),& \PZ{\mathcal{T}}(\mathcal{X_\mu}) &\xrightarrow{\:Z_\mathcal{T}(\mathcal{X}_\varsigma)X_\mathcal{T}(s)\:} \mathcal{X}_{{\mu}}(\overline{s})\PZ{\mathcal{T}}(\mathcal{X_\mu}),
\label{eq:Pauli gates are Clifford}\\
\PX{\mathcal{T}}(a)&\xrightarrow{\:U_\alpha\:}  \PX{\mathcal{T}}(\alpha(a)), &  \PZ{\mathcal{T}}(\mathcal{X_\mu})&\xrightarrow{\:U_\alpha\:}  \PZ{\mathcal{T}}(\mathcal{\PX{\alpha^{-*}(\mu)}} ), \label{eq:Automorphism gates are Clifford} \\
\PX{\mathcal{T}}(a)&\xrightarrow{\:D_\xi\:}  \xi(a) \PX{\mathcal{T}}(a) \PZ{\mathcal{T}}(\beta{(a)}), &   \PZ{\mathcal{T}}(\mathcal{\PX{\mu}})&\xrightarrow{\:D_\xi\:}   \PZ{\mathcal{T}}(\mathcal{\PX{\mu}}), \label{eq:Quadratic Phase gates are Clifford}\\
\PX{\mathcal{T}}(a)&\xrightarrow{\:\mathrm{QFT}\:}  \PZ{\mathcal{T}^*}(a), &  \PZ{\mathcal{T}}(\mathcal{X_\mu})&\xrightarrow{\:\mathrm{QFT}\:} \PX{\mathcal{T}^*}(\overline{\Hchi_\mu}).\label{eq:QFTs are Clifford}
\end{align}
When $\Comp_{\mathcal{T}}=\Comp_{\mathcal{T}_1}\otimes\cdots \otimes \Comp_{\mathcal{T}_m}$, the partial QFT over $\mathcal{T}_i$ simply transforms the $i$th tensor factor of the  Pauli operators according to  (\ref{eq:QFTs are Clifford}). In (\ref{eq:Quadratic Phase gates are Clifford}),  $\beta$ is a homomorphism from $\mathcal{T}$ to the subhypergroup of invertible characters $\mathcal{T}_\mathrm{inv}^*$ that depends on $\xi$; in (\ref{eq:Automorphism gates are Clifford}), $\alpha^{-*}$ is the inverse of the \emph{dual automorphism $\alpha^*$} \cite{McMullen79_Algebraic_Theory_Hypergroups}:

\begin{definition}[\textbf{Dual automorphism \cite{McMullen79_Algebraic_Theory_Hypergroups}}]  \label{eq:Dual Automorphism DEFINITION}
The  \emph{dual automorphism} of $\alpha$, denoted  $\alpha^*$, is the automorphism of $\mathcal{T}^*$  that takes $\mathcal{X}_\mu$ to the  character  $\mathcal{X}_{\alpha^{*}(\mu)} := \mathcal{X}_{\mu} \circ \alpha$  for fixed $\mathcal{X}_{\mu}$.\footnote{This is a morphism because $\mathcal{X}_{\alpha^*(\mu)}\mathcal{X}_{\alpha^*(\nu)}{=}\left(\mathcal{X}_{\mu}\circ\alpha\right)\left(\mathcal{X}_{\nu}\circ\alpha\right){=}\sum_{\gamma}m_{\mu\nu}^\gamma \left(\mathcal{X}_{\gamma}\circ \alpha\right){=} \sum_{\gamma}m_{\mu\nu}^\gamma\,\mathcal{X}_{\alpha^*(\gamma)}$.}
\end{definition}
Proving (\ref{eq:Pauli gates are Clifford}-\ref{eq:QFTs are Clifford}) involves  bulky yet beautifully structured hypergroup calculations that are carried out in appendix \ref{app:A}.
\end{proof}
It is worth noting  that  normalizer gates transform Pauli operators over $\mathcal{T}$ into Pauli operators over  $\mathcal{T}$ if they are not QFTs and into Pauli operators over  $\mathcal{T}^*$ otherwise\footnote{Recall from section \ref{sect: Circuit Model} that the hypergroup label $\mathcal{T}$ keeps track of the basis in which basis (\ref{eq:Quantum Fourier Transform over Hypergroup T}) measurements are performed and indicates which Pauli operators are diagonal in each basis.}.

\subsection{A normal form for stabilizer states and examples}
\label{sect:Normal Form CSS hypergroup Stablizier States}
In this final subsection we give examples and a normal form for a class of hypergroup states that generalize the well-known notion of Calderbank-Shor-Steane (CSS) stabilizer states \cite{CalderbankShor_good_QEC_exist,Steane1996_Multiple_Particle_Interference_QuantumErrorCorrection,Calderbank97_QEC_Orthogonal_Geometry}:

\begin{definition}[\textbf{CSS stabilizer state}]\label{def:CSS state}
A hypergroup stabilizer state $\ket{\psi}$ over $\mathcal{T}$  is said of CSS type if it is uniquely stabilized by two mutually commuting stabilizer hypergroups $\mathcal{S}_Z^{\lambda_z}$, $\mathcal{S}_X^{\lambda_x}$  consisting only of Z and X Pauli operators respectively.
\end{definition}
The standard definition of CSS state is recovered by setting $\mathcal{T}=\Integers_2^{n}$ (chapter \ref{chapterC}). For the sake of brevity, we assume  $\mathcal{S}_Z^{\lambda_z}$ and  $\mathcal{S}_X^{\lambda_x}$ to be maximal mutually commuting hypergroups in this section\footnote{This maximality assumption is not necessary in our derivation but it shortens the proofs.}. With these requirements, lemma \ref{lemma:Commutativity of Paulis}  imposes that $\mathcal{S}_Z^{\lambda_z}$,  $\mathcal{S}_X^{\lambda_x}$ must be of the  form 
\begin{align}
\mathcal{S}_X^{\lambda_x}&=\{\PX{\mathcal{T}}(a), a\in \mathcal{N}\}, & \lambda_x(\PX{\mathcal{T}}(a))&=\mathcal{X}_\varsigma({a}) \notag \\ \mathcal{S}_Z^{\lambda_z}&=\{\PZ{\mathcal{T}}(\mathcal{X}_\mu), \mathcal{X}_\mu\in \mathcal{N}^\perp\}, & \lambda_z(\PZ{\mathcal{T}}(\mathcal{X}_\mu))&= \mathcal{X}_\mu(s),\label{eq:Stabilizer Hypergroup CSS State}
\end{align}
where $s\in\mathcal{T}$, $\mathcal{X}_\varsigma\in\mathcal{T}^*$, $\mathcal{N} \le \mathcal{T}$ is a subhypergroup, and $\mathcal{N}^\perp$ is the \emph{annihilator} of $\mathcal{N}$ (\ref{eq:annhilator}).

We are now ready to prove the main technical result of this section, theorem \ref{thm:Normal Form CSS States}, which  characterizes the set of hypergroup stabilizer states of CSS type and leads to specific examples.

\begin{theorem}[\textbf{Normal forms for CSS-type hypergroup stabilizer states}]\label{thm:Normal Form CSS States}
$ $\\
\textbf{(a)} The quantum states stabilized by  $\{\mathcal{S}_X^{\lambda_x},\mathcal{S}_Z^{\lambda_z}\}$ from (\ref{eq:Stabilizer Hypergroup CSS State}) are those in the subspace
\[\mathcal{V_S}:=\mathrm{span}\left\{\sum_{x\in s\mathcal{N}}\psi_y(x)\ket
x,\,y{\,\in\,} s\mathcal{N}\right\}=\mathrm{span}\left\{\sum_{\mathcal{X}_\mu\in \mathcal{X_\varsigma N^\perp}}\widehat{\psi}_\nu(\mu)\ket{\Hchi_\mu},\,\mathcal{X}_\nu{\,\in\,}\mathcal{X_\varsigma N^\perp}\right\},\]
where $\psi_y$ and $\widehat{\psi}_\nu$ are functions supported on $s\mathcal{N}$ and $\mathcal{X_\varsigma N^\perp}$, respectively, and defined by
\begin{equation}\label{eq:Consistency CSS}
\psi_y(x):= \sqrt{\w{x}} \left(\sum_{\substack{b}\in\mathcal{N} }n_{ x, {\overline{y}}}^{ b}\mathcal{X}_{\overline{\varsigma}}(b)\right)\ \text{ and }\ \ \widehat{\psi}_\nu(\mu):=\sqrt{\w{\mathcal{X}_\mu}} \left(\sum_{\substack{\beta}\in\mathcal{N}^\perp }m_{ \mu, {\overline{\nu}}}^{ \beta}\mathcal{X}_{{\beta}}(s)\right).
\end{equation}
A state $\ket{\psi}$ is \emph{uniquely} stabilized by $\{\mathcal{S}_X^{\lambda_x},\mathcal{S}_Z^{\lambda_z}\}$ iff $\mathrm{dim}(\mathcal{V_S})=1$. \textbf{(b)} Furthermore, if either $s$ or $\mathcal{X}_\varsigma$ is invertible, then $\{\mathcal{S}_X^{\lambda_x},\mathcal{S}_Z^{\lambda_z}\}$  stabilizes a \emph{unique} state  of form\footnote{Cf.\ section \ref{sect:Subhypergroups, Quotients} for definitions of  $\w{\Hchi_{\overline{\varsigma}} \mathcal{N}^\perp}, \ws{s \mathcal{N}},\varpi_{s\mathcal{N}},\varpi_{\varsigma\mathcal{N}^\perp}$.} \begin{align}\label{eq:Normal Form CSS state}
\ket{\psi}&=\sum_{x\in s \mathcal{N}} \sqrt{\frac{\w{x} \w{\Hchi_{\overline{\varsigma}} \mathcal{N}^\perp} }{\varpi_{s\mathcal{N}}}}\, \mathcal{X}_{\overline{\varsigma}}(x) \ket{x}, \qquad \Fourier{\mathcal{T}}\ket{\psi} = \sum_{\mathcal{X}_\mu \in \mathcal{X}_\varsigma \mathcal{N}^\perp} \sqrt{\frac{\w{\mathcal{X}_\mu} \ws{s\mathcal{N}}}{\varpi_{\varsigma \mathcal{N}^\perp}}}\, \mathcal{X}_\mu(s) \ket{\mathcal{X}_\mu}.
\end{align}
\end{theorem}
Theorem \ref{thm:Normal Form CSS States} is proven at the end of the section after mentioning a few main applications.

We highlight that, despite our focus on stabilizer \emph{states}, theorem \ref{thm:Normal Form CSS States} can be easily extended to study hypergroup stabilizer \emph{codes}. For instance, in case \textbf{(b)}, we could choose a smaller stabilizer hypergroup $\mathcal{S}_X^{\lambda_x}=\{\PX{\mathcal{T}}(a), a\in \mathcal{K}\}$ with $\mathcal{K\subsetneq N}$ to obtain a stabilizer code $\mathcal{V_S}$, whose dimension is easy to compute with our techniques\footnote{With minor modifications of our proof of theorem \ref{thm:Normal Form CSS States}, we get that the dimension is the number of cosets of $\mathcal{K}$ inside $s\mathcal{N}$, if $\mathcal{X_\varsigma}$ is invertible, and the number of cosets of $\mathcal{N}^\perp$ inside $\mathcal{X_\varsigma K^\perp}$, if $s$ is invertible.}. Similarly, one could shrink $\mathcal{S}_Z^{\lambda_z}$ or both stabilizer hypergroups at the same time. 

\paragraph{Open question}  The hypergroup CSS code construction we outlined clearly mimics the standard qubit one \cite{CalderbankShor_good_QEC_exist,Steane1996_Multiple_Particle_Interference_QuantumErrorCorrection,Calderbank97_QEC_Orthogonal_Geometry,nielsen_chuang}. Interestingly, there could also be  hypergroup CSS codes with no qubit/qudit analogue if there exist groups (or even hypergroups $\mathcal{T}$ that do not arise from groups) for which $\mathcal{V_S}$ in theorem \ref{thm:Normal Form CSS States}\textbf{(a)} can be degenerate. Though this is never the case for abelian groups because the Pauli operators in (\ref{eq:Stabilizer Hypergroup CSS State}) generate a maximal linearly independent set of commuting operators (with  rank one common eigenprojectors), it can happen in our setting.\footnote{Products of hypergroup Pauli operators  in (\ref{eq:Stabilizer Hypergroup CSS State})  can be linearly dependent (choose $\mathcal{N}= \{\pm1,C_i\}$ for the quaternion group, section \ref{sect: Circuit Model}) and their cardinality $|\mathcal{N}||\mathcal{N^\perp}|$ may not match the dimension of $\mathcal{H}_{\overline{G}}$ \cite{Ichihara_thesis_Hypergroup_Extensions}.} We leave open the question of whether such codes exist. 

\subsubsection*{Examples and applications of theorem \ref{thm:Normal Form CSS States}}\label{sect:Examples Normal Form CSS States}

Theorem \ref{thm:Normal Form CSS States} is an important technical contribution of this work that will be used three times within the scope of the chapter: firstly, in the examples below,  to construct efficient classical descriptions for new kinds of complex many body states; secondly, in section \ref{sect:Simulation}, to devise classical  algorithms for simulating hypergroup normalizer circuits (theorems \ref{thm:simulation}); and finally, in section \ref{sect:Quantum Algorithms}, in the development of an efficient quantum algorithm for hidden subhypergroup problems (theorems \ref{thm:easy2}--\ref{thm:easy3}). 

We now give examples of CSS hypergroup stabilizer states  of the simpler form (\ref{eq:Normal Form CSS state}).

\paragraph{Example 1: standard basis states}

We show now that conjugacy-class states and character states (\ref{eq:Quantum Fourier Transform over Hypergroup T}) are instances of hypergroup CSS states (of type (b)), as anticipated above (claim 1). Consider, first, an arbitrary $\ket{a}$ with $a\in \mathcal{T}$. Eq.\ (\ref{eq:Pauli Z Properties}) implies that $\ket{\psi}$ is stabilized by    $\mathcal{S}_Z^{\lambda_z}=\{Z_\mathcal{T}(\mathcal{X}_\mu), {X}_\mu\in\mathcal{T^*}\}$ with maximal $\mathcal{N^\perp}=\mathcal{T^*}$ and $\lambda_z(Z_\mathcal{T}(\mathcal{X}_\mu))=\mathcal{X}_\mu(a)$. Letting $\mathcal{S}_X^{\lambda_x}$, $\mathcal{N}$, and $\lambda_x$  be trivial, the state can written in the form given in theorem \ref{thm:Normal Form CSS States}.{(b)} (note that $\lambda_x$ is an invertible character) and, hence, $\ket{a}$ is a uniquely stabilized  CSS state. An almost identical argument, using  (\ref{eq:Pauli X Properties}) instead, shows that any character state $\ket{\mathcal{X_\nu}}$ is uniquely stabilized by  $\mathcal{S}_X^{\lambda_x} = \{X_\mathcal{T}(a),a\in\mathcal{T}\}$ with $\lambda_x(X_\mathcal{T}(a))=\mathcal{X}_\nu(a)$. (Note that in both cases equation (\ref{eq:Normal Form CSS state}) reproduces (\ref{eq:Quantum Fourier Transform over Hypergroup T}) consistently.)

\paragraph{Example 2: hypergroup coset states}

The states in example 1 are always product states. Yet theorem \ref{thm:Normal Form CSS States} also implies that highly entangled states such as the abelian group coset states that appear in the abelian HSP quantum algorithms \cite{childs_vandam_10_qu_algorithms_algebraic_problems},
\begin{equation}\label{eq:Abelian Group Coset State}
\ket{x+H}=\sum_{h\in H} \frac{1}{\sqrt{|H|}}\ket{x+h}, \quad \textnormal{$H$ is a subgroup of a finite abelian group $G$},
\end{equation}
as well as the abelian hypergroup coset states used in the quantum algorithm \cite{Amini_hiddensub-hypergroup},
\begin{equation}\label{eq:Hypergroup Coset State}
\ket{s\mathcal{N}}=\sum_{x\in s \mathcal{N}} \sqrt{\frac{\w{x}}{\varpi_{s\mathcal{N}}}}\, \ket{x}, \quad \textnormal{$\mathcal{N}$ is a subhypergroup of a finite abelian hypergroup $\mathcal{T}$},
\end{equation}
are all instances of CSS hypergroup states of type (a) with trivial $\mathcal{X}_\varsigma$, uniquely stabilized by $\mathcal{S}_X^{\lambda_x}=\{X_\mathcal{T}(a),a\in\mathcal{N}\}$,  $\mathcal{S}_Z^{\lambda_z}=\{Z_\mathcal{T}(\mathcal{X}_\mu),\mathcal{X}_\mu\in\mathcal{N}^\perp\}$, $\lambda_z(Z_\mathcal{T}(\mathcal{X}_\mu))= \mathcal{X}_\mu(s)$, and trivial, invertible $\lambda_x$.

In the special case of  abelian group coset states  (\ref{eq:Abelian Group Coset State}), theorem \ref{thm:Normal Form CSS States} recovers a result by Van den Nest \cite{VDNest_12_QFTs} who identified the latter with generalized abelian group stabilizer states (see also theorem \ref{thm Normal form of an stabilizer state}). Theorem \ref{thm:Normal Form CSS States} extends the latter result demonstrating the existence of complex many-body states---hypergroup coset states (\ref{eq:Hypergroup Coset State}) and more (\ref{eq:Consistency CSS}-\ref{eq:Normal Form CSS state})---that can be  described within the present Hypergroup Stabilizer Formalism but not within the standard PSF nor the GSF (chapters \ref{chapterF}-\ref{chapterB}), or even (to the  best of our knowledge\footnote{The authors are not aware of any method to express hypergroup coset states in terms of monomial unitary stabilizers as in \cite{Gottesman_PhD_Thesis,Gottesman99_HeisenbergRepresentation_of_Q_Computers,Gottesman98Fault_Tolerant_QC_HigherDimensions,VDNest_12_QFTs,BermejoVega_12_GKTheorem,nest_MMS,NiBuerschaperVdNest14_XS_Stabilizers}. We doubt such a  description could exist and, at the same time, be easy to track under the action of hypergroup normalizer circuits as in theorem \ref{thm:Evolution of Stabilizer States}.}) or within other generalizations of the PSF such as the Monomial Stabilizer Formalism \cite{nest_MMS} and the X-S Stabilizer Formalism \cite{NiBuerschaperVdNest14_XS_Stabilizers}.

\subsubsection*{Proof of theorem \ref{thm:Normal Form CSS States}}

We finish this section by proving theorem \ref{thm:Normal Form CSS States} and giving a method for preparing coset states as a corollary  (corollary \ref{corollary:Coset State Preparations}). As announced in the previous section, the proof of this result relies on new technical ideas based on  hypergroup methods, which are needed to handle the \emph{non-unitary, non-monomial} stabilizers of theorem \ref{thm:Normal Form CSS States}.
 
\begin{proof}[Proof of theorem \ref{thm:Normal Form CSS States}] First note that the properties discussed in section \ref{sect:Pauli Operators} show that both stabilizer  hypergroups $\mathcal{S}_X^{\lambda_x}$, $\mathcal{S}_X^{\lambda_z}$ are well-defined. To prove the theorem, we will use some basic hypergroup theoretic results described in the following lemma.
\begin{lemma}[{\cite[2.4.15,2.4.16]{BloomHeyer95_Harmonic_analysis}}]
\label{lemma:Quotient-Annihilator Isomorphisms} For any subhypergroup $\mathcal{N}$, the hypergroup isomorphisms  $\mathcal{N}^*\cong\mathcal{T}^*/\mathcal{N}^\perp$ and $(\mathcal{T/N})^*\cong \mathcal{N}^\perp$ (cf.\ section \ref{sect:Glossary}), can be canonically realized as follows.
\begin{itemize}
\item[\textbf{(i)}] All characters of $\mathcal{N}$ are obtained via \emph{restriction} of characters of $\mathcal{T}$, and two characters $\mathcal{X}_\alpha,\mathcal{X}_\beta\in\mathcal{T}^*$ act \emph{equally} on $\mathcal{N}$ if and only if $\mathcal{X}_\alpha\in\mathcal{X}_\beta\mathcal{N}^\perp$. 
\item[\textbf{(ii)}]  All quotient characters are obtained by letting characters in $\mathcal{N}^\perp$ act on cosets $x\mathcal{N}$, and this map is well-defined because  the former act \emph{constantly} on the latter. 
\end{itemize}
\end{lemma}
Next, we identify  necessary and sufficient  conditions for a state $\ket{\psi}$ to be uniquely stabilized by $\{\mathcal{S}_X^{\lambda_x},\mathcal{S}_Z^{\lambda_z}\}$. First, condition (\ref{eq:Stabilizer State Condition}) says  that  $\ket{\psi}$ is stabilized by $\mathcal{S}_Z^{\lambda_z}$ iff
\begin{align}
\PZ{\mathcal{T}}(\mathcal{X}_{\mu})\ket{\psi_{}} &= \mathcal{X}_{\mu}( s )\ket{\psi_{}} =\lambda_{z}(\PZ{\mathcal{T}}(\mathcal{X}_{\nu}))\ket{\psi} \quad\textnormal{for every } \mathcal{X}_\mu \in \mathcal{N}^\perp.\label{inproof:CSS state is stabilized 1}
\end{align}
Due to lemma \ref{lemma:Quotient-Annihilator Isomorphisms}(ii), this holds iff the wavefunction $\psi$ is supported on a subset of the coset $s\mathcal{N}$. On the other hand, we show that $\ket{\psi}$ is further stabilized by $\mathcal{S}_X^{\lambda_x}$ iff $\psi$ belongs to the   image of the following operator:
\begin{equation}
 P_X:=\varw{\mathcal{N}}^{-1}\sum_{ b \in \mathcal{N}}\w{\Hchi_{\overline{\varsigma}}\mathcal{N}^\perp} \w{b}\mathcal{X}_{\overline{\varsigma}}({b}) \PX{\mathcal{T}}( b).
\end{equation}
The ``only if'' follows from the fact that $X_{\mathcal{T}}(b)\ket{\psi} = \Hchi_\varsigma(b)$ and the orthogonality relationship (\ref{eq:Character Orthogonality Subhypergroups}). The ``if'' follows from the calculation
\begin{align}
\PX{\mathsmaller{\mathcal{T}}}(a)P_X &=\sum_{ b, c\in  \mathcal{N}} \frac{\w{\mathcal{X}_{\overline{\varsigma}}\mathcal{N}^\perp}\w{b}}{\varw{\mathcal{N}}} n_{ a b}^{ c}\mathcal{X}_{\overline{\varsigma}}(b)\PX{\mathsmaller{\mathcal{T}}}( c)=\sum_{ c\in  \mathcal{N}} \frac{\w{\mathcal{X}_{\overline{\varsigma}}\mathcal{N}^\perp}\w{c}}{\varw{\mathcal{N}}} \left(\sum_{b\in  \mathcal{N}}n_{ {\overline{a}} c}^{ b}\mathcal{X}_{\overline{\varsigma}}(b)\right)\PX{\mathsmaller{\mathcal{T}}}( c)\\&=\sum_{c\in  \mathcal{N}} \frac{\w{\mathcal{X}_{\overline{\varsigma}}\mathcal{N}^\perp}\w{c}}{\varw{\mathcal{N}}} \mathcal{X}_{\overline{\varsigma}}(\overline{a})\mathcal{X}_{\overline{\varsigma}}(c)\PX{\mathsmaller{\mathcal{T}}}(c)=\mathcal{X}_{{\varsigma}}({a}) P_X,\label{eq:inproof:Projector Pauli X}
\end{align}
which implies with (\ref{eq:Character Orthogonality Subhypergroups}) that $P_X$ is a \emph{projector}, and consequently,  we get $\PX{\mathcal{T}}( a)\ket{\psi} = \PX{\mathcal{T}}(a)P_X\ket{\psi}=\mathcal{X}_\varsigma(a) P_X\ket{\psi}=\mathcal{X}_\varsigma(a)\ket{\psi}$ as desired.

As a result, we obtain that the stabilized states of $\{\mathcal{S}_X^{\lambda_x},\mathcal{S}_Z^{\lambda_z}\}$ are the quantum states in the vector space $\mathcal{V_S} :=\mathrm{span}\{P_X\ket{y}:y\in s\mathcal{N}\}$, where
\begin{align}\label{eq:inproof_P y}
P_X \ket{y} &= \sum_{\substack{ b\in \mathcal{N}\\
  x\in   s  \mathcal{N} } } \frac{\w{\Hchi_{\overline{\varsigma}}\mathcal{N}^\perp} \w{b}}{\varw{\mathcal{N}}} \sqrt{\frac{\w{y}}{\w{x}}} n_{ b, y}^{ x} \mathcal{X}_{\overline{\varsigma}}(b)\ket{ x} \propto \sum_{\substack{ x\in   s  \mathcal{N}} } \sqrt{\w{x}}  \left(\sum_{\substack{b}\in\mathcal{N} }n_{ x, {\overline{y}}}^{ b}\mathcal{X}_{\overline{\varsigma}}(b)\right)\ket{ x},
\end{align}
and that $\ket{\psi}$ is uniquely stabilized  iff this space is one dimensional. The proof for the RHS of (\ref{eq:Consistency CSS}) is the same: due to duality, we can apply a QFT (\ref{eq:QFTs are Clifford}) and reach this equality by  repeating the whole  proof from the beginning with exchanged roles for $\mathcal{T}$ and $\mathcal{T^*}$. This proves  case \textbf{(a)}.

Finally, we prove  \textbf{(b)}. First, in the simplest case, $s = e$, we can see that $\psi_y(x) = \sqrt{w_x} \Hchi_{\overline{\varsigma}}(x \overline{y})= \sqrt{w_x} \Hchi_{\overline{\varsigma}}(x)\Hchi_{\overline{\varsigma}}(\overline{y})=\psi_1(x)\Hchi_{\overline{\varsigma}}(\overline{y})$ by (\ref{eq:Character DEFINITION}), since $x,y\in\mathcal{N}$ and $\Hchi_{\overline{\varsigma}}$ is a character of $\mathcal{N}$. When $x, y \in s\mathcal{N}$, for $s \not = e$, we can, in general, have $n_{x,\overline{y}}^z \not= 0$ for $z \not\in \mathcal{N}$, so we cannot apply (\ref{eq:Character DEFINITION}). However, when $s$ is invertible (so $s \overline{s} = 1$), we can get the same result as in the simplest case by a simple change of variables.

For any $x,y\in s\mathcal{N}$, we define  $x':=\overline{s}x$ and $y':=\overline{s}y$. Since $x' \overline{y}' = \overline{s} x \overline{y} s = s \overline{s} x \overline{y} = xy$, we have $n_{x'\overline{y}'}^b = n_{x\overline{y}}^b$ and, taking $b=1$ and $y=x$, we have $w_{x'} = w_x$ from the definition of $w_x$. As these are the only appearances of $x$ and $y$ in $\psi_y(x)$, this shows that $\psi_y(x)=\psi_{y'}({x'})$. This combined with the previous easy case (for $x', y' \in \mathcal{N}$), shows that all $\psi_y$'s are  proportional to the non-zero function $\psi_1(x)$, which shows that the space is one-dimensional and contains $\ket{\psi}$. Finally, it is easily checked, in the case that $\mathcal{X}_\varsigma$ is invertible, that the normalization constant in (\ref{eq:Normal Form CSS state}, LHS) is  $(\w{\mathcal{X}_\varsigma \mathcal{N}^\perp}/\varpi_{s\mathcal{N}})^{-1/2}$; otherwise, it follows from (\ref{eq:Character Orthogonality Subhypergroups}). As in case \textbf{(a)}, duality lets us repeat the argument to get (\ref{eq:Normal Form CSS state}, RHS).
\end{proof}

As a final remark, we highlight that theorem \ref{thm:Normal Form CSS States} introduces many new states that we are not aware to be  preparable  by standard or character basis inputs and normalizer gates (the ingredients of the computational model in section \ref{sect: Circuit Model}), in general. However, we point out that the CSS states of type theorem \ref{thm:Normal Form CSS States}.(b) can always be prepared by measuring Pauli operators.
\begin{corollary}[\textbf{Coset state preparations}]\label{corollary:Coset State Preparations} Let $\mathcal{C}$ be a circuit  takes  the standard basis state  $\ket{\mathcal{X}_1}$ as input, performs $\Fourier{\mathcal{T}}^\dagger$ (an inverse QFT) or $\Fourier{\mathcal{T}^*}$, and then performs a syndrome measurement of the Pauli operators in a stabilizer hypergroup  $\mathcal{S}_Z^{\lambda_z}$ of form (\ref{eq:Stabilizer Hypergroup CSS State})\footnote{This is the canonical measurement defined by the common eigenprojectors that may be implemented, e.g., by measuring a poly-size set  generating set of $\mathcal{S}_Z^{\lambda_z}$, which exists if there is one for $\mathcal{N}^\perp$ (section \ref{sect:Stabilizer Formailsm}).}. Then $\mathcal{C}$ prepares a coset state. Specifically, it prepares $\ket{s\mathcal{N}}$  with probability $\varpi_{s\mathcal{N}}/\varpi_{\mathcal{T}}$. Furthermore, if $\ket{x_0}$ is given, $\Fourier{\mathcal{T}}$ or $\Fourier{\mathcal{T}^*}^\dagger$ is applied,  and $\mathcal{S}_X^{\lambda_x}$  (\ref{eq:Stabilizer Hypergroup CSS State}) is measured, then the outcome is a  coset state  $\ket{\mathcal{X}_\varsigma \mathcal{N}^\perp}$ with probability $\varpi_{\varsigma\mathcal{N}^\perp}/\varpi_{\mathcal{T}^*}$.
\end{corollary}
All states of form (\ref{eq:Normal Form CSS state}) can further be prepared from a coset state by applying Pauli gates.
\begin{proof} 
If we prove the first case, the second  holds  due to hypergroup duality. Lemma \ref{lemma:Quotient-Annihilator Isomorphisms}(ii) implies that measuring $\mathcal{S}_Z^{\lambda_z}$ is equivalent to performing a projective measurement with projectors $\{P_{s\mathcal{N}}=\ket{s\mathcal{N}}\bra{s\mathcal{N}}\}$. The claim follows by rewriting $\Fourier{\mathcal{T}}^\dagger\ket{\mathcal{X}_1}=\Fourier{\mathcal{T^*}}\ket{\mathcal{X}_1}=\sum_{a\in\mathcal{T}}\sqrt{\tfrac{\w{a}}{\varpi_{\mathcal{T}}}}\ket{a}=\sum_{s\mathcal{N}\in\mathcal{T/N}}\sqrt{\tfrac{\varpi_{s\mathcal{N}}}{\varpi_{\mathcal{T}}}}\ket{s\mathcal{N}}$.
\end{proof}

\section{Classical simulation of hypergroup normalizer circuits}\label{sect:Simulation}

In  chapter \ref{chapterF}, we gave efficient classical algorithms for simulating  normalizer circuits over  finite abelian groups, which recovered the standard  Gottesman-Knill theorem \cite{Gottesman_PhD_Thesis,Gottesman99_HeisenbergRepresentation_of_Q_Computers}. Together with the original GK theorem, these results demonstrate the existence of quantum computations that fail to give exponential quantum speed-ups despite usage of several powerful ingredients, such as  maximally entangled states \cite{nest06Entanglement_in_Graph_States}, quantum Fourier transforms and abelian-group coset states. Our next theorem extends a variant of the original Gottesman-Knill to normalizer circuits over arbitrary abelian hypergroups.
\begin{theorem}[\textbf{Simulation}]
\label{thm:simulation}
Let $\mathcal{C}$ be a normalizer circuit over a finite abelian hypergroup $\mathcal{T}$ containing \emph{global} QFTs, automorphism gates, and Pauli gates (but no quadratic phase gates) followed by a final measurement in the standard basis (cf.\ section \ref{sect: Circuit Model}). Then, given certain computability assumptions about $\mathcal{T}$ and its characters (section \ref{sect:Assumptions on Hypergroups}), there exists an efficient classical algorithm for sampling the measurement outcomes of $\mathcal{C}$.
\end{theorem}
The proof of the theorem is given at the end of the section. Our simulation result greatly expands the number of known families of quantum circuits that
can be classically simulated and it also adds yet more evidence to support the
idea that, for the HSP, quantum efficiency may go hand-in-hand with classical
simulability.

We highlight that our theorem generalizes the so-called CSS-preserving Gottesman-Knill theorem \cite{Delfosee14_Wigner_function_Rebits} without intermediate measurements, where the only normalizer gates allowed are those that send CSS states to CSS states (definition \ref{def:CSS state}). Our result also extends the CSS-preserving non-adaptive case of the theorems in chapter \ref{chapterF}.  Yet, theorem \ref{thm:simulation}  does not fully extend the ones in (chapter \ref{chapterF}-\ref{chapterI}, \cite{Gottesman_PhD_Thesis,Gottesman99_HeisenbergRepresentation_of_Q_Computers,VDNest_12_QFTs,BermejoVega_12_GKTheorem}), which altogether cover simulations of partial QFTs, quadratic phase gates, and intermediate Pauli-operators  measurements interspersed along the circuit. Simulating these extended cases is much harder in our nonabelian setting because of the  \emph{non-unitarity and non-monomiality} of hypergroup Pauli operators (cf.\ discussion in section \ref{sect:Stabilizer Formailsm}), which do not let us apply any  existing techniques for manipulating stabilizer codes \cite{Gottesman_PhD_Thesis,Gottesman99_HeisenbergRepresentation_of_Q_Computers,Gottesman98Fault_Tolerant_QC_HigherDimensions,VDNest_12_QFTs,BermejoVega_12_GKTheorem,BermejoLinVdN13_Infinite_Normalizers,BermejoLinVdN13_BlackBox_Normalizers,nest_MMS,AaronsonGottesman04_Improved_Simul_stabilizer,dehaene_demoor_coefficients,dehaene_demoor_hostens,VdNest10_Classical_Simulation_GKT_SlightlyBeyond,deBeaudrap12_linearised_stabiliser_formalism}; instead, the simulation method we give is based on the new hypergroup stabilizer techniques of section \ref{sect:Stabilizer Formailsm}. 

We stress that CSS normalizer operations can be highly nontrivial, as the quantum algorithms for  abelian HSP we investigated in chapter \ref{chapterB} are normalizer circuits with CSS structure. Hence, theorem \ref{thm:simulation} could be used to simulate, e.g., Shor's discrete-log quantum algorithm gate-by-gate if the information about the hidden subgroups was not manifestly hidden and groups were presented in a factorized form (cf.\ chapter \ref{chapterB} for an extended discussion). This means, for instance, that the entanglement present in a CSS-preserving circuit can be quite substantial.

Lastly, we  conjecture that our simulation result can be extended to all normalizer gates despite the non-monomiality/non-unitarity issues  we discuss.
\begin{conjecture}[\textbf{Conjecture}]\label{conj:Hypergroups}
There exist nontrivial families of abelian hypergroups for which the normalizer circuits of theorem \ref{thm:simulation} can still be efficiently classically simulated if they are supplemented with partial QFTs and quadratic phase gates acting at arbitrary circuit locations, and even if operations are chosen adaptively depending on the outcome of intermediate measurements of hypergroup Pauli operators.
\end{conjecture}
In conjecture \ref{conj:Hypergroups}, to qualify as ``nontrivial'', an abelian hypergroup family should not just consist of abelian groups and the weights of the elements of these hypergroups should be allowed to grow asymptotically with the number of bits needed to represent them (in order for the hypergroups not to be excessively ``group-like''). The measurement of a hypergroup Pauli operator is defined via its eigenvalue decomposition\footnote{Note that the following operators are manifestly diagonalizable: any $Z(\mathcal{X}_\mu),\mathcal{X}_\mu\in\mathcal{T^*}$ (\ref{eq:Pauli operators DEFINITION}); any $X(a),a\in\mathcal{T}$ (\ref{eq:QFTs are Clifford}); any product of commuting diagonalizable Paulis such as those in (\ref{eq:Stabilizer Hypergroup CSS State}) and in theorem \ref{thm:Normal Form CSS States}; any operator that is unitary equivalent to any of the latter. Unlike abelian-group Pauli operators, which are unitary (chapter \ref{chapterF}), it is not a priori obvious whether any \emph{product} of hypergroup Paulis always remains diagonalizable (this question was not investigated in this thesis); only diagonalizable products define measurements in conjecture \ref{conj:Hypergroups}.} as in section \ref{sect:PauliMeauserementsC1}. We consider examples of these measurements in theorem \ref{thm:AKR is normalizer}.

\myparagraph{Discussion: extensions of theorem \ref{thm:simulation}}\label{sect:Extensions theorem simulation:} In the light of our conjecture, we mention  a few simpler extensions of theorem \ref{thm:simulation} that we are aware of.

First, note that an efficient classical simulation is still possible if the main circuit is followed by another one  $\mathcal{C}'$ that contains any monomial normalizer gate (including quadratic-phase gates), which is then followed by a measurement in the standard basis\footnote{One can simply absorb those gates in the measurements \cite{nest_weak_simulations}.}: such circuits can prepare more types of entangled stabilizer states like (\ref{eq:Entangled State}) and the quaternionic cluster state (section \ref{sect:Quaternionic circuits}).

Second, the theorem can be easily extended to allow arbitrary CSS state/stabilizer state inputs with one further minimal assumption, namely, that  their corresponding wavefunctions can be sampled both in the  hypergroup element basis $\mathcal{B}_{\mathcal{T}}$ and in  the character basis $\mathcal{B}_{\mathcal{T}^*}$, which lets us, in particular, simulate QFTs acting on the state  (see section \ref{sect:Assumptions on Hypergroups}, condition (ii) and section \ref{sect:simulation proof}). Furthermore, if this holds for the simple CSS states of theorem \ref{thm:Normal Form CSS States}, then theorem \ref{thm:simulation} can be extended to circuits that can, e.g.,  prepare coset states as in  corollary \ref{corollary:Coset State Preparations} and/or accept coset states as inputs.

\subsection{Computability assumptions on hypergroups}\label{sect:Assumptions on Hypergroups}

In theorem \ref{thm:simulation}, we must restrict ourselves to  hypergroups with sufficient structure to let us efficiently compute within them and their character  hypergroups. Note that  assumptions of this kind are typically made in the HSP literature: for instance, in order for the HRT quantum algorithm for the HNSP \cite{Hallgren00NormalSubgroups:HSP} to be efficient one needs to be given the ability to intersect characters kernels.  The assumptions needed for theorem \ref{thm:simulation} are  listed next, followed by examples of hypergroups  that meet them.

First, in theorem \ref{thm:simulation} we assume that the hypergroup $\mathcal{T}$ as well as its dual $\mathcal{T}^*$ are \emph{efficiently computable}: we  say that a hypergroup $\mathcal{T}$ is {efficiently computable}\footnote{Efficiently computable hypergroups generalize  the black-box groups \cite{BabaiSzmeredi_Complexity_MatrixGroup_Problems_I} explored in chapter \ref{chapterB}.} if its elements can be uniquely represented with $n=O(\mathrm{polylog}{|\mathcal{T}|})$ bits and there are $O(\mathrm{poly}(n))$-time classical subroutines to perform the hypergroup multiplication, i.e.,\ given two elements $x_i$, $x_j\in \mathcal{T}$ and an index $k$, we can efficiently compute the coefficient $n_{ij}^k$ for any $i,j,k$.

In theorem \ref{thm:simulation},  we  further need to assume that the involved hypergroups are what we call  \emph{doubly efficiently computable}: a hypergroup $\mathcal{T}$ is doubly efficiently computable if both $\mathcal{T}$ and $\mathcal{T}^*$ are efficiently computable and, furthermore,  if the structure of their associated character tables is sufficiently well-known that we are able to efficiently perform the following tasks classically:
\begin{itemize}
\item[(i)] \textbf{Computable characters.} For any $a\in \mathcal{T}$, any character function $\Hchi_\mu(a)$ can be efficiently computed classically\footnote{For simplicity, we will assume that this can be done with perfect precision in this chapter. Our results readily extend if characters can be computed within an arbitrarily small error.}.
\item[(ii)] \textbf{Simulable input states.} Quantum Fourier transforms of allowed input states can be efficiently \emph{sampled} classically, or equivalently,  the distributions $\{p_a\}$ and $\{q_\mu\}$, with $p_a \defeq \tfrac{\w{a}\w{\mathcal{X}_\mu}}{\varpi_{\mathcal{T}}}|\mathcal{X}_\mu(a)|$ for fixed $\Hchi_\mu \in \mathcal{T}^*$ and $q_\mu \defeq \tfrac{\w{a}\w{\mathcal{X}_\mu}}{\varpi_{\mathcal{T}}}|\mathcal{X}_\mu(a)|$ for fixed $a \in \mathcal{T}$, can be efficiently sampled.
\item[(iii)] \textbf{Computable dual morphisms.} For any efficiently computable hypergroup automorphism $\alpha:\mathcal{T}\rightarrow \mathcal{T}$,  its inverse $\alpha^{-1}$ and its dual automorphism (definition \ref{eq:Dual Automorphism DEFINITION}) $\alpha^*:\mathcal{T}\rightarrow \mathcal{T^*}:\chi\rightarrow f^*_{\chi}$,  can both be efficiently determined and computed. Duals of computable hypergroup homomorphisms $f:\mathcal{T}\rightarrow \mathcal{T}'$ can also be computed\footnote{Applying def.\ \ref{eq:Dual Automorphism DEFINITION} to a  homomorphism $f:\mathcal{T}\rightarrow \mathcal{T}'$ one gets a dual morphism  $f^*:\mathcal{T}'\rightarrow \mathcal{T}$  \cite[1.6.(ii)]{McMullen79_Algebraic_Theory_Hypergroups}.}.
\end{itemize}

\paragraph{Examples and remarks}

Both computability notions presented are preserved by  taking direct products $\mathcal{T}_1\times \mathcal{T}_2$. Furthermore, the notion of doubly efficiently computable hypergroup is preserved under  taking duals $\mathcal{T} \leftrightarrow \mathcal{T}^*$.

Any hypergroup of the form $\mathcal{T}_1\times \cdots \times \mathcal{T}_m$ is doubly efficiently computable if  homomorphisms are restricted to be of a product form $f_1 \times \cdots \times f_m$, where $m$ is constant, or if they act nontrivially only in a constant number of sites. As a result,  \emph{normalizer circuits} over hypergroups of the from $\mathcal{T}_1\times \cdots \times \mathcal{T}_m$ with constant-size $\mathcal{T}_i$ will always turn out to be efficiently simulable if they contain  at most $k$-local entangling gates, for any constant $k$ (theorem \ref{thm:simulation}). The examples given in section \ref{sect:Quaternionic circuits} over the quaternions were of this form.

As another example, for any finite abelian group $G$, all problems in (i-ii-iii)  can be solved  in  $O(\polylog |G|)$ time given that $G$ is explicitly given in the form $G=\Integers_{N_1}\times \cdots\times \Integers_{N_m}$. Condition (ii) holds for any abelian group stabilizer state with known stabilizer group. These results are invariant of the bit-size of any $N_i$ (chapter \ref{chapterF}). 

For arbitrary finite abelian hypergroups finding simple bounds like those in chapter \ref{chapterF} is likely to be an ``impossible'' problem, since the question cannot even be addressed without classifying all conjugacy class and character hypergroups of all finite groups, whereas classifying the latter is regarded as a (so-called) ``wild'' problem  \cite{MO_Classification_Finite_Groups,MO_Classication_Problem_Wild}. It is easier to prove polynomial-time  bounds for particular hypergroup/group families  that fulfill the minimal computability requirements (i-ii-ii), like in the two examples given above.

Finally, we highlight that some efficiently computable hypergroups are provably not  \emph{doubly} efficiently computable unless efficient classical algorithms for believed-to-be-hard problems like computing discrete-logarithms exist (see appendix \ref{app:Discret Log}). In the abelian group case, the associated normalizer circuits can realize  Shor's \cite{Shor} algorithms  and lead to exponential quantum speed-ups (chapter \ref{chapterB}). In section \ref{sect:Quantum Algorithms}, we will develop new quantum algorithms based on normalizer circuits over such ``black-box'' hypergroups.

\subsection{Proof of theorem \ref{thm:simulation}}
\label{sect:simulation proof}

We finish this section proving theorem \ref{thm:simulation} by giving an explicit classical algorithm for sampling the outcome distribution after measuring the final state  of the computation $\mathcal{C}\ket{\psi_0}$,
being $\ket{\psi_0}$ the input state. Our algorithm is efficient given that the hypergroup $\mathcal{T}$ is doubly efficiently computable (section \ref{sect:Assumptions on Hypergroups}). The key technique that we exploit in our simulation is a normal form for CSS normalizer circuits.
\begin{lemma}[\textbf{Normal form}]\label{thm:Normal form} Let $\mathcal{C}$ be a normalizer circuit over a $\mathcal{T}$ as in theorem \ref{thm:simulation}. Then, $\mathcal{C}$ can be put in a layered normal form $\mathcal{C}=M  F$, where $F$ is either trivial or a QFT and $M$ is a monomial circuit\footnote{That is, a circuit whose transformation in matrix form has one entry per row and column.} of automorphism gates and Pauli gates. Furthermore, classical descriptions of $M$ and $F$ can be computed classically efficiently  if  $\mathcal{T}$ is  doubly efficiently computable.
\end{lemma}
\begin{proof}[Proof of lemma \ref{thm:Normal form}]
First, note that the Pauli gates in the circuit (which are   of the form $\PX{\mathcal{T}}(C)$,  $\PZ{\mathcal{T}}(\mathcal{X})$ in the conjugacy-class basis and of the form  $\PZ{\mathcal{T}^*}(C)$,  $\PX{\mathcal{T}^*}(\mathcal{X})$ in the character basis)  can be conjugated with all the other normalizer gates using the update rules in theorem \ref{thm:Evolution of Stabilizer States}(\ref{eq:Automorphism gates are Clifford},\ref{eq:Quadratic Phase gates are Clifford}). As a result, if $U$ is a Pauli gate at an intermediate circuit position $\mathcal{C}=\mathcal{C}_2 U \mathcal{C}_1$, it can be removed from its location by adding a new Pauli-correction term $U'=\mathcal{C}_2 U \mathcal{C}_2^\dagger$ at the beginning of the circuit. By doing this, we put $\mathcal{C}$ in an intermediate two-layered normal form $\mathcal{C}= P \mathcal{C'}$, where $P$ is a circuit of Pauli gates  and $\mathcal{C}'$ collects all  QFTs and all automorphism gates that were present in $\mathcal{C}$, in the same temporal order.

We finish the proof of the lemma by showing that  $\mathcal{C}'$ can be put in a normal form $A S$ where $S$ is either the identity gate or a QFT, and $A$ is a circuit of automorphism gates. Once we have that, we can combine $P$ and $A$ into a single layer $M$ and obtain $\mathcal{C}=M S$. Because every gate in $M$ is either a permutation or a diagonal unitary, $M$ is manifestly monomial.

To this end, we use that group automorphism gates can be conjugated with  Fourier transforms in an elegant way, by replacing them with dual automorphism gates. Specifically, we have $\mathcal{F_T} U_\alpha = U_{\alpha^*}^{-1} \mathcal{F_T} = U_{\alpha^{-*}} \mathcal{F_T}$ and $\mathcal{F_{T^*}} U_{\alpha^{-*}}=U_\alpha \mathcal{F_{T^*}}$, where $\alpha^{-*}$ is the dual of $\alpha^{-1}$ (definition \ref{eq:Dual Automorphism DEFINITION}). This follows by simple calculation, using (\ref{eq:Quantum Fourier Transform over Hypergroup T}) and the fact that automorphisms cannot change the weights of elements. Furthermore, we in fact have
\begin{equation}\notag
U_\alpha\ket{\mathcal{X}_\mu}=\sum_{a\in\mathcal{T}}\sqrt{\tfrac{\w{a} \ws{\overline{\Hchi_\mu}}}{\varw{\mathcal{T}}}} \, \overline{\Hchi_{\mu}}(a)\ket{\alpha(a)}\stackrel{b:=\alpha(a)}{=}\sum_{b\in\mathcal{T}}\sqrt{\tfrac{\w{b} \ws{\overline{\Hchi_{\alpha^{-*}(\mu)}}}}{\varw{\mathcal{T}} }}  \, \overline{\Hchi_{\alpha^{-*}(\mu)}}(b)\ket{b}=U_{\alpha^{-*}}\ket{\Hchi_\mu}.
\end{equation}
Hence, it follows that $U_\alpha = U_{\alpha^{-*}}$ as gates, which means that can implement the latter since we can implement the former by assumption.

Applying these rules, we can move the QFTs before the automorphisms. Furthermore, since the effect of each QFT is just to change the designated of the circuit, any product of QFTs is equivalent to a single QFT gate, $F$, that changes the basis once (or not at all). By this process, we get $\mathcal{C'}=AF$, where $A$ is a product of automorphism gates.

Finally, the assumption that both $\mathcal{T}$ and $\mathcal{T^*}$ are efficiently computable (and, hence, so is any hypergroup of form in (\ref{eq:Hypergroups})) means that we can efficiently compute the conjugations to determine the gates in $P$ using (\ref{eq:Automorphism gates are Clifford}, \ref{eq:QFTs are Clifford}). The assumption that $\mathcal{T}$ is doubly efficiently computable also implies that we can find $\alpha^{-*}$ efficiently for any $\alpha$, so that we can compute the automorphisms in $A$ efficiently as well. Finally, we efficiently obtain a classical description for $M$ (resp.\ $F$) by listing the Pauli and automorphism gates (resp.\ QFTs) it contains.
\end{proof}

To prove the theorem, we first apply lemma \ref{thm:Normal form} to put $\mathcal{C}$ in normal form, $M F$. Next, we note that $M$ acts as $M\ket{a}=\gamma_a\ket{\pi(a)}$, where $\gamma_a$ has unit modulus and $\pi$ is some permutation on the elements of the basis for the image of $F$. Thus, measuring in the final basis, after applying $M$, is equivalent to measuring in the basis after $F$ and then applying $\pi$. Because $\mathcal{T}$ is doubly efficiently computable (as defined in section~\ref{sect:Assumptions on Hypergroups}), we can, first, compute $M$ and $F$ by lemma~\ref{thm:Normal form}; second, simulate a measurement after applying $F$ by assumption (ii); and, finally, compute the action of $\pi$ on the obtained samples: for the latter step, we can  infer a poly-size boolean circuit implementing $\pi$ from the automorphism and Pauli $X$-gates in $M$---via assumption (iii) and  via the circuit that implements the hypergroup multiplication. Q.E.D.

\paragraph{Remark}

As mentioned in the discussion after theorem \ref{thm:simulation},  one can straightforwardly extend this simulation method to any input $\ket{\psi_0}$ if measurements on $\ket{\psi_0}$ on the hypergroup element and character bases are easy to simulate (because we only use that information about the state).

\section{Quantum algorithms for HNSP and abelian HSHP}
\label{sect:Quantum Algorithms}

In this last section, we apply the hypergroup methods of previous sections to the development new  quantum algorithms for the HNSP and for the CC-HSHP.

We give three quantum algorithms for the HNSP of increasing generality (and complexity). These algorithms are interesting because they are fundamentally different from the one of Hallgren et al. \cite{Hallgren00NormalSubgroups:HSP}, as they exploit the hypergroup structure of the HNSP:  to solve the problem, we turn it into a CC-HSHP (using theorems \ref{thm:reduction1}-\ref{thm:reduction2}, section \ref{sect:HNSP}) and solve the resulting abelian HSHP instead. Our results show that, in the cases considered here, the HNSP is easy because the CC-HSHP is easy, which gives an explanation for why the HNSP is easy in terms of the presence of an \emph{abelian} algebraic structure.

Our quantum algorithms for the CC-HSHP are interesting in their own right (beyond their use in solving the HNSP) as no provably efficient algorithms were previously known.

As we will see shortly, all of the quantum algorithms we consider here fit within our normalizer circuit model. This means that they can be analyzed using the stabilizer formalism of section \ref{sect:Stabilizer Formailsm}. The results of that section, especially theorem \ref{thm:Normal Form CSS States}, will be critical to all of our analysis below.

We begin in section \ref{sect:Comparison HRT vs AKR} by looking at a previously proposed algorithm for the HSHP \cite{Amini2011fourier}. We show that, in one subclass of cases, not only can we reduce the HNSP to the CC-HSHP (as we saw in section~\ref{sect:HNSP}), but in fact, the algorithm of \cite{Hallgren00NormalSubgroups:HSP} for the HNSP is identical, under a simple vector space isomorphism, to the proposed algorithm of \cite{Amini2011fourier} applied to the CC-HSHP. This demonstrates an even deeper connection between the HNSP and the CC-HSHP than that demonstrated by the reductions of section~\ref{sect:HNSP}.

In section~\ref{sect:AKR Analysis}, we use our stabilizer formalism to analyze the proposed algorithm of \cite{Amini2011fourier}. We describe instances where it does and does not work correctly. This analysis leads us to a new algorithm, described in the same section, which works correctly for all groups.

In section~\ref{sect:New Algorithms}, we develop new quantum algorithms, taking advantage of the unique structure of abelian hypergroups. The resulting algorithms work for all nilpotent (hyper)groups (along with some non-nilpotent groups) and requires fewer assumptions than those of section~\ref{sect:AKR Analysis}. As with the algorithms of section~\ref{sect:AKR Analysis}, the stabilizer formalism remains key to our analysis.

Finally, in section~\ref{sect:Open Problems}, we mention a few further results and some open problems.

\subsection{A comparison of two simple algorithms for the HNSP}
\label{sect:Comparison HRT vs AKR}

To illustrate ideas we use later and introduce the quantum algorithm proposed in \cite{Amini2011fourier} for the CC-HSHP, we begin by discussing it and comparing it to the standard algorithm for the HNSP \cite{Hallgren00NormalSubgroups:HSP} in the case when the oracle $f: G \rightarrow \{0,1\}^*$ is a class function. This happens if and only if $G/N$ is abelian, where $N$ is the hidden normal subgroup. We show that, for this case, the two algorithms actually \emph{coincide}: the HNSP becomes an instance of the CC-HSHP and the same algorithm solves both problems.

As we saw in section~\ref{sect:HNSP}, such an $f$ is easily transformed into an oracle $\overline{f} : \Conj{G} \rightarrow \{0,1\}^*$ for the CC-HSHP since each coset $xN$ is in a conjugacy class by itself. This allows us to turn algorithms for the CC-HSHP into algorithms for the HNSP (and vice versa). We now compare two algorithms designed for this common problem via two different perspectives.

The quantum algorithm of Hallgren, Russell, and Ta-Shma for HNSP \cite{Hallgren00NormalSubgroups:HSP}, henceforth referred to as ``HRT'', operates as follow:
\begin{enumerate}
\item Initialize a workspace register in a quantum state $\ket{\chi_1}$. 

\item Apply an inverse QFT in order to obtain a superposition $\sum_{g\in
G}\ket{g}$.

\item Evaluate the oracle on an ancillary register to obtain $\sum_g
\ket{g,f(g)}$. Measure the second register to project the state of the first
onto a coset state $\ket{xN}$, for some $x$ drawn uniformly at random.

\item Apply a QFT to $\ket{xN}$ and measure the label $\mu$ of an
irreducible representation.

\item Repeat the experiment $T$ times and record the outcomes $\mu_1, \ldots,
\mu_T$.

\item Determine the subgroup $\bigcap_i \ker \chi_{\mu_i}$. With exponentially
high probability $1-O(\tfrac{1}{2^T})$, the subgroup $\bigcap_i \ker
\chi_{\mu_i}$ is the hidden subgroup $N$.
\end{enumerate}
The quantum part of this algorithm (steps 1--5) can be implemented efficiently
if we have an efficient implementation of the QFT. However, the complexity of
the classical post-processing (step 6) is unknown, in general.

The quantum algorithm of Amini, Kalantar, and Roozbehani applies to the hidden
subhypergroup problem (HSHP). When applied to the conjugacy class hypergroup,
$\Conj{G}$, we refer to this algorithm as ``AKR''. It takes as input an oracle
$\overline{f} : \overline{G} \rightarrow \{0,1\}^*$ and operates as follows:
\begin{enumerate}
\item Initialize a workspace register in a quantum state $\ket{\Hchi_1}$. 

\item Apply an inverse QFT in order to obtain a superposition $\abs{G}^{-1/2}
\sum_{C_x\in \Conj{G}} \sqrt{\w{C_x}} \ket{C_x}$.

\item Evaluate the oracle on an ancillary register to obtain $\abs{G}^{-1/2}
\sum_{C_x} \sqrt{\w{C_x}} \ket{C_x,\overline{f}(C_x)}$. Measure the second register to
project the state of the first onto a hypergroup coset state $\ket{C_x
\subhyp{N}{G}}$, for some $x$ drawn uniformly at random.\footnote{Note that we
are working in the Hilbert space with basis $\set{\ket{C_x}}{C_x \in \Conj{G}}$ and
the state $\ket{C_x \subhyp{N}{G}}$ is a superposition of those conjugacy
classes that make up $C_x \subhyp{N}{G}$.}

\item Apply a QFT to the state $\ket{C_x \subhyp{N}{G}}$ and measure the label
$\Hchi_\mu$ of a character.

\item Repeat the experiment $T$ times and record the outcomes $\Hchi_{\mu_1},
\ldots, \Hchi_{\mu_T}$.

\item Determine the subhypergroup of classes in the kernels of all the
$\Hchi_{\mu_i}$'s.
\end{enumerate}
Like HRT, steps 1--5 can be implemented efficiently while the complexity of
step 6 is unknown, in general.

As the reader can see, the two algorithms perform the same steps. First,
they apply an inverse Fourier transform to prepare a superposition over the
entire basis on which they operate. Next, they apply their respective oracles
to an additional register, measure, and throw away its value. Finally, they
apply a Fourier transform and measure in the new basis. For AKR, this is
measuring a character label, while for HRT, this is measuring the label of an
irrep. It is critical to note that HRT does not use the value of the matrix
index register, which is part of the output of the QFT for the group.

When $f$ is a class function (so that $G/N$ is abelian), we have $C_x\subhyp{N}{G} = xN$
since each coset $xN$ is in its own conjugacy class. Hence, we can see that the
two algorithms are in matching states after steps 1--3. In particular, the
state of HRT after step 3 is conjugation invariant, and since the QFT preserves
conjugation invariance, so is the state of HRT after step 4. In fact, it is
easy to check that the QFT of $G$ applied to the conjugation invariant subspace
is exactly the QFT of $\Conj{G}$. More precisely, we have the following
result, which is also easy to check.
\begin{proposition}[\textbf{HRT = AKR}]
If $f : G \rightarrow \{0,1\}^*$ is a class function, then HRT operates
entirely within the conjugation invariant subspace $\Comp_{\overline{G}}$. Furthermore,
within this subspace, HRT is identical to AKR.
\end{proposition}
The first part of the following lemma simplifies our analysis of the algorithm. (And the second part will be useful to us later on.)
\begin{lemma}[\cite{Roth75_Character_Conjugacy_Hypergroups}]\label{lemma:Roth}
For any normal subgroups $N,K$ such that $N\trianglelefteq K \trianglelefteq G$, the  hypergroup $\overline{G}/\subhyp{N}{G}$ is isomorphic to $\overline{G/N}$ (the class hypergroup of $G/N$) and $\subhyp{K}{G}/\subhyp{N}{G}$ is a subhypergroup of $\overline{G}/\subhyp{N}{G}$ isomorphic to  $\subhyp{K/N}{G/N}$ (which is a subhypergroup of $\overline{G/N}$).
\end{lemma}
Since $G/N$ is abelian, the quotient hypergroup $\Conj{G} / \Conj{N}_G \cong \Conj{G/N}$ is actually a group. Hence, it follows immediately from the standard results on Fourier sampling of abelian groups \cite{kitaev_phase_estimation} that both algorithms are correct in this case as they are actually just performing Fourier sampling of an abelian group.

\begin{theorem}[\textbf{CC-HSHP is easy, I}]
\label{thm:easy1}
Let $G$ be a group. Suppose that we are given a function $\overline{f} :
\Conj{G} \rightarrow \{0,1\}^*$ hiding the subhypergroup corresponding to a
normal subgroup $N \lhd G$ such that $G/N$ is abelian, and suppose that we can
efficiently compute the QFT for $\Conj{G}$. Then there is an efficient
quantum algorithm for the CC-HSHP.
\end{theorem}
 
As a corollary of theorem \ref{thm:easy1}, we can see that there is an efficient hypergroup-based
algorithm for solving HNSP in the case when the oracle $f : G \rightarrow
\{0,1\}^*$ is a class function.

\begin{corollary}[\textbf{HNSP is easy, I}]
Let $G$ be a group. Suppose that we are given a hiding function $f : G
\rightarrow \{0,1\}^*$ that is  also a class function.  If we can efficiently
compute the QFT for $G$ and compute efficiently with conjugacy classes of $G$,
then we can efficiently solve this HNSP.
\end{corollary}

\begin{proof}
This follows since we can efficiently reduce the HNSP to a CC-HSHP by theorem \ref{thm:reduction1}, we can use the QFT of $G$ to implement the QFT of $\Conj{G}$ as described in appendix~\ref{app:CC implementation details}, and we can solve this CC-HSHP efficiently by theorem \ref{thm:easy1}.
\end{proof}

\subsection{Analysis of the algorithm of Amini et al.}
\label{sect:AKR Analysis}

In this subsection, we analyze the AKR algorithm in more detail. Our analysis will benefit from our hypergroup stabilizer formalism tools (section \ref{sect:Stabilizer Formailsm}), namely, our normal forms in theorem \ref{thm:Normal Form CSS States}, which will let us characterize the outcome probability distribution of the quantum algorithm.\footnote{This is a new result. No formula for this probability was given in \cite{Amini_hiddensub-hypergroup,Amini2011fourier}.} This is possible due to the following connection with our normalizer circuit framework.
\begin{theorem}[\textbf{AKR is normalizer}]\label{thm:AKR is normalizer}
For any finite group $G$, the AKR quantum algorithm for the CC-HSHP over $\overline{G}$ is a normalizer circuit  with intermediate Pauli measurements. Furthermore, all of its intermediate quantum states are  CSS stabilizer states of form (\ref{eq:Normal Form CSS state}).\footnote{Although we do not discuss the full AKR algorithm, this theorem also holds for all abelian hypergroups.}
\end{theorem}
\begin{proof}
Steps 1-3 implement a coset state preparation scheme as in corollary \ref{corollary:Coset State Preparations}, using  the oracle as a black box to perform a syndrome measurement of type $\mathcal{S}_Z^\lambda$ (cf.\  the proofs of corollary \ref{corollary:Coset State Preparations}, theorem \ref{thm:Normal Form CSS States} for details). Step 4  takes the QFT of a coset state of form (\ref{eq:Normal Form CSS state}).
\end{proof}
Using this connection, we can apply the tools developed in the previous
sections to compute the probability of measuring a given character label
$\Hchi_\mu$ at step 5.

The mixed state of AKR after the oracle is called and its value discarded is
given by
\[ \rho = \sum_{C_x\subhyp{N}{G} \in
      \overline{G}/\subhyp{N}{G}} \frac{\w{C_x\subhyp{N}{G}}}{\varw{\overline{G}/\subhyp{N}{G}}}
          \ket{C_x\subhyp{N}{G}}\bra{C_x\subhyp{N}{G}} \]
since the probability of measuring each coset $C_x \subhyp{N}{G}$ is proportional to its weight.\footnote{The initial superposition has probability of measuring $C_x$ proportional to $w_{C_x}$, so the probability of measuring the coset $C_x \subhyp{N}{G}$, which contains all the elements with the same value from the oracle as $C_x$, is proportional sum of all of their weights, $\varpi_{C_x \subhyp{N}{G}}$. These sums are proportional to the weights $w_{C_x \subhyp{N}{G}}$, so we get the form in the equation above after normalizing the probabilities to sum to 1.} 
Each coset state $\ket{C_x\subhyp{N}{G}}$ is a stabilized by $X_{\overline{G}}(C_y)$
for any $C_y \in \subhyp{N}{G}$ and any $Z_{\widehat{G}}(\Hchi_\mu)$ for any $\Hchi_\mu
\in \subhyp{N}{G}^\perp$, which means that it is a CSS stabilizer state of the form shown
in equation (\ref{eq:Stabilizer Hypergroup CSS State}) with $s=C_x$ and
$\Hchi_\varsigma = \Hchi_1$, the trivial character. Thus, we can read off the Fourier
transform of this state directly from theorem \ref{thm:Normal Form CSS States}.

Our application of  theorem \ref{thm:Normal Form CSS States} is greatly
simplified by the fact that $\Hchi_{\overline{\varsigma}}$ is the trivial character $\Hchi_{1}$. This means
 that
$\w{\Hchi_{\overline{\varsigma}}\subhyp{N}{G}^\perp} = 1$ since the weight of a character is the
dimension of the underlying representation $\mu$.\footnote{Note that
$\Hchi_{\overline{\varsigma}}\subhyp{N}{G}^\perp$ is an element of $\widehat{G}/\subhyp{N}{G}^\perp \cong
\subhyp{N}{G}^*$ (section \ref{sect:Glossary}), so this is the weight of $\Hchi_{\overline{\varsigma}}$ viewed as a
character of $\subhyp{N}{G}$, where it remains trivial.} This means that the state
$\ket{\psi}$ for $s = C_x$ and $\mathcal{N} = \subhyp{N}{G}$ is proportional to
$\sum_{C_y \in C_x\subhyp{N}{G}} \sqrt{\w{C_y}} \ket{C_y}$, which is proportional to
$\ket{C_x\subhyp{N}{G}}$, and thus, we have $\ket{\psi} = \ket{C_x\subhyp{N}{G}}$
since both states are normalized. Thus, theorem \ref{thm:Normal Form CSS
States} tells us
\[ \Fourier{\overline{G}} \ket{C_x\subhyp{N}{G}} = \sum_{\Hchi_\mu \in \subhyp{N}{G}^\perp}
   \sqrt{\frac{\w{\mu} \w{C_x\subhyp{N}{G}}}{\varpi_{\subhyp{N}{G}^\perp}}} \Hchi_\mu(C_x)
   \ket{\Hchi_\mu},  \]
and hence, the Fourier transform of the mixed state of AKR is
\begin{eqnarray*}
\Fourier{\overline{G}} \rho \Fourier{\overline{G}}^\dag
&=& \sum_{C_x\subhyp{N}{G}\in\overline{G}/\subhyp{N}{G}} \frac{ \w{C_x\subhyp{N}{G}} }{\varw{\overline{G}/\subhyp{N}{G}}}
 \left( \sqrt{\w{C_x\subhyp{N}{G}}} \sum_{\mu \in \subhyp{N}{G}^\perp}
    \sqrt{\frac{\w{\mu}}{\varpi_{\subhyp{N}{G}^\perp}}} \Hchi_\mu(C_x) \ket{\Hchi_\mu} \right)
  \times \\
&& \qquad \qquad \qquad \ 
   \left( \sqrt{\w{C_x\subhyp{N}{G}}} \sum_{\nu \in \subhyp{N}{G}^\perp}
     \sqrt{\frac{\w{\nu}}{\varpi_{\subhyp{N}{G}^\perp}}} \Hchi_\nu(C_x) \bra{\Hchi_\mu}
     \right) \\
&=& \sum_{\mu, \nu \in \subhyp{N}{G}^\perp} \sqrt{\w{\mu} \w{\nu}}
    \left( \sum_{C_x\subhyp{N}{G} \in \overline{G}/\subhyp{N}{G}}
      \frac{\w{C_x\subhyp{N}{G}}^2}{\varw{\overline{G}/\subhyp{N}{G}}^2}
      \Hchi_\mu(C_x) \overline{\Hchi_{\nu}}(C_x) \right)
    \ket{\Hchi_\mu} \bra{\Hchi_\nu}, \\
\end{eqnarray*}
where, in the last step, we have used the fact that $\subhyp{N}{G}^\perp \cong 
\left(\overline{G}/\subhyp{N}{G}\right)^*$. Finally, using lemma \ref{lemma:Roth}, we conclude that the probability of measuring the outcome $\Hchi_\mu$ is
\begin{equation}
\label{eq:AKR-prob1}
\Pr(\Hchi_\mu) =
\w{\mu} \sum_{C_x\subhyp{N}{G} \in \overline{G}/\subhyp{N}{G}} \frac{\w{C_x\subhyp{N}{G}}^2}{\varw{\overline{G}/\subhyp{N}{G}}^2}
      \Hchi_\mu(C_x) \overline{\Hchi_{\mu}}(C_x)=
      \w{\mu} \sum_{C_{xN} \in \overline{G/N}} \frac{\w{C_{xN}}^2}{\varw{\overline{G/N}}^2}
            \Hchi_\mu(C_x) \overline{\Hchi_{\mu}}(C_x).
\end{equation}
This formula is the key to our analysis of AKR in this section and the next.

\subsubsection{Non-convergence of AKR for simple instances}

We begin examining these probabilities by looking at an example.

\begin{example}[\textbf{AKR Counterexample}]
\label{ex:non-convergence}
The Heisenberg group over $\Integer_p$ (with $p$ prime) is the set
$\Integer_p^3$ with multiplication defined by $(x,y,z)\cdot(x',y',z') =
(x+x',y+y',z+z'+xy')$. For a nice review of its representation theory, see
the article of Bacon \cite{Bacon_Clebsch-Gordon}.

The center of this group is the normal subgroup $Z(G) = \set{(0,0,z)}{z \in
\Integer_p}$. Hence, any element of $Z(G)$ is in a conjugacy class of its own.
For any $(x,y,z) \not\in Z(G)$ (i.e., with $(x,y) \not= (0,0)$), it is easy to
check that its conjugacy class is the coset $(x,y,z) Z(G)$. Hence, the weight
of the former classes are 1 and those of the latter are $\abs{Z(G)} = p$.

Let us consider the case when the hidden subgroup is trivial $N = \{e\}$. For
any $(a,b) \not= (0,0)$ there is a 1-dimensional irrep of the Heisenberg group,
which we denote $\Hchi_{a,b}$ given by $\Hchi_{a,b}(x,y,z) = \omega_p^{ax+by}$,
where $\omega_p$ is a $p$-th root of unity. We can apply
equation~(\ref{eq:AKR-prob1}) to compute the probability of measuring this
irrep in AKR. To do this, we first note that this probability would be the
inner product of $\Hchi_\mu$ with itself except that weights have been squared.
Since there are only two sizes of conjugacy classes, it is not difficult to
rewrite this expression in terms of that inner product as follows.
\begin{eqnarray*}
\Pr(\Hchi_{a,b})
 &=& \sum_z \frac{1}{p^6} \abs{\Hchi_{a,b}(0, 0, z)}^2 +
     \sum_{(x,y) \not= (0,0)} \frac{p^2}{p^6} \abs{\Hchi_{a,b}(x,y,0)}^2 \\
 &=& \frac{1}{p^6} \sum_z 1 +
     \frac{p^2}{p^6} \sum_{(x,y) \not= (0,0)} 1 \\
 &=& \frac{1}{p^6} p + \frac{p^2}{p^6} (p^2 - 1) \\
 &=& \frac{p^4 - p^2 + p}{p^6} \\
\end{eqnarray*}
where we have used the fact that $\abs{\Hchi_{a,b}(x,y,z)} = 1$ for all $x,y,z
\in \Integer_p$.

Finally, the probability of measuring any of the 1-dimensional irreps is 
\[ \sum_{(a,b)\not=(0,0)} \Pr(\Hchi_{a,b})
    = (p^2-1) \frac{p^4 - p^2 + p}{p^6}
    = \frac{p^6-2p^4+p^3+p^2-p}{p^6}
    = 1 - O\left(\frac{1}{p^2}\right). \]
This means that if $p$ is exponentially large, we are unlikely to ever see
an irrep other than the $\Hchi_{a,b}$'s, and since $Z(G)$ is in the kernel of
all such irreps, the intersection of the kernels of polynomially many irreps
will include $Z(G)$ with high probability.

Since the AKR algorithm returns the intersection of the kernels of the sampled
irreps as its guess of the hidden subgroup, this demonstrates that the AKR
algorithm will fail to find the hidden subgroup with high probability for the
Heisenberg group with $N=\{e\}$. Indeed, this shows that AKR will fail to
distinguish between $N = Z(G)$ and $N = \{e\}$, despite the fact that the
former is exponentially larger than the latter.
\end{example}
The probability distribution over irreps established by AKR in this example
(and many others) favors the small dimensional irreps, whereas the distribution
established by HRT favors the large dimensional irreps. In this example, that
fact prevents AKR from ever seeing the irreps needed to uniquely determine $N$.
(Paradoxically, when finding non-normal hidden subgroups, it is often the small
irreps that are most useful and HRT that struggles to find them.\footnote{While
the AKR distribution would be better, AKR does not apply to finding non-normal
hidden subgroups since the hypergroup structure is no longer present.})

\subsubsection{An application of AKR: a 2nd hypergroup algorithm for the HNSP}

While AKR may fail to uniquely determine $N$ from its samples, the following
lemma tells us that it will, with high probability, learn something about $N$.
\begin{lemma}
If $N \not= G$, then the intersection $K$ of the kernels of the irreps sampled by AKR is a strict subgroup of $G$ --- i.e., we will have $N \le K \lneq G$ --- with high probability.
\end{lemma}
\begin{proof}
The intersection of the kernels will be smaller than $G$ provided that at least
one of the samples has a kernel smaller than $G$. In other words, the
intersection will be a strict subgroup provided that at least one of the irreps
sampled is not the trivial irrep.

We can use eq.~(\ref{eq:AKR-prob1}) to calculate the probability of sampling
the trivial irrep. Since $\Hchi_1(C_x) = 1$ for any $C_x$, the probability for the
trivial irrep is just $\sum_{C_{xN} \in \overline{G/N}} \w{C_{xN}}^2 /
\varw{\overline{G/N}}^2$. If we let $G'$ denote the group $G/N$, then we can also
write this as $\sum_{C_{x'} \in \overline{G'}} \w{C_{x'}}^2 / \varw{\overline{G'}}^2$. Now,
our job is to determine how close this can be to 1.

Let $C_1, \dots, C_m$ be the conjugacy classes of $G'$, and define $s_i =
\w{C_i} / \varw{G'} = \abs{C_i} / \abs{G'}$. The $s_i$'s satisfy $\sum_i s_i = 1$
(since every element of $G'$ is in some conjugacy class). Since the size of a
conjugacy class divides the size of the group, we also know that $s_i \le 1/2$
for every $i$. Forgetting everything but these constraints, we can upper bound
the probability of sampling the trivial irrep by the solution of the
optimization problem
\begin{equation*}
\begin{aligned}
\text{maximize} \quad   \sum_i s_i^2,          
\qquad \text{subject to}  \qquad \sum_i s_i = 1, \qquad  s_i \le \tfrac{1}{2}  \qquad \text{for}\ i=1 \dots m.
\end{aligned}
\end{equation*}
The latter problem can be solved by ordinary methods of calculus. In
particular, it is easy to check that the objective is increased if $s_i$ and
$s_j$ are replaced by $s_i+\epsilon$ and $s_j-\epsilon$ provided that $s_i >
s_j$. (I.e., the derivative in this direction is positive.) Hence, the
objective will be maximized when the two largest $s_i$'s have value $1/2$ and
all other $s_i$'s are 0. At that point, the objective is $(1/2)^2 + (1/2)^2 =
1/2$.

This tells us that the probability of sampling the $\Hchi_1$ is at most $1/2$ on
each trial. Hence, the probability that all the samples are the trivial irrep
is exponentially small.
\end{proof}
This lemma tells us that AKR will find, with high probability, a strict
subgroup $K \lneq G$ such that $N \le K$. This means that we can reduce the
problem of finding $N$ hidden in $G$ to the problem of finding $N$ hidden in
$K$, which is a strictly smaller group. Provided that we understand the
representation theory of $K$ as well and, in particular, have a QFT for it,
then we can recursively solve this problem. In more detail, we have:
\begin{theorem}[\textbf{CC-HSHP is Easy, II}]
\label{thm:easy2}
Let $G$ be a group and $N$ a normal subgroup. Suppose that, for each normal subgroup $K$ satisfying $N \le K \le G$, we have a function $\overline{f}_K : \overline{K} \rightarrow \overline{H_K}$ that hides $\subhyp{N}{K}$.\footnote{Alternatively, we may assume that, for any $K \lhd G$, we have a  function $\overline{f}_K$ that hides $\subhyp{(N \cap K)}{K}$.} If we can efficiently compute the QFT for each such $\Conj{K}$, then there is an efficient quantum algorithm for the CC-HSHP.
\end{theorem}
As a corollary, we can see that there is an efficient hypergroup-based
algorithm for solving HNSP in the case when the oracle $f : G \rightarrow H$ is
a group homomorphism.
\begin{corollary}[\textbf{HNSP is easy, II}]\label{cor:HNSP easy II}
Let $G$ be a group. Suppose that we are given a hiding function $f : G
\rightarrow H$ for $N \lhd G$ that is a group homomorphism. If we can
efficiently compute the QFT for any normal subgroup $K$ satisfying $N \le K \le
G$ and compute efficiently with conjugacy classes for any $K$ and $H_K$,
where $H_K$ is the image of $f|_K$, then we can efficiently solve this HNSP.
\end{corollary}
\begin{proof}
The restriction of $f$ to $K$, $f|_K : K \rightarrow H_K$, is itself a group
homomorphism, so by the assumptions of the corollary and
theorem \ref{thm:reduction2}, we can efficiently compute from it a CC-HSHP
oracle $\overline{f}_K : \overline{K} \rightarrow \overline{H_K}$. Hence, the
result follows from theorem \ref{thm:easy2} and the results of appendix~\ref{app:CC implementation details} on implementing the QFT of a hypergroup with the QFT of the group.
\end{proof}
We finish this subsection comparing the quantum and classical requirements of our quantum algorithms in theorem \ref{thm:easy2}, corollary \ref{cor:HNSP easy II} to those of HRT's. 

\myparagraph{(i) Post-processing requirements:}
Similarly to HRT's, the quantum algorithms in theorem \ref{thm:easy2}, corollary \ref{cor:HNSP easy II} are information theoretic and are only fully-efficient if certain  subroutines are given (e.g., as  oracles) to carry out classical post-processing tasks. Specifically, the HRT quantum algorithm (step 6 above), relies on a subroutine to compute kernels of group irreps. In our case, it is easy to show that the \emph{same}  kernel-intersection subroutine is necessary and sufficient in order to compute a description of $\overline{K}$ (resp.\ $K$) in every iteration; hence, our algorithms and HRT's have \emph{identical} post-processing requirements\footnote{The proof of our claim is straightforward. Our algorithms compute kernels of measured irreps of $G$  when restricted to a normal subgroup in every iteration. An algorithm that intersects kernels of irreps can be used to find $\Ker(\nu|_K)$ since the latter equals
$\Ker \nu \cap K$ and $K$ itself can be written as a irrep-kernel intersection by
induction. Conversely, given a list of irreps, one can find the
intersection of their kernels  by repeatedly computing the
kernel of the next irrep restricted to the intersection of the kernels of those
previous.}.

\myparagraph{(ii) Reduction requirements (only for corollary \ref{cor:HNSP easy II})}: Our HNSP quantum algorithm in corollary \ref{cor:HNSP easy II} requires the  ability to \emph{efficiently} convert an oracle $f:G\rightarrow X$ for the HNSP into an oracle $\overline{f}:\overline{G}\rightarrow X'$ for the CC-HSHP by some procedure. E.g.\ we know that  such a procedure exists if $f$ is a group homomorphism and we can  compute with  conjugacy classes (section \ref{sect:HNSP}). 

It is important to note that any HNSP oracle can always be converted into an HSHP. Indeed, all one needs to convert $f$ into $\overline{f}$ is to make $f$ \emph{worse} by \emph{forgetting}  how to \emph{distinguish} two conjugate cosets $xN$, $x^a N$. Of course, this can be done classically in (at most)  exponential time; an open question (which we leave open to future research) is whether a \emph{polynomial-time} reduction between these problems always exists.

\myparagraph{(iii) Quantum-circuitry requirements:} The quantum steps of our algorithms above rely on 
\begin{itemize}
\item[(iii.a)] (In both algorithms) our ability to implement  QFTs over normal subgroups of $G$;
\item[(iii.b)] (In the CC-HSHP case) our ability to construct the intermediate oracle $\overline{f}$ for any $K$.
\end{itemize}
Requirement (iii.a) is not needed in the original HRT algorithm \cite{HallgrenRusselTaShma2003_Normal_Subgroup} but appears in our setting  because we use recursion. Requirement (iii.b) may be prohibitive for groups, yet, when fulfilled, it gives us a general quantum algorithm that works \emph{for any group~$G$}.

\emph{We highlight that Requirements (iii.a-iii.b) are not fundamental.} In next section, we give improved quantum algorithms for \emph{specific} groups and hypergroups (including nilpotent ones), exploiting  additional algebraic structure to bypass these two assumptions. Yet, our quantum algorithms in theorem \ref{thm:easy2}, corollary \ref{corollary:Coset State Preparations} are \emph{conceptually} simpler and group-independent at the cost of having these two extra assumptions.

\subsection{Efficient quantum algorithm for the nilpotent group HNSP and CC-HSHP}
\label{sect:New Algorithms}

In this section, we present our most sophisticated hypergroup-based quantum algorithms for the HNSP and the CC-HSHP. Unlike those in previous sections, our new algorithm does not require any extra assumptions about subgroups (such as hiding functions for them or the ability to compute efficiently with their conjugacy classes). Remarkably, our algorithm for the HNSP is fundamentally different from HRT's, showing that this central problem can be solved efficiently via a hypergroup approach if the hidden-subgroup oracle is a group homomorphism of a nilpotent group.

Our approach takes advantage of unique properties of hypergroups. In
particular, as we are looking to reduce our problem on $G$ to a subproblem, we
note that, for any normal subgroup $K \le G$, there are actually two smaller
hypergroups associated to it. The first is the hypergroup $\Conj{K}$ that we
get by looking at $K$ as a group separate from $G$. The second is the
subhypergroup $\Conj{K}_G$, where $\Conj{K}_G$ contains the conjugacy classes
$C_x \in \Conj{G}$ such that $x \in K$.  Above, when we recursively solved a
problem in $K$, we were using the hypergroup $\Conj{K}$. However, as we will
see in this section, it is also possible to solve the subproblem on the
subhypergroup $\Conj{K}_G \le \Conj{G}$.

The subhypergroup $\Conj{K}_G$ has two advantages over $\Conj{K}$. The first is
that a CC-HSHP oracle for $\Conj{G}$ is also a CC-HSHP oracle for $\Conj{K}_G$:
since the conjugacy classes of the two hypergroups are the same, the condition
that the oracle is constant on conjugacy classes of $\Conj{G}$ means that the
same is true for $\Conj{K}_G$. The second advantage of this subhypergroup is
given in the following lemma.
\begin{lemma}[\textbf{Subhypergroup QFT}]
\label{lma:subhypergroup-qft}
Let $K \lhd G$. If we can efficiently compute $\Fourier{\overline{G}}$, the QFT   over $\overline{G}$, then we can also efficiently compute  $\Fourier{\Conj{K}_G}$, i.e., the QFT over $\Conj{K}_G$.
\end{lemma}
\begin{proof}
In fact, the QFT for $\Conj{K}_G$ is implemented by the same QFT as for
$\Conj{G}$ provided that we choose the \emph{appropriate basis} for the dual of $\Conj{K}_G$.

To describe this basis, we first note that every character on $\Conj{G}$ is a character on $\Conj{K}_G$ simply by restricting its domain to $\Conj{K}_G$. This map of $\Conj{G} \rightarrow \Conj{K}_G$ is in fact surjective with kernel $\subhyp{K}{G}^\perp$ \cite{BloomHeyer95_Harmonic_analysis}. This means that the characters of $\Conj{K}_G$ are in 1-to-1 correspondence with the cosets of $\subhyp{K}{G}^\perp$ in $\Repr{G}$ (i.e., $\Conj{K}_G^* \simeq \Repr{G} / \subhyp{K}{G}^\perp$), so our basis for characters of $\Conj{K}_G$ should be a basis of cosets of $\subhyp{K}{G}^\perp$ in $\Repr{G}$. As described in the example of section~\ref{sect:Examples Normal Form CSS States}, the cosets of $\subhyp{K}{G}^\perp$ are of the form $\ket{\Hchi_\nu \subhyp{K}{G}^\perp} = \sum_{\Hchi_\mu \in \Hchi_\nu \subhyp{K}{G}^\perp} \sqrt{w_\mu / \varpi_{\Hchi_\nu \subhyp{K}{G}^\perp}} \ket{\Hchi_\mu}$. Note that $\varw{\Hchi_{\mu}\subhyp{K}{G}^\perp}=w_{\Hchi_{\mu}\mathcal{K}^\perp}\varw{\subhyp{K}{G}^\perp}=\varw{\Hchi_{\mu}\subhyp{K}{G}^\perp}\varw{\left(\overline{G}/K\right)^*}=\varw{\Hchi_{\mu}\subhyp{K}{G}^\perp}\varw{\overline{G}/K}=\varw{\Hchi_{\mu}\subhyp{K}{G}^\perp}\varw{\overline{G}}/\varw{\subhyp{K}{G}}$, using the definitions in section~\ref{sect:Hypergroups}. This means that we can write the state instead as \[ \ket{\Hchi_\nu \subhyp{K}{G}^\perp} = \sum_{\Hchi_\mu \in \Hchi_\nu \subhyp{K}{G}^\perp} \sqrt{\frac{w_\mu\, \varw{\subhyp{K}{G}}}{\w{\Hchi_{\mu}\subhyp{K}{G}^\perp} \varw{G}}} \ket{\Hchi_\mu}, \] which is the definition we will use below.

With that basis chosen, we can now calculate the Fourier transform of a conjugacy class state $\ket{C_x}$ with $C_x \in \Conj{K}_G$. The key fact we will use below is that $\Hchi_\mu(C_x) = \Hchi_\nu(C_x)$ whenever $\Hchi_\mu \in \Hchi_\nu \subhyp{K}{G}^\perp$ since they differ only by multiplication with a character that is identity on $C_x$ (lemma \ref{lemma:Quotient-Annihilator Isomorphisms}.(i)). Hence, we can define $\Hchi_{\nu}\subhyp{K}{G}^\perp(C_x)$ to be this common value.
\begin{eqnarray*}
\Fourier{\overline{G}} \ket{C_x}
 &=& \sqrt{\frac{\w{C_x}}{\varw{G}}} \sum_{\Hchi_\mu \in \Repr{G}} \sqrt{\w{\mu}} \Hchi_\mu(C_x) \ket{\Hchi_\mu} \\
 &=& \sqrt{\frac{\w{C_x}}{\varw{G}}} \sum_{\Hchi_\nu \subhyp{K}{G}^\perp \in \Repr{G}/\subhyp{K}{G}^\perp}  \Hchi_\nu \subhyp{K}{G}^\perp(C_x) \sum_{\mu \in \Hchi_\nu \subhyp{K}{G}^\perp} \sqrt{\w{\mu}} \ket{\Hchi_\mu} \\
 &=& \sqrt{\frac{\w{C_x}}{\varw{G}}} \sum_{\Hchi_\nu \subhyp{K}{G}^\perp \in \Repr{G}/\subhyp{K}{G}^\perp } \Hchi_\nu \subhyp{K}{G}^\perp(C_x) \sqrt{\frac{\w{\Hchi_{\nu}\subhyp{K}{G}^\perp} \varw{G}}{\varw{\subhyp{K}{G}}}} \ket{\Hchi_{\nu}\subhyp{K}{G}^\perp} \\
 &=& \sqrt{\frac{\w{C_x}}{\varw{\subhyp{K}{G}}}} \sum_{\Hchi_\nu \subhyp{K}{G}^\perp \in \Repr{G}/\subhyp{K}{G}^\perp} \sqrt{\w{\Hchi_{\nu}\subhyp{K}{G}^\perp}} \Hchi_\nu \subhyp{K}{G}^\perp(C_x) \ket{\Hchi_{\nu}\subhyp{K}{G}^\perp}
\end{eqnarray*}
Because $\subhyp{K}{G}^*\cong \Repr{G}/\subhyp{K}{G}^\perp$ and  $\varw{\subhyp{K}{G}}=\varw{\subhyp{K}{G}^*}$, this last line is, by definition, the QFT for $\Conj{K}_G$, so we have seen
that the QFT for $\Conj{G}$ implements this QFT as well.
\end{proof}
The lemma shows that the assumption that we have an efficient QFT for the whole hypergroup $\Conj{G}$ is sufficient to allow us to recurse on a subproblem on $\Conj{K}_G$ without having to assume the existence of another efficient QFT specifically for the subproblem, as occurred in our last algorithm.

Our next task is to analyze the AKR algorithm applied to this hypergroup and,
in particular, determine the probability distribution that we will see on
character cosets when we measure. Recall that the algorithm starts by preparing
 a weighted superposition over $\subhyp{K}{G}$ (which is a uniform distribution over $K$) and  invoking the oracle.
The result is
\[ \rho = \sum_{C_x\subhyp{N}{G} \in (\subhyp{N}{G}/\subhyp{K}{G})}\frac{ \w{C_x\Conj{N}_G} }{\varw{(\subhyp{K}{G}/\subhyp{K}{G})}}
\ket{C_x\Conj{N}_G} \bra{C_x\Conj{N}_G}. \]

As with AKR, we can find the Fourier transform of this state directly from
theorem~\ref{thm:Normal Form CSS States}. Following the same argument as
before, we see that the $\ket{\psi}$ from part (b) with $\Hchi_\varsigma = \Hchi_1$,
the trivial character, and $s=C_x$ is precisely the state
$\ket{C_x\overline{N}_G}$. Note that, since we are no longer working in $\overline{G}$ but the
subhypergroup $\overline{K}_G$, we do a QFT over $\overline{K}_G$ (lemma \ref{lma:subhypergroup-qft}) and the subhypergroup $\mathcal{N}^\perp=\subhyp{N}{G}^\perp$ in theorem~\ref{thm:Normal Form CSS States} belongs to
$\overline{K}_G^*$. 
\[ \Fourier{\overline{K}_G} \ket{C_x\overline{N}_G} =
    \sum_{\Hchi_\mu \in \subhyp{N}{G}^\perp \le \subhyp{K}{G}^*}
   \sqrt{\frac{\w{\Hchi_\mu} \w{C_x\overline{N}_G} }{\varw{\subhyp{N}{G}^\perp}}}
    \Hchi_\mu(C_x) \ket{\Hchi_\mu}, \]
By a similar calculation to before, we have
\begin{eqnarray*}
\Fourier{\overline{K}_G} \rho \Fourier{\overline{K}_G}^\dag &=&
   \sum_{\Hchi_\mu, \Hchi_\nu \in \subhyp{N}{G}^\perp}
   \sqrt{\w{\Hchi_\mu} \w{\Hchi_\nu}} 
 \sum_{C_x\overline{N}_G \in (\subhyp{K}{G}/\subhyp{N}{G})}
   \frac{\w{C_x\overline{N}_G}^2}{\varw{\subhyp{N}{G}^\perp}^2} \Hchi_\mu(C_x) \Hchi_{\bar{\nu}}(C_x)
   \ket{\Hchi_\mu}\bra{\Hchi_\nu}.
\end{eqnarray*}
Finally, using $\subhyp{N}{G}^\perp\cong(\subhyp{K}{G}/\subhyp{N}{G})^*\cong (\subhyp{K/N}{G/N})^*$ (lemma \ref{lemma:Roth}),  we conclude that the probability of measuring $\Hchi_\mu\in\subhyp{N}{G}^\perp$ is
\begin{align}
\Pr(\Hchi_\mu) &= \w{\Hchi_\mu} \sum_{C_x\subhyp{N}{G} \in (\subhyp{K}{G}/\subhyp{N}{G})}
    \frac{\w{C_x\overline{N}_G}^2}{\varw{ (\subhyp{K}{G}/\subhyp{N}{G})}^2}
    \Hchi_\mu(C_x) \overline{\Hchi_{\mu}}(C_x)\notag\\
    &= \w{\Hchi_\mu} \sum_{C_{xN} \in ({\subhyp{K/N}{G/N}})}
    \frac{\w{C_{xN}}^2}{\varw{\subhyp{K/N}{G/N}}^2}
    \Hchi_\mu(C_{xN}) \overline{\Hchi_{\mu}}(C_{xN}),\label{eq:AKR-prob2}
\end{align}
which is analogous to what we saw in equation~(\ref{eq:AKR-prob1}).

With this in hand, we can now prove the following result for \emph{$p$-qroups} \cite{Humphrey96_Course_GroupTheory}, i.e., groups whose order is a power of a prime number $p$.
\begin{lemma}[\textbf{HSHP over \emph{p}-groups}]\label{lma:p-group}Let $G$ be a \emph{$p$-group} (where $p$ is prime). Suppose that we are given a hiding function $\overline{f} : \Conj{G} \rightarrow \{0,1\}^*$. If we can efficiently compute the QFT for $\overline{G}$ and we can efficiently compute the kernels of irreps when restricted to subgroups, then there is an efficient quantum algorithm for the CC-HSHP.
\end{lemma}
\begin{proof}
We follow a similar approach to before, applying AKR to subhypergroups
$\Conj{K}_G$ (starting with $K=G$) until we measure an irrep $\nu$ with
kernel smaller than $K$ and then recursing on $\subhyp{J}{G} \le \subhyp{K}{G}$, where $J =
\Ker(\nu|_K)$. By assumption, we can compute $\Ker(\nu|_K)$ efficiently. If we
fail to find such a $\nu$ in polynomially many samples, then we can conclude
that $K$ is the hidden subgroup with high probability.

By lemma \ref{lma:subhypergroup-qft} and the notes above it, our assumptions
imply that we have an oracle and an efficient QFT for each subproblem. To
implement AKR, we also need the ability to prepare a uniform superposition over
$K$. This was implemented earlier using the inverse Fourier transform. In this
case, that would require us to prepare a complicated coset state in
$\Repr{G}$. However, we can instead just prepare the superposition directly
by the result of Watrous \cite{Watrous_solvable_groups}.\footnote{This result
holds for black-box groups, so we only need to assume that we know the kernel
of each irrep not that we can intersect the kernels of arbitrary irreps.}

Finally, it remains to prove that we have a good probability (e.g., at least
$1/2$) of measuring a nontrivial irrep or, equivalently, that the probability
of measuring the trivial irrep is not too large (e.g., at most $1/2$). As
before, this amounts to putting a bound on $\sum_{C_{xN} \in(\subhyp{K/N}{G/N})}
(\w{C_{xN}} / \varw{\subhyp{K/N}{G/N}})^2$, this time by equation~(\ref{eq:AKR-prob2}). By
the same argument as before, this will hold if we can show that $\w{C_x\subhyp{N}{G}} /
\varw{\Conj{K/N}}$ is bounded by a constant less than 1.  Earlier, we showed this
by using the fact that the size of a conjugacy class divides the size of the
group.  Unfortunately, that does not help us here because we are comparing
$\w{C_x\subhyp{N}{G}}=|C_{xN}|$ not to the size of $G/N$ but to the size of the subgroup $K/N$. These
two need not be related by even a constant factor. In extreme cases, $K/N$ may
contain only one group element that is not in $C_{xN}$
\cite{math-exchange-one-non-identity-element}.

Instead, we will use properties of $p$ groups. Since $G$ is a $p$-group, so is
$K/N$, and the size of $K/N$ must be $p^k$ for some $k$. Since the size of a
conjugacy class divides the size of a group (this time $G/N$), the size of
$C_{xN}$ in $K/N$ must be $p^j$ for some $j$. Now, we must have $j \le k$ since $C_{xN} \subset \subhyp{K/N}{G/N}$. However, we cannot have $j=k$ unless $N = K$, which we have assumed
is not the case, since $K/N$ must contain at least two classes (one identity
and one non-identity). Hence, we can conclude that 
$\w{C_{xN}}=\abs{C_{xN}}$ is smaller than
$\varw{\subhyp{K/N}{G/N}}=|K/N|$ by at least a factor of $p \ge 2$. We conclude as before that the
probability of measuring the trivial irrep is at most $1/2$.
\end{proof}
\begin{theorem}[\textbf{CC-HSHP Is Easy, III}]
\label{thm:easy3}
Let $G$ be a nilpotent group. Suppose that we are given a hiding function
$\overline{f} : \Conj{G} \rightarrow \{0,1\}^*$. If we can efficiently compute
the QFT for $G$ and we can efficiently compute the kernels of irreps when
restricted to subgroups, then there is an efficient quantum algorithm for the
CC-HSHP.
\end{theorem}
\begin{proof}
A nilpotent group is a direct product of $p$-groups for different primes
\cite{Dummit_abstract_algebra}. This means that any subgroup must be a direct
product of subgroups, one in each of the $p$-groups.\footnote{This is, for
example, an immediate consequence of Goursat's Lemma \cite{Lang_algebra} (since
a quotient of a $p$-group and a quotient of a $q$-group, with $q \not= p$, can
only be isomorphic if they are both trivial groups).} Hence, it suffices to
solve the CC-HSHP in each of these $p$-groups.
\end{proof}
Finally, we can apply this result to the HNSP if we are given an oracle that
can be converted into a CC-HSHP oracle.
\begin{corollary}[\textbf{HNSP is easy, III}]\label{cor:HNSP easy III} Let $G$ be a nilpotent group. Suppose that we are given a hiding function $f :
G \rightarrow H$ that is a homomorphism. If we can efficiently implement the QFT
for $G$, compute with conjugacy classes of $G$ and $H$, and compute  kernels
of irreps when restricted to subgroups, then there is an efficient quantum
algorithm for the HNSP.
\end{corollary}
We note that the class of nilpotent groups includes the Heisenberg group,
which, as we saw in Example~\ref{ex:non-convergence}, is a case where the
original AKR algorithm does not find the hidden normal subgroup with
polynomially many samples. Our last algorithm, however, solves the problem
efficiently in this case.

\paragraph{Minimal requirements:} We highlight that our final quantum algorithms (theorem \ref{thm:easy3}, corollary \ref{cor:HNSP easy III}) work under a minimal amount of assumptions compared to those in last  section:
\begin{itemize}
 \item Our final quantum algorithm for CC-HSHP (theorem \ref{thm:easy3}) is provably efficient given \textbf{(a)} a \emph{single} circuit to implement the  QFT over $G$, due to lemma \ref{lma:subhypergroup-qft}, and \textbf{(b)} an HRT kernel-intersection  subroutine---requirement (i) in last section). In particular, this algorithm does not need the additional  requirements (iii.a-iii.b) of theorem \ref{thm:easy2}, hence,  runs efficiently under the same assumptions as HRT's.
 
 \item Our final quantum algorithm for HNSP (corollary \ref{cor:HNSP easy III}) is provably efficient given requirement \textbf{(ii)} (the HNSP oracle is efficiently convertible into an HNSP oracle); and, again,  \textbf{(a)} an \emph{single} circuit to implement the  QFT over $G$, due to lemma \ref{lma:subhypergroup-qft}, and \textbf{(b)} an HRT kernel-intersection  subroutine. Requirements (iii.a-iii.b) are no longer needed.
\end{itemize}

\myparagraph{Conclusion about the HNSP:} Altogether, the results in this section show that one can solve the HNSP by reducing it to CC-HSHP in many settings under reasonable assumptions, the most significant one being (in the view of the authors)  the need to find an oracle conversion protocol (which we handled restricting to group-homomorphism hiding functions). In such scenarios, we show that  the fact that the HNSP can be solved efficiently depends crucially on the fact that the hypergroups in the CC-HSHP is abelian. Thus, the results of the last two sections give us an explanation for why the HNSP is easy (in a wide range of settings) because of the presence of an \emph{abelian} algebraic structure, the abelian hypergroup of conjugacy classes.

Remarkably, though  both HRT's quantum algorithm and ours rely on equivalent  assumptions in the classical post-processing step,  the quantum parts of the two algorithms are \emph{fundamentally different}. Hence our  algorithm demonstrates a new way in which this important problem can be solved efficiently by quantum computers.

Finally, we mention that because of the very special mathematical structure that is common to both the HRT and AKR oracles---see requirement (ii) and section \ref{sect:AKR Analysis}---we are optimistic about the possibility of extending our results beyond the homomorphism-oracle setting\footnote{In fact, similarly to the abelian HSP setting (cf.\ chapter \ref{sect:Abelian HSPs}, theorem \ref{thm:HSP}) it can easily be shown that hiding-subgroup promise of the HRT oracle  $f:G\rightarrow X$ induces a group structure on $X$---an isomorphism onto $G/N$---such that $f:G\rightarrow X$ is a group homomorphism. A potential approach to extend our results would be to search for a method to exploit this hidden group-homomorphism structure.}.

\subsection{Further results and open problems}
\label{sect:Open Problems}

We finish this section with some discussion on whether this last result can be extended further. The most natural next step beyond nilpotent groups would be to show that the algorithm works for super-solvable groups. We start with a positive example in that direction.

\begin{example}[\textbf{Dihedral Groups}]
As we saw above, the algorithm will work correctly provided that the
probability of measuring the trivial irrep is not too close to 1. By
equation~(\ref{eq:AKR-prob2}), this is given by $\sum_{C_{xN} \in(\subhyp{K/N}{G/N})}
(\w{C_{xN}} / \varw{\subhyp{K/N}{G/N}})^2$.

Without loss of generality, we may assume $N=\{e\}$ by instead looking at the
group $G/N$. For a dihedral group, such a quotient is either dihedral or
abelian. Since abelian groups are (trivially) nilpotent, we know the algorithm
works in that case already.

By our earlier arguments, the probability $\sum_{C \in (\Conj{K})_G} (\w{C} /
\varw{\Conj{K}})^2$ is bounded by a constant below one provided that the
fractional weights $\w{C} / \varw{\Conj{K}}$ are bounded by a constant below one. In
other words, our only worry is that there is a normal subgroup $K$ containing a
conjugacy class that is nearly as large as $K$.

Let us consider the dihedral group of order $2n$, generated by a rotation $a$
of order $n$ and a reflection $r$ of over 2. Most of the normal subgroups are
contained in the cyclic subgroup $\langle a \rangle$. These are subgroups of
the form $\langle a^d \rangle$ with $d$ dividing $n$. Every conjugacy class in
this subgroup contains either 1 or 2 elements (since $r^{-1} a^j r = a^{-j}$
and hence $r^{-1} a^{-j} r = a^j$). Since a nontrivial normal subgroup cannot
consist of one conjugacy class, the worst case would be when $K$ has 3 elements
and contains a conjugacy class with 2 elements. In that case, the probability
of measuring the trivial irrep could only be as large as $2/3$, which still a
constant (independent of $n$) less than one\footnote{Note that if $\w{C_{x}} / \varw{\subhyp{K}{G}}\leq c$, then character orthogonality  (\ref{eq:Character Orthogonality Subhypergroups}) lets us bound the probability of measuring  a character $\Hchi_\mu\in\mathcal{T}^*$  (\ref{eq:AKR-prob2}) since $\mathrm{Pr}(\Hchi_\mu)/\ws{\Hchi_\mu}= \sum_{C_{x} \in\subhyp{K}{G}}
(\w{C_{x}} / \varw{\subhyp{K}{G}})^2 |\mathcal{\Hchi}_\mu(C_x)|^2\leq c (\sum_{C_{x} \in\subhyp{K}{G}}
 \w{C_{x}} / \varw{\subhyp{K}{G}}|\mathcal{\Hchi}_\mu(C_x)|^2)=c/\ws{\Hchi_\mu\subhyp{K}{G}^\perp}$. Specifically, for any invertible character we get $\mathrm{Pr}(\Hchi_\mu)\leq c$.}.

If $n$ is odd, then any normal subgroup $K$ containing $r$ is the whole group,
and the largest conjugacy class contains every $a^j r$ for $j \in \Integer_n$,
which is half the elements, so we get a bound of $1/2$ in that case. If $n$ is
even, then there are two more normal subgroups, one containing $a^{2j} r$ for
each $j$ and one containing $a^{2j+1} r$, but both also contain all rotations
of the form $a^{2j}$, so at least half of the elements in these subgroups
are contained in 1--2 element conjugacy classes, and once again we get a bound
of $1/2$.

All together, this shows that the probability of measuring the trivial irrep is
at most $2/3$ for the dihedral groups, so the algorithm will succeed with high
probability.
\end{example}
On the other hand, we also have a negative example.

\begin{example}[\textbf{Super-solvable group} \cite{math-exchange-one-non-identity-element}]
We will consider the group of simple affine transformations over $\Integer_p$.
These are transformations of the form $x \mapsto ax + b$ for some $a \in
\Integer_p^\times$ and $b \in \Integer_p$, which we denote by $(a,b)$. These
form a group under composition.  In particular, applying $(a,b)$ and then
$(c,d)$ gives $acx + bc + d$, which shows that $(c,d) \cdot (a,b) = (ac, bc +
d)$. A simple calculation shows the formula for the commutator \[[(a,b),(c,d)] = (1,
c^{-1}(1-a^{-1})d - a^{-1}(1 - c^{-1})b).\] This implies that the commutator
subgroup $[G,G]$ is contained in the set $\set{(1,b)}{b \in \Integer_p}$. On
the other hand, taking $a = 1$, $c = 2^{-1}$, and $d = 0$ in this formula gives
the result $(1,b)$, so $[G,G]$ must contain all the elements of this set. If
we mod out $[G,G]$, then we are left with the  abelian group
$\Integer_p^\times$. We have proven that the group is super-solvable.

On the other hand, for any element $(1,d)$, taking $a=2^{-1}, c=1$ and $b=0$ in the
formula above gives the result $(1,-d)$, which is not the identity $(1,0)$.
This means that the group has a trivial center, and thus, it cannot be
nilpotent.

Another simple calculation shows that conjugating $(1,b)$ by $(c,0)$ gives us
$(1,c^{-1}b)$. Hence, the conjugacy class of $(1,b)$ with $b\neq 0$ contains every $(1,b')$ with $b' \not= 0$. This is all of the subgroup $[G,G]$ except for the identity
element $(1,0)$. Hence, once we have $K=[G,G]$, we can see by
equation~(\ref{eq:AKR-prob2}) with $N=\{e\}$ that the algorithm will get the
trivial irrep with high probability, so we can see that the algorithm will fail
to find $N$ in this case.
\end{example}

Put together, these results show that our last algorithm works for some
non-nilpotent, super-solvable groups (like the dihedral groups\footnote{It is
super-solvable since it is a semi-direct product of abelian   groups, and it is
easy to check that it is not nilpotent unless $n$ is a power of 2.}), but not
all super-solvable groups since it fails on the affine linear group.
Determining exactly which super-solvable groups the algorithm does succeed on
is an open problem.

%% file: appendix2_infinite_chapter.tex
\chapter{Supplement to  chapter \ref{chapterGT}}\label{aGT}

\section{Supplementary material for section \ref{sect:Homomorphisms and matrix representations}}\label{appendix:Supplement to section Homomorphisms}

\subsection*{Proof of lemma \ref{lemma:properties of matrix representations}}

First we prove (a). Note that it follows from the assumptions that $\alpha(g+h)=A(g+h)=Ag+Ah\pmod{H}$, $\beta(x+y)=B(x+y)= Bx+By \pmod{J}$, for every $g,h\in G$, $x,y\in H$. Hence, $\beta\circ\alpha(g+h)\allowbreak=\beta(Ag+Ah+\textnormal{zero}_H)=BAg+BAh+ \textnormal{zero}_J\pmod{J}$, where zero$_X$ denotes some string congruent to the neutral element $0$ of the group $X$. As in the last equation zero$_J$ vanishes modulo $J$, $BA$ is a matrix representation of $\beta\circ\alpha$.

We prove (b). From the definitions of character, bullet group and bullet map it follows that
\begin{equation}\label{inproof:Properties Matrix Reps 1}
\chi_{\mu}(\alpha(g))=\exp\left(2\pii\sum_{ij} \mu^\bullet(i)A(i,j)g(j)\right)=\exp\left(2\pii(A^\transpose \mu^\bullet)\cdot g\right) \:\textnormal{for every $g\in G$.}
\end{equation}

Let $f$ be the function $f(g):=\exp\left(2\pii(A^\transpose \mu^\bullet)\cdot g\right)$. Then  it follows from (\ref{inproof:Properties Matrix Reps 1})  that $f$ is continuous and that $f(g+h)=f(g)f(h)$, since the function $\chi_\mu\circ \alpha$ has these properties. As a result,  $f$ is a continuous character \ $f=\chi_\nu$, where $\nu\in G^*$ satisfies $\nu^\bullet=\Upsilon_G \nu = A^\transpose \mu^\bullet\pmod{G^\bullet}$. Moreover, since $f=\chi_\mu\circ\alpha = \chi_{\alpha^*(\mu)}$ it follows that  $\alpha^*(\mu)=\nu\pmod{G^*}$ and, consequently, 
\begin{equation}
 \alpha^*(\mu)= \Upsilon_G^{-1} (A^\transpose \mu^\bullet)\pmod{G^*}= \Upsilon_G^{-1} A^\transpose \Upsilon_H\mu\pmod{G^*}. 
\end{equation}
Finally, since $\chi_\mu(\alpha(g))=\chi_x (\alpha(g))$ for any $x\in\R^n$ congruent to $\mu$, we  get that $\alpha^*(\mu)=\Upsilon_G^{-1} A^\transpose \Upsilon_Hx\pmod{G^*}$ for any such $x$, which proves the second part of the lemma.\qed

\subsection*{Proof of lemma \ref{lemma:existence of matrix representations}}\label{proof:lemma existence of matrix representations}

We will show that each of the homomorphisms $\alpha_{XY}$ as considered in lemma \ref{corollary:block-structure of Group Homomorphisms} has a matrix representation, say $A_{XY}$. Then it will follow from (\ref{eq:block decomposition of a homomorphism}) in lemma \ref{corollary:block-structure of Group Homomorphisms} that 
\begin{equation}
A:= \begin{pmatrix}
      A_{\Z\Z} & 0 & 0 & 0 \\
      A_{\R\Z} & A_{\R\R} & 0 & 0 \\
      A_{F\Z} & 0 & A_{FF} & 0 \\
      A_{\T\Z} & A_{\T\R} & A_{\T F} & A_{\T\T}.
    \end{pmatrix},
\end{equation}
as in (\ref{eq:block-structure of matrix representations}), is a matrix representation of $\alpha$.

First, note that if the group $Y$ is finitely generated, then the tuples  $e_i$ form a generating set of $Y$. It is then  easy to find a matrix representation $A_{XY}$ of $\alpha_{XY}$:  just choose the $j$th column of $A_{XY}$ to be  the element $\alpha(e_j)$ of $X$. Expanding  $g=\sum_i g(i)e_i$ (where the coefficients $g(i)$ are integral), it easily follows that $A_{XY}$ satisfies the requirements for being a proper matrix representation as given in definition \ref{def:Matrix representation}. Thus, all homomorphisms $\alpha_{XY}$ with $Y$ of the types $\Z^a$ or $F$ have matrix representations; by duality and lemma \ref{lemma:properties of matrix representations}(b), all homomorphisms $\alpha_{XY}$ with $X$ of type $\T^a$ or $F$ have matrix representations too.

The only non-trivial  $\alpha_{XY}$ left to consider is $\alpha_{\R\R}$. Recall that the latter is a continuous map from $\R^m$ to $\R^n$ satisfying $\alpha_{\R\R}(x+y)= \alpha_{\R\R}(x) + \alpha_{\R\R}(y)$ for all $x, y\in \R$. We claim that every such map must be linear, i.e. in addition we have \be \label{linearity}\alpha_{\R\R}(rx)= r\alpha_{\R\R}(x)\ee for all $r\in \R$. To see this, first note that $d \alpha_{\R\R}(kx/d )=k \alpha_{\R\R}(x)$, where $k/d$ is any  fraction ($k$, $d$ are integers). Thus  (\ref{linearity}) holds for all rational numbers $r=k/d$.  Using that $\alpha_{\R\R}$ is continuous and that the rationals are dense in the reals then implies that (\ref{linearity}) holds for all $r\in \R$. This shows that $\alpha_{\R\R}$ is a linear map; the existence of a matrix representation readily follows. \qed

\subsection*{Proof of lemma \ref{lemma:Normal form of a matrix representation}}\label{proof:lemma Normal form of a matrix representation}

It suffices to show (a), that any matrix representation $A$ of $\alpha$ must be an element of $\mathrm{Rep}$ and fulfill the consistency conditions (\ref{eq:Consistency Conditions Homomorphism}); (b), that  these consistency conditions imply that $A$ is of the form (\ref{eq:block-structure of matrix representations}) and fulfills propositions 1-4; and (c), that  every such matrix defines a group homomorphism.

We will first prove (a). Let $H_i$ is of the form $\Z$ or $\Z_{d_i}$. Then, for every $j=1,\ldots,m$, the definition of matrix representation \ref{def:Matrix representation} requires that  $(Ae_j)(i)=A(i,j)\pmod{H_i}$ must be an element of $H_i$. This shows that the $i$th row of $A$ must be integral and, thus, $A$ belongs to $\mathrm{Rep}$. Moreover, since $x:=c_je_j\equiv 0\pmod{G}$ and $y:=d^*_i e_i \equiv 0 \pmod{H^*}$, (due to the definition of characteristic) it follows that $Ax=0\pmod{H}$ and $Ay=0\pmod{G^*}$, leading to the consistency conditions (\ref{eq:Consistency Conditions Homomorphism}). 

Next, we will now prove (b).

First, the block form (\ref{eq:block-structure of matrix representations}) almost follows from (\ref{eq:block decomposition of a homomorphism}) in lemma \ref{corollary:block-structure of Group Homomorphisms}: we only have to show, in addition, that the zero matrix is the only valid matrix representation for any trivial group homomorphism $\alpha_{XY}=\mathpzc{0}$ in (\ref{eq:block decomposition of a homomorphism}). It is, however, easy to check case-by-case that, if $A_{XY}$ is a matrix representation of $\alpha_{XY}$ with $A_{XY}\neq 0$, then $\alpha_{XY}$ cannot be trivial.

Second, we prove propositions 1-4. In proposition 1, $A_{\Z\Z}$ must be integral since  $A_{\Z\Z}e_j(i)\in\Z$  (where, with abuse of notation, $i,\,j$ index the rows and columns of $A_{\Z\Z}$). By duality the same holds for $A_{\T\T}$ (it can be shown using  lemma \ref{lemma:properties of matrix representations}(b)). In proposition 2 the consistency conditions (including dual ones) are vacuously fulfilled and tell us nothing about  $A_{\R\Z}$, $A_{\R\R}$. In proposition 3, both matrices have to be integral to fulfill that $A_{XY}(e_i)\pmod{Y}$ is an element of $X$, which is of type $\Z^a$ or $F$; moreover, for $Y=F$, the  consistency conditions directly impose that the coefficients must be of the form (\ref{eq:coefficients of Matrix Rep for nonzero characteristic groups}), due to basic properties of linear congruences (see e.g.\ lemma 11 in \cite{BermejoVega_12_GKTheorem} for a similar derivation.) Lastly, in proposition 4, all consistency conditions associated to $A_{\T \Z}$ and $A_{\T\R}$ are, again, vacuous and tell us nothing about the matrix; however, the first consistency condition tells us that $A_{\T F}$ has rational coefficients of form $\alpha_{i,j}/c_j$.

Finally we will show (c). First, it is manifest that if $A$ fulfills 1-4 then $A\in \mathrm{Rep}$. Second, to show that $A$ is a matrix representation of a group homomorphism it is enough to prove that every $A_{XY}$ fulfilling 1-4 is the matrix representation of a group homomorphism from $Y$ to $X$.  This can be checked straightforwardly for the cases  $A_{\Z\Z}$,  $A_{\R\Z}$, $A_{\R\R}$,  $A_{F\Z}$, $A_{\T\Z}$,  $A_{\T\R}$ applying properties 1-4 of $A$ and using that, in all cases, there are no non-zero real vectors congruent to the zero element of $Y$. Obviously, for the cases where $A_{XY}$ must be zero the proof is trivial. It remains to consider the cases $A_{FF}$, $A_{\T F}$, $A_{\T\T}$. In all of this cases, it holds due to properties 1,3,4 that the  first consistency condition in (\ref{eq:Consistency Conditions Homomorphism}) is fulfilled. We prove the remaining cases in a single step, by letting $G'=G_F\times G_\T$, $H'=H_F\times H_\T$ and $A'=\begin{pmatrix}
A_{FF} & 0 \\ A_{\T F} & A_{\T\T}
\end{pmatrix}$  and showing that $A':G'\rightarrow H'$ is a homomorphism given (\ref{eq:Consistency Conditions Homomorphism}). To this end, we let $a_i'$ denote the $i$th column of $A'$,  $m'$ be the total number of columns; with this notation, we evaluate the action of $A'$  on   $g, h, g+h\in G'$ \emph{without} taking remainders to be
\begin{equation}\label{xxx proof matrix representation - homomorphism part}
A'g + A'h = \sum [g(i)+h(i)]a_i' ,  \qquad\qquad  A'(g+h)=\sum (g+h)(i) a_i'.
\end{equation}
Recalling associativity of $H$ and $G$, the latter expression shows that $A'h$ defines a function from $G'$ to $\Z^{m'}$, and, thus, $A'h \pmod{H'}$ is a function from $G'$ to $H'$. Last, it holds for every $i$ that $g(i)+h(i)=q_i c_i+(g+h)(i)$ for some integers  $q_i$ , since (by definition of the group $G$)  $(g+h)(i)$ is the remainder obtained when $g(i)+h(i)$ is divided by $c_i$ ($q_i$ is  the quotient). It follows, subtracting modularly, that $A'(g)+A'(h)-A'({g+h})=\sum_i q_i c_i a_i' = 0 \pmod{H'}$ for every $g$, $h$, using (\ref{eq:Consistency Conditions Homomorphism}); it follows that $A'$ (hence $A_{FF}$, $A_{\T F}$, $A_{\T\T}$) are group homomorphisms.
\qed

\section{Existence of general-solutions of systems of the form (\ref{eq:Systems of linear equations over groups})}\label{appendix:Closed subgroups of LCA groups / existence of general-solutions of linear systems}

In this section we  show that general-solutions of systems of linear equations over elementary abelian groups always exist (given that the systems admit at least one solution). 

We start by recalling an important property of elementary abelian groups.
\begin{lemma}[{See theorem 21.19 in \cite{Stroppel06_Locally_Compact_Groups} or section 7.3.3 in \cite{Dikranjan11_IntroTopologicalGroups}}]\label{lemma:Elementary_LCA_group property extends to subgroups, quotients, products}
The class of elementary abelian groups\footnote{Beware that in  \cite{Stroppel06_Locally_Compact_Groups} the class of elementary groups is referred as ``the category CGAL'', which stands for Compactly Generated Abelian Lie groups.} is closed with respect to forming closed subgroups, quotients by these, and finite products.\footnote{In fact, as mentioned in \cite{Stroppel06_Locally_Compact_Groups}, corollary 21.20 elementary LCA groups constitute the smallest subclass of LCA containing $\R$ and fulfilling all these properties.}
\end{lemma}
In our setting, the kernel of a continuous group homomorphism $A:G\rightarrow H$ as in (\ref{eq:Systems of linear equations over groups}) is always closed: this follows from the fact that the singleton $\{0\}\subset H$ is closed (because elementary abelian groups are Hausdorff  \cite{Stroppel06_Locally_Compact_Groups}), which implies that $\ker{A}=A^{-1}(\{0\})$ is closed (due to continuity of $A$). Hence, it follows from lemma \ref{lemma:Elementary_LCA_group property extends to subgroups, quotients, products} that $\ker{A}$ is topologically isomorphic to some elementary abelian group  $H':=\R^a\times\T^b \times \Z^c \times \DProd{N}{c}$; consequently, there exists a continuous group isomorphism  $\varphi$ from $H'$ to $H$.

Next, we write the group $H'$ as a quotient group $X/K$ of the group $X:=\R^{a+b}\times \Z^{c+d}$ by the subgroup $K$ generated by the elements of the form $\mathrm{char}(X_i)e_i$. The quotient group $X/K$ is the image of the quotient map $q:X\rightarrow X/K$ and the latter is a continuous group homomorphism \cite{Stroppel06_Locally_Compact_Groups}. By composing  $\varphi$ and $q$ we obtain a continuous group homomorphism $\mathcal{E}:=\varphi\circ  q$ from $X$ onto $H$. The map $\mathcal{E}$ together with any particular solution $x_0$ of (\ref{eq:Systems of linear equations over groups}) constitutes a general solution of (\ref{eq:Systems of linear equations over groups}), proving the statement.

\section{Proof of theorem \ref{thm:General Solution of systems of linear equations over elementary LCA groups}}\label{appendix:Systems of Linear Equations over Groups}

In this appendix, we prove theorem giving efficient classical algorithms for tasks (1-4).

\subsection{Algorithms for tasks (1-2)}

We show how to decide the existence of and find general solutions of the system $Ax=b\pmod{H}$. Our first step is to show that systems of linear equations over groups (\ref{eq:Systems of linear equations over groups}) can be reduced to systems of mixed real-integer linear equations (\ref{eq:System of Mixed Integer linear equations}). This is proven next.

Start with two elementary groups of general form $G$, $H$. First, notice that we can write $G$ and $H$ as $G=G_1\times\cdots\times G_m$, $H=H_1\times \cdots \times H_n$ where each factor $G_i$, $H_j$ is of the form $G_i=\textbf{X}_i/c_i\Z$, $H_j=\textbf{Y}_i/d_i\Z$ with $\textbf{X}_i$, $\textbf{Y}_i\in \{\Z, \R\}$; the numbers $c_i$, $d_j$ are the characteristics of the primitive factors. We assume w.l.o.g.\ that the primitive factors of $G$, $H$ are \emph{ordered} such that both groups are of the form $\Z^a\times F\times \T^b$: in other words, the finitely generated factors come first.

We now define a new group $\textbf{X}:=\textbf{X}_1\times\cdots\times \textbf{X}_m$ (recall that with the ordering adopted $X$ is of the form $\Z^a\times \R^b$) which will play the role of an \emph{enlarged solution space}, in the following sense. Let $\textbf{V}$ be the subgroup of $\mbf{X}$ generated by the elements $c_1e_1,\ldots, c_me_m$. Observe that the group $G$---the solution space in system (\ref{eq:Systems of linear equations over groups})---is precisely the quotient group $\mbf{X}/\mbf{V}$, and thus can be \emph{embedded} inside the larger group $\textbf{X}$ via the quotient group homomorphism $\textbf{q}:\textbf{X}\rightarrow G= \mbf{X}/\mbf{V}$:
\begin{equation}
\textbf{q}(\textbf{x}):=(\textbf{x}(1)\bmod{c_1}, \ldots, \textbf{x}(m)\bmod{c_m})=\textbf{x}\pmod{G};
\end{equation}
remember also that $\ker{\textbf{q}}=\textbf{V}$. Now let $\alpha:\textbf{X}\rightarrow H$ be the group homomorphism defined as $\alpha:=A\circ \textbf{q}$. Then it follows from the definition that $\alpha(\mathbf{x})=A\mathbf{x}\pmod{H}$, and $A$ is a  matrix representation of $\alpha$. (This is also a consequence of the composition property of matrix representations (lemma \ref{lemma:properties of matrix representations}.(a), since the $m\times m$ identity matrix $I_m$ is a matrix representation of $\textbf{q}$.) We now consider the relaxed\footnote{Notice that the new system is less constrained, as we look for solutions in a larger space than beforehand.} system of equations
\begin{equation}\label{eq:Enlarged system of equations 1}
\alpha(\textbf{x})=b\pmod{H},\quad  \textnormal{where } \textbf{x}\in \textbf{X}=\Z^a\times \R^b.
\end{equation}
Note that the problem of solving (\ref{eq:Systems of linear equations over groups}) reduces to solving (\ref{eq:Enlarged system of equations 1}), which  looks closer to a system of mixed real-integer linear equations. Indeed, let   $\textbf{X}_{\textnormal{sol}}$ denote the set  of all solutions  of system  (\ref{eq:Enlarged system of equations 1}); then\footnote{It is easy to prove $G_{\textnormal  {sol}}=\mathbf{q}(\mathbf{X}_{\textnormal{sol}})$ by showing $G_{\textnormal  {sol}}\supset \mathbf{q}(\mathbf{X}_{\textnormal{sol}})$ and the reversed containment for the preimage $\textbf{q}^{-1}(G_{\textnormal  {sol}})\subset\mathbf{X}_{\textnormal{sol}}$; then surjectivity of $\textbf{q}$ implies  $G_{\textnormal{sol}}=\textbf{q}(\textbf{q}^{-1}(G_{\textnormal{sol}}))\subset \textbf{q}(\textbf{X}_\textnormal{sol})$.}
\begin{equation}\label{eq:Relationship between solutions of original and enlarged system}
G_{\textnormal  {sol}}=\mathbf{q}(\mathbf{X}_{\textnormal{sol}})\quad\Longrightarrow\quad  G_{\textnormal  {sol}}=\textbf{q}(\textbf{x}_0)+\textbf{q}(\ker{\alpha})= \textbf{x}_0+\ker{\alpha}\pmod{G},
\end{equation}
Hence, our original system (\ref{eq:Systems of linear equations over groups}) admits solutions iff (\ref{eq:Enlarged system of equations 1}) also does, and the former can be obtained from the latter via the homomorphism $\mathbf{q}$. We further show next that (\ref{eq:Enlarged system of equations 1}) is equivalent to a system of  form (\ref{eq:System of Mixed Integer linear equations}). First, note that the matrix $A$ has a block form $A=\begin{pmatrix}
A_\Z & A_\R
\end{pmatrix}$ where $A_\Z$, $A_\R$ act, respectively, in integer and real variables. Since the constraint (mod ${H}$) is equivalent to the modular constraints mod ${ d_1},\ldots,$ mod ${ d_n}$, it follows that $\textbf{x}=\begin{pmatrix}
\mathbf{x_\Z} & \mathbf{x_\R}\end{pmatrix}\in\textbf{X}_{\textnormal{sol}}$ if and only if
\begin{equation}\label{eq:Enlarged system of equations 2}
A_\Z \mathbf{x_\Z} + A_\R\mathbf{x_\R} + D\mathbf{y}= c, \quad \textnormal{where } D=\textnormal{diag}(d_1,\ldots,d_n),\: \mathbf{y}\in \Z^n.
\end{equation}
Clearly, if we rename $A':=\begin{pmatrix}
A_\Z & D
\end{pmatrix}$, $\mathbf{x}':=\begin{pmatrix}
\mathbf{x_\Z} &  \mathbf{y}
\end{pmatrix}$, $B=A_\R$ and $\mathbf{y}':=\mathbf{x_\R}$, system (\ref{eq:Enlarged system of equations 2}) is a system of mixed-integer linear equations as in (\ref{eq:System of Mixed Integer linear equations}). Also,  system (\ref{eq:System of Mixed Integer linear equations}) can be seen as a system  of linear equations over abelian groups: note that in the last step the solution space $\mathbf{X}$ is increased by introducing new extra integer variables $\mathbf{y}\in\Z^n$. If we let $\mathbf{G}$ denote the group $\mbf{X}\times \Z^n$ that describes this new space of solutions, then  (\ref{eq:Enlarged system of equations 2}) can be rewritten as
\begin{equation}\label{eq:Enlarged system of equations 3}
\mathbf{A}\mathbf{g}:=
\begin{pmatrix}
A & D 
\end{pmatrix}\mathbf{g}=c,\quad \textnormal{where } \mathbf{g} \in \mathbf{G}
\end{equation}
and $c$ represents an element of $\mathbf{Y}$. 

Mind that (\ref{eq:Enlarged system of equations 2}) (or equivalently (\ref{eq:Enlarged system of equations 3})) admits solutions if and only if both of (\ref{eq:Enlarged system of equations 1}) and (\ref{eq:Systems of linear equations over groups}) admit solutions. Indeed, the solutions of (\ref{eq:Enlarged system of equations 2}) and (\ref{eq:Enlarged system of equations 1}) are---again---related via a surjective group homomorphism  $\pi:\mathbf{X}\times\Z^n\rightarrow \mbf{X}:(\mathbf{x}, \mathbf{y})\rightarrow \mathbf{x}$. It follows from the derivation of (\ref{eq:Enlarged system of equations 2}) that $\pi(\mathbf{G}_{\textnormal{sol}})=\mathbf{X}_{\textnormal{sol}}$ and, consequently, $\mathbf{q}\circ \pi(\mathbf{G}_{\textnormal{sol}})=G_{\textnormal{sol}}$; these relationships show that either all systems admit solutions or none of them do.\\

In the second step of the proof, we use existing algorithms to find a general solution $(\mathbf{g}_0, \mathbf{P})$ of system (\ref{eq:Enlarged system of equations 3}) and show how to use this information to compute a general solution of our original problem (\ref{eq:Systems of linear equations over groups}).

First,  we recall that  algorithms presented in \cite{BowmanBurget74_systems-Mixed-Integer_Linear_equations} can be used to: (a) \emph{check} whether a system of the form (\ref{eq:System of Mixed Integer linear equations},\ref{eq:Enlarged system of equations 3}) admits a solution; (b)  \emph{find} a particular solution $\mathbf{g}_0$ (if there is any) and a matrix $\mathbf{P}$ that defines a group endomorphism of $\textbf{G}=\mbf{X}\times \Z^n$ whose image $\textnormal{im}\,\mathbf{P}$ is precisely the kernel\footnote{In fact, the matrix $P$ is also idempotent and defines a projection map  on $\textbf{G}$ and $\ker{\mathbf{A}}$ is the image of a projection map: subgroups satisfying this property are called \emph{retracts}. Though the authors never mention the fact that $\textbf{P}$ is a projection, this follows immediately from their equations (10a,10b).} of  $\mathbf{A}=
\begin{pmatrix}
A & D 
\end{pmatrix}$ (for details see theorem 1 in \cite{BowmanBurget74_systems-Mixed-Integer_Linear_equations}). 

Assume now that (\ref{eq:Enlarged system of equations 2}) admits solutions and that we have already found a general solution $(\mbf{g}_0=(\mathbf{x}_0,\mathbf{y}_0), \mathbf{P})$. We show next how a general solution $(x_0,P)$ of (\ref{eq:Systems of linear equations over groups}) can be computed by making use of the map $\mathbf{q}\circ \pi$. We also discuss the overall worst-case running time we need to compute $(x_0,P)$, as a function of the sizes of the matrix $A$ and the tuple $b$ given as an input in our original problem (\ref{eq:Systems of linear equations over groups}) (the bit-size  or simply \emph{size} of an array of real numbers---tuple, vector or matrix---is defined as the minimum number of bits needed to store it with infinite precision), size($G$) and size($H$):
\begin{itemize}
\item First, note that $(\mbf{g}_0=(\mathbf{x}_0,\mathbf{y}_0), \mathbf{P})$ can be computed in polynomial-time in  $\textnormal{size}(A)$, $\textnormal{size}(b)$, size($G$) and size($H$), since there is only a polynomial number of additional variables and constrains in (\ref{eq:Enlarged system of equations 2}) and the worst-time scaling of the algorithms in \cite{BowmanBurget74_systems-Mixed-Integer_Linear_equations} is also polynomial in the mentioned variables. (We discussed the complexity of these methods in section~\ref{sect:Systems of linear equations over groups}.)
\item Second, a particular solution $x_0$ of (\ref{eq:Systems of linear equations over groups}) can be easily computed just by taking $x_0:=\mathbf{q}\circ\pi((\mathbf{x}_0,\mathbf{y}_0))=\pi(\mathbf{x}_0)\pmod{G}$: this computation is clearly efficient in size($\textbf{x}_0$) and size($G$). 
\item Third, note that the composed map $P:=\mathbf{q}\circ\pi\circ \mathbf{P}$ defines a group homomorphism  $P:\mathbf{G}\rightarrow G$ whose image is precisely the subgroup $\ker A$; a matrix representation of $P$ (that we denote with the same symbol) can be efficiently computed, since
\begin{equation}
\textnormal{if}\quad \mathbf{P}=\begin{pmatrix}
\mathbf{P}_{\mathbf{X} \mathbf{X}} & \mathbf{P}_{\mathbf{X} \Z}\\
\mathbf{P}_{\Z \mathbf{X}} & \mathbf{P}_{\Z \Z}
\end{pmatrix}
\quad
\textnormal{then} \quad P:=\begin{pmatrix}
\mathbf{P}_{\mathbf{X} \mathbf{X}} & \mathbf{P}_{\mathbf{X} \Z}
\end{pmatrix}
\end{equation}
is a matrix representation of $\mathbf{q}\circ\pi\circ \mathbf{P}$ that we can take without further effort. 
\end{itemize}
The combination of all steps above yields a deterministic polynomial-time algorithm to  compute a general solution $(x_0, P;\: \mathbf{G})$ of system (\ref{eq:Systems of linear equations over groups}), with  worst-time scaling as a polynomial in the variables $m$, $n$, $\|A\|_{\mathbf{b}}$, $\|b\|_{\mathbf{b}}$, $\log{c_i}$, $\log{d_j}$. This proves theorem \ref{thm:General Solution of systems of linear equations over elementary LCA groups}.

\subsection{Algorithm for problem (3-4)}

First, we show that problem 4 can be solved by via the algorithm for  problems 1-2-3. First, we can use algorithms 1-2 to decide if a general solution exists and (in the affirmative case) find  $(x_0, \mathcal{E})$ such that $G_\mathrm{sol}=x_0+\mathrm{im}\mathcal{E}$. Moreover, we can run our algorithm 3 to find (a) an elementary group $Q$ isomorphic to $\mathrm{im}\mathcal{E}$, which must necessarily be finite and of form $Q=\DProd{D}{m}$, so that the total number of solutions is $|G_\mathrm{sol}|=|\mathrm{im}\mathcal{E}|=|Q|=D_1\ldots D_m$, which is efficiently computable; and (b) a matrix representation of the isomorphism $\mathcal{E}_\mathrm{iso}:Q\rightarrow \mathrm{im}\mathcal{E}$, the columns of which form a generating set of $\mathrm{im}\mathcal{E}$; hence, we can simply set the sought elements $x_1,\ldots,x_r$ to be the columns of  $\mathcal{E}_\mathrm{iso}$.

The above reduction shows that we can finish our proof if we give an efficient classical algorithm for problem 4. We address this question next.

Recall that, in problem 4, we are given $G=\Z^a\times\DProd{N}{b}$,  $\mathcal{X}=\Z^{\alpha+\beta}$ and $(x_0,\mathcal{E})$ as an input. Our task is to devise an algorithm to compute a primitive-group decomposition of the quotient $Q=\mathcal{X}/\ker{\mathcal{E}}$ and a matrix representation of the isomorphism $\mathcal{E}_\mathrm{iso}:Q\rightarrow\mathrm{im}\mathcal{E}$. To this end, we first apply our algorithm in theorem \ref{thm:General Solution of systems of linear equations over elementary LCA groups} to obtain a $(\alpha+\beta)\times\gamma$ matrix representation $A$ of a group homomorphism\footnote{Lemma \ref{lemma:Normal form of a matrix representation} ensures that real factors do not appear in the  domain of  $A$ because there are no non-trivial continuous group homomorphisms from products of $\R$ into products of $\Z$.} $A:\Z^{\gamma} \rightarrow \mathcal{X}$ such that  $\mathrm{im}\,A = \ker{\mathcal{E}}$ (where $\gamma=\alpha+\beta+m$). We can represent these maps in a diagram:
\begin{equation}
\begin{tikzcd}
\Z^{\gamma}\arrow{r}{A}
&\Z^{\alpha+\beta}\arrow{r}{\mathcal{E}}
&\mathrm{im}\mathcal{E}
\end{tikzcd}
\end{equation}
The worst-case time complexity needed to compute $A$ is polynomial in the variables  $m$, $\alpha$, $\beta$, $ \log{N_i}$, $\|\mathcal{E}\|_{\mathbf{b}}$. 
Next, we compute two integer  invertible matrices $U$, $V$ such that $A=USV$ and $S$ is in Smith normal form (SNF). This can be done in  $O\left(\ppoly{m,\alpha,\beta,\log{N_i}, \|\mathcal{E}\|_\mathbf{b}), \log{\frac{1}{\varepsilon}}}\right)$ time with existing  algorithms to compute the SNF of an integer matrix (see e.g.\ \cite{Storjohann10_Phd_Thesis} for a review). Each matrix $V$, $S$, $U$ is the matrix of representation of some new group homomorphism, as illustrated in the following diagram.
\begin{equation}\label{eq:Commutative diagram}
\begin{tikzcd}
\Z^{\gamma}\arrow{r}{A}\arrow{d}{V}
&\Z^{\alpha+\beta}\arrow{r}{\mathcal{E}}
&\mathrm{im}\mathcal{E}\\
\Z^\gamma \arrow{r}{S}&\Z^{\alpha+\beta}\arrow{u}{U}
\end{tikzcd}
\end{equation}
Since $V$, $U$ are invertible integer matrices the maps $V:\Z^\alpha\rightarrow\Z^\alpha$ and $U:\Z^{\alpha+\beta}\rightarrow\Z^{\alpha+\beta}$ are continuous group isomorphisms and, hence, have trivial kernels. As a result, $\mathrm{im}\, S = \mathrm{im}\,{U^{-1}AV^{-1}} = \mathrm{im}\,U^{-1}A = U^{-1}(\mathrm{im}\,A)=U^{-1}(\ker{\mathcal{E}})$, which shows that $\ker{\mathcal{E}}$ is isomorphic to $\mathrm{im}\, S$ via the isomorphism $U^{-1}$. These facts together with  lemma \ref{lemma:properties of matrix representations}.(a) show that   $\mathcal{E}_{\mathrm{iso}}:=\mathcal{E}U$ is a matrix representation of a \emph{group isomorphism} from the group  $Q:=\mathcal{X}/\mathrm{im}\, S$ into  $\mathrm{im}\mathcal{E}$. 

Finally, we show that $Q$ can be written explicitly as a direct product of primitive groups of type $\Z$ and $\Z_{d}$. We make crucial use of the fact that $S$ is Smith normal form, i.e.\
\begin{equation}\label{eq:Smith normal form}
S=\left(\begin{matrix}
s_1 &     &         &     &\vline   \\
    & s_2 &         &     &\vline  \\
    &     & \ddots  &     &\vline  \\
    &     &         & s_{(\alpha+\beta)} &\vline \\
\end{matrix}\begin{array}{c}
\quad\mbox{\huge 0}\end{array}\:\right)
=\left(\begin{matrix}
I_{\mathbf{a}} &     &         &     \vline   \\
    &  
    \begin{matrix}
    \sigma_1 & &\\
             &\ddots &\\
             &       &\sigma_{\textnormal{\textbf{b}}}
    \end{matrix} &     &   \vline\\
    &   &  \mbox{\Large 0}&\vline  
\end{matrix}\begin{array}{c}
\quad\mbox{\Huge 0}
\end{array}\:\right),
\end{equation}
where the coefficients $\sigma_i$ are strictly positive. It follows readily that $\mathrm{im}\,S = \Z^\mathbf{a}\times \sigma_1\Z\times\cdots\sigma_\mathbf{b}\Z\times \{0\}^{\mathbf{c}}$, and therefore
\begin{equation}\label{inproof:Sampling Algorithm Decomposition of a Quotient}
Q=\Z^{a+b}/\mathrm{im}\,S = \{0\}^\mathbf{a}\times \DProd{\sigma}{\mathbf{b}}\times \Z^{\mathbf{c}}.
\end{equation}

\section{Efficiency of Bowman-Burdet's algorithm}\label{appendix:Efficiency of Bowman Burdet}

In this appendix we briefly discuss the time performance of Bowman-Burdet's algorithm \cite{BowmanBurget74_systems-Mixed-Integer_Linear_equations} and argue that, using current algorithms to compute certain matrix normal forms (namely, Smith normal forms) as subroutines), their algorithm can be implemented in worst-time polynomial time.

An instance of the problem $Ax+By = C$, of the form (\ref{eq:System of Mixed Integer linear equations}), is specified by the rational matrices $A$, $B$ and the rational vector $C$. Let $A$, $B$, $C$ have $c\times a$, $c\times b$ and $c$ entries. Bowman-Burdet's algorithm (explained in \cite{BowmanBurget74_systems-Mixed-Integer_Linear_equations}, section 3) involves different types of steps, of which the most time consuming are (see equations 8-10 in \cite{BowmanBurget74_systems-Mixed-Integer_Linear_equations}):
\begin{enumerate}
\item the calculation of a constant number of certain types of generalized inverses introduced by Hurt and Waid \cite{HurtWaid70_Integral_Generalized_Inverse};
\item a constant number of matrix multiplications.
\end{enumerate}
A Hurt-Waid generalized inverse $M^\#$ of a rational matrix $M$ can be computed with an algorithm given in \cite{HurtWaid70_Integral_Generalized_Inverse}, equations 2.3-2.4. The  worst-case running time of this procedure is dominated by the computation of a Smith Normal form  $S=UMV$ of $M$ with pre- and post- multipliers $U$, $V$. This subroutine becomes the bottleneck of the entire algorithm, since existing algorithms for this problem are slightly slower than those for multiplying matrices (cf.\ \cite{Storjohann10_Phd_Thesis} for a slightly outdated review). Furthermore, $S$, $U$ and $V$ can be computed in polynomial time (we refer the reader to \cite{Storjohann10_Phd_Thesis} again). 

The analysis above shows that Bowman-Burdet's algorithm runs in worst-time polynomial in the variables $\|A\|_{\mathbf{b}}$, $\|B\|_{\mathbf{b}}$, $\|C\|_{\mathbf{b}}$, $a$, $b$, $c$, which is enough for our purposes.

\section{Proof of lemma \ref{lemma:Computing Inverses}}\label{appendix:Computing inverses}

As a preliminary, recall  that group homomorphism form an abelian group with the point-wise addition operation. Clearly, matrix representations inherit this group operation and form a group too. This follows from the following formula,
\begin{equation}\label{eq:Homomorphisms form a group}
(\alpha+\beta)(g)=\alpha(g)+\beta(g)=Ag + Bg = (A+B)g\pmod{G},
\end{equation}
which also states that the sum $(A+B)$ of the  matrix representations $A$, $B$  of two homomorphisms $\alpha$, $\beta$ is a matrix representation of the homomorphism $\alpha+\beta$. The group structure of the matrices is, in turn, inherited by their  \emph{columns}, a fact that will be exploited in the rest of the proof; we will denote by  $X_j$ the abelian group formed by the $j$th columns of all matrix representations with addition rule inherited from the matrix addition operation. 

A consequence of lemma \ref{lemma:Normal form of a matrix representation} is that the group $X_j$ is always an elementary abelian group, namely
\begin{align}
G_j&=\Z  &\Rightarrow\quad\qquad & X_j =  G= \Z^{a}\times \R^{b} \times \DProd{N}{c} \times \T^{d};\notag\\
G_j&=\R   &\Rightarrow\qquad\quad & X_j = \{0\}^{a}\times \R^{b} \times \{0\}^{c} \times \R^{d};\notag\\
G_j&=\Z_{N_j}  &\Rightarrow\qquad\quad & X_j =  \{0\}^{a}\times \{0\}^{b} \times \left(\eta_{1,j} \Z \times\cdots \times \eta_{c,j}\Z\right)\times \left(\tfrac{1}{N_j}\Z\right)^{d};\notag\\
G_j&=\T  &\Rightarrow\qquad \quad& X_j = \{0\}^{m_z}\times \{0\}^{m_r} \times \{0\}^{m_f} \times  \Z^{m_t};\label{eq:Columns of M representations form a group}
\end{align}
where   $\eta_{i,j}:=N_i/\gcd{(N_i, N_j)}$.

We will now prove the statement of the lemma.

First, we reduce the problem of computing a valid matrix representation $X$ of $\alpha^{-1}$ to that of solving   the equation $\alpha\circ\beta=\mathrm{id}$ ($\alpha$) is now the given automorphism) where $\beta$ stands for any continuous group homomorphism $\beta:G\rightarrow G$. It is easy to show that this equation admits $\beta=\alpha^{-1}$ as unique solution, since
\begin{equation}
\alpha\circ\beta=\mathrm{id}\quad\Longrightarrow\quad\beta=(\alpha^{-1}\circ\alpha)\circ\beta = \alpha^{-1}\circ(\alpha\circ\beta) = \alpha^{-1}.
\end{equation}
Hence, our task will be to find a matrix $X$ such that $g\rightarrow Xg\pmod{G}$ is a continuous group \emph{homomorphism} and such that $AX$ is a matrix representation of the identity automorphism. The latter condition reads  $AX g = g\pmod{G}$ for every $g\in G$ and is equivalent to
\begin{align}\label{inproof:Computing inverses 0}
AX\left( \sum_j g(j)e_j\right) = \sum_j g(j) Ax_j= \sum_j g(j)e_j \pmod{G} ,\,  \textnormal{ for every $g \in G,$}
\end{align}
where $x_j$ denotes the $j$th column of $X$. Since (\ref{inproof:Computing inverses 0}) holds, in particular, when all but  one number $g(j)$ are  zero, it can be re-expressed as an equivalent system of equations:
\begin{equation}\label{inproof:Computing Inverses 1}
  g(j)Ax_j=  g(j)e_j \pmod{G} ,\,  \textnormal{ for any $g(j) \in G_j$, for $j=1,\ldots,m$}.
\end{equation}
Finally, we will reduce each individual equation in (\ref{inproof:Computing Inverses 1}) to  a linear system of equations of the form (\ref{eq:Systems of linear equations over groups}). This will let us apply the algorithm in theorem \ref{thm:General Solution of systems of linear equations over elementary LCA groups} to compute every individual column $x_j$ of $X$. 

We begin by finding some simpler equivalent form for (\ref{inproof:Computing Inverses 1}) for the different types of primitive factors:
\begin{itemize}
\item[(a)] If $G_j=\Z$ or  $G_j=\Z_{N_j}$ the coefficient $g(j)$ is integral and can take the value 1. Hence, equation (\ref{inproof:Computing Inverses 1}) holds iff $Ax_j=  e_j \pmod{G}$.
\item[(b)] If $G_j=\R$ or  $G_j=\T$ we show that (\ref{inproof:Computing Inverses 1}) is equivalent to $Ax_j = e_j\pmod{X_j}$. Clearly, (\ref{inproof:Computing Inverses 1}) implies $g(j)Ax_j = g(j)e_j + \mathit{zero}$ where $\mathit{zero}= 0\pmod{G}$ and where  we fix a value of $g(j)\in G_j$. Since $G_j$ is divisible,  $g(j)'=g(j)/d$ is also an element of $G_j$ for any positive integer $d$. For this value we get $\tfrac{g(j)}{d} Ax_j   = \tfrac{g(j)}{d}e_j  + \mathit{zero}'$. These two equations combined show that  $\mathit{zero}= d\,\mathit{zero}'$ must hold for every positive integer $d\in \Z$. Since both $\mathit{zero}$ and $\mathit{zero}'$ are integral, it follows that the entries of $\mathit{zero}$ are \emph{divisible} by all positive integers; this can only happen if $\mathit{zero}=0$ and, consequently, (\ref{inproof:Computing Inverses 1}) is equivalent to $Ax_j = e_j$. Since both  $Ax_j$ and $e_j$  are $j$th columns of matrix representations, the latter equation can be  written as $Ax_j = e_j\pmod{X_j}$ with $X_j$  as in (\ref{eq:Columns of M representations form a group}).  
\end{itemize}
Finally, we argue that the final systems (a) $Ax_j=  e_j \pmod{G}$ and (b) $Ax_j=  e_j \pmod{X_j}$ are linear systems of the form (\ref{eq:Systems of linear equations over groups}).  First notice that for any two homomorphisms $\beta$, $\beta'$ with matrix representations $X$, $Y$, it follows from (\ref{eq:Homomorphisms form a group}) and lemma \ref{lemma:properties of matrix representations}.(a)  that $A(X+Y)=AX+AY$ is a matrix representation of the homomorphism $\alpha\circ(\beta+\beta')=\alpha\circ\beta+\alpha\circ\beta'$. Consequently,
\begin{equation}\label{inproof:Computing Inverses 3}
A(X+Y)g = (AX + AY) g\pmod{G}, \textnormal{for every }g\in G.
\end{equation}
The argument we used to reduce $AXg=g\pmod{G}$ to the cases (a) and (b) can be applied again to find a simpler form for (\ref{inproof:Computing Inverses 3}). Applying the same procedure step-by-step (the derivation is omitted), we obtain that, if $G_j=\Z$ or  $G_j=\Z_{N_j}$, then  (\ref{inproof:Computing Inverses 3}) is equivalent to $A(x_j+y_j)=Ax_j + Ay_j \pmod{G}$; if $G_j=\R$ and  $G_j=\T$, we get $A(x_j+y_j)=Ax_j + Ay_j \pmod{X_j}$ instead. It follows that the map $x_j\rightarrow Ax_j$ is a group homomorphism from $X_j$ to $G$ in case (a) and from $X_j$ to $X_j$ in case (b). This shows that systems  (a) and  (b) are  of the form (\ref{eq:Systems of linear equations over groups}).

\section{Supplementary material for section \ref{sect:quadratic_functions}}\label{app:Supplement Quadratic Functions}

\subsection*{Proof of lemma \ref{lemma:Normal form of a bicharacter 1}}

The lemma is a particular case of proposition 1.1 in \cite{Kleppner65Multipliers_on_Abelian_Groups}. We reproduce a shortened proof of the result in \cite{Kleppner65Multipliers_on_Abelian_Groups} (modified to suit our notation) here.

If $\beta$ is an continuous homomorphism from $G$ into $G^*$ then $ B(g, h) = \chi_{\beta(g)}(h)$ is continuous, since composition preserves continuity. Also, it follows using the linearity of this map and  of the character functions that  $B(g, h)$ is bilinear, and hence a bicharacter.  Conversely, consider an arbitrary bicharacter $B$. The condition that $B$ is a character on  the second argument says that for every $g$ the function $f_g: h\to B(g, h)$ is a character. Consequently $f_g(h)=B(g,h)=\chi_{\mu_g}(h)$ for all $h\in G$ and some $\mu_g\in G^*$ that is determined by $g$. We denote by $\beta$ be the map which sends $g$ to $\mu_g$. Using that $g\to B(g, h)$ is also a character  it follows that $\chi_{\beta{(g+g')}}(h)=\chi_{\beta{(g)}}(h)\chi_{\beta{(g')}}(h)$ for all $h\in G$, so that $\beta:G\rightarrow G^*$ is a group homomorphism. It remains to show that $\beta$ is continuous; for this we refer to the proof in \cite{Kleppner65Multipliers_on_Abelian_Groups}, where the author analyzes how  neighborhoods are transformed under this map.

\subsection*{Proof of lemma \ref{lemma symmetric matrix representation of the bicharacter homomorphism}}

We obtain (a) by combining (\ref{eq:definition of bullet map2}) with the normal form (\ref{eq:first normal fomr of a bicharacter}): the matrix $M$ is  of the form $\Upsilon X$ where $X$ is a matrix representation of $\beta$;  (b) follows from this construction. (c) follows from the normal form in lemma \ref{lemma:Normal form of a bicharacter 1}, property (a) and  lemma \ref{thm_extended_characters}.

To prove (d) we bring together (a) and the relationship $B(h,g)=B(g,h)$, and derive
\begin{equation}\label{eq:symmetry of the starred-matrix representation of a bicharacter MODULO Z}
 g^\transpose M h =  g^\transpose M^\transpose  h\mod{\Z},\quad\text{for every $g$, $h\in G$.}
\end{equation}
Write $G=G_1\times\dots\times G_m$ with $G_i$ of primitive type. If $G_i$ is either finite or equal to $\Z$ or $\R$ then the canonical basis vector $e_i$ belongs to $G$. If $G_i=\T$ then $te_i\in G$ for all $t\in[0, 1)$. If neither $G_i$ nor $G_j$ is equal to $\T$, taking  $g= e_i$, $h=e_j$ in equation (\ref{eq:symmetry of the starred-matrix representation of a bicharacter MODULO Z}) yields $M(i, j)\equiv M(j, i)$ mod $\Z$. If $G_i$ and $G_j$ are equal to $\T$, setting $g= te_i$ and $h=se_j$ yields $stM(i, j)\equiv stM(j, i)$ mod $\Z$ for all $s, t\in [0, 1)$, which implies that \be st (M(i, j)- M(j, i))\in \Z  \ee for all $s, t\in [0, 1)$. This can only happen if $M(i, j)= M(j, i)$. The other cases are treated similarly. In conclusion, we find that $M$ is symmetric modulo $\Z$. This proves (d). 

Lastly, we prove (e). Note that we have just shown that $M(i,j)=M(j,i)$ if  $G_i=G_j=\T$; the same argument can be repeated (with minor modifications) to show $M(i,j)=M(j,i)$ if either one of $G_i$ or $G_j$ is  of the form $\R$ or $\T$. 
Hence, $M(i,j)\neq M(j,i)$ can only happen if $G_i$, $G_j$ are of the form   $\Z$ or $\Z_d$. In this case, we denote by $\Delta_{ij}$ the number such that $M(j,i)=M(i,j)+\Delta_{i,j}$. (d) tells us that $\Delta_{ij}$ is an integer. Moreover, by choosing  $g=g(i)e_i$, $h=h(j)e_j$ in (\ref{eq:symmetry of the starred-matrix representation of a bicharacter MODULO Z}) it follows that
\begin{equation}\label{inproof: Delta ij}
M(j,i)g(i)h(j)=M(i,j)g(i)h(j) + \Delta_{i,j}g(i)h(j) \mod{\Z},
\end{equation}
As $g(i)$ and $h(j)$ are integers the factor $\Delta_{i,j}g(i)h(j)$  gets canceled modulo $\Z$ and produces no effect. Finally, we  define  a new symmetric matrix  $M'$ as $M'(i,j)=M(i,j)$ if $i\geq j$, and $M'(i,j)=M(j,i)$ if $i< j$. It follows from our discussion that  $g^\transpose M'h=g^\transpose Mh\bmod{\Z}$ for every $g,h\in G$, so that $M'$ manifestly fulfills (a).

It remains to show that $M'$ fulfills (b)-(c). Keep in mind that $h\rightarrow Mh\pmod{G^\bullet}$ defines a group homomorphism into $G^\bullet$. From our last equations, it follows that either $M(i,j)h(j)=M'(i,j)h(j)$ or $M(i,j))h(j)=M'(i,j)h(j)\bmod{\Z}$ if both $G_i$ and $G_j$ are discrete groups. From the definition of bullet group (\ref{eq:Bullet Group}), it is now easy to derive that $Mh=M'h\pmod{G^\bullet}$ for every $h$, and to extend this equation to all tuples $x$ congruent to $h$  (this reduces to analyzing all possible combinations of primitive factors). As a result, $M'$ is a matrix representation that defines the same map as $M$, which implies (b). The fact that $M'$ satisfies (c) follows using the same argument we used for $M$.

\subsection*{Proof of lemma \ref{lemma quadratic B-representations differ by a character}}

We prove that the function $f(g):=\xi_1(g)/\xi_2(g)$ is a character, implying that there exists $\mu\in G^*$ such that $\chi_{\mu}=f$:
\begin{equation}
f(g+h):=\frac{\xi_1(g)}{\xi_2(g)}\frac{\xi_1(h)}{\xi_2(h)}\frac{B(g,h)}{B(g,h)}=f(g)f(h).
\end{equation}

\subsection*{Proof of lemma \ref{lemma:B-representations can always be constructed}}

Define the function $q:G\rightarrow \R$ as
\begin{equation}
q(g):= g^{\transpose} M g + C^\transpose g.
\end{equation}
We prove that $q(g)$ is a quadratic form modulo $2\Z$ with associated bilinear form $b_q(g,h):=2 g^\transpose M h$; or, in other words, that the following equality holds for every $g,h\in G$:
\begin{equation}\label{eqinproof:quadratic form over 2Z}
q(g+h)=q(g)+q(h)+ 2 g^\transpose M h \pmod{2\Z}.
\end{equation}
Assuming that (\ref{eqinproof:quadratic form over 2Z}) is correct, it follows readily that the function $Q(g)=\exp{\left(\pii q(g)\right)}$ is quadratic and also a $B$-representation, which is what we wanted:
\begin{equation}
Q(g+h)=Q(g)Q(h)\exp{\left(2\pii\, g^\transpose M h\right)}
\end{equation}
We prove  (\ref{eqinproof:quadratic form over 2Z}) by direct evaluation of the statement. First we define $q_M(g):=g^{\transpose} M g$ and $q_C(g):=C^{\transpose} g$, so that $q(g)=q_M(g)+q_C(g)$. We will also (temporarily, i.e. only within the scope of this proof) use the notation $g\oplus h$ to denote the group operation in $G$ and reserve $g+h$ for the case when we sum over the reals. Also, denoting $G=G_1\times\dots G_m$ with $G_i$ primitive, we define $c:=(c_1,\ldots,c_m)$ to be a tuple containing all the characteristics $c:=\charac{G_i}$. With these conventions we  have $g\oplus h = g+h + \lambda\circ c$, where $\lambda$ is a vector of integers and $\circ$ denotes the entrywise product: $\lambda\circ c=(\lambda_1 c_1, \dots, \lambda_m c_m)$.  Note that $\lambda\circ c$  is the most general form of any string of real numbers that is congruent to $0\in G$ (the neutral element of the group).
We then have (using that $M=M^T$):
\begin{align}\label{eqinproof quadratic form derivation 1}
q_M(g\oplus h) = \: & q_M (g) + q_M (h) + 2 g^\transpose M h  \nonumber\\ &+ 2g^\transpose M (\lambda\circ c) + 2h^\transpose M (\lambda\circ c) + (\lambda\circ c)^\transpose M (\lambda\circ c),\\
q_C(g\oplus h) =\: & q_C(g) + q_C(h) +  \sum_i M(i,i) \lambda(i)c_i^2.
\end{align}
Consider an $x\in \R^m$ for which there exists  $g\in G$ such that $x\equiv g$ mod $G$. Then $x^\transpose M (\lambda\circ c)$ with $x\in G $ must be an integer. Indeed, we have \be \label{inproof:B(g,0)} 1 = B(g,0)=\exp{\left(2\pii x^\transpose M (\lambda\circ c) \right)},\ee where in the second identity we used lemma \ref{lemma symmetric matrix representation of the bicharacter homomorphism} together with the property $\lambda\circ c\equiv 0$ mod $G$.   This shows that $x^\transpose M (\lambda\circ c)$ is an integer. It follows that the fourth and fifth terms on the right hand side of eq.\ (\ref{eqinproof quadratic form derivation 1}) must be equal to an even integer and thus cancel modulo $2\Z$. Combining results we end up with the expression
\begin{equation}
q(g\oplus h) = q (g) + q (h) + 2 g^\transpose M h  + \Delta \pmod{2\Z},
\end{equation}
where
\begin{equation}
\Delta := (\lambda\circ c)^\transpose M (\lambda\circ c) + \sum_i M(i,i)\lambda(i) c_i^2.
\end{equation} We finish our proof by showing that $\Delta$ is an even integer too, which proves (\ref{eqinproof:quadratic form over 2Z}).

First, we note that, due to the symmetry of $M$, we can expand $(\lambda\circ c)^\transpose M (\lambda\circ c)$  as
\begin{equation}\label{Delta}
(\lambda\circ  c)^\transpose M (\lambda\circ c) = \sum_{i,j\: :\: i<j} 2  M(i,j) \lambda(i)\lambda(j) c_i c_j + \sum_i M(i,i)\lambda(i)^2 c_i^2.
\end{equation}
Revisiting (\ref{inproof:B(g,0)}) and  choosing $x=e_i$ and $\lambda = e_j$ for all different values of $i$, $j$, we obtain the following consistency equation for $M$
\begin{equation}\label{eqinproof:consistency conditions of the BICHARACTER matrix}
c_j M(i,j) = c_i M(i,j) = 0 \pmod{\Z}
\end{equation}
It follows that all terms of the form $2M(i,j) \lambda(i)\lambda(j) c_i c_j$  are \emph{even} integers. We can thus remove these terms from (\ref{Delta}) by taking modulo $2\Z$, yielding
 \begin{align}
 \Delta &=  \sum_i M(i,i)\lambda(i)^2 c_i^2 + \sum_i M(i,i) \lambda(i) c_i^2 \pmod{2\Z}\\
 &=  \sum_i M(i,i) c_i^2 \lambda(i)(\lambda(i)+1) =0  \pmod{2\Z},
 \end{align}
where in the last equality we used the fact that $\lambda(i)(\lambda(i)+1)$ is  necessarily even.

\subsection*{Proof of lemma \ref{lemma:Quadratic Function composed with Automorphism}}

The fact that $\xi_{M,v}\circ \alpha$ is quadratic follows immediately from the fact that $\xi_{M,v}$ is quadratic and that $\alpha$ is a homomorphism. Composed continuous functions lead to continuous functions. As a result, theorem \ref{thm:Normal form of a quadratic function} applies and we know $\xi_{M',v'}=\xi_{M,v}\circ \alpha$ for some choice of $M'$, $v'$. 

Let $B_M(g,h)=\exp\left(2\pii g^\transpose M h\right)$ be the bicharacter associated with $\xi_{M, v}$. One can show by direct evaluation (and using lemma \ref{lemma:properties of matrix representations}(a) and lemma  \ref{lemma symmetric matrix representation of the bicharacter homomorphism}) that $B_{M'}$ with $M':=A^\transpose M A$ is the bicharacter associated to $\xi_{M,v}\circ \alpha$. Let $Q_{M'}(g):=\exp(\pii \, (  g^{\transpose} M' g + C_{M'}^\transpose g )),$ be the quadratic function in lemma \ref{lemma quadratic B-representations differ by a character}. By construction,  both $\xi_{M,v}\circ \alpha$ and $Q_{M'}$ are $B_{M'}$-representations of  the bicharacter $B_{M'}$. As a result, lemma \ref{lemma quadratic B-representations differ by a character} tells us that the function $f(g):=\xi_{M,v}\circ \alpha(g)/ Q_{M'}(g)$ is a character of $G$, so that there exists $v'\in G^*$ such that $\chi_{v'}(g)=f(g)$. We can compute $v'$ by direct evaluation of this expression:
\begin{equation}
\chi_{v'}(g)= \exp\left( \pii \left( A^\transpose C_{M} - C_{A^\transpose M A}\right)g \right) \exp\left( 2\pii \left( A^\transpose v \right)\cdot  g \right).
\end{equation}
It can be checked that the function $\exp \left( 2\pii \left( A^\transpose v \right) g \right) $ is a character, using that  it is the composition of a character $\exp(2\pi v^\transpose g)$  (theorem \ref{thm:Normal form of a quadratic function}) and a continuous group homomorphism $\alpha$. Since $\chi_{v'}$ is also a character, the function  $\exp\left( \pii \left( A^\transpose C_{M} - C_{A^\transpose M A}\right)g \right)$ is a character too (as characters are a group under multiplication), and it follows that  $v_{A,M}=( A^\transpose C_{M} - C_{A^\transpose M A} )/2$ is congruent to some element of $G^\bullet$ \footnote{This statement can also be proven (more laboriously) by explicit evaluation, using arguments similar to those in the proof of lemma \ref{lemma:B-representations can always be constructed}.}; we obtain that $v'= A^\transpose v + v_{A,M}$ is an element of $G^\bullet$. Finally, we obtain that  $\xi_{M',v'}$ is a normal form of $\xi_{M,v}\circ \alpha$, using the relationship $\xi_{M,v}\circ \alpha (g)=Q_{M'}(g) f(g)=Q_{M'}(g) \chi_{v'}(g)=\xi_{M',v'}(g)$.

%% file: appendix1_finite_chapter.tex
\chapter{Supplement to chapter \ref{chapterF}}\label{aF}

\section{Proof of lemma \ref{LEMMA REDUCTION TO SYSTEMS OF LINEAR EQUATIONS}}\label{Appendix:A1}

In the following, we define $A_H$, $A_K$ to be integer matrices whose columns are the elements of the sets $\{h_1,\ldots,h_r\}$ and $\{k_1,\ldots,k_s\}$; the latter generate respectively $H$ and $K$. Also, we  denote by $d$  the least common multiplier of $d_1,\ldots,d_m$. One can  use lemma \ref{lemma:Normal form of a matrix representation} to check that the matrices $A_H$, $A_K$ and $[A_H|A_K]$ define group homomorphisms from $\Z_d^{t}$ to $G$, if the value of $t$ is respectively chosen to be $r$, $s$ and $r+s$.

We will  show how to turn the problems (a-c) into system  of the form $Ax = b\pmod{\mbf{G}}$ such that $\mbf{G}$ equals the original group $G$;  $\mbf{G}_{sol}$ is chosen to be $\Z_d^{t}$, for some $t$; and $A$ is an integer matrix that defines a group homomorphism from $\mbf{G}$ to $\mbf{G}_{sol}$: \\

\textbf{(a)} $b$ belongs to $H$ if and only if $b$ can be obtained as a linear combination of elements of $H$, i.e.,  if and only if $A_H x = b \pmod G$ has at least one solution $x\in \Z_d^{r}$. Moreover, if one finds a particular solution $w$, this element fulfills $b=A_H w = \sum w(i)h_i \pmod{G}$.\\

\textbf{(b)} The order of $H$ is the number of distinct linear combinations of columns of $A_H$, which coincides with the order of the \textit{image} of the group homomorphism  $A_H:\Z_d^{r}\rightarrow G$. With this knowledge, it suffices to count the number of solutions of $A_H x = 0 \pmod G$, which equals $|\ker A_H|$. Then, one can compute  $|H|=|\textnormal{im}A_H|=d^r/|\ker A_H|$, where the latter identity comes from the first isomorphism theorem ($\text{im}{A_H}\cong \Z_d^{r}/\ker{A_H}$).\\

\textbf{(c)}  $g$ belongs to  $H\cap K$ iff it can be simultaneously written as $h=\sum x(i) h_i = \sum y(i) k_i$ for some  $(x,y)\in \Z_d^{r}\times\Z_d^{s}$; or, equivalently, iff there exist an element $(x,y)$ of the kernel of $\left[A_H | A_K\right]: \Z_d^{r}\times\Z_d^{s}\rightarrow G$ such that $h=A_H x=-A_Ky\pmod{G}$. Thus, given a generating-set $\{(x_i,y_i)\}$ of  $\ker\left[A_H | A_K\right]$, the elements $g_i:=A_H x_i\pmod{G}$ generate $H\cap K$, and, owing to, the problem reduces to finding  solutions of $\left[A_H | A_K\right]\binom{x}{y}=0\pmod{G}$.\\

\textbf{(d-e)} Note  that problem (d) reduces to (e) by setting all $a_i$ to be $0$---this yields the system (\ref{orthogonal group EQUATION}), whose solutions are the elements of the annihilator subgroup. Therefore, it will be enough to prove the (e)th case. Moreover, since the equations $\chi_{h_i}(g)=\gamma^{a_i}$ can be fulfilled for some $g\in G$ only if all $\gamma^{a_i}$ are $\Inputsize$th-roots of the unit, this systems can only have solutions if all  $a_i$ are even numbers. As we can determine it efficiently whether these numbers are even, we assume from now on that it is the case.

Now define a tuple of integers $b$ coefficient-wise as $b(i):=a_i/2$; use the later to  re-write $\gamma^{a_i}=\exp{(2\uppi\textnormal{i}\, b(i)/\Inputsize)}$. Also, denote by $H$ the group generated by the elements $h_i$. By letting $|G|$ multiply numerators and denominators of all fractions in (\ref{Character Functions DEFINITION}), the system of complex exponentials $\chi_{h_i}(g)=\gamma^{a_i}$ can be turned into an equivalent  system of  congruences $\sum_j (|G|/d_j)\, h_i(j) g(j)=b(i) \mod |G|$. Finally, by defining a matrix $\Omega$ with coefficients $\Omega(i,j):=({|G|}/{d_j})\, h_i(j)$ the system can be written as
\begin{equation}
\Omega g = b \pmod{\mbf{G}},
\end{equation}
where  $b$ belongs to $\mbf{G}=\Z_{|G|}^{r}$, being $r$ the number of generators $h_i$, and we look for solutions inside $\mbf{G}_{sol}=G$. Moreover, the coefficients of $\Omega$ fulfill $d_j \Omega(i,j)=0\bmod{|G|}$; hence, condition (\ref{eq:Consistency Conditions Homomorphism}) is met and $\Omega$ defines a homomorphism.\\

\textbf{(h)}  Note that the group homomorphism $\omega(g):=\Omega(g)\pmod{\Z_{|G|}^{r}}$ fulfills $\ker{\omega}=H^{\perp}$. Therefore, if we substitute $H$ with $H':=H^{\perp}$ in the procedure above, given   $s=\polylog{\Inputsize}$ generators of $H'$, we would obtain an $s\times m$ integer matrix $\Omega'$ that defines a second group homomorphism $\varpi:G\rightarrow \Z_{|G|}^{s}$ such that
\begin{equation}\label{Oracle function definition}
\varpi(g)=\Omega'(g)\pmod{G}\qquad\textnormal{and}\qquad\ker\varpi=H'^{\perp}=H.
\end{equation}
As a result, our classical algorithm for problem (d) can be efficiently adapted to solve (h). \\

\textbf{(Remarks:)} Finally, note that $\Omega$ can be computed in $O(\polylog{\Inputsize})$ using standard algorithms to multiply and divide integers (chapter \ref{sect_CompComp_FiniteAbelianGroups}). It is now routine to check, using the concepts developed thus far, that both $\log{|\mbf{G}_{sol}|}$ and $\log{|\mbf{G}|}$ are $O(\polylog{\Inputsize})$;  as a result,   the input-size of the new problem, as well as the memory needed to store $\Omega$ and $b$, are all $O(\polylog{\Inputsize})$.

Finally, notice that $r$, $s$ and $r+s$ are $O(\polylog{\Inputsize})$ due to the initial assumption that the generating-sets are poly-size, and that $d$ is $O(d_1 d_2\cdots d_m)=O(|\mathbf{G}|)$; as a consequence, $\log|\mbf{G}_{sol}|$, $\log|\mbf{G}|$ are also $O(\polylog{\Inputsize})$; and, thus, we need $O(\polylog{|G|})$ memory to store the matrix $A$. It follows that the input-size of the new problem is $O(\polylog{|G|})$ and, therefore, we have reduced all problems (a-c) to systems of linear equations over finite abelian groups in polynomial time.

%% file: appendix3_blackbox.tex
\chapter{Supplement to  chapter \ref{chapterB}}\label{aB}

\section{Proof of theorem \ref{thm:ModExp requires Z}}\label{app:ModExp requires Z}

\newcommand{\ganc}{\Z_M}

To prove the result we can assume that we know a group isomorphism $\varphi:\mathbf{B}\rightarrow G$ that decomposes the black-box group as a product of cyclic factors $G= \DProd{N}{d}$. Let $U_{\varphi}:\mathcal{H}_{\mathbf{B}}\rightarrow \mathcal{H}_G$ be the unitary that implements the isomorphism  $U_\varphi\ket{b}=\ket{\varphi(b)}$ for any $b\in\mathbf{B}$. It is easy to check that $\mathcal{C}$ is a normalizer circuit over $\ganc\times G$ if and only if $(I\otimes U_\varphi) \mathcal{C} (I\otimes U_\varphi)^\dagger $ is a normalizer circuit over $\ganc\times \mathbf{B}$: automorphism (resp. quadratic phase) gates get mapped to automorphism (resp. quadratic phase) gates and vice-versa; isomorphic groups have isomorphic character groups \cite{Morris77_Pontryagin_Duality_and_LCA_groups}, and therefore Fourier transforms get mapped to Fourier transforms.
  
As a result, it is enough to prove the result in the basis labeled by elements of $\ganc\times G$. The advantage now is that we can use results from  chapter \ref{chapterF}, \cite{VDNest_12_QFTs}. In fact, the rest of the proof will be similar to the proof of theorem 2 in \cite{VDNest_12_QFTs}.
  
The action of $U_\mathrm{me}$ in the group-element basis reads  $U_\mathrm{me}\ket{m,g}=\ket{m,m\alpha+g}$, in additive notation. Define a function $F(m,g)=(m,m\alpha+g)$.  We now assume that the order $M$ of $1$ as an element of $\Z_M$ is not divisible by $|a|$ and that there exists a normalizer circuit $\mathcal{C}$ such that $\|\mathcal{C}-U_\mathrm{me}\|< \delta$ with $\delta=1-1/\sqrt{2}$ and try to arrive to a contradiction. This property implies that $\|\mathcal{C}\ket{m,g}-U_\mathrm{me}\ket{m,g}\|< \delta$ for any standard basis state, and consequently
  \begin{equation}\label{inproof:Ume is no Clifford}
  |\bra{F(m,g)}\mathcal{C}\ket{m,g}|> 1- \delta=\tfrac{1}{\sqrt{2}}
  \end{equation}
We now from, e.g., theorem \ref{thm Normal form of an stabilizer state}  that $\mathcal{C}\ket{m,g}$ is a uniform superposition over some subset $x+K$ of $\Z_M\times G$, being $K$ a subgroup. If $K$ has more than two elements, then $\mathcal{C}\ket{m,g}$ is a uniform superposition over more than two computational basis states. It follows that $\bra{m',g'}\mathcal{C}\ket{m,g}\leq\frac{1}{\sqrt{2}}$ for any basis state $\ket{m',g'}$ in contradiction with (\ref{inproof:Ume is no Clifford}), so that we can assume $K=\{0\}$ and  that $\mathcal{C}\ket{m,g}$ is a standard basis state. Then (\ref{inproof:Ume is no Clifford}) implies that $\ket{F(m,g)}$ and $\mathcal{C}\ket{m,g}$ must coincide for every $(m,g)\in \ganc\times G$, so that $\mathcal{C}$ must perfectly realize the transformation $\ket{(m,g)}\rightarrow\ket{F(m,g)}$; however, the only classical functions that can be implemented by normalizer circuits of this form are affine maps \cite{VDNest_12_QFTs}, meaning that $F(m,g)=f(m,g)+b$ for some group automorphism $f:\ganc\times G\rightarrow\ganc\times G$ and some $b\in \ganc\times G$. 
  
  Finally, we arrive to a contradiction showing that if $F(m,g)$ is  affine then  $M$ need to be a multiple of $|a|$. First, by evaluating $F(m,g)=f(m,g)+b=(m,m\alpha+g)$ at $(0,0)$,$(1,0)$ and elements of the form $(0,g)$, we check that $b=0$, so that $F(m,g)$ must be an automorphism. Because of $M$ is the order of $(1,0)$ in $\ganc \times G$, it follows that  $(0,0)=F((0,0))=F(M(1,0))=MF((1,0))=M(1,\alpha)$ modulo $\ganc\times G$, which holds only if $M$ is  a multiple of the order of $\alpha$.

\section{Quantum algorithm for discrete logarithms over elliptic curves}\label{sect:Elliptic Curve}

In this appendix, we show that a quantum algorithm given by Proos and Zalka  \cite{ProosZalka03_Shors_DiscreteLog_Elliptic_Curves} to compute \emph{discrete logarithms over elliptic curves} can be implemented with black-box normalizer circuits. This generalizes our result from section \ref{sect:Discrete Log}, where we saw  that black-box normalizer circuits can compute discrete logarithm in $\Z_p^\times$ and break the Diffie-Hellman key exchange protocol: specifically, we showed that Shor's algorithm for this problem decomposes naturally in terms of normalizer gates over $\Z_{p-1}^2\times \Z_p^\times$, where $\Z_p^\times$  is treated as a black-box group. Unlike the previous setting,  our implementation of Proos and Zalka algorithm requires either  normalizer gates over an \emph{infinite} group $\Z^2\times E$ (similarly to Shor's factoring algorithm, section \ref{sect:Factoring}) or an order-finding oracle.

\subsubsection*{Basic notions}

To begin with, we review some  rudiments of the theory of elliptic curves. For simplicity, our survey focuses only on the particular types of elliptic curves that were studied in \cite{ProosZalka03_Shors_DiscreteLog_Elliptic_Curves}, over fields with characteristic different than 2 and 3. Our discussion applies equally to the (more general) cases considered in  \cite{Kaye05_optimized_Quantum_Elliptic_Curve,CheungMaslovMathew08_Design_QuantumAttack_Elliptic_CC}, although the definition of the  elliptic curve group operation becomes more cumbersome in such settings\footnote{Correspondingly, the complexity of performing group multiplications in \cite{Kaye05_optimized_Quantum_Elliptic_Curve,CheungMaslovMathew08_Design_QuantumAttack_Elliptic_CC} is greater.}. For more details on the subject, in general, we refer the reader to \cite{childs_vandam_10_qu_algorithms_algebraic_problems, lorenzini1997invitation}.

Let $p> 3$ be prime and let $K$ be the field defined by endowing the set $\Z_p$ with the addition and multiplication operations modulo $p$.  An \emph{elliptic curve} $E$ over the field $K$ s a finite abelian group formed by the solutions  $(x,y)\in K\times K$ to an equation
\begin{equation}\label{eq:Elliptic Curve}
C:y^2 = x^3 + \alpha x + \beta
\end{equation}
together with a special element $O$ called the ``point at infinity''; the coefficients $\alpha$, $\beta$ in this equation live in the field $K$. The discriminant  $\Delta:=-16(4\alpha^3 + 27 \beta^2)$ is assumed to be nonzero, ensuring that  the curve is non-singular. The elements of $E$ are endowed with a commutative group operation. If $P\in E$ then $P+O= O + P = P$. The inverse element $-P$ of $P$ is obtained by the reflection of $P$ about the $x$ axis. Given two elements $P=(x_P,y_P)$ and $Q=(x_Q,y_Q)\in E$, the element $P+Q$ is defined via the following rule:
\begin{equation} \label{eq:Elliptic curve group operation}
P+Q= 
\begin{cases}
O & \textnormal{if $P=(x_P,y_P)=(x_Q,-y_Q)=-Q$,}  \\
 -R& \textnormal{otherwise (read below).} 
\end{cases}
\end{equation}
In the case $P\neq Q$, the point $R$ is computed as follows:

\begin{minipage}{0.4\textwidth}
\begin{align}
x_{R} &=\lambda^2 - x_P - x_Q  \notag \\
y_{R} &= y_P- \lambda(x_P-x_{R})  \notag 
\end{align}
\end{minipage}
\begin{minipage}{0.4\textwidth}
\begin{equation} 
\lambda:=\notag
\begin{cases}
\tfrac{y_Q-y_P}{x_Q-x_P} & \textnormal{if $P\neq Q$} \\
 \tfrac{3 x_P^2+\alpha}{2 y_P} & \textnormal{if $P= Q$ and $y_P \neq 0$}
\end{cases}
\end{equation}
\end{minipage}\\

\noindent  $R$ can also be defined, geometrically, to be the ``intersection between the elliptic curve and the line through $P$ and $Q$'' (with a minus sign) \cite{childs_vandam_10_qu_algorithms_algebraic_problems}. 

It is not hard to check form the definitions above that the elliptic-curve group $E$ is finite and abelian; from a computational point of view, the elements of $E$ can be stored with $n\in O(\log |K|)$ bits and the group operation can be computed in $O(\poly{n})$ time. Therefore, the group $E$ can be treated as a \textbf{black box group}.

Finally, the \textbf{discrete logarithm problem} (DLP) over an elliptic curve is defined in a way analogous to the $\Z_p^\times$ case, although now we use additive notation: given $a$, $b\in E$ such that $xa=b$ for some integer $x$; our task is to find the least nonnegative integer $s$ with that property. The elliptic-curve DLP is believed to be intractable for classical computers and can be used to define cryptosystems  analog to Diffe-Hellman's \cite{childs_vandam_10_qu_algorithms_algebraic_problems}.

\subsubsection*{Finding discrete logarithms over elliptic curves with normalizer circuits}

In this section we review Proos-Zalka's quantum approach to solve the DLP problem over an elliptic curve \cite{ProosZalka03_Shors_DiscreteLog_Elliptic_Curves}; their quantum algorithm is, essentially, a modification of Shor's algorithm to solve the DLP over $\Z_p^\times$, which we covered in detail in section \ref{sect:Discrete Log}. 

Our \textbf{main contribution} in this appendix is a proof that Proos-Zalka's algorithm can be implemented with normalizer circuits over the group $\Z\times \Z \times E$. The proof reduces to combining ideas from sections \ref{sect:Discrete Log} and \ref{sect:Factoring} and will be sketched in less detail.
\begin{algorithm}[\textbf{Proos-Zalka's \cite{ProosZalka03_Shors_DiscreteLog_Elliptic_Curves}}]\label{alg:ProosZalka} $\quad $
\begin{alg_in}
An elliptic curve with associated group $E$ (the group operation is defined as per (\ref{eq:Elliptic curve group operation})), and two points $a,b \in E$. It is promised that $sa = b$ for some nonnegative integer $s$.
\end{alg_in}
\begin{alg_out}
Find the least nonnegative integer $s$ such that $sa=b$.
\end{alg_out}
\begin{enumerate}
\item We use a register $\mathcal{H}_E$, where $E$ is the group associated with the elliptic curve (\ref{eq:Elliptic Curve}), and two ancillary registers $\mathcal{H}$ of dimension $N=2^n$, associated with the group $A=\Z_N\times\Z_N$. The computation begins in the state $\ket{0,0,O}$, where $(0,0)\in A$ and $O\in E$.
\item Fourier transforms are applied to the ancillas to create the superposition $\sum_{(x,y)\in A}\ket{x,y,O}$. 
\item The following transformation is applied unitarily:
\begin{equation}
\sum_{(x,y)\in A}\ket{x,y,O}\quad  \xrightarrow{\quad c\mbox{\,-}U\quad } \quad \sum_{(x,y)\in A}\ket{x,y,xa + yb}.
\end{equation}
\item Fourier transforms are applied again over the ancillas and then measured, obtaining an outcome of the form $(x',y')$. These outcomes contain enough information to extract the number $s$, with similar post-processing techniques to those used in Shor's DLP algorithm.
\end{enumerate}
\end{algorithm}
Algorithm \ref{alg:ProosZalka} is not a normalizer circuit over $\Z_N\times \Z_N \times E$. Similarly to  the factoring case, the algorithm would become a normalizer circuit  if the classical transformation in step 3 was an automorphism gate; however, for this to occur, $N$ needs to be  a common multiple of the orders of $a$ and $b$ (the validity of these claims follows with similar arguments to those in section \ref{sect:Factoring}). In view of our results in sections \ref{sect:Discrete Log} and \ref{sect:Factoring}, one can easily come up with two approaches to implement algorithm \ref{alg:Discrete Log} using normalizer gates.
\begin{itemize}
\item[(a)] The first approach would be to use our normalizer version of Shor's algorithm (theorem \ref{thm:Order Finding}) to find the orders of the elements $a$ and $b$: normalizer gates over $\Z\times E$ would be used in this step. Then, the number $N$ in algorithm \ref{alg:ProosZalka} can be set so that all the gates involved become normalizer gates over $\Z_N\times \Z_N \times E$.
\item[(b)] Alternatively, one can choose not to compute the orders by making the ancillas infinite dimensional, just as we did in algorithm \ref{alg:Order Finding infinite precision}. The algorithm becomes a normalizer circuit over $\Z\times \Z \times E$: as in algorithm \ref{alg:Order Finding infinite precision},  the ancillas are initialized to the zero Fourier basis state, and the discrete Fourier transforms are replaced by QFTs over $\T$ (in step 2) and $\Z$ (in step 4). A finite precision version of the algorithm can be obtained in the same fashion as we derived algorithm \ref{alg:Order Finding infinite precision}. Proos-Zalka's original algorithm could, again, be interpreted as a discretization of the resulting normalizer circuit.
\end{itemize}

\section{Proof of theorem \ref{thm:Simulation}} \label{app:Simulation proof}
In this section we will prove theorem \ref{thm:Simulation}. The proof uses results of Section \ref{sect:Group Decomposition}; the reader may wish to review that section before proceeding with this proof.

A key ingredient of our proof will be the main  simulation result of chapter \ref{chapterI} (\textbf{theorem \ref{thm:Main Result}}). We recall that this theorem can be applied given the following conditions: (i) $G$ is given in an explicitly decomposed form; (ii)  any group automorphism gate is specified as a rational matrix $A$, as in the normal form of theorem \ref{lemma:Normal form of a matrix representation}; (iii) any quadratic phase gate is specified as $(M,v)$, where rational $M$ is a matrix and $v$ is a rational vector, as in the normal form of theorem \ref{thm:Normal form of a quadratic function}; (iv)  (partial) quantum Fourier transforms are specified by the elementary subgroups it acts on. Note that, in the black-box normalizer-circuit setting, only condition (iv) is granted by assumption. 

Hence, given a black-box normalizer circuit acting on a black-box group $\mathbf{G}= \Z^a \times \T^b \times \DProd{N}{c} \times \mathbf{B}$, there are two things we need to do to ``de-black-box" it, so that the circuit can be classically simulated via  {theorem \ref{thm:Main Result}}:
\begin{enumerate}
\item Decompose the black-box portion of $\mathbf{G}$, $\mathbf{B}$: i.e., find  $\Z_\mathbf{B} := \Z_{N_{c+1}} \times \cdots \times \Z_{N_{c+d}}$, isomorphic to  $\mathbf{B}$, and matrix representations of the isomorphisms $\varphi,\varphi^{-1}$ that relate these groups.
\item Calculate \emph{normal forms} for each of the normalizer gates in the computation.
\end{enumerate}
Since we are given  access to an oracle for Group Decomposition,  we do not to show step 1. In the rest of the paper we show how to tackle task 2.

\subsection{Switching from black-box encoding to decomposed group encoding.}\label{sect:Encodings}

Throughout the rest of the appendix, we fix  $G=G_1\times \cdots \times G_m$  be the decomposed group
$$G= \Z^a \times \T^b \times \DProd{N}{c} \times  \Z_{N_{c+1}} \times \cdots \times \Z_{N_{c+d}}$$
where  $G_i$ corresponds to the $i$th primitive factor in the above equation, $m:=a+b+c+d$ and  $\Z_\mathbf{B} := \Z_{N_{c+1}} \times \cdots \times \Z_{N_{c+d}}$ is the group given in step 1. above.  We recall now that our classical algorithms to decompose  our black box group $\mathbf{B}$ output a set of linearly independent generators $b_{1},\cdots,b_{k'}$ of $\mathbf{B}$ such that $\mathbf{B} =\langle \beta_1\rangle\oplus\cdots\oplus\langle \beta_\ell\rangle$, as well as the order $N'_{i} = N_{c+i}$ of $\beta_i$. 

In our proof below, we will need to be able to convert elements back and forth from the original black-box encoding and this decomposed group encoding. To change between encodings, we need show how to perform the following tasks:
\begin{enumerate*}
\item[(a)] Our first task is to map an element from the decomposed group $\Z_{\mathbf{B}}$ to the black box group $\mathbf{B}$. In other words, we need to be able to compute the following group homomorphism $\varphi$:
\begin{equation*}
\varphi:\Z_{\mathbf{B}}\rightarrow\mathbf{B},\qquad\varphi(g)=b_1^{g(1)}\cdots b_{d}^{g(d)},\quad\textnormal{for any $g\in \Z_{\mathbf{B}}$.}
\end{equation*}
\item[(b)]Our second task is to convert elements from the original black-box group encoding to the new encoding defined by $\Z_{\mathbf{B}}$. In other words, given an arbitrary $\mathbf{b}\in \mathbf{B}$, we need to be able to compute $\varphi^{-1}{(\mathbf{b})}$.
\end{enumerate*}
Note that it is always possible to compute $\varphi(g)=b_1^{g(1)}\cdots b_d^{g(d)}$ for any $g\in \Z_\mathbf{B}$, since this can be done using a polynomial number of queries to the black-box group oracle (using repeated squaring if necessary for the exponentiation). Task (a) is therefore immediate.

As for Task (b), we note that computing $\varphi^{-1}(\mathbf{b})$ for an element $\mathbf{b}\in \mathbf{B}$ is equivalent to finding a list of integers $(g(1),\cdots,g(d))$ such that $b_1^{g(1)}\cdots b_{d}^{g(d)} = \mathbf{b}$. This is a special case of the multivariate discrete logarithm problem, defined in lemma \ref{lemma:Multivariate Discrete Log}; from lemma \ref{lemma:Multivariate Discrete Log} we see that Task (b) can be solved efficiently with a polynomial number of calls to the Group Decomposition oracle.

\subsection{Step (i): Group automorphism gates}

Recall that, by assumption, we have  access to a group automorphism oracle $\alpha:\: \mathbf{G}\rightarrow \mathbf{G}$ which, by the change of encoding of section \ref{sect:Encodings}, can be efficiently turned into a classical rational automorphism $f:G\rightarrow G$. Furthermore, $f$ can be efficiently evaluated by using the oracle $\alpha$ and switching the input and output of $\alpha$ from the black-box encoding (where the group action is implemented as a black-box circuit) to the decomposed group encoding (where elements of the group are given as a list of numbers, and the group action is simply addition of vectors), and vice versa (see previous subsection for details).

Our next step is to find a matrix representation  $A$ for $f$. We will assume (for the efficiency of this algorithm) that the size and precision of the coefficients are upper bounded by a known parameter $D$, i.e. each element of $M$ can be written as $A_{i,j} = \alpha_{i,j}/\beta_{i,j}$ for integers $\alpha_{i,j},\beta_{i,j}$ with absolute value no more than $D$.\footnote{Note that $D$ can be inferred from the precision bound $n_\textrm{out}$ (s.\ \ref{sect:Normalizer circuits over blackbox groups}) of an automorphism gate: because the output of $\alpha$ can only be $n_\textrm{out}$  bits larger than its input, it follows that the size of the denominator/numerator of every matrix element increases  at most by $D=2^{n_\textrm{out}}$. A similar argument will hold for quadratic phase gates.}

We now  show how to obtain the matrix representation $A$ from by evaluating $f:\: \mathbb{Q}^{a+b+c+d} \rightarrow \mathbb{Q}^{a+b+c+d}$ (which we view as a function sending rational inputs to  rational outputs). Because this function is a group automorphism we know that it further fulfills
\be
f(x) \equiv f(x') \mbox{ mod } G  \quad \text{if } x \equiv x' \mbox{ mod } G,
\ee
where two vectors are equal modulo $G$ if each pair of corresponding entries are equal modulo $\text{char}(G_i)$. Recalling that any matrix representation $A$ for $f$ has a specific block-structure of characterized by lemma \ref{lemma:Normal form of a matrix representation}(\ref{eq:block-structure of matrix representations}), we show how find a matrix representation $A$ for $f$ coefficient-by-coefficient. 

Let $c_i = \text{char}(G_i)$ be the characteristic of the $i$th primitive-group factor $G_i$ of $G$ as in lemma \ref{lemma:Normal form of a matrix representation}. Then, for most entries of $A$ this is trivial: note that we have
\be
A_{i,j} \equiv f(e_j)_i \mbox{ mod } c_i.
\ee
Hence by evaluating $f$ on the unit vectors $e_i$, we can determine $A_{i,j}$ modulo $c_i$. Thus we can evaluate $A_{\T F}$ exactly, the coefficients of the $i$-th rows of $A_{F\Z}$ and $A_{FF}$ modulo $\Z_{N_i}$, and the coefficients of $A_{\T\Z}$ and $A_{\T F}$ modulo $1$. This is sufficient for the cases listed above; the only case we still need to treat is $A_{\T\T}$, whose entries are arbitrary integers (and $c_i=\text{char}(\T)=1$ in this case). We can instead evaluate $f(e_j / \Delta)$ for some large integer $\Delta$:

\be
A_{i,j}/\Delta \equiv f(e_j/\Delta)_i \mbox{ mod } c_i
\ee
which allows us to determine $A_{i,j}$ modulo $\Delta c_i$ for our choice of $\alpha$. Choosing $\Delta > 2D$ then allows us to determine $A_{i,j}$ exactly for the case of $A_{\T\T}$.

\subsection{Step (ii): quadratic phase gates}\label{sect:QuadraticPhaseMatrices}

Next, we consider  a quadratic phase gate $\xi$, implemented as a classical circuit family $q:\: G \rightarrow \mathbb{Q}$ such that
\be
\xi(g) = \euler^{2\pii q(g)} \quad \forall g \in G.
\ee
For simplicity, we assume that we have changed from the original encoding $\mathbf{G}$ to $G$ using the same technique as in previous section and treat the elements of $G$ as a vector in $\mathbb{Q}^{a+b+c+d}$. Our next goal is  to write the quadratic function $\xi(g)$ in the normal form given by theorem \ref{thm:Normal form of a quadratic function}, i.e. find $M,v$ as in theorem  such that
\begin{equation}
\xi(g)=\euler^{\pii \,\left(g^{\transpose} M g \: +  \: C^{\transpose} g  \: +  \: 2v^\transpose g\right)}.
\end{equation}
Here, $q$, $M$, $C$, and $v$, are rational by the assumptions  and we do not need to find $C$, which is determined by $M$. Furthermore,  due to theorem \ref{thm:Normal form of a quadratic function}, lemmas \ref{lemma symmetric matrix representation of the bicharacter homomorphism}-\ref{lemma:Normal form of a matrix representation}, $M$, $v$ have some additional structural features, namely: $v$ is an element of the bullet group $G^\bullet$; $M$ is the matrix representation of a group homomorphism from $G$ to $G^\bullet$; and, up to a permutation, $M$ has the following upper triangular structure
\begin{equation}
M:= \begin{pmatrix}
      M_{\T\Z} & M_{\T F} & M_{\T\T} \\
      M_{F^{\bullet}\Z} & M_{F^{\bullet}F} & 0 \\
      M_{\Z\Z} & 0 & 0
    \end{pmatrix}
\end{equation}
where (i) $M_{\Z\Z}$ and $M_{\T\T}$ are arbitrary integer matrices; (ii) $M_{F^\bullet\Z}$ and $M_{\T F}$ have rational entries, the former with the form $M(i,j) = \alpha_{i,j}/N_i$ and the latter with the form $M(i,j) = \alpha_{i,j}/N_j$, where $\alpha_{i,j}$ are arbitrary integers, and $N_i$ is the order of the $i$-th cyclic subgroup $\Z_{N_i}$; (iii) $M_{F^\bullet F}$ is a rational matrix with coefficients of the form $M(i,j)= \frac{\alpha_{i,j}}{\gcd{(N_i, N_j)}}$  where $\alpha_{i,j}$ are arbitrary integers, and $N_i$ is the order of the $i$-th cyclic subgroup $\Z_{N_i}$; (iv)
$M_{\T\Z}$ is an arbitrary real matrix. The entries of $M_{F^\bullet\Z}$, $M_{\T F}$, $M_{F^\bullet F}$, and $M_{\T\Z}$ can be assumed to lie in the interval $[0,1)$.
Moreover, $M$ can be assumed to be symmetric, i.e. $M_{\Z\Z}^\transpose = M_{\T\T}$, $M_{F^\bullet\Z}^\transpose = M_{\T F}$, $M_{F^\bullet F}^\transpose = M_{F^\bullet F}$, and $M_{\T\Z}^\transpose = M_{\T\Z}$.

We now show how to compute $M$ and $v$, To this end, we assume, as before, that the size and precision of the coefficients are upper bounded by some known constant $D$, i.e. each element of $M$ can be written as $M_{i,j} = \alpha_{i,j}/\beta_{i,j}$ for integers $\alpha_{i,j},\beta_{i,j}$ with absolute value no more than $D$.

To do this, let us first determine the entries of $M$. This can be done in the following manner: it should be straightforward to verify that
\begin{equation}
\xi(x+y)=\xi(x)\xi(y)\euler^{2\pii \,x^{\transpose} M y}
\end{equation}
for any $x,y \in G$, and therefore
\begin{equation}
x^{\transpose} M y \equiv q(x+y) - q(x) - q(y) \mbox{ mod } \Z.
\end{equation}
We can use this method to determine nearly all the entries of $M$ exactly, by taking $x$ and $y$ to be unit vectors $e_i$ and $e_j$; this would determine $M_{ij}$ up to an integer, i.e.
\begin{equation}
M_{i,j} = e_i^{\transpose} M e_j \equiv q(e_i+e_j) - q(e_i) - q(e_j) \mbox{ mod } \Z.
\end{equation}

This determines all entries of $M$ except for those in $M_{\Z\Z}$ and $M_{\T\T}$ (the other entries can be assumed to lie in $[0,1)$). To deal with $M_{\Z\Z}$ we take $x = \Delta^{-1}e_i$, and $y = e_j$, such that the coefficient $M(i,j)$ is in the submatrix $M_{\Z\Z}$ and $1/\Delta$ is an element of the circle group with $\Delta>2D$, where $D$ is the precision bound. We obtain an analogous equation
\begin{equation}
\left(\frac{e_i^{\transpose}}{\Delta} M e_j\right) \equiv \frac{M_{i,j}}{\Delta}  \equiv  q(\Delta^{-1}e_i+e_j) - q(\Delta^{-1}e_i) - q(e_j) \mbox{ mod } \Z,
\end{equation}
which allows us to determine $M_{i,j}$: since the number $M_{i,j}/\Delta$ is smaller than $1/2$ in absolute value, the coefficient is not truncated modulo 1. One can apply the same argument to obtain the coefficients of $M_{\T\T}$, choosing  $x = e_i$, and $y = \Delta^{-1} e_j$.

Once we determine all the entries of $M$ in this manner, we get immediately the vector $C$ as well (since $C(i) = c_iM(i,i)$) (theorem \ref{thm:Normal form of a quadratic function}). It is then straightforward to calculate the vector $v$. Thus we can efficiently find the normal form of $\xi(g)$ through polynomially many uses of the classical function $q$.

\section{Extending theorem \ref{thm:Simulation}  to the abelian HSP setting}\label{app:Extending}

In this appendix, we briefly discuss that theorem \ref{thm:Simulation} (and some of the results that follow from this theorem) can be re-proven in the general hidden subgroup problem oracular setting that we studied in section \ref{sect:Abelian HSPs}.  This fact supports our view (discussed in the main text) that the oracle models in the HSP and in the black-box setting are  very close to each other.

Recall that the main result  in this section (theorem \ref{thm:HSP}) states that the quantum algorithm abelian HSP is a normalizer circuits over a group of the form $\DProd{d}{m}\times \mathcal{O}$, where $\mathcal{O}$ is a group associated with the abelian HSP oracle $f$ via the formula (\ref{eq:Oracular Group Operation}). The group $\mathcal{O}$ is not a black-box group, because no oracle to multiply in $\mathcal{O}$ was provided. However, we discussed at the end of section \ref{sect:Abelian HSPs} that one can use the hidden subgroup problem oracle to perform certain multiplications implicitly.

We show next that theorem \ref{thm:Simulation} can be re-casted in the HSP setting as ``\emph{the ability to decompose the oracular group $\mathcal{O}$  renders normalizer circuits over $\DProd{d}{m}\times \mathcal{O}$ efficiently classically simulable}''. To see this, assume a group decomposition table $(\alpha, \beta, A, B, c)$ is given. Then we know $\mathcal{O}\cong\Z_{c_1}\times \cdots \times \Z_{c_m}$. Let us now view the function  $\alpha(g)=(g,f(g))$ used in the HSP quantum algorithm as a  group automorphism of $G\times \mathcal \Z_{c_1}\times \cdots \times \Z_{c_m}$, where we decompose $\mathcal{O}$. Then, it is easy to check that $\begin{pmatrix}
1 & 0 \\
B & 1 \\
\end{pmatrix}$ is  a matrix representation of this map. It follows that the group decomposition table can be used to ``de-black-box'' the HSP oracle, and this fact allows us to adapt the proof of theorem \ref{thm:Simulation} step-by-step to this case.

We point out further that the extended Cheung-Mosca algorithm can be adapted to the HSP setting, showing that normalizer circuits over $G\times \mathcal{O}$ can be used to decompose $\mathcal{O}$. This follows from the fact that the function $f$ that we need to query to decompose $\mathbf{B}$ using the extended Cheung-Mosca algorithm (algorithm \ref{alg:group decomposition}) has precisely the same form as the HSP oracle. Using the HSP oracle as a subroutine in algorithm \ref{alg:group decomposition} (which we can query \emph{by promise}), the algorithm computes a group decomposition tuple for $\mathcal{O}$.

Finally, we can combine these last observations with theorem \ref{thm:Hidden Kernel Problem} and conclude that the problem of decomposing groups of the form $\mathcal{O}$ is classically polynomial-time equivalent to the abelian hidden subgroup problem. The proof is analogous to that of theorem \ref{thm:Hidden Kernel Problem}.

%% file: appendix4_hypergroups.tex
\chapter{Supplement to  chapter \ref{chapterH}}\label{aH}

\section{Proof of theorem \ref{thm:Evolution of Stabilizer States}, part II}\label{app:A}

In this appendix, we derive equations (\ref{eq:Pauli gates are Clifford}-\ref{eq:QFTs are Clifford}) finishing the proof of theorem \ref{thm:Evolution of Stabilizer States}.  We treat the different types of normalizer gates separately below.
\begin{enumerate}
\item \textbf{Automorphism Gates.}
It follows from the definition in section \ref{sect:Hypergroups} that any hypergroup automorphism $\alpha$ fulfills  $n_{\alpha(a),\alpha(b)}^{\alpha(c)}=n_{a,b}^{c}$  and $\w{\alpha(a)}=\w{a}$ for all $a,b,c\in \mathcal{T}$. Combining these properties with  (\ref{eq:Pauli operators DEFINITION}) we derive (\ref{eq:Automorphism gates are Clifford}).
\item \textbf{Quadratic phase gates.} The RHS\ of  (\ref{eq:Quadratic Phase gates are Clifford}) follows because $D_\xi$ is diagonal, hence, commutes with $Z_\mathcal{T}(\mathcal{X}_\mu)$.

The LHS can be derived by explicitly evaluating the action of $D_\xi \PX{\mathcal{T}}(a)D_\xi^\dagger$  on basis states in $\mathcal{B}_\mathcal{T}=\{\ket{b},b\in\mathcal{T}\}$  using that, for any $c, c'\in ab$ and any quadratic function $\xi$ with associated bicharacter $B$, the following identities holds:
 \begin{align*}
  (\mathrm{i})\quad  \xi(c)=\xi(c')=\xi(ab),  \quad  (\mathrm{ii}) \quad \xi(c)= \xi(a)\xi(b)  B(a,b), \quad (\mathrm{iii}) \quad  B(a,b)= \Hchi_{\beta(a)}(b)
 \end{align*}
  for some  homomorphism $\beta$ from $\mathcal{T}$ onto $\mathcal{T}_\mathrm{inv}^*$. Above, (i) follows from the triangle inequality and given properties, namely, $\xi(ab)=\sum_{c\in ab} n_{ab}^{c}\xi(c)$,  $|\xi(c)|=1$, and $\sum_{c} n_{ab}^c{=}1$; (ii) follows from (i) and the definition of quadratic function; and last, the normal form for bicharacters (iii) can be obtained by extrapolating the group-setting argument given in \cite{VDNest_12_QFTs}, lemma 5. 
\item \textbf{Quantum Fourier transforms.} We derive (\ref{eq:QFTs are Clifford}) by explicitly computing the action of Pauli operators on the states  $\Fourier{\mathcal{T}}^\dagger\ket{\mathcal{X}_\mu}$ (which form a basis)  using  (\ref{eq:Quantum Fourier Transform over Hypergroup T}): 
  \begin{align}
  \PX{\mathcal{T}}(a)\Fourier{\mathcal{T}}^\dagger\ket{\mathcal{X}_\mu}&=\sum_{b\in \mathcal{T}} \sqrt{\tfrac{\w{b}\ws{\overline{\Hchi_{\mu}}}}{\varpi_{\mathcal{T}}}} \overline{\Hchi_{\mu}}(b)\PX{\mathcal{T}}(a)\ket{b} \\&\stackrel{(\ref{eq:Pauli operators DEFINITION})}{=}\sum_{b\in \mathcal{T}}
  \sqrt{\tfrac{\w{b}\ws{\overline{\Hchi_{\mu}}}}{\varpi_{\mathcal{T}}}} \overline{\Hchi_{\mu}}(b)\left(\sum_{c} \sqrt{\tfrac{\w{b}}{\w{c}}}\,  n_{a,b}^{c} \ket{c}\right)\notag \\
  &\stackrel{(\ref{eq:Reversibility Property})}{=}\sum_{c\in \mathcal{T}} \sqrt{\tfrac{\w{c}\ws{\overline{\Hchi_{\mu}}}}{\varpi_{\mathcal{T}}}} \left(\sum_{b}   n_{{\overline{a}},c}^{b} \overline{\Hchi_{\mu}}(b) \right)\ket{c}=\sum_{c\in \mathcal{T}} \sqrt{\tfrac{\w{c}\ws{\overline{\Hchi_{\mu}}}}{\varpi_{\mathcal{T}}}} \overline{\Hchi_{\mu}}({\overline{a} } ) \overline{\Hchi_{\mu}}(c) \ket{c}\notag\\  &= \mathcal{X}_{{\mu}}({{a} } )\Fourier{\mathcal{T}}^\dagger\ket{\mathcal{X}_\mu}= \Fourier{\mathcal{T}}^\dagger\PZ{{\mathcal{T}^*}}(a)\ket{\Hchi_\mu}
  \end{align} 
 \begin{align}
 \PZ{\mathcal{T}}(\mathcal{X}_\mu)\Fourier{\mathcal{T}}^\dagger\ket{\mathcal{X}_\nu}&\notag\stackrel{(\ref{eq:Pauli operators DEFINITION})}{=}\sum_{b\in \mathcal{T}} \sqrt{\frac{\w{b} \ws{\overline{\Hchi_{\nu}}}}{\varpi_{\mathcal{T}}}} \, \mathcal{X}_{\mu}(b)\overline{\Hchi_\nu}(b)\ket{b}
\\&= \sum_{b\in \mathcal{T}} \sqrt{\frac{\w{b} \ws{\overline{\Hchi_{\nu}}}}{\varpi_{\mathcal{T}}}} \,  \left(\sum_{\overline{\Hchi_\gamma}\in{\mathcal{T}^*}}m_{\mu\overline{\nu}}^{\overline{\gamma}} \mathcal{X}_{{\overline{\gamma}}}(b)\right)\ket{b}\\
&= \sum_{\overline{\Hchi_\gamma}\in{\mathcal{T}^*}} \sqrt{\frac{\ws{\overline{\Hchi_{\nu}}}}{\ws{\Hchi_{\overline{\gamma}}} }}  m_{\mu\overline{\nu}}^{\overline{\gamma}}\left(\sum_{b\in \mathcal{T}} \sqrt{\frac{\w{b} \ws{\Hchi_{\overline{\gamma}}}}{\varpi_{\mathcal{T}}}} \,  \mathcal{X}_{{\overline{\gamma}}}(b)\ket{b}\right)\notag\\
& =\Fourier{\mathcal{T}}^\dagger \sum_{\overline{\Hchi_\gamma}\in{\mathcal{T}^*}} \sqrt{\frac{\ws{\overline{\Hchi_{\nu}}}}{\ws{\Hchi_{\overline{\gamma}}}}}  m_{\mu\overline{\nu}}^{\overline{\gamma}}\ket{\mathcal{X}_{{\gamma}}}= \Fourier{\mathcal{T}}^\dagger \PX{{\mathcal{T}^*}}(\overline{\Hchi_{\mu}}) \ket{\Hchi_\nu},
 \end{align} 
where we used   $\ws{\Hchi_{\overline{\gamma}}}=\ws{\Hchi_{\gamma}}$ and $m_{\mu\overline{\nu}}^{\overline{\gamma}}=m_{\overline{\mu}\nu}^\gamma$ from section \ref{sect:Hypergroups}. The analogous statement for partial QFTs follows straightforwardly using that character hypergroup of $\mathcal{T}_1\times \cdots \times \mathcal{T}_m$ is $\mathcal{T}_1^*\times \cdots \times \mathcal{T}_m^*$ \cite{BloomHeyer95_Harmonic_analysis} and the tensor-product structure of Pauli operators (section \ref{sect:Pauli Operators}).
\item\textbf{Pauli gates.} We can now use  (\ref{eq:QFTs are Clifford}) to get   
$Z_\mathcal{T}(\mathcal{X}_\varsigma)\PX{\mathcal{T}}(a) Z_\mathcal{T}(\mathcal{X}_\varsigma)^\dagger = \mathcal{X}_{{\varsigma}}(a) \PX{\mathcal{T}}(a)$ and  $Z_\mathcal{T}(\mathcal{X}_\varsigma)\PZ{\mathcal{T}}(\mathcal{X_\mu})Z_\mathcal{T}(\mathcal{X}_\varsigma)^\dagger=\PZ{\mathcal{T}}(\mathcal{X_\mu})$
since invertible characters are quadratic functions with trivial $B$ and $\beta$. Moreover, we can apply   (\ref{eq:QFTs are Clifford}) and repeat the argument in the character basis, obtaining $X_\mathcal{T}(s)\PX{\mathcal{T}}(a) X_\mathcal{T}(s)^\dagger=\PX{\mathcal{T}}(a)$,  $X_\mathcal{T}(s)\PZ{\mathcal{T}}(\mathcal{X_\mu}) X_\mathcal{T}(s)^\dagger = \mathcal{X_\mu}(\overline{s})\PZ{\mathcal{T}}(\mathcal{X_\mu})$. Equation (\ref{eq:Pauli gates are Clifford}) is derived combining these expressions.
\end{enumerate}

\section{Quadratic functions}\label{app:Quadratic functions}

We prove that the functions $\xi_{i}$, $\xi_{j}$, $\xi_{k}$ and $\xi$   defined in section \ref{sect:Quaternionic circuits} are quadratic. The quadraticity of $\xi_{x}$, with $x=i,j,k$, follows from the fact that the function  can be obtained by composing the quotient map $\overline{Q}_8\rightarrow \overline{Q}_8/\{\pm 1, \pm x\} \cong \Integers_2$, with the isomorphism $\overline{Q}_8/\langle x\rangle \rightarrow  \Integers_2$ and the map $\Integers_2\rightarrow \C:a \rightarrow i^a$; since the latter is a quadratic function of $\Integers_2$ \cite{VDNest_12_QFTs}, it follows easily that $\xi_x$ is a quadratic function of $\overline{Q}_8$. Note that in this derivation we implicitly use that  $\{\pm 1, \pm x\}$ is a subhypergroup of $\overline{Q}_8$  \cite{Roth75_Character_Conjugacy_Hypergroups}, that the quotient $\overline{Q}_8/S$ is an abelian hypergroup for any subhypergroup $S$, and that the quotient map $\overline{Q}_8\rightarrow \overline{Q}_8/S$ is a  hypergroup homomorphism \cite{Roth75_Character_Conjugacy_Hypergroups}.

To show that $\xi:\overline{Q}_8\times \overline{Q}_8\rightarrow\C$  is quadratic, we use the fact, prove below, that the function $B(C_x,C_y):=f_{C_x}(C_y)$ is a symmetric bi-character of $\overline{Q}_8$. Given that property as a  promise and recalling that $\xi((C_x,C_y))=B(C_x,C_y)$ (by definition), we can see that
\begin{align}
\xi\left((C_a,C_b)\cdot (C_c,C_d)\right)&=B(C_{a}C_c,C_b C_d)=B(C_a,C_b C_d)B(C_{c},C_b C_d))\notag\\
&=B(C_a,C_b)B(C_a,C_d)B(C_c,C_b)B(C_c,C_d)\notag\\
&=\xi\left((C_a,C_b)\right)\xi\left( (C_c,C_d)\right)B(C_a,C_d)B(C_c,C_b)\notag\\
&=\xi\left((C_a,C_b)\right)\xi\left( (C_c,C_d)\right)B'\left((C_a,C_b),(C_c,C_d)\right),
\end{align}
where we define $B'\left((C_a,C_b),(C_c,C_d)\right)=B(C_{a},C_d)B(C_{c},C_b)$. The latter is easily seen to be a bi-character of $\overline{Q}_8\times\overline{Q}_8$, so that $\xi$ is indeed quadratic.

It remains to show that  $B(C_x,C_y)$ is a symmetric bi-character. To see this, note,  that both the quotient hypergroup  $\overline{Q}_8/\{\pm 1\}$ and the subhypergroup of linear characters $\widehat{Q_8}_\ell$ of $Q_8$ are isomorphic to the Klein four group $\Integers_2\times \Integers_2$. Observe next that the map $C_x\rightarrow f_{C_x}$ is a homomorphism $\overline{Q}_8\rightarrow \widehat{Q_8}_\ell$, as it can be obtained composing the quotient map $\overline{Q}_8\rightarrow \overline{Q}_8/\{\pm 1\}$ with a chain of isomorphisms $\overline{Q}_8/\{\pm 1\} \rightarrow \Integers_2\times \Integers_2 \rightarrow \widehat{Q_8}_\ell$.  This latter fact implies that $B(C_x,C_y)=f_{C_x}(C_y)$ is a character in both arguments, hence, a bi-character. Finally, it is routine to check that $B(C_x,C_y)=B(C_y,C_x)$   by explicit evaluation, which completes the proof.

\section{Efficient vs.\ doubly efficient computable hypergroups}\label{app:Discret Log}

We  give an example of an efficient computable abelian hypergroup that cannot be   doubly efficiently computable unless we are given the ability to compute discrete logarithms over $\Integers_p^\times$. In chapter \ref{chapterB}, we discussed that this problem is believed to be hard for classical computers (being the basis of the Diffie-Hellman public-key cryptosystem \cite{DiffieHellman}), yet it can be solved via  Shor's discrete-log quantum algorithm \cite{Shor}. This problem reduces to the so-called hidden subgroup problem over $\Integers_{p-1}^2$ \cite{Nielsen02UniversalSimulations} for a certain hiding function $f$, which defines a group homomorphism from $\Integers_{p-1}^2$ to $\Integers_p^\times$ (chapter \ref{sect:Discrete Log}).

Considering now the group $\mathcal{T}=\Integers_{p-1}^2\times\Integers_p^\times$, which is manifestly efficiently computable following our definition, we can define an efficiently computable group automorphism $\alpha:\mathcal{T}\rightarrow \mathcal{T}: (m,x)\rightarrow(m,f(m)x)$.  We show that  $\mathcal{T}$ cannot be doubly efficiently computable unless the initial hidden subgroup problem and, hence, the discrete logarithm problem, can be solved in probabilistic polynomial time (which, up to date, is not possible).

First, we show that, if we are able to compute\footnote{We are implicitly assuming that there are efficient unique classical encodings for representing the characters of $\Integers_p^\times$, which is a strong yet \emph{weaker} assumption that $\Integers_p^\times$ being doubly efficiently computable.} $\alpha^*$, we must also be able to compute $f^*:\widehat{\Z}_p^\times \rightarrow \widehat{ \Z}_{p-1}^2$ (the dual of $f$, which is defined analogously to $\alpha^*$), since for any $\Hchi_{\mu,\nu}:=\Hchi_\mu\otimes\Hchi_\nu$ we have
\begin{equation}\notag
\Hchi_{\alpha^*(\mu,\nu)}(m,x)=(\Hchi_\mu\otimes\Hchi_\nu)(m,f(m)x)=\Hchi_\mu(m)\Hchi_\nu(f(m))\Hchi_\nu(x) = \Hchi_\mu(m){\Hchi_{f^*(\nu)}}(m) \Hchi_\nu(x),
\end{equation}
consequently, $\Hchi_{\alpha^*(\mu,\nu)} = \left(\Hchi_{\mu}\cdot \Hchi_{f^*(\nu)}\right) \otimes \Hchi_{\nu}$. Hence, if we can evaluate $\alpha^*$ on any character  $\Hchi_1\otimes \Hchi_{\mu}$, then we can determine  $\Hchi_{f^*(\nu)}$, the value of $f^*$ on $\nu$, for any $\Hchi_{\nu}$. If we now evaluate $f^*(\mu_i)$ on all  elements of a $O(\log p)$-sized randomly-obtained generating set $\{\Hchi_{\mu_i}\}$ of $\widehat{\Z}_{p}^\times$  and use our classical algorithms (theorem \ref{thm:General Solution of systems of linear equations over elementary LCA groups}) to solve the system of equations $\{[f^*(\mu_i)](x)=\Hchi_{\mu_i}(f(x))=1, x\in \Integers_{p-1}^2 \}$, whose solutions are those $x$ for which $f(x)=e$, we have found (in these $x$'s) generators of the hidden subgroup. This finishes the reduction.

\section{Implementing normalizer circuits over $\Conj{G}$}
\label{app:CC implementation details}

In this section, we present more details on how to efficiently implement normalizer circuits over the hypergroup $\Conj{G}$ when we are working in the Hilbert space $\mathcal{H}_G = \set{\ket{g}}{g \in G}$ labeled by elements of the group. As described in section \ref{sect:Circuits nonabelian group}, normalizer circuits over $\Conj{G}$ can be thought of as operating entirely within the subspace $\mathcal{I}_G \le \mathcal{H}_G$ of conjugation invariant wavefunctions; however, we will describe these operations in this section in terms of how they operate on the entire Hilbert space. In section~\ref{sect:character basis operations}, we discuss operations applied in the character basis, and in section~\ref{sect:class basis operations}, we discuss the same in the conjugacy class basis.

\subsection{Working in the character basis}
\label{sect:character basis operations}

Normalizer circuits allow of the following operations to be performed in the character class basis: preparation of initial states; Pauli, automorphism, and quadratic phase gates; and measurement of final states. It should be easy to understand how each of these could be implemented efficiently if we worked in a basis $\set{\ket{\mu}}{\mu \in \mathrm{Irr}(G)}$ of irrep labels. However, the Hilbert space $\mathcal{H}_G$ is only naturally labeled by group elements, which is why, in section~\ref{sect:Circuits nonabelian group}, we defined the character basis states $\set{\ket{\Hchi_\mu}}{\Hchi_\mu \in \Repr{G}}$ in the element basis.

Below, we will describe how to implement an isometry $\ket{\Hchi_\mu} \stackrel{\tau}{\mapsto} \ket{\mu}$ and its inverse, using a readily available choice for the basis $\set{\ket{\mu}}{\mu\in\mathrm{Irr}(G)}$. It should then be clear that we can implement each of the above gates by applying $\tau$, performing the operation in the irrep label basis, and then applying $\tau^{-1}$. To prepare an initial state $\ket{\Hchi_\mu}$, we prepare $\ket{\mu}$ in the irrep label basis and then apply $\tau^{-1}$. Finally, to measure in the character basis, we apply $\tau$ and then read the irrep label.

Our definition of the irrep label basis $\set{\ket{\mu}}{\mu\in\mathrm{Irr}(G)}$ comes from the definition of the QFT over the group $G$. Recall  that the QFT over any finite group $G$ \cite{childs_vandam_10_qu_algorithms_algebraic_problems}, denoted $\mathcal{F}_G$, is a unitary gate that sends an element state $\ket{g}$, for any $g\in G$, to a weighted superposition $\abs{G}^{-1/2} \sum_{\mu\in\mathrm{Irr}(G)} d_\mu \ket{\mu, \mu(g)}$, where  $\ket{\mu}$ is a state  that labels the irrep $\mu$ and $\ket{\mu(g)}$ is a $d_\mu^2$ dimensional state defined via
\begin{equation}\label{eq:QFT over NAB group}
\ket{\mu(g)}=\left(\mu(g)\otimes I_{d_\mu}\right) \sum_{i=1}^{d_\mu} \frac{\ket{i,i}}{\sqrt{d_\mu}}= \sum_{i,j=1}^{d_\mu} \frac{[\mu(g)]_{i,j}}{\sqrt{d_\mu}}\ket{i,j}.
\end{equation}
This transformation $\Fourier{G}$ has been extensively studied in the HSP literature and efficient quantum implementations over many groups are currently known (including  the symmetric group,  wreath products of  polynomial-sized groups and metabelian groups \cite{childs_vandam_10_qu_algorithms_algebraic_problems}). 

To see how we can use this, let's look at what $\Fourier{G}$ does to a character class state. In (\ref{eq:Bases Nonabelian Group}), we defined the state $\ket{C_x}$, when living inside the Hilbert space $\mathcal{H}_G$, to be a uniform superposition over the elements in the class $C_x$. If we apply $\mathcal{F}_G$ to this state, the result is
\begin{eqnarray*}
\mathcal{F}_G \ket{C_x}
 &=& \frac{1}{\sqrt{\abs{C_x}}} \sum_{g \in C_x} \mathcal{F}_G \ket{g} \\
 &=& \frac{1}{\sqrt{\abs{C_x}}} \sum_{g \in C_x} \frac{1}{\sqrt{\abs{G}}} \sum_{\mu \in \mathrm{Irr}(G)} d_\mu \ket{\mu} \otimes \sum_{i,j=1}^{d_\mu} \frac{[\mu(g)]_{i,j}}{\sqrt{d_\mu}} \ket{i,j} \\
 &=& \frac{1}{\sqrt{\abs{C_x} \abs{G}}} \sum_{\mu \in \mathrm{Irr}(G)} d_\mu \ket{\mu} \otimes \sum_{i,j=1}^{d_\mu} \frac{[\sum_{g \in C_x} \mu(g)]_{i,j}}{\sqrt{d_\mu}} \ket{i,j}.
\end{eqnarray*}
To simplify further, we need to better understand the sum in the numerator on the right.

The sum $\sum_{g \in C_x} \mu(g)$ is more easily analyzed if we write it as $(\abs{C_x}/\abs{G}) \sum_{h \in G} \mu(x^h)$: by standard results on orbits of group actions \cite{Lang_algebra}, each $\mu(g)$, for $g \in C_x$, arises the same number of times in the sum $\sum_{h \in G} \mu(x^h)$, which hence must be $\abs{G}/\abs{C_x}$ times for each, so we have $(\abs{C_x}/\abs{G}) \sum_{h \in G} \mu(x^h) = \sum_{g \in C_x} \mu(g)$. The sum $(1/\abs{G})\sum_{h \in G} \mu(x^h)$ may be familiar, as it is well known to be $\tfrac{1}{d_\mu} \chi_\mu(x) I$ \cite{Serre_representation_theory}.\footnote{This is a simple application of Schur's lemma. This sum is a $G$-invariant map $\mathcal{H}_G \rightarrow \mathcal{H}_G$, so it must be a constant times the identity. The constant is easily found by taking the trace of the sum.}

Putting these parts together, we can see that
\begin{eqnarray*}
\mathcal{F}_G \ket{C_x}
 &=& \frac{1}{\sqrt{\abs{C_x} \abs{G}}} \sum_{\mu \in \mathrm{Irr}(G)} d_\mu \ket{\mu} \otimes \sum_{i=1}^{d_\mu} \frac{\abs{C_x} \chi_\mu(x)}{d_\mu \sqrt{d_\mu}} \ket{i,i} \\
 &=& \sqrt{\frac{\abs{C_x}}{\abs{G}}} \sum_{\mu \in \mathrm{Irr}(G)} d_\mu \frac{\chi_\mu(x)}{d_\mu} \left(\frac{1}{\sqrt{d_\mu}} \sum_{i=1}^{d_\mu} \ket{\mu,i,i} \right),
\end{eqnarray*}
which is rewritten in our usual hypergroup notation as
\begin{equation}\label{eq:group Fourier on class}
\mathcal{F}_G \ket{C_x} = \sum_{\Hchi_\mu \in \Repr{G}} \sqrt{\frac{w_{C_x} w_{\Hchi_\mu}}{w_\Conj{G}}} \Hchi_\mu(C_x) \left(\frac{1}{\sqrt{d_\mu}} \sum_{i=1}^{d_\mu} \ket{\mu,i,i} \right).
\end{equation}
This precisely mirrors the definition of $\mathcal{F}_\Conj{G}$ with $\ket{\Hchi_\mu}$ replaced by $d_\mu^{-1/2} \sum_{i=1}^{d_\mu} \ket{\mu,i,i}$. We will denote the latter state below by $\ket{\mu_\mathrm{diag}}$. Thus, it follows by (\ref{eq:Character Orthogonality}) that $\Fourier{G} \ket{\Hchi_\mu} = \ket{\mu_\mathrm{diag}}$.

To implement the operation $\tau$, we apply $\Fourier{G}$ and then \emph{carefully} discard the matrix index registers.\footnote{In full detail, we do the following. First, apply the map $\ket{i,j} \mapsto \ket{i,j-i}$, which gives $\ket{i,0}$ when applied to $\ket{i,i}$. Next, write down $d_\mu$ in a new register and then invoke the Fourier transform over $\Integer_{d_\mu}$ on the first index register. The result of this will always be $\ket{0}$, so after uncomputing $d_\mu$, we are left with the state $\ket{0,0}$ in the index registers regardless of the value of $\mu$. At that point, they are unentangled and can be safely discarded.} By the above discussion, we can see that this maps $\ket{\Hchi_\mu}$ to the state $\ket{\mu}$, so this implements the operation $\tau$ correctly for any conjugation invariant state.

To implement the operation $\tau^{-1}$, we do the above in reverse. Starting with a state $\ket{\mu}$, we adjoin matrix index registers, prepare a uniform superposition over $\ket{1}, \dots, \ket{d_\mu}$ in the first index register using the inverse Fourier transform over the abelian group $\Integer_{d_\mu}$, and then copy the first index register to the second\footnote{Or rather, apply the map $\ket{i,j} \mapsto \ket{i, i+j}$, which gives $\ket{i,i}$ when applied to $\ket{i,0}$.} to get the state $\ket{\mu_\mathrm{diag}}$. Finally, we apply $\Fourier{G}^\dagger$ to get the state $\ket{\Hchi_\mu}$ per the calculations above.

As discussed earlier, the operations $\tau$ and $\tau^{-1}$ are all that we need in order to implement each of the required operations of normalizer circuits over $\Conj{G}$ in the character basis.

\subsection{Working in the character class basis}
\label{sect:class basis operations}

Most of the time, gates applied in the conjugacy class basis arise from operations on the whole group. For example, automorphisms of conjugacy classes often arise from automorphisms of the group. Likewise, Pauli Z operators in the conjugacy class basis are applications of characters, which are defined on the whole group, and Pauli X operators can also be implemented using multiplication in the group. Hence, it remains only discuss how to prepare initial states and measure in the conjugacy class basis.

For this, we need to assume that we can perform certain operations on conjugacy classes, as described in the following definition.

\begin{definition}\label{def:compute with CC}
Let $C_1, \dots, C_m$ be the conjugacy classes of $G$. Consider the following operations for working with conjugacy classes:
\begin{itemize}
\item Given a conjugacy class label $i$, produce the size of this class, $\abs{C_i}$.
\item Given an $x \in G$, produce the pair $(i,j)$, where $x = x_j$ in the class $C_i = \{x_1, \dots, x_t\}$.
\item Given a pair $(i,j)$, produce the element $x_j$ from $C_i$.
\end{itemize}
If each of these operations can be performed efficiently, then we say that we can \emph{compute efficiently with conjugacy classes} of $G$.
\end{definition}

We note that this assumption is trivial for abelian groups since each element is in its own conjugacy class. For some common examples of nonabelian groups, such as the dihedral and Heisenberg groups (and their higher nilpotent generalizations), elements are normally encoded in this manner already, so no additional assumption is actually required. For other common examples like the symmetric group, while elements are not always encoded directly in this manner, it is easy to see how the above calculations can be performed efficiently. In general, while we must formally make this assumption, we are not aware of any group for which these calculations cannot be performed efficiently.

If we can compute efficiently with conjugacy classes of $G$, then we can prepare initial states as follows. Starting with the conjugacy class label $i$ in a register, we first compute the size $\abs{C_i}$ into a new register. Next, we adjoin another new register and invoke the inverse Fourier transform over the abelian group $\Integer_{\abs{C_i}}$. After uncomputing the size $\abs{C_i}$, we are left with the superposition $M^{-1/2} \sum_{j=1}^M \ket{i,j}$, where $M = \abs{C_i}$. Finally, we apply the operation that turns pairs into group elements to get $M^{-1/2} \sum_{j=1}^M \ket{x_j}$, where $x_1, \dots, x_M$ are the elements of $C_i$, which is the desired initial state.

To perform a measurement in the conjugacy class basis, we can do the reverse of how we prepared the initial states in order to produce a conjugacy class label $\ket{i}$ in a register. Alternatively, we can simply measure in the group element basis and then, afterward, compute the conjugacy class of this element. These two approaches will give identical measurement probabilities.

Finally, we note that the two operations just described are the equivalent of the operations $\tau$ and $\tau^{-1}$ from section~\ref{sect:character basis operations} for the conjugacy class basis.\footnote{Indeed, the separation of a group element label into a conjugacy class label and an index label is analogous to how, in the space of irreducible representations, we separate each basis element into an irrep label and matrix index labels. It is frequently assumed that we can separate the latter into different registers whenever convenient, so our assumption that we can do the same for conjugacy classes is only affording the same convenience for the hypergroup $\Conj{G}$ that is often assumed for $\Repr{G}$.} As a result, if we do have gates that can be easily implemented on conjugacy classes but do not extend easily to the whole group, then we can implement these gates in the same manner as in the character basis: apply $\tau$ to convert into a basis of conjugacy class labels $\set{\ket{i}}{C_i \in \Conj{G}}$, apply the gate in this basis, and then apply $\tau^{-1}$ move back to the conjugacy class basis in $\mathcal{H}_G$.

Thus, we can see that the ability to compute efficiently with conjugacy classes of $G$ allows us to fully implement normalizer circuits operations applied in the conjugacy class basis. If we also have an efficient QFT for $G$, then as we saw in the previous section, we can implement normalizer circuits operations applied in the character basis as well. Together, these two assumptions allow us to fully implement normalizer circuits over $\Conj{G}$ when working in the Hilbert space $\mathcal{H}_G$.

%% file: appendix23_normalizer=gaussian.tex
\chapter{Normalizer circuits over $\mathbb{R}$ generate all bosonic Gaussian unitaries}\label{aG}

To complement our discussion in sections \ref{sect:chapter1-chapter2}-\ref{sect:Introduction2},  we show in this appendix that  normalizer circuits over  real groups of the form $\R^m$ (which were not considered in this thesis) coincide with the well-known definitions of (bosonic) Gaussian unitaries, which  are central in continuous-variable quantum information processing  \cite{Braunstein98_Error_Correction_Continuous_Quantum_Variables,Lloyd98_Analog_Error_Correction,Gottesman01_Encoding_Qubit_inan_Oscillator,Bartlett02Continuous-Variable-GK-Theorem,BartlettSanders02Simulations_Optical_QI_Circuits,Barnes04StabilizerCodes_for_CV_WEC,LloydBraunstein99_QC_over_CVs,BraunsteinLoock05QI_with_CV_REVIEW,GarciaPatron12_Gaussian_quantum_information}. This appendix is based on unpublished joint work with Geza Giedke \cite{BermejoGeza}.

To begin with, we expand on an earlier comment in section \ref{sect:Introduction2}, where we mentioned that the infinite-group normalizer circuit model (chapter \ref{sect:Normalizer Gates Infinite Group}) is well-defined for \emph{any} abelian group that has a locally compact Hausdorff topology and and, in particular, for groups of the  form $\Z^a \times \T^b \times \Z_{N_1}\times\cdots \times \Z_{N_c}$ with additional $\R^m$ factors. Here, we discuss how to define normalizer gates and Pauli operators over $\R^m$ groups. In fact, this turns out to be  slightly easier than for the groups in chapter \ref{chapterI}, since $\R^m$ groups are always isomorphic to their own character groups and have several other benign algebraic  features: namely, they are  vector spaces, have a well-defined inner product and, because $\R$ is a field of zero characteristic, do not contain zero divisors. In fact, because   $\R^m = (\R^m)^* \cong \widehat{\R^m}$ (section \ref{sect:characters}) all designated bases (\ref{group_labels_basis}) and Pauli operators (\ref{eq:Pauli operator type X, over G, eigenket definition}-\ref{eq:Pauli operator type Z}) are labeled by the same index group $\R^m$; hence, the distinction between $G$ and $G^*$ can be dropped from our formalism, similarly to the the finite abelian group setting of chapter \ref{chapterF}). 

We now relate normalizer circuits over $\R^m$ to unitary gates that act on harmonic oscillators.  Note, first, that  the Hilbert space of the computation   $\mathcal{H}_{\R^m}=\mathcal{H}_\R^{\otimes m}$ has a group element basis $\{\ket{x}_{X},x\in\R^m\}$ and  a character basis $\{\ket{p}_P,p\in\R^m\}$ that are related through the quantum Fourier transform over $\R^m$ as
\begin{equation}\label{eq:QFT over Reals}
\ket{p}_{P}=\int_{\R^m} \mathrm{d}x\, \overline{\euler^{2\pii p x}} \ket{x}_{X}.
\end{equation}
Unlike  for $\Z^n$ integer groups, there exists a natural 1-to-1 mapping between the elements and characters of $\R^m$ that lets us implement  the QFT over $\R^m$ as the unitary gate $\mathcal{F}_{\R^m}$ that implements the map $\ket{x}_{X}\rightarrow \int_{\R^m} \mathrm{d}x\, {\euler^{2\pii p x}} \ket{p}_{X}$. Realize that, w.l.o.g, we can identify $\mathcal{H}_{\R^m}$ with the Hilbert space   $\mathcal{H}_{\mathrm{osc}}^{\otimes m}$ of $m$ harmonic oscillators and  the states $\ket{x}_X$ (respectively  $\ket{p}_P$) with the joint eigenstates of all position operators $\hat{X}_i$ (respectively, all  momentum ones $\hat{P}_j$) of the  $m$-mode harmonic-oscillator. With this identification, it follows from \cite[2.80,2.83]{KokLovett10Intro_to_Optical_Quantum_Information_Processing} that the generalized Pauli operators $Z_{\R^m}(p)$, $X_{\R^m}(x)$  over $\R^m$ coincide with tensor-products of so-called position and momentum shift operators in the Gaussian formalism: 
\begin{equation}
X_{\R^m}(x)=\bigotimes_{i=1}^m X(x_i):=\bigotimes_{i=1}^m \exp\left(\imun\, x_i \hat{P}_i\right), \quad Z_{\R^m}(p)=\bigotimes_{i=1}^m Z(p_i):=\bigotimes_{i=1}^m \exp\left(\imun\, p_i \hat{X}_i\right). 
\end{equation}
Pauli operators are, hence, examples of normalizer circuits\footnote{Recall that Pauli X gates can be implemented by a  normalizer circuit with 3 normalizer gates due to lemma~\ref{lemma:Fourier transform diagonalizes the Regular Representation} (the proof of which applies to any locally compact abelian group).} over $\R^m$ that are \emph{Gaussian} unitaries, i.e., gates that can be generated by  Hamiltonians that are (at most) quadratic polynomials of  position and momentum operators.
Furthermore, it follows from \cite[2.87,2.89]{KokLovett10Intro_to_Optical_Quantum_Information_Processing} that single-\emph{mode} (in our notation, \emph{mode} is synonym of register) partial quantum Fourier transforms over $\R^m$  (\ref{eq:QFT over Reals}) are instances of  so-called single-mode Gaussian \emph{phaseshifters}
\begin{equation}
U(\theta):=\exp\left(   \theta\hat{a}^\dagger \hat{a}\right)=\exp\left( \theta\tfrac{\hat{X}^2 +\hat{P}^2}{2}\right),
\end{equation}
with angle  $\theta=3\uppi /4$, 
(here and below we fix physical units so that $1=\hbar = \omega$, where $\omega$ is the oscillator frequency); it follows that any QFT over $\R^m$ and, in particular, the global $m$-modes QFT  $\mathcal{F}_{\R^m}=\mathcal{F}_{\R}\otimes \cdots \otimes \mathcal{F}_{\R} = U(3\uppi/4)^{\otimes m}$ is Gaussian. Our next result shows that \emph{all} other normalizer gates can also be implemented via Gaussian unitaries and vice-versa.
\begin{theorem}[\textbf{Bosonic Gaussian = Normalizer over $\R^m$}]\label{thm:Gaussian=Normalizer} Let $\mathcal{H}_{\R^m}=\mathcal{H}_{\mathrm{osc}}^{\otimes m}$ be the Hilbert space of  $m$ harmonic oscillators. Then, any $m$ mode Gaussian unitary $U$ can be  approximated up to error $\varepsilon$ by circuit $\tilde{U}$ of $O(\polylog{\tfrac{1}{\varepsilon}})$ two-mode  normalizer gates over $\R^m$. Moreover,  any normalizer circuit $V$ over $\R^m$ can be approximated up  to error $\varepsilon$ by a circuit $\tilde{V}$ of $O(\polylog{\tfrac{1}{\varepsilon}})$ two-mode Gaussian unitaries\footnote{These error bounds are the standard ones that come from  quantum gate synthesis algorithms \cite{Dawson06_Solovay_Kitaev}}.
\end{theorem}
To prove the result, we give an explicit classical algorithm that  outputs classical descriptions for $\tilde{U}$, $\tilde{V}$. This classical algorithm is \emph{efficient} if certain maps that describe the action of $U$ $V$ on Pauli operators--see (\ref{inproof:Normalizer = Gaussian}) below---can be efficiently computed.
\begin{proof}
First, note that both normalizer circuits and Gaussian unitaries send Pauli operators (over $\R^m$) to Pauli operators under conjugation: for the former, this follows from theorem \ref{thm:Normalizer gates are Clifford} (the proof of which holds for any locally compact abelian  $G$); for the later, it follows from the fact that Gaussian unitaries send shift operators to shift operators \cite{KokLovett10Intro_to_Optical_Quantum_Information_Processing}. We prove our result by showing that \emph{any} unitary $\mathcal{C}$  that sends Paulis to Paulis can be efficiently implemented by circuits $U$, $V$ built of Gaussian unitaries and normalizer gates, respectively. The results follows by choosing the circuit $\mathcal{C}$ to be either Gaussian or normalizer.

Our next step is to write the action of $\mathcal{C}$ on Pauli operators as
\begin{equation}\label{inproof:Normalizer = Gaussian}
\mathcal{C}Z(p)X(x)\mathcal{C}^\dagger=\gamma^\mathcal{C}(p,x)Z(\alpha_Z^\mathcal{C} (p,x))X(\alpha_X^\mathcal{C} (p,x))
\end{equation}
for some functions $\gamma^\mathcal{C},\alpha^\mathcal{C}$. To prove our claim we will  assume that $\gamma^\mathcal{C},\alpha^\mathcal{C}$ can be computed efficiently for any target $\mathcal{C}$ that we may want to implement. This choice will lead to efficient algorithms to decompose $\mathcal{C}$ in terms of simpler Gaussian and normalizer gates. We highlight that  if we drop this assumption then our algorithm becomes inefficient but still outputs poly-size approximations for $\mathcal{C}$: the latter claim, though weaker, is already enough to show that normalizer circuits over $\R^m$ and Gaussian unitaries define the same families of unitary gates.

Now, note that, because $\mathcal{C}\sigma_1 \sigma_2 \mathcal{C}^\dagger =(\mathcal{C}\sigma_1 \mathcal{C}^\dagger)(\mathcal{C}\sigma_2 \mathcal{C}^\dagger)$  for any two Pauli operators $\sigma_1:=Z(p_1)X(x_1), \sigma_2:=Z(p_2)X(x_2)$, the map  $\alpha^\mathcal
C$ needs to be a  group automorphism of $\R^m\times\R^m$ and, necessarily, a continuous one, because $U$ is continuous. Hence, $\alpha^\mathcal{C}$ has a matrix representation $A^\mathcal{C}$  (lemma \ref{lemma:existence of matrix representations}) whose entries  $A^\mathcal{C}(e_i,e_j)$, where we denote $e_i:= (0, \dots, 0, 1_i, 0, \dots, 0)$, can be efficiently inferred by evaluating  $\alpha^\mathcal{C}$. Furthermore, because of the identity
\begin{equation}
\mathcal{C}\sigma_1 \sigma_2 \mathcal{C}^\dagger = \mathcal{C}\sigma_2 \sigma_1 \euler^{2\pii [(p_1,x_1),(p_2,x_2)]} \,\mathcal{C}^\dagger =(\mathcal{C}\sigma_2 \mathcal{C}^\dagger)(\mathcal{C}\sigma_1 \mathcal{C}^\dagger) \euler^{2\pii [A^{\mathcal{C}}(p_1,x_1),A^{\mathcal{C}}(p_2,x_2)]}
\end{equation}
for any $x_1,p_1,x_2,p_2\in \R^m$, it follows that  $A^\mathcal{C}$  must be a \emph{symplectic} matrix, i.e., it must preserve the symplectic product defined as $[(p_1,x_1),(p_2,x_2)]=p_1\cdot x_2 - x_1\cdot p_2$.

We will now show that $\mathcal{C}$ can be efficiently approximated both by Gaussian-unitary and normalizer-gate circuits.  For this, we use the following lemma, which says that if we can find  (Gaussian-gate or normalizer-gate) circuits  that act like $\mathcal{C}$  on Pauli operator labels (i.e., on \emph{phase space} in the CV QIP jargon) then we are done.
\begin{lemma}[\textbf{Phase space actions}]\label{lemma:PhaseSpaceActionsA23}
Let  $U$ be any  unitary that sends Pauli operators to Pauli operators  and whose action on (phase space) Pauli operator labels (\ref{inproof:Normalizer = Gaussian}) is identical to that of $\mathcal{C}$: i.e., such that $\alpha^U=\alpha^\mathcal{C}$ but $\gamma^U$ might differ from $\gamma^\mathcal{C}$. Then, $U$ coincides with $\mathcal{C}$ up to a correction term $W := Z_{\R^m}(b)X_{\R^m}(a)$ that is a Pauli operator over $\R^m$. Moreover, if $\alpha^{U},\gamma^{U}$ and their inverses are given to us as oracles, then $W$ can  be efficiently determined classically.
 \end{lemma}
 Note that in lemma \ref{lemma:PhaseSpaceActionsA23} an oracle could be any arbitrary poly-size circuit that approximates $\alpha^{U},\gamma^{U}$. In particular, a valid oracle could be given by a circuit $\tilde{U}$ of $O(\polylog{ \tfrac{1}{\varepsilon}})$ (not necessarily nearest neighbor) $k$-mode gates (with constant $k$) that fulfill (\ref{inproof:Normalizer = Gaussian}) and implements  $U$ up to  error $\varepsilon$, as studied below: from such a description, one can efficiently infer classical boolean circuits that compute these maps.
 
 Lemma \ref{lemma:PhaseSpaceActionsA23} is important because it tells us that if we can find an efficient circuit $U$ that implements the action of $\mathcal{C}$ in phase space then we can also implement $\mathcal{C}$ efficiently by performing a Pauli correction. Since the latter can be implemented with either Gaussian unitaries or normalizer gates, we can reduce our original problem to finding good Gaussian and normalizer approximations of $\mathcal{C}$ in phase space.
 \begin{proof}[Proof of lemma \ref{lemma:PhaseSpaceActionsA23}] 
For  $\mathcal{C}':= \mathcal{C}U^\dagger$ and any $\alpha,\beta\in \R^m$, we   consider the  stabilizer  groups $$\mathcal{S}_Z=\{\overline{\euler^{2\pii \alpha\cdot p}}Z_{\R^m}(p),p\in \R^m\} ,\qquad\mathcal{S}_X=\{\euler^{2\pii \beta\cdot x}X_{\R^m}(x),x\in \R^m\},$$ which are easily seen to uniquely stabilize, respectively, the states $\ket{\alpha}_{X}$ and $\ket{\beta}_{P}$ (cf.\ proof of lemma \ref{lemma:Stabilizer Group for Standard Basis State}) . By   assumption, the gate  $\mathcal{C'}$ sends Paulis to Paulis (via conjugation)  with $\alpha_{\mathcal{C}'}=\mathrm{id}$, being the identity. Hence, $\mathcal{C'}$ transforms the stabilizer groups  as  $\mathcal{S}_Z\rightarrow\mathcal{S}_{Z'}, \mathcal{S}_X\rightarrow\mathcal{S}_{X'}$ where
$$\mathcal{S}_Z'=\{\gamma^{\mathcal{C}'}(p,0)\overline{\euler^{2\pii \alpha\cdot p}}Z_{\R^m}(p),p\in \R^m\} ,\qquad\mathcal{S}_X=\{\gamma^{\mathcal{C}'}(0,x)\euler^{2\pii \beta\cdot x}X_{\R^m}(x),x\in \R^m\}.$$
Moreover, using that  $Z_{\R^m}(p_1+p_2)=Z_{\R^m}(p_1)Z_{\R^m}(p_2)$, $X_{\R^m}(x_1+x_2)=X_{\R^m}(x_1)X_{\R^m}(x_2)$ it follows that $\gamma^{\mathcal{C}'}$ restricts to a character of $\R^m$ on the subgroups $\R^m\times\{0\}$ and $\{0\}\times \R^m$. Therefore, there exist $\alpha',\beta'\in \R^m$ such that $\gamma^{\mathcal{C}'}(p,0)=\overline{\euler^{2\pii \alpha'\cdot p}}$ and $\gamma^{\mathcal{C}'}(0,x)=\euler^{2\pii \beta'\cdot x}$. From these equations, we derive that the action of $\mathcal{C'}$ is 
\begin{equation}\label{inproof:CSSnosymp=Pauli}
\mathcal{C'}\ket{\alpha}_{X}=\varphi(\alpha)\ket{\alpha+\alpha'}_{X}, \qquad  \mathcal{C'}\ket{\beta}_{P}=\vartheta(\beta)\ket{\beta+\beta'}_{P},
\end{equation}
where the additional   terms  $\varphi(\alpha)$, $\vartheta(\beta)$ are introduced since stabilizer states are only well-defined up to an (arbitrary) phase. Finally, we show that the complex function $\varphi$ (resp.\ $\vartheta$) is proportional (as a vector), to the character function $\chi_{(-\beta')}$ (resp.\ $\chi_{\alpha'}$) of $\R^m$: for  $\varphi$ this follows from the identity
$$\mathcal{C}'\ket{\beta}_P\stackrel{\textnormal{(\ref{eq:QFT over Reals})}}{=}\int_{\R^m} \mathrm{d}\alpha\, \overline{\euler^{2\pii \beta\cdot \alpha}} \varphi(\alpha)\ket{\alpha+\alpha'}_{X}\stackrel{\textnormal{(\ref{inproof:CSSnosymp=Pauli},\ref{eq:QFT over Reals})}}{=}\vartheta(\beta)\int_{\R^m} \mathrm{d}\alpha''\, \overline{\euler^{2\pii (\beta+\beta')\cdot \alpha''}} \ket{\alpha''};$$ the proof for $\varphi$ is analogous. These last equations fully determine $\mathcal{C'}$ as a unitary gate to be of form  $\mathcal{C}'\propto W=Z_{\R^m}(b)X_{\R^m}(a)$ with $a:=\alpha',b:=-\beta'$,  up to a (neglectable) global phase.

Finally, note that  $W$ can be efficiently identified because that the vectors $a,b$ can be inferred by evaluating the function $\gamma_{\mathcal{C'}}=(\gamma_{\mathcal{C}}\circ \alpha^{U^{\inverse}})(\overline{\gamma_{U}})$ on basis vectors $e_i$,  which can be done efficiently by computing the oracles that we are given.
\end{proof}
Next, we recall that  for any  (necessarily invertible)  symplectic matrix $A^\mathcal{C}$, there exist efficient classical algorithms \cite{Braunstein05_Squeezing_Irreducible_Resource,KokLovett10Intro_to_Optical_Quantum_Information_Processing} that can be used to find a circuit of \emph{Gaussian} unitaries $U$ such that  $\alpha^U=\alpha^\mathcal{C}$; here $\gamma^U$ might differ from $\gamma^\mathcal{C}$. Hence, adding a Pauli correction, $U$ provides an exact Gaussian implementation of $\mathcal{C}$.  Moreover, this result still holds if we restrict our Gaussian gates to belong to simple gate sets: first, $U$ can be written as a circuit of $O(m)$ Gaussian single-mode squeezers, which are gates of the form
\begin{equation}
S(r):=\exp{ \left(\imun r \left(\hat{X}\hat{P}+\hat{P}\hat{X}\right)\right)}
\end{equation}
where $r$ is a real parameter and two global Gaussian passive\footnote{Passive unitaries are those that preserve the energy eigenspaces of the global $m$-mode oscillator hamiltonian.} transformations\footnote{In the standard approaches \cite{Braunstein05_Squeezing_Irreducible_Resource,KokLovett10Intro_to_Optical_Quantum_Information_Processing}, this step is implemented by computing the so-called Euler decomposition of a $2m\times 2m$ symplectic matrix \cite{Arvind95RealSymplecticGroups_Q_Mechanics_Optics} using efficient   classical algorithms for the Singular Value Decomposition.}; furthermore,  techniques from \cite{Bouland14_Universal_Linear_Optics_Beamsplitter} let us approximate the latter global operations  to any error $\varepsilon$ by Gaussian circuits comprising $O(\polylog{\tfrac{1}{\varepsilon}})$ (non nearest-neighbor) two-mode beamsplitters 
\begin{equation}
U_{BS}=\exp\left(i\frac{\uppi}{4} \left(\hat{X}_1\hat{X}_2 + \hat{P}_1\hat{P}_2\right)\right)
\end{equation} and single-mode  QFTs (ie.\ phaseshifters of the form $U(3\uppi/4)$). Combining these two facts, we obtain a poly-size circuit of two-local\footnote{Here we used non nearest neighbor interactions. Of course, these could be decomposed into linear-size circuits of two-mode nearest neighbor ones using two-mode swap operations (which are both  Gaussian and normalizer).} Gaussian gates $\tilde{U}$ approximating $U$. 

We complete our proof showing that all gates in the Gaussian circuit $\tilde{U}$ are normalizer gates over $\R^m$ up to a Pauli correction. Since, we have already discussed that all QFTs and Pauli gates are normalizer, we just need to show this for $S(r)$ and $U_{BS}$. Moreover,  due to lemma \ref{lemma:PhaseSpaceActionsA23}, it is enough to show that the action in phase space of these gates coincides with that of some normalizer gates. Writing our phase space points as $(p_1,p_2, x_1,x_1)\in \R^{2m}$, the symplectic matrices associated to these gates are \cite{GarciaPatron12_Gaussian_quantum_information}
\begin{equation}\label{inproof:SqueezerBS}
A^{S(r)}=\begin{pmatrix}
 \euler^{-r} & 0 \\ 0  & \euler^r
\end{pmatrix}
,\qquad \qquad A^{U_{BS}} = \frac{1}{\sqrt{2}}
\begin{pmatrix}
\begin{matrix}
  1 & 1 \\ -1 & 1 
  \end{matrix} & \begin{matrix}
 0 & 0 \\ 0 & 0 
 \end{matrix}\\ \begin{matrix}
  0 & 0 \\ 0 & 0 
  \end{matrix}  & \begin{matrix}
    1 & 1 \\ -1 & 1 
    \end{matrix}
\end{pmatrix}.
\end{equation}
It follows that, up to a Pauli, $S(r)$, $U_{BS}$ can be implemented as the one-mode and two-mode normalizer automorphism gates $$U_{\alpha_{S(r)}}\ket{x}=\frac{1}{\sqrt{ \euler^{r} }}\ket{\euler^{r} x}, \quad x\in \R,\qquad  U_{\alpha_{BS}}\ket{x_1, x_2}=\ket{\frac{1}{\sqrt{2}}\begin{pmatrix}
  1 & 1 \\ -1 & 1 
  \end{pmatrix} (x_1,x_2)}, \quad (x_1,x_2)\in \R^2$$
which implement the classical maps $\alpha_{S(r)}:x\to \euler^{r} x$ and $(x_1,x_2)\to \frac{1}{\sqrt{2}}\begin{pmatrix}
  1 & 1 \\ -1 & 1 
  \end{pmatrix}(x_1,x_2)$. These are easily seen to be continuous group automorphisms of $\R$ and $\R^2$ whose actions in phase space are also described by the  matrices (\ref{inproof:SqueezerBS}): the proof of the latter fact is analogous to the one of eq. (\ref{inproof:LabelUpdateAutomorphismGate}) in lemma \ref{lemma:Evolution of Pauli labels}. 
  
  \subsection*{Discussion: renormalization factor for automorphism gates}
  
 Finally, we mention that there is a subtle (albeit negligible) difference between the automorphism gates over $\R^m$ compared to those in chapters \ref{chapterC}-\ref{chapterI}: unlike earlier,  for the $\R^m$,  a  normalization factor is needed in the definition of  automorphism gate in order that they are always unitary. This is exemplified in our formulas for $U_{\alpha_{S(r)}}$ and $U_{\alpha_{BS}}$ above. In general, for any automorphism gate $U_\alpha$, a re-scaling by a factor of $1/\sqrt{|\det A|}$ is needed because of the well-known change-of-variable formula used in integration by substitution
 $$\int_{\R^m} \mathrm{d}x |\psi^2(x)|=\int_{\R^m} \mathrm{d}y |\det{A}|\, |\psi^2(\alpha(y))| $$
 where $\mathrm{det} A$ is the determinant of a matrix representation of $\alpha$.
 
 More generally, for an arbitrary locally compact abelian group $G$, the re-scaling factor needed for an automorphism gate $U_\alpha$ (with respect to earlier chapters) is $1/\sqrt{\mathrm{mod}\alpha}$, where $\bmod\,\alpha$ denotes the so-called \emph{module function}  of $G$ \cite{Stroppel06_Locally_Compact_Groups}. The latter equals $|\det{A}|$ for $G=\R^m$ and is defined  via the analogous integration-by-substitution formula: 
 $$\int_{G} \mathrm{d}g |\psi^2(g)|=\int_{G} \mathrm{d}h  \,\mathrm{mod} \alpha\,|\psi^2(\alpha(h))|.$$
The existence of this function follows from properties of the Haar measure of an LCA group and guarantees that re-scaled automorphism gates do not increase volumes locally, at the level of group-element labels, hence,  preserve inner products at the quantum-state level: it follows that they are always \emph{unitary gates}. On the other hand, this re-normalization was not needed for the groups $\DProd{D}{a}\times \Z^b\times \T^b$ considered earlier  because their associated module function is always trivial (intuitively, because continuous invertible  endomorphisms cannot locally increase/shrink volumes  on  toruses and discrete groups).

Finally, we highlight that the presence of these normalization factors is meaningless from a stabilizer formalism perspective. This is because  $\bmod\,\alpha$ is a group homomorphism $\mathrm{Aut}(G)\rightarrow \R^{+}$, which readily  implies $\bmod\,\alpha^{-1}= 1/\bmod\,\alpha$ and  $U_{\alpha^{-1}}=U_\alpha^\dagger$. As a result, these factors always get canceled when automorphism gates act by conjugation on the Heisenberg picture.

\end{proof}